\documentclass[11pt, a4paper, twoside, BCOR10mm, DIV14, liststotoc, bibtotoc, headsepline, footsepline, pointlessnumbers]{scrbook} %
\usepackage{a4wide}           %
\usepackage[T1]{fontenc}      %
\usepackage[latin1]{inputenc}
\usepackage[english,ngerman]{babel}   %
\usepackage{mathpazo}  %
\usepackage[sumlimits,intlimits,namelimits]{amsmath}   %
\usepackage{amssymb}
\usepackage{units}     %
\usepackage{graphicx}
\usepackage{icomma}    %
\usepackage{float}     %
\usepackage{scrpage2}  %
\usepackage{url}       %
\usepackage{xspace}    %
\usepackage{amsthm}    %
\usepackage{dcolumn}   %
\usepackage{booktabs}  %
\usepackage[usenames,dvipsnames]{color} %
\usepackage{ifpdf}     %
\usepackage[a4paper,bookmarks,plainpages=false,pdfpagelabels,bookmarksopen=false,bookmarksnumbered=true,pdfauthor={},pdftitle={},urlcolor=black,linkcolor=black,breaklinks=true,citecolor=black,colorlinks]{hyperref} %
\usepackage{overpic}

\usepackage{listings}  %
\usepackage{booktabs}
\newcommand{\arraystretchdefault}{\renewcommand{\arraystretch}{1.2}}
\arraystretchdefault

\newcommand{\resettabcolsep}{\setlength{\tabcolsep}{6pt}}

\usepackage{threeparttable}
\renewcommand{\TPTtagStyle}{\underline}

\usepackage[printonlyused, withpage]{acronym}
\renewcommand{\bflabel}[1]{\normalfont{\normalsize{#1}}\hfill} 
\newcommand{\acslink}[1]{\acsu{#1}\label{acro:#1}} %
\newcommand{\acsplink}[1]{\acsp{#1}\acused{#1}\label{acro:#1}}

\usepackage{dcolumn}
\newcolumntype{d}[1]{D{.}{.}{#1}}

\usepackage{pdflscape}

\usepackage{multirow}
\usepackage{bigdelim}

\usepackage{longtable}

\usepackage{rotating}
\def\Vert#1{\rotatebox{90}{#1}}

\usepackage{afterpage}

\usepackage[table]{xcolor}
\xdefinecolor{lightblue}{rgb}{0.9,0.9,1}

\usepackage{collcell,array}

\renewcommand{\thefootnote}{\underline{\alph{footnote}}}

\usepackage[normalem]{ulem}

\usepackage{textcomp}

\usepackage{makeidx}
\makeindex{}

\usepackage{placeins}

\usepackage{bibtopic}

\usepackage{cite}

\usepackage{currvita}

\usepackage{mathtools}

\usepackage[absolute]{textpos}
\textblockorigin{0mm}{0mm}

\usepackage{numprint}       %
\npstyleenglish             %
\npthousandsep{\,}          %

\usepackage{datetime}

\setheadsepline{.4pt}
\setfootsepline{.4pt}

\addtokomafont{captionlabel}{\bfseries}  %
\setcapindent{0em}                       %

\newcommand{\abovecaptionskipdefault}{\renewcommand{\abovecaptionskip}{3pt}} %
\abovecaptionskipdefault
\makeatletter
\g@addto@macro\bfseries{\boldmath}
\makeatother

\makeatletter
\renewcommand{\@pnumwidth}{2.1em}
\makeatother

\renewcommand{\div}{\text{div}\,}

\newcommand{\mustbe}{\stackrel{!}{=}}

\newcommand{\ten}[1]{\cdot 10^{#1}}
\newcommand{\imag}{\dot{\imath}}

\newcommand{\setR}{\mathbb{R}}

\newcommand{\as}{\inmath{\alpha_s}}
\newcommand{\expectation}[1]{\langle #1 \rangle}   %
\newcommand{\definedas}{\ensuremath{:=}}           %
\newcommand{\asdefined}{\ensuremath{=:}}           %
\newcommand{\intd}{\ensuremath{\,\mathrm{d}}}                     %
\newcommand{\folgt}{\ensuremath{\Rightarrow}\xspace} %
\newcommand{\sub}[1]{\ensuremath{_\text{#1}}\xspace} %
\newcommand{\eff}[1]{\inmath{\epsilon\sub{#1}}} %
\newcommand{\percent}[1]{{#1}\,\%}

\newcommand{\inmath}[1]{\ensuremath{#1}\xspace}
\newcommand{\calL}{\inmath{{\cal L}}} %
\newcommand{\Linst}{\inmath{\calL_\text{inst}}} %
\newcommand{\Lint }{\inmath{\calL_\text{int}}}  %
\newcommand{\instlumi}[1]{\inmath{\unit[#1]{Hz/cm^2}}}                %
\newcommand{\meff}{\inmath{m_\text{eff}}}
\newcommand{\mttwo}{\inmath{m_\text{T2}}}

\newcommand{\pt}{\inmath{p_T}}
\newcommand{\TeV}[1]{{\unit[#1]{TeV}}}
\newcommand{\GeV}[1]{{\unit[#1]{GeV}}}

\newcommand{\ipb}[1]{\inmath{\unit[#1]{pb^{-1}}}} %
\newcommand{\ifb}[1]{\inmath{\unit[#1]{fb^{-1}}}} %

\newcommand{\onehalf      }{\inmath{{\nicefrac12}}}
\newcommand{\mzero        }{\inmath{{m_0}}}
\newcommand{\moh          }{\inmath{{m_\onehalf}}}
\newcommand{\signmu       }{\inmath{{\operatorname{sign}\mu}}}
\newcommand{\tanbeta      }[1]{\inmath{\tan\beta = #1}}
\newcommand{\avgmu}   {\inmath{\langle\mu\rangle}} %
\newcommand{\seventev}{\inmath{\sqrt{s} = \TeV{7}}}
\newcommand{\widebar}[1]{\ensuremath{\overline{#1}}}
\newcommand{\micro}{\ensuremath{\textnormal{\textmu}}}
\newcommand{\nreco}{\inmath{n_\text{pv,reco}}}

\newcommand{\Poisson}[2]{\ensuremath{\operatorname{Pois}(#1|#2)}\xspace}

\newcommand{\Normal}[1]{\ensuremath{\operatorname{Normal}(#1)}\xspace}

\makeatletter
\DeclareRobustCommand*{\trigger}[1]{%
  \begingroup\@activeus\scantokens{#1 }\endgroup}
\begingroup\lccode`\~=`\_\relax
   \lowercase{\endgroup\def\@activeus{\catcode`\_=\active \let~\_}}
\makeatother

\makeatletter
\renewcommand\paragraph{\@startsection{paragraph}{4}{\z@}%
  {-3.25ex\@plus -1ex \@minus -.2ex}%
  {1.5ex \@plus .2ex}%
  {\normalfont\normalsize\bfseries}}
\makeatother

\newcommand{\shiftdown}[1]{\smash{\raisebox{-.5\normalbaselineskip}{#1}}}
\newcolumntype{S}{>{\collectcell\shiftdown}c<{\endcollectcell}}

\let\origappendix\appendix %
\renewcommand\appendix{\cleardoublepage\pagenumbering{roman}\renewcommand{\thepage}{A-\arabic{page}}\origappendix}

\newcommand{\jetmet}{{jet~+~\met}\xspace}
\newcommand{\Jetmet}{{Jet~+~\met}\xspace}
\newcommand{\jetmetj}[1]{EF\_j75\_{\allowbreak}jetNoEF\_{\allowbreak}EFxe#1\_noMu\xspace} %
\newcommand{\jetmetf}[1]{EF\_j50\_{\allowbreak}jetNoEF\_{\allowbreak}xe#1\_noMu\xspace} %

\newcommand{\jetmetfall}{\jetmetf{*}}
\newcommand{\jet}[1]{EF\_j#1\_jetNoEF}       %
\newcommand{\metcomponent}[1]{\me{x,y}^\text{#1}}
\newcommand{\miss   }[1]{\inmath{{\not\mathrel{#1}}}}
\newcommand{\met    }{\inmath{\miss{E}_T}}
\newcommand{\me     }[1]{\inmath{\miss{E}_{#1}}}
\newcommand{\mex    }{\me{x}}
\newcommand{\mey    }{\me{y}}

\newcommand{\alphas }{\inmath{\alpha_\text{S}}}

\newcommand{\sumet  }{\inmath{\sum{E_T}}}
\newcommand{\Antikt }{{Anti-k$_\textsf{t}$}\xspace} %
\newcommand{\antikt }{{anti-k$_\textsf{t}$}\xspace} %
\let\OldEta\eta
\renewcommand{\eta}{\ensuremath{\OldEta}\xspace} %
\newcommand{\seta}{\ensuremath{|\eta|}} %
\newcommand{\cerenkov}{\v{C}erenkov\xspace}
\newcommand{\betadist}{$\beta$-distribution\xspace}
\newcommand{\Bi}{\inmath{\operatorname{Bi}}} %
\newcommand{\CL}{C.~L.\xspace}
\newcommand{\cls}{\inmath{\text{CL}_s}} %
\newcommand{\clsb}{\inmath{\text{CL}_{s+b}}} %
\newcommand{\clb}{\inmath{\text{CL}_b}} %
\newcommand{\kfactor}{$k$-factor\xspace}
\newcommand{\kfactors}{$k$-factors\xspace}
\newcommand{\EMJES}{EM+JES\xspace}
\newcommand{\Zjets}{{$Z$\,+ jets}\xspace}
\newcommand{\Wjets}{{$W$\,+ jets}\xspace}
\newcommand{\ZJets}{{$Z$~+ Jets}\xspace}
\newcommand{\WJets}{{$W$~+ Jets}\xspace}
\newcommand{\hypo}[1]{\inmath{{\cal H}_{#1}}} %
\newcommand{\hypob}{\hypo{B}}
\newcommand{\hyposb}{\hypo{S+B}}
\newcommand{\ssqrtb}{\inmath{s/\sqrt{b}}}
\newcommand{\soverb}{\inmath{s/b}}
\newcommand{\tp }[2]{\inmath{ \begin{array}{c} #1 \\ #2 \end{array} }}  %
\newcommand{\tpk}[2]{\inmath{ \begin{pmatrix} #1 \\ #2 \end{pmatrix} }} %
\newcommand{\Athena}{\mbox{\propername{Athena}}\xspace}
\newcommand{\ATLAS} {\mbox{\propername{ATLAS}}\xspace}
\newcommand{\ALICE} {\mbox{\propername{ALICE}}\xspace}
\newcommand{\CDF}   {\propername{CDF}\xspace}
\newcommand{\CMS}   {\propername{CMS}\xspace}
\newcommand{\COOL}  {\propername{COOL}\xspace}
\newcommand{\Dzero} {\mbox{D\O}\xspace}
\newcommand{\LEP}   {\propername{LEP}\xspace}
\newcommand{\LHC}   {\propername{LHC}\xspace}
\newcommand{\LHCb}  {\propername{LHCb}\xspace}
\newcommand{\LHCf}  {\propername{LHCf}\xspace}
\newcommand{\MoEDAL}{\propername{MoEDAL}\xspace}
\newcommand{\ROOT}  {\propername{ROOT}\xspace}
\newcommand{\TOTEM} {\propername{TOTEM}\xspace}
\newcommand{\Tevatron}{Tevatron\xspace}

\newcommand{\neutralinon}[1]{\inmath{{\widetilde{\chi}_{#1}^0}}}
\newcommand{\neutralino }{\neutralinon{1}}
\newcommand{\charginon  }[1]{\inmath{{\widetilde{\chi}_{#1}^\pm}}}
\newcommand{\charginopn }[1]{\inmath{{\widetilde{\chi}_{#1}^+}}}
\newcommand{\charginomn }[1]{\inmath{{\widetilde{\chi}_{#1}^-}}}
\newcommand{\gluino     }{\inmath{{\widetilde{g}}}}
\newcommand{\squark     }{\inmath{{\widetilde{q}}}}

\newcommand{\fig}[1]{Figure~\ref{#1}\xspace} %
\newcommand{\Fig}[1]{Figure~\ref{#1}\xspace}
\newcommand{\Figs}[1]{Figures~\ref{#1}\xspace}
\newcommand{\Eq}[1]{Equation~\eqref{#1}\xspace}
\newcommand{\Eqs}[1]{Equations~\eqref{#1}\xspace}
\newcommand{\Tab}[1]{Table~\ref{#1}\xspace}
\newcommand{\Tabs}[1]{Tables~\ref{#1}\xspace}
\newcommand{\Sec}[1]{Section~\ref{#1}\xspace}
\newcommand{\Secs}[1]{Sections~\ref{#1}\xspace}

\newcommand{\eg}{e.\,g.\@\xspace}
\newcommand{\ie}{i.\,e.\@\xspace}
\makeatletter %
\newcommand*{\etc}{%
    \@ifnextchar{.}%
        {etc}%
        {etc.\@\xspace}%
}
\makeatother

\newcommand{\widthtwoplots}  {7.5cm}
\newcommand{\widthsingleplot}{ 10cm}
\newcommand{\widthwideplot}  {0.618\paperwidth}

\newcommand{\kommentar}[1]{}

\newcommand{\revised}[1]{#1}                        %
\newcommand{\latein}[1]{\textit{#1}}     %
\newcommand{\define}[1]{\emph{#1}}       %
\newcommand{\propername}[1]{\textsc{#1}} %
\newcommand{\code}[1]{\texttt{#1}}       %
\newcommand{\tagname}[1]{\texttt{#1}}    %

\let\inindex\index %
\def\index#1{\def\x##1##2{\MakeUppercase{##1}{##2}}#1\inindex{#1@\x#1}}

\newcommand{\removesection}[1]{}

\newcommand{\incgraphics}[2]{\includegraphics[#1]{#2}}
\newcommand{\incgraphicsdraft}[2]{
  \begin{overpic}[#1]{#2}
    \put(10,10){\huge NOT FINAL PLOT}
  \end{overpic}
}
\newcommand{\incgraphicsclose}[2]{
  \begin{overpic}[#1]{#2}
    \put(10,10){\huge CLOSE TO FINAL}
  \end{overpic}
}

\graphicspath{{figures/}{plots/}}

\begin{document}

  \frontmatter %

\extratitle{
  \begin{flushleft}
    \vspace*{2.5cm}
    {\scshape \large Alexander Mann \\}
    \vspace*{1cm}
    {
  \scshape \LARGE
  Calorimeter-Based Triggers \\
  at the \ATLAS Detector \\
  for Searches for Supersymmetry \\
  in Zero-Lepton Final States \\
    }
    \vspace*{1cm}
    {\scshape \large Dissertationsschrift}

  \begin{textblock}{20}( 10,10.2)
    \incgraphics{height=4.5cm}{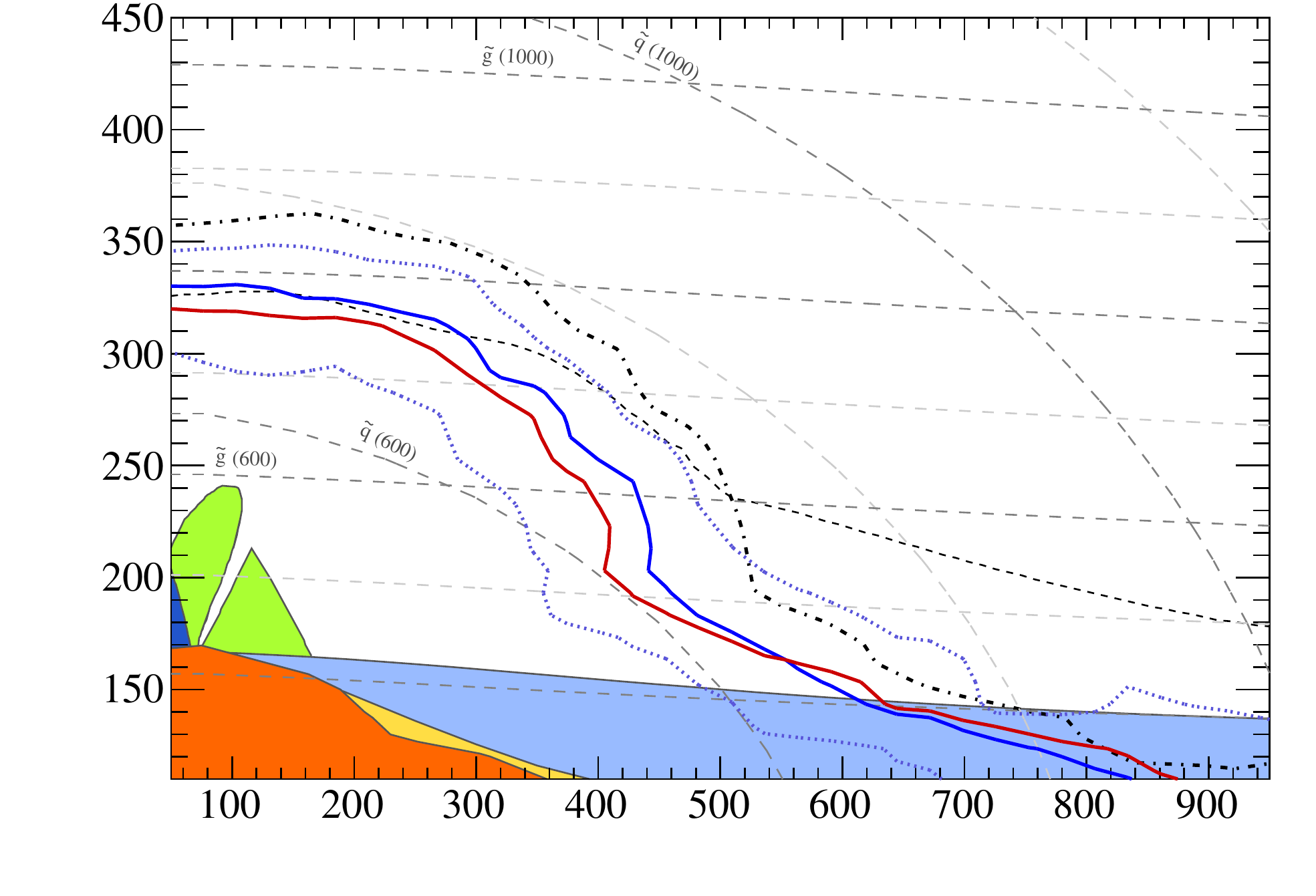}
  \end{textblock}
  \begin{textblock}{20}(  6,10.2)
    \incgraphics{height=4.5cm}{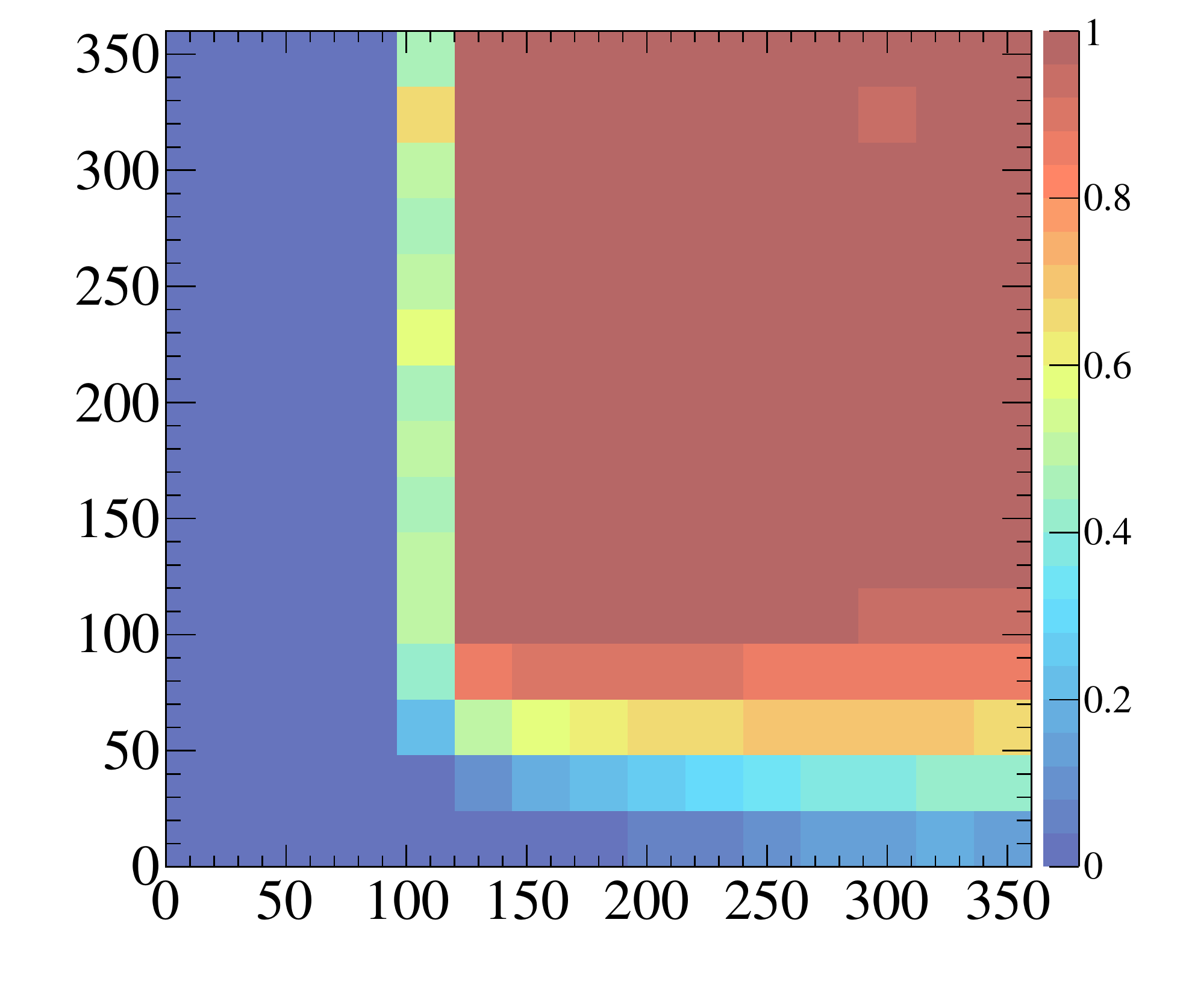}
  \end{textblock}
  \begin{textblock}{20}(0.8,10.2)
    \incgraphics{height=4.5cm}{{{COOL.combine_rates_own2010_rate2010IHG_avg2_EF_xe15_noMu.pdf-rate2010IHG_avg2_EF_xe35_noMu.pdf_182726-189425_titlepage}}}
  \end{textblock}

  \end{flushleft}

}

\title{
  Calorimeter-Based Triggers \\
  at the \ATLAS Detector \\
  for Searches for Supersymmetry \\
  in Zero-Lepton Final States \\[3cm]
  \selectlanguage{ngerman} %
  \large
  \textnormal{ %
  Dissertation\\
  zur Erlangung des mathematisch-naturwissenschaftlichen Doktorgrades\\
  "`Doctor rerum naturalium"'\\
  der Georg-August-Universität Göttingen\\[2cm]
  }
}
\author{
  vorgelegt von\\[0cm]
  Alexander Mann\\[0cm]
  aus Göttingen\\[2cm]
}
\date{
  Göttingen, Januar 2012 \\
}
\selectlanguage{english}

\lowertitleback{
  \revised{
	Referent: Prof. Dr. Arnulf Quadt, Universität Göttingen\\
	Korreferent: Dr. Carsten Hensel, Universität Göttingen\\
	Tag der mündlichen Prüfung: 16. Februar 2012\\
	\\
	Referenznummer: II.Physik-UniGö-Diss-2012/01
	}
}

\dedication{
\begin{quote}
  The existence of Supersymmetry is almost proven:\\
  Nearly half of the particles of the Minimal Supersymmetric Standard Model have been observed already.
  \flushright{\textit{--anonymous}}
\end{quote}
}

\maketitle

\setcounter{tocdepth}{1} %
\setcounter{tocdepth}{2} %
\tableofcontents

\chapter*{Abstract}
\chaptermark{Abstract}
\label{sec:abstract}

This thesis consists of three closely related parts.
An analysis of data recorded by the \ATLAS detector in 2010
in proton-proton collisions at a center-of-mass energy of \seventev
with an integrated luminosity of $\unit[33.4]{pb^{-1}}$ is performed,
searching for supersymmetric final states containing jets and missing transverse energy and no electrons or muons (zero-lepton channel).
No excess over the Standard Model background expectation is observed.
Using the \cls and PLR methods,
exclusion limits are set in a minimal supergravity model with \tanbeta{3}, $A_0 = 0$ and $\mu > 0$
and in a simplified supergravity model.
These considerably extend the excluded parameter ranges from earlier experiments.
The analysis validates the official analysis carried out within the \ATLAS Supersymmetry group
and additionally takes into account the uncertainty from pile-up effects.

The rates and efficiencies of triggers based on combined signatures with jets plus missing transverse energy in \ATLAS are studied,
which are the primary triggers for the search for Supersymmetry in the zero-lepton channel.
Methods to measure the efficiencies on data 
are tested and optimized to obtain sufficient statistics.
For the measurement of the efficiencies in data collected in 2010 and 2011,
the bootstrap method is applied
to correct for the sample trigger bias.
Different sample triggers based on jets and missing transverse energy are compared
and their efficiencies are measured.
A reweighting approach is studied and used to correct
for the bias from the propagation of the uncertainties in the bootstrap method.
The resulting efficiency estimates for the primary triggers
allow to determine the onset of the plateau of the two-dimensional turn-on curves
and are input to the official analyses in the \ATLAS Supersymmetry group in 2010 and 2011.

A universal model is developed to describe the contribution
of fake missing transverse energy from resolution effects
to the rates of missing transverse energy triggers
as function of the level of in-time pile-up, \ie the number of concurrent proton-proton interactions.
The input parameters to tune the model to the properties of the \ATLAS triggers are determined,
and the model predictions are compared to measurements %
of trigger rates
in \ATLAS.
Good agreement is found for missing transverse energy triggers with low thresholds %
for which the rates are dominated by resolution effects,
whereas the rates for higher thresholds are underestimated
due to additional sources of fake and real missing transverse energy which are not incorporated in the model.

\selectlanguage{ngerman}
\newcommand{\englisch}[1]{\textit{#1}}     %

\section*{Kurzfassung}
\label{sec:kurzfassung}
Diese Arbeit besteht aus drei eng zusammenhängenden Teilen.
Es wird eine Analyse von Daten des \ATLAS-Detektors aus dem Jahr 2010 durchgeführt,
die in Proton-Proton-Kollisionen bei einer Schwerpunktsenergie von \seventev aufgezeichnet wurden
und eine integrierte Luminosität von $\unit[33.4]{pb^{-1}}$ umfassen,
mit dem Ziel, supersymmetrische Endzustände mit Jets, fehlender transversaler Energie und ohne Elektronen und Myonen nachzuweisen.
Dabei wird kein über den erwarteten Standardmodelluntergrund hinausgehender Überschuss an Ereignissen beobachtet.
Auf die Massenparameter von zwei supersymmetrischen Szenarien mit gravitationsvermittelter Symmetriebrechung,
von mSUGRA mit \tanbeta{3}, $A_0 = 0$ und $\mu > 0$ und von einem vereinfachten Modell,
werden Ausschlussgrenzen mithilfe der \cls- und der PLR-Methode gesetzt,
welche die Ausschlussgrenzen früherer Experimente deutlich übertreffen.
Die Analyse validiert die offiziellen Ergebnisse der \ATLAS-Supersymmetrie-Gruppe
und berücksichtigt zusätzlich Unsicherheiten infolge der Überlagerung mehrerer gleichzeitiger Proton-Proton-Wech\-sel\-wir\-kun\-gen. %

Die Raten und Effizienzen der im \ATLAS-Detektor verwendeten Trigger,
die auf kombinierten Signaturen mit Jets und fehlender transversaler Energie beruhen, %
werden untersucht.
Diese Trigger sind die Primärtrigger für die vorgestellte Suche nach Supersymmetrie im Null-Lepton-Kanal.
Methoden zur Bestimmung der Triggereffizienzen auf Daten werden getestet und auf größtmögliche Statistik hin optimiert. %
Für die Messung der Effizienzen auf Daten aus 2010 und 2011
wird die Bootstrap-Methode verwendet,
um den Bias der Sample-Trigger zu korrigieren.
Verschiedene, auf Jets und fehlender transversaler Energie basierende Sample-Trigger werden verglichen
und ihre Effizienzen werden bestimmt.
Der Bias, der der Fehlerfortpflanzung in der Bootstrap-Methode inhärent ist,
wird untersucht und korrigiert.
Die hierdurch mögliche Bestimmung der Lage des Plateaus in den erhaltenen zweidimensionalen Triggereffizienzkurven
ist Bestandteil der offiziellen Analysen der \ATLAS-Supersymmetrie-Gruppe 2010 und 2011.

\begin{sloppypar} %
Um den Anteil der durch Auflösungseffekte hervorgerufenen scheinbaren feh\-len\-den trans\-ver\-sa\-len Energie
an den Triggerraten als Funktion der Anzahl der gleichzeitigen Proton-Proton-Wechselwirkungen beschreiben zu können,
wird ein allgemeingültiges Modell entwickelt.
Die Modellparameter werden an die Eigenschaften des \ATLAS-Triggers angepasst
und die erhaltenen Vorhersagen mit Messungen der entsprechenden Raten des \ATLAS-Triggers verglichen.
Eine gute Überstimmung zeigt sich für Trigger mit niedrigen Schwellwerten,
deren Raten von Auflösungseffekten dominiert werden,
während die Triggerraten für höhere Schwellwerte unterschätzt werden.
Dies wird begründet mit dem Auftreten zusätzlicher Quellen von echter und scheinbarer fehlender transversaler Energie,
die im Modell nicht abgebildet werden.
\end{sloppypar}

\selectlanguage{english}

\chapter*{Curriculum Vitae}
\chaptermark{}

\begin{cv}

\begin{cvlist}{Personal Details}
  \item[Name] Alexander Mann
  \item[Date of Birth] 26 December 1981
  \item[Place of Birth] Göttingen, Deutschland
  \item[Nationality] German %
\end{cvlist}

\newcommand{\PhD}{PhD\xspace} %
\begin{cvlist}{Education}
  \item[2008 -- \revised{2012}]{\PhD (in Physics), Georg-August-Universität Göttingen}
  \item[2004 -- 2008]{Diplom (in Physics), Georg-August-Universität Göttingen}
  \item[2002 -- 2004]{Vordiplom (in Physics), Georg-August-Universität Göttingen}
  \item[2001 -- 2002]{Civilian Service, Krankenhaus Göttingen-Weende e.V.}
  \item[1995 -- 2001]{Abitur, Felix-Klein-Gymnasium Göttingen}
\end{cvlist}

\end{cv}

\chapter*   {Acknowledgements} %
\chaptermark{Acknowledgements} %
\label{sec:acknowledgements}

{\parindent0mm %

It is my pleasure to thank the many people who made this thesis possible. %

In particular, I wish to express my gratitude to my supervisor,
Dr. Carsten Hensel,
head of our Emmy Noether young researcher group at Göttingen,
which focuses on the search for supersymmetric new physics models. %
He gave me the opportunity to work on the very interesting, experimentally challenging and theoretically appealing topic of Supersymmetry.
I'd like to thank him for his continued encouragement and invaluable suggestions during my work.

I would like to thank
Prof. Dr. Arnulf Quadt,
chair of the II. Institute of Physics at Göttingen,
as co-referee of my thesis and member of my thesis committee.
He managed, with great ambitions, to bring particle physics
to this city reknown for its excellent physicists.
Moreover, I thank Prof. Dr. Detlev Buchholz as the third member of my thesis committee.

I am very grateful to Dr. Julien Morel.
It has been a great pleasure to work with him during my time at CERN and afterwards.

The results presented in this thesis are based on the efforts of a huge number of people,
who designed, built and operate the \LHC machine and the \ATLAS detector.
I wish to extend my gratitude to everybody in our collaboration who contributed to making the \LHC and \ATLAS projects a great success,
and in particular to my colleagues for the friendly and warm atmosphere
in our group at Göttingen
and the working groups I was allowed to be part of.

For many useful comments and remarks I would like to thank everybody
who was so kind to read one of the various draft versions of parts of my thesis.

I would like to express my deep and sincere gratitude to all my friends who accompanied me on my way,
for the fun we had together, their help and inspiration.

Lastly, and most importantly, I wish to thank my parents for their support.

This thesis has been supported by the German Research Foundation (DFG)
and the Federal Ministry of Education and Research (BMBF).
The financial support is gratefully acknowledged.

\chaptermark{}

}

  \mainmatter %
\cleardoublepage
\chapter{Introduction} %
\label{sec:introduction}

Elementary particle physics deals with the building blocks of matter on the smallest scale of size.
It describes the constituents of matter and how they interact with one another.
The theoretical framework of elementary particle physics is the Standard Model.
The Standard Model is a %
very satisfactory %
theory of the electromagnetic, weak and strong interactions of particles,
which allows to make predictions and to compute quantities like cross sections and decay rates at high precision and in agreement with experimental findings.
However, there is a number of questions it cannot answer,
and it is believed to be the low-energy limit of a deeper theory %
which would at some point also incorporate a quantum theory of the fourth fundamental interaction, gravity.
Moreover, the Standard Model has theoretical imperfections. %
One of these is the fine-tuning or hierarchy problem,
which requires a remarkably precise cancellation of physical parameters
in order to protect the Higgs mass from quadratic divergences.
To address these shortcomings, an extension of the Standard Model is needed.
A very promising candidate %
is Supersymmetry.
Supersymmetric extensions of the Standard Model
introduce a new symmetry which unifies fermions, the constituents of matter,
and bosons, the particles which mediate the forces between them.
This not only solves the fine-tuning problem in a very elegant way,
but there are also other quantitative indications
in favor of Supersymmetry,
including gauge coupling unification and the existence of natural candidates for cold dark matter. %
The introduction of Supersymmetry doubles the particle content of the Standard Model:
every fermion is accompanied by a bosonic partner and vice versa.
Inevitably, this leads to the question how all of these new particles could have evaded observation so far.
A possible explanation is that Supersymmetry is a broken symmetry
so that the supersymmetric partners are much heavier than the Standard Model particles.
They can then only be produced at collider experiments which reach sufficiently high energies.

All previous searches for evidence of Supersymmetry have yet been unsuccessful,
but with the recent start-up of the Large Hadron Collider at CERN,
with its unprecedented center-of-mass energy of \unit[7]{TeV},
it is now possible to search for new physics in an energy regime which has never been accessible before at a particle collider.
The confrontation of theory with the upcoming new experimental results
will either confirm or disprove %
the expectations of theorists
and may reshape the foundations of particle physics.
In this thesis, data will be evaluated which is collected with the \ATLAS detector,
one of the two large multi-purpose detectors at the Large Hadron Collider.
The \ATLAS detector is a device to detect the particles which are produced in high-energy collisions of protons.
Comparing its data output to simulations will allow to
draw conclusions on what has happened in the collision events
and to search for signatures of the production of supersymmetric particles and their decay products.

In many supersymmetric models, the lightest supersymmetric particle has a large mass and evades detection
due to its weak interactions,
leading to an imbalance in the sum of all measured particle momenta.
Therefore, a typical supersymmetric signature involves
a considerable amount of missing transverse energy from the lightest supersymmetric particle,
plus high-energetic jets from decays of heavy supersymmetric particles.
One of the challenges in a collider experiment is
to spot these signatures among the collision events,
the overwhelming majority of which
contains only well-known Standard Model processes.
In order to collect a sufficient number of the rare new physics events,
very high collision rates need to be employed.
This yields an enormous data output, many orders of magnitudes larger than what can be stored.
Hence, a dedicated detector component, the trigger, is needed,
to do fast online measurements
and to decide automatically in real-time
whether or not a given collision event is interesting, and thus kept, or discarded. %
As in the resulting dataset
only events can be examined which have been accepted by the trigger,
whereas everything else is irretrievably lost,
it must be ensured that the trigger actually does what it is supposed to do.
A vital part of data analysis are therefore trigger studies.
These are concerned with the efficiency of the trigger selection,
but also with the trigger rates and their dependencies.

The final goal and unifying element %
of the work documented 
in this thesis is the search for Supersymmetry,
which is carried out on data taken with the \ATLAS detector.
In accordance with the explanations from above,
the analysis of this data is complemented by detailed studies of the efficiency of the trigger selection
which is used to collect the dataset being analyzed. %
A further aspect under study in this thesis
is the output rate of the trigger
for a certain type of collision events with missing transverse energy.
The outline of this thesis is the following.
In Chapter~\ref{ch:theory},
the theoretical background is given.
Both the Standard Model of Particle Physics and the fundamentals of Supersymmetry are covered,
and the resulting particle content and the phenomenological implications are introduced.
Afterwards, in Chapter~\ref{ch:experimentalsetup},
the experimental setup is presented,
in particular the collider and the detector
which are used to produce the particle collisions and to record the data evaluated in this thesis.
As studies concerned with the trigger system constitute an important part,
Chapter~\ref{sec:tdaq} will be dedicated to the trigger and data-acquisition system of the \ATLAS detector.
In this context, also the processing of data and event reconstruction are described,
thereby complementing the previous description of the hardware of the \ATLAS detector.
After these introductory texts on the setting,
Chapter~\ref{sec:triggerstudies} comprises the studies of trigger rates and trigger efficiencies,
which together quantify the selection quality of the trigger system.
The focus of these studies lies on the trigger selection
which is based on a combined signature of jets and missing transverse energy.
Both parts, the jets and the missing transverse energy, are reconstructed and computed from calorimeter measurements. %
In addition to the rates, the efficiencies and the methods to measure those,
the benefits of using this combined signature for the selection of collision events by the trigger are discussed.
The trigger rates based on measurements of the missing transverse energy exhibit a non-trivial behavior
as function of the activity in the collision events.
A simple model to explain this behavior is developed in Chapter~\ref{sec:results_met_model},
and its predictions are compared to the measured rates of the \ATLAS trigger.
In the final Chapter~\ref{sec:analysis_susysearch},
the search for Supersymmetry is presented.
After applying a suitable selection of events,
a comparison of event counts in data and simulations is employed
to set limits on the parameters of two supersymmetric models.
Two different methods are used to compute these limits
and the results are compared to results from earlier experiments.
\newcommand{\natunitssecond}[2]{#2} %

\chapter{Theory}
\label{ch:theory}

This chapter describes the theoretical foundations and motivations of the studies which are carried out in this thesis.
First of all, a review of the Standard Model of Particle Physics,
the theoretical framework that comprises the current understanding of experimental results in elementary particle physics,
will be given in three steps.
In the first step, the theoretical background of the Standard Model is provided.
Afterwards, the particle content of the Standard Model is summarized,
and the fundamental interactions among the particles and their phenomenology are described in the third step.
The ensuing discussion %
of open questions and problems
will make clear the need for an extension of the Standard Model.
Two important alternative extensions, technicolor and extra dimensions, will be briefly discussed, %
before Supersymmetry is introduced as a theoretically appealing model,
which can provide a number of answers.
Supersymmetry is the physics model which is searched for in the analysis of collision data in this thesis.
The theory of Supersymmetry,
the particle content of the Minimal Supersymmetric extension of the Standard Model
and its implications for the phenomenology form the last part of this chapter.

\section{The Standard Model of Particle Physics}

\subsection{From Classical Mechanics to the Standard Model}
\label{sec:theory_sm_lagrangians}

The following summary of the theory of the Standard Model is given in terms of gauge theories.
It starts with principles of classical mechanics and leads up to the Standard Model,
including the Higgs mechanism to generate the masses of the particles consistent with observation.
No derivations will be given for the sake of brevity.
It is a very fast-paced introduction\footnote{
  A very clear presentation can be found in \cite{Griffiths},
  on which this summary is based in large parts, together with \cite{HalzenMartin}.
  Quantum Field Theory is covered \eg in \cite{PeskinSchroeder,Maggiore}.
},
needed as a foundation
for the following motivation and explanation of the supersymmetric extensions of the Standard Model.

\subsubsection{Lagrangians in Classical Mechanics}
In classical mechanics, the \index{Lagrangian formalism} is based on the Lagrangian function $L$ (often called Lagrangian for short),
which is comprised of the kinetic energy $T$ and the potential energy $U$ of the system,
\begin{equation}
  L = T - U.
\end{equation}
Using Hamilton's principle, also known as the principle of stationary action,
the variational principle
\begin{equation}
  \delta S[\vec q, \, \delta{\vec q}] \mustbe 0
\end{equation}
is applied to the action $S$,
\begin{equation}
  S[\vec q] = \int_{t_i}^{t_f} L(\vec q, \, \dot{\vec q}, \, t) \intd t.
\end{equation}
$S[\vec q]$ is the integral over the Lagrangian and thus a functional of the generalized coordinates $q_i$ spanning the configuration space of the mechanical system.
Conventionally, their derivatives with respect to time $t$ are written as $\dot q_i$.
Assuming the variation $\delta \vec q$ to vanish at the integration boundaries $t_i$ and $t_f$,
the variational principle gives the Euler-Lagrange equations
\begin{equation} \\
  \frac{\intd}{\intd t} \left( \frac{\partial L}{\partial \dot q_i} \right) = \frac{\partial L}{\partial q_i} \quad \forall i.
\end{equation}
They relate the generalized coordinates $q_i$ and their derivatives $\dot q_i$ and yield the equations of motion.

In field theory, the point-like particles of classical mechanics with trajectories $\vec {x_i}(t)$ in three-dimensional space
are replaced by fields which are functions $\phi_i(\vec x, t)$ of position and time.
In this generalization, the Lagrangian becomes a \index{Lagrangian density} \calL,
which, integrated over space, gives the Lagrangian function.
Therefore, integration over all four spacetime coordinates $x^\mu$ gives the action
\begin{equation}
  S[\phi_i] = \int \calL(\phi_i, \, \partial_\mu \phi_i, \, x_\mu) \intd^4x. %
\end{equation}
The Lagrangian density $\calL$ fully describes the physics of the theory.
This is why theoretical models are often specified by stating their Lagrangian density,
which is also called Lagrangian for short.
The Euler-Lagrange equations become
\begin{equation}
  \partial_\mu \left( \frac{\partial\calL}{\partial (\partial_\mu \phi_i)} \right) = \frac{\partial\calL}{\partial \phi_i},
  \label{eq:theory_euler_lagrange_for_fields}
\end{equation}
where space and time are now treated on equal footing, as befits for a relativistic theory.
The Lagrangian (density) in field theory is %
constructed axiomatically to give the desired field equations when evaluating the Euler-Lagrange equations.
If the constructed $\calL$ describes a Lorentz scalar field, %
the resulting equations of motion are automatically covariant.

Some important Lagrangians for free (non-interacting) fields
commonly met in elementary particle physics describe particles with spin 0, \onehalf and 1.
For a single scalar (spin-$0$) field $\phi$ describing a free particle of mass $m$,
the Lagrangian is given by
\begin{equation}
\natunitssecond{
  \calL_\text{Klein-Gordon} \definedas \frac{1}{2} (\partial_\mu \phi) (\partial^\mu \phi) - \frac12 \left(\frac{mc}{\hbar}\right)^2 \phi^2,
}{
  \calL_\text{Klein-Gordon} \definedas \frac{1}{2} (\partial_\mu \phi) (\partial^\mu \phi) - \frac12 m^2 \phi^2,
}
  \label{eq:theory_lagrangian_klein_gordon}
\end{equation}
yielding the \index{Klein-Gordon equation}
\begin{equation}
\natunitssecond{
  \partial_\mu \partial^\mu \phi + \left(\frac{mc}{\hbar}\right)^2 \phi = 0
}{
  \partial_\mu \partial^\mu \phi + m^2 \phi = 0
}
  \label{eq:theory_eom_klein_gordon}
\end{equation}
by \Eq{eq:theory_euler_lagrange_for_fields}.
The Dirac Lagrangian for a spinor (spin-\onehalf) field $\psi$ describing a particle of mass $m$ is
\begin{equation}
\natunitssecond{
  \calL_\text{Dirac} \definedas \imag (\hbar c) \widebar\psi \gamma^\mu \partial_\mu \psi - (mc^2) \widebar\psi \psi,
}{
  \calL_\text{Dirac} \definedas \imag \widebar\psi \gamma^\mu \partial_\mu \psi - m \widebar\psi \psi,
}
  \label{eq:theory_lagrangian_dirac}
\end{equation}
which leads to the \index{Dirac equation}
\begin{equation}
\natunitssecond{
  \imag \gamma^\mu \partial_\mu \psi - \left( \frac{mc}{\hbar} \right) \psi = 0.
}{
  \imag \gamma^\mu \partial_\mu \psi - m \psi = 0.
}
  \label{eq:theory_eom_dirac}
\end{equation}
Finally, the Lagrangian of a vector (spin-$1$) field $A^\mu$,
\begin{equation}
\natunitssecond{
  \calL_\text{Proca} \definedas -\frac{1}{16\pi} F^{\mu\nu} F_{\mu\nu} + \frac{1}{8\pi} \left( \frac{mc}{\hbar} \right)^2 A^\nu A_\nu
}{
  \calL_\text{Proca} \definedas -\frac{1}{16\pi} F^{\mu\nu} F_{\mu\nu} + \frac{m^2}{8\pi} A^\nu A_\nu
}
  \label{eq:theory_lagrangian_proca}
\end{equation}
with
\begin{equation}
  F^{\mu\nu} \definedas \partial^\mu A^\nu - \partial^\nu A^\mu,
  \label{eq:theory_lagrangian_proca_F}
\end{equation}
leads to the corresponding \index{Proca equation} for a particle of spin 1 and again mass $m$,
\begin{equation}
\natunitssecond{
  \partial_\mu F^{\mu\nu} + \left( \frac{mc}{\hbar} \right)^2 A^\nu = 0.
}{
  \partial_\mu F^{\mu\nu} + m^2 A^\nu = 0.
}
  \label{eq:theory_eom_proca}
\end{equation}
In the limit $m\to0$, this yields Maxwell's equations for the electromagnetic field in empty space. %

\subsubsection{Symmetries and Noether's Theorem}
\label{sec:theory_sm_noether_theorem}
Starting from the above Lagrangians for free fields,
the principle of \index{local gauge invariance},
\ie the symmetry of the free Lagrangian under local phase transformations,
will lead to the appearance of additional fields mediating fundamental interactions.
Symmetries in general play an important role in %
physics.
In this context, a symmetry is understood to be an operation that leaves a system invariant,
\ie carries it into a configuration which cannot be distinguished from the original one.
The properties of symmetry operations (closure, associativity and existence of the identity and inverse)
are the defining properties of a mathematical group.
This motivates the importance of group theory for %
physics,
and in particular Lie groups in case of continuous symmetries.
Every group can be represented by a group of matrices, %
which can then be used to carry out concrete computations.
The \index{Poincaré group} \cite{Greiner3a} includes all transformations of type
\begin{equation}
  x'^\mu = {\Lambda^\mu}_\nu x^\nu + a^\mu,
\end{equation}
\ie translations by a vector $a^\mu$ and Lorentz transformations ${\Lambda^\mu}_\nu$.
Poincaré transformations have ten independent parameters, four for the translation, three for rotations in space and three describing Lorentz boosts.
They leave the distance between two points in Minkowski space invariant.
\inindex{Noether's Theorem}
Noether's (first) Theorem, published 1918 by Emmy Noether \cite{Noether1918}, states that
every \revised{continuous} symmetry of nature yields a conservation law,
and every conservation law reflects an underlying symmetry. %
In a more precise formulation,
if a system (or its Lagrangian density) exhibits a continuous symmetry of the action,
then there are corresponding quantities whose values are conserved in time. %
Noether's Theorem connects invariance under translations in space and time with the conservation of momentum and energy, respectively,
and rotational invariance with the conservation of angular momentum.
In addition to these, there are internal symmetries in elementary particle physics, %
which are independent of the spacetime coordinates,
\ie the transformations commute with the spacetime components of the wave functions. %
The simplest example for such an internal symmetry is the invariance of a wave function $\psi$ under a global phase shift $\theta \in \setR$,
\begin{equation}
  \psi \to \exp( \imag \theta) \, \psi,
  \label{eq:theroy_global_phase_transformation}
\end{equation}
which also holds for the Dirac Lagrangian in \Eq{eq:theory_lagrangian_dirac}.
The family of transformations described by \Eq{eq:theroy_global_phase_transformation}
form a unitary Abelian (commutative) group $U(1)$,
and the corresponding invariance, a global gauge invariance,
through Noether's theorem gives rise to the conservation of the electromagnetic charge. %

\subsubsection{Quantum Electrodynamics}
In contrast to the global phase shift,
the Dirac Lagrangian is no longer invariant
under a local phase transformation, where the phase depends on the spacetime coordinates~$x^\mu$,
\begin{equation}
  \psi \to \exp\!\left( \imag \theta(x) \right) \psi.
  \label{eq:theory_local_phase_transformation}
\end{equation}
The local gauge invariance of the Lagrangian can be restored by replacing the normal derivative $\partial_\mu$ by the covariant derivative
\begin{equation}
\natunitssecond{
  {\cal D}_\mu \definedas \partial_\mu + \imag \frac{q}{\hbar c} A_\mu,
}{
  {\cal D}_\mu \definedas \partial_\mu + \imag q A_\mu,
}
  \label{eq:theory_covariant_derivative_qed}
\end{equation}
which entails the introduction of an additional gauge field $A^\mu$.
A constant factor $q$ has been pulled out. %
This vector field requires that an additional free field term given by \Eq{eq:theory_lagrangian_proca} be included in the Lagrangian,
describing the kinetic energy of the field.
As $A^\nu A_\nu$ is not invariant under the local phase transformation \eqref{eq:theory_local_phase_transformation},
from \Eq{eq:theory_lagrangian_proca} it can be seen that the particle described by $A^\mu$ must be massless.
$A^\mu$ can be identified with the electromagnetic potential, %
\ie the photon field, and transforms as
\begin{equation}
  A_\mu \to A_\mu - \frac1q \partial_\mu \theta.
\end{equation}
Imposing the local phase invariance on the free Dirac Lagrangian thus leads to the Lagrangian of \ac{QED},
\begin{equation}
\natunitssecond{
  \calL_\text{QED} =   \widebar{\psi} \left[ \imag \hbar c \gamma^\mu \partial_\mu - mc^2 \right] \psi
                     - q \widebar{\psi} \gamma^\mu A_\mu \psi
                     - \frac{1}{16\pi} F^{\mu\nu} F_{\mu\nu},
}{
  \calL_\text{QED} =   \widebar{\psi} \left[ \imag \gamma^\mu \partial_\mu - m \right] \psi
                     - q \widebar{\psi} \gamma^\mu A_\mu \psi
                     - \frac{1}{16\pi} F^{\mu\nu} F_{\mu\nu},
}
  \label{eq:theory_lagrangian_qed}
\end{equation}
where $q$ is the charge of the Dirac particle. %

\subsubsection{Quantum Chromodynamics}
The same strategy can be applied to derive the Lagrangian of \ac{QCD},
starting from three color fields,
which are combined into one $\psi$ vector to simplify the notation,
$\psi = (\psi_r, \psi_g, \psi_b)^T$.
Invariance under local $SU(3)$ gauge transformations,
\begin{equation}
  \psi \to S\psi \quad\text{with}\quad S \definedas \exp\!\left(\imag \vec a(x) \cdot \vec \lambda \right),
\end{equation}
is imposed on the Dirac Lagrangian, with eight real numbers $a_1, \dots, a_8$ as parameters
and the Gell-Mann $3\times3$ matrices $\lambda_1, \dots, \lambda_8$ as generators of the transformation.
The invariance can again be restored by introducing the covariant derivative
\begin{equation}
\natunitssecond{
  {\cal D}_\mu = \partial_\mu + \imag \frac{g}{\hbar c} \vec \lambda \cdot \vec {A_\mu}, %
}{
  {\cal D}_\mu = \partial_\mu + \imag g \vec \lambda \cdot \vec A_\mu,
}
\end{equation}
which now involves eight gauge fields $\vec A_\mu$ corresponding to the gluons and one new coupling constant $g$.
Due to the non-commutativity of $SU(3)$, the transformation behavior of $\vec A_\mu$ is different.
For infinitesimal transformations it is given by
\begin{equation}
\natunitssecond{
  {\vec A_\mu} \to {\vec A_\mu} %
                  - \frac{\hbar c}{g} \partial_\mu \vec a %
                  - 2 f_{abc} \vec e_a a_b A_\mu^c %
                  , %
}{
  {\vec A_\mu} \to {\vec A_\mu}
                  - \frac1g \partial_\mu \vec a
                  - 2 f_{abc} \vec e_a a_b A_\mu^c
                  ,
}
  \label{eq:theory_gluon_transformation}
\end{equation}
with the (real) structure constants $f_{abc}$ of $SU(3)$, %
$\vec e_i$ being the $i$th unit vector
and summing over repeated indices implied. %
This leads to the \ac{QCD} Lagrangian,
which describes quarks and gluons and their interactions:
\begin{equation}
\natunitssecond{
  \calL_\text{QCD} =   \widebar{\psi} \left[ \imag \hbar c \gamma^\mu \partial_\mu - mc^2 \right] \psi
                     - g \left( \widebar{\psi} \gamma^\mu \vec \lambda \psi \right) \cdot \vec A_\mu
                     - \frac{1}{16\pi} \vec F^{\mu\nu} \cdot \vec F_{\mu\nu}
                     .
}{
  \calL_\text{QCD} =   \widebar{\psi} \left[ \imag \gamma^\mu \partial_\mu - m \right] \psi
                     - g \left( \widebar{\psi} \gamma^\mu \vec \lambda \psi \right) \cdot \vec A_\mu
                     - \frac{1}{16\pi} \vec F^{\mu\nu} \cdot \vec F_{\mu\nu}
                     .
}
  \label{eq:theory_lagrangian_qcd}
\end{equation}
Like the photon, the eight gluon fields have to be massless
because the Proca mass term proportional to $\vec A^\nu \cdot \vec A_\nu$ is excluded by local gauge invariance. %
Due to the different form of \Eq{eq:theory_gluon_transformation},
also $\vec F^{\mu\nu}$ takes a slightly more complicated form than in \Eq{eq:theory_lagrangian_proca_F},
\begin{equation}
\natunitssecond{
  \vec F^{\mu\nu} \definedas \partial^\mu \vec{A^\nu} - \partial^\nu \vec{A^\mu} - \frac{2g}{\hbar c} f_{abc} \vec e_a {A^\mu_b} {A^\nu_c}.
}{
  \vec F^{\mu\nu} \definedas \partial^\mu \vec{A^\nu} - \partial^\nu \vec{A^\mu} - 2g f_{abc} \vec e_a {A^\mu_b} {A^\nu_c}.
}
\end{equation}
This has the important consequence that the kinetic energy term in \Eq{eq:theory_lagrangian_qcd} is not purely kinetic,
but includes (three- and four-point) %
self-interaction terms for the gluons.
Gluons can thus not only interact with quarks, but also among themselves.

\subsubsection{Higgs Mechanism}
\label{sec:theory_sm_higgs_mechanism}
The derivation of the \ac{QED} and \ac{QCD} Lagrangian %
is straightforward,
using the Dirac Lagrangian for the participating particles
and imposing the principle of local gauge invariance.
In this procedure, the gauge fields turn out to be massless,
which is adequate for the gluons and the photon, %
but the mediators of the weak interaction,
the vector bosons $W^\pm$ and $Z^0$, are massive.
A possible mechanism to allow for massive gauge fields
is \index{spontaneous symmetry-breaking}, %
or, specifically, the \index{Higgs mechanism}, %
which is the spontaneous symmetry-breaking of a local gauge invariance\footnote{
  It is not possible to put in the mass terms by hand,
  thereby breaking local gauge invariance,
  because the resulting theory would be unrenormalizable. %
}.

The main difference is that,
unlike for the Lagrangians formulated above,
the potential energy now is chosen such that the ground state (or vacuum state) no longer is the trivial one of vanishing fields $\phi = 0$.
A common choice for such a potential is the ``Mexican hat'' potential.
For a complex scalar field $\phi$, it is given by
\begin{equation}
  V(\phi) = - \frac{\rho^2}{2} \phi^* \phi + \frac{\lambda^2}{4} (\phi^* \phi)^2, %
  \label{eq:theory_higgs_potential}
\end{equation}
with $\rho^2 > 0$ and $\lambda^2 > 0$. $V(\phi)$ has a circle of minima in the complex $\phi$ plane at
\begin{equation}
  \phi^* \phi = \frac{\rho^2}{\lambda^2}, %
\end{equation}
which is easier to recognize writing $\phi \definedas \phi_1 + \imag\phi_2$, thus $\phi^*\phi = \phi_1^2+\phi_2^2$.
The corresponding Lagrangian
\begin{equation}
    \calL_\text{Mexican} = \frac12 \left( \partial_\mu \phi \right)^* \left( \partial_\mu \phi \right) + \frac{\rho^2}{2} \phi^* \phi - \frac{\lambda^2}{4} (\phi^* \phi)^2
  \label{eq:theory_lagrangian_mexican}
\end{equation}
possesses a global $U(1)$ symmetry  $\phi \to \exp(\imag \theta) \phi$.
In order to find the masses of the participating fields, %
the Lagrangian needs to be rewritten in different field variables and to be expanded about a fixed vacuum ground state. %
Choosing a particular vacuum state breaks the symmetry
and allows to identify the mass terms of the fields. %
This is called spontaneous symmetry-breaking.
Nevertheless, both Lagrangians, before and after rewriting in different field variables, describe the same physical system,
and the manifest symmetry only gets hidden by rewriting the Lagrangian.
Moreover, according to a theorem by 't Hooft \cite{tHooft1971}, %
the theory remains renormalizable. %

Breaking the global symmetry of the Lagrangian \eqref{eq:theory_lagrangian_mexican} is not sufficient,
because the Goldstone theorem states that massless scalars (called \index{Goldstone boson}s) must occur
whenever a continuous global symmetry is spontaneously broken. %
No massless scalars have been found in nature,
and thus breaking the global symmetry does not describe reality.
Instead, the Lagrangian \eqref{eq:theory_lagrangian_mexican} must be made invariant under a local $U(1)$ gauge symmetry, %
following the recipe from above for the derivation of the \ac{QED} and \ac{QCD} Lagrangian.
This implies the introduction of an additional gauge field.
Again, a vacuum ground state is fixed and the Lagrangian rewritten in different field variables
in an expansion around the chosen ground state.
The local $U(1)$ symmetry can be exploited to pick a particular gauge, %
in which it becomes apparent that the Lagrangian now describes two interacting massive fields,
one of which is a massive vector gauge boson and the other a massive scalar,
which is named the Higgs particle.
The massless Goldstone boson has been turned into the longitudinal polarization of the massive gauge particle \cite{Higgs1964}. %
This mechanism is called the Higgs mechanism \cite{HalzenMartin}. %
Note that the precise form of the Higgs potential in \Eq{eq:theory_higgs_potential} is unknown,
but it has to be quartic in the fields for the theory to be renormalizable. %
The Higgs particle has not yet been found,
but the Higgs mechanism, which allows to avoid the appearance of massless particles,
is believed to be the true mechanism for generating masses in the Standard Model.

\subsubsection{Weinberg-Salam Model}

To obtain three massive gauge fields for the charged $W^\pm$ bosons and the neutral $Z^0$ boson %
in the Standard Model,
while retaining a massless photon field,
the electroweak Lagrangian is obtained from \Eq{eq:theory_lagrangian_qed}
by replacing the $U(1)$ electromagnetic interaction with two interaction terms. %
They contain the generator $T$ of the $SU(2)_L$ gauge group of the weak isospin
and the generator $Y$ of the $U(1)_Y$ gauge group of the weak hypercharge.
The underlying symmetry group of the electroweak interactions is thus $SU(2) \otimes U(1)$. %
The interaction terms involve an isospin triplet %
of weak vector fields $\vec W^\mu$ %
and a fourth single vector field $B^\mu$
coupling to the weak isospin and hypercharge current, respectively \cite{HalzenMartin}. %
These fields mix to give the physical fields $W^\pm$, $Z^0$ and the photon $\gamma$.

The Higgs mechanism is applied by introducing four scalar fields
and choosing a suitable vacuum expectation value
which leaves the $U(1)_\text{em}$ symmetry with generator $Q = T_3 + Y$ unbroken, %
so that the photon remains massless.
This gives the Weinberg-Salam\inindex{Weinberg-Salam model} or minimal model of electroweak interactions.
It includes one neutral scalar Higgs field
as a remainder of the four scalar fields originally introduced as the Higgs isospin doublet. %
The other three degrees of freedom have been absorbed
by the now massive weak gauge bosons.
The Higgs doublet can also be used to give masses to all fermions appearing in the Standard Model, quarks and leptons.
Mass terms like $- m_e (\widebar e_R e_L + \widebar e_L e_R)$ %
for the left- and right-chiral electron fields $e_{L,R}$, %
which would otherwise break gauge-invariance,
can be included in the Lagrangian in a gauge-invariant way through the Higgs mechanism \cite{HalzenMartin}. %
This leads to additional terms representing couplings of the fermions to the Higgs field $H$ itself, \eg
\begin{align}
  - \frac{m_e}{v} (\widebar e_R e_L + \widebar e_L e_R) H = - \frac{m_e}{v} \, \widebar{e} e \, H, %
  \label{eq:theory_em_Higgs_fermion_coupling}
\end{align}
where the vacuum expectation value of the Higgs field
\natunitssecond{
  $\sqrt{\hbar c}v = \GeV{246}$ %
}{
  $v = \GeV{246}$ %
}
can be calculated from the mass of the $W$ \cite{Griffiths}.
These couplings are chosen to be proportional to the mass of the fermion\footnote{
  This will play an important role later on when discussing problems of the Standard Model.
}.
The mass of the Higgs $m_H$ itself is given by
\begin{equation}
  m_H = v \sqrt{\frac {\lambda^2} 2} %
  \label{eq:theory_sm_Higgs_mass}
\end{equation}
at tree-level,
where $\lambda^2$ is the strength of the Higgs self-interaction in \Eq{eq:theory_higgs_potential}.
Like all other particle masses, the mass of the Higgs boson is therefore not predicted by the theory.
The resulting Lagrangians after spontaneous symmetry breaking are quite lengthy
already for the simple model with the Lagrangian given in \Eq{eq:theory_lagrangian_mexican},
and therefore are not spelled out here.
The complete Lagrangian of the Standard Model,
before and after spontaneous symmetry-breaking, is given in \cite{Langacker2003}.

The above has lead naturally to the Higgs mechanism
as a \revised{procedure} for generating the masses of the particles which appear in the Standard Model.
In the next section, these particles and the fundamental interactions will be described.
Afterwards, it will be necessary to discuss some of the problems with the Higgs mechanism.
This leads to the need for an extension of the Standard Model
and to the introduction of Supersymmetry.

\subsection{Particle Content}
\label{sec:theory_particle_content_sm}
\inindex{Particle Content!Standard Model}

\arraystretchdefault
\begin{table}
  \centering
  \begin{tabular}{l*{5}{c}}
    \toprule
    Field & Spin & $SU(3)_C$ & $SU(2)_L$ & $U(1)_Y$ & $T_3$ \\
    \midrule
    $q_L = \tpk{u_L}{d_L}$, $\tpk{c_L}{s_L}$, $\tpk{t_L}{b_L}$ & $\frac12$ & $\bf 3$ & $\bf 2$ & $+\frac16$ & \tp{+\frac12}{-\frac12}\\
    $(u_R)^c = \widebar{u}_L$, $(c_R)^c$, $(t_R)^c$ & $\frac12$ & $\bf\widebar3$ & $\bf 1$ & $-\frac23$ & 0\\
    $(d_R)^c = \widebar{d}_L$, $(s_R)^c$, $(b_R)^c$ & $\frac12$ & $\bf\widebar3$ & $\bf 1$ & $+\frac13$ & 0\\
    $l_L = \tpk{\nu_{e L}}{e_L^-}$, $\tpk{\nu_{\mu L}} {\mu_L^-}$, $\tpk{\nu_{\tau L}}{\tau_L^-}$ & $\frac12$ & $\bf 1$ & $\bf 2$ & $-\frac12$ & \tp{+\frac12}{-\frac12}\\
    $(e_R)^c = \widebar{e}_L$, $(\mu_R)^c$, $(\tau_R)^c$ & $\frac12$ & $\bf 1$ & $\bf 1$ & $+1$ & 0\\
    $g$ & 1 & $\bf 8$ & $\bf 1$ & 0 & 0\\
    $W$ = $\begin{pmatrix} W^+ \\ W^0 \\ W^- \end{pmatrix}$ & 1 & $\bf 1$ & $\bf 3$ & 0 & $\begin{array}{c} +1 \\ 0 \\ -1 \end{array}$\\
    $B^0$ & 1 & $\bf 1$ & $\bf 1$ & 0 & 0\\
    \bottomrule
  \end{tabular}
  \caption{
    Field content of the Standard Model of particle physics \cite{Stock,Aitchison}.
    $T_3$ is the third component of the weak isospin corresponding to the $SU(2)_L$ symmetry.
    For $U(1)_Y$, the weak hypercharge $Y$ quantum number is given,
    for $SU(3)_C$ and $SU(2)_L$, the dimension of the representation.
    The Higgs field is not included.
  }
  \label{tab:theory_particle_content_sm}
\end{table}

Depending on their spin quantum number, particles belong to one of two distinct groups:
Fermions with spin \onehalf or in general half-numbered spin,
which are the building blocks of matter,
and bosons with integer spin,
which are the mediators of the interactions in the Standard Model of particle physics. %
There are two groups of fermions within the Standard Model, quarks and leptons, which all have spin \onehalf.
Six quarks $q$ \revised{are} known to exist,
which bear the names up $u$, down $d$, charm $c$, strange $s$, top $t$ and bottom (or beauty) $b$.
These six flavors\inindex{Quark flavor} are organized in three generations (also called families) in the order of increasing masses.
Each generation has a quark with charge $+\nicefrac23$ ($u$, $c$, $t$) and one with charge $-\nicefrac13$ ($d$, $s$, $b$) in units of the electron charge magnitude $e$.
Quarks carry color charge and can therefore participate in strong interactions,
in contrast to leptons.
There are six leptons $l$ in the Standard Model,
which are also grouped into three generations,
each with one charged and one neutral particle.
The charged leptons are, with increasing mass,
the electron $e$, muon $\mu$ and tau $\tau$,
which have negative charge of $-e$.
Every charged lepton is associated with a neutral partner,
the electron neutrino $\nu_e$, muon neutrino $\nu_\mu$ or tau neutrino $\nu_\tau$, respectively.
The twelve fermions in the Standard Model are complemented by twelve antiparticle partners with opposite charge\footnote{
  \revised{Neutrino and antineutrino are both neutral, of course.}
},
which reflects the fact that the Dirac equation has positive and negative energy solutions.
Two additive conserved quantities are associated with quarks and leptons.
The baryon number $B$ is $\nicefrac13$ for quarks, $-\nicefrac13$ for antiquarks and $0$ for all other particles.
The lepton number $L$ is $1$ for leptons, $-1$ for antileptons and $0$ for all other particles.
All observations are consistent with $B$ and $L$ conservation in Standard Model processes. %
Except for neutrino mixing, $L$ is conserved for the three lepton families separately.
The bosons in the Standard Model are the massless neutral photon $\gamma$,
the heavy charged $W^\pm$ bosons and the heavy neutral $Z^0$ boson,
and the eight massless neutral gluons $g$.
All of them have spin $1$ and are called gauge bosons
because they arise from gauge symmetries, as discussed in the previous section.
In addition, there is also the hypothesized Higgs boson $H$, which has spin $0$.

\Tab{tab:theory_particle_content_sm} summarizes the observed particle content of the Standard Model in terms of fields.
The indices $L$ and $R$ give the chirality of the fields with respect to the weak interaction,
which only acts on left-chiral particles.
The left-chiral quarks and leptons which belong to the same generation are written as isospin doublets.
Right-chiral particles are singlets under $SU(2)_L$,
\ie they do not transform under the weak interaction.
$(u_R)^c$ denotes the charge conjugate of the right-handed up quark field,
which is identical to the left-handed anti-up quark field $\widebar{u}_L$ and so on.
Following the usual convention,
antiparticles are indicated by a bar over the symbol for the respective particle.
The table also lists the $W$ and $B$ bosons,
which yield the photon and weak bosons as linear combinations.
The electric charge can be calculated as $Q = T_3 + Y$, %
where $T_3$ is the third component of the weak isospin,
and $Y$ is the weak hypercharge quantum number.
(Note that other conventions use twice the value for~$Y$.) %
The mass spectrum of the Standard Model particles is summarized in \Tab{tab:theory_particle_masses}.
For the neutrinos, the upper limit on the mass from tritium decay is given.
More details are given in the discussion of neutrinos in the Standard Model at the end of \Sec{sec:theory_neutrino_masses_sm}. %

\renewcommand{\arraystretch}{1}
\begin{table}
  \centering
  \begin{tabular}{llcrrr@{}l@{ }rr}
    \toprule
     & Name & Symbol & \multicolumn{2}{c}{Charge\hspace*{-3mm}} & \multicolumn{3}{c}{Mass}  \\
    \midrule
    Fermions: Leptons & electron & $e$ & -1 &  & 0 & .5110 & MeV\\
     & electron neutrino & $\nu_e$ & 0 &  & \multicolumn{2}{c}{$<2$} & eV  \\
     & muon & $\mu$ & -1 &  & 105 & .7 & MeV\\
     & muon neutrino & $\nu_\mu$ & 0 &  & \multicolumn{2}{c}{$<2$} & eV  \\
     & tau (lepton) / tauon & $\tau$ & -1 &  & 1776 & .82(16) & MeV\\
     & tau neutrino & $\nu_\tau$ & 0 &  & \multicolumn{2}{c}{$<2$} & eV  \\
    Fermions: Quarks & up quark & $u$ & $+2/3$ &  & \multicolumn{2}{c}{$1.7$ -- $3.3$} & MeV  \\
     & down quark & $d$ & $-1/3$ &  & \multicolumn{2}{c}{$4.1$ -- $5.8$} & MeV  \\
     & charm quark & $c$ & $+2/3$ &  & 1270 & $^{+70}_{-90}$ & MeV\\
     & strange quark & $s$ & $-1/3$ &  & 101 & $^{+29}_{-21}$ & MeV\\
     & top quark & $t$ & $+2/3$ &  & 172 & $\pm0.9\pm1.3$ & GeV\\
     & bottom quark & $b$ & $-1/3$ &  & 4190 & $^{+180}_{-60}$ & MeV\\
    Bosons & Photon & $\gamma$ & 0 &  & \multicolumn{2}{c}{$< 10^{-18}$} & eV  \\
     & Gluon & $g$ & 0 &  & \multicolumn{3}{c}{(assumed to be \unit[0]{eV})} \\
     & $W$ boson & $W^\pm$ & $\pm1$ &  & 80 & .399(23) & GeV\\
     & $Z$ boson & $Z^0$ & 0 &  & 91 & .1876(21) & GeV\\
    \bottomrule
  \end{tabular}
  \caption{
    Masses of the Standard Model particles
    \cite{PDB2010}.
    Charges are given in terms of the electron charge magnitude. %
    Masses of quarks except for the top quark are given in the $\widebar{\text{MS}}$ scheme (see text). %
    The masses of the electron and muon are known better than to \unit[1]{ppm},
    thus no uncertainty is given.
  }
  \label{tab:theory_particle_masses}
\end{table}
\arraystretchdefault

\subsection{Fundamental Interactions}

Only three of the four fundamental interactions of physics are so far included in the Standard Model of Particle Physics:
the weak, the electromagnetic and the strong interaction.
The fourth, gravitation, has therefore not been mentioned in the previous section,
but for interactions of elementary particles it can be safely ignored
due to its relative weakness at the energy scales which are within reach at current collider experiments.

\subsubsection{Electromagnetic Interaction}
\acf{QED} describes the interactions among charged particles like leptons and quarks,
via the exchange of photons. %
The photon is electrically neutral and therefore cannot interact with other photons at tree level.
The coupling constant that describes the strength of electromagnetic interactions is \cite{Griffiths}
\begin{equation}
\natunitssecond{
  g_e = -q\sqrt{4\pi / \hbar c} = \sqrt{4\pi \alpha} \quad\text{ (for $q = -e$)}, %
}{
  g_e = -q\sqrt{4\pi} = \sqrt{4\pi \alpha} \quad\text{ (for $q = -e$)},
}
\end{equation}
in Gaussian units, %
where $q$ is the electric charge of the particle and $\alpha$ is the fine-structure constant.
In fact, $q$ and $\alpha$ are not constant,\inindex{Running (of coupling constants)}
but %
depend on the energy scale $|k^2|$ of the interaction that is considered,
where $|k^2| = -k^2$ and $k$ is the virtual 4-momentum transferred by the photon.
$\alpha$ is roughly $1/137$ at $|k^2| = 0$ and increases slowly to $1/128$ at energy scales
that correspond to the mass of the $W$ boson, $|k^2| = m_W^2$ \cite{PDB2010}. %
This running of the coupling constant can be explained in terms of a vacuum polarization effect,
leading to a screening of the charge.

\removesection{ %
This running of the coupling constant can be explained in terms of elementary particles %
splitting for a short time into two particles,
\eg giving rise to electron loops within photon propagator lines. %
In consequence, there is a vacuum polarization effect around an electric charge in analogy to a charge
which is immersed in a polarizable (dielectric) medium.
A probe at a large distance will sense a smaller effective charge due to the polarization.
This charge screening becomes less effective close to the charge,
so that the effective charge increases close to the charge.
Summing up the loop contributions from vacuum polarization one obtains in the leading log approximation %
\begin{equation}
\natunitssecond{
  \alpha(|k^2|) = \frac{\alpha(0)}{1 - \frac{\alpha(0)} { 3\pi} \ln\left( \frac{|k^2|}{m^2c^2} \right) }  %
}{
  \alpha(|k^2|) = \frac{\alpha(0)}{1 - \frac{\alpha(0)} { 3\pi} \ln\left( \frac{|k^2|}{m^2} \right) }  %
}
\quad
\text{for}
\quad
\natunitssecond{
  |k^2| \gg (mc)^2.
}{
  |k^2| \gg m^2.
}
\end{equation}
} %

\subsubsection{Weak and Electroweak Interaction}

The \index{weak interaction} is mediated by three massive vector bosons,
two of which, the $W^+$ and $W^-$ bosons, are electrically charged 
and one, the $Z^0$ boson, is neutral.
Weak interactions respect the lepton generations,
\ie there is no cross-generational coupling of the type $e^- \to \nu_\mu + W^-$. %
However, it is possible to change a quark into a quark from another generation in a vertex involving a $W$ boson\footnote{
  In the Standard Model,
  the couplings of the $Z$ boson to fermions are flavor diagonal at tree level %
  and suppressed at higher orders due to the so-called GIM mechanism \cite{Glashow1970}. %
  Flavor changing neutral currents,
  \ie processes which change the flavor of a fermion without altering its electric charge, %
  are however predicted in many new physics models including Supersymmetry \cite{Langacker}. %
}.
This is called quark mixing.
The Cabibbo-Kobayashi-Maskawa (CKM) matrix, %
a complex unitary $3\times 3$ matrix $V$ that can be parametrized using three angles and one phase factor,
relates the physical quark states to the weak eigenstates:
\begin{equation}
  \begin{pmatrix} d' \\ s' \\ b' \end{pmatrix}
  = V
  \begin{pmatrix} d \\ s \\ b \end{pmatrix}
\quad
\text{with}
\quad
  |V_{ij}| =
  \begin{pmatrix}
    0.974 & 0.225 & 0.003 \\
    0.225 & 0.973 & 0.041 \\
    0.009 & 0.040 & 0.999 \\
  \end{pmatrix}
\end{equation}
The uncertainties on the magnitudes of the values in the CKM matrix are about one per mille or smaller \cite{PDB2010}.
The non-zero off-diagonal elements of $V$ cause transitions between generations of quarks in weak processes.
The values in the CKM matrix have to be determined experimentally and are not predicted by the Standard Model.
Weak processes violate parity conservation (in case of $W$ bosons maximally), %
\ie the symmetry of physical processes under inversion of spatial coordinates, %
because the weak interaction only couples to left-chiral fermions and right-chiral antifermions.
The observed small violation of invariance under the combined charge conjugation and parity (CP) operation
can be accommodated by a complex phase factor in the CKM matrix.
Violation of invariance under time reversal and CP (TCP) is theoretically impossible within the current framework of \ac{QFT}.

In the Glashow-Weinberg-Salam (GWS) model of unified weak and electromagnetic (electroweak) interactions, %
the weak and electromagnetic coupling constants and the mass ratio of the $W$ and $Z$ bosons
are completely determined by the weak mixing angle $\theta_W = 28.75^\circ$ \cite{PDB2010}. %
This angle is again a parameter that is not predicted by the Standard Model.
It defines the mixing of the neutral $W$ and $B$ fields,
which give the photon field %
and the $Z$ boson field.
For the coupling constants of the weak and the electromagnetic interaction one obtains %
\begin{equation}
  g_W \sin \theta_W = g'\cos\theta_W = g_e \quad\text{ and }\quad g_Z = \frac{g_e}{\sin \theta_W \cos \theta_W}, %
\end{equation}
where $g_W$, $g_Z$ and $g_e$ are the coupling constants for the weak and electromagnetic couplings, respectively,
and $g'$ is the coupling to the weak hypercharge.
The Standard Model with its Higgs doublet also predicts the weak mixing angle $\theta_W$
to appear in the ratio of the masses of the weak bosons,  %
\begin{equation}
  \cos\theta_W = \frac{m_W}{m_Z}.
\end{equation}

\subsubsection{Strong Interaction}

\acf{QCD} describes the interactions of particles which carry color charge by the exchange of gluons.
Only quarks and gluons carry color charge %
and can \revised{participate} in \index{strong interaction}s.
Unlike photons, the eight massless gluons can also interact among themselves.
This self-interaction leads to two intimately related properties of the strong interaction,
which are very different from QED: asymptotic freedom and confinement. %
\index{Asymptotic freedom} describes the fact that the strong interaction gets weaker at short distances.
This is the opposite of the running of the coupling constant in QED.
\removesection{ %
This is just the opposite of the screening effect in QED, but it is again due to virtual loop corrections.
The antiscreening effect arising from gluon loops gives a larger effect than the screening caused by the quark loops,
so that in the end antiscreening dominates
and the strong coupling constant \alphas decreases with increasing energy scale.
The virtual loop corrections can be summed up to give
\begin{equation}
   \alphas(|k^2|) = \frac{12\pi}{(11N_c-2N_f)\ln(|k^2|/\Lambda^2)} %
\quad
\text{for}
\quad
  |k^2| \gg \Lambda^2.
\end{equation}
Here, the number of colors $N_c = 3$
and the number of quark flavors $N_f \leq 6$ with $4m_q^2 < |k^2|$ enters. %
$\Lambda$ is a scale parameter that needs to be determined from experiment, $\Lambda = \unit[0.2(1)]{GeV}$ \cite{MartinShaw}.
Note that \alphas varies much stronger with $|k^2|$ than $\alpha$.
} %
While the strong interaction is weak at short distances,
at distances above approximately \unit[1]{fm}
quarks experience a confining potential $V(r) \propto \alpha_s r$ \cite{Langacker}. %
The energy stored in the color field (flux tube) when separating quarks against this non-diminishing force %
at some point becomes large enough to create new quark-antiquark pairs,
which then form hadronic states with the original quarks. %
Color is an internal three-valued quantum number.
The possible values of the color degree of freedom are labelled red, green and blue.
Gluons carry one unit of color plus one unit of anticolor, %
quarks carry only one unit of color and antiquarks carry one unit of anticolor. %
Due to confinement, no isolated free quarks have been observed so far,
and all observed particles are colorless. %
This rule limits the possibilities to combine quarks and antiquarks into hadrons,
composite particles made of quarks.
Colorless combinations of quarks are mesons, $q_r\widebar{q}_r$, $q_g\widebar{q}_g$, $q_b\widebar{q}_b$,
and baryons,
where a red, a green and a blue quark, $q_r q_g q_b$,
or three antiquarks with different anticolors,
$\widebar q_r\widebar q_g\widebar q_b$, come together.
As quarks so far have not been observed isolated,
their masses,
with the exception of the top quark as explained below,
can only be inferred indirectly from hadronic properties.
To define quark masses,
typically one uses the
modified minimal subtraction ($\widebar{\text{MS}}$) scheme \cite{Langacker,PeskinSchroeder}, %
a renormalization scheme for QCD perturbation theory
which yields the running mass $\overline{m}(\mu)$,
where $\mu$ is a dimensionful scale parameter \cite{PDB2010}.
Quarks that are assigned a mass of less than \GeV{1}
conventionally are called \index{light quark}s. This includes the up, down and strange quark.
Quarks above this threshold are called \index{heavy quark}s,
including thus the charm, bottom and top quark.
The top quark is the heaviest known elementary particle with a mass of \GeV{172} \cite{PDB2010}. %
It is special among the six quarks
because unlike all of the others it decays by electroweak decay %
before it can hadronize and form a bound state.
From its decay products,
one can therefore deduce properties of the top quark, %
and in particular measure its mass relatively precisely,
with an uncertainty of less than \percent{1} (see \Tab{tab:theory_particle_masses}).

\paragraph{The Parton Model, Factorization, Hadronization}
\label{sec:theory_parton_model}

According to the simple quark parton model \cite{Langacker}, %
a proton consists of three quarks, $uud$, which yields a proton charge of $2 \cdot \frac23-\frac13 = +1$.
An antiproton consists of three antiquarks, $\bar u \bar u \bar d$, yielding a charge $-(2\cdot \frac23-\frac13) = -1$.
Going one step beyond this simple description,
in addition to the three valence quarks in the proton,
a sea of quark-antiquark pairs and gluons exists,
which are produced by soft QCD processes. %
Due to quark-mass effects, the probability for heavy quark pairs to occur is smaller than for lighter pairs.
The momentum of the proton is shared amongst its constituents,
\begin{equation}
  \sum_i \int_0^1 x[q_i(x) + \bar{q}_i(x) ] \intd x + \int_0^1 x G(x) = 1,
\end{equation}
where the sum runs over all quark flavors,
and $q_i$ and $\bar{q}_i$ are the \index{parton distribution function}s (\acs{PDF}s).
$q_i(x)$ is the probability density for finding a quark with flavor $i$
carrying a longitudinal momentum fraction $x$ of the proton (or more generally nucleon) momentum.
$G(x)$ is the probability density for a gluon with a longitudinal momentum fraction $x$ of the proton.
The parton distribution functions can be measured by deep-inelastic scattering of leptons with nucleons \cite{PDB2010}. %
The above is written in terms of parton distribution functions that are independent of the transferred momentum $|k^2|$.
In fact, the parton distribution functions depend on both $x$ and $|k^2|$,
and the evolution of the PDFs with $|k^2|$ %
is described by the Dokshitzer-Gribov-Lipatov-Altarelli-Parisi (DGLAP) equations \cite{Ellis1996}. %
The evolution kernel in the DGLAP equations is determined by splitting functions,
which give the probability for a parton to emit another parton carrying a fraction of the original parton's momentum.
The parton distribution functions have been derived by various groups.
The most prominent are CTEQ \cite{CTEQ6.6} and MRS/MRST/MSTW \cite{MSTW2008}.
They are an important input to simulations of particle reactions and computations of cross sections.
\revised{As they are universal,} the parton distribution functions for proton-proton collisions are the same as for deep inelastic scattering.
The cross section can therefore be calculated through the factorization theorem,
according to which for a process
in which two hadrons $H_{1,2}$ produce a specific observed final state $F$ plus proton rests $X$,
the cross section reads \cite{Ellis1996}
\begin{equation}
  \sigma(H_1 H_2 \to F+X) = \sum_{i,j \in \{q,\bar q,g\}} \int \!\! \intd x_1 \intd x_2 \, f_i^1(x_1, \mu^2) \, f_j^2(x_2, \mu^2) \, \hat\sigma_{ij \to F} (x_1 P_1, x_2 P_2, \alphas(\mu^2), Q^2/\mu^2),
  \label{eq:theory_factorization_theorem}
\end{equation}
where the sum runs over all quark and antiquark flavors and gluons.
The cross section thus factors into a partonic cross section $\hat\sigma$ averaged over spin and color for the relevant subprocess,
and the distribution functions $f^{1,2}$ for the respective parton defined at a factorization scale $\mu$.
The factorization scale is a scale parameter that separates the long- and short-distance physics, %
\ie the low transverse momentum physics below $\mu$ that is considered as part of the hadron structure,
and the high transverse momentum physics that goes into the partonic cross section.
$\mu$ is often chosen to be equal to the renormalization scale, %
and often both are set to the characteristic scale of the hard scattering process $Q$ \cite{Ellis1996}. %
Quarks and gluons, the production of which can be described by \Eq{eq:theory_factorization_theorem},
will subsequently hadronize into mesons and baryons, %
which in the detector appear as clustered energy depositions or jets (cf. \Sec{sec:software_hadronization}). %
The hadronization is a long-distance (low-energy) process,
\ie it cannot be described by perturbative QCD due to the large values of the coupling constant \alphas at these energies. %
Instead, there are phenomenological models tuned to fit experimental data and non-perturbative lattice QCD models \cite{Langacker}.
The computation of the cross sections from the factorization theorem
and the phenomenological treatment of the hadronization
gives a separation of perturbative and non-perturbative effects,
which is often used in simulations in high-energy physics (cf. \Sec{sec:software_MC}).

\subsubsection{Neutrinos in the Standard Model}
\label{sec:theory_neutrino_masses_sm}

In many descriptions of the Standard Model, neutrinos are taken to be massless,
although by now it is established experimentally that neutrinos can oscillate between different flavors, $\nu_e$, $\nu_\mu$ and $\nu_\tau$,
which is only possible if they have different masses.
The values of the neutrino masses are unknown.
As the neutrino oscillations are only sensitive to differences in neutrino masses
via the conversion probability
\begin{equation}
\natunitssecond{
  P(\nu_e \to \nu_\mu) = \sin^2(2\theta) \sin^2\left( \frac{(m_2^2-m_1^2)c^3}{4\hbar E} d \right),
}{
  P(\nu_e \to \nu_\mu) = \sin^2(2\theta) \sin^2\left( \frac{(m_2^2-m_1^2)}{4E} d \right),
}
  \label{eq:theory_two_neutrino_mixing}
\end{equation}
they can only be used to set experimental constraints on the mass splittings.
Here, $E$ is the neutrino energy, $m_{1,2}$ are the masses,
$d$ is the distance traveled by the neutrino,
and $\theta$ is the mixing angle relating the mass and flavor eigenstates.
\Eq{eq:theory_two_neutrino_mixing} is written for two-neutrino mixing,
but in general all three neutrino flavors mix. %
Direct measurements from the high energy cut-off of the beta-decay spectrum of tritium
can only set upper limits on neutrino masses.
The upper limit for the electron neutrino mass is $\unit[2]{eV}$ \cite{Stock}.
In addition, lower bounds from atmospheric neutrino oscillations exist. %

It is not yet clear whether neutrinos are their own antiparticles (Majorana particles).
This could be proven by establishing the existence of the neutrinoless double beta decay \cite{Griffiths}.
Majorana neutrinos would be required by the ``see-saw'' mechanism,  %
which predicts very heavy neutrino partners at high mass scales \cite{Binetruy}. %

\subsection{Open Questions and Problems} %

The Standard Model of particle physics has been in place and constantly improved over many decades now,
and it describes and explains many observations at high precision,  %
in particular in the domain of QED,
for example from the agreement of different measurements of the fine-structure constant which are able to test QED \cite{Langacker}. %
It is thus a very successful theoretical framework.
However, there are several indications which suggest that the Standard Model needs to receive modifications or extensions,
which can roughly be grouped into two classes:

First, there are experimental observations which the Standard Model fails to describe
or which directly contradict the respective theoretical prediction from the Standard Model.
One example is the anomalous magnetic moment of the muon,
which can be measured very precisely \cite{Bennett2003}
and differs from the Standard Model expectation by $3.4\sigma$ \cite{Amsler2008}. %
Using a different determination (from hadronic $\tau$ decays)
for one of the correction terms reduces the discrepancy to $0.9\sigma$.
A possible explanation are additional contributions as they are found in supersymmetric extensions of the Standard Model \cite{Langacker}.

Second, despite its success, the Standard Model cannot provide answers to a large number of open questions\footnote{
  A comprehensive overview of interpretations and suggestions of new physics models that provide possible explanations is given in \cite{Langacker}.
}:
\begin{itemize}
  \item Why is charge quantized in thirds of the charge of the electron?
    Why is (only) the weak interaction chiral?
    Why is the underlying symmetry group a complicated product of three groups, $SU(3)\times SU(2)_L \times U(1)$?
    Is there a unification of these interactions?
  \item Why are there three generations of quarks and leptons?
    Why are the masses of the elementary particles so different,
    going from a fraction of an electronvolt for neutrinos to almost \unit[$2\ten{11}$]{electronvolts} for the top quark?
    Why can quark generations mix, whereas the lepton number,
    apart from neutrino oscillations, is conserved for each generation individually?
  \item What has caused the prevalence of matter over antimatter as observed today?
\end{itemize}
In addition to these open questions,
the Standard Model has a large number of free parameters
that are not predicted by theory.
Another two open questions will be discussed in more detail in the following.

\subsubsection{The Hierarchy Problem of the Higgs Mass}

One of the most unsatisfactory aspects of the Standard Model
is known as the fine-tuning\inindex{Fine-tuning problem} or \index{hierarchy problem} \cite{Aitchison,SUSYPrimerMartin1997} %
and concerns the mass of the Higgs particle, as predicted in the Standard Model.
The Higgs mass, as given at tree level in \Eq{eq:theory_sm_Higgs_mass},
receives contributions from loop corrections.
For a Dirac fermion $f$ which couples to the Higgs through a term $-\lambda_f H \widebar{f} f$ %
in the Lagrangian (cf. \Eq{eq:theory_em_Higgs_fermion_coupling}),
this contribution turns out to be \cite{SUSYPrimerMartin1997}
\begin{equation}
  \Delta m_H^2 = - \kappa |\lambda_f|^2 \Lambda_\text{UV}^2, %
  \label{eq:theory_bsm_higgs_correction_fermion}
\end{equation}
with a constant prefactor $\kappa = 1/8\pi^2$.
$\Lambda_\text{UV}$ is a cut-off representing the scale at which new physics appears
and is believed to be indicated by the Planck mass
\begin{equation}
  m_P \simeq \GeV{1.22\ten{19}} %
\end{equation}
at most, which is the scale at which quantum gravity is expected to become important. %
The loop corrections to the Higgs mass are thus many orders of magnitude larger
than the phenomenologically required value of the Higgs mass,
which is of the order of a few hundreds of \GeV{}\footnote{
  Electroweak precision data favor rather low values of the Higgs mass,
  below \GeV{167} at \percent{95} confidence level \cite{Langacker}. %
  The even lower current limits on the Higgs mass will be discussed below.
}.
The required low value for the Higgs mass can still be retained
by adjusting the parameter $-\rho^2$ of the (unknown) Higgs potential in \Eq{eq:theory_higgs_potential}
with a one-loop corrected physical value, %
but this would rely on a very precise cancellation of two very large numbers.
Although not excluded in principle, this fine-tuning is considered very unlikely and therefore unnatural.
Note that these quadratic divergences of the loop integrals are a problem that only affects scalar fields.
The fermion masses in the Standard Model are protected by the chiral symmetry,
reducing the quadratic divergences to logarithmic divergences \cite{Aitchison}. %
The fact that the hierarchy problem can be solved by introducing Supersymmetry,
as shown below,
is one of the main indications in favor of Supersymmetry,
although Supersymmetry historically has not been introduced to solve this problem. %

\subsubsection{Dark Matter and Dark Energy}

Another unsatisfactory aspect is that the Standard Model only covers a few percent of the energy and matter content of the universe.
Measurements of velocities of galaxies first done %
in 1933 using the Doppler effect \cite{Zwicky1933,Zwicky1937},
and later more precise measurements of the rotation curves of galaxies \cite{Rubin1970,Rubin1980},
\ie the rotational velocity as function of the distance from the center of the galaxy,
revealed that there must be a lot more matter in galaxies than the matter in visible stars.
This additional non-luminous and non-absorbing matter,
which is evident only through its gravitational effects,
is called \index{dark matter}.
Today, from a number of cosmological observations like %
the cosmic microwave background radiation,
galactic lensing
or the acceleration and large scale distribution of matter in the universe,
it is found that the universe is flat
and about \percent{21} of its energy content are dark matter.
Only about \percent{5} are ordinary matter as it is described by the Standard Model,
and the remaining \percent{74} are attributed to dark energy \cite{Langacker}. %
This hypothesized \index{dark energy}, the nature of which is unknown \cite{Griffiths}, %
may explain, via its negative pressure, the observed acceleration of the expansion of the universe. %

The term dark matter implies that this substance is thought to be made up from elementary particles.
Most of the dark matter content must be non-relativistic, so-called \acf{CDM},
so that it can clump together and become gravitationally bound on large scales. %
This rules out neutrinos, which are also too light to contribute more than a small fraction to dark matter. %
In general, no Standard Model particle has the right properties to explain cold dark matter, %
which is supposed to be comprised of \acfp{WIMP} \cite{Baer2009}. %
In physics models which go beyond the Standard Model,
there are many candidates for \acp{WIMP},
for example the lightest supersymmetric particles,
similar stable particles in little Higgs or universal extra dimension models,
or axions associated to a potential solution of the strong CP problem \cite{Langacker}. %
\section{Possibilities for Beyond Standard Model Physics}

Before discussing \acf{SUSY}, this section gives an overview of two important theoretical models,
which, like Supersymmetry, are capable of addressing shortcomings of the Standard Model,
in particular the hiearchy problem of the Higgs mass.
The two models are technicolor and (universal) extra dimensions.
In some cases, they may give phenomenological signatures which are very similar to Supersymmetry.

\subsection{Technicolor}

\index{Technicolor} \cite{TechnicolorReview} provides a natural solution to the hierarchy problem
by not introducing a scalar Higgs boson in the first place.
Instead, a dynamical breaking of electroweak and flavor symmetry is employed, %
in which a bilinear of fermions %
acquires a vacuum expectation value. %
To achieve this, in technicolor models a new non-Abelian gauge group is introduced,
which is modelled on QCD:
asymptotically free at very high energies and strong and confining at low energies.
In addition, massless fermions called technifermions are introduced.
The technifermions can form a condensate,
which spontaneously breaks the global chiral symmetry of the fermions.
Following the Goldstone theorem,
the breaking of the symmetry procures Goldstone bosons,
which are referred to as technipions. %
Three of them become longitudinal components of $W$ and $Z$,
which thereby acquire mass. %

Additional interactions are needed to give mass to the Standard Model quarks and leptons, %
which cannot have bare mass terms. %
This is achieved in \index{extended technicolor} by extending the new gauge group
to include color, technicolor and flavor symmetries, %
thereby allowing technifermions to couple to quarks and leptons. %
After the breaking of the larger gauge symmetry,
the resulting massive gauge bosons
can mediate transitions between Standard Model and technicolor fermions,
giving rise to couplings that make the Standard Model fermions massive. %
The interactions of the technicolor massive gauge bosons can also raise the masses of technipions,  %
so that they can have escaped detection so far, %
mediate their decays to Standard Model fermions,
and induce mixing of quarks.  %
A successful technicolor model would not only predict the masses and mixings of quarks and leptons,
but also why there are three families of each. %
Technicolor can also provide dark matter candidates %
in terms of the lowest-lying bound state of technifermions \cite{Bagnasco1994}. %

Similar in spirit to technicolor models,
in the sense that they avoid introducing Higgs particles,
are a number of other Higgsless models which use extended gauge groups. %
There are also little Higgs models with composite Higgs states,
which use extended gauge groups combined with novel mechanisms for breaking the gauge symmetry
and naturally give rise to light Higgs bosons without relying on Supersymmetry.
An overview and discussion of these models, detector signatures and searches for technicolor is given in \cite{PDB2010} and \cite{Hill2003}.

\subsection{Extra Dimensions}

The concept of \index{extra dimensions} was first formulated by Kaluza and Klein in the 1920s
in an attempt to unify gravitation and electromagnetism, known as the \index{Kaluza-Klein theory}. %
It was revived in the 1980s in the context of string theory.
Extra dimensions are needed in consistent quantum theories of strings,
which predict ten or eleven spacetime dimensions~\cite{Quiros2006}, %
as new degrees of freedom for the fundamental objects: the one-dimensional strings. %
Strings can be thought of as a generalization of point-like particles. %
The extra dimensions must be compactified,
\ie instead of being infinite, they are curled up in a circle of small radius,
thus having periodic boundary conditions and a finite length of the order of inverse TeV, %
so that they can evade direct observation. %
The scale of the extra dimensions,
which is given by the compactification radius,
is constrained by experimental data.
In the simplest possible compactification scheme, %
the extra-dimensional part of spacetime is a toroidal structure with one characteristic radius $R$~\cite{Stock}.
The periodic boundary conditions in compactified dimensions imply %
that fields can be expanded as a Fourier series along this dimensional direction,
which results in an infinite tower of massive states, the Kaluza-Klein modes,
collectively called the Kaluza-Klein tower.

The theory of \acf{UED} \inindex{Universal extra dimensions} is a phenomenological model
based on the concept of extra dimensions
similar to the first string theories, %
with the assumption that the extra dimensions have a flat (rather than a warped) metric,
and that all fields can propagate in the compact dimensions.
Momentum conservation %
in the extra dimensions translates %
into the conservation of the number of Kaluza-Klein modes at tree level (KK-parity).
Each interaction vertex involves at least two Kaluza-Klein excitations,
so that the direct production of Kaluza-Klein states is only possible in pairs
and the lightest KK-state is absolutely stable,
making it a good candidate for cold dark matter. %
The implications of KK-parity for the phenomenology of \ac{UED} at particle colliders
are therefore very similar to supersymmetric models with conserved $R$-parity,
as will become clear below.
The mass bounds for the lowest excited modes are relatively low. %
The limits on $R^{-1}$ for one extra dimension are $300$ to $\GeV{500}$, depending on the Higgs mass.
They lie within the region up to $R^{-1} \sim \TeV{1.5}$ which is experimentally accessible at the LHC \cite{Stock}.
By introducing large compact dimensions,
where the term large refers to any dimension which is larger than the Planck scale,
the hiearchy problem can be traded for the problem of a large volume of the extra dimensions,
which relates the Standard Model scale and the Planck scale. %

Mixed extra dimensions are interpolations between the two extreme cases
of universal extra dimensions,
where the Standard Model fields can propagate in all compact dimensions,
meaning that the Standard Model particles also acquire a tower of KK-excitations, %
and scenarios, where the Standard Model fields are confined to a D3-brane, %
\ie a region of spacetime on which open strings can end with three spatial dimensions. %
In mixed extra dimensions, only some fields are constrained to a brane,
\eg the gauge bosons can propagate in the extra dimensions,
whereas the fermions are confined to the Standard Model D3-brane,
which gives different collider bounds.
Minimal \ac{UED} is a five-dimensional model with only one additional flat dimension,
where all particles can propagate in all dimensions.
The experimental signatures of universal extra dimensions at hadron colliders are very similar to those of Supersymmetry (cf. \Sec{sec:theory_supersymmetry}) \cite{Cheng2002,Macesanu2006}:
\begin{itemize}
  \item Both UED and \ac{SUSY} have additional massive particles
    with the same couplings and quantum numbers as their Standard Model partners,
    only the spin being different by half a unit for the supersymmetric partners of the Standard Model particles.
    In UED, there is a tower of additional states, but phenomenologically only the lowest is of interest at this stage. %
  \item In both UED and SUSY, a multiplicatively conserved quantum number can be defined,
    KK-parity in UED and $R$-parity in SUSY.
    In both cases, this implies that the partner particles can only be produced in pairs,
    and that the lightest partner particle at the end of the decay chain is stable
    so that it is a candidate for cold dark matter.
\end{itemize}
This has two important consequences:
First, analyses optimized for Supersymmetry may also be sensitive to UED.
Second, this shows that detector signatures are not unique and can be interpreted in many different models,
so that when indications for beyond Standard Model physics are found,
more detailed studies of the produced new particle states are imperative.

An important difference between UED and SUSY lies in the degeneracy of the masses,
which for Standard Model particles and their supersymmetric partners are very different after breaking of Supersymmetry,
whereas in universal extra dimensions,
the masses of the first levels of KK-excitations at tree level are almost degenerate.
This is because their splitting is provided by Standard Model mass terms,
and most Standard Model particles are relatively light.
Loop corrections can lift the degeneracy in UED,
but the mass differences will still be small,
so that the decay products will be difficult to identify at hadron colliders.
The situation can be improved by allowing the violation of KK-parity through gravity,
which can yield final states with one or more heavy jets and missing transverse energy,
similar to the signature searched for in the analysis in this thesis in \Sec{sec:analysis_susysearch} \cite{Macesanu2006}.

\section{Supersymmetry}
\label{sec:theory_supersymmetry}

This section summarizes the ideas and concepts needed as theoretical background
for the evaluation of data from particle colliders in analyses concerned with \index{Supersymmetry}.
Supersymmetry is theoretically appealing
because it is based on a simple and natural extension
of the existing theoretical framework of particle physics.
Moreover, it follows in many string models,
and it can solve a number of problems of the Standard Model outlined above,
without entailing predictions which are inconsistent with existing experimental observations.

\subsection{General Idea}

The basic idea of Supersymmetry is to introduce a new discrete symmetry
beyond those already included in the Standard Model.
This symmetry relates fermions and bosons,
and thus every fermion has a bosonic supersymmetric partner particle and vice versa.
Supersymmetry therefore doubles the Standard Model particle spectrum,
with the exception of the Higgs sector,
where the (still elusive) scalar and neutral Higgs boson from the Standard Model %
is replaced by five Higgs particles in minimal supersymmetric models. %
The supersymmetric particles have, apart from spin, the same quantum numbers and masses as their Standard Model partners.
Experimental constraints therefore require that Supersymmetry is broken
so that the masses of the supersymmetric partners of the Standard Model particles are shifted to much higher values.
Otherwise the superpartners would have been observed already.
Different models for breaking of Supersymmetry exist and are explained in \Sec{sec:theory_supersymmetry_breaking}.

\subsection{Motivations}
\label{sec:theory_bsm_susy_motivations}
Supersymmetry provides a natural solution to the hierarchy problem of the Higgs mass
by introducing supersymmetric partner particles,
because the loop corrections to the Higgs mass arising from bosons and fermions are of opposite sign:
The correction for a scalar particle $S$ coupling to the Higgs field $H$ via $-\lambda_S|H|^2|S|^2$ is given by \cite{SUSYPrimerMartin1997}
\begin{equation}
  \Delta m_H^2 = \frac\kappa2 \lambda_S \Lambda_\text{UV}^2.
  \label{eq:theory_bsm_higgs_correction_scalar}
\end{equation}
If the coupling constants %
in \Eqs{eq:theory_bsm_higgs_correction_fermion} and \eqref{eq:theory_bsm_higgs_correction_scalar} were equal,
$\lambda_S = |\lambda_f|^2$, %
the quadratic divergences would exactly cancel to all orders in perturbation theory due to the opposite sign\footnote{
  The additional factor $2$ accounts for the fact
  that for each Dirac fermion two complex scalar partners are introduced. %
}.
This would, of course, be an extraordinary coincidence if not enforced by a symmetry~\cite{Wess1974}.
In fact, Supersymmetry does not only cancel the quadratic divergences,
but also divergences which are logarithmic in $\Lambda_\text{UV}$
and contribute to the %
mass term in the Higgs potential \cite{Aitchison}. %
After the necessary (``soft'', see below) breaking of Supersymmetry, the quadratic divergences still cancel,
but there are finite contributions to the Higgs mass of the order of the Supersymmetry breaking scale \cite{Langacker}.

Another motivation for Supersymmetry is that it provides a natural candidate for dark matter
in terms of a weakly interacting massive particle,
which does not exist in the Standard Model.
Most supersymmetric models postulate the conservation of a new quantum number which is called $R$-parity.
This conservation makes the lightest supersymmetric particle (\acs{LSP}) absolutely stable.
If, as is the case in most supersymmetric scenarios,
the LSP is colorless, neutral and heavy,
which are all requirements met by the supersymmetric neutralinos,
the LSP is an attractive candidate for dark matter.

\begin{figure}
  \centering
  \incgraphics{width=\textwidth}{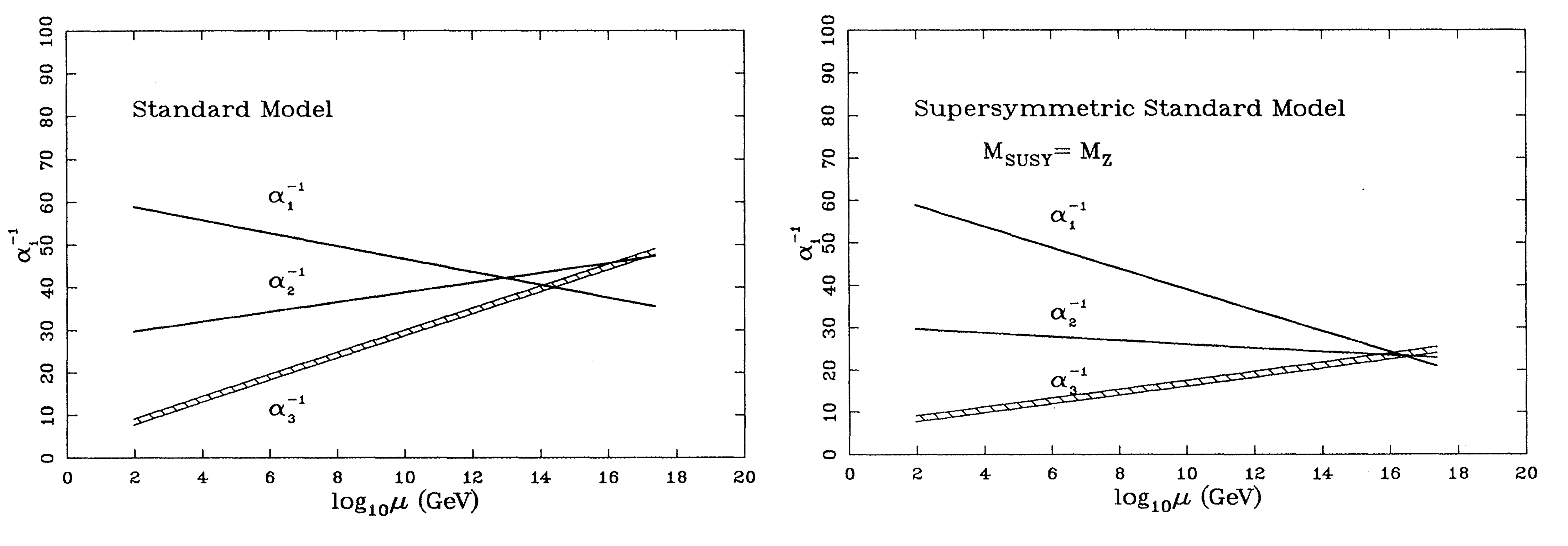}
  \caption{
    Comparison of the extrapolation of the gauge couplings to the GUT scale, %
    assuming the Standard Model (left) and the MSSM (right).
    The vertical axis is $\alpha_i^{-1}$,
    where $\alpha_i$ is the coupling of the group $U(1)_{Y/2}$, $SU(2)_L$, $SU(3)_C$ for $i=1,\,2,\,3$, respectively. %
    $\mu$ is the energy scale.
  }
  \label{fig:theory_gut_unification} %
\end{figure}

In addition, there is another motivation in favor of a supersymmetric extension of the Standard Model,
which comes from the extrapolation of the three gauge coupling constants
to the energy scale of \acp{GUT} %
around $\unit[10^{16}]{GeV}$. %
At this scale, the inverse gauge couplings are predicted to meet in GUTs,
but assuming the Standard Model they \revised{only approximately do so.}
By introducing the superpartners of Standard Model particles, %
the energy dependence of the running coupling constants is modified
such that, in the \acf{MSSM}, it is possible to make them meet exactly in one point at the GUT scale \cite{Langacker}.
This behavior is demonstrated in \Fig{fig:theory_gut_unification}
in a comparison between the Standard Model and the MSSM evolution
of the running of the coupling constants of the three gauge groups $U(1)_{Y/2}$, $SU(2)_L$ and $SU(3)_C$~\cite{LangackerPolonsky1993}.

\subsection{Theoretical Background}
\label{sec:theory_bsm_susy_theory}

As was explained in \Sec{sec:theory_sm_lagrangians}, symmetries play a very important role in physics,
because they allow to deduce physical laws
from very simple assumptions about the invariance of a physical system under a symmetry operation
(cf. Noether's theorem). %
The introduction of a new symmetry beyond those already implemented in the Standard Model seems to fail in the first place,
because the Coleman-Mandula theorem states
that the only conserved charges which transform as tensors under the Lorentz group %
are $P_\mu$ %
and $M_{\mu\nu}$,
the generators of the Poincaré group. %
This means that charges associated with symmetries besides those of the Poincaré group
are Lorentz scalars.
However, the Coleman-Mandula theorem can be circumvented by using fermionic generators \cite{Haag1975}, %
which under Lorentz transformations transform as spinors. %
This is required for generators of supersymmetric transformations
because they transform an integer spin field into a spinor field,
and therefore must carry a spinorial index.
It follows that they do not commute with Lorentz transformations,
and follow anticommutation relations due to the (anti-)commutation rules of bosons and fermions. %
Restrictions from the Coleman-Mandula theorem fix the algebra of the fermionic generators,
making Supersymmetry unique in this sense \cite{Binetruy}. %

The fermionic Supersymmetry generators %
of the Supersymmetry transformations (``supertranslations'') are denoted $Q_a$.
Their algebra for the case of one generator%
\footnote{
  One can also consider representations with more than one SUSY generator,
  but making a supersymmetric extension of the Standard Model
  then becomes difficult
  because the different behavior of the left- and right-chiral components of the lepton fields cannot be easily accommodated \cite{Aitchison}.
} is given by \cite{Aitchison}
\begin{equation} 
  \{ Q_a, Q_b^\dagger \} = 2 \left(\sigma^\mu\right)_{ab} P_\mu, \\ %
  \label{eq:theory_bsm_SUSY_algebra}
\end{equation}
where the brackets denote the anticommutator
and $\sigma^\mu$, $\mu \in\{0,\,\dots,\,3\}$, are the Pauli matrices~\cite{Messiah2},
including the $2\times2$ identity matrix as $\sigma^0$. %
$P_\mu$ is again the generator of translations. %
This is written in the Weyl formalism, %
where the Supersymmetry charges $Q_a$ %
are anticommuting quantum field operators with two components\footnote{
  The Hermitian conjugation $^\dagger$ converts left-chiral spinors into right-chiral spinors,
  which are conventionally also denoted with dotted indices and a bar over the symbol,
  \ie $ Q_b^\dagger =  \widebar{Q}_{\dot b}$.
}, \revised{
$a \in \{1,\,2\}$.
}
They transform under Lorentz transformations as spinors, %
\ie
\begin{equation}
  [Q_a, M_{\mu\nu}] = {(\sigma_{\mu\nu})_a}^b Q_b, %
\end{equation} %
with $M_{\mu\nu}$ being the generators of the Lorentz transformations \revised{
and $\sigma^{\mu\nu} = \frac\imag2\left[\gamma^\mu, \gamma^\nu\right]$.}
Further\revised{more}, being a symmetry operator, they commute with the Hamiltonian of the system. %
The anticommutation relation in \Eq{eq:theory_bsm_SUSY_algebra} is complemented by the relation
\begin{equation}
  [Q_a, P_\mu] = [Q_a^\dagger, P_\mu] = 0,
\end{equation}
stating that the SUSY generators and the generators of translations commute,
and so does~$P^2$.
This means that the states in a supermultiplet
which are connected by the actions of the Supersymmetry generators
must all have the same 4-momentum,
and thus in particular the same mass. %
When the generators $Q_a$ and $Q_a^\dagger$ act on a single particle state,
they either create a state with spin differing by half a unit,
or they annihilate it
if it is at the ``end'' of a supersymmetric multiplet. %

Three different massless supersymmetric multiplets (supermultiplets) are of interest:
\begin{itemize}
  \item The \index{chiral supermultiplet} contains a massless complex scalar field and a massless Weyl fermion field.
    For example, the left-chiral electron and neutrino form an $SU(2)_L$ doublet,
    which is partnered by a corresponding doublet of scalars to form a left-chiral supermultiplet.
    Similarly, the right-chiral electron field, which is an $SU(2)_L$ singlet,
    will be partnered by a corresponding supersymmetric state to form a right-chiral supermultiplet.
  \item The vector\inindex{Vector supermultiplet} or \index{gauge supermultiplet} consists of a massless spin-1 (vector) state
    and a massless Weyl spin-\onehalf fermion partner.
    The gauge bosons of the Standard Model are assigned to gauge supermultiplets. %
  \item The \index{gravity supermultiplet} combines a spin-2 graviton and its spin-$\nicefrac{3}{2}$ gravitino partner.
    This supermultiplet is needed in theories of supergravity.
\end{itemize}
\inindex{Minimal Supersymmetric Standard Model}
Only the first two, the massless chiral and vector supermultiplets, are needed to construct the %
Minimal Supersymmetric (extension of the) Standard Model, %
which is the supersymmetric theory including all Standard Model particles with the minimal additional particle content.
To account for the fact that the supersymmetric transformations mix states with different spin,
it is useful to define a notational extension of the four-dimensional spacetime,
which includes two additional anticommuting coordinates in terms of Grassmann variables. %
The operators which are functions of the coordinates in this \index{superspace} are then called \index{superfield}s \cite{Langacker}. %

\subsection{The Minimal Supersymmetric Standard Model}
\label{sec:theory_bsm_susy_mssm}

\begin{table}
  \centering
  \begin{tabular}{l*{6}{c}}
    \toprule
    Name & Symbol & Spin 0 & Spin \onehalf & SU(3)$_C$ & SU(2)$_L$ & U(1)$_Y$ \\
    \midrule
    Squarks, Quarks & $Q$ & $\tpk{\tilde{u}_L}{\tilde{d}_L}$ & $\tpk{u_L}{d_L}$ & $\bf 3$ & $\bf 2$ & $+\frac16$ \\
    & $\bar{u}$ & $\tilde{\bar{u}}_L = \tilde{u}_R^\dagger$ & $\bar{u}_L = (u_R)^c$ & $\bf\bar3$ & $\bf 1$ & $-\frac23$\\
    & $\bar{d}$ & $\tilde{\bar{d}}_L = \tilde{d}_R^\dagger$ & $\bar{d}_L = (d_R)^c$ & $\bf\bar3$ & $\bf 1$ & $+\frac13$\\
    Sleptons, Leptons & $L$ & $\tpk{\tilde{\nu}_{e L}}{\tilde{e}_L}$ & $\tpk{\nu_{eL}}{e_L}$ & $\bf 1$ & $\bf 2$ & $-\frac12$\\
    & $\bar{e}$ & $\tilde{\bar{e}}_L = \tilde{e}_R^\dagger$ & $\bar{e}_L = (e_R)^c$ & $\bf 1$ & $\bf 1$ & $+1$\\ %
    Higgs, Higgsinos & $H_u$ & $\tpk{h_u^+}{h_u^0}$ & $\tpk{\widetilde{h}_u^+}{\widetilde{h}_u^0}$ & $\bf 1$ & $\bf 2$ & $+\frac12$\\
    & $H_d$ & $\tpk{h_d^0}{h_d^-}$ & $\tpk{\widetilde{h}_d^0}{\widetilde{h}_d^-}$ & $\bf 1$ & $\bf 2$ & $-\frac12$\\
    \midrule
     & & Spin \onehalf & Spin 1 & && \\
    \midrule
    Gluinos, Gluons   & & $\tilde{g}$ & $g$ & $\bf 8$ & $\bf 1$ & 0\\
    Winos, $W$ bosons & &
      $\widetilde W$ = $\begin{pmatrix} \widetilde{W}^+ \\ \widetilde{W}^0 \\ \widetilde{W}^- \end{pmatrix}$ &
      $W$ = $\begin{pmatrix} {W}^+ \\ {W}^0 \\ {W}^- \end{pmatrix}$ &
      $\bf 1$ & $\bf 3$ & 0\\
    Bino, $B$ boson & & $\widetilde B$ & $B$ & $\bf 1$ & $\bf 1$ & 0\\
    \bottomrule
  \end{tabular}
  \caption{
    Field content of the Minimal Supersymmetric extension of the Standard Model~\cite{Aitchison,SUSYPrimerMartin1997}.
    The fields with tildes are the supersymmetric partners of the Standard Model fields.
    Note that only the first generation is shown for the quarks and leptons.
    The Higgs of the Standard Model is replaced by two Higgs doublets in the \acs{MSSM}.
    $\tilde{u}_L$ and $\tilde{u}_R$ are independent scalar fields.
    The index refers to the chirality of their superpartners.
  }
  \label{tab:theory_particle_content_mssm}
\end{table}

\inindex{Particle Content!MSSM}
\Tab{tab:theory_particle_content_mssm} shows the field content of the \acf{MSSM}.
Every Standard Model particle is accompanied by a superpartner,
conventionally denoted with a tilde.
The fermions in the Standard Model are the lepton fields,
\eg the $SU(2)_L$ doublet consisting of the neutrino $\nu_{eL}$ and electron field $e_L$,
and the quark fields,
which are triplets under the $SU(3)_C$ color gauge group.
The left-chiral lepton fields are accompanied by a doublet of scalar fields, %
\eg the sneutrino $\tilde{\nu}_{eL}$ and selectron $\tilde{e}_L$,
where the prefix ``s'' stands for scalar rather than for supersymmetric.
The quarks have corresponding squark fields,
where the $SU(2)_L$ doublet $(u_L, d_L)$ is partnered by $(\tilde{u}_L, \tilde{d}_L)$.
The same holds for the right-chiral fields,
which are $SU(2)_L$ singlets, %
$\tilde{e}_R$ partners $e_R$,
$\tilde{u}_R$ partners $u_R$ and so on.
The indices of the scalar fields, \eg of $\tilde q_L$,
refer to the chirality of their superpartners
and show what their $SU(2)_L\times U(1)$ quantum numbers are. %
For the spin $1$ particles of the Standard Model,
the gluons, $W$ bosons and the $B$ boson,
there is an $SU(3)_C$ octet of Weyl fermions %
called gluinos (\gluino)\footnote{
  In models with supergravity,
  the symbols for gluons and gluinos are often chosen to be $G$ and $\widetilde G$
  to reserve $g$ and $\widetilde g$ for the graviton and gravitino with spin $3/2$.
},
an $SU(2)_L$ triplet of Weyl fermions called winos ($\widetilde{W}^{\pm,0}$),  %
and finally a $U(1)_Y$ bino ($\widetilde{B}$). %

The Higgs sector in supersymmetric models needs to be extended with respect to the Standard Model
to include two Higgs doublets instead of one
because Supersymmetry does not allow the needed Yukawa couplings to generate masses for both $u$ and $d$ quarks with only one Higgs doublet. %
The two independent Higgs chiral supermultiplets
are $H_u = (H_u^+ \, H_u^0)^T$ and $H_d = (H_d^0 \, H_d^-)^T$
with scalar and fermionic components such as $h_u^+$ and $\tilde h_{uL}^+$ for $H_u^+$. %
After symmetry breaking,
this leaves three neutral and one conjugate pair of charged Higgs fields. %
In analogy to the mixing of the three Standard Model $W$ bosons and the $B$ boson,
in the MSSM %
their supersymmetric partners mix to give the winos, a zino $\widetilde Z^0$ and a massless photino $\widetilde \gamma$. %
But also the Higgsinos can mix with the winos and the bino to produce %
two mass eigenstate Dirac charginos ($\tilde\chi_{1,2}^\pm$, so four particles in total)
and four mass eigenstate Majorana neutralinos ($\neutralinon{1,\dots,4}$ with $m_\neutralinon{1} < \dots < m_\neutralinon{4}$). %

The supersymmetric interactions are fixed by specifying the \index{superpotential},
which is a holomorphic function of the left-chiral superfields only %
and thus itself a left-chiral superfield.
In the MSSM, the superpotential $W$ is \cite{Aitchison}
\begin{equation}
  W = y_u^{ij} \widebar{u}_i Q_j \cdot H_u - y_d^{ij} \widebar{d}_i Q_j \cdot H_d - y_e^{ij} \widebar{e}_i L_j \cdot H_d + \mu H_u \cdot H_d, %
  \label{eq:theory_bsm_superpotential_mssm}
\end{equation}
where the symbols for the chiral supermultiplet fields from \Tab{tab:theory_particle_content_mssm} have been used\footnote{
  Note that $Q = \tpk{U}{D}$,
  with \eg $\tilde u_L$ being the scalar and $u_L$ the fermion spinor component of the left-chiral supermultiplet $U$.
}.
The color indices for the terms involving quarks are not written out.
The dimensionless couplings are given in terms of $3\times 3$ matrices $y$ in generation space,
which are exactly the same as the couplings in the Standard Model.
$\mu$ is thus the only additional parameter needed in this supersymmetric extension of the Standard Model.
From the superpotential, the interaction terms of the Lagrangian can be obtained from the so-called $F$ component,
which automatically makes the Lagrangian invariant under supersymmetric transformations. %
Note that the full Lagrangian of the \ac{MSSM} will contain auxiliary complex fields $F$, %
which do not correspond to physical particles,
but which are needed to ensure a supersymmetric action. %
When the Higgs fields acquire vacuum expectation values,
the first three terms with $y$ couplings give masses to the quarks and leptons and their supersymmetric partners. %
The last term in the superpotential yields supersymmetric masses for the Higgs and Higgsino fields. %
It is connected with the $\mu$-problem \cite{KimNilles1984},
which is a problem of naturalness within supersymmetric models.

\subsubsection{\texorpdfstring{$R$-Parity} {R-parity}}

The superpotential of the MSSM in \Eq{eq:theory_bsm_superpotential_mssm} is a specific choice
which does not include all gauge-invariant and renormalizable terms.
The terms which have been left out are~\cite{Aitchison}
\begin{equation}
  W_{\Delta L = 1} = \lambda_{e}^{ijk} L_i \cdot L_j \widebar e_k +
                     \lambda_{L}^{ijk} L_i \cdot Q_j \widebar d_k +
                     \mu_{L}^{i}       L_i \cdot H_u, %
\end{equation}
which would violate lepton number conservation, and
\begin{equation}
  W_{\Delta B = 1} = \lambda_{B}^{ijk} \, \widebar u_i \widebar d_j \widebar d_k, %
\end{equation}
which would violate bayron number conservation.
$\lambda$ and $\mu$ represent all possible coupling constants.
In the Standard Model, %
no renormalizable terms in the Lagrangian are possible,
that would violate lepton or baryon number conservation.
In the MSSM, however, such terms are possible.
Including both non-zero couplings $\lambda_{L}^{ijk}$ of the leptoquark term
and $\lambda_{B}^{ijk}$ of the diquark term
would allow the proton to decay. %
This is in conflict with observation,
because the lower limit on the lifetime of the proton is of the order of $10^{31}$ years \cite{PDB2010}. %
It can thus be ruled out that both terms appear with non-vanishing couplings,
but there is no fundamental reason not to include any of these terms.
To forbid lepton number and baryon number violating terms,
an additional symmetry is postulated \cite{SUSYPrimerMartin1997}.
It is called $R$-parity\inindex{R-parity@$R$-parity},
\begin{equation}
  R = (-1)^{2s+L+3B},
\end{equation}
where $s$ is the spin of the particle, $L$ the lepton number and $B$ the baryon number,
yielding $R=+1$ for Standard Model particles and $R=-1$ for supersymmetric particles.
$R$-parity is multiplicatively conserved in supersymmetric processes
and forbids all terms in $W_{\Delta L = 1}$ and $W_{\Delta B = 1}$.
It has furthermore two important phenomenological consequences:
First, supersymmetric particles are always produced in pairs at collider experiments, %
which means that in supersymmetric events there will always be at least two overlaid decay chains.
Second, the \ac{LSP} is absolutely stable,
because there are no lighter supersymmetric particles,
and a decay to only Standard Model particles would violate $R$-parity.
In many supersymmetric models, the \ac{LSP} is a neutralino,
and therefore, if it is stabilized by $R$-parity,
it gives an attractive candidate for non-baryonic cold dark matter.
Due to their weakly interacting nature,
these \acp{LSP} would be detected as missing momentum in collider experiments.

\subsection{Breaking of Supersymmetry}
\label{sec:theory_supersymmetry_breaking}

\inindex{Supersymmetry breaking}

Because no degenerate supermultiplets are observed,
some form of breaking of Supersymmetry is inevitable
in order to bring the masses of the sparticles
into the not yet excluded range of several hundred gigaelectronvolts.
It can also be shown that the Lagrangian of the \ac{MSSM},
being invariant under Supersymmetry,
cannot accommodate electroweak symmetry breaking. %
Like other symmetries, Supersymmetry can be broken explicitly or spontaneously.
In any case, the Supersymmetry breaking needs to be ``soft''
in order to avoid the reintroduction of the hierarchy problem. %
``Soft'' means that the breaking only appears via terms in the Lagrangian with positive mass dimension, %
and in general that it maintains the cancellation of quadratically divergent radiative corrections to masses of scalar fields. %
As a consequence, Supersymmetry breaking terms can only
be mass terms for scalars or gauginos or cubic scalar couplings. %
Spontaneous symmetry breaking is favored for a number of reasons \cite{Langacker}, %
and can occur for example as so-called $D$-type or $F$-type breaking:
either one of the auxiliary fields $D$,
which need to be introduced for the gauge supermultiplets, %
or one of the auxiliary fields $F$ mentioned above
acquires a symmetry-breaking non-zero vacuum expectation value. %

A generic feature of spontaneous Supersymmetry breaking is the appearance of a Weyl fermion, the Goldstino\footnote{ %
  The Goldstino is not the fermionic superpartner of a scalar Goldstone boson,
  but itself a fermionic Goldstone mode \cite{Aitchison}. %
},
which in supergravity becomes a component of the gravitino in the super-Higgs mechanism (see below).
Furthermore, spontaneous symmetry breaking in general implies a sum rule
for tree-level mass squares of fermionic and bosonic degrees of freedom.
This sum rule requires some superpartners to be light,
which means that tree-level spontaneous symmetry breaking cannot be phenomenologically viable for the MSSM. %
To evade these constraints, the breaking of Supersymmetry must %
occur in a sector which is only weakly coupled to the chiral supermultiplets of the MSSM,
and therefore called \index{hidden sector}. %
The mechanism responsible for the mediation
from the hidden sector to the MSSM particles can be of various types
which are outlined in \Sec{sec:theory_bsm_susy_models}.
Independently of the exact mechanism,
the parametrization of the possible gauge invariant terms
in the now effective Lagrangian at low energy scales %
introduces a large number of additional free parameters in the MSSM.
In the MSSM with conserved $R$-parity,
there are 124 free parameters \cite{Dimopoulos1995}, %
without including neutrino masses and right-handed scalar neutrinos,
\ie 105 new parameters compared to the Standard Model \cite{Aitchison}. %
This makes the model flexible,
but also makes it difficult to transcribe results from searches for Supersymmetry
into conclusions for the MSSM.
Thus, simplified models are constructed within the MSSM,
which enforce phenomenological constraints, %
and in the same step reduce the number of free parameters of the breaking mechanism significantly. %
Some examples are presented in \Sec{sec:theory_bsm_susy_models}.

\subsection{Higgs Mass Bounds and the MSSM}

Assuming an appropriate form of Supersymmetry breaking,
the known Standard Model particle content seems to be fully compatible
with a possible extension to a supersymmetric model.
The question is whether this does also hold for the Higgs boson,
which is believed by many to exist,
but at the time of writing,
this is not yet experimentally proven.
In the Standard Model, there is one Higgs doublet with four real scalar degrees of freedom. %
After electroweak symmetry breaking, three of these become longitudinal modes of the massive vector bosons of the weak interaction,
while the remaining fourth becomes the neutral scalar Higgs boson.
In the Minimal Supersymmetric Standard Model,
there are two complex Higgs doublets with thus eight %
scalar degrees of freedom.
Again, three of them are needed as longitudinal modes of the massive vector bosons,
which leaves five Higgs bosons in the MSSM,
three of which are neutral ($A^0$, $h^0$, $H^0$) and two are charged ($H^\pm$). %
If the existence of a Higgs boson is established experimentally,
this does not rule out that there are more Higgs bosons and therefore does not exclude the MSSM.
However, the opposite case of a light Higgs boson being excluded might be fatal for the MSSM,
which sets quite stringent upper bounds on the mass of the lightest Higgs boson $h^0$,
\begin{equation}
  m_{h_0} \leq m_Z |\cos(2\beta)| \leq m_Z \quad\text{ (tree-level)},
  \label{eq:theory_bsm_lightest_higgs_lower_bound}
\end{equation}
where $m_Z$ is the mass of the $Z$ boson and $\tan\beta \definedas v_u/v_d$ %
is the ratio of the vacuum expectation values of the $H_u^0$ and $H_d^0$ Higgs fields \cite{SUSYPrimerMartin1997}. %
The experimental lower bound on the mass of the (Standard Model) Higgs boson
from the \acf{LEP} is \GeV{114.4} at \percent{95} confidence level (\CL) \cite{LEPh02003}.
However, the lower bound in \Eq{eq:theory_bsm_lightest_higgs_lower_bound} receives loop-corrections
which may be of sufficient size to shift the theoretical upper bound on $m_{h_0}$ close to \GeV{140},
safely above the lower bound from \ac{LEP} \cite{Degrassi2003}. %
Recent results of Higgs searches in \ATLAS and \CMS
exclude (Standard Model) Higgs masses between $130$ and \GeV{235} \cite{ATLAS-COM-CONF-2011-195}
and $127$ and \GeV{600} \cite{CMS-PAS-HIG-11-032}
at \percent{95} \CL, respectively,
which still leaves a gap that would allow for an MSSM Higgs boson.

\subsection{Important Supersymmetric Models}
\label{sec:theory_bsm_susy_models}

This section concludes the introduction to Supersymmetry
by presenting \acf{mSUGRA} and several other prominent supersymmetric models.
The attractiveness of these models
lies in the small number of parameters compared to the 124 parameters of the full MSSM.
They rely on breaking of Supersymmetry in a hidden sector
and mediation of this breaking to the visible sector of MSSM particles via different mechanisms.

\subsubsection{Minimal Supergravity}
\label{sec:theory_bsm_model_minimal_supergravity}

The relevant supersymmetric model for the studies in this thesis is the \ac{mSUGRA} model\inindex{Minimal supergravity model},
which is also one of the most frequently encountered models in Supersymmetry analyses,
and often used for investigations of the phenomenological consequences of weak scale Supersymmetry \cite{Chamseddine1982, Hall1983}. %
As pointed out in~\cite{Binetruy},
making the theory of special relativity invariant under local coordinate transformations %
leads to the theory of general relativity,
and along the same lines,
local Supersymmetry is expected to lead to a theory of gravity called \index{supergravity}.
Indeed, when making the Supersymmetry transformations local,
\ie when the parameter in the transformations generated by the supersymmetric charges become spacetime dependent,
this necessitates the introduction of a graviton field and its superpartner, the gravitino, and thus gravity,
in analogy to other localized gauge symmetries. %
Supergravity is not a full quantum theory of gravity,
but an effective theory which is nonrenormalizable.
Again, in analogy to spontaneous breaking of a continuous gauge symmetry (cf. \Sec{sec:theory_sm_higgs_mechanism}), %
a~Goldstone mode in terms of the Goldstino provides the missing degrees of freedom of the gravitino. %
This is known as the \index{super-Higgs mechanism}. %
The graviton remains massless. %
In supergravity, the breaking of Supersymmetry in the hidden sector
is mediated by gravitational couplings,  %
\ie the effective Lagrangian contains nonrenormalizable terms for the mediation,
which are suppressed by powers of the Planck mass \cite{SUSYPrimerMartin1997}. %
Models of minimal supergravity have five free parameters,
of which four are continuous and one is discrete,
and which are conventionally denoted with the following symbols \cite{Binetruy}:
\begin{itemize}
  \item \moh: a common mass for all gauginos at the GUT scale, %
  \item \mzero: a common mass for all scalar partners of leptons and quarks and the Higgs doublets at the GUT scale, %
  \item $\tan \beta$: the ratio of Higgs vacuum expectation values,
  \item \signmu: the sign of the Higgs mixing parameter and %
  \item $A_0$: a common value for the soft SUSY-breaking trilinear couplings at the GUT scale.
\end{itemize}
Subsets of these parameters can be traded for others,
but this set is commonly used for the parametrization of \ac{mSUGRA}.
Thus, all squarks and sleptons are degenerate in mass at the \ac{GUT} scale,
which allows to eliminate mixings of states with the same electroweak quantum numbers by unitary transformations.
The couplings and masses evolve according to \acfp{RGE} %
from the \ac{GUT} scale down to the electroweak scale.
One prediction of the \ac{RGE} evolution common to most supersymmetric models
is that the gluino is expected to be heavier than the states from the electroweak sector \cite{Aitchison}. %
The mass of the gravitino $m_{\nicefrac32}$ in minimal supergravity models
is of the order of the masses of the other supersymmetric particles \cite{SUSYPrimerMartin1997} %
and may be set equal to $m_0$ \cite{Olive2011}. %
As its couplings are of gravitational strength,
it is not relevant for collider physics. %

\subsubsection{Other Possibilities}

In \cite{Abdus2011SUSYBenchmark},
an overview of a number of models is given,
which all have fewer parameters than the full MSSM and can be tested at current collider experiments.
What is referred to in this thesis as the mSUGRA model is called the Constrained MSSM (CMSSM) there,
and mSUGRA is defined instead as a subset of the CMSSM models with additional assumptions.
Another related model which is introduced in \cite{Abdus2011SUSYBenchmark} is the Non-Universal Higgs Mass Model (NUHM).
It generalizes the CMSSM by relaxing the boundary conditions at the unification scale,
thus introducing additional %
parameters to allow for non-universal Higgs masses~$m_H$
or $m_{H_u}$ and $m_{H_d}$ at the GUT scale.
Such extensions of mSUGRA are needed, for example,
to make mSUGRA predictions compatible with the observed neutralino relic density \cite{Baer2009}. %

Furthermore, there are a number of models,
where instead of gravity other mediators communicate between the visible MSSM and the hidden sector.
In \acf{GMSB} models, %
new chiral supermultiplets are introduced as messenger fields %
that interact with the hidden sector.
They are also charged under the Standard Model gauge group
and thus couple to MSSM particles through gauge boson and gaugino interactions. %
In GMSB, the gravitino often is the \ac{LSP}.
This makes the phenomenology quite different from gravity-mediated models, %
and typically the \ac{NLSP} decays to a photon and the gravitino.
The simplest supersymmetric extension of the Standard Model with a scale-invariant superpotential beyond the MSSM is the \acf{NMSSM}: %
It solves the $\mu$-problem by introducing an additional singlet superfield.
In addition to an extended Higgs sector with five neutral Higgs states, %
the NMSSM also has an extended neutralino sector with five neutralinos including the \index{singlino}. %
This leads to significant modifications of all sparticle decay cascades compared to the MSSM
if the \ac{LSP} is singlino-like.
Again, constraints can be made to reduce the now even larger number of free parameters to a smaller set,
as is done in the models derived from the MSSM.
\inindex{R-parity violation@$R$-parity violation}
As said above, the superpotential of the MSSM does not include terms
which violate the conservation of the multiplicative quantum number $R$,
which was introduced to forbid terms which would lead to a decay of the proton.
There are possibilities, though,
to replace the $R$-parity by other symmetries like baryon-triality or lepton-parity \cite{Abdus2011SUSYBenchmark},
which then allow for violation of the conservation of either the lepton number $L$ or baryon number $B$,
while still guaranteeing stability of the proton. %
In these \ac{RPV} models, additional couplings are introduced in the superpotential,
which open up new phenomenological signatures.
In general, $R$-parity violation also means that the LSP no longer is stable and may decay,
potentially giving displaced vertices in the detector.

\chapter{Experimental Setup}
\label{ch:experimentalsetup}

The data evaluated in this thesis has been collected with the \ATLAS detector.
The \ATLAS detector is one of the four large-scale detectors
used to probe proton-proton collisions that are produced at the \acf{LHC} at \acs{CERN}.
The first part of this section gives a general overview of particle detectors,
with a focus on selected topics which are particularly relevant for this thesis,
like calorimeters, missing transverse energy and pile-up.
It is intended to provide definitions of the fundamental notions which are used in the following.
In the second part,
after a short introduction with respect to the rôle of CERN in particle physics,
a description of the \acl{LHC}, the largest collider ring currently in operation, is given.
Afterwards, the hardware of the \ATLAS detector is described.

\section{Conventions and Units}

In particle physics, it is common to set $c = \hbar = k_B = 1$,
where $c$ is the speed of light in vacuum, %
$\hbar \definedas h/{2\pi}$ is the reduced Planck constant %
and $k_B$ is the Boltzmann constant.
This is called ``\index{natural units}'' and will be adopted in the following,
in particular with respect to masses and momenta of particles,
which will be given in electron volts as a unit of energy ($\unit[1]{eV} = \unit[1.602\ten{-19}]{J}$).
The additional factors of $c$, $\hbar$ and $k_B$ will not be written explicitly,
but can be found from dimensional analysis,
\eg the unit of momentum is $\unit[]{eV}/c$,
for masses it is $\unit[]{eV}/c^2$ %
and for lengths it is $\hbar c / \unit[]{eV}$.
The typical energy scale for particles produced at current colliders is GeV.
In nuclear physics, lower energy ranges are studied (MeV range),
astroparticle physics covers a broad spectrum up to very high energies in cosmic rays with $\unit[10^{20}]{eV}$. %
An important unit in particle physics and nuclear physics is \index{barn},
which is a unit of area and used to express cross sections
and, by its inverse, luminosities (cf. \Sec{sec:definition_luminosity}). %
It is defined to be $\unit[10^{-28}]{m^2}$. %
In spite of the large negative value of the exponent,
this is still a very large cross section
and therefore typically used together with the SI prefixes
for pico ($\unit[1]{pb} = \unit[10^{-40}]{m^2}$)  %
or femto ($\unit[1]{fb} = \unit[10^{-43}]{m^2}$). %

The following symbols are conventionally used:
$E$ is the energy of a particle and $\vec p$ its vector three-momentum,
which together constitute the contravariant four-momentum $P = (P^\mu) = (E, \, \vec p)$, $\mu = 0,\,1,\,2,\,3$. %
The transverse component of the three-momentum,\inindex{Transverse momentum}
given by its projection onto the $x$-$y$ plane,
is denoted by $\vec p_T$.
Its absolute value is \pt.
In analogy, the \index{transverse energy} $E_T$ is defined as $E_T \definedas E \cdot \pt/|\vec p|$. %
The \index{invariant mass} of a system of $N$ particles is given by the sum of their energies and momenta as
\begin{equation}
  m^2 = \left(\sum_i^N E_i\right)^2 - \left(\sum_i^N \vec p_i\right)^2. %
  \label{eq:invariant_mass}
\end{equation}
The invariant mass of a decaying particle is equal to the invariant mass of its decay products, %
and in the center-of-mass frame,
the invariant mass is the total available energy
and therefore gives the kinematic boundary for particle production.
The center-of-mass energy in high-energy physics is usually denoted by the square root of $s$, %
one of the Lorentz-invariant Mandelstam variables,
defined as
\begin{equation}
  \sqrt{s} = \sqrt{P_\mu P^\mu},
\end{equation}
where $P = P_1+P_2$ is the sum of the four-vector momenta of the two initial particles in a two-body reaction
and summation over repeated indices is understood (Einstein convention).
In the center-of-mass system,
the total momentum is zero and $\sqrt{s}$ is again the energy available for particle production.
In a fixed target experiment,
the center-of-mass energy scales as $\sqrt{s} \sim \sqrt{2 E m_t}$, %
where $m_t$ is the mass of the resting target particle
and $E$ is the energy of the beam particle in the laboratory frame  \cite{MartinShaw}. %
In a collider experiment, where both of the colliding particles are accelerated to have an energy $E$ and collided head-on,
much higher energies can be reached,
because here $\sqrt{s} = 2 E$. %

To account for the rotational symmetry around the beam axis in collisions of elementary particles at a zero crossing angle,
usually a cylindrical or spherical coordinate system is used, which will be introduced in \Sec{sec:atlas}.

\section{Collision Detectors in High-Energy Physics}

The purpose of the detector is to determine the kinematic properties of the particles produced in high-energy particle collisions.
The final goal of the measurement is to correctly identify as many of the particles as possible.
Typical quantities that are used to describe the particles produced in collision experiments are
their positions, momenta and energies, and the impact times of the particles.
The impact time is normally used to group particles according to the collisions from which they emerged.

Basically, there are two types of instruments in a detector.
Tracking devices measure the momentum $\vec p$ of charged particles from their deflection in a magnetic field.
Calorimeters measure the energy $E$ of particles,
which is only possible by stopping them.
They must thus be optimized to maximize the energy loss and surround the tracking devices.
Unlike all other charged Standard Model particles,
the energy loss of muons is usually too small for them to be stopped in the calorimeters.
Muons can then be identified in muon spectrometers surrounding the calorimeters,
where a supplementary measurement of their momentum can be done.
The resulting sequence of instrumentation, tracking devices, calorimeters, muon spectrometers,
which is typical of general multi-purpose particle detectors,
is also reflected in the design of the \ATLAS detector,
which is described in detail in \Sec{sec:atlas}.

\subsection{Tracking Devices}

Charged particles can be detected by the ionization they produce when passing through a medium,
or by the production of photons as scintillation or \cerenkov light or as transition radiation.
Tracking devices\inindex{Tracking device} aim at determining the space and time coordinates of charged particles
and are mostly used in magnetic spectrometers,
which additionally allow a measurement of the momentum of the particle via
\begin{equation}
  p = |\vec p| = 0.2998 \frac{\text{GeV}}{\text{T} \, \text{m}} \cdot q \, B \, r,
  \label{eq:experimental_setup_momentum_measurement}
\end{equation}
if the strength of the magnetic field $B$ %
and the bending radius $r$ of the particle trajectory are known. %
\Eq{eq:experimental_setup_momentum_measurement} is written down for a particle with charge $q$
in units of the electron charge magnitude $e$.
From the sign of the right-hand side in \Eq{eq:experimental_setup_momentum_measurement},
the sign of the charge of the particle can be determined.
The relative error of the momentum measurement asymptotically scales as $\sigma(p)/p \sim p$ \cite{Kleinknecht}. %
For very fast or heavy particles,
the deflection of the track in the magnetic field is small and thus the curvature of the track difficult to measure.
In addition to the momentum measurement,
tracking devices are needed for the reconstruction of primary and secondary interaction vertices,
where the measurement of the distance of the secondary vertices allows
the determination of lifetime and thus identity of sufficiently slowly decaying particles.
Tracking devices are usually the innermost subdetector close to the beam pipe and interaction point,
which requires a high radiation hardness.
They also need to be optimized for an energy loss of the particles as small as possible
so that they do not interfere with the energy measurement in the calorimeters.

The two main techniques for tracking devices are gaseous detectors and solid state detectors \cite{Stock}.
When an ionizing particle passes through a gas-filled detector,
it creates electron-ion pairs.
If the electric field applied to the gas volume is sufficiently strong,
the ejected electrons can further ionize the medium and create an avalanche and thus a high signal gain.
For small electric fields, the electrons only drift along field lines to be captured by anodes. %
By choosing the voltage appropriately,
gaseous detectors can be run in proportional mode,
in which the signal amplitude is proportional %
to the primary charge produced within the sensitive volume. %
Multiwire proportional chambers consist of a planar cathode
and tautened anode wires in a plane parallel to the cathode.
A segmentation of the cathode gives additional information along the wire direction.
Alternatively, the charge division over resistive sense wires can be employed.
In a drift chamber, the constant drift velocity in the region with a homogeneous
electic field is exploited to find the position orthogonal to the cathode plane.
Straw-tube chambers are long cylindrical chambers with a central anode wire
and are often combined in modules or layers. %
Resistive plate chambers are low-cost, non-proportional detectors
with flat, high-resistivity electrodes,
between which violent discharges are induced by ionizing particles.
They have an excellent timing-resolution and provide short pulses that can be used for triggering. %

Solid state detectors\inindex{Solid state detector} use a semiconductor material, usually silicon or germanium,
which is doped to have a $pn$-junction.
Through application of a bias voltage, a carrier-free depletion zone is created,
which acts as a kind of ionization chamber,
in which charged particles create electron-hole pairs.
For this to happen, much less energy is needed than in gaseous detectors,
the energy being given by the band gap of the semiconductor material,
which is of the order of a few electron volts.
The charge is collected in an external electric field.
Spatial information can be obtained from a segmentation of the electrodes,
which may be reflected in the doped regions as well.
If the electrodes are approximately quadratic,
the device is called \revised{a} pixel detector;
if it is a very elongated rectangular shape,
it is called a strip detector.
Using orthogonal strips yields two-dimensional track information.
The irradiation of the doped semiconductor material
will at some point lead to an $n$ to $p$ type inversion,
and thus a reversed and with time increasing bias voltage is needed.
Recent research on solid state detectors includes
using diamond as semiconductor material
because of its much higher radiation tolerance
and the development of three-dimensional detectors.
These have electrodes going through the bulk of material,
which allows to improve the charge collection efficiency
and to lower the depletion voltage without changing the sensor thickness \cite{Stock}. %

\subsection{Calorimeters}
\label{sec:calorimeters_in_general}

A \index{calorimeter} is basically an instrumented block of dense absorbers,
in which the produced particles are stopped and (ideally) all of their energy is converted in a very short time
into a measurable quantity (light or charge),
which is then detected by embedded sensors.
This is thus a destructive measurement.
It proceeds in a cascade of interactions
and the depth needed for a complete absorption grows logarithmically with the energy of the particle.

Calorimeters are important for two reasons:
They can measure charged and neutral particles by detecting the charged secondaries,
and the asymptotic behavior of the uncertainty of the energy measurement for high energies is in principle
$\sigma(E)/E \sim E^{-1/2}$ due to the stochastic nature of the absorption process.
This means the more energy the particle carries, the smaller the relative uncertainty
because of the higher number of particles produced in the cascade\footnote{
  See below for a full discussion of the parametrization.
}.
This is complementary to the momentum measurement.
Not only the energy deposited in calorimeters is of interest,
but also the timing %
and the pulse shape as a measure of the quality of the energy determination.
Depending on the segmentation and granularity of the calorimeter,
also the direction and impact point of the incident particle can be measured.

In general, highly energetic particles will have enough energy to initiate the production of secondary particles
and thus to start the creation of an avalanche or shower of particles.
The shower continues until the energy of the particles produced in the showering process
falls below the threshold of energy needed to create further particles,
or until other processes of energy dissipation become dominant.
The energy of the particles that have been produced up to this point is then absorbed via low-energy processes.

It is common to equip detectors with two types of calorimeters,
one specialized to electrons and photons in front of one specialized to hadrons.
Photons, electrons and positrons require less material to be stopped compared to hadrons,
which interact mainly through the strong interaction.
This is also expressed by the radiation length $X_0$, %
which is the characteristic length for the development of an electromagnetic shower by pair production and bremsstrahlung,
and is smaller than the nuclear absorption or interaction length $\lambda$, %
which gives the mean free path %
or, equivalently, the probability of an inelastic collision over a certain distance for a hadron passing through the material.

Electromagnetic showers are initiated by electrons, positrons or photons
and are alternating cascades of photon emission via bremsstrahlung and creation of electron-positron pairs.
Electrons and positrons above a critical energy %
mainly lose energy via bremsstrahlung,
and at low energies through collisions with material, leading to ionization and excitation.
Photons at high energies %
mainly lose energy via $e^+ e^-$-pair production,
and at low energies through Compton scattering and the photoelectric effect. %
The expected energy loss for an electron above the critical energy %
is given by the radiation length $X_0$, $\expectation{E(x)} = E_0 \exp(-x/X_0)$. %
For photons, the intensity goes as $\expectation{I(x)} = I_0 \exp(-7x/9X_0)$ %
due to pair production.
The scale of electromagnetic showers is therefore determined by $X_0$.
Hadronic showers are initiated by hadrons like pions, kaons or protons
and consist of secondary hadrons produced via strong interactions in inelastic scattering with the nuclei of the calori\-meter material. %
The scale of the hadronic cascade is governed by the nuclear absorption length~$\lambda$.
The charged component of the cascade can be measured through its ionization and excitation effects, %
and photons from the de-excitation of excited nuclei can be detected directly.

The signal response of a calorimeter to hadrons will on average be lower
than that to an electron with the same energy due to several effects \cite{Kleinknecht}:
In a hadronic shower,
some of the produced particles like neutrinos or muons simply escape from the calorimeter, %
other particles like neutrons, fragments from spallation or recoiling nuclei are only partially detectable. %
A significant fraction %
of the energy in hadronic showers is removed by the production of neutral pions,
which will decay predominantly electromagnetically, $\pi^0\to\gamma\gamma$,
thus leading to a one-way conversion to an electromagnetic shower component,
which will give the same signal yield as an electron.
The ratio of the signal responses to pions (as representative of hadronic particles) and electrons
is usually denoted $\pi/e$ and of the order of $\pi/e \approx 0.8$. %
If $\pi/e \neq 1$, this also makes the calorimeter response to hadrons a non-linear function of the energy,
because the electromagnetic component increases with larger shower lengths \cite{Stock}.

Calorimeters fall into two groups:
In a sampling calorimeter,
an absorber material as passive component is sandwiched with active layers containing signal sensors.
Sampling calorimeters in general suffer from increased signal fluctuations due to their structure (sampling fluctuations). %
On the other hand,
in a sampling calorimeter it is possible to tune several variables, %
and by increasing the hadronic and decreasing the electromagnetic response to bring the response ratio $\pi/e$ back to one.
Calorimeters designed to this aim are called %
``compensating'' calorimeters.
In the simpler but often larger or more expensive homogeneous calorimeter,
the absorber material also serves as detecting medium,
for example in the calorimeter of the \CMS detector at the \LHC.
For the energy measurement itself,
there are two primary detection techniques.
In inorganic or organic scintillators,
molecules are excited through interactions with the incident particle
and subsequently emit a small fraction of the absorbed energy in form of photons,
which can then be detected using photodetectors, usually photomultipliers. %
In liquefied noble gases or semiconductors, %
the ionization in terms of electron-ion or electron-hole pair creation %
can be measured directly by applying an external electric field to collect the charge created by a particle passing through the material.

The conventional parametrization for the relative energy resolution of a calorimeter consists of three terms
with different scaling behavior with $E$ \cite{PDB2010},
\begin{equation}
  \frac{\sigma(E)}{E} = \frac a {\sqrt{E}} \oplus \frac b E \oplus c,
  \label{eq:experimental_setup_general_calorimeter_resolution_parametrization}
\end{equation}
where $\oplus$ stands for the quadratic sum, $x\oplus y \definedas \sqrt{x^2+y^2}$.
The parameter $a$ gives the stochastic term,
accounting for intrinsic stochastic fluctuations in the number of particles produced in the shower process,
which are proportional to $\sqrt{N} \sim \sqrt{E}$,
but also \eg for sampling fluctuations.
The constant term $b$, dominating the resolution for low energies, %
is the noise term.
The systematic term $c$, dominating for high energies,
comes from instrumental contributions to the energy resolution
like detector non-uniformities, %
incomplete shower containment (leakage) or calibration uncertainties which scale as $\sigma(E) \sim E$.
For measurements of hadronic shower energies,
non-compensation also contributes to this term.

\subsubsection{Jets}
\label{sec:experimental_setup_general_jets}

\begin{figure}
  \centering
  \incgraphics{width=\widthsingleplot}{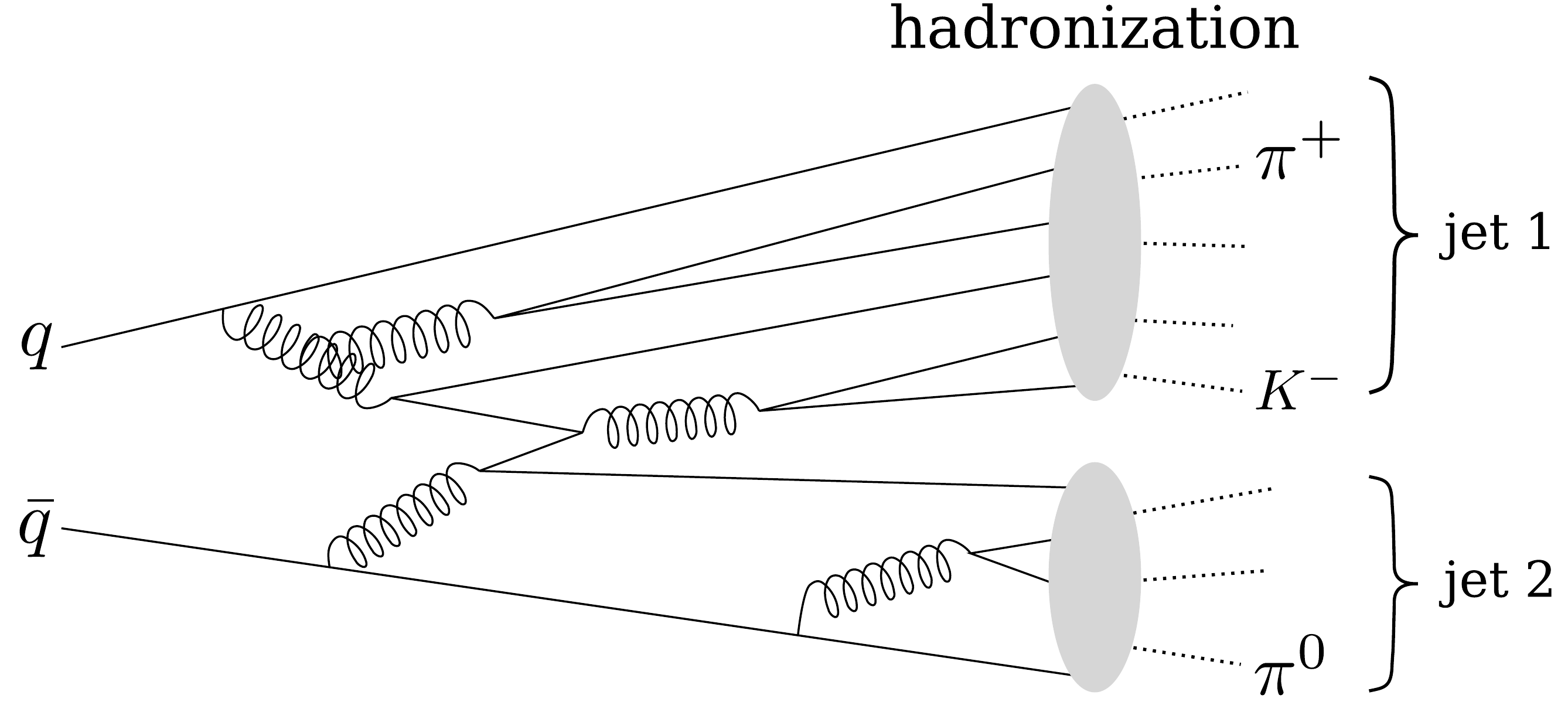} %
  \caption{
    Schematic view of the hadronization and jet formation process
    in a style similar to a Feynman diagram.
    The curly lines represent gluons, the straight lines quarks and antiquarks.
    The exact process of hadronization is unknown and therefore drawn as grey blobs,
    from which hadrons emerge as dotted lines.
    Some examples for hadrons have been put.
  }
  \label{fig:schematic_jet_formation}
\end{figure}

The \index{hadronic jet}s, which are measured as clustered energy depositions in the calorimeter,
are initiated by partons, gluons and quarks,
but the particles which interact with the detector material are hadrons and not the partons themselves,
which due to confinement cannot leave the interaction point as isolated entities (cf. \Sec{sec:theory_parton_model}).
A jet can thus be defined at different levels:
At parton level, where every high-energy (hard) parton corresponds to a jet,
or at particle level as the collection of hadrons produced from a parton in the hadronization process.
Finally, from the point of view of the calorimeter measurement,
a jet is defined as a clustered energy deposition in the cells of the calorimeter,
which is due to one original particle,
be it a prompt electron or photon in case of electromagnetic jets
or a parton from the parton level jet in case of hadronic jets.
The algorithms that are used to reconstruct the original jet structure
from the energy depositions in the calorimeters are described in \Sec{sec:software_jet_reconstruction}.

\subsubsection{Missing Transverse Energy}
\label{sec:general_missing_transverse_energy}

Many models of physics beyond the Standard Model like Supersymmetry
predict the existence of stable and heavy particles which only interact weakly (and gravitationally) with matter.
Besides the precise measurement of the particles like electrons, muons or hadrons in terms of jets,
the detection of particles which only very rarely interact with the detector material is therefore of great importance.
The direct detection of these particles usually is not possible,
unless the detector is specifically designed for this purpose and uses very large volumes and sophisticated methods for background suppression.
But although these weakly interacting particles are invisible\inindex{Invisible particle} to general purpose particle detectors like \ATLAS,
in the sense that they almost always evade direct detection,
they show up through the momentum and energy they carry away.
Due to the principle of momentum conservation,
in symmetric particle collisions the sum of all momenta must be zero
because the total momentum of the initial particles is zero.
If a particle leaves the detector unnoticed,
its momentum will be missing in a measurement of the total momentum computed as a vectorial sum of the momentum of all detected particles,
and lead to a non-zero result.
Of course, a lot of detector effects like noise and mismeasurements will also contribute to the apparent deviation in the sum from zero as discussed below.
At hadron colliders, there is an additional complication.
As the partons in the initial state of the interaction carry only a fraction of the proton momentum (cf. \Sec{sec:theory_parton_model}), %
their momentum along the beam axis direction is unknown,
and only the conservation of the transverse component of the momenta can be used,
by assuming that the transverse momentum of the initial state partons is zero.
The \index{missing transverse momentum} is then defined as
\begin{equation}
  \miss{\vec p}_T \definedas - \sum_i \vec p_T(i),
\end{equation}
where the sum runs over all visible particles in the final state \cite{PDB2010}.
The missing transverse momentum is identified with the sum of the transverse momenta of all invisible particles.

The momentum of a particle can be measured in tracking devices only for charged particles,
therefore instead of the momentum usually the energy is employed.
In the high energy approximation,
according to which particle masses are negligible,
the magnitude of the momentum of a particle is equal to its energy,
which can be inferred from a measurement in the calorimeter.
The \index{missing transverse energy} (\met, also abbreviated MET rather than MTE)
is then defined as the absolute value of the vectorial sum
of all contributions to the transverse energy,
\begin{equation}
  \met \definedas \left| - \sum_i \vec E_T(i) \right|.
\end{equation}
In practice, this is computed from a weighted projection of all calorimeter cells onto the transverse plane (cf. \Sec{sec:software_reconstruction_met}).
Muons as minimum ionizing particles\footnote{
  A minimum ionizing particle can be defined as a charged particle
  with a such a velocity that its energy loss is near the minimum
  of the Bethe equation %
  describing the stopping power $\langle -\intd E/\intd x\rangle$ as function of $\beta\gamma = p/mc$~\cite{PDB2010}. %
} typically only deposit a small fraction of their energy in the calorimeters.
To account for this, the measurement of the calorimeter \met can be corrected by using the momentum information from the muon spectrometers.
A quantity related to \met is the sum of the absolute values of the transverse energies,
\begin{equation}
  \sumet \definedas \sum_i \left| \vec E_T(i) \right|.
\end{equation}
The value of \sumet gives an indication of the total activity in an event.

In collisions in which no invisible particles are produced,
an ideal detector would measure $\met = 0$.
In reality however, there are many detector effects which contribute to the \met measurements.
In particular due to the definition of \met as the sum of squares of the $x$ and $y$ components of the missing energy,
\begin{equation}
  \met = \sqrt{{\me{x}}^2 + {\me{y}}^2},
\end{equation}
although the mean of the noise of the energy measurement in the calorimeter is zero for both components individually,
it will give a non-zero contribution to \met.
This is discussed in detail in \Sec{sec:results_met_model}.
Additional contributions to \met come from
non-instrumented parts of the detector, %
``dead material''\inindex{Dead material} %
such as cooling, cables for read-out and powering and support structures
and crack regions in the calorimeter, %
where deposited energy cannot be measured and therefore is lost,
and mismeasurements of the \met from the limited resolution and calibration effects.
All these detector effects will be referred to as fake \met\inindex{Fake MET@Fake \met} in the following as opposed to real \met\inindex{Real MET@Real \met},
which is induced by non-interacting particles like the Standard Model neutrinos or neutralinos from Supersymmetry
and is the quantity which is of interest with respect to physics analyses.
In addition, there is also the notion of true \met\inindex{True MET@True \met},
which refers to the Monte Carlo information about the actual amount of real \met in simulated events.
Note that the presumed conservation of the transverse momentum neglects a number of small effects,
giving rise to a non-zero transverse momentum even if all particle momenta were known exactly.
Such effects are, for example,
that the partons in the initial state of the collisions
have a small, but non-zero \pt,
or the non-zero crossing angle between the colliding beams (cf. \Sec{sec:experimental_setup_LHC}),
leading to a residual momentum in the $y$ direction of the order of a GeV \cite{ATL-DAQ-PUB-2011-001}. %

\subsection{Muon Spectrometers}

\index{Muon spectrometer}s are tracking devices, too, but specifically designed for muons.
As said above, they typically lie outside the calorimeter,
and they usually cover a large volume with low material density to avoid scattering of the muons.
Different types of muon chambers are described
in the context of the muon spectrometer of the \ATLAS detector in \Sec{sec:experimental_setup_atlas_muon_system}.

\subsection{Luminosity}
\label{sec:definition_luminosity}

The \index{luminosity} governs the rate of interactions in a collider experiment.
It has unit of a flux, \unit[]{Hz/cm$^2$}, often also written as \unit[]{cm$^{-2}$s$^{-1}$}.
The \index{instantaneous luminosity} in a collider experiment is given by
\begin{equation}
  \Linst = f n \frac{N_1 N_2} {A},
  \label{eq:def_inst_lumi_1}
\end{equation}
where $N_i$ is the number of particles in the bunches of the two beams which are brought to collision,
$f$ the revolution frequency
and $n$ the number of particle bunches in each of the beams.
While the number of particles in each bunch can differ,
especially if the two beams consist of two different types of particles,
$n$ usually is the same for both beams.
$A$~is the cross-sectional area of the beams at the interaction point,
which can be expressed as $A = 4\pi\sigma_x\sigma_y$. %
The transverse beam profile can in many cases be assumed to be Gaussian \cite{PDB2010} %
and is characterized by the widths $\sigma_{x,y}$ in the horizontal and vertical direction.
In order to make the dependence on the collider machine parameters explicit,
this can be written as \cite{LHC_TDR_CERN_Vol3}
\begin{equation}
  \Linst = f n \frac{N_1 N_2 \gamma_r}{4\pi\epsilon_T\beta^*}, %
  \label{eq:def_inst_lumi_2} \\
\end{equation}
where $\epsilon_T$ is the transverse emittance of the beams,
$\beta^*$ is the wavelength of the betatron oscillations of the beams \cite{Ellis1996}, %
and $\gamma_r$ %
is the relativistic factor of the colliding particles.

The integral of the instantaneous luminosity over time gives the \index{integrated luminosity}
\begin{equation}
  \Lint = \int_{t_1}^{t_2} \! \Linst \, \rm dt.
  \label{eq:def_int_lumi}
\end{equation}
The amount of data collected over a certain interval of time is specified in terms of the integrated luminosity.
The number of events of a given type $N$,
which is expected to be contained in a data sample,
is related to the corresponding integrated luminosity via the total inclusive cross section $\sigma$,
\begin{equation}
  \langle N \rangle = \sigma \cdot \Lint.
  \label{eq:lumi_and_expected_number_of_events}
\end{equation}
The product of the instantaneous luminosity
and the cross section for the relevant scattering process
gives the reaction rate $\langle R \rangle = \sigma \cdot \Linst$.
\Eq{eq:lumi_and_expected_number_of_events} is important for the normalization of samples
with Monte Carlo simulated events %
to the number of events expected in the sample of real detector data.
The number of events $N_0$ in the Monte Carlo sample
needs to be scaled to the integrated luminosity \Lint of the data sample,
using the respective cross section $\sigma$ computed by the Monte Carlo generator.
The event weight serving as scaling factor is given by $f =  \langle N \rangle / N_0$.
\begin{figure}
  \centering
  \incgraphics{width=\widthsingleplot}{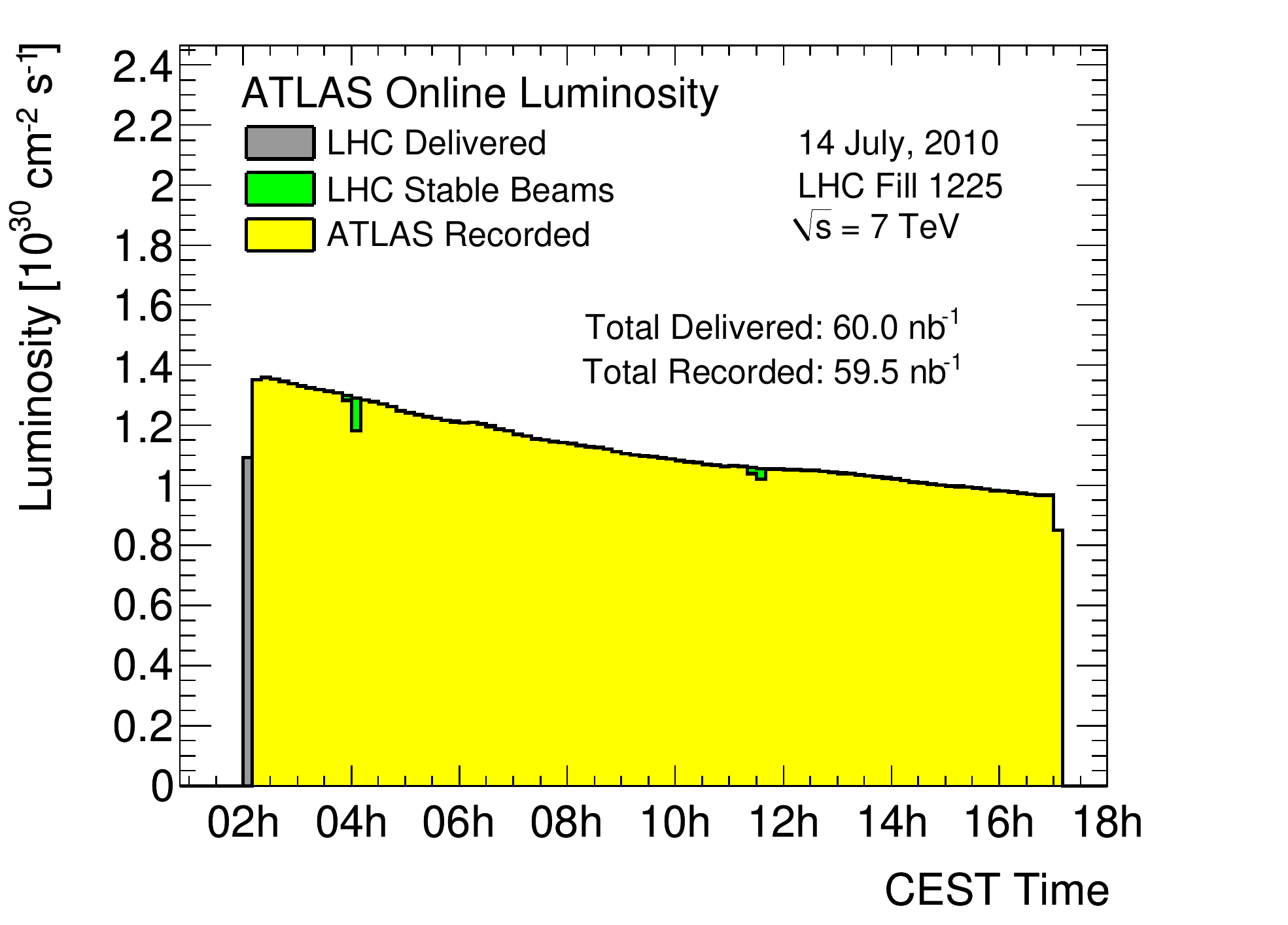}
  \caption{
    Plot of the instantaneous luminosity as recorded by \ATLAS,
    for fill number 1225 of the LHC.
  }
  \label{fig:example_instant_lumi} %
\end{figure}
The instantaneous luminosity decreases exponentially with time during a fill of the collider
due to several effects with different time constants.
Normally, the decrease of the instantaneous luminosity should be dominated
by the reduction of the number of particles in the bunches due to collisions
rather than due to any other losses like beam-gas interactions (cf. \Sec{sec:experimental_setup_LHC}). %
\Fig{fig:example_instant_lumi} shows exemplary the exponential decay of the instantaneous luminosity
for one of the first long runs of the \acf{LHC} and the \ATLAS detector in 2010. %
The plot shows in three different colors the luminosity delivered by the LHC,
the luminosity delivered during the time when the beams were declared to be stable by the LHC operators,
and the luminosity that was actually recorded by the \ATLAS detector.
After about 15 hours, the beams were dumped and the data taking ended.

\subsection{Pile-up}
\label{sec:define_pileup}

In the context of trigger rates,
and in particular in the discussion of the \met model in \Sec{sec:results_met_model},
the notion of \index{pile-up} will be important.
Pile-up means that the particles produced in more than one particle-particle collision
reach the detector at the same time,
or more generally that their signals are overlaid in a way that they cannot be disentangled.
It does, however, not refer to the effect
that due to the large extent of the detector, the small temporal bunch spacing and the finite speed of light,
the products emerging from several subsequent bunch crossings coexist in the detector volume,
as will become clear below from the explanation of the two different types of pile-up.
Beforehand, it is useful to introduce the expected average number of interactions per bunch-crossing.

\subsubsection{Average Number of Interactions per Bunch-crossing}
\label{sec:experimental_setup_general_define_avgmu}

When bunches of particles are collided, \ie they run through each other,
the probability for an interaction is proportional to the particle densities,
or better, including the particle velocity, the flux,
which is expressed in terms of the instantaneous luminosity.
The number of actual particle collisions taking place when two particle bunches cross (dubbed a \define{\index{bunch crossing}})
is a random variable that follows a Poisson distribution.
For low instantaneous luminosities, in most bunch crossings nothing will happen;
for high instantaneous luminosities, in most bunch crossings several particle collisions will take place at the same time.
In the following, \avgmu is defined as the (expected) average number of (inelastic) proton-proton interactions per bunch-crossing.
It is an important quantitative measure for the activity within an event %
and can be calculated from the instantaneous luminosity as shown in \Sec{sec:results_trigger_rates_describe_COOL_plots}.

Like the instantaneous luminosity,
\avgmu depends on the number of protons in the colliding bunches,
which is not necessarily the same for all bunches in the collider ring
because it may vary depending on the injection quality conditions.
Nevertheless, in the following it is a good approximation to assume for simplicity that it is the same for all bunches.

\subsubsection{In-time Pile-up}
\inindex{In-time pile-up}

Depending on the value of \avgmu, in a significant fraction of bunch crossings
more than one hard proton-proton interaction will take place concurrently,
so that the particles produced in these collisions reach the surrounding detector at the same time.
Due to its large inclusive cross section, most of the interactions will be minimum-bias events (cf. \Sec{sec:define_minimum_bias}), %
and only very rarely will two processes with small cross sections (\eg $W$ production, cf. \Fig{fig:lhc_cross_sections})
take place in the same bunch crossing.
In Monte Carlo simulations, in-time pile-up is thus incorporated by overlaying
the signals from the physics process of interest
with signals corresponding to a given number of additional independent minimum-bias collisions.

Depending on the subdetector and type of measurement,
it may or may not be possible to
distinguish between particles coming from different concurrent proton-proton interactions. %
For the measurement of \met and in the trigger,
it is in general not possible.
For charged particles, it may be possible to trace back their origin to spatially separated production vertices.
The impact of in-time pile-up on the rates of the \met trigger
is studied in detail in \Sec{sec:results_met_model}.

\subsubsection{Out-of-time Pile-up}
\inindex{Out-of-time pile-up}

\begin{figure}
  \centering
  \incgraphics{width=\widthsingleplot}{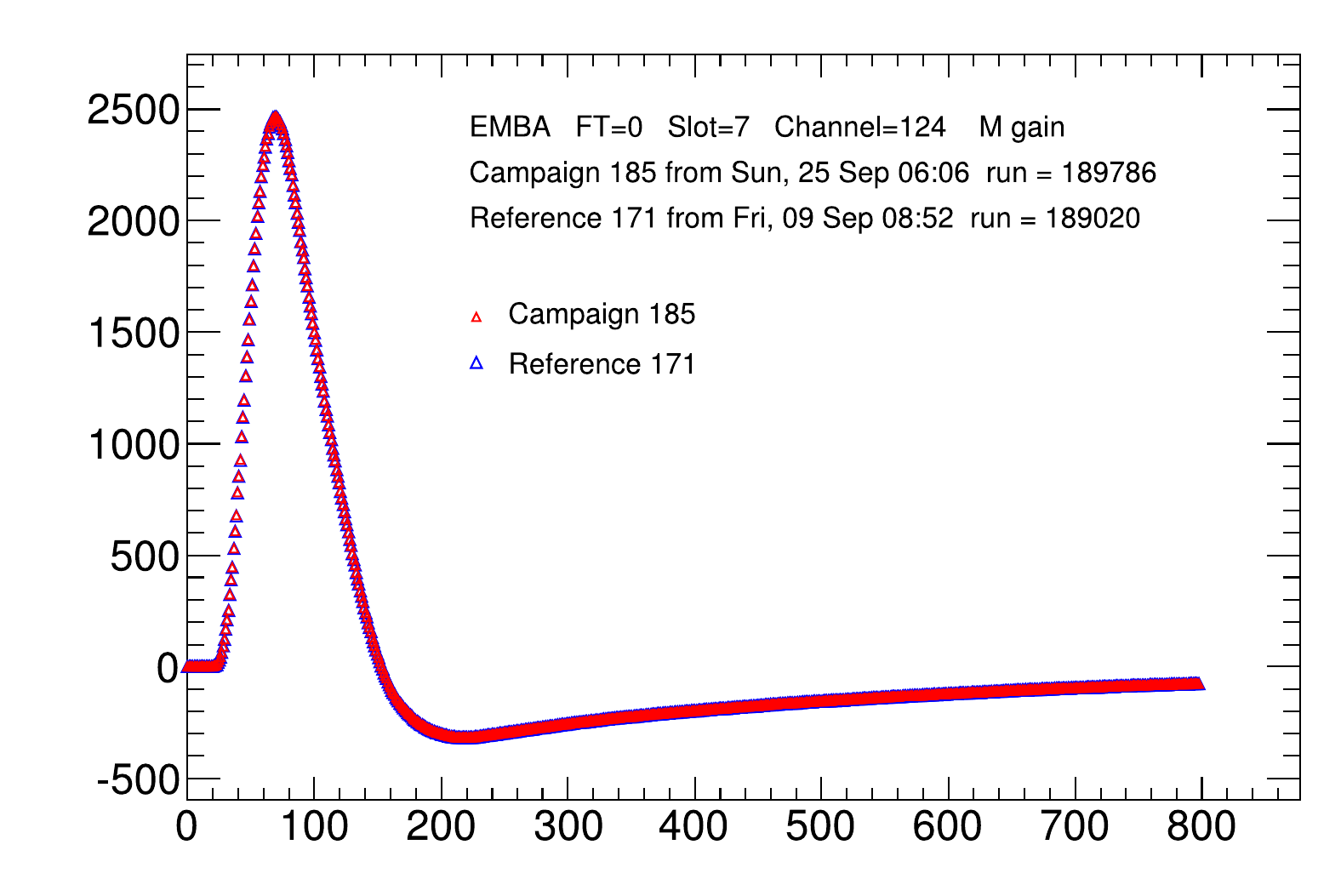}
  \caption{
    Calibration plot from September 2011
    showing the pulse shape of one read-out channel of the electromagnetic barrel calorimeter of \ATLAS (side A).
    The horizontal axis is the time in nanoseconds,
    the vertical axis the amplitude of the pulse in Analog-to-Digital Converter (ADC) counts.
  }
  \label{fig:example_pulse_shape} %
\end{figure}

Out-of-time pile-up subsumes effects
which originate from the time the detector needs to return to its stand-by state being longer than the temporal bunch spacing.
This means that signals from particles produced in a subsequent collision
reach the detector before it is ready to start the next measurement.
While in-time pile-up will always lead to an increase of the activity in the event,
to more particles appearing in the detector
and to more energy being deposited in the calorimeter,
out-of-time pile-up may also lead to the opposite effect in calorimeter measurements.
Due to the pulse shape of the calorimeter signals, %
which exhibit an undershoot before coming back to the zero baseline,
energy depositions which occur at the same position in the calorimeter,
but which are separated by a certain time interval,
will lead to a destructive interference in the measurement
if the second pulse starts during the undershoot of the first pulse.
This may lead to a considerable negative bias in the energy measurements
that needs to be taken into account when running at high luminosities
with short bunch distances that allow for out-of-time pile-up.
\Fig{fig:example_pulse_shape} gives an example of a typical pulse shape in the electromagnetic liquid-argon calorimeter of \ATLAS.
It can be seen that the peak of the signal is followed by a long undershoot,
which spans tens of bunch crossings at the nominal bunch spacing of \unit[25]{ns} (see below).

\subsubsection{Pile-up and Luminosity}

\Eqs{eq:def_inst_lumi_2} and \eqref{eq:def_int_lumi} show what possibilities there are to increase the luminosity.
These have different implications for the amount of pile-up.
Increasing the integration time (``physics time'' with stable beam conditions)
or the number of colliding bunches $n$ in the collider does not increase the level of in-time pile-up
because the average number of interactions per bunch crossing stays the same.
However, changes in any of the other parameters
which govern the instantaneous luminosity per bunch, 
\ie an increase in the number of particles in the bunches $N_{1,2}$
or a reduction of the beam parameters $\epsilon_T$ and $\beta^*$,
all do increase in-time pile-up.
(The revolution frequency $f$ and relativistic factor $\gamma$ are fixed.)

\section{CERN and the LHC}
\label{sec:cernlhc}

\subsection{L'Organisation Européenne pour la Recherche Nucléaire}
\acslink{CERN} is an international organization with 20 member states
and is one of the largest centers for scientific research.
It is situated near Geneva on the border of Switzerland and France and
operates the world's largest particle physics laboratory.
Its official name today is ``European Organization for Nuclear Research''
or \guillemotleft~L'Organisation Européenne pour la Recherche Nucléaire~\guillemotright\xspace in French.
The acronym CERN stood for the French ``Conseil Européen pour la Recherche Nucléaire'' (European Council for Nuclear Research),
which was the name of the council founded in February 1952
with the mandate to establish a fundamental physics research organization in Europe.
At the end of September 1954, after the ratification by France and Germany,
the European Organization for Nuclear Research officially came into being %
and the council was dissolved, but the name CERN was retained.

A variety of particle accelerators and colliders have been built on CERN grounds since its foundation,
some of which, like the Proton Synchrotron (PS) started up in 1959, are still in use despite their old age in terms of a machine used for research.
The PS is an important part of the accelerator chain which feeds the LHC (see below).
Besides accelerating particles, also slowing down particles is a scientific occupation,
for example in the ATRAP experiment,
which aims at comparing hydrogen atoms with their antimatter equivalents,
for which the ingredients, antiprotons and positrons need to be cooled down so that they can form antihydrogen.
So far, five Nobel Prizes were awarded to CERN physicists, %
in connection to the discovery of the weak bosons, the $J/\psi$ meson, the muon neutrino and detector technology and methods. %
CERN also claims to be the place where the World Wide Web has been invented,
originally for sharing information between scientists working in different places all over the world.

\subsection{The Large Hadron Collider}
\label{sec:experimental_setup_LHC}

The \acf{LHC} \inindex{Large Hadron Collider} %
is a circular (synchrotron) proton-proton accelerator and collider,
designed for a peak instantaneous luminosity of \instlumi{10^{34}} and an energy of \unit[7]{TeV} per beam,
giving a center-of-mass energy of \revised{$\sqrt{s} = \unit[14]{TeV}$} available in collisions \cite{LHC_TDR_Published2008}. %
It is currently the world's largest particle accelerator
and located about \unit[100]{m} underground near Geneva,
where it spans the border between Switzerland and France. %
Its primary experimental goals are to study the mechanism of electroweak symmetry breaking,
find the Higgs boson and search for indications of physics beyond the Standard Model.
A summary of relevant parameters of the \ac{LHC} is given in \Tab{tab:experimental_setup_lhc_parameters}.

The \ac{LHC} occupies a tunnel with a circumference of \unit[27]{km}
which was originally excavated for the \acf{LEP}
that ran successfully until the end of 2000. %
The \ac{LEP} required such a large radius due to the synchrotron radiation of the electrons and positrons. %
The major differences with respect to the older Tevatron,
the hadron collider with the second highest energy in the world operated in the U.S. at Fermilab,
are the higher center-of-mass energy and the fact that at the LHC proton pairs are collided,
whereas at the Tevatron protons and antiprotons are collided.
This leads to significant differences in the composition of the initial states.
In addition to colliding protons with protons,
there are also (shorter) phases in the operation of the LHC
with dedicated heavy ion runs,
in which lead ions (Pb$^{82+}$) are collided with lead ions.
This allows to give an even higher total center-of-mass energy,
although the energy per \revised{nucleon} of \unit[2.76]{TeV} is lower than in proton-proton collision mode.

\begin{figure}
  \centering
  \incgraphics{width=\widthsingleplot}{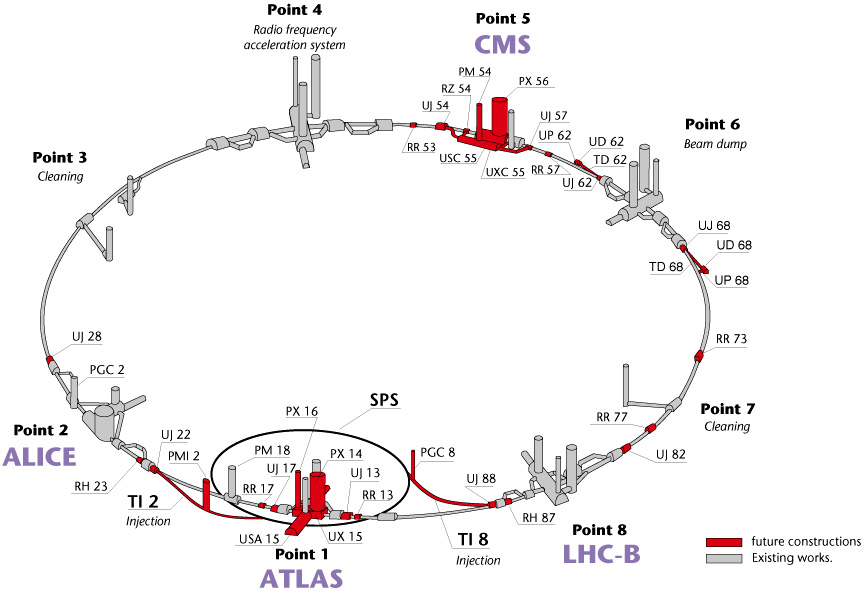}
  \caption{
    Layout of the LHC infrastructures.
    The \ATLAS detector is housed at Point~1,
    one of the four interaction points.
    (Note that this image is from 1993 and the ``future constructions'' by now have been completed.)
  }
  \label{fig:experimental_setup_lhc_layout} %
\end{figure}

The center-of-mass energy at a hadron collider like the \LHC
is currently limited by the available magnet technology.
A short calculation (cf. \Eq{eq:appendix_bending_magnets}) shows that per \TeV{} of beam energy,
the magnetic field needs to be increased by about \unit[1]{Tesla}
to bend the protons onto a track with the radius of the \LHC tunnel.
In fact, the magnetic field actually needs to be higher,
as the \LHC is not a perfect circle,
but sections equipped with dipole magnets are interspersed with straight segments.
The effective radius is thus smaller and the bending force needs to be larger.
In total, the LHC has 9300 magnets with superconducting Niobium-titanium (NbTi) coils,
which are cooled with liquid nitrogen and superfluid helium, in which they are immersed,
to an operating temperature of \unit[1.9]{K}. %
The magnetic field of the 1232 main bending magnets is designed to be \unit[8.33]{T}
at the maximum current through the magnets of \unit[11.8]{kA}.
They make up about two thirds of the length.
In the remaining length, there are quadrupole magnets for focusing,
the beam injection points, the beam dump, beam cleaning stations and long straight sections,
at which also the four interaction regions with the main detectors
\ALICE, \ATLAS, \CMS and \LHCb are located. %
The interaction regions,
in which the two proton beams cross,
are distributed over the eight octants\footnote{
  Correspondingly, there are eight special locations, one in the middle of each octant, %
  which are numbered clockwise around the ring and called Point~1 through Point~8.
  Points~1, 2, 5 and 8 house the \ATLAS, \ALICE, \CMS and \LHCb detector, respectively.
}
into which the \LHC is divided (cf. \Fig{fig:experimental_setup_lhc_layout}).
In \ATLAS and \ALICE, the beams cross in the vertical plane,
in \CMS and \LHCb in the horizontal plane. %
At Points~2 and~8, the proton beams from the SPS, the last step of the accelerator chain, are injected into the LHC.

The \index{accelerator chain} for the injection of protons %
into the LHC is the following (for lead ions the chain it is slightly different).
Protons for each beam are accelerated in the Linac2, a \unit[30]{m} long linear accelerator, to \unit[50]{MeV},
then transferred into the Proton-Syn\-chro\-tron-Booster (PSB) to be accelerated to \unit[1.4]{GeV},
extracted to be accelerated to \unit[25]{GeV} in the Proton-Syn\-chro\-tron (PS),
and then to \unit[450]{GeV} in the Super-Proton-Synchrotron (SPS),
from which they are injected into the LHC to be ramped up to the final center-of-mass energy.
The acceleration of the protons in the LHC is done using a sinusoidal electric field at a frequency of \unit[400.8]{MHz}
in eight superconducting radio frequency cavities per proton beam at Point~4 with a maximum power of \unit[4.8]{MW}.

The beam crossing frequency of the LHC is $f_\text{LHC} = \unit[40.079]{MHz}$,
which results in an approximate \unit[25]{ns} width
of the 3564 \index{time bucket}s\inindex{LHC time bucket} %
and a revolution frequency of \unit[11.245]{kHz}. %
The number of time buckets is directly related to the %
relations of the circumferences of the preceding accelerators, the PS and SPS \cite{Stock}. %
Due to hardware restrictions, in particular from the rise times of the injection kicker magnets, %
after subtracting abort and injection gaps,
only a maximum of 2808 of the 3564 time buckets can be filled with proton bunches. %
Buckets which contain a proton bunch are often called \index{filled bunch}es in the following,
buckets without protons are called \index{empty bunch}es.
Finally, proton bunches in buckets that are brought to collision at the interaction point in \ATLAS are called \index{colliding bunch}es.
The distribution of the proton bunches over buckets is usually given in terms of \index{bunch train}s,
which describe a certain repeated pattern of filled and empty buckets.

The operation of the LHC requires a very good vacuum to reduce beam-gas interactions.
Otherwise, proton bunches would be rapidly depleted and detector measurements would suffer from high background noise levels.
The vacuum also determines the single-beam lifetime,
which is the time in which the number of particles in stable orbit is reduced by a factor of $1-\exp(-1)$ due to beam-gas interactions,
and is designed to be significantly larger than the luminosity lifetime of 45 hours due to the proton-proton collisions \cite{LHC_TDR_CERN_Vol1}. %
The required equivalent hydrogen density to achieve this is \unit[$10^{15}$]{m$^{-3}$},
close to the detectors it will be even a 100 times better.

\begin{figure}
  \centering
  \incgraphics{width=\widthsingleplot}{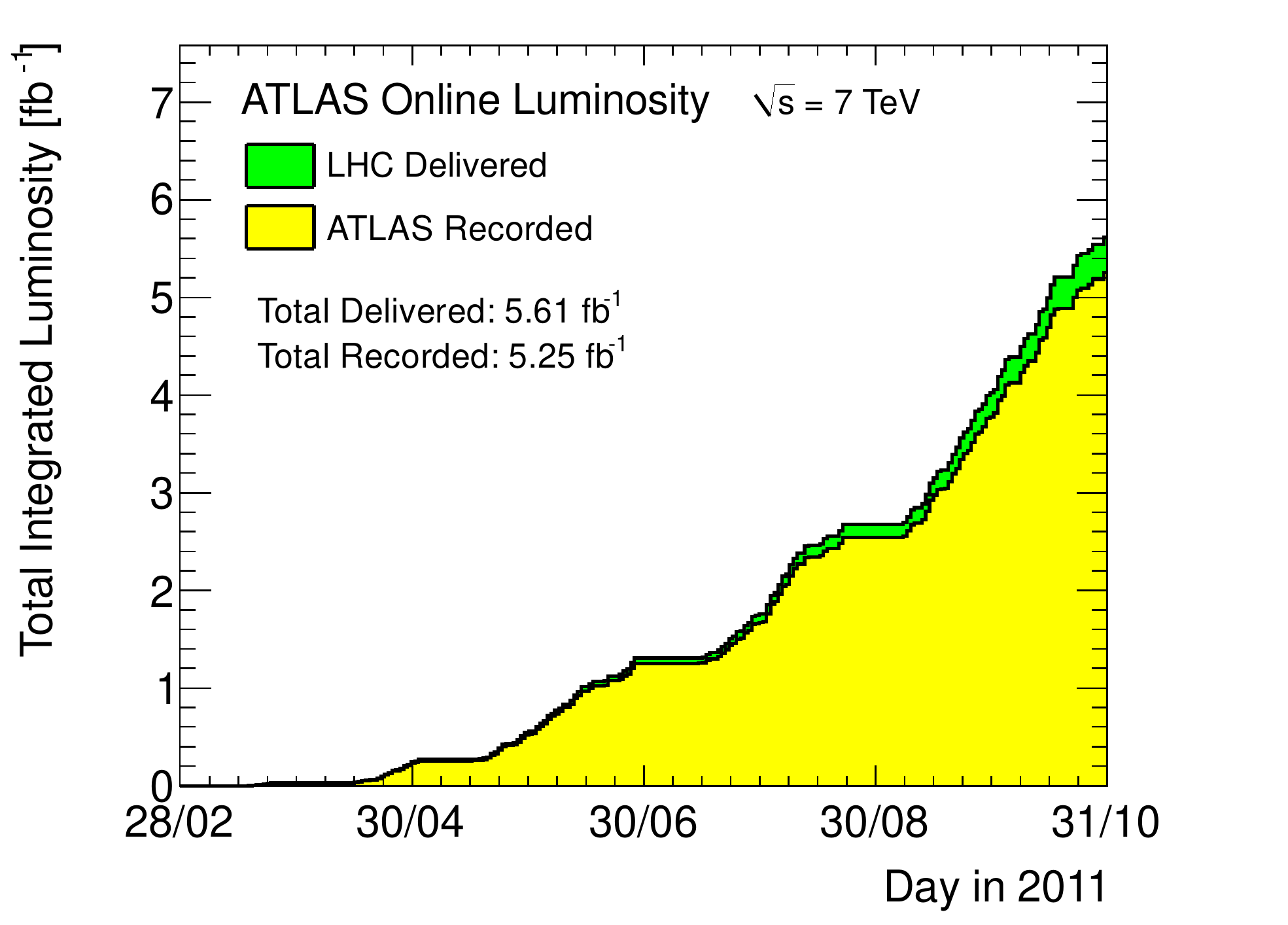}
  \caption{
    Integral of the delivered and recorded luminosity for \ATLAS in 2011 as function of time.
    Over \unit[5]{fb$^{-1}$} have been collected in 2011 at \seventev.
  }
  \label{fig:example_total_lumi} %
\end{figure}

After an incident in September 2008\inindex{Incident of 2008} in the commissioning phase of the LHC \cite{URL_LHCIncident},
which revealed problems with the bus bars, electrical interconnectors between the magnets,
it was decided that it is only safe to run the LHC at a reduced center-of-mass energy of \TeV{7}, \TeV{3.5} per beam,
until all the superconducting interconnect splices will have been overhauled,
which will require a long consolidation shutdown \cite{URL_LHCIncident2}.
In March 2010, for the first time proton-proton beams were collided at a center-of-mass energy of \TeV{7} in the LHC.
Since then, the instantaneous luminosity has been constantly increasing.
At the time of writing (November 2011),
the peak instantaneous luminosity is already above \unit[$3.5\ten{33}$]{Hz/cm$^2$} \cite{URL_Records},
so there is only little needed to reach the design luminosity of the LHC.
Part of this can be achieved by reducing the bunch spacing from \unit[50]{ns} to \unit[25]{ns}, which will double the instantaneous luminosity.
In 2010, the LHC has delivered \unit[49]{pb$^{-1}$} of integrated luminosity,
in 2011 over \unit[5.6]{fb$^{-1}$}.
A plot of the evolution of the integrated luminosity over time is shown in \Fig{fig:example_total_lumi} for 2011.
The increase in the instantaneous luminosity is reflected by the increase of the slope in the integrated luminosity.

\renewcommand{\arraystretch}{1} %
\begin{table}
  \centering
  \begin{tabular}{ll}
    \toprule
    Parameter & Value \\
    \midrule
    Proton energy  &  \TeV{7}  \\
    Relativistic $\gamma$ factor of protons &  7461  \\
    
    Number of proton bunches  &  2808  \\ %
    Number of protons per bunch  &  $1.15\ten{11}$  \\
    Transverse emittance & \unit[3.75]{\micro m} \\ %
    $\beta$ function at IP1 and IP5 & \unit[0.55]{m} \\ %
    Half crossing-angle &  \unit[$\pm142.5$]{\micro rad} \\ %

    Peak luminosity at IP1 and IP5  &  $\unit[1\ten{34}]{Hz/cm^2}$ \\ %
    \quad per bunch crossing & $\unit[3.56\ten{30}]{Hz/cm^2}$ \\
    $pp$ collisions per bunch crossing  &  19  \\
    
    Revolution frequency  &  \unit[11.245]{kHz}  \\
    Bunch spacing & \unit[24.97]{ns} \\ %
    Beam crossing frequency & \unit[40.079]{MHz}  \\
    RF frequency  &  \unit[400.790]{MHz}  \\
    
    Stored energy per beam  &  \unit[362]{MJ}  \\
    Circulating beam current  &  \unit[0.582]{A}  \\
    Beam current  lifetime  &  \unit[14.9]{hours}  \\
    
    Ring circumference  &  \unit[\numprint{26658.883}]{m}  \\
    Number of main bends  &  1232  \\
    Field of main bends  &  \unit[8.33]{T}  \\
    Bending radius  &  \unit[2803.95]{m}  \\
    \bottomrule
  \end{tabular}
  \caption{
    Summary of relevant LHC parameters \cite{LHC_TDR_CERN_Vol1}.
    The parameters given in this table are the design parameters.
    At the moment (2011), the energy per beam is only \TeV{3.5} and the design luminosity not yet reached (see text).
    IP1 and IP5 are the interaction points at the center of the \ATLAS and \CMS detector (cf. \Fig{fig:experimental_setup_lhc_layout}).
  }
  \label{tab:experimental_setup_lhc_parameters}
\end{table}
\arraystretchdefault

\subsubsection{Cross Sections at the LHC}
\begin{figure}
  \centering
  \incgraphics{width=\widthsingleplot}{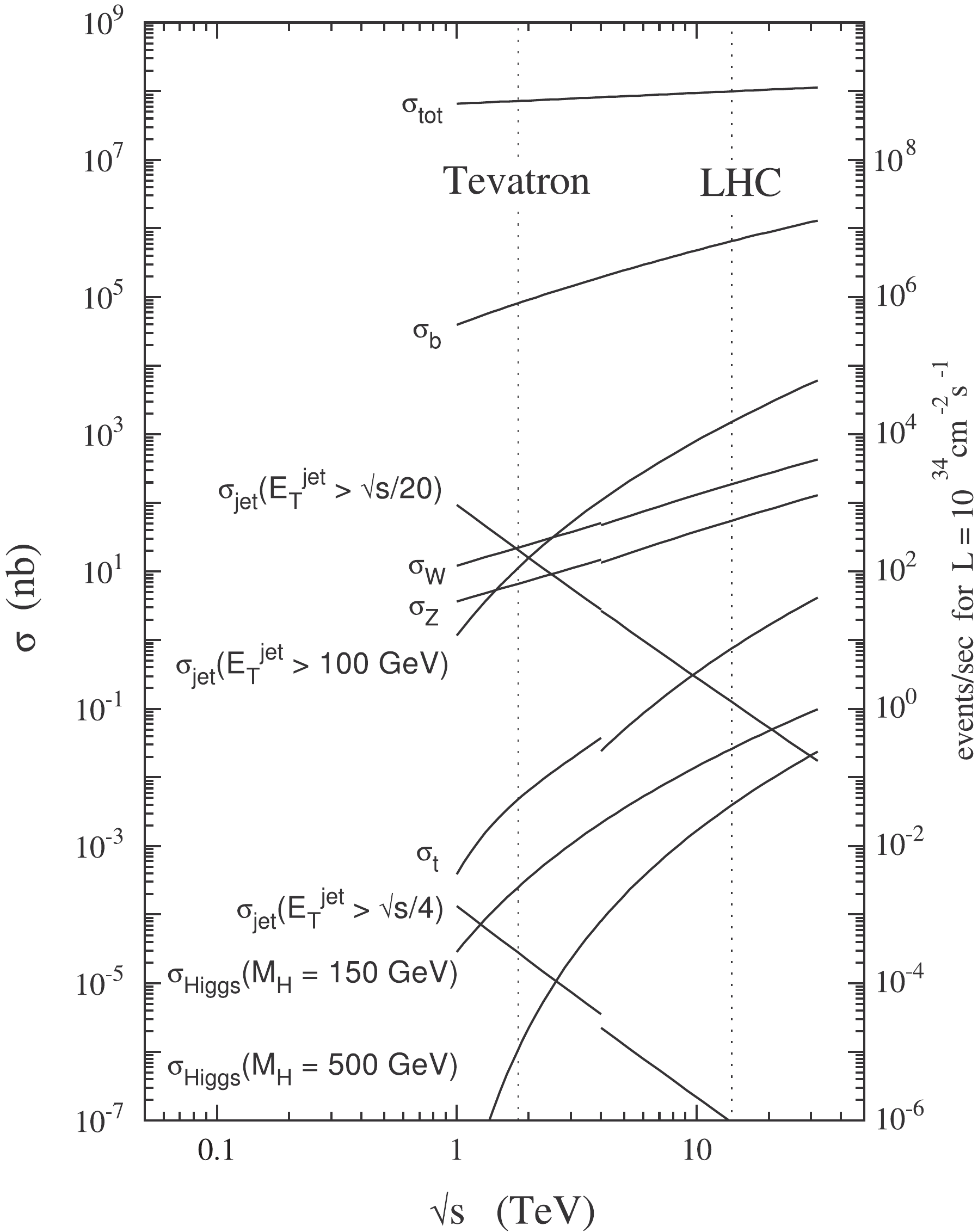}
  \caption{
    Cross sections for several important processes at the LHC compared to the Tevatron,
    the center-of-mass energies of which are indicated by the dashed vertical lines.
    (For the LHC the design center-of-mass energy of \TeV{14} is assumed.)
    The number of events per second on the right vertical axis
    is computed for the design luminosity of the LHC.
  }
  \label{fig:lhc_cross_sections} %
\end{figure}

\Figs{fig:lhc_cross_sections} compares the cross sections of several important processes 
as function of the center-of-mass energy.
The center-of-mass energies of the LHC and the Tevatron are indicated by the dashed vertical lines.
Some of the curves in the plot exhibit a discontinuity, which has nothing to do with the center-of-mass energy being \TeV{4}.
Instead, these are processes which have different cross sections at the Tevatron, being a proton-antiproton collider,
and the LHC, being a proton-proton collider, due to the nature of the dominant production mode.
For processes where gluon fusion is the dominant production mode, like for Higgs bosons, %
the cross section curves have a smooth transition between the Tevatron and LHC.
The vector bosons $W$ and $Z$ get significant contributions to their production from quark-antiquark annihilation,
and therefore the cross section at a fixed center-of-mass energy is slightly larger at the Tevatron
because the antiquark here can be a valance quark from the antiproton,
whereas at the LHC antiquarks only appear as sea quarks,
which carry smaller fractions of the proton momentum.

\subsubsection{Plans for LHC Upgrades}
As the LHC is not yet running at its design parameters,
in particular only at half the center-of-mass energy $\sqrt{s}$ it is designed for,
the next step naturally is to ramp up $\sqrt{s}$ to $\unit[14]{TeV}$,
which is planned for a two-year shutdown in 2013 and 2014 \cite{URL_LHCUpgrade}.
It may be that only $\sqrt{s} = \unit[13]{TeV}$ will be reached.
After three years of running at design luminosity,
the peak instantaneous luminosity shall be increased to $2$ or \instlumi{3\ten{34}},
to collect about \ifb{300} of integrated luminosity until 2021.

Afterwards, the \ac{LHC} may be replaced by the \index{Super Large Hadron Collider} (SLHC) or High Luminosity LHC,
raising the peak instantaneous luminosity to \instlumi{10^{35}}.
Among the possibilities to achieve this (cf. \Sec{sec:definition_luminosity}),
current plans foresee to increase the beam intensities by a factor 5
and by a stronger squeezing of the beams at the interaction points 1 and 5
to decrease $\beta$ by a factor 2 \cite{Stock}. %
With the increased luminosity and higher event rates,
also the radiation doses will be much higher,
which will be a challenge especially for the detector systems close to the beam pipe.
The SLHC shall deliver $200$ to \ifb{300} per year to collect a total of \ifb{3000}.

\subsubsection{After the LHC Era}
Considering the long time-scales involved in the research and development for a machine as large and complex as the LHC ---
from the first discussions in 1984 \cite{Gianotti2007} %
to the official start in 2008 more than 20 years passed --- it is already time to think about what will come afterwards.
Some of the projects being discussed in the scientific community are outlined here.

There are two competing projects for building an electron-positron collider, the ILC and the CLIC project.
As the energy loss per revolution due to synchrotron radiation for cyclic collider rings goes as $E^4 \, r^{-1} \, m_0^{-4}$ %
with the energy $E$, orbital radius $r$ and rest mass $m_0$ of the accelerated particle
and is thus much more severe for electron colliders than for proton colliders,
both projects are based on linear machine layouts.
The \acf{ILC}\inindex{International Linear Collider}
is a proposed \unit[31]{km} long linear electron-positron collider \cite{URL_ILCFacts},
which is supposed to have a total energy of \GeV{500}, with an option to upgrade to \TeV{1}.
It would use 16,000 superconducting RF cavities with an accelerating gradient of \unit[31.5]{MeV/m},
allowing to scan a center-of-mass energy range of $200$ up to \unit[500]{GeV}.
The \acf{CLIC}\inindex{Compact Linear Collider} is a proposed electron-positron collider
with a design center-of-mass energy of \unit[3]{TeV}.
To reach this energy with a machine of reasonable physical dimensions,
CLIC is supposed to use very high accelerating gradients of \unit[150]{MeV/m}
at room-temperature using a two-beam technology with a high-current, low-energy drive beam parallel to the main beam.
Although the above electron-positron colliders enter the \TeV{} regime,
the highest center-of-mass energies are still reserved to hadron colliders.
The \index{Very Large Hadron Collider} (VLHC) is a (very) theoretical proton-proton collider,
with a foreseen center-of-mass energy of \unit[100]{TeV} \cite{Stock}.
Another LHC-related project is lepton-nucleon scattering in the LHeC,
colliding the existing LHC proton or heavy ion beam with an electron beam,
possibly from an additional beamline in the LHC tunnel and synchronous to the proton-proton collisions,
or from a linear accelerator which could give a lower instantaneous luminosity but a higher center-of-mass energy \cite{URL_LHeC}. %

\subsection{The 7 Detectors at the LHC}

In addition to \ATLAS, which will be the subject of a section of its own,
there are three other large-scale detectors hosted at the \LHC: \ALICE, \CMS and \LHCb;
and three smaller ones, \LHCf, \MoEDAL and \TOTEM. %
\inindex{ALICE@\ALICE}
\ALICE (A Large Ion Collider Experiment) is specifically designed to probe lead-ion collisions,
in which a quark-gluon plasma is expected to be created,
like it is supposed to have existed a very short time after the Big Bang.
In this plasma, the energy density is high enough so that quarks and gluons are deconfined.
The aim of the \ALICE detector is to study the properties and cooling of the quark-gluon plasma
and the formation of particles which constitute the matter content of the universe today.
\ALICE is \unit[26]{m} long, \unit[16]{m} high and wide and weighs \unit[10,000]{tonnes}.
It has an asymmetric geometry with a central barrel and a dimuon spectrometer in the forward direction.
\ALICE will record data at a much lower rate than \ATLAS,
but the size of the recorded data per collision is much larger, %
giving a challenging data flux of \unit[1.25]{GB/s}.

\inindex{LHCb@\LHCb}
The \LHCb (Large Hadron Collider beauty) detector is another specialized detector.
Its main purpose is to answer the question,
why there is such an asymmetry between the matter and antimatter content in the universe.
To do so, it studies the properties of $b$ quarks in $B$ mesons and their decay products.
\LHCb is \unit[21]{m} long, \unit[10]{m} high and \unit[13]{m} wide and weighs $5600$ tonnes.
Its design is very different from general-purpose detectors
in that it does not enclose the interaction region,
but is a forward spectrometer with a number of planar detectors stacked one after the other on one side of the interaction point.
Like \ATLAS, the \acf{CMS} detector\inindex{CMS@\CMS} is a general-purpose detector
with the typical onion-like layout of its subdetectors.
\CMS shares many of the scientific goals of the \ATLAS detector,
and having these two detectors will allow to do cross-checks of the results of the respective other detector.
The structure of \CMS is quite different from the \ATLAS detector though.
A huge solenoid magnet with a cylindrical coil of superconducting cable
generates a magnetic field of \unit[4]{T} %
and surrounds both the silicon tracker as well as the electromagnetic and hadronic calorimeters.
The electromagnetic calorimeter uses crystals of lead tungstate, PbWO$_4$,
which are very dense but optically clear,
whereas the hadronic calorimeter is a sampling calorimeter made of brass and steel interleaved with plastic scintillators.
Outside the solenoid,
a huge iron return yoke interspersed with muon chambers guides and confines the magnetic field.
\CMS is smaller, but much heavier than \ATLAS due to the weight of the iron yoke.
At a size of \unit[21]{m} in length and \unit[15]{m} diameter,
it weighs \unit[12,500]{tonnes}.
The trigger system of \CMS is basically a two-level system with respect to its hardware,
consisting of a Level~1 and a High Level Trigger,
which roughly correspond to the hardware-based Level~1 and software implementation of the Event Filter of the \ATLAS detector.
Like in \ATLAS, only muon and calorimeter information are available at Level~1.
The second level at \CMS can take a much higher rate than the Event Filter in \ATLAS. %
However, the processing in the High Level Trigger of \CMS is also subdivided into three steps,
of which the first one uses only calorimeter and muon information,
and the second only partial tracker information \cite{CMSHLT2006}.

\inindex{LHCf@\LHCf}
\LHCf (Large Hadron Collider forward) are two very small detectors ($\unit[0.3\times0.8\times 0.1]{m^3}$),
sharing the interaction region at Point~1 with the \ATLAS detector,
but at a distance of \unit[140]{m} to either side of the interaction point
exactly at an angle of zero degrees in a Y-shaped transition of the beam tube \cite{LHCf2011}. %
They measure the number and energy of neutral pions produced in the forward directions in the proton-proton collisions,
with the aim of understanding ultrahigh-energy cosmic ray events which happen in the upper atmosphere.
\inindex{TOTEM@\TOTEM}
\TOTEM (TOTal Elastic and diffractive cross section Measurement) %
also studies forward particles.
It consists of three sub-detectors extending over \unit[440]{m} in length at Point~5 and is \unit[5]{m} high and wide and weighs \unit[20]{tonnes}.
Using Roman pots (specially designed detectors housed in cylindrical vessels), %
gas electron multipliers and cathode strip chambers, %
its physics goals are a precise measurement of the proton-proton interaction cross section, %
as well as the study of the proton structure.
Also, all of the \LHC detectors will use \TOTEM's measurement to calibrate their luminosity monitors.
\inindex{MoEDAL@\MoEDAL}
\MoEDAL (Monopole and Exotics Detector At the LHC) sits in the cavern of \LHCb
and is comprised of an array of plastic nuclear track detectors.
It searches for highly ionizing stable massive particles such as magnetic monopoles
or the lightest stable states in $R$-parity conserving Supersymmetry
or models of universal extra dimensions with Kaluza-Klein parity.

\section{The ATLAS Detector} %
\label{sec:atlas}

The \ATLAS\inindex{ATLAS@\ATLAS} detector\footnote{
  \ATLAS is an acronym standing for ``A Toroidal LHC ApparatuS''.
} is a general-purpose detector for particle physics,
with a forward-backward symmetric cylindrical geometry and near $4\pi$ coverage in solid angle,
built for probing proton-proton and heavy-ion collisions.
It is a unique machine of \unit[25]{m} in height and \unit[44]{m} in length,
weighing approximately \unit[7000]{tonnes} in total. %
The \ATLAS detector is housed at Point~1, %
one of the interaction points of the \ac{LHC} ring adjacent to the CERN main entrance,
in a cavern about \unit[100]{m} below the surface.

In order to be able to describe the location of the different subsystems in the following,
first the detector \index{coordinate system} will be introduced,
which will be used throughout this thesis.
The right-handed \ATLAS coordinate system has its origin at the nominal interaction point at the center of the detector.
The direction of the beam pipe defines the $z$-axis %
and the $x$-$y$ plane is transverse to the beam direction.
The positive $x$-axis points from the interaction point to the center of the \ac{LHC} ring,
and the positive $y$-axis points upwards.
The side of the \ATLAS detector with positive $z$ is side ``A'', the other side is side ``C''.
Polar coordinates with a radial distance $r$ and an azimuthal angle $\phi$ are used in the transverse plane.
In spherical coordinates,
the third coordinate can either be expressed through the polar angle $\theta$ with respect to the $z$-axis %
or the \index{pseudorapidity}
\begin{equation}
  \eta \definedas -\ln\tan(\theta/2).
  \label{eq:define_pseudorapidity}
\end{equation}
The $x$-$y$ plane at $z=0$ is defined by $\eta=0$.
In the directions along the beam pipe the pseudorapidity diverges.
For particles at very high energies so that their mass is negligible,
the pseudorapidity is equal to the \index{rapidity} $y$,
which is defined as
\begin{equation}
  y \definedas \frac{1}{2} \ln \left(\frac{E+p_z}{E-p_z}\right).
  \label{eq:define_rapidity}
\end{equation}
Geometrical distances\inindex{Delta R@$\Delta R$} of reconstructed objects
are often expressed in terms of
\begin{equation}
  \Delta R \definedas \sqrt{\Delta\eta^2 + \Delta\phi^2}.
\end{equation}
The following description of the design of the \ATLAS detector is based on \cite{ATLASTDR1999,ATLASDetector2008,CSCNotes}.

\subsection{Detector Components}

\begin{figure}
  \centering
  \incgraphics{width=\widthwideplot}{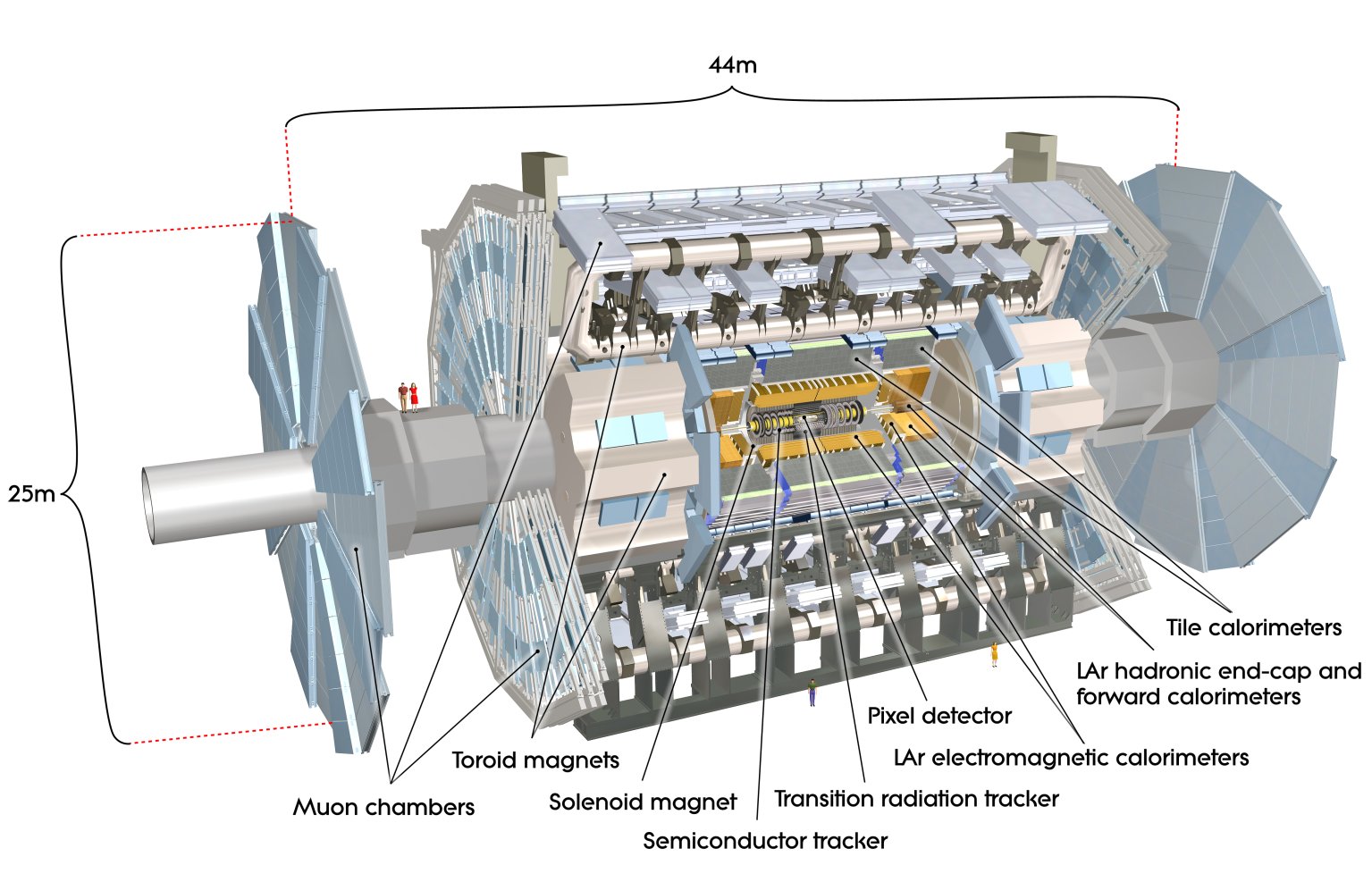}
  \caption{
    The \ATLAS detector (computer-generated cut-away view).
    The different subsystems are explained in the text.
  }
  \label{fig:experimental_setup_atlas_full_detector_view} %
\end{figure}

\ATLAS has a cylindrical geometry with the beam pipe as axis of its rotational symmetry
and the nominal interaction point at its very center,
where the proton beams delivered by the LHC are brought to collision.
The beam pipe has a diameter of \unit[58]{mm} and is constructed from \unit[0.8]{mm} thick beryllium.
The detector components are described as being part of the barrel
if they are located in the central pseudorapidity region
or as part of one of the two end-caps if they are in the forward or backward region.
In the following, the detector components will be described,
going from the center to the outermost components.
Three main components can be distinguished,
which give complementary information and consist of several subsystems each.
Closest to the interaction point at the center of the detector is the Inner Detector,
which is a tracking device used to measure precisely the tracks of charged particles
in the pseudorapidity region up to $|\eta| < 2.5$.
It is immersed in a \unit[2]{T} solenoidal magnetic field and surrounded by the calorimeter system,
which measures the energy of easily stopped particles.
The overall dimensions of the detector are defined by the muon spectrometer,
which consists of a toroidal magnet system and several different types of muon chambers.
They measure the tracks of particles such as the highly penetrating muons,
which cannot be stopped in the calorimeter.
A computer-generated view of the detector and its components is shown in \Fig{fig:experimental_setup_atlas_full_detector_view}.

\subsection{Inner Detector and Solenoid Magnet}

The \index{Inner Detector} (ID) consists of three independent and complementary tracking systems.
The innermost is the silicon pixel detector (Pixel),
surrounded by another precision silicon tracker, the \acf{SCT}, %
and the \acf{TRT}.
The Pixel and SCT cover the region $|\eta| < 2.5$,
while the TRT, being only the outer part of the cylinder, %
covers $|\eta| < 2$,
as can be seen in \Fig{fig:experimental_setup_atlas_inner_detector_view},
which shows a detailed overview of the Inner Detector only.
In the barrel region, the components are arranged as concentrical cylinders,
in the endcap region as disks perpendicular to the beam axis.
The design goals of the Inner Detector are to provide a high momentum resolution, %
precisely determine the primary and secondary vertex position and detect charged tracks with \pt above \GeV{0.5}.

\begin{figure}
  \centering
  \incgraphics{width=\widthwideplot}{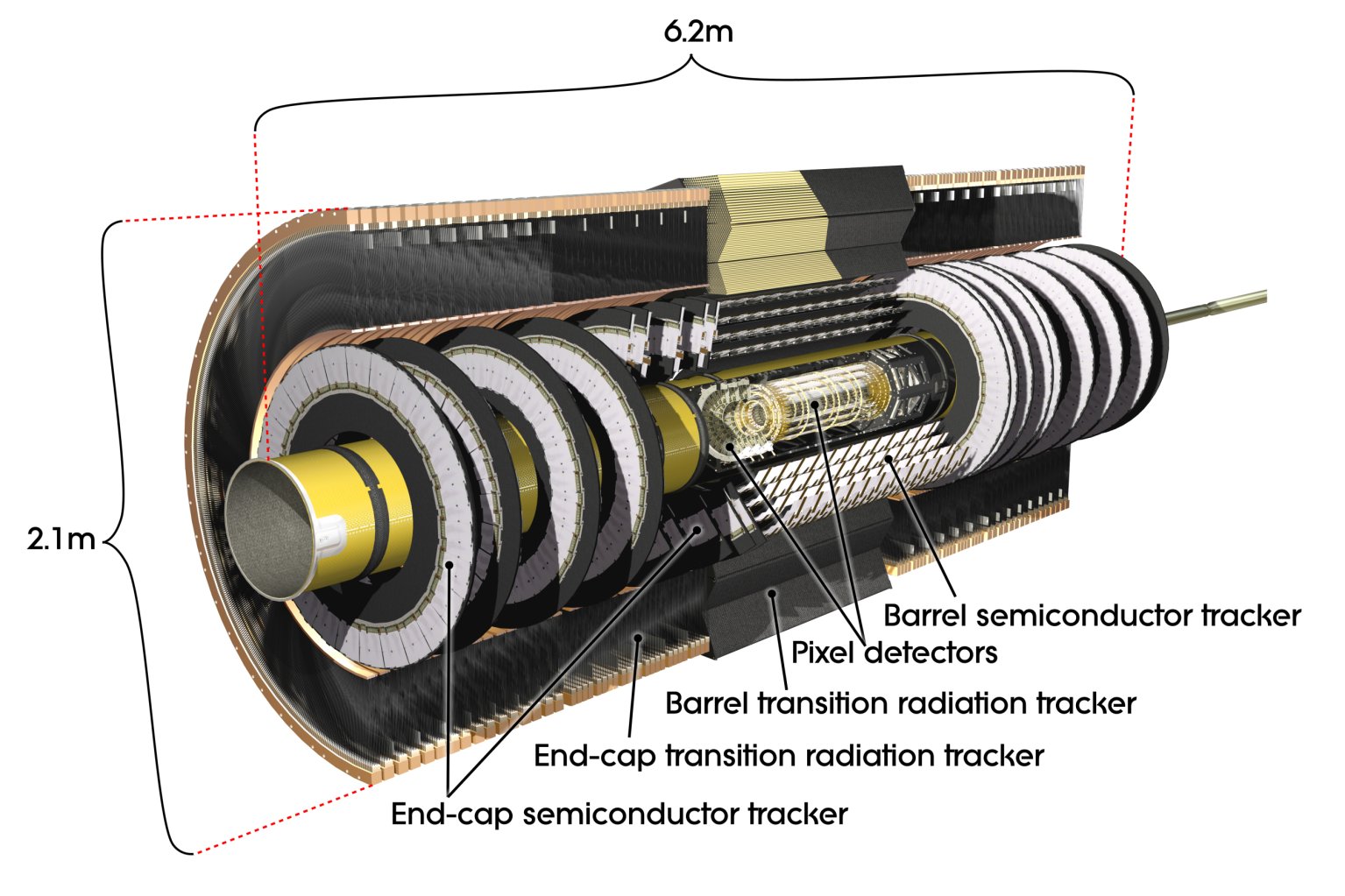}
  \caption{
    The \ATLAS Inner Detector (computer-generated cut-away view).
  }
  \label{fig:experimental_setup_atlas_inner_detector_view} %
\end{figure}

The silicon pixel detector needed for the vertex detection has the highest granularity.
It consists of three layers in the barrel and three disks in each of the endcaps,
with a minimum pixel size of $\unit[50]{\micro m} \times \unit[400]{\micro m}$.
The intrinsic accuracies of the layers in the barrel are \unit[10]{\micro m} in $R$--$\phi$
and \unit[115]{\micro m} in the $z$-direction and vice versa in the disks.
The pixel detector is operated at a temperature of $-5$ to \unit[$-10$]{\textdegree C} to reduce electronic noise
and has $80.4$ million readout channels in total.
The silicon microstrip tracker has four coaxial cylindrical layers in the barrel and nine disks in each endcap,
which are arranged such that a straight track with $|\eta| < 2.5$ crosses at least four modules,
giving four space-point measurements with two hits per module.
The two times two sensors on each module are rotated against each other
and have a mean pitch of the strips of \unit[80]{\micro m}.
The rotation by a small angle allows to do stereo measurements of both coordinates in the plane.
The intrinsic accuracies per module in the barrel are $\unit[17]{\micro m}$ in $R$--$\phi$
and $\unit[580]{\micro m}$ in the $z$-direction and vice versa in the disks.
In total the SCT has $6.3$ million readout channels.
The \acf{TRT} is a combination of a straw tracker and a transition radiation detector.
It is made of gaseous straw tube elements with \unit[4]{mm} diameter,
which are thin-walled proportional drift tubes,
interleaved with foils and fibres of transition radiation material.
The straw tubes are filled with a Xe/CO$_2$/O$_2$ gas mixture. %
In constrast to the semiconductor trackers,
the TRT is designed to operate at room temperature.
Its intrinsic accuracy is \unit[130]{\micro m} per straw. %
In the barrel, the straws are mounted parallel to the beamline with a length of \unit[144]{cm} in up to 73 layers.
In the end-cap, the straws run radially and are only \unit[37]{cm} long, organized in wheels of 160 planes.
The TRT has $351,000$ readout channels in total.
Straw hits in outer layers contribute significantly to the momentum measurement,
in spite of their lower resolution due to the long lever arm of the measurement.
In addition to the tracking abilities,
the TRT provides for electron identification by the detection of transition radiation photons:
If a charged particle travels through a medium at a velocity
higher than the local speed of light $c/n$,
where $n$ is the index of refraction of the medium,
the medium radiates photons at a characteristic angle. %
This \cerenkov transition radiation is used to identify electrons with energies between $0.5$ and $\unit[150]{GeV}$. %
The distinction between transition radiation and tracking signals is made
using separate low and high thresholds in the front-end electronics. %

The Inner Detector is fully contained in the \index{solenoid magnet}
consisting of one thin superconducting coil with a length of \unit[5.8]{m} and a diameter of \unit[2.56]{m}. %
It weighs \unit[5.7]{tonnes} and provides a \unit[2]{T} axial field,
in which the Inner Detector components are immersed,
to bend the tracks of charged particles and thereby make possible the momentum measurements.
At nominal current,
the energy stored in the magnet field is \unit[40]{MJ}.
The flux is returned by the steel of the \ATLAS hadronic calorimeter (see below) and its girder structure.
It can be cooled down to the operating temperature of \unit[4.5]{K} in one day.
In order not to deteriorate the calorimeter performance,
the construction of the solenoid magnet had to be as light-weight and thin as possible.
It contributes approximately $0.66$ radiation lengths.

A major problem for the Inner Detector components is that,
being so close to the beam-pipe,
they are subject to a high radiation dose
which is absorbed by the material over the running time of the experiment.
It therefore needs to have a good radiation hardness. %
The pixel inner vertexing layer must therefore be replaced after approximately three years of running at the design luminosity.
Also, the bias voltage will need to be increased from initially \unit[150]{V} to up to \unit[600]{V}
to achieve a good charge collection efficiency after several years of running.

\subsection{Calorimeters}

\begin{figure}
  \centering
  \incgraphics{width=\widthwideplot}{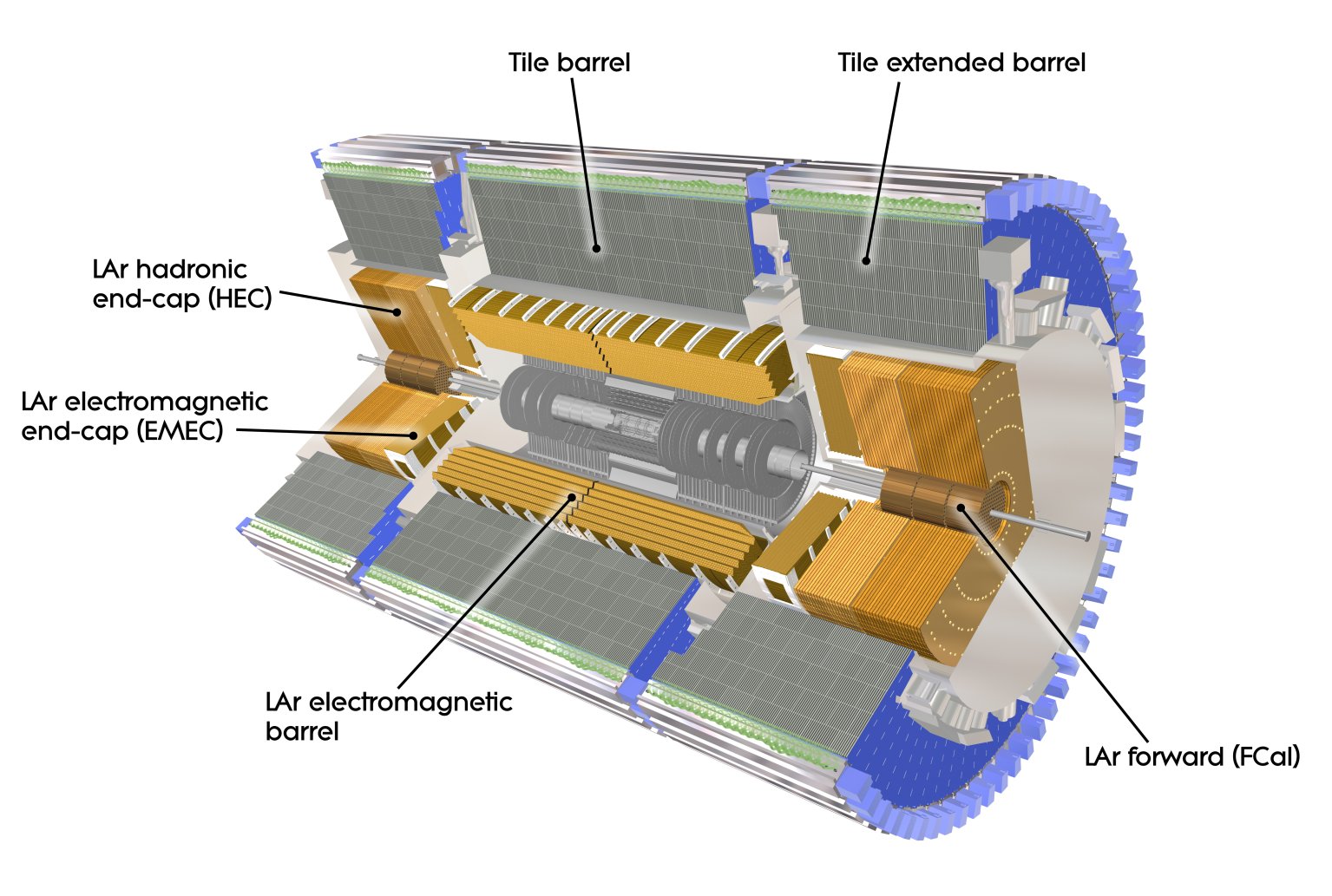}
  \caption{
    The \ATLAS Calorimeters (computer-generated cut-away view).
  }
  \label{fig:experimental_setup_atlas_calorimeters} %
\end{figure}

The \ATLAS calorimeter system covers the region up to $|\eta| < 4.9$
and consists of electromagnetic calorimeters and hadronic calorimeters
which are located directly behind (``downstream'') the electromagnetic calorimeters.
The calorimeter system is based on different materials,
but all components are sampling calorimeters.
In general, the inner cylinder is made up of calorimeters
which use liquid argon (\acslink{LAr}) as active medium,
surrounded by an outer concentric cylinder of tile calorimeters,
as shown in \Fig{fig:experimental_setup_atlas_calorimeters}.
The calorimeter system has a full $\phi$-symmetry and coverage around the beam axis.
The thickness of the electromagnetic calorimeters is $> 22$ radiation lengths in the barrel and $> 24$ radiation lengths in the end-cap, %
and for the hadronic calorimeters the thickness amounts to $9.7$ interaction lengths in the barrel and about $10$ in the end-caps. %
Over the $\eta$ region that is covered by the Inner Detector,
the electromagnetic calorimeter has the highest granularity
to provide precision measurements of the energy deposited by electrons and photons.

The LAr \index{electromagnetic calorimeter} is a sampling calorimeter
with accordion-shaped kapton electrodes %
and lead absorber plates and LAr as the active medium.
It consists of a barrel part which extends up to $|\eta| < 1.475$,
with a small gap of \unit[4]{mm} at $z=0$ between two symmetric halves (EMA and EMC),
and two end-cap components in the region $1.375 < |\eta| < 3.2$.
The end-cap components (EMEC) are again \revised{mechanically} divided into two wheels with $1.375 < |\eta| < 2.5$ and $2.5<  |\eta| < 3.2$.
The barrel and the inner wheels have together at least three segmentations in depth (sampling layers),
while the outer wheels have only two.
The granularity of the electromagnetic calorimeter in $\eta$ is very fine,
down to $\Delta \eta = 0.025/8$ in the first layer in the range up to $|\eta| < 1.8$. %
The granularity in $\phi$ is mostly $0.1$ in the first and $0.025$ in the second and third layers,
in which most of the energy is collected for high energy electrons (see \cite{ATLASDetector2008} for details). %
A presampler,
consisting of an instrumented active LAr layer, %
is installed in front of the electromagnetic calorimeter
and covers the region up to $|\eta| < 1.8$.
It corrects for the energy lost by electrons and photons before reaching the calorimeter.

The \index{hadronic calorimeter} uses liquid argon as well as scintillating tiles as active components.
It is comprised of three parts:
  the tile calorimeter,
  the LAr hadronic calorimeters in the end-caps
  and the hadronic part of the LAr forward calorimeter.
(Note that there is also an electromagnetic calorimeter in the forward region, see below.)
The tile calorimeter (Tile) consists of three parts, one barrel and two extended \revised{barrel parts},
which cover the region up to $|\eta| < 1.7$.
It uses steel as absorber and scintillating tiles as active component, %
and has three segments in depth. %
The LAr hadronic calorimeters in the end-caps (HEC)
cover $1.5 < |\eta| < 3.2$ to have a slight overlap with the tile and forward calorimeters.
They use copper plates interleaved with LAr as active medium
and consist of two independent wheels per end-cap, each divided into two segments in depth.

The LAr \index{forward calorimeter} (FCal)
covers the forward and backward regions closest to the beam with $3.1<|\eta|<4.9$.
It consists of three modules in each end-cap,
of which the first uses copper and is optimized for electromagnetic measurements,
whereas the second and third use tungsten and measure predominantly energy from hadrons.
Again, liquid argon is the active medium flowing within the metal matrix.
The longitudinal segmentation helps to do electromagnetic shower identification.

All calorimeter systems together have roughly $1.9\ten{5}$ readout channels.
There are two separate readout paths:
One with coarse granularity,
for which calorimeter cells occupying the same area in \eta and $\phi$ are combined to trigger towers.
This is used by the Level~1 trigger (cf. \Sec{sec:tdaq}).
The other has a fine granularity and is used for the final readout by the \acl{HLT} and in the offline reconstruction.
The performance in the transition region between the barrel and end-cap electromagnetic calorimeters, $1.37 < |\eta| < 1.52$,
is expected to suffer from the larger amount of material in front of the first active calorimeter layers.
The shapes of the calorimeter signals are compared to nominal shapes,
allowing to determine a $\chi^2$-like quality factor.
Also the timing of signal is measured and can be used in analyses.

\subsubsection{LAr~Hole}
\label{sec:experimental_setup_lar_hole}
The ``\index{LAr~hole}'' is the commonly used term within the \ATLAS collaboration
for a detector problem
which has occured in the first half of 2011 data taking\footnote{
  30.04.2011, starting with the beginning of data-taking period 2011~E. %
}
and which affects part of the electromagnetic liquid-argon calorimeter in the barrel.
Due to the failure of a crate controller,
the information from six front-end boards is lost,
four of which could be recovered starting with period 2011~I.
The region affected by the LAr~hole is \revised{$[0.0,\,1.45] \times [\numprint{-0.78847},\,\numprint{-0.59213}]$}
in $\eta \times \phi$\footnote{
  No official documentation on this issue appears to be publicly available at the time of writing. %
}. %
For physics analyses, a way to deal with this problem
is to reject events which include jets that fall into the LAr~hole region.
Due to the high number of jets in QCD events, this may make up a significant fraction of events
and therefore refined veto strategies are recommended.
The analysis presented in \Sec{sec:analysis_susysearch} uses only 2010 data and is thus not affected by the problem.
The impact on the efficiencies of \jetmet triggers in 2011 is studied in \Sec{sec:results_trigger_performance_2011_consistency}.

\subsection{Muon Spectrometers and Toroidal Magnet System}
\label{sec:experimental_setup_atlas_muon_system}

\begin{figure}
  \centering
  \incgraphics{width=\widthsingleplot}{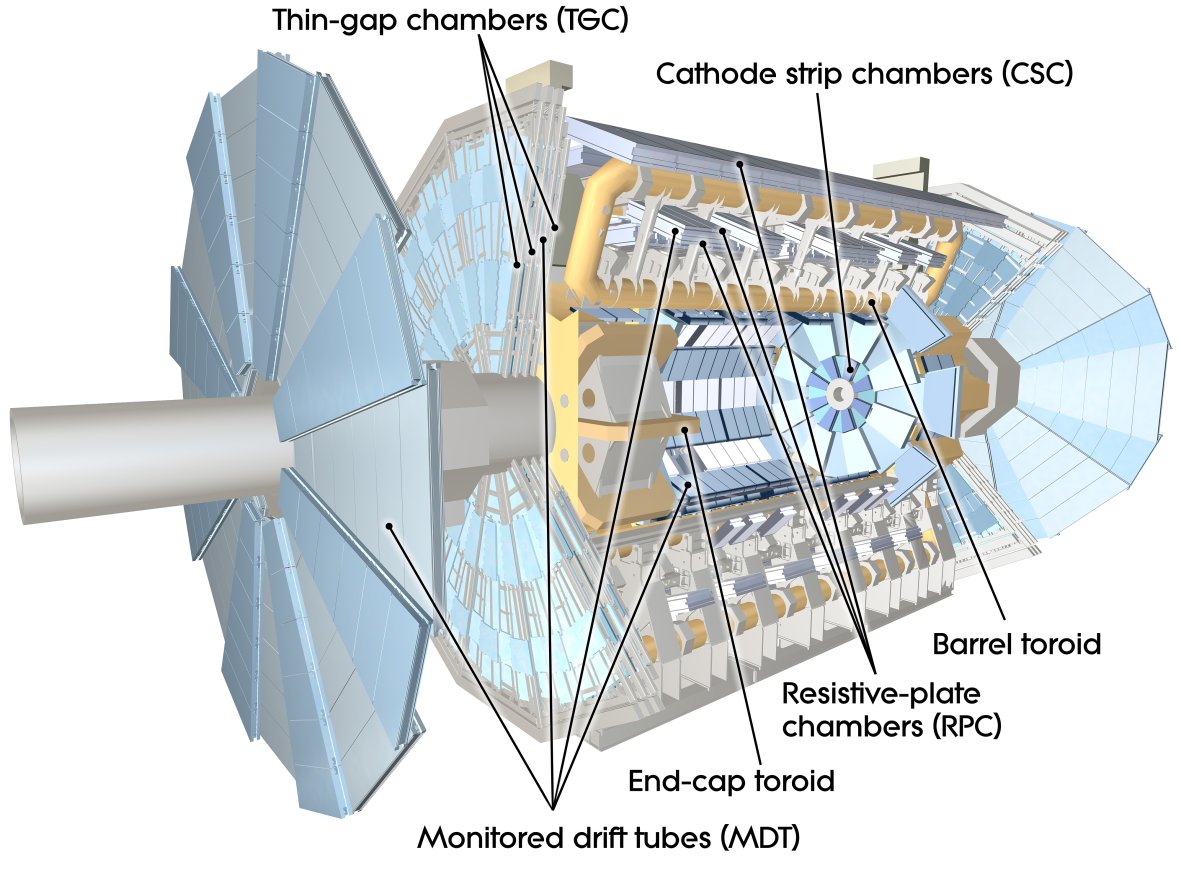}
  \hfill
  \caption{
      The muon system of the \ATLAS detector (computer-generated cut-away view).
  }
  \label{fig:experimental_setup_atlas_muon_systems} %
\end{figure}

The muon spectrometer of the \ATLAS detector is a large tracking system
and measures the charge and momentum of charged particles leaving the calorimeters.
It also serves for the identification of particles as muons.
All electromagnetically or strongly interacting Standard Model particles
except for the muon are expected to be stopped within the calorimeters,
unless they have exceedingly high energies such that they can punch through the calorimeters.

The four subsystems of the muon spectrometer are mounted in and around air-core to\-roids\inindex{Toroid magnet},
which generate the necessary magnetic bending field as can be seen in \Fig{fig:experimental_setup_atlas_muon_systems}.
They use four different technologies
and can be classified into two subsystems which do precision tracking
and another two which are specialized to provide information to the Level~1 trigger at a high time-resolution.
The systems are organized in three concentric cylindrical layers at approximately $5$, $7.5$ and \unit[10]{m} in the barrel.
In the transition and end-cap region, they form large wheels installed orthogonal to the beam direction,
as can be seen in \Fig{fig:experimental_setup_atlas_muon_systems_schematic}.

\begin{figure}
  \centering
  \incgraphics{width=\widthwideplot}{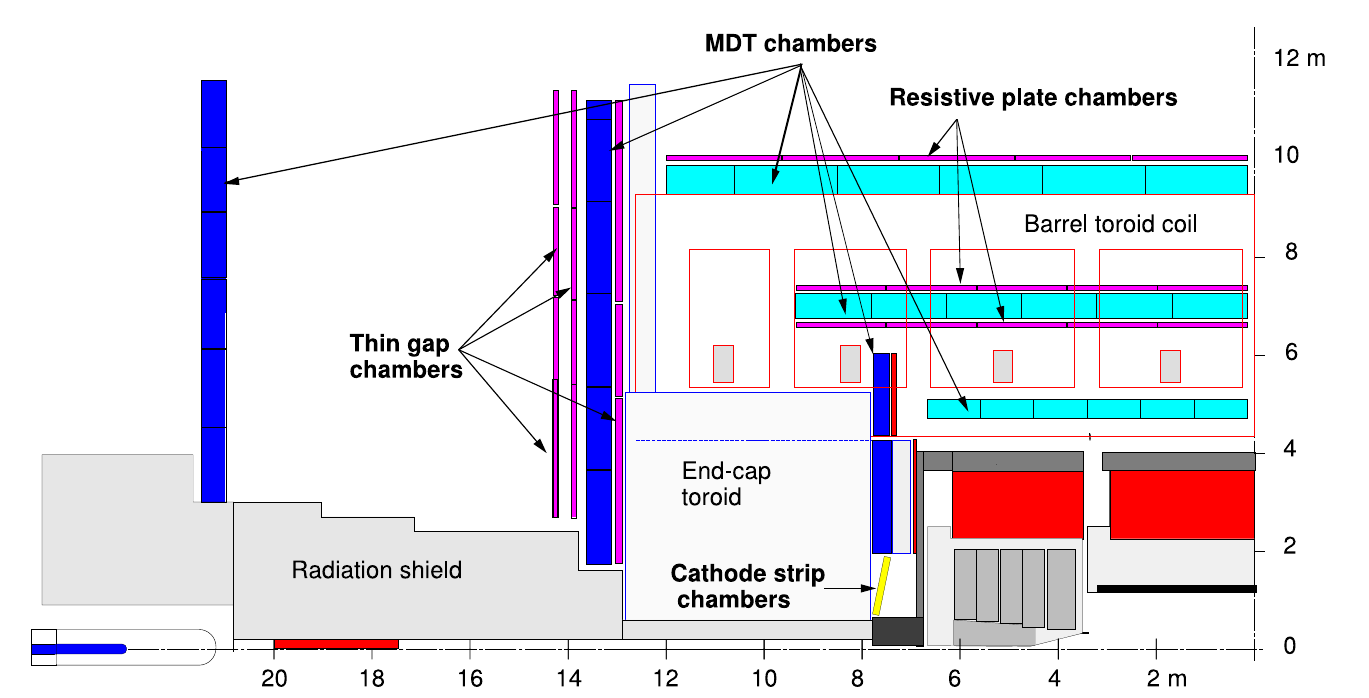}
  \hfill
  \caption{
    Cross-sectional view in the $y$-$z$ plane showing the positions
    of the four muon systems and the coils of the toroid magnet
    in one quadrant of the \ATLAS detector.
  }
  \label{fig:experimental_setup_atlas_muon_systems_schematic} %
\end{figure}

The high-precision tracking chambers measure the track coordinates in the bending plane.
The \acfp{MDT}, covering $|\eta|< 2.7$ (2.0 for the innermost of the three concentric layers),
consist of up to eight layers of pressurized drift tubes per chamber,
organized in two multilayers,
and achieve an average resolution of \unit[80]{\micro m} per tube and about \unit[35]{\micro m} per chamber.
They are called monitored,
because their deformations and relative positions are monitored by $12,000$ alignment sensors based on optical monitoring.
The \acp{MDT} are complemented by \acfp{CSC},
which cover $2.0<|\eta|<2.7$
and have a higher rate capability up to counting rates of \unit[1000]{Hz/cm$^2$},
compared to \unit[150]{Hz/cm$^2$} for the \acp{MDT}. %
The \acp{CSC} are multiwire proportional chambers with cathodes segmented into strips in orthogonal directions, %
thus allowing to measure both coordinates.
Their resolution is \unit[40]{\micro m} in the bending plane and about \unit[5]{mm} in the transverse plane.

The muon trigger chambers have a time resolution of $1.5$ to \unit[4]{ns}.
Two different technologies are used as trigger chambers,
\acfp{RPC} for $|\eta|<1.05$
and \acfp{TGC} for $1.05<|\eta|<2.7$ ($2.4$ for triggering).
They are also used for measuring the muon coordinate in the direction orthogonal %
to that determined by the precision chambers with a resolution of several millimeters.
The \acp{TGC} are also multiwire proportional chambers,
with the wire-to-wire distance being smaller than the wire-to-wall distance.
This allows for running in a saturated mode with a very high gas gain. %
The $\phi$-symmetry of the muon chamber systems follows the structure of the toroids, consisting of eight octants.
There are gaps in the coverage of the muon system
to allow for services to the solenoid magnet,
the calorimeters and the Inner Detector and the detector support structure.
The total number of readout channels of the muon systems is $1.1\ten{6}$. %
For details see \cite{ATLASDetector2008}.

\subsubsection{Toroid Magnets}

The muon spectrometer relies on a set of large superconducting air-core toroidal magnets.
The light and open structure of the magnets is needed
to minimize multiple scattering effects,
which would degrade the muon identification performance and momentum resolution.
The toroidal magnet system consists
of a long barrel toroid in the region up to $\seta<1.4$
and two smaller end-cap magnets inserted into both ends of the barrel toroid which cover $1.6 < \seta < 2.7$.
The strength of the magnetic field is approximately $0.5$ and \unit[1]{T} in the central and end-cap regions, respectively.
Each of the three toroids consists of eight coils,
the end-cap toroid coil system being rotated by \unit[$\pi/8$]{rad} with respect to the barrel toroid coil system
to improve the overlap of the magnetic fields.
The toroid magnets have an axial length of $25.3$ and \unit[5.0]{m}
and an outer diameter of $20.1$ and \unit[10.7]{m} for the barrel and end-cap parts, respectively.
The total weight is 830 plus $\unit[2\times239]{tonnes}$,
and the total energy stored in the magnetic field is \unit[1.58]{GJ},
the region where the magnetic field exceeds \unit[50]{mT} being \unit[12,000]{m$^3$}.
The strength of the magnet field is monitored by $1800$ Hall sensors,
which are distributed throughout the spectrometer volume.
Note that the direction of deflection for charged particles
is different from the one effected by the field of the solenoidal magnet,
in which the Inner Detector is immersed.

\subsection{Performance Goals}

\begin{table}
  \centering
  \begin{tabular}{ll}
    \toprule
    Detector component & Required resolution \\
    \midrule
    Tracking & $\sigma(\pt) / \pt  = \percent{0.05} \, \pt \oplus \percent{1}$ \\
    Muon spectrometer & $\sigma(\pt) / \pt  = \percent{10} $ at $\pt =  \TeV{1}$ \\
    EM calorimetry & $\sigma(E) / E = \percent{10} / \sqrt{E} \oplus \percent{0.7} $ \\
    Hadronic calorimetry (jets) \\ 
    \quad  barrel and end-cap & $\sigma(E) / E = \percent{50} / \sqrt{E} \oplus \percent{3}  $ \\
    \quad  forward & $ \sigma(E) / E = \percent{100} / \sqrt{E} \oplus \percent{10} $ \\
    \bottomrule
  \end{tabular}
  \caption{
    Design goals of the \ATLAS detector with respect to energy and momentum resolution of the different subdetectors.
    $E$ and \pt are given in GeV.
    The numbers are taken from \cite{ATLASDetector2008} and may differ from the resolutions actually achieved.
  }
  \label{tab:experimental_setup_atlas_goals}
\end{table}

In \Tab{tab:experimental_setup_atlas_goals},
an overview is given of the performance goals
with respect to the design of the \ATLAS detector components that have been described above.
The numbers are taken from \cite{ATLASDetector2008} and mainly for illustrational purposes.
The resolutions actually achieved may differ, cf. \Sec{sec:software_reconstruction_event_reconstruction}.

\subsection{Minimum-Bias Trigger Scintillators}
\label{sec:atlas_minimum_bias_trigger_scintillators}
\inindex{Minimum-bias trigger scintillators}
For the initial phase of data taking at low instantaneous luminosities,
the \acf{MBTS} can be used %
to trigger on a minimum collision activity during proton-proton collisions \cite{ATLAS-CONF-2010-060}.
The MBTS detector consists of 32 scintillator paddles,
which are organized in two disks perpendicular to the beam pipe
and installed at either side of the detector at a distance of $z = \pm\unit[356]{cm}$ from the interaction point.
Each disk covers an \eta range of $[2.09, 3.84]$
and is split into two rings with eight independent sectors each.
The light from the scintillators is collected
and converted into an electrical signal in photomultiplier tubes,
which is passed on to the \ac{CTP} (see \Sec{sec:tdaq}),
where the number of hits per scintillator is evaluated
by comparing the signal to a given discriminator threshold.
Triggering on minimum-bias events can be done by requiring either single, multiple or coincident hits in both disks.

\subsection{Forward Detectors}

The forward detectors are a group of special detectors \cite{ATLASDetector2008} located relatively
far away from the interaction point at very high pseudorapidities.
The \acslink{LUCID}\inindex{LUCID} detector,
where LUCID stands for luminosity measurement using a \cerenkov integrating detector,
is a detector which is primarily dedicated to relative luminosity measurements.
It is located at $\unit[\pm17]{m}$ from the interaction point
and can record the luminosity for each bunch crossing separately.
The \acf{ZDC}\inindex{ZDC} is $\unit[\pm140]{m}$ away from the interaction point
and detects forward neutrons in heavy-ion collisions.
The \acslink{ALFA} (\acl{ALFA}) detector\inindex{ALFA} at a distance of $\unit[\pm240]{m}$
consists of scintillating-fibre trackers which are located inside Roman pots.

\chapter{Data Acquisition and Data Processing}
\label{sec:tdaq}

Following the description of the \ATLAS detector hardware,
this chapter describes the acquisition and the processing of data in \ATLAS,
and is mainly concerned with software issues.
The central part of the data-acquisition system is the trigger.
Its purpose in general and the necessary terminology are discussed first,
before the \ATLAS trigger system itself is described.
When the data has been recorded,
it needs to be suitably processed.
This is called event reconstruction and is explained in the following part.
In the last part of this chapter,
a brief introduction to Monte Carlo event generation is given
and important related notions are introduced.
Additional technical background can be found in \Sec{sec:computing}.

\section{Purpose of a Trigger}
\label{sec:triggering}

The trigger system of a detector can be thought of as a filter that,
out of the large number of collision events taking place in a collider experiment,
selects those that are stored for later processing and analysis.
The need for such a filter becomes evident when looking exemplarily at the following numbers:
The \ac{LHC} is designed to deliver proton-proton collisions at a rate of up to \unit[40]{MHz}, %
\revised{so that every \unit[25]{ns} a collision of proton bunches takes place at all interaction points.}
The output rate, at which the measurements of the collision events can be written out,
is limited mainly by the allocated storage capacity in long terms
and the output bandwidth of data-taking system in short terms.
For \ATLAS, the latter amounts to a few hundred megabytes per second.
Assuming an average event size of \unit[1.5]{MB},
this yields \revised{an allowed output rate} of $200$ to \unit[300]{Hz},
which may be raised up to \unit[600]{Hz} for limited periods of time \cite{TriggerPerf2010}.
(Of course, there are also limitations from the speed at which the detector systems can be read out at full granularity,
but those are less severe.)
This means that a rate reduction of the order of $10^5$ is needed,
implying that most of the collision data will be discarded.
In fact, it is not a problem to achieve this reduction from the point of view of analysis,
as most of the collisions will only procure events
which contain processes that are well known and studied.
Of interest are those events which contain rare processes,
like the production and decay of some so far unknown particle.
The task is therefore to identify those events among the huge number of collisions taking place every second,
which are interesting in some sense.
This is done by the trigger system:
It uses a suitably reduced set of information to identify,
in real time during data taking (``online''), the most interesting events to retain for detailed analysis (``offline''),
ideally by rejecting all uninteresting events while keeping the interesting ones at a high efficiency.
The trigger system will usually comprise several more or less independent triggers running in parallel.
If any of these triggers fires, \ie decides to accept the collision event, the event is stored.

A well-designed trigger system ought to meet a number of requirements:\inindex{Trigger requirements}
\begin{itemize}
  \item Robustness:
    The operation of the trigger should be reliable and stable.
    Technical problems leading to missing, incomplete or invalid data should not lead to a crash of the whole trigger system or parts of it.
  \item Redundancy:
    Redundancy is needed both with respect to online data taking and offline aspects.
    During data-taking, it may become necessary to disable a particular trigger %
    because its output rate is too high,
    or because it is misconfigured and disturbs the rest of the data-taking process. %
    With respect to offline, it may be found afterwards that a particular trigger has selected other types of events than it was intended to.
    In both cases, loss of data can only be prevented if the task of the disabled or malfunctioning triggers can be taken over by another.
    This redundancy normally should not increase the output rate a lot.
  \item Inclusiveness:
    Keeping the selection criteria a trigger applies to the data rather general
    allows to use the data taken with this trigger for a lot of different analyses.
    It also reduces the risk of retrieving only a biased subsample of all events
    or to miss unexpected but interesting collision events.
  \item Simplicity:
    Defining the triggers in such a way that their decisions are easy to comprehend
    simplifies the commissioning and the maintenance of the trigger system.
    It also helps understanding the trigger (in-)efficiencies,
    an aspect that is discussed in detail in \Sec{sec:triggerefficiencies}.
\end{itemize}

In order to be able to run within the constraints of the trigger system,
in addition to these requirements,
the triggers need to have a precision that is high enough to make an efficient decision
and at the same time be fast enough to process all events within the given time.
The performance of a trigger can be quantified in terms of its rate and efficiency, \ie the quality of its selection.
The optimization of the trigger is then done by finding the best working point
as a compromise between smaller rates and higher efficiency. %
This is discussed in \Sec{sec:results_trigger_rates_benefits_combined_triggers} using the example of \jetmet triggers.  %
The main challenge is that all events which are rejected by the trigger are lost forever.
Thus, it is possible to confirm that the triggered events are interesting,
but not that there are no interesting events among those which have been discarded.
The solution to this problem is to keep a few additional events,
regardless of the trigger decision (pass-through events, see below),
which allows to search a small sample of discarded events for anything interesting that may have been missed.

\section{Terminology of Triggers}

\revised{Before the \ATLAS trigger system will be described in the next section,}
some basic notions and concepts will be introduced,
which are needed in the discussion of triggers in general.

\label{sec:definition_prescales}
The rate of a particular trigger can be adjusted,
for example to follow the constant decay of the instantaneous luminosity within a fill of the collider,
by applying different \define{prescales}\inindex{Prescale factor}.
Applying a prescale factor $N$ to a trigger
means that out of $N$ events in which this trigger has given a positive decision
only one randomly selected event is actually stored.
The rate at which the trigger fires is therefore directly proportional to the inverse of the prescale.
A trigger is called prescaled if it has a prescale larger than one.
In some sense the opposite of the prescale factor is the \define{pass-through} factor\inindex{Pass-through factor}.
A pass-through factor of $N$ means that one in $N$ events is treated
as if it had been accepted by the trigger regardless of the actual trigger decision.
A use case for pass-through factors is to obtain a control sample of discarded events as mentioned above.
Triggers can also be run in \define{\index{transparent mode}},
which means that the event is accepted without executing the trigger.
Finally, an error within the trigger system may lead to a \define{\index{forced accept}}
of the event without further execution of triggers.
The accept signal of a trigger may not only be inhibited by prescales,
but also due to a dead time veto or busy veto of the trigger system.
\define{\index{Dead time}} occurs always after an event has been accepted (see below). %
A busy signal can be issued when some part of the trigger or data-acquisition system cannot cope with the current trigger rate.

Triggers can be categorized with respect to their type and their function.
With respect to their type,
a distinction is made between
single object triggers, multiple object triggers, combined triggers and topological triggers.
\define{Single} and \define{multiple object triggers} are used for final states with at least one or two or more characteristic objects of the same type.
\define{Combined triggers} select events with two or more objects of different types,
whereas the decision of \define{topological triggers} is based on the combined properties of objects of different types.
With respect to their purpose,
there are primary, supporting and monitoring or calibration triggers.
\define{Primary triggers}\inindex{Primary trigger} are used to select events for physics analysis,
which potentially contain the searched for signal.
\define{Supporting triggers}\inindex{Secondary trigger} are also used in physics analysis,
but have a supplementary function.
They select events, for example, for studies of the background.
The data collected with \define{monitoring} and \define{calibration triggers}
is used to ensure the correct operation of the trigger and detector.
The configuration of the trigger system is stored in the \define{\index{trigger menu}},
which includes information about which triggers are executed and their parameters.
Different triggers are rarely exclusive,
so that some events will be triggered, \ie accepted, by more than one trigger.
The consequence of this overlap between triggers is that deactivating a particular trigger
will not reduce the total output rate of the trigger system
by the full rate of this trigger, but only by the \define{\index{unique rate}} of this trigger.
This is the rate of events in which exclusively this trigger fires.
The overlap of triggers makes it difficult to predict the impact of changes in the trigger menu on the total rate of the system.

\section{The ATLAS Trigger System}
\label{sec:tdaq_ATLAS_trigger}

\subsection{General Architecture}

\begin{figure}
  \centering
  \incgraphics{width=\widthwideplot}{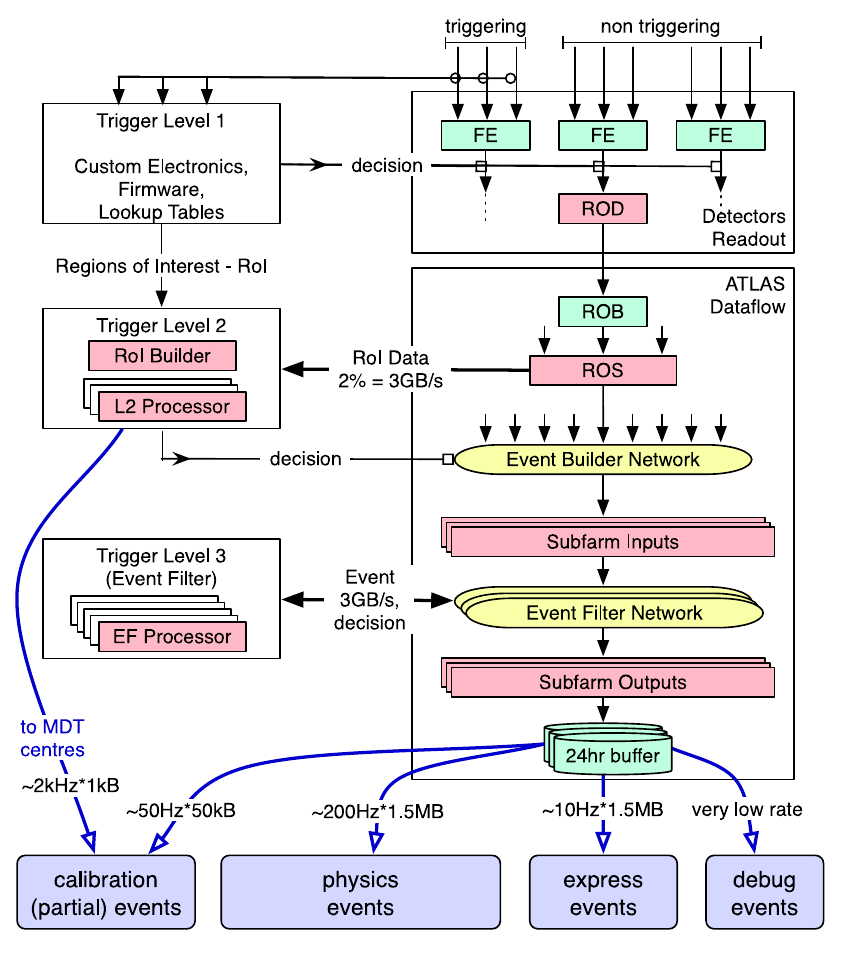}
  \caption{
    Schematic diagram of the \ATLAS trigger and data-acquisition system.
    On the left, the data-processing components are shown,
    on the right, the data-acquisition components.
  }
  \label{fig:tdaq_system_overview_schematic} %
\end{figure}

The \ATLAS detector is equipped with a three-level trigger system \cite{CSCNotes},
filtering events in three subsequent steps,
where each level confirms and refines the decision of the previous level.
The three trigger levels are called \acf{L1}, \acf{L2} and \acf{EF}.
A~schematic diagram of the \ATLAS trigger and data-acquisition system is shown in
\Fig{fig:tdaq_system_overview_schematic}.
The dataflow between the different components
and the propagation of the trigger decisions are indicated by arrows.
Details can be found in \cite{ATL-DAQ-PROC-2010-017}.

\index{Level~1} is implemented as a hardware-based system \cite{CERN-LHCC-98-14},
using fast custom electronics and algorithms which are programmed into the firmware.
This part of the trigger is located underground close to the detector
and only uses coarse information from the calorimeters and the dedicated muon trigger chambers,
to be able to make a decision within an average time of \unit[25]{ns}.
The main element of Level~1 is the \acf{CTP}\inindex{Central Trigger Processor},
which computes the trigger decision,
using as input up to 256 different trigger signal bits from the calorimeter and muon trigger processors. %
In addition, bunch group requirements are applied in order to restrict some of the triggers at Level~1 to only filled or empty bunch crossings.
The maximum latency of Level~1 is \unit[2.5]{$\mu$s}. %
As this time spans up to 100 bunch crossings,
the detector data needs to be kept in pipeline memories
until the decision for the respective collision event has been computed.
Upon a Level~1 Accept signal,
the data is transferred to the detector-specific \acp{ROB},
where it is stored until requested by Level~2 algorithms or by the \index{Event Builder} upon a positive Level~2 decision.
The maximum designed output rate of Level~1 is \unit[75]{kHz}, depending on the trigger menu.
In 2010, the maximum rate was \unit[30]{kHz}.
In addition to the rate reduction,
the main purpose of Level~1 is to define \acp{RoI},
which describe geometrical regions of the detector in \eta-$\phi$ space and are used in the next step by Level~2.

\index{Level~2} and the \index{Event Filter} constitute the \acf{HLT}\inindex{High Level Trigger} which is implemented in software.
It runs on a cluster of commercially available computers and networking hardware
near the detector, but at the surface.
In 2010, the computer farm comprises 1100 nodes in total,
each node having 8 \acfp{CPU}, most of which are running at \unit[2.4]{GHz} \cite{TriggerPerf2010}.
For the design luminosity, the number of nodes will be increased to 500 for L2 and 1800 dedicated to the EF.

Level~2 is seeded by the \acp{RoI} defined at Level~1.
It can read out the detector data with full granularity,
but only in a limited region around the defined \acp{RoI}.
Level~2 can also access the Inner Detector data in addition to the calorimeter and muon spectrometer information,
which is not used before, because no track finding can be done within the time available at Level~1.
In 2010, the output rate of Level~2 was \unit[3]{kHz} and the average processing time \unit[50]{ms} per event.
If an event is accepted by Level~2,
the full detector data is collected into an event record by the Event Builder
and passed to the Event Filter.
At the Event Filter level, it is possible to evaluate the full data in the events
with more sophisticated software algorithms.
These algorithms basically are the same algorithms which are used in the offline processing of data.
The maximum design output rate of the Event Filter is \unit[200]{Hz}.
In 2010, the average processing time was about \unit[0.4]{s} per event.
The maximum time available to the Event Filter to process an event
depends on the number of nodes in the computer farm and the Level~2 output rate.
The trigger configuration of the \ATLAS trigger system is organized in terms of {chains}.\inindex{Trigger chain}
Each chain spans all three trigger levels and specifies the sequence of reconstruction and selection steps at each level.
(Often the terms trigger and trigger chain are used interchangeably.)
The process that controls the execution of the trigger chains
is referred to as the {\index{trigger steering}} \cite{HLTSteering2008}.
To minimize the average time spent on the computation of the trigger decision for every event,
the trigger steering is designed to reject events as early as possible ({\index{early-rejection principle}}).
In order to achieve this,
each chain is processed in a step-wise and seeded manner and all chains are processed in parallel.
Higher levels use the output of lower levels and from previous steps within the chains at the different levels.
To be accepted and stored, an event needs to pass at least one of the trigger chains.

The HLT algorithms which the trigger chains are comprised of
can be separated into two classes with different tasks.
The {\acf{FEx}} algorithms\inindex{Feature-extraction algorithm} perform the time-con\-sum\-ing %
unpacking of data and reconstruction of trigger objects like calorimeter clusters.
These objects are called features in the trigger, hence the name.
The {hypothesis} algorithms\inindex{Hypothesis algorithm} perform selection cuts based on the reconstructed objects,
which is comparably fast.
Therefore, each chain at all three levels typically consists of one or more feature-extraction algorithms,
followed by a hypothesis algorithm which computes the actual decision of the chain at that level.
The caching mechanism of features allows objects which have been reconstructed by one chain
to be reused by all other chains.

\ATLAS uses an inclusive streaming model \cite{ATL-DAQ-PROC-2010-017,EventStreamingATLAS},
where every trigger chain can be associated to one or more \index{data stream}s.
Every event which has been accepted by the trigger is written to all streams
for which at least one of the trigger chains associated with the stream has fired.
There are four primary streams for storing events for physics analyses,
called
\tagname{Egamma} (for photon and electron triggers),
\tagname{Muons} (for muon triggers),
\tagname{JetTauEtmiss} (for jet, tau-lepton and \met triggers)
and \tagname{MinBias} (for minimum-bias and random triggers),
\inindex{Egamma stream@\tagname{Egamma} stream}
\inindex{Muons stream@\tagname{Muons} stream}
\inindex{JetTauEtmiss stream@\tagname{JetTauEtmiss} stream}
\inindex{MinBias stream@\tagname{MinBias} stream}
according to the type of triggers that feed events into the stream.
Muon chains would trigger the event to be written to the \tagname{Muons} stream,
an event which fires, for example, both a muon and an~\met trigger would be copied
to both the \tagname{Muons} and \tagname{JetTauEtmiss} stream.
In addition to writing complete events to a stream,
there are also calibration streams,
to which only partial information from one or more sub-detectors is written.
The inclusive streaming model implies that there will be overlaps
between sets of events from different streams,
which needs to be taken into account when combining them in physics analyses,
and also means that a fraction of the bandwidth is wasted
on writing out the same data redundantly.

In \ATLAS, prescales can also be used to deactivate a given trigger by assigning a negative prescale to it\footnote{
  The absolute value of a negative prescale does not matter,
  but different negative values may be found
  when a trigger is deactivated which is normally prescaled.
  The absolute value then usually corresponds to the prescale the trigger would have if it were active.
}.
In contrast to the composition of the trigger chains or their thresholds,
the prescales can be changed at any time within a data-taking run at the beginning of a new luminosity block.
Primary triggers are usually not prescaled,
whereas the other triggers, in particular those for monitoring and calibration purposes, in general are.
As said above, the prescales can be changed independently from the trigger menu.
In total, three keys are needed to retrieve the configuration of the \ATLAS trigger from the trigger database (cf. \Sec{sec:software_ATLAS_databases}),
the SuperMasterKey (SMK) which selects the trigger menu
and the Level~1 and HLT Prescale Keys (L1PSK and HLTPSK)
which determine the set of prescales that is applied \cite{TriggerConfig2008}.
Besides, there is the Bunch Groups (BGS) Key which specifies the bunch configuration.
The keys are unique integers by which the complete configuration for all recorded runs can be accessed.

Dead time of the trigger, during which the accept signal is vetoed,
is introduced by the \ac{CTP} after every Level~1 Accept %
to allow the read-out systems to finish processing and to return to normal operation,
and %
to protect the front-end read-out buffers from overflowing due to short-scale rate fluctuations (preventive dead time) \cite{TriggerPerf2010}. %
This type of dead time mainly affects the first trigger level. %
At the subsequent levels, the trigger rates can be assumed to be low enough
so that dead time of detector components does not play a role under normal running conditions.
There are other sources of trigger vetoes,
like dead time of the data-acquisition system itself coming from backpressure,
\ie if the trigger rate is higher than the rate at which the events can be processed or written out,
or when the data acquisition is paused by the operator.

To study and understand the trigger performance,
it is important to know how the trigger algorithms work.
Therefore, in the following a summary of the missing transverse energy and jet triggers will be given,
complemented by a brief overview of other important triggers.

\subsection{\texorpdfstring{\met Trigger} {MET Trigger}}
\label{sec:tdaq_met}

The algorithms which compute \met in the trigger \cite{ATL-DAQ-INT-2010-004,ATL-DAQ-PUB-2011-001} run unseeded
because the \met is a global quantity and therefore no \aclp{RoI} can be defined.
In fact, the \met in the trigger is computed for all events at all levels
if any trigger has accepted the event at the respective lower level.
This makes it possible to emulate the trigger decision for all recorded events,
whether or not they have been accepted by an \met trigger chain (cf. \Sec{sec:tdaq_explain_trigger_emulation})
and to study combined triggers $\met + X$.
In order to compute \met (and \sumet), the full detector needs to be read out,
which currently is only possible at Level~1 and at the Event Filter, but not at Level~2.
The \met at Level~2 therefore is the same as \met at Level~1,
plus the muon correction (see below),
which in turn is not possible to include at Level~1.

At Level~1, \met and \sumet are computed from trigger towers,
which are comprised of cells occupying the same area in \eta-$\phi$ space
and spanning all of the different calorimeter samplings (both from \ac{EM} and hadronic calorimeters).
The approximately 7200 trigger towers with a granularity of approximately $0.1\times0.1$ in \eta-$\phi$ space
cover the region up to $|\eta| \leq 4.9$ in pseudorapidity. %
The analog signals from the trigger towers are transmitted to the Level~1 system,
where they are digitised and associated with a bunch crossing and the pedestal subtraction is performed.
The Level~1 system also applies a noise suppression threshold per trigger tower %
and extracts the transverse components. %
Actually, \met is not computed at Level~1,
but instead a look-up table mapped by the vector sums of the components \mex and \mey is used
to check whether any of the defined thresholds have been exceeded.
If the measurement saturates the look-up table,
a flag is set to indicate an overflow
so that the event is passed to Level~2 regardless of the configured \met thresholds.

If any chain accepts the event at Level~1,
the \mex and \mey values computed at Level~1 are passed to Level~2,
where \met is computed as the square root of the sum of the squares of \mex and \mey, %
rather than using a look-up table.
The decision of Level~1 and Level~2 may therefore differ slightly if \met lies close to a trigger threshold.
Additionally, the \pt of muons reconstructed at Level~2 are included in the computation of \met and \sumet at this trigger level,
but both computations, with and without the muon contribution, are stored separately.
The decision of the \met and \sumet triggers at Level~2 and Event Filter, however,
is based on the computation of \met and \sumet without the muon contributions.
Note that muons in general do not deposit much of their energy in the calorimeter,
so that the contributions from muons measured in the muon spectrometer to \met are not negligible.

At Event Filter, a complete recomputation of \met and \sumet is done
by looping over all calorimeter cells,
which gives a better resolution than possible by using the trigger tower results.
Only cells with energies above a given threshold
which corresponds to three times the noise \ac{RMS} of the cell %
are included in the summation.
This one-sided cut leads to a small bias in the \sumet computation of about $10$ to \GeV{20} (cf. \Fig{fig:results_met_model_sumet0}).
For the muon contribution, the result of the Level~2 computation is reused.
All measurements of \met in the trigger are done at the uncalibrated electromagnetic scale.

\subsection{Jet Trigger}
\label{sec:tdaq_jet}

The jet triggers start at Level~1 by constructing jet elements
which are comprised of $2\times 2$ blocks of the trigger-towers mentioned above.
These are then combined into proto-jets using a sliding-window algorithm,
which selects high-energy depositions in a square of programmable size of
$0.4\times0.4$, $0.6\times0.6$ or $0.8\times0.8$ in $\eta\times\phi$.
If the multiplicity and energy of the Level~1 trigger jets
which have been constructed in this way
is sufficient to pass the trigger thresholds,
the geometrical positions of the \aclp{RoI} of the Level~1 jets are passed to Level~2.
The Level~1 jet algorithm identifies jets within $|\eta| < 3.2$.
\revised{Forward jet trigger chains,}
which are independent from the normal trigger chains,
use the forward calorimeter outside this region with a coarser granularity,
but they are not used in this thesis.
The Level~2 algorithm is a cone clustering algorithm,
which runs a limited number of iterations on calorimeter clusters
in a rectangular region centered around the \acp{RoI} identified at Level~1
with higher calorimeter granularity.

The Event Filter basically uses the same algorithms to reconstruct jets
that are also used in the offline processing of data.
Throughout 2010, all of the jet chains have been run with the Event Filter level in pass-through mode,
until at the beginning of 2011, the new trigger algorithms based on \antikt were commissioned.
An important point is that the jet reconstruction algorithms in the trigger at Event Filter level,
starting from 2011, run in an unseeded full-scan mode,
\ie they are executed when Level~2 has accepted the event,
but they do not rely on the \aclp{RoI} or jets identified by the lower chains.
Moreover, another change in 2011 with respect to 2010 is that the jet trigger chains with the lowest thresholds at Event Filter
are seeded by random triggers and not by jet triggers at Level~1 and Level~2,
which can be seen in \Tab{tab:results_trigger_performance_jet_chains_2011} and will be discussed later.
Note that the description of the Level~1 and Level~2 jet algorithms is based on \cite{CSCNotes}
and may differ from the current implementation.

\subsection{Other Types of Triggers}

Besides the jet and \met triggers,
there are many more types of triggers
among the approximately 500 triggers which are defined in the \ATLAS trigger menu.
The ones which are relevant in the context of this thesis are described below.
Other groups of triggers that are not used and therefore not discussed here
are electron, photon, tau lepton and $b$-physics triggers. %

\subsubsection{Muon Triggers}

Muons in \ATLAS can be triggered on within the range $|\eta| < 2.4$.
Two of the four muon subsystems have a timing resolution that is good enough
to be used for triggering, at the price of lower precision.
In the barrel region, $|\eta| < 1.05$, these are the Resistive Plate Chambers (RPC),
and in the endcap region, $1.05 < |\eta| < 2.4$, the Thin Gap Chambers (TGC).
While the trigger coverage in the endcap is close to \percent{100},
in the barrel it is significantly lower because of the lower geometrical acceptance
due to a crack around the mirror symmetry axis of the detector at $\eta = 0$,
the magnet support structure, and two small elevators in the bottom part of the spectrometer \cite{CSCNotes}. %
The Level~1 trigger works by finding hit coincidences between the different muon detector layers.
Six \pt thresholds can be configured using look-up tables.
The Muon CTP Interface (MuCTPI) collects the information from the muon trigger chambers (RPC and TGC) and does overlap removal.
Level~2 starts from the \aclp{RoI} defined at Level~1 and runs three algorithms successively.
The first algorithm fits tracks using the MDT measurements, %
and for the determination of \pt uses look-up tables.
The next algorithm %
combines the previously found tracks from the muon spectrometer with Inner Detector tracks.
In the final step, %
a discrimination between isolated and non-isolated muon candidates from the previous steps is done,
taking into account the energy depositions in the calorimeter using two concentric cones around the muon.
The inner, smaller cone is used to exclude the energy depositions by the muon itself,
and the energy content within the outer cone defines the isolation.
The Event Filter can again basically use the same algorithms as offline,
seeded by muons found at Level~2
and combining reconstructed tracks from the muon spectrometer and from the Inner Detector.

\subsubsection{Minimum-Bias and Random Triggers}
\label{sec:tdaq_minimum_bias_triggers}

Minimum-bias\inindex{Minimum-bias trigger} and \index{random trigger}s are triggers
which are supposed to give an unbiased sample of all possible collision events taking place in the detector.
In particular the random triggers, if they are correctly implemented, give a truely unbiased sample.
They trigger, as can be easily guessed from their name, randomly on any bunch crossing without taking any event properties into account.
There are different types of random triggers,
which either only trigger on crossings of filled bunches,
where proton-proton interactions are expected to take place (\eg \trigger{EF_rd0_filled_NoAlg}), %
or only on empty bunches, where no collisions are expected (\eg \trigger{EF_rd0_empty_NoAlg}).
As the prescales implemented within the trigger steering are sufficient for randomly triggering events,
no additional algorithm needs to be implemented at Level~2 and Event Filter.
At Level~1, the random triggers require the \trigger{RNDM0} item to accept an event,
which in normal running it does at an unprescaled rate of $f_\text{LHC} / 8$,
$f_\text{LHC} \approx \unit[40]{MHz}$ being the beam crossing frequency at the interaction points.
Furthermore, the items BGRP0 and BGRP1 or BGRP3 are required,
which select non-vetoed filled or empty bunch crossings, respectively.

The minimum-bias triggers differ from the random triggers
in that they are based on the output from some detector component and not purely random.
There are many different triggers which are all subsumed as minimum-bias triggers,
having in common that they trigger on events in which some minimum activity of any kind has been detected.
The minimum-bias trigger used in \Sec{sec:results_trigger_performance_measurements_data_2010} is the \trigger{EF_mbMbts_1_eff} chain
which is based on the counts which have been detected in the minimum-bias trigger scintillators (cf. \Sec{sec:atlas_minimum_bias_trigger_scintillators}).

\subsubsection{Combined Triggers}

The single triggers from above can be combined in many ways to give \index{combined trigger}s.
The reconstruction of the trigger objects, which has been described above, stays the same,
only the hypothesis algorithms, which act on the reconstructed trigger objects,
are replaced by new ones which implement a conjunction (logical ``AND'' condition) of several signatures.
The most important example in the context of this thesis
are the combined \jetmet triggers which are studied in \Secs{sec:results_trigger_rates_benefits_combined_triggers} and \ref{sec:triggerefficiencies}. %
They combine requirements on the online measurement of jets and \met in the trigger and exist in many variants.
They have undergone a number of changes between 2010 and 2011 as explained in the sections referenced above.
These changes are partly due to how they combine the jet and \met trigger objects and what thresholds are applied,
but also due to changes in how the underlying trigger objects are reconstructed.

Closely related to the combined triggers are multi-object triggers,
which require a given multiplicity of objects of the same type,
for example $n$ jets above given thresholds,
and special triggers,
which combine existing trigger objects in a more sophisticated fashion.
An example of the latter are $\Delta\phi$ triggers,
which require a certain angular separation between two or more trigger objects that have been reconstructed online.
Special triggers are designed to trigger on very specific physics signatures,
and as such will become more important when first indications for new physics have been found
which allow to narrow down the search,
but for the first years of data taking,
presumeably most of the trigger bandwidth
will be spent on inclusive trigger chains rather than exotic signatures.

\subsection{Trigger Nomenclature}
\label{sec:tdaq_trigger_nomenclature}

In this thesis,
trigger chains will be referred to by giving their official names as they appear in the trigger menu.\inindex{Trigger nomenclature}
These names are composed according to the following conventions.
The general structure of the trigger names is
\begin{center}
  \trigger{(Trigger Level)_(Multiplicity if > 1)(Threshold Type)(Threshold Value)_(Postfix)}
\end{center}
The prefix indicating the trigger level is always L1, L2 or EF, followed by an underscore.
For triggers which require a multiplicity of trigger objects larger than 1,
the multiplicity is then given in front of the threshold type.
The threshold type is an abbreviation stating the type of the trigger.
The most important abbreviations for object triggers are
e for electrons,
g for photons,
mu for muons,
tau for tau leptons,
j for jets and
fj for forward jets.
Triggers which use global quantities have
mb for minimum bias,
rd for random triggers,
te for the sum of the transverse energy,
je for the total energy in jets,
xe for the missing transverse energy and,
introduced in the middle of 2011,
xs for the significance of missing transverse energy (cf. \Sec{sec:results_met_model_predictions_xs_triggers}).
In addition, at Level~1, there is EM,
standing for the item triggering on electron/photon objects,
which cannot be distinguished at L1.
The threshold value is usually given in GeV.
Threshold type and value may appear several times for combined triggers.

The names for \ac{L1} items always consist only of uppercase letters,
whereas for \ac{L2} and \ac{EF} usually lowercase letters are used, but here the postfixes may vary.
Common postfixes specify additional trigger features,
for example isolation, ``filled'' or ``empty'' for the random triggers or
``loose'' or ``tight'' for variants of the same chain differing by the thresholds at lower levels.
Important postfixes of the chains studied here are ``noMu'',
which is common to all configured \met trigger chains in 2010 and 2011 at L2 and EF
and says that no muon contributions are taken \revised{into} account in the trigger decision.
Furthermore, ``jetNoEF'' or ``jetNoCut''
appear in jet and combined \jetmet trigger chains in the 2010 trigger menu,
to make clear that no cuts are \revised{made} at Event Filter level on trigger jets.
``EFFS'' marks jet and combined \jetmet trigger chains which use the full-scan algorithm at EF.

Some examples for illustration are:
\begin{itemize}
  \item \trigger{L1_MU20}
    = L1 muon trigger item, requiring a muon with a \pt of at least \GeV{20}.
  \item \trigger{L2_j45_xe20_noMu}
    = \jetmet trigger at L2, requiring a jet with at least \GeV{45} and \met of at least \GeV{20} excluding muon contributions.
  \item \trigger{EF_xe30_loose_noMu}
    = \met trigger, requiring at least \GeV{30} at EF, which has looser thresholds at L1 and / or L2.
\end{itemize}
In the last example, the difference with respect to \trigger{EF_xe30_noMu}
becomes clear only from the full specification of the chain,
including the thresholds at L1 and L2.
Therefore, in the trigger efficiency studies presented in this thesis,
for the relevant chains the thresholds are listed for all trigger levels explicitly.
Note also that giving a trigger name like ``EF\dots'',
unless otherwise specified,
usually refers to the whole trigger chain,
not only to the cut at Event Filter level.

\section{Event Reconstruction}
\label{sec:software_reconstruction_event_reconstruction}

In this section, the reconstruction of events from the raw detector output is described.
The \index{event reconstruction} is the act of turning the pattern of signals from the detector into physics objects,
thereby reducing the large amounts of raw data into a form which is more suitable for physics analysis.
Due to the large dimensions of the \ATLAS detector and the high interaction rate,
combined with the finite velocity of the particles produced in the proton-proton collisions,
reconstructed particles which are detected at the same time but at different distances from the collision point at the center of the detector
may have been produced in different collisions.
This needs to be taken into account when reconstructing event records from the detector output.
The same reconstruction algorithms run on real data and simulated Monte Carlo pseudodata,
which is both a test for the reconstruction algorithms
as well as a device to study the signatures left in the detector by specific particles or reactions
and thereby to develop and refine physics analyses.

The reconstruction is done centrally in a post-processing step of the data which has been stored during data taking
and uses algorithms which are implemented within the \Athena framework.
In the following, the reconstruction of the different types of physics objects will be discussed separately.
The physics objects relevant in the context of this thesis
with respect to the trigger studies and the search for Supersymmetry
are leptons, in particular electrons and muons, jets,
the missing transverse energy and sum of transverse energy,
and the primary vertices.
Not discussed are \eg photons or the tagging of jets arising from bottom quarks.
At the end of the section, also the measurement of the luminosity
and the computation of derived quantities
which are used in the search for Supersymmetry such as the effective mass are described.
There are two important figures of merit for the quality of the reconstruction of a certain type of physics object.
The probability that a given object,
which has been produced in a particle collision or subsequent decays,
is actually reconstructed is called the efficiency of the reconstruction\inindex{Reconstruction efficiency}.
In the common nomenclature,
the \index{fake rate} gives the fraction in which a certain object is reconstructed from the detector data
where in reality no such object has been.
The \index{purity} gives the fraction of correctly identified objects in all reconstructed objects.

The description of the reconstruction algorithms is limited to the information relevant for the scope of this thesis.
Like for the \ATLAS detector itself, a detailed documentation can be found in~\cite{ATLASTDR1999} and~\cite{CSCNotes}.
Note that the reconstruction algorithms are subject to a constant process of evolution and refinement,
and therefore the description here may differ from the algorithms in the current release of the \Athena software.

\subsection{Leptons}

Charged leptons from the first two generations, electrons and muons, are relatively easy to identify,
because the muons in particular leave a unique signature in the detector.
Identification of tau leptons is more complicated
because due to their high mass they have a lot of different possibilities to decay
and their lifetime is so short that they will decay close to the interactions point even at large relativistic factors.
Neutrinos cannot be reconstructed as individual objects and only be detected by the energy they carry away,
which becomes apparent as a violation of the conservation of momentum in the transverse plane in terms of a non-zero value of \met.

\subsubsection{Electrons (and Photons)}
\label{sec:software_electron_reconstruction}
The reconstruction of electrons can be subdivided into two tasks:
The reconstruction of electron candidates and the subsequent identification of electrons by rejecting fake electrons.
One of the challenges when reconstructing electrons from clustered energy deposits in the calorimeter
is to decide whether the cluster is due to an electron,
which would leave a track in the Inner Detector due to its electric charge,
or a photon.
(In fact, in the trigger such a distinction is not even attempted nor possible at Level~1,
thus there are only \code{egamma} objects in the L1 trigger.)

To reconstruct \acf{EM} clusters,
a sliding-window algorithm is used,
which builds rectangular clusters with a fixed size
which are positioned so that the amount of energy within the cluster is maximised.
Alternatively, clusters can be formed by connecting neighbouring cells
until the cell energy falls below a threshold as is done for jets,
but this is not used by the default electron reconstruction algorithms.
The optimal cluster size depends on the calorimeter region and is different for electrons and photons,
thus different collections of clusters are reconstructed with different window sizes.
The calibrated cluster energy is the sum of four contributions:
The energy deposited in the material in front of the EM calorimeter,
in the calorimeter inside the cluster,
outside the cluster (lateral leakage) and behind the EM calorimeter (longitudinal leakage).

An electron is reconstructed from the sliding-window clusters
if it matches a track with $\pt > \GeV{0.5}$.
If several tracks are available,
the one lying closest in \eta-$\phi$ space to the (weighted) center of the cluster %
is chosen. %
If there is no track or if there is a reconstructed conversion vertex matching the cluster,
a photon is reconstructed.
Electron candidates with $|\eta| > 2.5$ are called forward electrons.
They lie outside the range of the tracking systems and therefore need to be reconstructed using an alternative algorithm.
Forward electrons are not used in the following.
To give a feeling for the reconstruction efficiencies and fake rates for electrons,
about \percent{0.9} of electrons with $E_T > \GeV{20}$ and $|\eta| < 2.5$ from $Z \to ee$ decays are not reconstructed at all
and \percent{2.1} are reconstructed as photons~\cite{ATLAS-CONF-2010-005}.
The other way round \percent{2.1} of photons from Higgs decays $H\to\gamma\gamma$ are reconstructed as electrons.

Actually, three algorithms are run in parallel to reconstruct electrons.
The default algorithm,
which is seeded by the EM clusters,
is dedicated to high \pt isolated electrons.
For low \pt electrons and electrons in jets a second algorithm is available,
which is seeded by tracks in the Inner Detector.
The third algorithm reconstructs forward electrons.
For the analysis in this thesis,
electrons reconstructed by only the cluster-based \code{AuthorElectron} algorithm starting from calorimeter seeds
or by both the track-based \code{AuthorSofte} algorithm and the \code{AuthorElectron} algorithm are used (cf. \Sec{sec:analysis_object_definitions_electrons}). %
An overlap removal to avoid that the same electron enters the collection of reconstructed electrons multiply is included in \Athena. %
The identification of electrons (and photons) in \ATLAS relies on independent rectangular cuts on variables
which have been found to give a good separation between isolated electrons (or photons) and fake signatures from QCD background.
These variables include information from the calorimeter and, in the case of electrons, the tracker,
and different sets of optimized cuts are provided by the Electron/Gamma combined performance group~\cite{ATLAS-CONF-2010-005,CSCNotes}.
In addition to the identification of electrons, which aims at rejecting fake electrons,
a procedure has been \revised{established} to reject electrons the reconstruction of which will suffer from detector problems
such as regions in the calorimeter with dead power supplies, faulty read-out electronics
(\eg Front End Boards with dead optical links~\cite{ATL-LARG-PROC-2011-010}) %
or channels with high noise.
It relies on so-called Object Quality maps,
which for a certain range of data-taking runs specify geometrical regions of the detector
which were affected by known detector problems,
so that electrons reconstructed in these regions can be treated appropriately in the physics analysis.

\subsubsection{Muons}
\label{sec:software_muon_reconstruction}

There is a number of different algorithms to reconstruct muons
from Inner Detector, muon spectrometer and calorimeter information~\cite{ATL-PHYS-PROC-2009-113,ATLAS-CONF-2010-064}.
Depending on the method which is used to identify and reconstruct them,
four kinds of muon candidates are distinguished:
The trajectory of \define{standalone muons} is reconstructed from the muon spectrometer only,
correcting the measured momentum for energy loss in the calorimeter
and extrapolating the track back to the interaction region.
\define{Combined muons} are muons which are identified by successfully combining
a full Inner Detector track and a full track reconstructed in the muon spectrometer.
The trajectory in the Inner Detector is used to determine
the impact parameter of the muon trajectory with respect to the primary vertex.
The other two types start from trajectories in the Inner Detector.
If the trajectory in the Inner Detector can be associated with an energy deposition in the calorimeters
that is compatible with the hypothesis of a minimum-ionizing particle,
it is used to reconstructed a \define{calorimeter-tagged muon}.
If its extrapolation to the muon spectrometer can be associated with a straight-track segment there,
it is used to reconstructed a \define{segment-tagged muon}.

There are two muon algorithm chains in \ATLAS, which are both considered suitable for physics analyses, %
and each is comprised of several algorithms covering all of the above strategies using different pattern recognition strategies.
They are called \tagname{MuID} and \tagname{Staco} after the name of their respective combined reconstruction algorithms. %
In this thesis, muons from the muon collection created by the \tagname{Staco} chain are used.\inindex{Staco muons@\tagname{Staco} muons}
Three different algorithms contribute to this collection:
\tagname{Staco} attempts to statistically merge the Inner Detector tracks with tracks in the muon spectrometer to produce combined muons.
\tagname{Muonboy} is a segment-finding code,
building a track starting from the outer and middle stations in the barrel and inner and middle \ac{MDT} stations in the end-cap, %
\ie where second coordinate chambers are available, then iteratively adding segments in the other layers
until the full track is obtained.
\tagname{Mutag} starts with Inner Detector tracks above a \GeV{3} cutoff and attempts to associate them with \tagname{Muonboy} segments, %
based on the $\eta - \phi$ location and angle where possible. %
It is used to complement \tagname{Staco} in difficult areas and produces segment-tagged muons.

Combined muons are the baseline recommendation~\cite{URL_MuonRecoRel15}. %
To recover efficiency at $|\eta| \sim 0$ and $|\eta| \sim 1.2$ and at low $\pt < \GeV{6}$,
in addition to this,
segment-tagged muons can be used in physics analyses.
Isolation is an efficient way to discriminate between primary muons,
\ie muons which have been produced in the hard \revised{scattering} process at the interaction point,
and muons which have been produced in a jet.
To this aim, the energy deposition in the calorimeter within a cone around the muons (excluding the energy deposited by the muon itself),
the number of tracks or the sum of momenta of reconstructed objects within such a cone can be used.
In studies with cosmic rays~\cite{CosmicRays2011}, a resolution of the relative transverse momentum
of about \percent{2} at $\pt = \GeV{10}$ and \percent{20} at $\pt = \TeV{1}$ has been measured.
The measurement also shows that the momentum resolution of Inner Detector and muon spectrometer complement each other,
because the momentum resolution in the muon spectrometer for small \pt is worse
due to uncertainties on the energy loss corrections associated with the extrapolation of the muon spectrometer track parameters to the perigee,
so that in this range the Inner Detector gives a more precise measurement of the momentum,
which improves the performance of combined muons compared to standalone muons.
Also, first performance studies on data show that the resolution of the fractional momentum resolution
being better than \percent{10} at $\pt = \TeV{1}$ expected from the simulation is not yet fully achieved~\cite{ATLAS-CONF-2011-046},
the fractional resolution in data being worse by a factor of about $1.5$ compared to the simulation. %
The \ATLAS Muon Combined Performance group therefore suggests to smear the resolution of muons in events from Monte Carlo simulations
to achieve a better agreement with the reconstruction of real data,
but this is not yet implemented in the analysis of 2010 data presented in this thesis.
The reconstruction efficiency has been found to be above \percent{98} for muons from $J/\psi$ decays with $\pt > \GeV{6}$~\cite{ATLAS-CONF-2011-021}. %

\subsubsection{Taus}
Tau leptons are much heavier than electron and muon, have a short lifetime and decay mostly hadronically,
which makes their identification more difficult than that of the other charged leptons.
If they decay leptonically, they are difficult to distinguish
from the prompt production of electrons or muons,
therefore often only hadronic tau decays are considered.
Tau leptons are produced \eg in decays of $W$ or $Z$ bosons,
but are also expected to appear in the decay of neutral Standard Model or MSSM Higgs bosons~\cite{ATLAS-CONF-2011-132} if these particles exist.
Their reconstruction is done starting from jets with nearby tracks as candidates,
with a subsequent identification step to distinguish tau from quark jets.
In the zero-lepton analysis presented in this thesis, tau identification is not used.

\subsection{Jets}
\label{sec:software_jet_reconstruction}

Jets can in principle be reconstructed at three different levels.
From data, jets are always reconstructed as calorimeter jets (detector level),
\ie the jet constituents are groups of calorimeter cells with energy deposits induced by particles.
In Monte Carlo, also truth jets can be reconstructed,
which are either built from the final stable particles (particle level) in the Monte Carlo event generation,
excluding neutrinos and usually also muons,
or from the partons in the final state of the hard \revised{scattering} process (parton level).
This is not directly possible in data,
because the information about the original partons
gets washed out by the process of hadronization.
In the following, only calorimeter jets will be considered.

Different inputs can be used to reconstruct jets.
The two main concepts are calorimeter towers and topoclusters~\cite{ATL-LARG-PUB-2008-002}.
Topological energy clusters or short topoclusters
are three-dimensional objects designed to follow the shower development,
taking advantage of the calorimeter segmentation of the \ATLAS detector.
Topoclusters are built around seeds,
which are calorimeter cells with a signal-to-noise ratio above a threshold of~$4$,
$|E_\text{cell}| > 4\sigma_\text{cell}$, 
where $\sigma_\text{cell}$ refers to the \ac{RMS} of the energy distribution for random events 
and is dependent on both the sampling layer in which the cell resides and the position along the calorimeter in~$\eta$~\cite{ATLAS-CONF-2010-053}.
In the next step, neighboring cells are iteratively included
if they have a signal-to-noise ratio of at least~$2$.
In the final step, one additional surrounding layer of all nearest-neighbor cells is added to the cluster,
regardless of their signal-to-noise ratio.
This approach was shown to improve energy resolution in single pion test beam studies. %
It may result in clusters covering large areas of the detector if sufficient energy
is present between incident particles.
Therefore, in a following step a cluster splitting algorithm
splits individual clusters based on local maxima in terms of energy content.
Energy from cells may then be shared between some of the new clusters~\cite{ATL-LARG-PUB-2008-002}.
Due to the subtraction of a %
pedestal term %
from the measured energy, some clusters may exhibit a negative energy.
These clusters are rejected from the jet reconstruction.
The other possible input to jet reconstruction algorithms are calorimeter towers.
They use a fixed geometrical grid of calorimeter cells with size $\Delta\eta \times \Delta \phi = 0.1 \times 0.1$,
where a tower is created from the energy in all cells within one grid cell in radial direction.
Noise-suppressed towers use only cells belonging to topological clusters,
thus applying the same type of noise suppression.
Also, towers used as inputs for jet reconstruction are required to have an energy above zero. %
\removesection{
  Another type of towers are simple (or ghost) towers which do not use noise suppression,
  but fulfill the requirement of non-negative energies by
  their energy being set to a very small positive value (\GeV{$10^{-6}$}) if its negative.
}

\subsubsection{Energy scale}
\label{sec:software_jet_energy_scale}
\inindex{Jet-energy scale}
The baseline measurement of the energy deposited by a jet is done at the electromagnetic energy scale (EM scale),
which is the energy scale measured directly by the calorimeters,
\ie the scale or calibration established using test-beam measurements
for electrons and muons in the electromagnetic and hadronic calorimeters.
This is the correct scale for the energy of photons and electrons,
but does not account for detector effects.
As explained in \Sec{sec:calorimeters_in_general},
there is a fundamental difference
between hadronic and electromagnetic showers for non-compensating calorimeters
so that the calorimeter signal from hadrons is lower than that from electrons.
Other effects, which concern both hadronic and electromagnetic jets in the same way, are:
\begin{itemize}
  \item Dead material: \inindex{Dead material} %
    Energy which is deposited by particles from the jet in inactive regions of the detector
    is simply lost because it is not measured.
    These uninstrumented regions of the detector are called dead material.
  \item Leakage: If particles from a jet escape from the calorimeter (\eg a ``punch-through''),
    the energy of the jet is also underestimated, because
    its energy is not completely absorbed within the calorimeter.
  \item Out-of-acceptence particles: Jets at high \eta may lose energy
    because they are partly out of the region covered by the calorimeter.
  \item Out-of-cone particles:
    Jets have a fixed size set by the reconstruction algorithm.
    Particles outside this cone are not included in the energy sum.
  \item Differences between truth and calorimeter jets: %
    Particles may fall out of the reconstructed calorimeter jet, but be included in the truth jet.
    This may lead to a mismatch between truth jets and reconstructed jets that affects Monte Carlo based calibrations. %
\end{itemize}
To obtain a better resolution in the jet energy measurement,
a \acf{JES} calibration is applied to the measurement at the \ac{EM} scale.
Different calibration schemes are available,
all aiming at correcting the energy and momentum of the jets measured in the calorimeter.
\inindex{EMJES calibration@\EMJES calibration}
The calibration that will be used in the analysis presented in this thesis (cf. \Sec{sec:analysis_susysearch}) %
to convert the energy measurements in the calorimeters
from electromagnetic calibration (EM scale) to the calibrated hadronic scale %
is a simple \pt- and $\eta$-dependent calibration scheme called \EMJES calibration.
It corrects for the non-linear correlation between the energy reconstructed in the
calorimeter and the energy of the particles forming jets,
and can be calculated from Monte Carlo simulations 
or from data using $\gamma$~+~jet and dijet balance techniques~\cite{ATLAS-CONF-2010-053}.
The current calibration is based on Monte Carlo simulations
and has been validated with collision data~\cite{ATLAS-CONF-2010-056}.
It is also known as jet numerical inversion correction~\cite{ATL-PHYS-INT-2011-009}.
Other calibration schemes developed in \ATLAS are
the Global Sequential (GS), Global Cell energy-density Weighting (GCW),
and Local Cluster Weighting (LCW calibration) calibration scheme (see below). %
These use additional information such as the longitudinal and transverse properties of the jet structure (GW)
or weights that compensate for differences between hadronic and electromagnetic jet responses and energy losses in dead material (GCW and LCW)
in order to reduce fluctuations in the jet energy measurement,
thereby improving the jet resolution between \percent{10} and \percent{30}~\cite{ATL-COM-PHYS-2011-240}.

\subsubsection{Resolution}
\label{sec:jet_resolution}
\inindex{Jet-energy resolution}
For the computation of the jet resolution,
two different methods have been applied to data and Monte Carlo in~\cite{ATLAS-CONF-2010-054},
using the dijet balance~\cite{Dijet2001} and the bi-sector techniques~\cite{Bisector1984}.
Both methods give consistent results,
and the resolution on Monte Carlo is found to agree with the resolution on data within \percent{14}
for jets with $\GeV{20} < \pt < \GeV{80}$ and  rapidities $|y|<2.8$.
In the updated version,
a better agreement within \percent{10} of the resolution in Monte Carlo and data
for jets with $\GeV{30} < \pt < \GeV{500}$ and  rapidities $|y|<2.8$
is stated~\cite{ATL-COM-PHYS-2011-240}, %
the resolution in data being higher than in Monte Carlo by an average \percent{4},
but consistent within the systematic uncertainties.
The parametrization used to fit the relative momentum resolution is
the same functional form as typically used for the parametrization of the calorimeter resolution (cf. \Eq{eq:experimental_setup_general_calorimeter_resolution_parametrization}).
The uncertainty on the resolution of jets as well as the jet-energy scale uncertainty
are input to physics analyses in terms of systematic uncertainties.

\subsubsection{Jet Algorithms}
\inindex{anti-kt@\Antikt algorithm}
The most important algorithm for jet reconstruction in \ATLAS is the \antikt algorithm~\cite{Cacciari2008},
which has superseded the previously used Cone algorithm~\cite{CSCNotes} as default algorithm in \ATLAS, %
starting with 2011 data taking also in the trigger at Event Filter level\footnote{
  At Level~1 and~2 of the trigger much simpler algorithms are used, cf. \Sec{sec:tdaq_jet}.
}.
Here, only the \antikt algorithm will be considered.

Several theoretical guidelines can be formulated
to make the reconstructed jets meaningful with respect to the underlying scattering process
in which the particles were produced which deposit energy in the calorimeter. %
According to these, a good jet algorithm should be \index{infrared safe} and \index{collinear safe},
\ie additional soft particles should not disturb the correct reconstruction of the jet
and the jet should be reconstructed independent of
whether a certain amount of transverse momentum is carried by one particle
or this particle is split into two collinear particles.
Finally, the same hard scattering should be reconstructed independently at parton, particle and detector level.
Moreover, the jet reconstruction should be independent of detector properties,
stable against changes in the rest of the event and fast with respect to its implementation.

Algorithms for the reconstruction of jets are typically iterative procedures,
which start from seeds that can be partons, particles %
or reconstructed detector objects with four-mo\-men\-tum representations such as calorimeter clusters,
and then repeat the same step a number of times,
iteratively combining seeds to form larger entities,
until a predefined stopping criterion is met and a jet has been formed.
The \antikt algorithm belongs to a more general class of sequential recombination algorithms
that differ in the exponent $p$ of the energy scale in the distance measure between two entities $i$ and $j$,
\begin{equation}
  d_{ij} = \min\{k_{ti}^{2p}, k_{tj}^{2p} \} \frac{\Delta^2_{ij}}{R^2},
\end{equation}
which is used to determine which entities are to be merged next,
and is a function of the radius parameter $R$, the transverse momentum $k_t$ and the geometrical distance $\Delta^2_{ij}$.
For $p=1$ this gives the inclusive $k_t$ algorithm, $p=0$ leads to the inclusive Cambridge~/~Aachen algorithm,
and $p=-1$ gives the \antikt algorithm introduced and discussed in~\cite{Cacciari2008}.
The advantage of the \antikt algorithm is that it shares the positive features of certain iterative cone algorithms,
in particular that the reconstructed jets have a regular shape and their boundaries are resilient with respect to soft radiation,
but without suffering from collinear unsafety found for iterative cone algorithms.

\subsubsection{Relevant Jet Collections}
The jets which are used in this thesis are taken from the jet collections \tagname{AntiKt4H1Topo} and \tagname{AntiKt4TopoNewEM},
which contain jets reconstructed with the \antikt algorithm with distance parameter $R=0.4$ in \eta-$\phi$ space, %
using topological clusters as inputs.
The energy of the jets is calibrated by applying a jet-energy scale factor
to the energy measurement at the electromagnetic scale,
which depends on the transverse jet momentum and the rapidity (\EMJES scale calibration).
An implicit minimum \pt of around \GeV{7} at the GCW calibrated scale is induced by the reconstruction algorithm
(cf. \Secs{sec:results_trigger_performance_dataselection_2010} and \ref{sec:appendix_MC_study_QCD_jets}),
and only central offline jets with $|\eta|<2.8$ (or smaller) are used here.

\subsection{Missing Transverse Energy and Sum of Transverse Energy}
\label{sec:software_reconstruction_met}

The missing transverse energy in \ATLAS
is calculated from the measurements of energy deposits in the calorimeter,
supplemented by additional information from other detector components, %
in particular from the measurement of muons in the muon spectro\-me\-ters~\cite{ATLAS-CONF-2010-057,METPerformance2010}.
The missing transverse energy basically is the sum of two terms from the calorimeter and the muon contribution,
\begin{equation}
  \me{x,y} = \metcomponent{calo} + \metcomponent{muon}.
\end{equation}
The two components in the transverse plane are obtained as the sum of the projected energies $E_i$ over all $N_\text{cells}$ cells in the range $|\eta| < 4.5$,
\begin{align}
  \me{x} &= - \sum_{i=1}^{N_\text{cells}} E_i \sin\theta_i\cos\phi_i, \\
  \me{y} &= - \sum_{i=1}^{N_\text{cells}} E_i \sin\theta_i\sin\phi_i,
\end{align}
where $\theta_i$ and $\phi_i$ are the polar and azimuthal angle.
The sum of the transverse energy \sumet is computed via
\begin{equation}
  \sumet = \sum_{i=1}^{N_\text{cells}} E_i \sin\theta_i,
\end{equation}
using the same calibration scheme which is described for \met in the following.

The \ATLAS calorimeters are non-compensating,
meaning that the response to hadrons is lower than the response to electromagnetic energy deposits.
Energy is also lost in dead material in front of and between the calorimeters,
so that dedicated calibration \revised{schemes} for the hadronic energy deposits in calorimeters need to be developed
which include corrections to account for the different calorimeter response in order to improve the resolution of the \met measurement.
The muon term can be computed from muons reconstructed with different algorithms in two different ways, 
either from the momentum reconstructed by the external spectrometer only
or from the combined measurement of the momentum in the spectrometer and the Inner Detector,
where then the energy deposited by the muon in the calorimeters has to be subtracted to avoid double counting.

The high granularity of the calorimeter necessitates to apply a noise suppression
which reduces the number of cells which are used in the computation.
This electronic noise suppression can be done either by applying an energy threshold to all cells,
or by using only cells within topological clusters constructed as described for the jet reconstruction above. %
The cell energies can be calibrated in three different ways: %
In the global cell energy-density weighting calibration scheme (global calibration or GCW), %
the different calorimeter response to hadronic and electromagnetic energy deposits
and the energy losses in dead material
are compensated for by applying cell-level signal weights obtained from Monte Carlo simulations.
In the local cluster weighting calibration scheme (local hadronic calibration or LCW),
properties of topological clusters are used to calibrate them individually
based on their classification as being of electromagnetic or hadronic nature,
using weights which are determined from Monte Carlo simulations of charged and neutral pions.
These weights also account for lost energy.

The third way is the refined calibration,
which applies different weights to calorimeter cells
based on reconstructed physics objects to which the cells can be associated.
As the response of the calorimeter differs between physics objects,
knowing which object is reponsible for the energy deposited in a given calorimeter cell
allows to improve the \met reconstruction
because an individual calibration for a particular physics object naturally gives
a more precise description and thus a better accuracy
than it is possible with only the averaged global calibration.
To find the mapping between calorimeter cells and physics objects,
calorimeter cells are associated with reconstructed and identified high-\pt objects in the following order:
electrons, photons, hadronically decaying tau leptons, jets, muons.
Navigating back from the reconstructed object to its component clusters and back again to the constituent calorimeter cells,
the calibration weights depending on the type of the reconstructed object 
are used to replace to initial global calibration of the associated cells.
If a cell belongs to several types of reconstructed objects,
only the first association is used, avoiding double counting of cells.
Cells which cannot be associated with any reconstructed object
are included in the \met calculation using the initial global calibration scheme. %

The full composition of the refined \met is given by the following sum:
\begin{equation}
  \metcomponent{calo} =
    \metcomponent{$e$} +
    \metcomponent{$\gamma$} +
    \metcomponent{$\tau$} +
    \metcomponent{jets} +
    \metcomponent{soft jets} +
    \metcomponent{calo,$\mu$} +
    \metcomponent{CellOut}
\end{equation}
Each term is calibrated independently~\cite{METPerformance2010} and computed as the negative sum of the calibrated cluster energies inside the corresponding objects.
The upper index specifies the type of object:
electrons, photons, tau jets from hadronically decaying tau leptons,
jets with calibrated $\pt > \GeV{20}$ and
soft jets with $\GeV{7} < \pt < \GeV{20}$.
$\metcomponent{calo,$\mu$}$ is the contribution from energy lost by muons in the calorimeter
and the $\metcomponent{CellOut}$ term is calculated from the cells in topological clusters
which are not included in any reconstructed objects.
Different choices for the calibrations and selections of the various terms can be done and adopted to the needs of the analysis.
A large number of different ``flavors'' (computation schemes) of the \met calibration are thus available.
In the following, the flavors that are used in this thesis will be further explained.

\tagname{MET\_Topo}
\inindex{MET_Topo@\tagname{MET\_Topo}}
is used as offline reference for the computation of trigger efficiencies in data 2010 in \Sec{sec:results_trigger_performance_measurements_data_2010}.
It is a very simple computation of \met,
not using the refined calibration described above,
but the uncalibrated \met from cells in topological clusters~\cite{ATLAS-CONF-2010-039} %
and comparable to what is done in the trigger (cf. \Sec{sec:tdaq_met}),
which makes it a useful offline reference for trigger studies.
For the trigger studies in 2011 data in \Sec{sec:results_trigger_performance_measurements_data_2011},
it has been replaced by \tagname{MET\_LocHadTopo}\inindex{MET_LocHadTopo@\tagname{MET\_LocHadTopo}},
which is also computed from topological clusters, but applying the local hadronic calibration.
The missing transverse energy that is used in the analysis presented in \Sec{sec:analysis_susysearch},
and also in the trigger efficiency studies on 2010 data,
is the \met defined by the \ATLAS Supersymmetry group based on a simplified refined calibration,
which is called \tagname{MET\_EMJES\_Ref\-Final\_CellOutEM}~\cite{ATL-PHYS-INT-2011-009}.
\inindex{MET_EMJES_RefFinal_CellOutEM@\tagname{MET\_EMJES\_RefFinal\_CellOutEM}}
It is a simplified version that differs from the full refined \met shown above
in that only for a subset of the objects listed above refined calibrations are used.
Contributions that do not enter this version of the refined \met are those %
from photons and tau leptons. %
As cells are not available in the \ac{AOD} file format, from which the SUSY \met\footnote{
  To avoid repeating the long name of the \met variant defined by the \ATLAS Supersymmetry group,
  it will be denoted ``SUSY \met'' in the following.
} is computed,
topological clusters (\tagname{TopoClusters} and \tagname{egClusters}) are used instead.
All electrons and muons that pass a given selection are included in the \met,
applying no additional overlap removal.
(As said above there is an overlap removal done at cell level when computing the refined \met.)
All electrons passing medium purity criteria are used, but as in the analysis an updated, different set of purity criteria 
(RobustMedium, cf. \Sec{sec:analysis_object_identification_and_overlap_removal}) is used,
this will necessitate to correct the \met for the different definition of electrons.
Contributions from all jets above \GeV{20} at \EMJES scale are included without any cut on~\eta.
Jets below \GeV{20} enter the \met via the CellOut term at electromagnetic scale.
For the trigger studies in 2011 data,
the updated definition of \met used by the SUSY group~\cite{ATL-PHYS-INT-2011-085} is called \tagname{MET\_Simplified20\_RefFinal}.
\inindex{MET_Simplified20_RefFinal@\tagname{MET\_Simplified20\_RefFinal}}
It is basically identical to the previous SUSY \met computation. %

Note that the names like \tagname{MET\_Topo} are names of data structures (containers),
which indicate the calibration or reconstruction variant of the measurements of the energy.
These containers do not only contain the \met values,
but \eg also its components, \me{x} and \me{y}, and the sum of the transverse energy.
Nevertheless, if it is not specified which quantity is used,
the use of the container name usually refers to the \met measurement.
When using \met as offline reference, \eg in trigger efficiency plots,
the offline \met variable is written as \eg MET\_Topo\_et in this thesis,
which refers to the missing transverse energy values taken from the \tagname{MET\_Topo} container.

\subsection{Primary Vertices}
\label{sec:software_vertex_reconstruction}

The reconstruction of the primary vertex or vertices is part of the track reconstruction~\cite{ATLAS-CONF-2010-027, PVReco2008}.
It is divided in two steps, although these steps often may be mixed in the actual implementation of vertex finding algorithms.
In the first step, the primary vertex finding, reconstructed tracks are associated to a particular vertex candidate.
In the second step, the vertex fitting, the actual vertex position is reconstructed and its quality estimated.
Candidates for primary vertices can for example be created from a pre-selection of tracks with a given minimum \pt,
which are compatible with the expected bunch-crossing region, and grouping these into clusters.
These clusters are then cleaned iteratively for outlying tracks based \eg on the $\chi^2$ of the fit of the track and the vertex.

\subsection{Effective and Stransverse Mass}
\label{sec:reconstruction_effective_contransverse_mass}

Two additional global event variables are employed in the analysis of \ATLAS data
in the search for Supersymmetry presented in this thesis to define the signal regions:
the effective mass \meff and the ''stransverse`` mass \mttwo,
which is a generalization of the transverse mass to pair decays.

The \index{effective mass} is a straightforward sum of energies and momenta
obtained from the reconstructed objects passing the baseline selection,
\begin{equation}
  \meff \definedas \met + \sum_i p_T^{(i)},
  \label{eq:reconstruction_definition_meff}
\end{equation}
where the sum with the symbolical index $i$ usually includes a given number of leading jets and all identified leptons,
but the concrete implementation may vary from analysis to analysis.
In the case of the zero-lepton channel which is relevant for this thesis,
no leptons are expected in the selected events and the sum runs only over the $n$ leading jets ($n=2,\,3$) defining the analysis channel.
The second term in \Eq{eq:reconstruction_definition_meff} is sometimes denoted $H_T$.
The effective mass provides an estimate of the overall energy scale of the event at hand
and can be used to discriminate Standard Model and supersymmetric events.
Selecting high \meff values will prefer events
in which heavy particles are produced and stand at the beginning of the decay chain,
which aims at giving sensitivity to events in which the final state of the proton-proton interaction
includes heavy supersymmetric particles.

The \index{transverse mass}, which was used for the $W$ boson discovery, is defined as~\cite{Barr2009}
\begin{equation}
  m_T^2 \definedas m_v^2 + m_i^2 + 2 \left( e_v e_i - \vec v_T \cdot \vec q_T \right),
  \label{eq:reconstruction_definition_mt}
\end{equation}
with the masses $m_v$ and $m_i$ and the transverse momenta $\vec v_T$ and $\vec q_T$ of the visible and invisible particles.
The transverse energies are defined to be
\begin{equation}
  e_v^2 \definedas m_v^2 + \vec v_T^2
  \quad \text{and} \quad
  e_i^2 \definedas m_i^2 + \vec q_T^2.
\end{equation}
The transverse momentum of the invisible particle is assumed to be equal to the missing transverse momentum %
in the event.
In decays of a single parent particle with mass $m_0$ to a visible and an invisible daughter like in decays of $W$ bosons,
it holds $m_T \leq m_0$, and the maximum value of the continuous spectrum of $m_T$ over many events can be taken as an estimation for the parent particle mass.

While the transverse mass $m_T$ is a function of the momentum of one visible particle and the missing transverse momentum,
the \index{stransverse mass} is a function of the momenta of two visible particles and the missing transverse momentum,
defined in the following way~\cite{Lester1999}:
\begin{equation}
  \mttwo^2 \definedas
           \min_{\sum \vec q_T = \miss{\vec p}_T}  \left[
           \max \left\{
             m_T(1), m_T(2)
           \right\}
           \right],
  \label{eq:reconstruction_definition_mt2}
\end{equation}
where $m_T(n)$ is the transverse mass computed from the momenta of the first and second visible particle and the hypothesized unobserved momenta, respectively.
Taking the minimum accounts for the fact that the transverse momenta of the invisible particles is not known individually.
Only their sum $\sum \vec q_T = \vec q_T(1) + \vec q_T(2)$ is constrained by the measurement of the missing transverse momentum $\miss{\vec p}_T$.
In events where two parent particles of mass $m_0$ decay,
\mttwo gives the lowest upper bound on $m_0$ consistent with the observed momenta and the hypothesized daughter masses.
The maximum of \mttwo over events can again be taken as an estimate of the mass of the parent particle.
Studies which use a cut on \mttwo have shown
that it is a good compromise between a combination of single-variable cuts with least computational effort but worst signal efficiency,
and computing the likelihood for all events for every signal and background hypothesis which is computationally very challenging~\cite{Barr2009}.
It can be used to suppress both physics and detector background~\cite{ATL-PHYS-INT-2011-009}. %
In that, its application is, of course, not limited to searches for signals from supersymmetric processes. %
The actual computation of $\mttwo$ in the analysis code is done using the \tagname{Basic\_Mt2\_332\_Calculator}
class from the ''Oxbridge MT2 / Stransverse Mass Library``~\cite{Lester1999,Barr2003,URL_mttwo}.

\subsection{Runs and Good Runs Lists}
\label{sec:software_good_runs_lists}
\label{sec:analysis_explain_grl}

The chronological subdivision of the data taken with the \ATLAS detector has several levels:
The largest sub-blocks of data which is taken within one year are the \index{data-taking period}s.
Every data-taking period spans a certain interval of time,
or, equivalently, a certain range of run numbers,
and is given a letter starting with ``A'' at the beginning of the year.
There may also be subperiods which are numbered with arabic numbers, \ie period 2011A consists of A1 and A2. %
The data from proton-proton collisions taken in 2010 is subdivided into nine periods, labelled with letters A~--~I.
The declaration of the a new data-taking period usually indicates
that there has been a major change with non-negligible impact on the detector performance.

Each period may consist of one or more physics runs.
A physics run usually starts after the beams in the LHC have been accelerated to the nominal center-of-mass energy and declared stable
and ends several hours later shortly before a scheduled beam dump,
unless there are problems with the detector or the data-taking infrastructure
which necessitate an abort of the current data-taking run.
Each run is comprised of \index{luminosity block}s (\acsplink{LB}) with a nominal length of one (for 2011 data taking) or two (for 2010) minutes,
which in turn consist of several thousand events each. %
A luminosity block is the smallest entity over which the detector and beam conditions are assumed to be stable.
Its length is somewhat arbitrary,
but when the data-taking conditions are changed by manual intervention,
in most cases due to a change in the trigger prescales (cf. \Sec{sec:tdaq_ATLAS_trigger}),
the beginning of a new luminosity block is enforced.

Not all of the data which has been recorded with the \ATLAS detector is suitable for physics analyses.
Problems with the detector may compromise the data and render part of the data unusable.
The requirements of individual physics analyses with respect to the quality of the data will differ depending on the type of analysis.
If, for example, there is a problem with the muon spectrometer deteriorating the muon momentum resolution,
analyses which do not use muons can still use the data from the run in which this problem occurred.
To accommodate for the needs of different analyses,
while providing a unified approach to assess the data quality,
the status of the detector is monitored throughout data taking and observed problems with the data are stored in a database.
From this information, \acfp{GRL}\inindex{Good Runs List} can be created~\cite{ATL-COM-GEN-2009-015}.
They are part of the input to every analysis and define which subset of the data taken with the \ATLAS detector
is suitable to be considered for physics or performance analyses.

The data-quality status flags are stored in the \ATLAS conditions database (\acs{COOL}).
They either directly describe the status of various subdetectors %
or at a more abstract level the quality of some aspect of the data,
\eg some reconstructed physics quantity or the performance of some reconstruction algorithm. %
The flags can either be green, yellow or red.
Green means that no significant problem was spotted with the data taken by this subdetector or subsystem.
Yellow in general means that the data is flawed and not fulfilling all quality requirements,
but may be usable for physics analysis after some reprocessing or action taken to fix the problem.
Red means that the data is probably lost and cannot be used for physics analysis.
(Additionally, they could be gray
\revised{if} the data quality has not been assessed
or black if some component has been taken out for this particular run.)
It has been found that this ``traffic light'' categorization in many cases is not flexible enough
to consistently reflect the data quality and possible problems,
and from the beginning of 2011 has been supplemented by a defect database. %

\acp{GRL} are stored as text files in the \acf{XML} file format
and contain lists of runs, and for each run a list of the luminosity blocks for which all quality requirements have been met.
\removesection{
  They are created from the a query of the database based on the set of status flags
  that have been defined as relevant by the respective physics groups for their analyses
  and other run requirements such as the beam energy.
  It may then be used in the analysis job to filter events.
}
The \ATLAS Supersymmetry group has defined their own set of criteria and provides \acp{GRL} based on these,
which are used in the analysis described in chapter \ref{sec:analysis_susysearch}.
Looser criteria might be sufficient for the other studies which are presented in this thesis,
but the gain in statistics is expected to be negligible,
and thus for consistency the same \acp{GRL} are used to select data for these studies, too.

\subsection{Luminosity Measurements}
\label{sec:luminosity_measurement}

\inindex{Luminosity measurement}
The accurate determination of the luminosity is crucial
for precision measurements of cross sections of particular physics processes,
which are an important part of the physics program of the LHC,
because the result for the cross section scales directly with the integrated luminosity.
For online measurements of the luminosity,
\ATLAS can make use of six subdetectors\footnote{
  \acs{LUCID}, a \cerenkov detector specifically designed for luminosity measurements in \ATLAS,
  FCAL, the forward calorimeter,
  \acs{ZDC}, the zero-degree calorimeter,
  BCM, the beam conditions monitors placed at the Inner Detector,
  \acs{MBTS}, the minimum-bias trigger scintillators located at the endcaps of the \acs{LAr} calorimeter,
  and vertex counting in the \ac{HLT}.
  \acs{ALFA} is not used according to \cite{LuminosityEPJC}.
},
giving a total of 16 independent methods for the luminosity determinations,
which differ in performance, sensitivity to background, acceptance and efficiency~\cite{Maettig:1307100}. %
The comparison between the measurements of the different subdetectors
allows to do cross-checks and helps estimating the systematic uncertainties.
The luminosity is measured online, \ie in real-time, using multiple detectors and multiple techniques,
and stored together with the \ATLAS data quality information in the \acs{COOL} database for each luminosity block (\acs{LB}).
In addition to the online measurements,
the results from offline algorithms are also available.
Offline algorithms can access more detailed information that is not easily accessible online,
but can operate only on the small fraction of events which is recorded and are thus less powerful~\cite{LuminosityEPJC}.
The luminosity information stored in the database is given in terms of delivered luminosity, without loss corrections.
This allows to compute the integrated luminosity for a subset of the data
based on the association of LB number and amount of integrated luminosity for the LB,
taking into account the trigger prescales,
dead time of the data-acquisition system, LBs marked as bad during the data quality assessment
and other sources of data loss, and also allows to use and compare different algorithms and calibrations.
Each time the calibration is updated,
a new database tag is created.

Measurements of the total inelastic $pp$ cross section that rely on Monte Carlo have been found to have relatively large systematic uncertainties.
An alternative is the determination of the absolute luminosity using van-der-Meer scans,
also called beam-separation scans,
which does not require knowledge of the cross section and gives smaller uncertainties,
dominated by the measurement of the LHC beam-current~\cite{ATLAS-CONF-2010-060}.
In the first measurements in 2010,
the systematic uncertainty on the absolute calibration of the luminosity via beam-separation scans
was found to be \percent{11} for the $pp$ collision runs taken at the LHC at a center-of-mass energy of \GeV{900} and \TeV{7}~\cite{LuminosityEPJC}. %
An update is presented in~\cite{ATLAS-CONF-2011-011}. %
It states a \percent{3.7} larger visible cross section for $pp$ collisions in 2010, compared to~\cite{ATLAS-CONF-2010-060},
with a reduced relative uncertainty of~\percent{3.4}.
This is the luminosity calibration that will be used in this thesis
to compute the total integrated luminosity (cf. \Sec{sec:analysis_luminosity_calculation}),
which is then used for the normalization of the Monte Carlo and setting exclusion limits.
Due to a number of changes in the \ATLAS luminosity detectors during the shutdown in winter 2010~/ 2011,
the 2010 calibration cannot be applied directly to 2011 data.
A new calibration was recently made available for the 2011 data
\cite{ATLAS-CONF-2011-116} which has a relative uncertainty of \percent{3.7}. %

In \ATLAS, the official tool to compute the integrated luminosity
is \propername{iLumiCalc},
which is also available as an online application \cite{URL_iLumiCalc}.
It relies on the online and offline information about the instantaneous luminosity that is stored in the \ac{COOL} database,
which strictly speaking is the instantaneous luminosity integrated over a very short time interval,
over which the instantaneous luminosity is assumed to be constant.
The calculation of the integrated luminosity for a given run or set of runs is based on \acp{GRL} (see above). %
The same \ac{GRL} that was used to select the events for the analysis processing is fed into the \propername{iLumiCalc} tool,
which then basically sums up the luminosity of all luminosity blocks that are declared good by the \ac{GRL}.

\removesection{ %
Two issues make this more complicated than a simple addition.
The first issue is the prescale of the trigger that is used to collect the data.
As explained in \Sec{sec:definition_prescales},
the aggregate trigger prescale is the product of the three prescales at \ac{L1}, \ac{L2} and \ac{EF}.
The prescale of a trigger may not only vary between different runs,
but also within a run at the boundaries of the \acp{LB}.
Correcting for the prescales is not difficult as the luminosity scales directly with the prescales.
A trigger prescaled by a factor $n$ on average takes one $n$th of the events it would take without prescales,
so the integrated luminosity from each LB needs to be divided by the prescale of the trigger for this LB.
The second issue is the dead time of the trigger,
\ie the time the trigger needs to recover after it has fired,
in which it is not able to trigger another event.
The dead time needs to be subtracted from the collected luminosity.
The fraction of time that the trigger system has been running and able to accept events
is estimated by \propername{iLumiCalc} as the live-fraction of a trigger, %
using a high rate trigger to minimize statistical uncertainties and fluctuations.
} %

\section{Monte Carlo Generation}
\label{sec:software_MC}
\inindex{Monte Carlo method}
\acf{MC} methods are a class of computational algorithms
which use random sampling as an essential part of their execution flow.
They only yield approximate solutions,
but can often provide sufficient accuracy for problems
for which exact algorithms would need prohibitively long running times.
Another important field of application are computations which include uncertainties on the input data
or simulations of physics processes which are of inherently stochastic nature.

In high-energy physics, the simulation of particle collisions
and the subsequent interaction of the produced particles with the detector material are the primary example.
Monte Carlo methods are used here to generate event data in as much detail and complexity
as they would be observed at a real collider and detector.
The computer codes simulate the physics processes taking place in the interaction of the colliding particles
and produce a simulated final state of an interaction, with particle identities and momenta.
Following the notion of a physics event, these codes are called \define{\index{event generator}s}.
They try to yield results as close as possible to real events,
within the bounds of the current understanding of the underlying physics processes, %
and allow to translate theoretical physics models into simulated physics events.
The simulated events can then be compared to the actual measurements
to test predictions of the Standard Model and new physics models.
Event generators as the first step of the Monte Carlo production chain are independent of the detector.
Their output is used as input to the following step,
in which the propagation of the produced particles and their interaction with the detector material is simulated.
Only in this step, the exact geometry and behavior of the detector and its subsystems \revised{enter},
and the simulated output of the virtual detector is created.
To make a clear distinction from real detector data,
events which are produced in Monte Carlo simulations are sometimes called (Monte Carlo) pseudodata.

A fundamental difference between real data and Monte Carlo pseudodata is
that in Monte Carlo events exact information about all particles is available,
including the identity and kinematics of particles, positions of decay vertices and the relationship between parent and daughter particles.
This information is called (Monte Carlo) \define{\index{truth information}}
and is vital for systematic studies
of the response of the detector to certain classes of particles or event types
and of how signals of new physics processes can be identified.
The Particle Data Book \cite{PDB2010} gives a numbering scheme for Monte Carlo particles,
which was introduced in 1988 to facilitate interfacing between event generators, detector simulation, and analysis packages,
and is used for example by the widely used event generators \propername{Pythia} and \propername{Herwig}.
It not only covers particles which are known to exist,
but also a range of hypothetical particles from extensions of the Standard Model like Supersymmetry or technicolor.

Ideally, real data and Monte Carlo pseudodata are fully interchangeable from a structural point of view
so that the same analysis code can be run on one or the other without any changes.
This makes the data processing transparent and thus less error-prone.
It is not always possible and desirable to achieve this though,
either because the Monte Carlo simulation may lack some important detail,
or because truth information is needed in the analysis which is not available in real data.

\label{sec:software_MC_split_MC}
The generation of Monte Carlo pseudodata is usually done separately for groups of closely related physics processes
rather than simulating inclusive samples which contain all conceivable processes.
In addition to the splitting at process levels,
there can be another splitting based on event kinematics.
The separation has a number of advantages.
There are huge differences in the cross sections of different physics processes (cf. \Fig{fig:lhc_cross_sections})
so that even in large Monte Carlo samples with millions of events,
for rare processes %
only few events would be included in the sample.
However, after applying the cuts implementing the analysis selection
precisely these rare events may become dominant.
Therefore, to achieve a better coverage of the available phase space and not to run out of statistics after harsh cuts,
the phase space can be divided into bins,
as is done \eg in dijet samples,
using the momentum exchanged between the two initial quarks,
and simulate a certain number of events for each bin.
To obtain physically meaningful results,
all subsamples need to be combined again later,
weighting events from the subsamples according to the corresponding cross sections.
Having different physics processes split up into different samples also
makes it possible to compare different Monte Carlo generators
or to study systematic variations,
\eg using different top masses as templates in an analysis aiming at a measurement of the top quark mass \cite{TopAntitopMass2011}.
When generating Monte Carlo samples which are binned in some kinematic property,
this is usually done by applying a filter which retains only the desired events.
This \index{filter efficiency} is a multiplicative factor
which needs to be taken into account when computing the cross section of the Monte Carlo sample.

\subsection{Hadronization and Jet Formation}
\label{sec:software_hadronization}

In the discussion of the parton model in \Sec{sec:theory_parton_model} %
and of the measurement of hadronic jets in \Sec{sec:experimental_setup_general_jets},
it was already explained that the partons hadronize and give particle showers
before being detected in the calorimeter as clustered energy depositions.
The simulation of jets in Monte Carlo event generators can be divided into several steps which reflect this evolution.

The first step is the simulation of the hard process by evaluating the corresponding matrix elements.
The hard process takes place at high energy scales,
which allows to use perturbative \ac{QCD} calculations
and compute the respective matrix elements exactly.
This, however, can only be done to a limited order in the strong coupling constant \alphas,
because the complexity increases roughly factorially with the order \cite{Ellis1996}. %
A simple example of a higher order process is the emission of a gluon, %
which eventually gives rise to an additional hadron jet.
This corresponds to electromagnetic bremsstrahlung,
but due to the larger strong coupling constant is more frequent.
Instead of doing an exact calculation of higher orders,
a parton shower algorithm can be employed,
which does an approximate perturbative treatment of the QCD dynamics
above some given threshold for the square of the momentum transfer. %
It is thereby possible to include higher order terms
which can be important in certain regions of phase space.
The parton shower proceeds in a sequential and probabilitistic manner according to splitting functions,
which describe the emission of gluons and creation of quark-antiquark pairs. %
These branchings in the parton shower can be efficiently implemented
in terms of the so-called \index{Sudakov form factor}, using Monte Carlo techniques. %
The parton shower uses as input the partons from the initial and final state of the hard process.
Some care has to be taken to avoid double counting of phase-space regions %
for partons produced in the hard process at higher orders %
and partons created in the following shower process.
The parton shower output, still consisting of partons, %
is then used as input to the hadronization which converts the partons into hadrons.

The hadronization of partons is an intrinsically non-perturbative process
in the low-mo\-men\-tum transfer, long-distance regime, %
which can only be modeled phenomenologically at present.
There are various models for the hadronization \cite{Ellis1996}. %
The simplest model assumes the independent fragmentation of every parton individually
by iteratively combining quarks with quark-antiquark pairs until the energy falls below a threshold,
and splitting gluons into quark-antiquark pairs.
An improvement on this model is the string fragmentation model,
where the colored partons are connected by extending flux tubes while they move apart,
which break when enough energy is stored in them,
creating $q\bar q$ pairs and forming colorless objects.
In cluster hadronization models,
after the parton branching colorless clusters of partons form and then decay into hadrons \cite{Langacker}.
Finally, short-lived hadrons may undergo decays before reaching detector material.
What is finally observed are their decay products.
The string model is used \eg in \propername{Pythia}, %
the cluster hadronization \eg in \propername{Herwig}. %

\subsection{\texorpdfstring{The \kfactor} {The k-factor}}

It may happen that Monte Carlo samples have been produced by a leading order generator,
which yields the \acf{LO} cross section,
but the \acf{NLO} cross section may be available from dedicated cross section calculations, too.
The difference between LO and NLO calculations is quantified in the \kfactor,
which is defined as the ratio of the cross section at NLO and LO,
\begin{equation}
  k = \frac{\sigma_\text{NLO}}{\sigma_\text{LO}}.
\end{equation}
The \kfactor can be used to correct the LO cross section produced by MC generators to compensate for missing higher-order terms.
It can become significantly larger than one, especially when new subprocesses appear at next-to-leading order.
Scaling all kinematic distributions by the same factor assumes
that these distributions are invariant under the inclusion of higher-order terms.
This is not necessarily always true,
and studies of this can be done by comparing the predictions for fundamental distributions
such as that of the \pt of the leading jets 
from LO generators to those of higher order where available.

\subsection{Minimum Bias}
\label{sec:define_minimum_bias}

Here, a few basic terms will be introduced that are used to classify hadron interactions.
A~distinction is made between \define{hard processes},
in which a large momentum is transferred between the interacting partons
and which are well described by perturbative QCD,
and \define{soft interactions} with a small momentum transfer,
which require phenomenological models as they cannot be treated perturbatively
due to the large values of the strong coupling constant in this regime.
Soft collisions will usually dominate,
whereas the production of final states with high-\pt particles and the creation of massive new particles can only happen in hard collisions.

The notion of a \define{\index{minimum-bias event}} is commonly used.
Minimum bias is an experimental definition and can be described
as everything that would be triggered by a totally inclusive trigger, %
including any single inelastic collision of two protons, %
therefore including very rare high-\pt scatters and very common low-\pt scatters.
The name comes from the fact that this type of trigger introduces the least possible bias by its selection.
Minimum-bias triggers attempt to be as inclusive as possible by \revised{making} very loose selections (cf. \Sec{sec:tdaq_minimum_bias_triggers}).
\removesection{ %
The relation between this experimental definition and physics in terms of a cross section is %
\begin{equation}
  \sigma_\text{total} = \sigma_\text{elastic} + \sigma_\text{sd} + \sigma_\text{dd} + \sigma_\text{nd},
\end{equation}
saying that the total cross section $\sigma_\text{total}$ consists of contributions
from elastic, single-diffractive (sd), double-diffractive (dd) and non-diffractive (nd) interactions.
In a single-diffractive event, one of the two interacting particles is still intact after the collision,
the other particle has been excited to a system of higher mass but with the same quantum numbers \cite{Alner1987445}.
In double-diffractive events, particles are excited to higher mass diffractive systems which are assumed to decay independently.
In non-diffractive events, particles are produced in a central region
with a distribution that is flat in rapidity, and in two fragmentation regions.
As nd and dd events are difficult to distinguish experimentally,
they are often combined in a class of non-single diffractive (nsd) events
which then comprise what is taken to be a minimum-bias event when doing Monte Carlo simulations. %
Some definitions also include sd events in minimum bias.
The cross sections for the LHC at \seventev are $48.5$, $13.7$ and \unit[9.17]{mb} for non-diffractive, single- 
and double-diffractive processes, respectively \cite{ATLAS-CONF-2010-024}.
} %

A related term is the \define{\index{underlying event}},
mostly defined to include all particles from a single particle collision except the process of interest
in terms of the hard physics process or all high-energy particles.
Still, the underlying event may contain high-\pt particles.
It is different from minimum-bias events %
due to correlations with the hard process.
Both a minimum-bias event and the underlying event are dominated by soft partonic interactions. %

\chapter{Trigger Studies}
\label{sec:triggerstudies}

The trigger is a vital part of \revised{any} detector operating at high interaction rates,
and trigger studies thus are a vital part of every analysis which evaluates data taken at such a detector.
Two major aspects concerning triggers are covered in this chapter:
the trigger rates and the trigger efficiencies of the \ATLAS trigger system.
Concerning the rates,
the bandwidth limits of the \acf{TDAQ} define how much data can be taken
in terms of the amount of data that can be stored per unit of time.
Within the resulting maximum total trigger rate,
the relative rates of the trigger chains
determine the composition of the sample of events that is obtained.
The efficiencies account for the fact
that the trigger decides in real-time for every event whether it should be kept or discarded.
This online decision has to be made within a time span
which is much shorter than the time available for the offline reprocessing of the data.
Therefore, only a fraction of the full information can be evaluated,
and the decision will in some cases differ from the decision
which would have been made based on the full information and offline processing
so that events which should have been kept are discarded.
The \revised{trigger mistag} rate induced by the necessary compromise between speed and accuracy of the online data-taking process
is expressed in terms of the efficiency of the trigger.
To know these efficiencies is crucial for the correct interpretation of the recorded data.

The context and the motiviation for the studies presented in the following %
is the search for Supersymmetry at the \ATLAS detector,
specifically in zero-lepton final states,
for which the combined \jetmet trigger is the primary trigger.
The measurements of the trigger rates will %
allow to demonstrate the advantages of using a combined trigger for this analysis
rather than the easier to utilize single object triggers.
This necessitates to understand the combined \jetmet trigger and to measure its efficiencies,
important studies
to which the largest part of this chapter is dedicated.

\section{Trigger Rates}
\label{sec:triggerrates}

\subsection{Measurements}
\label{sec:results_trigger_rates_rate_measurements}

The measurements and plots of trigger rates presented in this thesis make use of the trigger counts
which are stored in the \COOL database.
These counts are online measurements, which are recorded at the time of data taking
and are available for each run and luminosity block.
They are written out by the trigger system for all triggers which are configured \revised{according to} the corresponding trigger menu.
On the one hand, the advantage of using this database is
that it allows to access a very large range of runs and different triggers easily,
without extracting the trigger decision run by run from the data in the different streams.
This would be a very laborious process and also occupy a lot of data-processing resources. %
On the other hand, using these trigger counts it is neither possible to \revised{apply} selection cuts based on physics objects
nor to determine the correlation between triggers or the relative rate of triggers with respect to other triggers
because the trigger counts are already summed over luminosity blocks and therefore an average over a large number of events.
However, at least the \ac{GRL} can be enforced
because it is a simple selection based on luminosity blocks,
and thus data suffering from bad detector conditions or configurations which differ from normal data taking can be masked out.

The trigger rates can be \revised{investigated} as function of different variables.
Here, the average number of interactions per bunch crossing \avgmu (as defined in \Sec{sec:experimental_setup_general_define_avgmu}) %
and the instantaneous luminosity \Linst are of interest
to study the behavior of the rates with changing beam conditions.
These quantities are also available per luminosity block in the \COOL database.
How the plots from \COOL are produced is explained in detail in \Sec{sec:results_trigger_rates_describe_COOL_plots} in the Appendix. %
The rates shown in the following plots are in general corrected for the trigger prescales which have been applied,
\ie the counts which are read from the database are multiplied by the prescale factors at all three trigger levels for the respective run and luminosity block.
One dot in the rate plots corresponds to a \revised{single luminosity block}
or an average over several luminosity \revised{blocks}.

The trigger rates can also be obtained by other methods.
An important alternative to predict trigger rates
is the measurement of trigger rates relative to another, looser trigger.
This allows to study the relative rate reduction introduced by raising trigger thresholds or additional cuts.
Measuring absolute rates of triggers is more complicated.
As with increasing instantaneous luminosity, 
the triggers evolve to harsher cuts and not the other way round,
the study of evolving trigger cuts relative to existing triggers should be always possible.
There are also special runs dedicated to measuring trigger rates for chains with low thresholds,
which are called \index{enhanced bias runs}. %
In these runs, triggers with relatively low thresholds are run,
and the rates of other triggers can then be measured relative to these.
In contrast to the rate measurements stored in \COOL,
this also allows to study correlations and overlap of triggers and unique rates as well as inclusive trigger rates.

\subsubsection{Example Plot: Rate of a Jet Trigger}

\begin{figure}
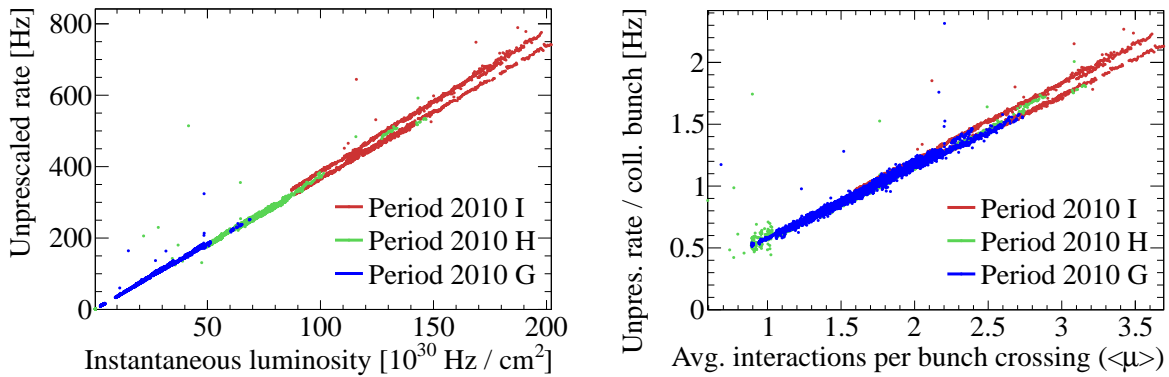

  \centering
  \incgraphics{width=\widthtwoplots}{{{COOL.plot_rates_L1_J30_2010I,2010H,2010G_withGRL_Xoffllumi}}}
  \hfill
  \incgraphics{width=\widthtwoplots}{{{COOL.plot_rates_L1_J30_2010I,2010H,2010G_withGRL_Xofflmu}}}
  \caption{
    Unprescaled rate of the \trigger{L1_J30} trigger in the last three periods of 2010 data taking (periods G, H and I)
    as function of the instantaneous luminosity (left) or the average number of interactions per bunch crossing \avgmu (right).
  }
  \label{fig:trigger_rate_example_L1_J30}
\end{figure}

An example of a trigger rate extracted from the \COOL database is given in \Fig{fig:trigger_rate_example_L1_J30}.
The plot shows the rate of the \trigger{L1_J30} trigger item before the application of prescales
in the periods with the highest instantaneous luminosity in 2010.
Every point in the plot corresponds to one luminosity block from one run
so that each point is an average of the trigger rate over approximately two minutes.
The \ac{GRL} of the \ac{SUSY} group has been applied
\revised{to mask out} data from bad luminosity blocks.
The rate is plotted as a function of the instantaneous luminosity
and shows a linear behavior over a range of almost two orders of magnitude in luminosity,
going from about $5\ten{30}$ to $\unit[2\ten{32}]{Hz/cm^2}$.
This linear behavior is advantageous
because it makes extrapolations %
of the rates easy.
It is an indication for the stability of this trigger,
in particular with respect to pile-up.
As the center-of-mass energy is constant,
and therefore the cross sections are constant, too,
the instantaneous luminosity is directly proportional to the number of events taking place in the detector in a given interval of time.
A direct proportionality between the trigger rates and the instantaneous luminosity is therefore consistent with the naive %
expectation.
For the \met triggers,
which are discussed in \Sec{sec:results_met_model_motivation},
this will turn out not to be true.

\subsection{Benefits of Combined Triggers}
\label{sec:results_trigger_rates_benefits_combined_triggers}

This section explains the motivation to use combined triggers
by demonstrating their benefits over single object triggers.
This is done first in terms of general considerations,
and in the following part by giving a concrete example for \jetmet triggers
with numbers from \ATLAS data taking in 2010.

At the first glance, using a combined trigger entails a number of disadvantages\revised{.}
The measurement of the trigger efficiencies becomes more complicated (cf. \Sec{sec:results_trigger_performance_terminology}).
At least a two-dimensional binning is needed to accommodate for the two (or more) offline variables,
and thus the turn-on curves have a higher dimensionality and more statistics is needed.
There may also be fewer possibilities to measure the efficiencies
because there are less options for orthogonal triggers,
and correlations between the components may give rise to very subtle and hard to understand effects.
\revised{The advantage of using a combined trigger is
that for single object triggers
the thresholds will \revised{typically} be much higher than for combined triggers.}
Combined triggers therefore allow a higher signal yield than single object triggers
within the constraints %
imposed by the trigger system.

The maximum trigger rate and bandwidth are fixed,
and the trigger thresholds need to be optimized within this affordable total rate,
\ie which events are recorded needs to be optimized.
The choice of the trigger can be optimized for a specific physics signal
by matching the trigger signature to the signature of the searched for signal.
Specifically in the context of this thesis
discussing the search for Supersymmetry in zero-lepton final states (cf. \Sec{sec:analysis_susysearch}),
this leads to a trigger which combines a jet signature with an \met signature,
considering the following prerequisite:
For a signal where only \met and \revised{none} or only soft jets,
or where only jets and little \met \revised{are} expected,
only single object triggers can be used.
But many Supersymmetry models have long decay chains,
giving hard jets due to the high masses of the supersymmetric partners of the Standard Model particles
and lots of \met due to the heavy \acp{LSP}, which evades detection and carries away a lot of energy.
The fundamental signature is thus a combination of both large \met and hard jets. %
This is illustrated in the exemplary plot to the left in \Fig{fig:jetmet_example_combined_trigger_overlap},
which shows a two-dimensional histogram with the \pt of the leading jet at \EMJES scale on the horizontal axis and \met %
on the vertical axis,
for a Monte Carlo sample implementing a \ac{mSUGRA} scenario with $\mzero = \GeV{360}$ and $\moh = \GeV{280}$.

\begin{figure}
  \centering
  \incgraphics{width=\widthtwoplots}{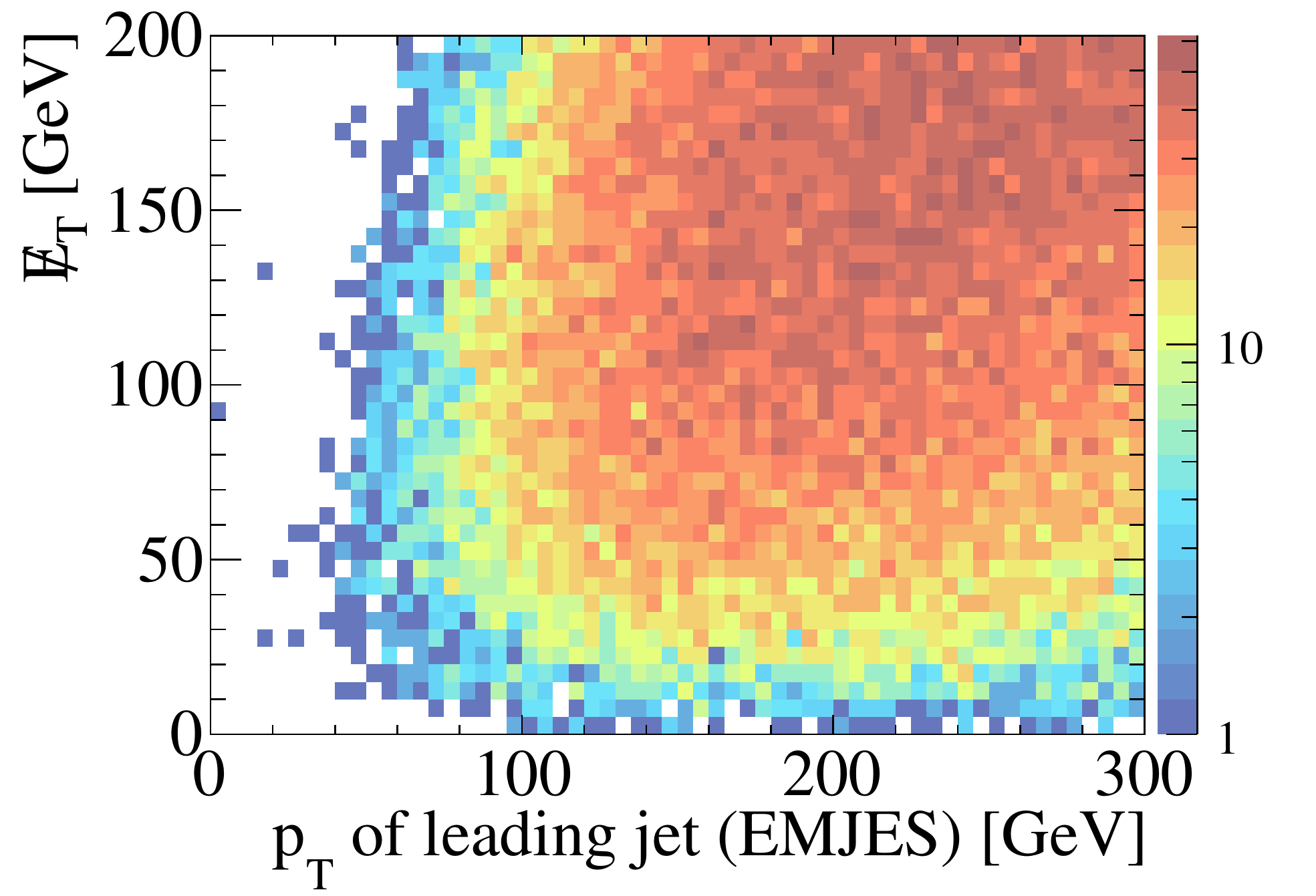} %
  \hfill
  \begin{overpic}[width=\widthtwoplots]{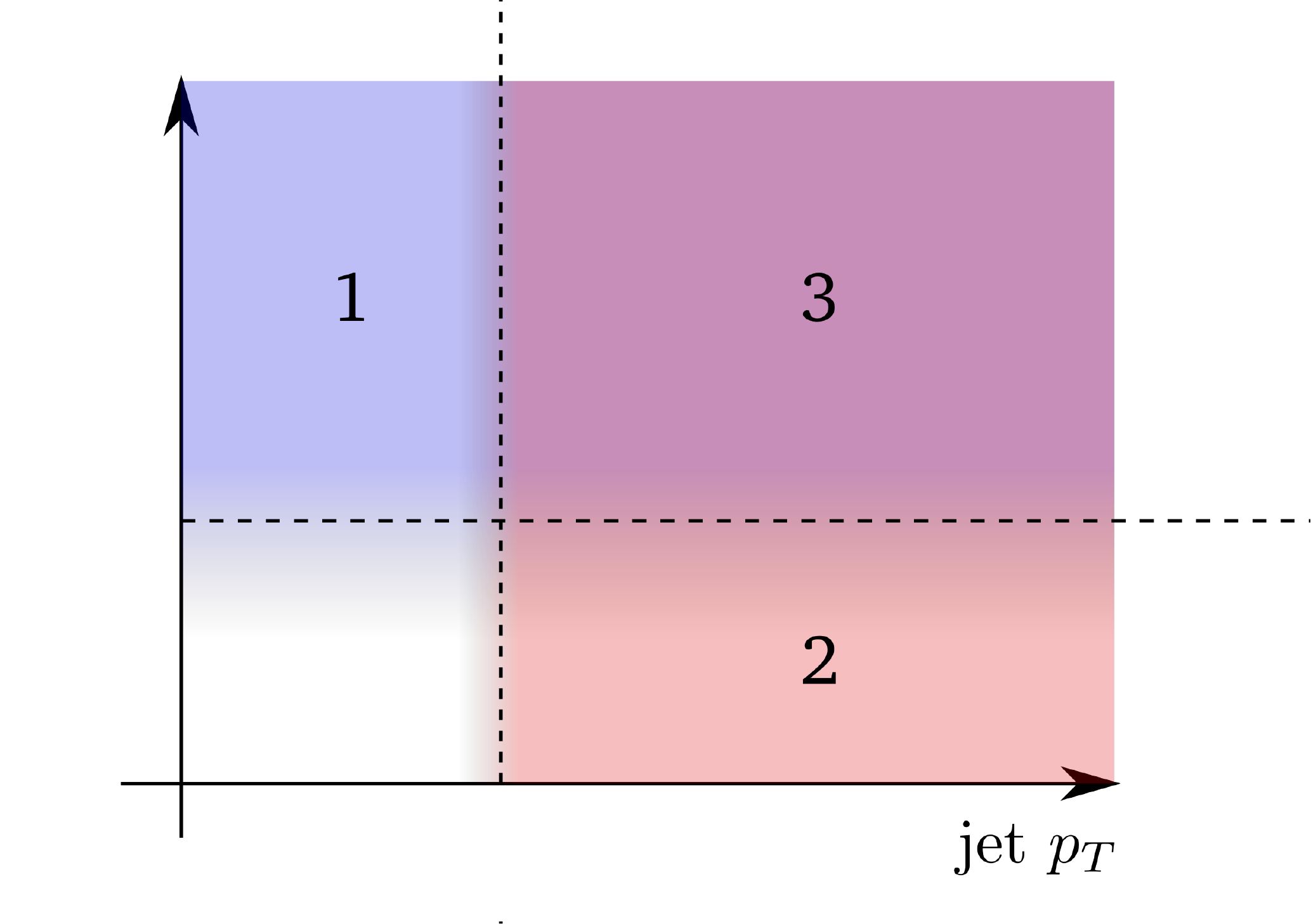}
    \put(4,60){\met} %
  \end{overpic}
  \caption{
    Left: Two-dimensional spectrum of the \pt of the leading jet and \met reconstructed offline as color-coded plot,
    for one point of the mSUGRA grid used in \Sec{sec:analysis_susysearch} with $\mzero = \GeV{360}$ and $\moh = \GeV{280}$.
    Every entry corresponds to one Monte Carlo event. %
    \newline
    Right: Defining three regions with different types of events in the phase space relevant for \jetmet triggers.
  }
  \label{fig:jetmet_example_combined_trigger_overlap}
\end{figure}

The right plot in \Fig{fig:jetmet_example_combined_trigger_overlap} shows a schematic view of the coverage of
different triggers in the plane spanned by \revised{leading} jet \pt and \met.
The red shaded region \revised{represents} the single jet trigger efficiency,
the blue shaded region \revised{represents} the \met trigger efficiency.
Events which have \met and jet \pt in region 1 will only appear in a sample taken with the \met trigger,
events in regions 2 will be triggered by the single jet trigger only,
and events in region 3 will be triggered by both the \met, the jet and a combined \jetmet trigger.
A general guideline is that using an combined trigger is advantageous
if the signal has few events in regions 1 and 2 and is concentrated in region 3.
The backgrounds, which drive the trigger rates, usually have a steeply falling spectrum
and populate in particular regions 1 and 2.
By \revised{rejecting events in} regions 1 and 2, a combined trigger gives a much better signal and background separation than single object triggers,
which helps to save bandwidth and enrich the recorded event sample with signal events.
To substantiate these considerations,
the rates and signal yield of the combined \jetmet triggers
in the \ATLAS trigger menu from 2010 will be discussed in the following. %

\subsubsection{\texorpdfstring{Benefits of \Jetmet Triggers: Rates} {Benefits of JetMET Triggers: Rates}}
\label{sec:results_trigger_rates_illustration_jetmet}

\begin{figure}
  \centering
  \incgraphics{width=\widthwideplot}{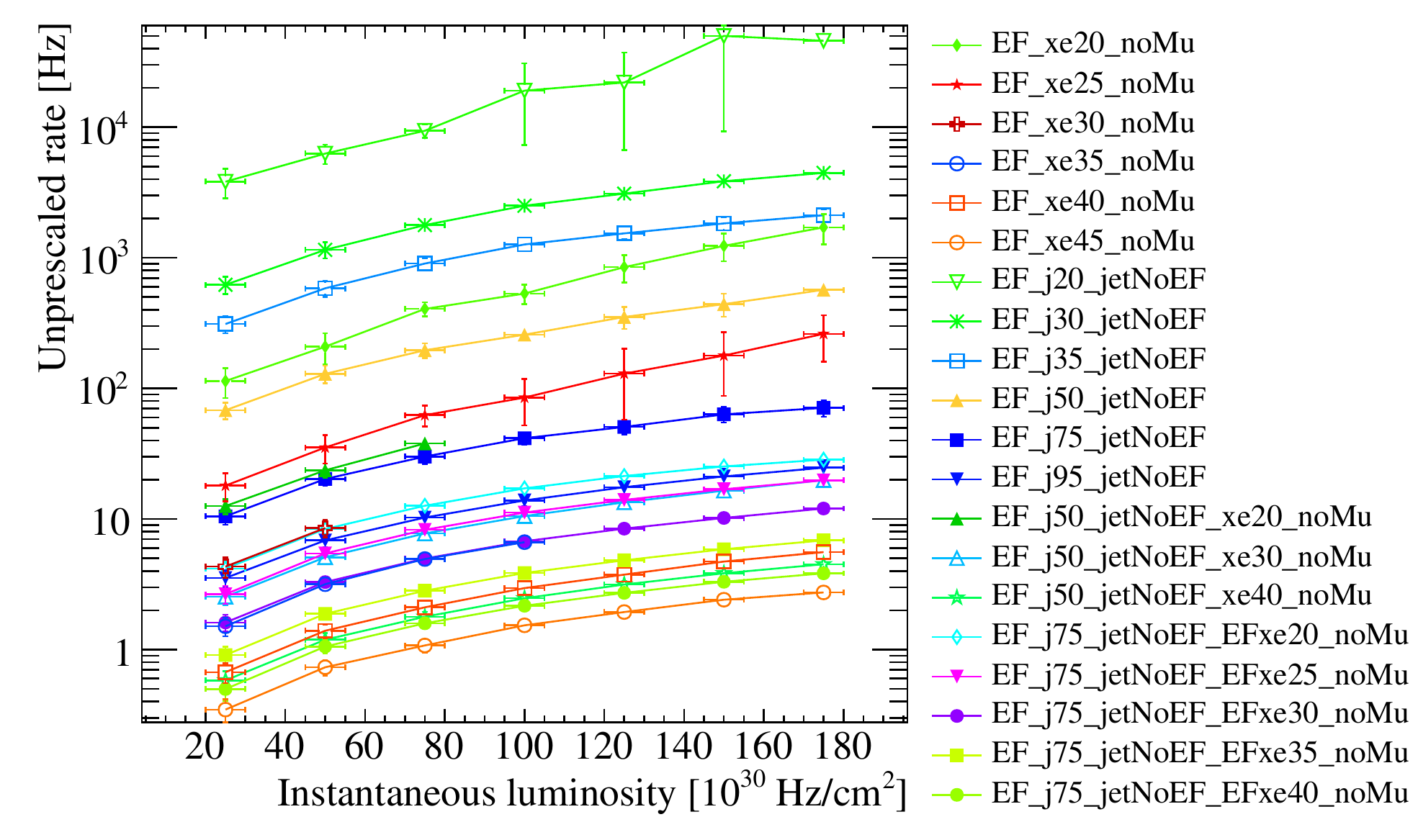}
  \caption{
    Unprescaled rates of several jet, \met and combined \jetmet triggers in 2010 as function of the instantaneous luminosity.
  }
  \label{fig:results_trigger_rates_overview_2010}
\end{figure}

\Fig{fig:results_trigger_rates_overview_2010} gives an overview of the unprescaled rates
of several jet, \met and combined \jetmet triggers in 2010 as function of the instantaneous luminosity.
In steps of $\unit[25\ten{30}]{Hz/cm^2}$ in the instantaneous luminosity,
all rate measurements from an interval of $\unit[\pm5\ten{30}]{Hz/cm^2}$
around the central value are collected and fitted with a normal distribution.
The fit result gives the rates and their uncertainties,
which are shown as vertical error bars in this plot.
The horizontal error bars indicate the interval of instantaneous luminosity,
from which the rate measurements for the fit are taken.
Some points are missing for triggers which were deactivated at the respective instantaneous luminosity.
The plot \revised{demonstrates} that the rates of these triggers cover several orders of magnitude,
from the lowest rate for the \met trigger \trigger{EF_xe45_noMu} with a threshold of \GeV{45}
up to the highest rate for the jet trigger \trigger{EF_j20_jetNoEF} with a threshold of \GeV{20}.
In particular the single-object jet and \met triggers with low thresholds have very high rates,
but going to higher thresholds strongly reduces the rates.
For low \met thresholds, already an increase of \GeV{5} may reduce the rate by a factor of $10$.

The primary physics trigger for the analysis of 2010 data in the search for Supersymmetry in zero-lepton final states,
which is described in this thesis,
is the combined \jetmet trigger \trigger{EF_j75_jetNoEF_EFxe25_noMu} (cf. \Sec{sec:analysis_susysearch} and \Tab{tab:analysis_luminosities2010}).
From the plot in \Fig{fig:results_trigger_rates_overview_2010}
it can be read off that this trigger has a rate of about \unit[20]{Hz}
at the highest instantaneous luminosity included in the plot,
which is $\unit[1.75\ten{32}]{Hz/cm^2}$.
This provides an estimate of the rate allotted to this analysis.
If, instead of this combined trigger, the single object triggers with the same thresholds were to be used,
this would \revised{result in} an (unprescaled) rate of about \unit[70]{Hz} for the jet trigger \trigger{EF_j75_jetNoEF}
and about \unit[200]{Hz} for the \met trigger \trigger{EFxe25_noMu},
resulting in a total rate of $200$ to \unit[270]{Hz}, depending on the overlap\footnote{
  A total rate of \unit[270]{Hz} would %
  imply that there is no overlap between these triggers,
  and then the combined trigger would never fire.
  This is not the case, of course.
} of the single object triggers.
Note that the rates given here are inclusive rates,
\ie they include events which are triggered by other triggers, too.
Therefore, switching off these chains does not reduce the total trigger rate by the full rate given here,
but having these triggers in the menu means that the total trigger rate will be at least that high.
These rates are unaffordable and thus these two triggers have to be prescaled.
Their prescales in one of the latest runs in 2010 (run 167776 from period 2010~I2)
vary between $36$ and $144$ for \trigger{EF_j75_jetNoEF} and between $750$ and $3000$ for \trigger{EFxe25_noMu},
depending on the instantaneous luminosity within the run.
\revised{In contrast,} \trigger{EF_j75_jetNoEF_EFxe25_noMu} ran unprescaled until the end of 2010 $pp$ data taking.

The bottom line is that using the combined \jetmet trigger can effect
considerable savings of trigger bandwidth compared to the single object triggers.
It reduces the rate roughly by a factor of $10$ %
in this example.
Of course, this argument does not take into account
how many signal events can be collected with the different triggers.
Using a trigger with \TeV{} thresholds would give much lower rates,
but it would probably also cut away most of the signal.
Therefore, in the following the signal yield of the triggers will be included in the considerations.

\subsubsection{\texorpdfstring{Benefits of \Jetmet Triggers: Signal Yield} {Benefits of JetMET Triggers: Signal Yield}}

\begin{figure}
  \centering
  \incgraphics{width=\widthwideplot}{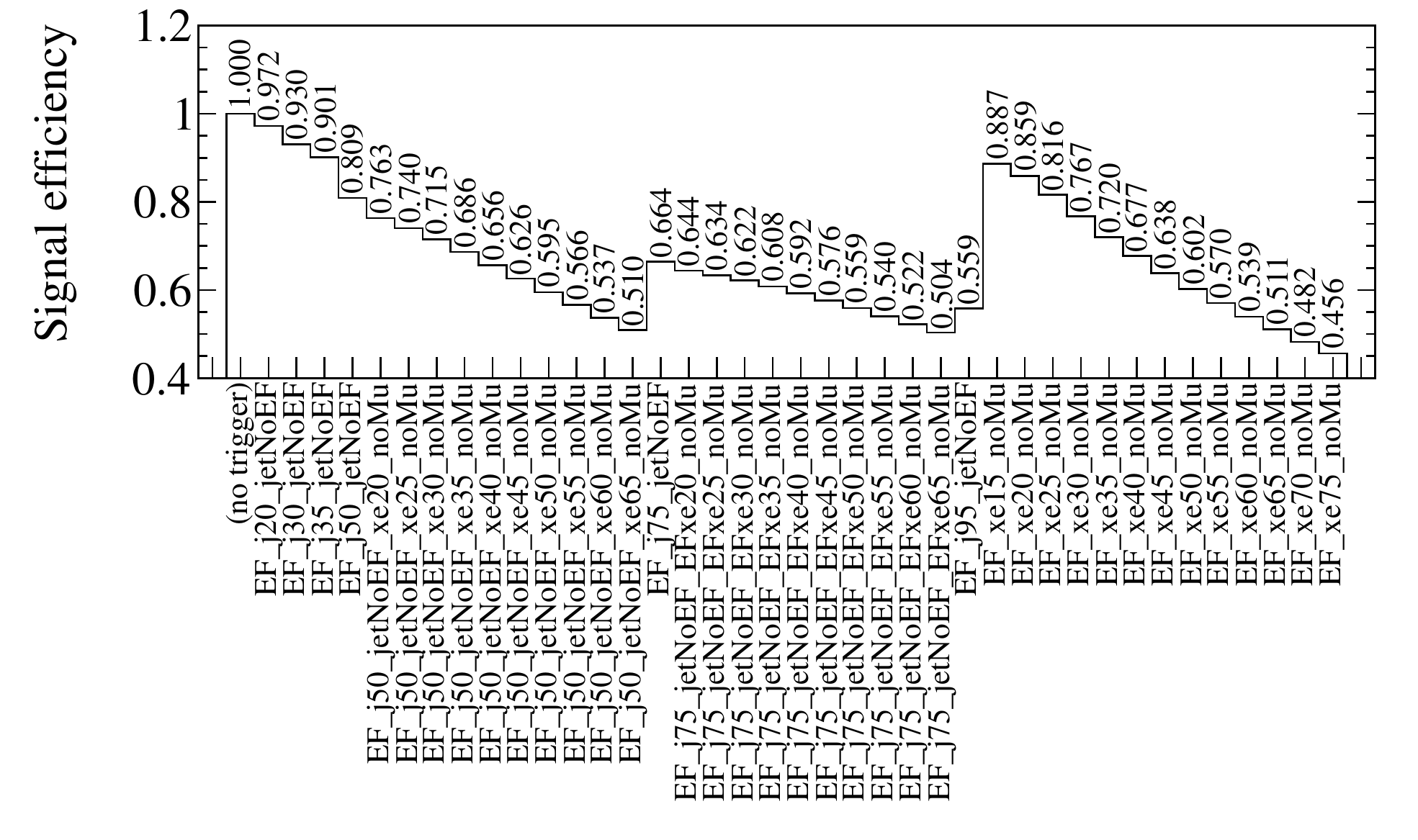} %
  \caption{
    Signal efficiency $\epsilon_T$  for a number of triggers for Monte Carlo events at the supersymmetric benchmark point SU4.
    Given is the fraction of events in this sample
    in which the respective trigger fires.
    (In the interest of better readability, histogram lines are drawn instead of markers.)
  }
  \label{fig:results_trigger_rates_trigger_yield_SU4_2010}
\end{figure}

To estimate the impact of using a combined trigger instead of the single object triggers on the signal yield,
the fraction of events which pass the trigger selection can be counted on Monte Carlo samples with supersymmetric physics processes.
This trigger-dependent \index{signal efficiency} is shown in \Fig{fig:results_trigger_rates_trigger_yield_SU4_2010}
for a sample of Monte Carlo events generated for the benchmark point SU4\inindex{SU4 benchmark point}.
SU4 is a supersymmetric benchmark scenario implementing \ac{mSUGRA} (cf. \Sec{sec:theory_bsm_model_minimal_supergravity}),
which has a large cross section of \unit[402.19]{pb} at \ac{NLO}
\revised{ and $\sqrt{s} = \TeV{14}$ \cite{CSCNotes}.} %
The \ac{mSUGRA} parameter values are
$\mzero = \GeV{200}$,
$\moh = \GeV{160}$,
$A_0 = \GeV{-400}$ and
$\tanbeta{10}$,
and $\mu$ is positive.
In the \mzero-\moh plane, SU4 lies in the low-mass region,
close to the exclusion bounds set by the Tevatron experiments.

The plot in \Fig{fig:results_trigger_rates_trigger_yield_SU4_2010} shows several triggers
and the fraction of events within the sample for which the respective trigger fires.
This fraction will depend on the spectrum of \met and \pt of the jets 
which are produced in the supersymmetric decay chains,
and therefore also on the masses of the supersymmetric particles.
Due to the low masses of the SUSY particles at SU4,
it is to be expected that the signal yield will \revised{strongly} depend on the trigger thresholds,
and \revised{decrease} rapidly for triggers with higher thresholds.
For comparison,
in \Fig{fig:results_trigger_rates_trigger_yield_114039_2010} the same plot for a different point in parameter space is given,
using events from one of the Monte Carlo samples of the mSUGRA grid that is introduced in \Sec{sec:analysis_SUSY_GRIDs}.
This point has much larger masses for the supersymmetric particles than the SU4 benchmark point,
the relevant mass parameters at the unification scale being $\moh = \GeV{340}$ and $\mzero = \GeV{1160}$.
(Mass spectra for points with similar values of \moh and \mzero can be seen in \Fig{fig:analysis_mass_spectrum_msugra}.)
\Fig{fig:results_trigger_rates_trigger_yield_114039_2010} clearly shows,
as expected,
that the loss in the signal yield when going to higher trigger thresholds is much lower in this case.

\begin{figure}
  \centering
  \incgraphics{width=\widthwideplot}{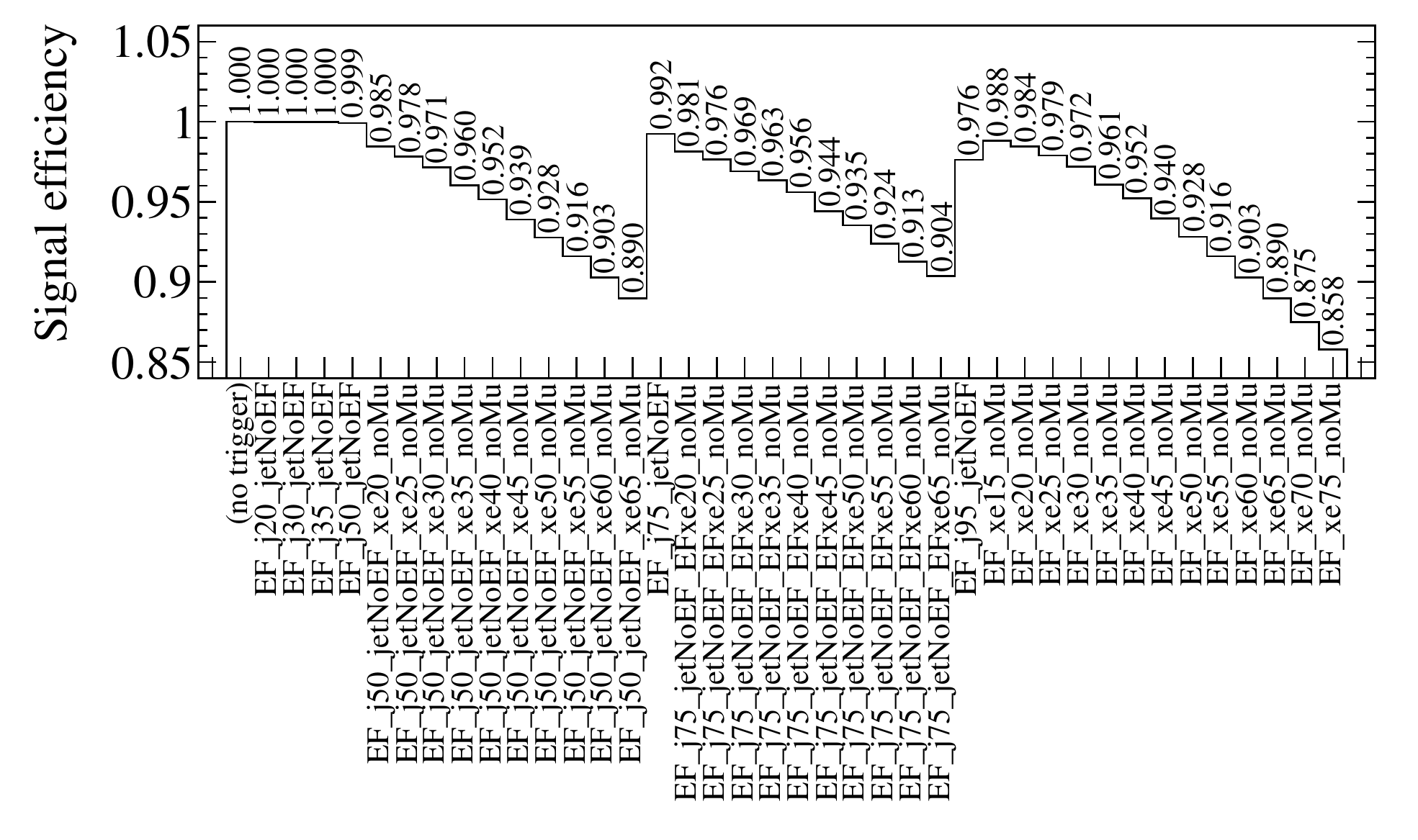} %
  \caption{
    Signal efficiency $\epsilon_T$ for a number of triggers for Monte Carlo events at one of the points of the mSUGRA grid 
    introduced for the analysis of 2010 data in \Sec{sec:analysis_SUSY_GRIDs}.
    This is a point in parameter space with very large masses, $\moh = \GeV{340}$ and $\mzero = \GeV{1160}$.
    \revised{The numbers give the fraction of events in this sample in which the trigger fires.}
    Note the different range on the vertical axis with respect to \Fig{fig:results_trigger_rates_trigger_yield_SU4_2010}.
  }
  \label{fig:results_trigger_rates_trigger_yield_114039_2010}
\end{figure}

A comparison of different types of triggers
will serve here as a second motivation
why the combined trigger \trigger{EF_j75_jetNoEF_EFxe25_noMu} is indeed a good choice within the bandwidth limits
and for the order of magnitude of the instantaneous luminosity in 2010.
Background efficiencies and offline cuts will be neglected for this,
and an in-depth study would consider a large number of grid points rather than only single points,
but the \revised{basic} argument in favor of combined \jetmet triggers will be the same.
The number of signal events which can be collected at an instantaneous luminosity \Linst
within a time $\Delta t$
with a particular trigger $T$ is given by
\begin{equation}
  \begin{aligned}
    N_T &= \Linst \cdot \Delta t \cdot \sigma_\text{SUSY}  \cdot \frac{\epsilon_T } {f^\text{pre}_T} \\
        &= c \cdot  \frac{\epsilon_T}{ f^\text{pre}_T },
    \label{eq:results_trigger_rates_signal_yield}
  \end{aligned}
\end{equation}
where $\epsilon_T$ is the signal efficiency of the trigger
and $\sigma_\text{SUSY}$ is the \revised{production} cross section \revised{for} the events in the Monte Carlo sample.
$f^\text{pre}_T$ is the prescale of the trigger introduced to \revised{limit} its rate if it is too high.
For simplicity, all quantities are assumed to be constant over~$\Delta t$.
$c$~collects factors which are common to all triggers.
$N_T / c$~will be referred to as the \index{signal yield} of the trigger~$T$.
This is the quantity to be optimized.

\begin{table}
  \centering
  \begin{tabular}{lrrc}
    \toprule
    Trigger name & \multicolumn{1}{c}{Rate [Hz]} & \multicolumn{1}{c}{$f^\text{pre}_T$} & \multicolumn{1}{c}{$N_T / c = {\epsilon_T} / { f^\text{pre}_T }$ } \\
    \midrule

    \trigger{EF_j20_jetNoEF} & \numprint{19043.4} & \numprint{1731.22} & 0.001 \\
    \trigger{EF_j30_jetNoEF} & \numprint{2505.3} & 227.76 & 0.004 \\
    \trigger{EF_j35_jetNoEF} & \numprint{1265.1} & 115.01 & 0.009 \\
    \trigger{EF_xe20_noMu} & 532.2 & 48.38 & 0.020 \\
    \trigger{EF_j50_jetNoEF} & 257.7 & 23.43 & 0.043 \\
    \trigger{EF_xe25_noMu} & 85.1 & 7.73 & 0.127 \\
    \trigger{EF_j75_jetNoEF} & 41.6 & 3.78 & 0.263 \\
    \trigger{EF_j75_jetNoEF_EFxe20_noMu} & 17.2 & 1.56 & 0.629 \\
    \trigger{EF_j95_jetNoEF} & 13.9 & 1.26 & 0.774 \\
    \trigger{EF_xe45_noMu} & 1.5 & 0.14 & 0.940 \\
    \trigger{EF_j50_jetNoEF_xe40_noMu} & 2.5 & 0.22 & 0.952 \\
    \trigger{EF_xe40_noMu} & 3.0 & 0.27 & 0.952 \\
    \trigger{EF_j75_jetNoEF_EFxe40_noMu} & 2.2 & 0.20 & 0.956 \\
    \rowcolor{lightblue} %
    \trigger{EF_j75_jetNoEF_EFxe25_noMu} & 11.2 & 1.02 & 0.959 \\
    \trigger{EF_xe35_noMu} & 6.6 & 0.60 & 0.961 \\
    \trigger{EF_j75_jetNoEF_EFxe35_noMu} & 3.9 & 0.35 & 0.963 \\
    \trigger{EF_j75_jetNoEF_EFxe30_noMu} & 6.8 & 0.61 & 0.969 \\
    \trigger{EF_j50_jetNoEF_xe30_noMu} & 10.5 & 0.96 & 0.971 \\

    \bottomrule
  \end{tabular}
  \caption{
    The rates of different triggers,
    measured at an instantaneous luminosity of $\unit[10^{32}]{Hz/cm^2}$,
    together with the resulting prescales needed to limit the trigger rate to an assumed allotted rate of \unit[11]{Hz}
    and the signal yield $N_T/c$ as defined by \Eq{eq:results_trigger_rates_signal_yield}.
    The row with the primary trigger of the zero-lepton Supersymmetry analysis is highlighted.
  }
  \label{tab:results_trigger_rates_yields}
\end{table}

The instantaneous luminosity, which determines the trigger rates,
will be assumed to be $\unit[10^{32}]{Hz/cm^2}$ for this example,
about half the peak instantaneous luminosity in 2010. %
The allotted rate is set to \unit[11]{Hz},
which is the approximate rate of %
the trigger \trigger{EF_j75_jetNoEF_{\allowbreak}EFxe25_{\allowbreak}noMu}
at this instantaneous luminosity.
\Tab{tab:results_trigger_rates_yields} lists the rates of the triggers shown in \Fig{fig:results_trigger_rates_overview_2010},
together with the prescale $f^\text{pre}_T$
which would be needed to bring the rate within the limit of \unit[11]{Hz}.
The resulting signal yield is $N_T / c$.
The numbers in the table demonstrate that indeed the combined triggers give the best signal yields.
It should be noted that the yield depends strongly on the allotted rate, %
and a lower rate limit will usually prefer triggers with higher thresholds,
indicating that the loss in signal yield from higher prescales is more severe
than the loss from harsher online cuts,
a behavior which is expected for signals with high-energetic objects.
Even if triggers with higher thresholds seem capable of higher yields,
in comparison to a trigger with lower thresholds which can also be run unprescaled,
they would not exhaust the allotted rate and thus give a smaller number of events in the end.
\trigger{EF_j75_jetNoEF_EFxe25_noMu} has the best signal yield among the triggers which fully exploit the rate,
\ie have prescales larger than or equal to one,
and therefore seems to perform best.
This result is biased to some extent
because the rate limit has been adjusted to the rate of this particular trigger.
It is interesting to see that pure \met triggers perform quite well,
but their rates do not scale linearly with increasing instantaneous luminosity,
whereas those of the \jetmet triggers do.
The results of the \met triggers can therefore not be directly extrapolated to higher luminosities.
\revised{This is discussed in detail in \Sec{sec:results_met_model}.}

\subsubsection{Conclusion}

Considering the savings in terms of trigger rate together with the signal yield,
\Fig{fig:results_trigger_rates_trigger_yield_SU4_2010} shows that even for a low-mass point
the signal efficiency of the combined \jetmet trigger \trigger{EF_j75_jetNoEF_EFxe25_noMu} is still \percent{74}.
Thus in the worst case, only a quarter of events is lost,
having on the other hand a factor $10$
of reduction in trigger rate compared to single object triggers.
To sum up,
the search for Supersymmetry in the zero-lepton channel
can benefit considerably from using \jetmet triggers instead of single object triggers,
even if these triggers are more complex,
and more work is required to understand their behavior. %
Therefore, the next chapter will be dedicated to the study of the efficiencies of combined \jetmet triggers.

\section{Trigger Efficiencies: Methodology}
\label{sec:triggerefficiencies}

\subsection{Introduction}

The efficiency of a trigger, or trigger efficiency for short, is the probability of a trigger to accept a given event\footnote{
  Note that in the following, the term trigger can have two different meanings:
  It can refer either to a specific trigger chain or to the trigger system as a whole.
  Which meaning is intended can be inferred %
  from the context.
}.
It depends on the detector hardware and implementation details of the trigger system,
and needs to determined individually for every trigger chain that is defined for the online data taking.
The trigger efficiency will, of course,
also depend on the properties of the events themselves. %
This dependency may be of arbitrary complexity,
although a well-defined trigger should be designed such that these dependencies are reduced to the least possible.

The total number of interactions $N$ in a collider experiment %
is per definition given by the instantaneous luminosity \Linst and the cross section $\sigma$ of the process under study,
integrated over time,
\begin{equation}
  N = \int \Linst(t) \, \sigma \intd t.
\end{equation}
The number of recorded events $n$, however, may be smaller,
\begin{equation}
  n = \epsilon \, N,
\end{equation}
due to the trigger efficiency $\epsilon \leq 1$,
which includes the acceptance and online reconstruction efficiency.
To retrieve the true number of events $N = n/\epsilon$,
the efficiency of the trigger needs to be known.

Taking an \met trigger in \ATLAS as an illustation, %
the efficiency of this trigger will depend on the calorimeter,
the calibration and the details of the online measurement of \met,
and needs to be measured specifically for the \ATLAS detector.
The efficiency in terms of the probability for a given event to fire a given trigger
depends on the \met threshold of the trigger and the actual amount of \met in the event.
As the actual amount of \met is unknown,
the \met value which is reconstructed offline
is taken as the best available estimate to study the efficiency of the trigger.
Plotting the probability of the trigger to fire as function of the online \met measurement would yield a step function:
If the measured value is below the trigger threshold, the probability is zero;
if it is above, the probability is one.
In the ideal case,
the relation between the online and the offline measurement of \met is a monotonous function,
and the correlation is one.
But as the two measurements are not exactly the same,
the probability as function of the offline \met variable
will be a convolution of the step function and the relation between the online and the offline measurement.
As said above, there may also be additional dependencies on other event properties.
Indeed, for the \met trigger it can be seen on Monte Carlo
that the efficiency strongly depends on the type of \met
and is different for events with fake \met and events with real \met\footnote{
  This can be seen in \Fig{fig:results_trigger_performance_measurements_2d_efficiency_MC_LocHadTopo_projections}, for example.
}.
In case of the \met trigger,
it is \revised{challenging} to distinguish between fake and real \met,
and the trigger efficiencies therefore can only be measured as an average over events with both sources of \met,
which will result in larger uncertainties of the trigger efficiency.
But there are other examples,
in which the influence of the additional dependencies on the trigger efficiency can be determined,
\eg a potential dependence on the isolation of a muon which fires a muon trigger,
or the geometrical distance of two jets in an event accepted by a jet trigger.

As there sometimes seems to be some confusion about what is a trigger efficiency,
the following is intended to clarify the interpretation which shall be adopted here:
The trigger efficiency by itself is a figure which is the same for all analyses
because the trigger does not know about the offline selection.
The decision which is made by the trigger system online
therefore is independent of what happens later in the analysis.
Still, care has to be taken
because folding in the offline selection will lead to a seemingly different trigger efficiency.
If, for example, the offline selection requires many hard jets,
this will yield a sample of events which receives large contributions to \met from mismeasurements of the jets
(which is proportional to the jet \pt),
so that the measured efficiency of the \met trigger on this sample will differ from its efficiency on an unbiased sample. %
This is related to the fact that the trigger by definition acts as a filter,
and thus does preferentially select certain event topologies.
Thus, if the efficiency of the trigger seems to depend on the offline selection,
this is due to the impact of the offline selection on the sample on which the efficiency is measured,
rather than the trigger efficiency being different itself.
The trigger efficiency is also independent of the type of events in terms of the underlying physics process,
in the sense that it only depends on the objects which are reconstructed by the trigger and used in the trigger decision.
Along the same lines, it should be noted furthermore
that if the offline variable, as function of which the trigger efficiencies are presented,
depends on the event topology or type of event in a different way than the online measurement,
this can lead to a distorted shape of the turn-on curves or to dependencies on the offline event selection\footnote{
  This effect can be seen, for example, in \Fig{fig:results_trigger_performance_measurements_met_turnon_susymet_met_turnon_muonthreshold}.
}.
Such a distortion can be avoided by using an offline variable which comes as close to the online measurement as possible.
It is therefore important when interpreting trigger efficiencies
to take into account on which sample of events and as function of which variable they are presented.

\subsection{Terminology of Trigger Efficiencies}
\label{sec:results_trigger_performance_terminology}

\begin{figure}
  \centering
  \incgraphics{width=\widthsingleplot}{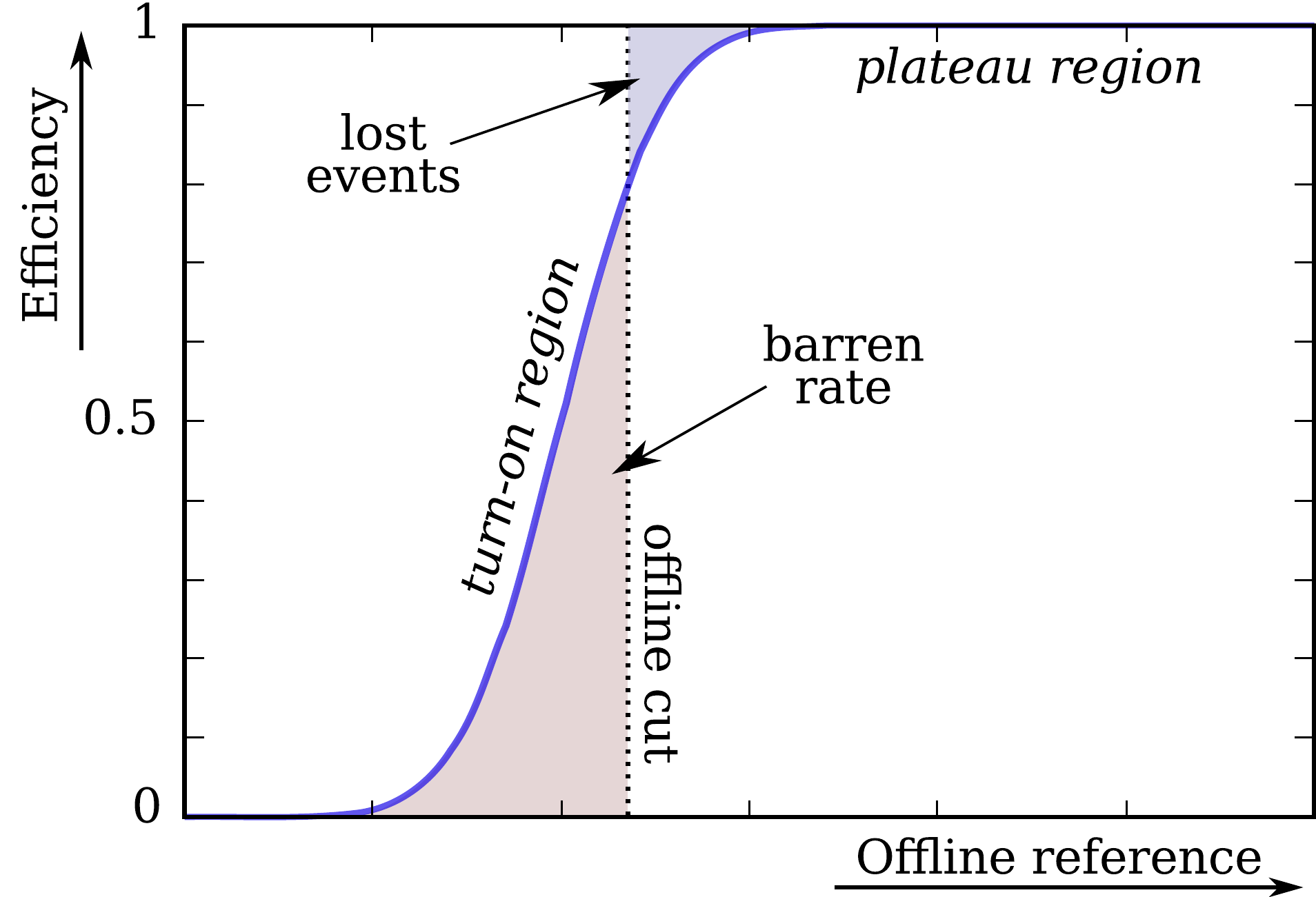} %
  \caption{
    Typical example of the efficiency of a trigger as function of some offline reference,
    explaining some of the terms which will be used in this text.
  }
  \label{fig:results_trigger_performance_schematic_turnon}
\end{figure}

In the following, the vocabulary needed for the discussion of trigger efficiencies will be introduced.
The plot of a trigger efficiency as function of the offline reference is referred to as the \define{\index{turn-on curve}} of a trigger.
A sketch of a one-dimensional turn-on curve is shown in \Fig{fig:results_trigger_performance_schematic_turnon}.
Turn-on curves often resemble a Gauss error function in shape
and have three distinct regions:
one where the efficiency is flat and close to zero for small values of the offline reference,
an intermediate region, in which the slope is large,
and the region at high values, where the efficiency flattens out again.
The latter is called the \define{plateau (region)}\inindex{Plateau region} of the efficiency curve.
The average efficiency in the plateau region is called the \define{\index{plateau efficiency}}.
For jet triggers, the plateau efficiency is usually close to one %
because jets are hard to miss if they have sufficiently high energy\footnote{
  Even in case parts of the calorimeter are malfunctioning,
  if this affects both online and offline measurement in the same way,
  the effect cancels out and is not visible in the trigger efficiency
  (cf. the impact of the \revised{so-called} \acs{LAr} hole on the efficiencies in \Sec{sec:results_trigger_performance_2011_consistency}).
}.
For similar reasons, the plateau efficiency of \met triggers is close to one,
whereas for \eg muon triggers it may well be below one
because muons may \revised{escape} through cracks in the muon spectrometer
or, being minimum ionizing particles, leave too weak a signal.
In any case, the efficiency for a well-defined trigger should be constant in the plateau region.
The region in which the trigger efficiency changes from zero to its plateau value is called the \define{\index{turn-on region}}.
If the turn-on region is broad, the turn-on is slow, if it is narrow, the turn-on is steep.
A steep or sharp turn-on is another indicator of a well-defined trigger
and of a high resolution of the online measurement with respect to offline.
The shaded regions in \Fig{fig:results_trigger_performance_schematic_turnon} highlight two different, undesired effects,
coming from the fact that the trigger efficiency is not a step function as function of the offline reference.
The ``barren'' rate is made up by events in which the trigger fires,
but which are not used in the analysis because they fall below the offline cut.
This increases the trigger rate,
but does not have a direct impact on the analysis and may be affordable or unavoidable.
``Lost'' events are events for which the trigger does not fire,
although they would have exceeded the offline threshold.
These events will be missing later in the analysis
and will lead to wrong conclusions if they are not accounted for.

\begin{figure}
  \centering
  \incgraphics{width=\widthsingleplot}{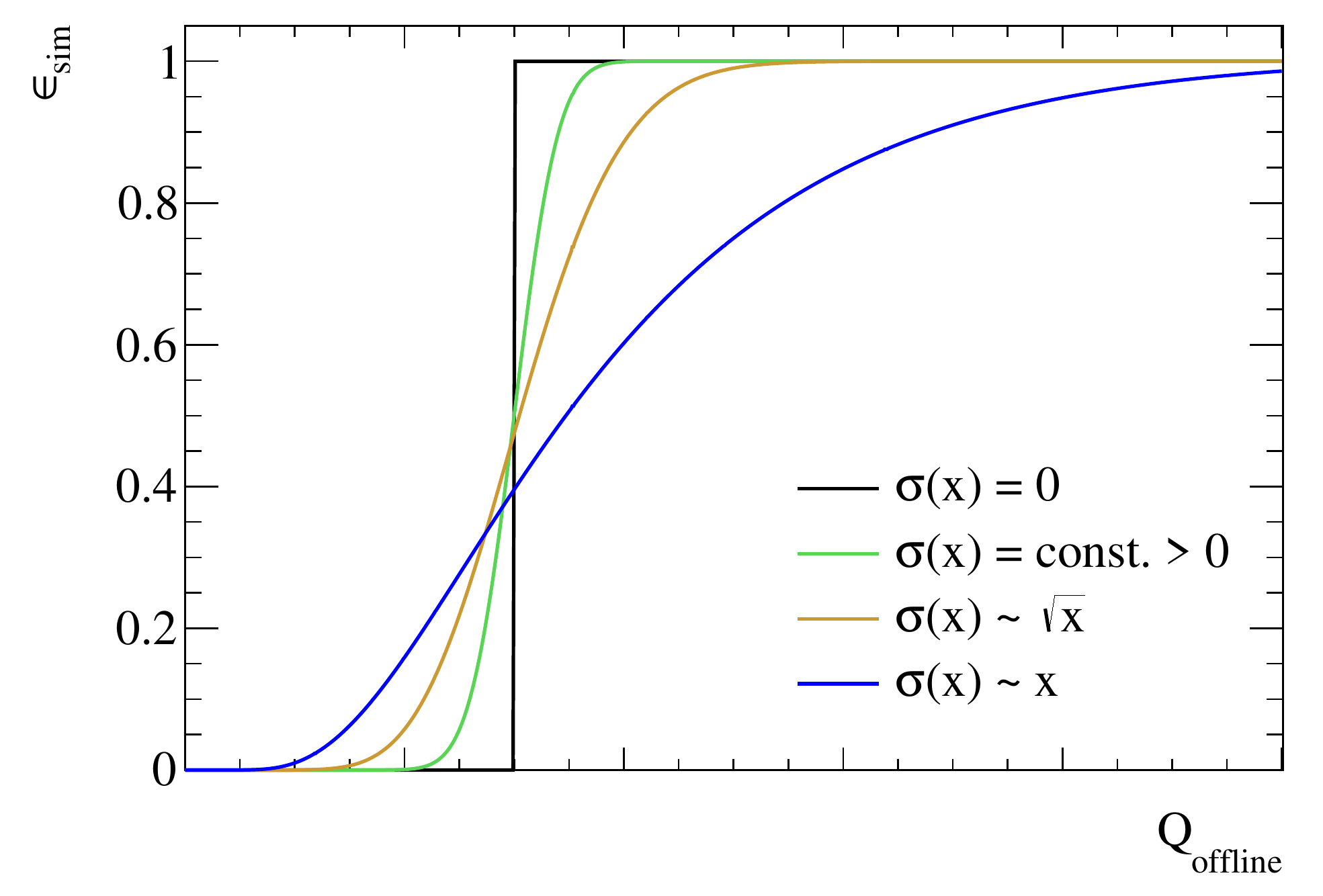}
  \caption{
    Schematic behavior of a typical trigger turn-on curve,
    showing the trigger efficiency $\epsilon_\text{sim}$ as function
    of the offline quantity $Q_\text{offline}$
    for different functional forms of the resolution $\sigma(Q_\text{offline})$
    of the online measurement with respect to offline.
  }
  \label{fig:results_trigger_performance_schematic_turnon_resolution_dependence}
\end{figure}

If the online measurement of the quantity the trigger cuts on
can be described as a smearing of the offline measurement with a Gaussian of constant width,
the trigger efficiency as function of the offline reference can be described by a Gauss error function.
In general, the resolution of the online measurement with respect to offline may change as function of the offline variable,
and then the trigger efficiency deviates from a pure Gauss error function.
This is demonstrated in the plot in \Fig{fig:results_trigger_performance_schematic_turnon_resolution_dependence},
where %
the trigger efficiency is computed for a resolution of the online measurement
which depends in different ways on the value of the offline reference.
Assuming that the resolution gets worse for higher values,
this makes the turn-on curve asymmetric and its upper part stretch more to the right.

\subsection{Estimation of Efficiencies}

The basic principle of the computation of a trigger efficiency is
counting the number of occurrences in which the trigger could have fired (the denominator of the efficiency estimate $n$),
and the number of occurrences in which it actually did (the enumerator $m$),
both as \revised{a} function of the offline variables which are used to \revised{parametrize} the trigger efficiency.
The ratio of the counts is used as the estimator of the trigger efficiency,
\begin{equation}
  \hat\epsilon = \frac{m}{n} \xrightarrow{n \to \infty} \epsilon,
\end{equation}
where $m$ obviously depends on $n$.
The hat, which is often used to denote an estimator, is omitted in the following,
identifying for practical applications the estimate of the trigger efficiency with the true efficiency.
The counting is usually done using two one- or higher dimensional histograms
with an appropriate binning in the relevant offline variables.
The difficult part is to determine which offline variables are relevant,
\ie in which variables the efficiency is not flat, %
and how to reliably do the actual counting,
in particular with respect to the choice of the event sample on which the counting is done.

So far, trigger efficiencies have been discussed in terms of probabilities which are assigned to every event as a whole.
In some cases, the probability for an event to fire a trigger can be broken down into object probabilities
which, appropriately combined, give an approximation of the \revised{total} event probability.
If the object probabilities are independent of the rest of the event,
this approximation is exact.
In many cases, the dependence is so small
that the approximation works well,
at least for low object multiplicities. %
If the trigger efficiency of an object described by a set of parameters $\vec o$ is given by $\epsilon_\text{obj}(\vec o)$,
the trigger efficiency of an event with $n$ \revised{such} objects,
\ie the probability of such an event to fire the trigger,
is given by
\begin{equation}
  \epsilon_\text{ev}(\vec o_1, \dots, \vec o_n) = 1 - \prod_{i=1}^n 1-\epsilon_\text{obj}(\vec o_i) ,
\end{equation}
under the assumption of independent object probabilities.

To be able to compute the object-wise trigger efficiencies,
a matching needs to be done
between the objects reconstructed in the processing of all online data by the trigger system,
which are called trigger objects,
and the respective objects which are reconstructed later in the offline processing of the recorded data.
In this matching, corresponding trigger objects and offline objects are identified,
usually based on \revised{their geometrical distance in $\eta \times \phi$ space}.
From the properties of a trigger object, it can be inferred whether this object has fired the trigger or not.
Alternatively, the information whether a given trigger object has issued the trigger signal %
is usually stored in
the event record.
For the binning of the trigger efficiency, not the value measured for the trigger object is relevant, %
but the value of the respective property of the corresponding offline object.
This is made explicit in the description of the concrete implementation of the computation of trigger efficiencies (cf. \Sec{sec:results_trigger_performance_measurements_introduction}). %

\subsection{Use Cases and Trigger Strategies}

Taking into account the trigger efficiency is an important aspect of every physics analysis
and needs to be done with care
because the large rate reduction accomplished by the trigger system,
which stands at the beginning of every analysis,
may have a considerable influence on the analysis results.
Generally speaking, the trigger efficiencies are needed to estimate the impact of the trigger selection
on the composition of the event sample on which the analysis is carried out.

To this end, the trigger efficiency can be exploited in different ways:
The simplest method,
for a trigger which attains near \percent{100} efficiency in its plateau region,
is to adjust the offline selection so that only events in the plateau region pass,
and then to assume the trigger to be fully efficient.
This approach is adopted in the official search for Supersymmetry in the zero-lepton channel  %
and the analysis presented in \Sec{sec:analysis_susysearch}. %
The \jetmet triggers used in this analysis fulfill the condition of
having a plateau efficiency near \percent{100} (cf. \Sec{sec:results_trigger_performance_measurements_data}). %
If changing the offline selection is undesirable or not possible for some reason,
the trigger efficiency can be taken into account by using the result of the trigger simulation in Monte Carlo
to accept and reject events in the same way as it is done in data.
This relies on the correctness of the trigger simulation,
and has the disadvantage that for low trigger efficiencies a considerable part of the Monte Carlo is lost,
namely all events which are rejected based on the (binary) simulated trigger decision.
For triggers with a maximum efficiency which is considerably below one,
like it is found \eg for the muon triggers in the barrel which only reach around \percent{75} \cite{TriggerPerf2010},
the fraction of discarded Monte Carlo events is sizeable.
In this case, it is better not to use the trigger simulation to reject these events,
but to assign a weight to each Monte Carlo event,
which reflects the trigger efficiency for this particular event
and thus the probability that such an event contributes to the data sample collected online.
This reweighting of events allows to extract the maximum possible information from the Monte Carlo.
Other analyses which, for example, aim at the determination of production cross sections,
need precise estimations of trigger efficiencies from data only
to be able to account for inefficiencies of the trigger
and obtain the true number of selected events.

\subsection{\texorpdfstring{Trigger Efficiencies and the $\beta$-Distribution} {Trigger Efficiencies and the Beta-Distribution}}
\label{sec:results_trigger_performance_introduce_beta}

The trigger efficiency is defined as the average probability of the trigger to fire
on a sample containing an infinite number of events.
For the finite case, the outcome of a series of statistically independent experiments
with success probability $p$ is in general \revised{described} by the Binomial distribution.
The probability of seeing $k$ successes in $n$ identical trials
according to this distribution is given by
\begin{equation}
  \Bi(k | n,p) = \binom{n}{k} p^k \left(1-p\right)^{n-k},
  \label{eq:definition_binomial_1} %
\end{equation}
where $\binom{n}{k}$ is the binomial coefficient \cite{Barlow1989}.
The measurement of trigger efficiencies yields two numbers for each bin $i$:
the total number of events or trials $n_i$ and the number of successes in which the trigger fired $k_i \leq n_i$.
From these two numbers, the probability distribution $P(\epsilon_i| k_i,n_i)$ for the true efficiency $\epsilon_i$ is to be computed.

In order to find the distribution $P(\epsilon_i| k_i,n_i)$
starting from the assumption that the distribution of $k_i$, given $n_i$ and $p_i$, is given by $\Bi(k_i| n_i,p_i)$,
Bayes' theorem can be employed.
In terms of (conditional) probabilities on a discrete set of events\footnote{
  ``Event'' being meant here in the general stochastic sense, not as the record of a particle collision.
} which includes \revised{the possible outcomes} $A$ and $B$,
it can be written as
\begin{equation}
  P(A|B) = \frac{P(B | A) \cdot P(A)}{P(B)}.
  \label{eq:results_trigger_performance_Bayes_discrete}
\end{equation}
For continuous probability distributions,
this translates into $P(\epsilon_i| k_i,n_i) \propto P(k_i| \epsilon_i, n_i) P(\epsilon_i| n_i)$.
Inserting the Binomial distribution and requiring the prior $P(\epsilon_i| n_i)$ distribution
for the efficiency to be independent of the sample size $n_i$ yields
\begin{equation}
  P(\epsilon_i| k_i,n_i) \propto \Bi(k_i| \epsilon_i, n_i) P(\epsilon_i).
  \label{eq:results_trigger_performance_epsilon_1}
\end{equation}
The conjugate prior distribution of the Binomial distribution is the \betadist, %
meaning that if a \betadist is inserted as the prior probability $P(\epsilon_i)$ in \Eq{eq:results_trigger_performance_epsilon_1},
the resulting distribution $P(\epsilon_i| k_i,n_i)$ belongs again to the family of {\betadist}s.

The {\betadist}s\inindex{beta distribution@\betadist}
$\beta(x| a, b)$ are a family of continuous probability distributions over the interval $x\in(0, 1)$
with two positive parameters $a$ and $b$.
They are given by %
\begin{equation}
  \beta(x|a,b) = \frac{x^{a-1}(1-x)^{b-1}} {\int_0^1 u^{a-1} (1-u)^{b-1} \intd u},
\end{equation}
and the following of their properties are relevant here:
\begin{alignat}{2}
  \label{eq:beta_dist_mean}
  \text{mean:}&& \quad \operatorname{E}(X) &= \frac{a}{a + b} \\
  \label{eq:beta_dist_var}
  \text{variance:}&& \quad \operatorname{V}(X) &= \frac{a b}{(a + b)^2 \, (a + b + 1)} \\
  \label{eq:beta_dist_mode}
  \text{mode:}&& \quad \operatorname{M}(X) &= \frac{a-1}{a + b -2}
\end{alignat}

From setting $P(\epsilon_i) = \beta(\epsilon_i|a,b)$ in \Eq{eq:results_trigger_performance_epsilon_1},
it follows that
\begin{equation}
  P(\epsilon_i| k_i,n_i) = \beta(\epsilon_i| a', b'),
\end{equation}
with $a' = a + k_i$ and $b' = b + n_i - k_i$ \cite{Casadei2009}.
In this thesis, a uniform (``flat'') prior $P(\epsilon_i)$ will be used,
because this makes the mode, the most likely value of the \betadist,
fall together with the ratio of the number of passed events $k_i$ over all events $n_i$.
This is the intuitive expectation for the trigger efficiency,
and it can also be motivated \latein{a priori} as the assumption of no value for the efficiency being more likely than any other.
The uniform distribution corresponds to parameters $a_1=b_1=1$,
so that the posterior probability of the trigger efficiency
is given by
\begin{equation}
  P(\epsilon_i| k_i,n_i) = \beta(\epsilon_i | k_i + 1, n_i - k_i + 1).
  \label{eq:results_trigger_performance_final_beta_with_uniform_prior}
\end{equation}
From \Eq{eq:beta_dist_mode}, it follows that the mode then is $k_i / n_i$ as expected.
The error bars which are given in the trigger efficiency plots in this thesis
indicate the smallest interval covering \percent{68.3}
of the \betadist obtained in the way described above.

\subsection{Overview of Methods}
\label{sec:results_trigger_methods}
In this section, an overview of methods is given that can be used to determine trigger efficiencies.
One way is to use Monte Carlo,
where the trigger efficiencies can be determined by direct counting.
This is sometimes referred to as \define{\index{Monte Carlo counting}}.
Monte Carlo counting relies on the correctness of the trigger simulation
and can be used to validate the other methods described below.
The efficiencies obtained from Monte Carlo are always specific to the type of events in the sample and thus biased.
This can be an advantage, for example to optimize the trigger on certain event topologies. %
However, it is not possible to measure the trigger efficiency for an unbiased sample in Monte Carlo
because statistics in Monte Carlo are too small to obtain meaningful results for triggers which select rare events.
A general advantage of data-driven techniques is
that they do not rely on the trigger simulation in Monte Carlo.

Therefore, the common objective of the methods presented here
to measure trigger efficiencies on data
is to obtain an unbiased sample of events
and thus an unbiased estimate of the trigger efficiency:
\begin{itemize}
  
  \item The \textit{tag \& probe} method exploits the fact that for a given two-body decay with known kinematics,
    from the observation of one of the decay products it can be inferred that also the other particle should be present in the event.
    Decays with two identical particles, usually leptons, in the final state are used.
    One of the particles is selected as the tag which is used to collect the event sample,
    the other particle is the probe which is used to perform the actual measurement of the trigger efficiency.
    The efficiency is computed as the number of probes firing the trigger divided by the total number of probes.
    The tag \& probe method can be used to study \eg
    electron \cite{CMS-PAS-EGM-07-001} %
    or muon \cite{ATLAS-CONF-2011-021}
    trigger efficiencies in decays of $Z$~bosons, $J/\psi$~mesons or $\Upsilon$~mesons. %
    A cut on the invariant mass of the reconstructed pair of particles
    \revised{and the back-to-back topology}
    can be used to obtain a well-defined sample of such decays.
    Note that in this method a sample dependence remains,
    in particular with respect to the kinematic distributions of the decay products.
    An important assumption is furthermore that the trigger efficiencies of the objects are uncorrelated,
    \ie the trigger efficiency for the probe is assumed to be independent of the fact
    that there is always another object of the same kind in the event which has fired the trigger.
    The tag \& probe method is also employed to compute (offline) reconstruction efficiencies~\cite{ATLAS-CONF-2011-063}.
  
  \item An event sample to compute an unbiased estimate of the trigger efficiency can also be collected using an \define{\index{orthogonal trigger}},
    which is a trigger that, in the ideal case, is completely independent of the trigger under study.
    It is not always possible to identify such a trigger and the orthogonality will often be only approximate.
    For example, for the computation of \met trigger efficiencies,
    muon triggers could be used to collect the event sample
    because the \met triggers in \ATLAS at the time of writing
    are independent of the measurements in the muon spectrometer.
    However, the muon selection will enhance the %
    fraction of events with $W\to\mu\nu_\mu$ decays,
    which have real \met from the neutrino,
    thereby changing the composition of \met in the event sample. %
  
    To obtain a sample with very little or no bias at all,
    \define{minimum-bias}\inindex{Minimum-bias trigger} or \define{\index{random trigger}s} can be used.
    For most studies, these triggers yield far too small statistics,
    which is obvious considering the magnitude of the rate reduction in the trigger system.
    To get down from the design collision rate of \unit[40]{MHz}
    to the rates of a typical physics trigger with a rate of a few Hertz,
    a rate reduction of the order of at least $10^6$ is necessary.
    This means that to collect a sample of a few thousands of events in which the target trigger actually fires,
    billions of events would have to be collected with random or minimum-bias triggers,
    and these triggers typically are run at rather low rates,
    so that even for the lowest thresholds direct studies on these samples are not feasible. %
    
  \item \define{Bootstrapping} overcomes the problem of insufficient statistics in samples taken with unbiased triggers
    by using a biased event sample instead, on which the trigger efficiencies are computed.
    The bias which is introduced by the sample trigger can then be corrected under certain conditions,
    using Bayes' theorem and the turn-on curve of the sample trigger,
    which must be known to perform the bias correction.

\end{itemize}
Bootstrapping is the main data-driven method used in this thesis to compute trigger efficiencies,
and is therefore explained in detail in \Sec{sec:bootstrapping}.
In this section and the following \Sec{sec:results_trigger_performance_addition_weighted_samples},
it is also described how to \revised{perform} the actual computation of the efficiency estimates from the event counts
and how to combine the trigger efficiencies from several weighted samples,
which is needed to correctly propagate the uncertainties of efficiencies computed on weighted Monte Carlo samples.

\subsection{Bootstrapping}
\label{sec:bootstrapping}

In this section, the notion of bootstrapping \revised{in the context of} measuring trigger efficiencies will be explained.
The trigger under study will be referred to as the target trigger in the following.
The trigger that is used to collect the sample of events on which the efficiency of the target trigger is measured
will be referred to as the sample trigger.
The basic idea has already been sketched in \Sec{sec:results_trigger_methods}. %
Instead of insisting to have an unbiased sample of events,
on which the efficiency of the target trigger is measured,
this sample may now also be collected using a sample trigger which introduces a bias.
This solves the problem that for triggers that are tailored towards selecting rare events,
often it is simply impossible to acquire an unbiased sample, comprised of minimum-bias or random events,
which contains enough of these rare events
so that it is possible to determine the efficiency of the target trigger.
Of course, measuring the trigger efficiency on a biased sample yields a biased measurement,
but this bias can be removed by applying the bootstrapping method.

\index{Bootstrapping} is based on Bayes' theorem\revised{,}
which is given by \revised{(cf. \Eq{eq:results_trigger_performance_Bayes_discrete})}
\begin{equation}
  P(T) = \frac{P(T|S) \cdot  P(S)}{P(S|T)}.
  \label{eq:results_trigger_performance_Bayes_discrete_repeat}
\end{equation}
The probabilities $P(S)$ and $P(T)$ can now be identified with the probability
that the sample trigger chain issues a trigger signal in a given collision event, $P(S)$,
and the probability that the target trigger issues a trigger signal in this collision event, $P(T)$.
This makes $P(T|S)$ the probability of the target trigger to fire in an event where the sample trigger has fired,
which is true for all events from the sample which have been collected using the sample trigger.

If the probability $P(S|T)$ in the denominator were known,
the unconditional (unbiased) probability $P(T)$ of the target trigger to fire could be computed from $P(T|S)$ and $P(S)$.
In practice, the sample and target trigger are chosen such that \inindex{Bootstrapping condition}
\begin{equation}
  P(S|T) = 1,
  \label{eq:bootstrap_requirement}
\end{equation}
\ie if in a given event the target trigger fires, the sample trigger fires with probability one, too.
It is then possible to calculate the unbiased target trigger efficiency $P(T)$
from the trigger efficiency of the sample trigger $P(S)$
and the biased efficiency of the target trigger on the sample of events collected with the sample trigger:
\begin{equation}
  P(T) = P(T|S) \cdot P(S).
  \label{eq:bootstrap_multiplication}
\end{equation}
This is the central equation of the bootstrapping method.
Although in this simple derivation, the necessity of \Eq{eq:bootstrap_requirement} for this method to work is obvious,
it should be stressed that it is vital to make sure that this condition really holds when applying \Eq{eq:bootstrap_multiplication}.
If both the sample trigger and the target trigger make their decision based on the same online trigger quantity,
e.g. on the \met reconstructed by the trigger,
and if the sample trigger has a threshold lower or equal to the target trigger,
then \Eq{eq:bootstrap_requirement} holds by definition.

\pagebreak
In short, the following three \revised{conditions must be met to apply} the bootstrapping method:
\begin{itemize}
  \item The unbiased sample trigger efficiency needs to be known,
    either from another method
    or by bootstrapping the sample trigger efficiency from a third (even more inclusive) trigger.
  \item The target trigger efficiency needs to be measured with respect to the sample trigger,
    \ie all events which fire the sample trigger go into the denominator
    and all events which fire the target trigger go into the enumerator.
  \item The sample trigger must be chosen such that \Eq{eq:bootstrap_requirement} holds.
    This will be discussed in \Sec{sec:results_trigger_performance_sample_triggers}.
\end{itemize}

\subsubsection{Propagation of Uncertainties}
\Eq{eq:bootstrap_multiplication} is written in terms of discrete probabilities,
rather than the probability distribution which appears in \Eq{eq:results_trigger_performance_final_beta_with_uniform_prior}
as the result of the computation of the trigger efficiency.
As such, it only holds in the case of infinite statistics,
\ie if the probabilities were known exactly.
The form of \Eq{eq:results_trigger_performance_final_beta_with_uniform_prior} expresses the fact that this is not the case,
but only estimates of the efficiencies can be derived from the measurement.
It accounts for the uncertainty on the efficiency estimates by describing the efficiency in terms \revised{of} a probability distribution.
The generalization of Bayes' theorem and \Eq{eq:bootstrap_multiplication} to probability distributions is straightforward,
and then implies the multiplication of two probability distributions.
Note that this multiplication does not depend in any \revised{way} on offline or online trigger quantities,
but solely and independently on the counts in the respective bins of the enumerator and denominator histograms.

This multiplication of two probability distributions suffers from a fundamental problem.
In case of low event counts, both distributions are relatively broad
so that, even if the estimator for both efficiencies is one %
and the probability distribution for both efficiency estimates is peaked at one,
the multiplication will yield a distribution that no longer is peaked at one due to the tails.
This leads to a systematic underestimation of the efficiency,
and in particular makes the efficiency appear to decrease for the bins at high \revised{values of the offline reference},
which \revised{typically} have less statistics.
It is then not clear whether this decrease stems from the decrease in statistics
or is due to \revised{an inherent deficiency of} the trigger algorithm.
To amend this problem, a reweighting of the product probability distribution can be done\revised{,
as it is explained in %
the next section.}
The actual multiplication of the probability distributions (which are {\betadist}s) may be done analytically \cite{Bhargava1981},
but for simplicity a numerical approach is preferred here.
It consists of randomly sampling numbers from the two distributions and filling a histogram with the product.
From this histogram\revised{ } the mode and the smallest interval covering \percent{68.3} \revised{are} computed
as the estimate of the efficiency and its lower and upper confidence intervals,
covering quantiles equivalent to one \revised{standard deviation} for a Gaussian distribution. %

\subsubsection{Accounting for Bias from Bootstrapping: Logarithmic Reweighting}
\label{sec:lnreweighting}

\begin{figure}
  \centering
  \incgraphics{width=\widthtwoplots}{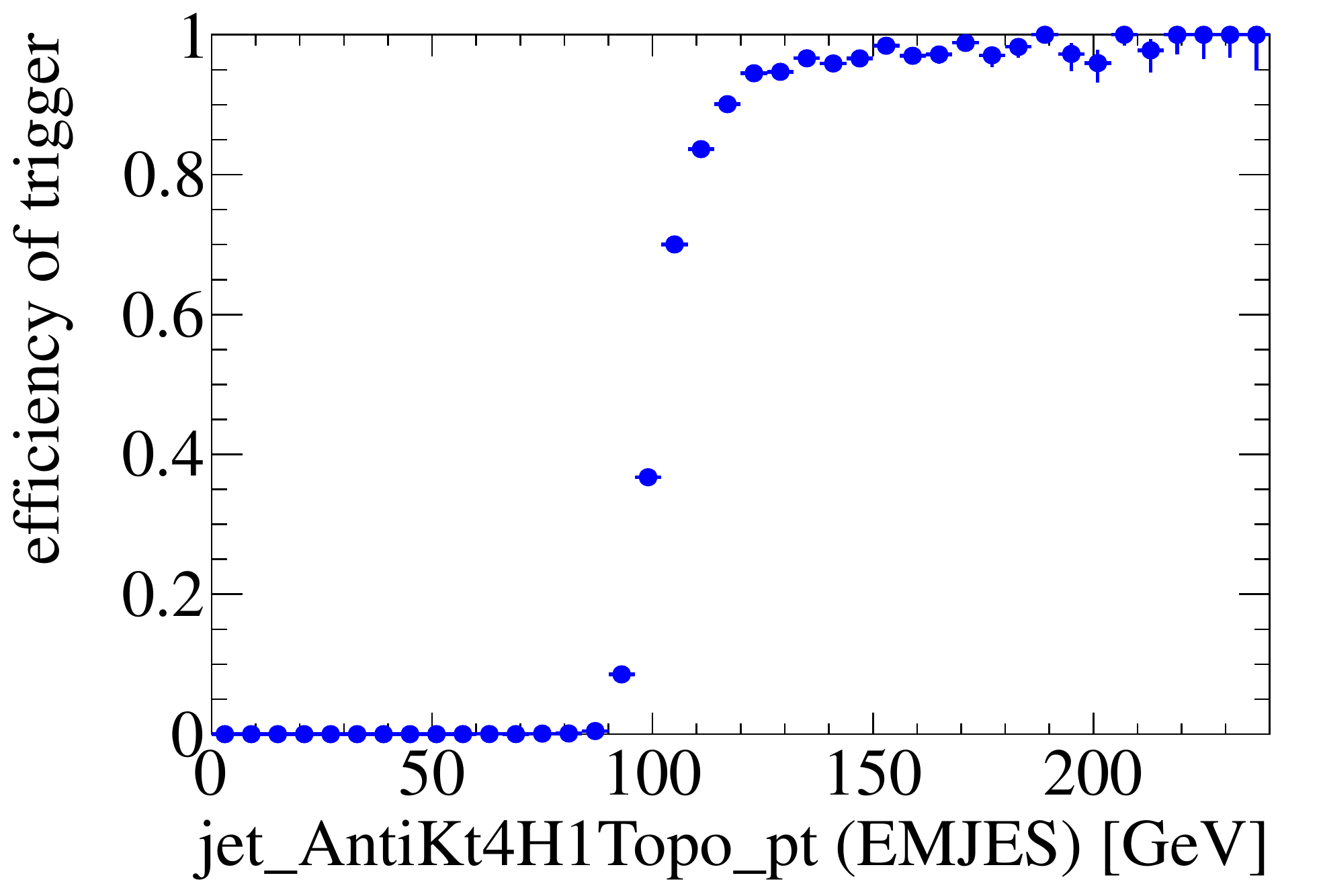}
  \hfill
  \incgraphics{width=\widthtwoplots}{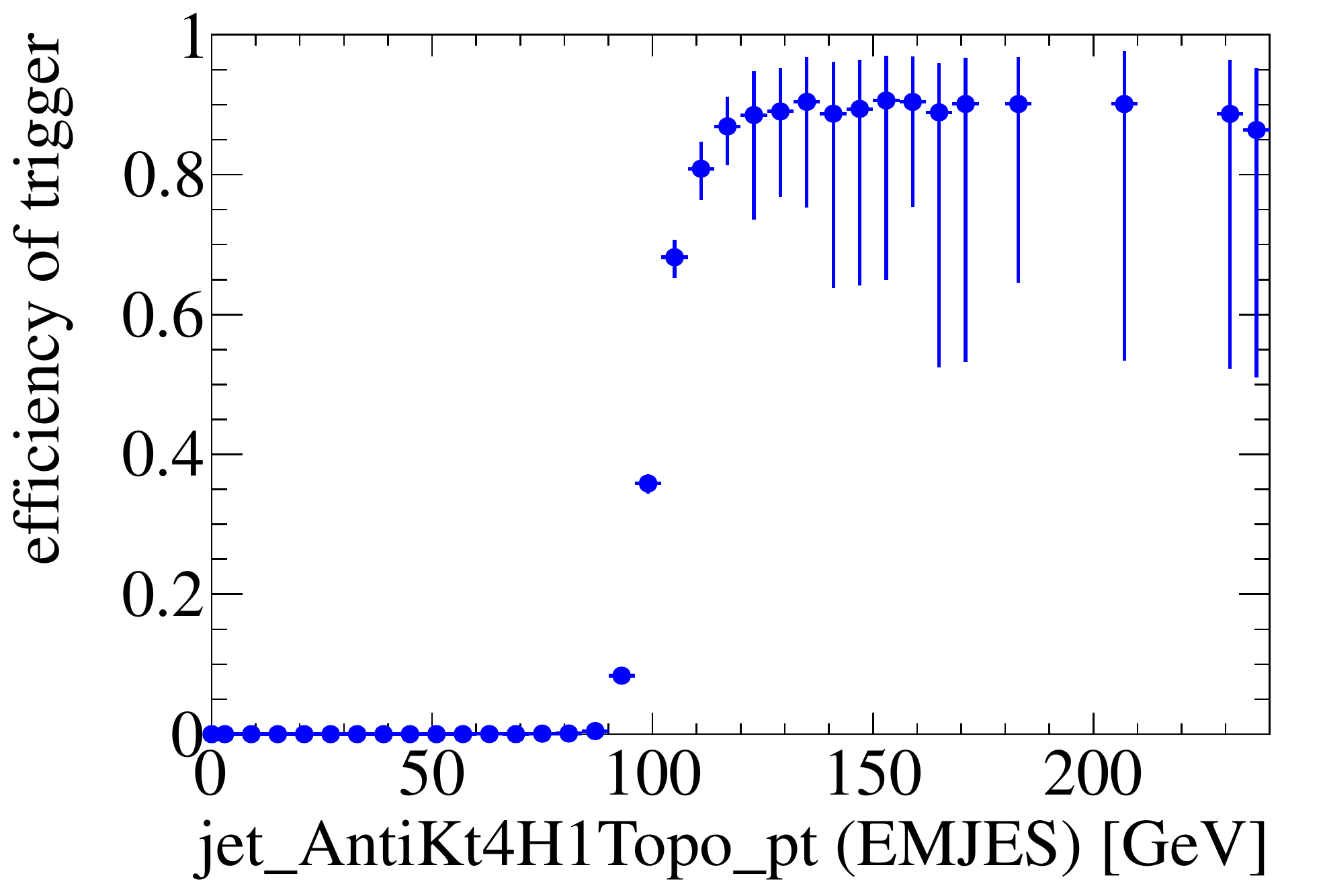}
  \\
  \incgraphics{width=\widthtwoplots}{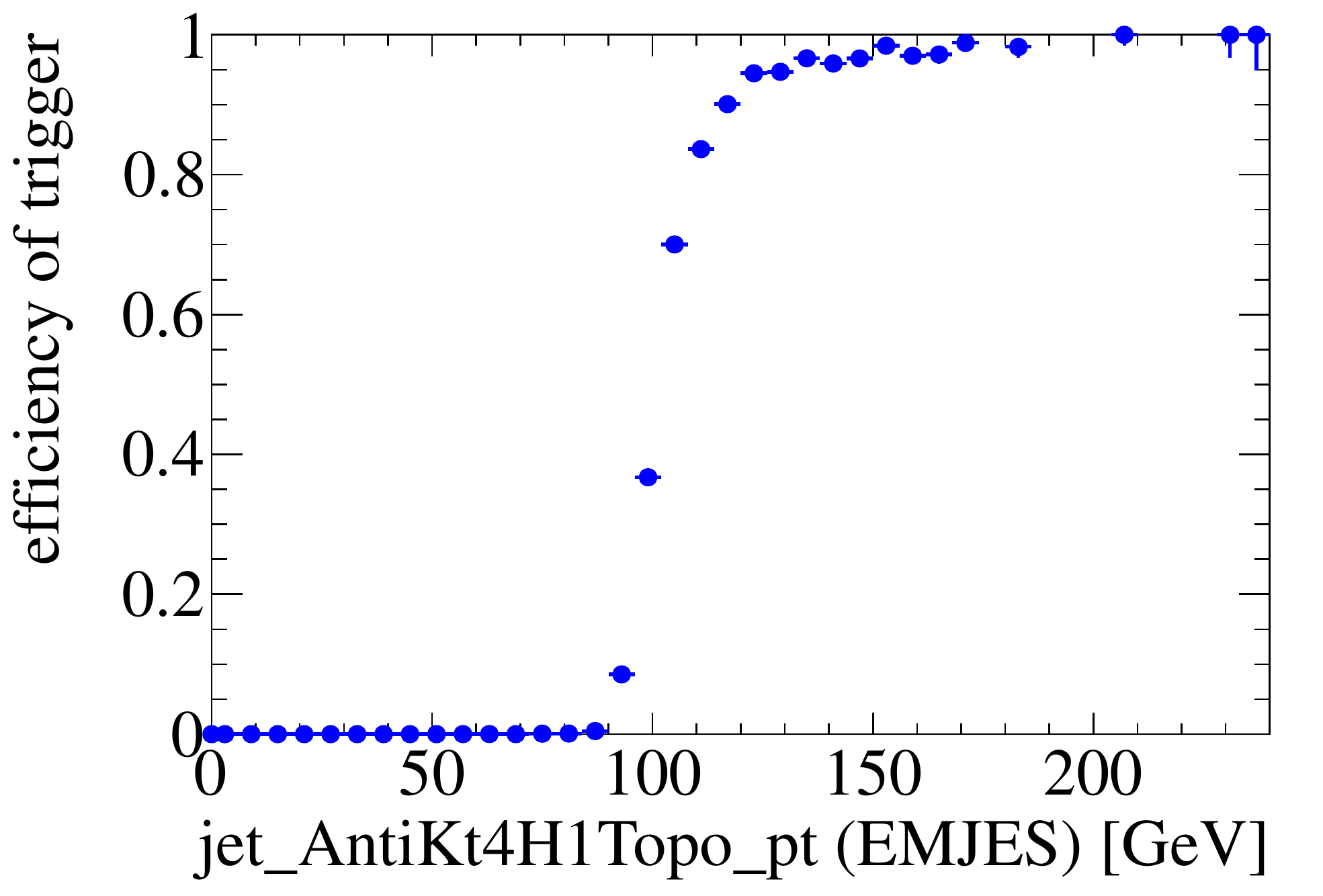}
  \hfill
  \incgraphics{width=\widthtwoplots}{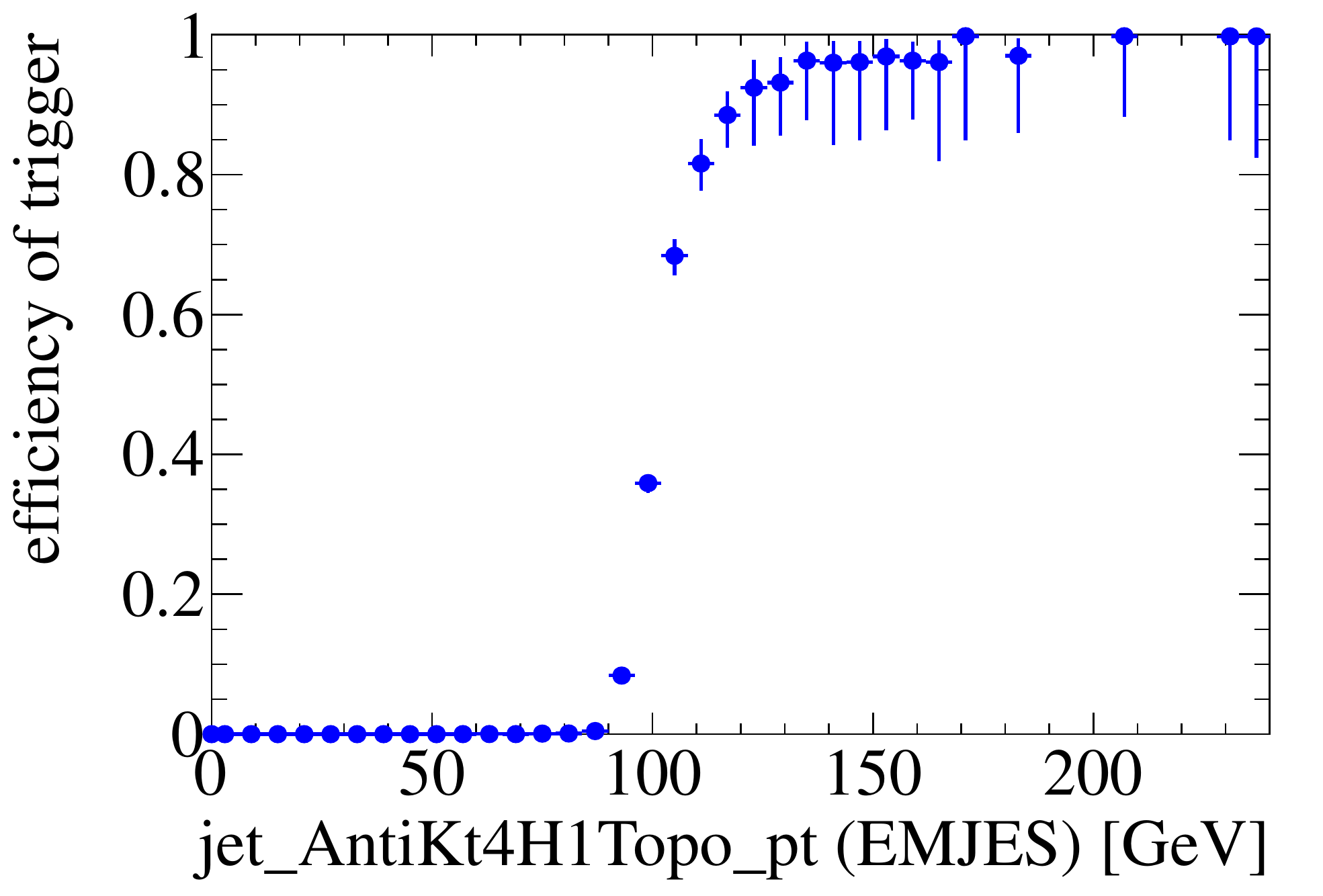}
  \caption{
    Turn-on of the jet trigger \trigger{EF_j75_jetNoEF} measured on events taken with \trigger{EF_j30_jetNoEF} in 2010 G~--~I with different methods:
    Biased turn-on before bootstrapping (upper left), after bootstrap correction of bias (upper right),
    after bootstrapping without error propagation (lower left) and with bootstrapping and reweighting (lower right).
    The offline reference is the \pt of jets from the \tagname{AntiKt4H1Topo} collection at \EMJES scale (cf \Sec{sec:software_jet_reconstruction}).
  }
  \label{fig:results_trigger_performance_jet_turnon_four_versions}
\end{figure}
The multiplication of two {\betadist}s, as is necessary in the propagation of uncertainties in the bootstrapping method,
will in general not leave the mode invariant.
In particular, the mode of the product distribution is lower than the product of the modes,
if the two distributions are peaked at one.
This will lead to an underestimation of the trigger efficiency,
if the probability distributions are broad as is the case for low statistics.
This is demonstrated in \Fig{fig:results_trigger_performance_jet_turnon_four_versions},
using the example of the efficiency of the jet trigger \trigger{EF_j75_jetNoEF} (target trigger),
which is measured on events taken with \trigger{EF_j30_jetNoEF} as sample trigger in 2010 G~--~I.
The four plots compare four different ways of computing the efficiency.
The upper left plot shows for comparison the biased efficiency of the target trigger
measured on the sample of events taken with \trigger{EF_j30_jetNoEF} before bootstrapping.
In the other three plots, bootstrapping is applied.
For the lower left plot, it was assumed that the estimate of the efficiency of the sample trigger efficiency is exact,
\ie the uncertainties of the estimate of the sample trigger efficiency are not propagated,
effectively multiplying with the mode instead of the full distribution of the sample trigger efficiency\footnote{
  Note that the two plots to the left are the same, except that in the lower version some points are missing.
  Bootstrapping does not change the mode of the efficiency estimate of the \trigger{EF_j75_jetNoEF} here,
  because \trigger{EF_j30_jetNoEF} reaches its plateau below the onset of the turn-on region of \trigger{EF_j75_jetNoEF}.
  This is typical for jet triggers which usually have comparably steep turn-on curves.
  The missing points stem from empty bins in the denominator of the sample trigger efficiency,
  where the efficiency of the sample trigger is thus undefined.
}.
This plot serves as a reference of what the bootstrapping method ought to return as the estimate of the target trigger efficiency
(but not for the uncertainties, which naturally are too small).
Comparing this to the upper right plot,
which shows normal bootstrapping using the full \revised{sample trigger efficiency} distribution,
the underestimation of the efficiencies is clearly visible in the plateau.

\inindex{Logarithmic reweighting}
The solution to this problem is a phenomenological approach,
in which the resulting product distribution is reweighted.
This approach has been introduced and described in \cite{Hensel:1306492},
and is motivated by considering the product $Y=X_1\cdot X_2$ of two independent random variables $X_1$ and $X_2$
with uniform distribution over the interval $[0,\,1]$. %
Its probability distribution is given by
\begin{equation}
  f_Y(y) = {\cal N} \iint f_{X_1}(x_1) \, f_{X_2}(x_2) \, \delta(x_1 \cdot x_2 - y) \intd x_2 \intd x_1,
  \label{eq:results_trigger_performance_factor_rev1}
\end{equation}
where ${\cal N}$ is a suitably chosen normalization factor,
$f$ denotes the probability density function of the respective random variable,
and the integrals are over the support of the probability distributions.
For $0\leq y\leq1$, the integrals evaluate to $f_Y(y) = -\ln y$. %
This factor can be understood from considering the extreme cases:
There is only one possibility to combine $x_1$ and $x_2$ to obtain $y=1$ ($= 1\cdot1$),
but infinitely many for $0$ ($= 0\cdot x \; \forall x\in[0,\,1]$).
Identifying $X_1$ and $X_2$ with the trigger efficiencies which are multiplied in \Eq{eq:bootstrap_multiplication},
a factor $-1/\ln(y)$ is used to reweight the resulting distribution.
This corrects for the combinatoric effect involved in the efficiency multiplication,
which would otherwise lead to the underestimation of the bootstrapped trigger efficiency discussed above.
This is not a stringent derivation,
but is shown in \cite{Hensel:1306492} and in examples in the following to work reasonably well,
in the sense that it can remove the bias without distorting the propagated uncertainties.

\begin{figure}
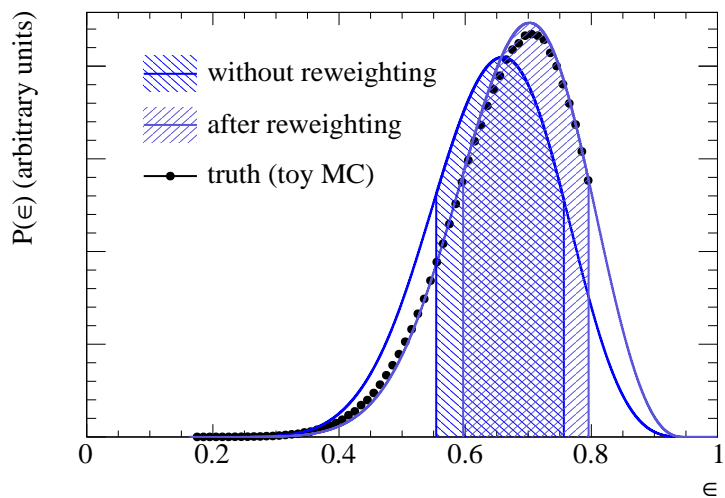

  \centering
    \incgraphics{width=\widthsingleplot}{{{test.trigeff.reweight_sample0.8_demo0.7shaded}}}
  \caption{
    Example of the impact of the reweighting correction for bootstrapped efficiencies,
    in a toy Monte Carlo study for a specific choice of parameters (see text),
    comparing the bootstrapping results with and without reweighting (blue lines)
    against the distribution of the true efficiency (black dots, the error bars are smaller than the marker size).
    The shaded regions give the smallest intervals covering \percent{68.3},
    the curves are normalized to have the same area in the range $[0,\,0.8]$.
  }
  \label{fig:results_trigger_performance_bootstrap_reweighting_toy_MC_example}
\end{figure}

First evidence is given in the lower right plot in \Fig{fig:results_trigger_performance_jet_turnon_four_versions},
in which the bootstrapping is combined with a subsequent reweighting.
It shows indeed that the efficiencies are consistent with the left plots,
and an improvement over the upper right plot which does not have the reweighting.
Moreover, the error bars appear to have the correct magnitude.
The plot in \Fig{fig:results_trigger_performance_bootstrap_reweighting_toy_MC_example} shows the result of a study on toy Monte Carlo.
It compares the distribution of the efficiency estimate obtained from bootstrapping,
with and without the reweighting factor applied,
\revised{to} the true distribution of the efficiency
and demonstates that here the reweighted distribution matches the true distribution \revised{much} better.
For this plot, the efficiency of the sample trigger was chosen to be $16/20 = 0.8$,
and the efficiency of the target trigger chosen to be $0.7$
so that the relative efficiency of the target trigger on the sample is $14/16 = 0.875$,
the fraction giving the event counts which dictate the uncertainties.
Note that \Eq{eq:bootstrap_requirement} limits the true efficiency to values below $0.8$.
To create the histogram of the true efficiency,
care has to be taken not to do the toy Monte Carlo the wrong way round:
In bootstrapping, the efficiencies are estimated
from the event counts in the enumerator and denominator histograms for the sample trigger and for the target trigger.
Thus, the toy Monte Carlo must yield the true distribution of the target trigger efficiency
for a given observed number of the event counts,
not for a given target trigger efficiency.

\subsubsection{Systematic Studies and Improved Reweighting Factor}

In the course of the trigger studies,
with the increase of statistics available for the computation of the combined trigger efficiencies,
it became clear that the preformance of the reweighting approach is not optimal under certain circumstances.
This affects in particular the computation of efficiencies of the combined \jetmet trigger
using a jet trigger as sample trigger,
where a one-dimensional turn-on curve describing the jet trigger efficiency as function of jet \pt is employed for the bootstrapping.
In that case,
it may happen for a bin with high \met in the plateau region
that the event counts for the efficiency of the target trigger are much smaller than for the sample trigger at the same offline jet \pt.
In particular, the denominator for the target trigger may be much smaller than the enumerator for the sample trigger,
which would otherwise be equal if no additional binning (here: in \met) were done.
This imbalance makes the probability distribution of the efficiency of the sample trigger very narrow,
whereas the probability distribution for the target trigger efficiency is broad,
and in this edge case,
the reweighting factor $-1/\ln(y)$ leads to an overestimation.

\begin{figure}
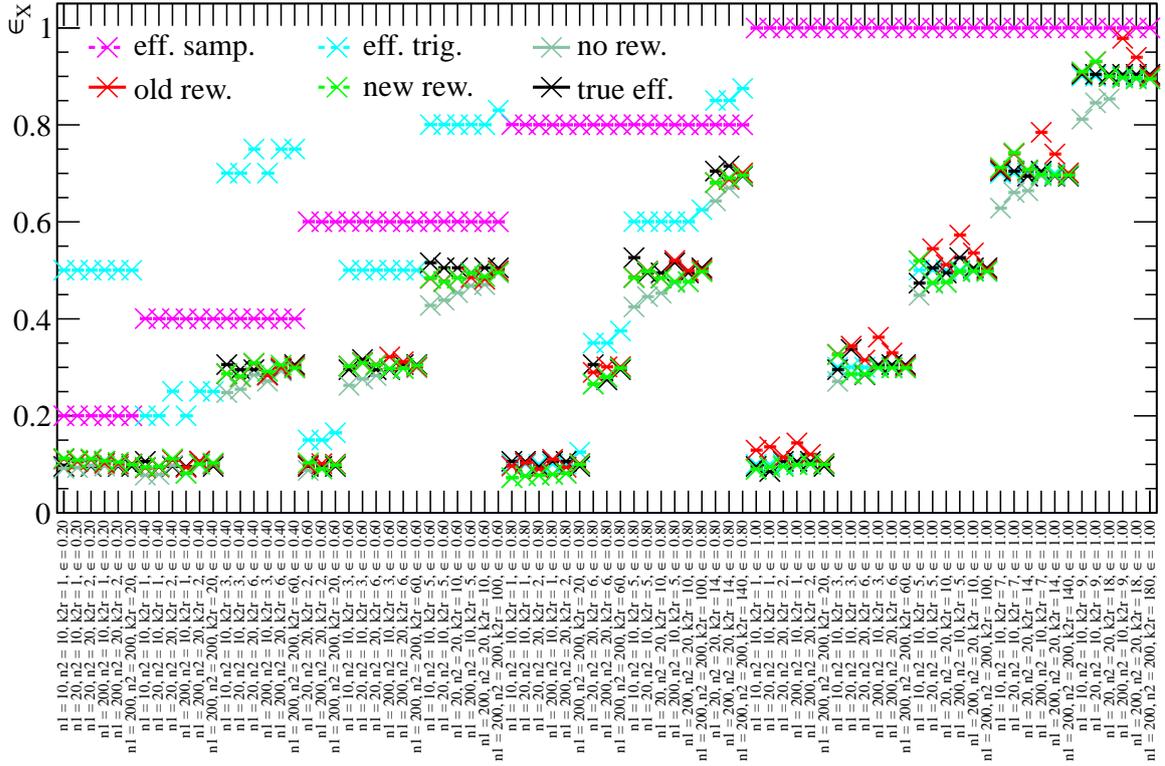

  \centering
    \incgraphics{width=\textwidth}{{{runner_test.trigeff_3_noe}}}
  \caption{
    Comparison of the estimates of the target trigger efficiency (eff. trig.) from bootstrapping
    when applying the new or old reweighting (new / old rew.),
    against the true target trigger efficiency (true eff.) in a toy Monte Carlo study.
    The plot also shows the sample trigger efficiency (eff. samp.)
    and the bootstrapped efficiency without reweighting (no rew.).
    The numbers in the labels on the horizontal axis are
      the event counts in the denominator for the efficiency measurement of the sample ($n_1$) and target trigger ($n_2$),
      the observed number of events in which the target trigger fires on the event sample taken with the sample trigger ($k_{2r}$)
      and the assumed sample trigger efficiency ($\epsilon$).
    $\epsilon_X$ on the vertical axis stands for the different efficiencies explained in the legend.
  }
  \label{fig:results_trigger_performance_bootstrap_reweighting_toy_MC_systematic}
\end{figure}

It turns out that an improvement of the reweighting factor can be achieved
by restricting the integral in \Eq{eq:results_trigger_performance_factor_rev1}
to the interval over which the numerical multiplication is done instead of the full interval $[0,\,1]\times[0,\,1]$ \cite{PCHamer1}.
The numerical multiplication effectively uses only a subset of $[0,\,1]$.
Due to discretization effects,
bins outside this subset do not contribute to the product in the sampling process,
because their weight is so small that they will never be hit in the random sampling.
Denoting these limits with $a_1 \leq x_1 \leq b_1$ and $a_2 \leq x_2 \leq b_2$,
\Eq{eq:results_trigger_performance_factor_rev1} turns into
\begin{align}
  f_Y(y) &= {\cal N} \int_{a_2}^{b_2} \int_{a_1}^{b_1} f_{X_1}(x_1) \, f_{X_2}(x_2) \, \delta(x_1 \cdot x_2 - y) \intd x_2 \intd x_1 \\
    &= {\cal N} \int_{a_2}^{b_2} \frac{1}{x_2} \left[ \Theta(b_1 x_2 - y) - \Theta(a_1 x_2 - y) \right] \intd x_2,
\end{align}
writing $\Theta(x)$ for the Heaviside step function. The expression in square brackets translates into the condition
\begin{equation}
  \frac{y}{b_1} \leq x_2 \leq \frac{y}{a_1},
\end{equation}
so that, writing $I \definedas [\min(b_2, y/a_1) ,\, \max(a_2, y/b_1)]$ for the domain of integration,
\begin{align}
  \folgt f_Y(y) &= {\cal N} \int_I \frac{1}{x_2} \intd x_2 \\
                &= {\cal N} \ln\left( \frac{\min(b_2, y/a_1)}{\max(a_2, y/b_1)} \right).
\end{align}
For $a_1 = a_2 = 0$, $b_1 = b_2 = 1$, this goes over into $f_Y(y) = -\ln y$ again.

\Fig{fig:results_trigger_performance_bootstrap_reweighting_toy_MC_systematic}
is the result of a toy Monte Carlo study of the same type as described above.
It shows a systematic comparison of the estimates of the target trigger efficiency from bootstrapping
when applying the new or old reweighting, %
and the true target trigger efficiency.
Each test is defined by
the event counts in the denominator for the efficiency measurement of the sample ($n_1$) and target trigger ($n_2$),
the observed number of events in which the target trigger fires on the event sample taken with the sample trigger ($k_{2r}$)
and the assumed sample trigger efficiency ($\epsilon$).
The plot \revised{conveys} three messages:
\begin{itemize}
  \item It confirms the underestimation of the efficiency without the reweighting.
  \item It demonstrates the improvement achieved by reweighting, where the old reweighting overestimates the efficiency under the conditions described above.
  \item It shows the better agreement of the new reweighting with the true value compared to the old reweighting in most cases.
\end{itemize}
The histograms in \Fig{fig:results_trigger_performance_bootstrap_reweighting_toy_MC_systematic_distribution}
have been created by repeating the toy Monte Carlo on a larger set of parameters.
They show the distribution of the discrepancies
between the true and the non-reweighted efficiencies,
and between the true efficiencies and the efficiencies with the old and new reweighting,
all scaled by the inverse of the symmetrized uncertainties.
It demonstrates the three items listed above in less detailed representation.
To quantify the differences between the three curves,
a fit with a Gaussian function is done,
which yields the fit parameters in \Tab{tab:results_trigger_performance_fit_gauss}.
The mean $x_0$ moves closer to zero, which shows the improvement of new reweighting over the old reweighting.
Note that the composition of the test parameters is arbitrary to a certain degree,
so that the slight downward bias of the new reweighting may stem from the dominance of high efficiencies.
Finally,
\Fig{fig:results_trigger_performance_bootstrap_reweighting_toy_MC_systematic_distribution_uncertainties} compares
the distribution of the error of the computed uncertainties on the efficiency estimates between the three approaches.
It shows that basically the reweighting does not change the distribution of the uncertainties,
which is important because the reweighting would be pointless if it changed the uncertainties:
bootstrapping without error propagation can be achieved without multiplying two distributions (see above).
In conclusion, for the efficiency plots which rely on bootstrapping in 2011
and for which the edge cases causing a slight overestimation have been observed for the first time,
the new reweighting is used by default.

\begin{table}[t]
  \centering
  \begin{tabular}{crrr}
    \toprule
                & \multicolumn{3}{c}{Fit parameter} \\
    \cmidrule{2-4}
    Reweighting & \multicolumn{1}{c}{$\cal N$} & \multicolumn{1}{c}{$x_0$} & \multicolumn{1}{c}{$\sigma$} \\
    \midrule
    none & 112(7) & -0.105(6) & 0.116(4) \\
    old  & 119(7) &  0.036(6) & 0.110(4) \\
    new  & 116(7) & -0.021(6) & 0.115(4) \\
    \bottomrule
  \end{tabular}
  \caption{
    Fit parameters from a fit of the three distributions in \Fig{fig:results_trigger_performance_bootstrap_reweighting_toy_MC_systematic_distribution}
    with a Gaussian function $f(x) = {\cal N} \exp\left(-\frac{(x-x_0)^2}{2\sigma^2}\right)$.
  }
  \label{tab:results_trigger_performance_fit_gauss}
\end{table}

\begin{figure}[t]
  \centering
    \incgraphics{width=\widthsingleplot}{{{runner_test.trigeff_3_pulls}}}
  \caption{
    \revised{Pull distribution} of the estimates of the target trigger efficiency from bootstrapping
    when applying no, the old or the new reweighting. %
    \Fig{fig:results_trigger_performance_bootstrap_reweighting_toy_MC_systematic} shows a subset of the data used to fill these histograms in detail.
  }
  \label{fig:results_trigger_performance_bootstrap_reweighting_toy_MC_systematic_distribution}
\end{figure}

\begin{figure}[t]
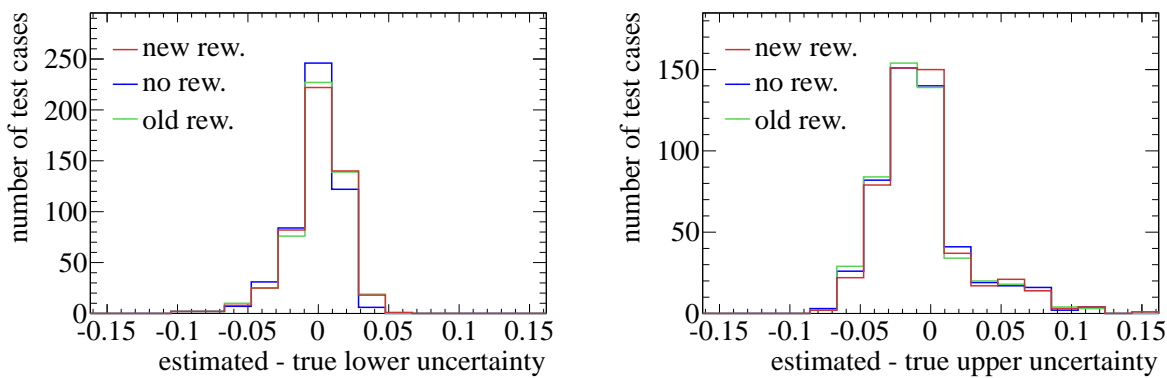

  \incgraphics{width=\widthtwoplots}{{{runner_test.trigeff_3_leerr}}}
  \hfill
  \incgraphics{width=\widthtwoplots}{{{runner_test.trigeff_3_heerr}}}
  \caption{
    Absolute error on the uncertainty intervals of the efficiency estimate.
    The plots show the differences of the lower (left plot) and upper (right plot) uncertainty
    to the true uncertainty from the toy MC experiments,
    for the new and the old reweighting method and without reweighting.
  }
  \label{fig:results_trigger_performance_bootstrap_reweighting_toy_MC_systematic_distribution_uncertainties}
\end{figure}

\subsubsection{Bootstrapping and Projections}
The efficiencies of combined triggers like the \jetmet triggers require at least a two-di\-men\-sio\-nal binning
to account for the two different offline quantities which are involved.
To study the turn-on behavior of one of the two components of the trigger in detail,
a projection onto one of the two axes is convenient.
This is done in many examples in \Secs{sec:results_trigger_performance_measurements_MC} and \Sec{sec:results_trigger_performance_measurements_data}.
If a one-dimensional turn-on of the sample trigger is used,
which is the case throughout all of the studies in these sections\footnote{
  The one exception being \Fig{fig:results_trigger_performance_measurements_turnons_jetmetf30_topo}.
}. %
The projection however can only be combined with bootstrapping
if the projection axis \revised{coincides} with the axis in which the binning of the sample trigger efficiency is done:
In this case, it is possible to add up all entries from the enumerator and denominator bins above a given threshold on the orthogonal axis,
and multiply the resulting probability distribution of the efficiency with the known probability distribution of the sample trigger efficiency for this bin.
In the opposite case,
when the axis in which the sample trigger efficiency is binned and the projection axis are orthogonal,
it is not possible to perform bootstrapping.
Only a weighted projection can be done,
which by multiplying with the efficiency of the sample trigger in the respective bin yields the correct efficiency estimate,
but cannot propagate the errors correctly.
This corresponds to the assumption that the error on the sample trigger efficiency can be neglected,
which in general is a valid assumption due to the cut being applied,
which, if it brings the trigger into its plateau,
also ensures that the sample trigger has reached the plateau of its efficiency.
\subsection{Computation of Trigger Efficiencies on Weighted Samples}
\label{sec:results_trigger_performance_addition_weighted_samples}
Often when generating Monte Carlo pseudodata,
a physics process is split into several subsamples to achieve a better coverage of the phase space.
This splitting can for example be done
in the number of additional \revised{partons} or the transferred momentum in the hard \revised{scattering} process.
To obtain physically meaningful results, the events from all subsamples have to be added up,
using the corresponding cross sections as weights.
The same needs to be done when computing trigger efficiencies,
so the number of events passing a given trigger $k$ and the total number of events $n$ need to be scaled properly.
For the expectation value,
the weighting is simple,
the estimated efficiency being the ratio of the weighted number of events passing the trigger and the weighed total number of events (cf. \Eq{eq:results_trigger_performance_weighted_expectation} below).

The uncertainty computation is more involved, as the error bars need to be computed from the \revised{\betadist}
representing the probability distribution of the efficiencies,
which is not known \latein{a priori}.
For a sample that has ten times as many events as another sample but only a tenth of the weight,
even if the values for $k$ and $n$ are the same,
the uncertainty on the efficiency estimate should be smaller due to the higher statistics,
but by just scaling $k$ and $n$ this is not accounted for.
Instead, one solution is to use the method of moments %
to obtain an approximation of the probability distribution of the efficiency $\epsilon$
in terms of a \betadist based on the values of the expectation value $E(\epsilon)$ and the variance $V(\epsilon)$ \cite{Casadei2009},
from which the asymmetric uncertainties can be determined, as will be shown in the following\footnote{
  Note that the expectation value $E(\epsilon)$ and variance $V(\epsilon)$ of course depend on the
  probability distribution $P(\epsilon)$ rather than on $\epsilon$,
  although this is not made explicit in the notation.
}.

Fixing the index $i$ of the binning in the offline variable
and using the index $j$ to run over the samples which are averaged,
the best estimate of the efficiency for the complete sample is given by the expectation value
\begin{equation}
  E(\epsilon_i) = \frac{\sum_j \omega_j \, k_i^j} {\sum_j \omega_j \, n_i^j}, %
  \label{eq:results_trigger_performance_weighted_expectation}
\end{equation}
where $\omega_j$ is the weight of the sample $j$.
The variance scales with the square of the weights,
\begin{equation}
  V(\epsilon_i) = \frac{\sum_j \omega_j^2 \, V(\epsilon_i^j)} {(\sum_j \omega_j)^2}.
  \label{eq:results_trigger_performance_weighted_variance}
\end{equation}

In the following, the parameter $\epsilon_i$ will be left out to simplify the notation,
\ie $E \equiv E(\epsilon_i)$ and $V \equiv V(\epsilon_i)$.
In the same way, all other quantities are understood to refer to a given bin $i$.
The estimates of $E$ and $V$ obtained from $\{k_i^j\}$ and $\{n_i^j\}$
in \Eqs{eq:results_trigger_performance_weighted_expectation} and \eqref{eq:results_trigger_performance_weighted_variance}
can be used to find a \betadist with this mean and variance. %
Solving \Eqs{eq:beta_dist_mean} and \eqref{eq:beta_dist_var} for $a$ and $b$
yields the method-of-moments estimates of the parameters \cite{URL_NIST1}
\begin{eqnarray}
  \begin{aligned}
  a &&=&&     E & \left( \frac{E (1-E)}{V} - 1 \right)           \qquad\text{and} \\
  b &&=&& (1-E) & \left( \frac{E (1-E)}{V} - 1 \right).
  \label{eq:mom_ab_mean}
  \end{aligned}
\end{eqnarray}
There is a problem though.
From \Eqs{eq:mom_ab_mean}, it can be seen that this does not work when $E=0$ or $E=1$
because then $a$ and $b$ take values (negative or zero) for which the \betadist is not defined.
This problem stems from the fact that there is no \betadist with mean $0$ or $1$.
Therefore, a similar approach is used here, but instead of the mean $E$ the mode $M$ is used. %
Solving \Eqs{eq:beta_dist_var} and \eqref{eq:beta_dist_mode} for $a$ and $b$
yields the slightly more complicated expression
\begin{equation}
  \label{eq:EVapprox_find_b_from_a}
  b = (a+2) + \frac{a - 1}{M},
\end{equation}
where $a$ is given by the solutions \revised{of} a cubic equation,
\begin{alignat}{2}
    &a^3&+& \nonumber\\
    \left(M^3/V - M^2/V + 7M - 3 \right) &a^2&+& \nonumber\\
    \left(-2 M^3/V + 16 M^2 + M^2/V -14M + 3 \right) &a^1&+& \nonumber\\
    \left(12 M^3-16 M^2+7 M - 1\right) & &=& \quad 0.
\end{alignat}
In the implementation, the solutions are found numerically using \texttt{gsl\_poly\_solve\_cubic} from the \propername{GNU Scientific Library}.
The case $E=1/2$, in which the coefficients of \revised{the monomials} $a^1$ and $a^0$ vanish, is treated specially.
If $a$ is found to be $1$, it follows $M=0$, and the result of \Eq{eq:EVapprox_find_b_from_a} is undefined.
\revised{In} that case, \Eq{eq:beta_dist_var} simplifies to
\begin{equation}
  V =\frac{b}{(b^2 + 2b + 1)(b+2)},
\end{equation}
yielding
\begin{equation}
  \Rightarrow b^3 + 4 b^2 + (5-1/V) b + 2 = 0,
\end{equation}
which is solved to find $b$ using \texttt{gsl\_poly\_solve\_cubic} again.
A closure test has been performed to ensure that the solutions for $a$ and $b$
using the largest root of the cubic equations yield a \betadist with the desired values of $M$ and $V$.

\section{Measurements of Trigger Efficiencies}
\label{sec:results_trigger_performance_measurements_introduction}

The methods which have been introduced in \Sec{sec:results_trigger_methods}
will now be applied to measure trigger efficiencies both in Monte Carlo and data.
The \revised{ultimate} goal of the trigger efficiency studies
is the measurement of the efficiencies of the combined \jetmet triggers,
which are used by the \ATLAS Supersymmetry group as the primary trigger for the zero-lepton analysis
and in the search for Supersymmetry which is presented in this thesis in \Sec{sec:analysis_susysearch}. %
Before the measurements of trigger efficiencies on data are presented,
on which these analyses are based,
the results of complementary studies of the efficiencies of \jetmet triggers on Monte Carlo samples are discussed in the next subsection.
The determination of the efficiencies on data follows afterwards in \Sec{sec:results_trigger_performance_measurements_data}.

Which method is employed to compute the efficiencies is specified for each turn-on curve \revised{individually}.
The efficiencies of triggers involving jets are in general determined object-wise,
\ie a matching between trigger jets and jets which were reconstructed offline is done
based on the geometrical distance in $(\eta,\phi)$ space with the usual metric $\Delta R = \sqrt{(\Delta\eta)^2 + (\Delta\phi)^2}$.
For this matching, the trigger jet at the highest active trigger level, Level~2 in 2010 or Event Filter in 2011, is used.
The \pt of all jets from the respective offline selection is filled into the denominator histograms,
the enumerator is filled using only the \pt of offline jets which have been matched to a trigger jet that has actually fired the trigger.
There may be more than one trigger jet in the same event firing a given trigger chain.
For combined \jetmet chains, an additional condition for filling the enumerator is, of course,
that the amount of \met computed by the trigger at all levels exceeds the respective thresholds.

\subsection{\texorpdfstring{Measurements of \Jetmet Trigger Efficiencies in Monte Carlo} {Measurements of JetMET Trigger Efficiencies in Monte Carlo}}
\label{sec:results_trigger_performance_measurements_MC}

\begin{figure}
  \centering
  \incgraphics{width=\widthtwoplots}{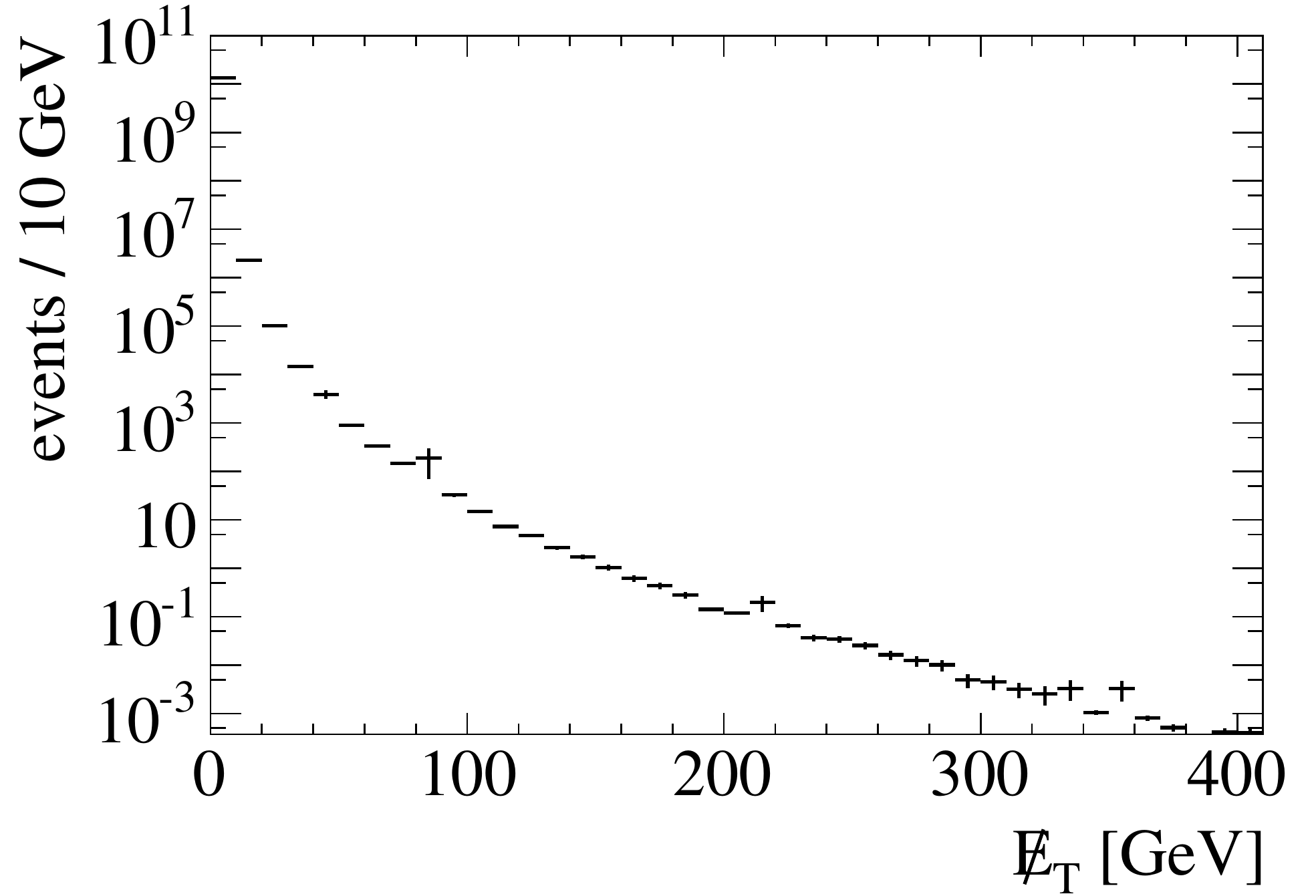}
  \hfill
  \incgraphics{width=\widthtwoplots}{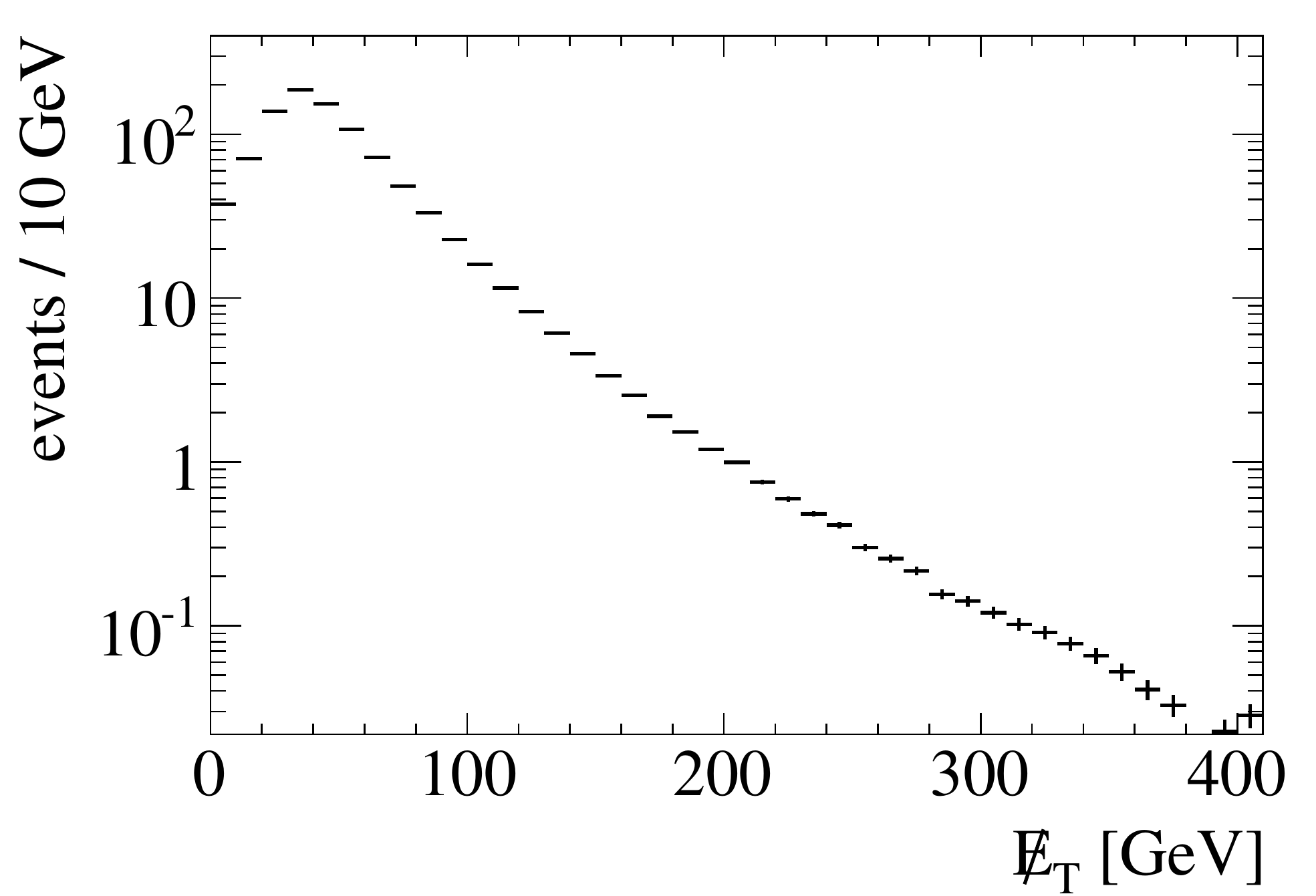}
  \caption{
    Spectrum of \met coming from non-interacting particles in \propername{Pythia} QCD (left) and \propername{Alpgen} $Z\to\nu\nu$~+ jets (right).
    The error bars indicate the statistical uncertainties due to limited Monte Carlo statistics.
    The number of events is normalized to an integrated luminosity of \ipb{1}.
  }
  \label{fig:results_trigger_performance_measurements_MC_mets}
\end{figure}

The efficiencies of the combined \jetmet triggers are evaluated on two different Monte Carlo samples
which differ strongly \revised{in} the composition of \met.
The first sample is the \propername{Pythia} dijet sample used later in the analysis to model the \acs{QCD} background,
and is intended to obtain an \revised{idea} of the performance of the \jetmet trigger on the majority of events the trigger will run on,
because the \ac{QCD} background is the dominant background at the \acs{LHC}.
(A measurement on minimum-bias Monte Carlo is not possible because of insufficient statistics.
On the other hand it is safe to assume that minimum-bias events indeed contain only very soft jets and low \met
and will therefore always be rejected by the combined \jetmet triggers.
This is in contrast to the dijet samples which contain events with hard jets which have a non-negligible cross section.)
Whilst the QCD sample is expected to contain jets with high \pt,
the main contribution to \met comes from jet mismeasurements
and does not include real \met,
which in QCD processes could only arise from neutrinos produced within jets.

The second sample contains $Z\to\nu\nu$ decays and jets from additional partons produced in the hard \revised{scattering} process.
If the neutrinos are produced back-to-back \revised{in the laboratory frame},
this again does not give \met,
because their \revised{transverse} momenta add up to zero.
\revised{However,} additional jets recoiling against the neutrino system can give rise to significant \met,
which can be seen in the right plot in \Fig{fig:results_trigger_performance_measurements_MC_mets}.
This figure shows the spectrum of \met coming from non-interacting particles for both samples, QCD and $Z\to\nu\nu$,
where the subsamples have been weighted according to their cross section as given in \Sec{sec:analysis_mc_sm_background}, %
and the \met spectrum is plotted from Monte Carlo Truth information.
As expected from the above explanations,
the real \met in %
\revised{the vast majority}
of the QCD events is negligible,
whereas most events in the $Z\to\nu\nu$~+ jets sample contain real \met of around \GeV{40}.

\begin{figure}
  \centering
  \incgraphics{width=\widthtwoplots}{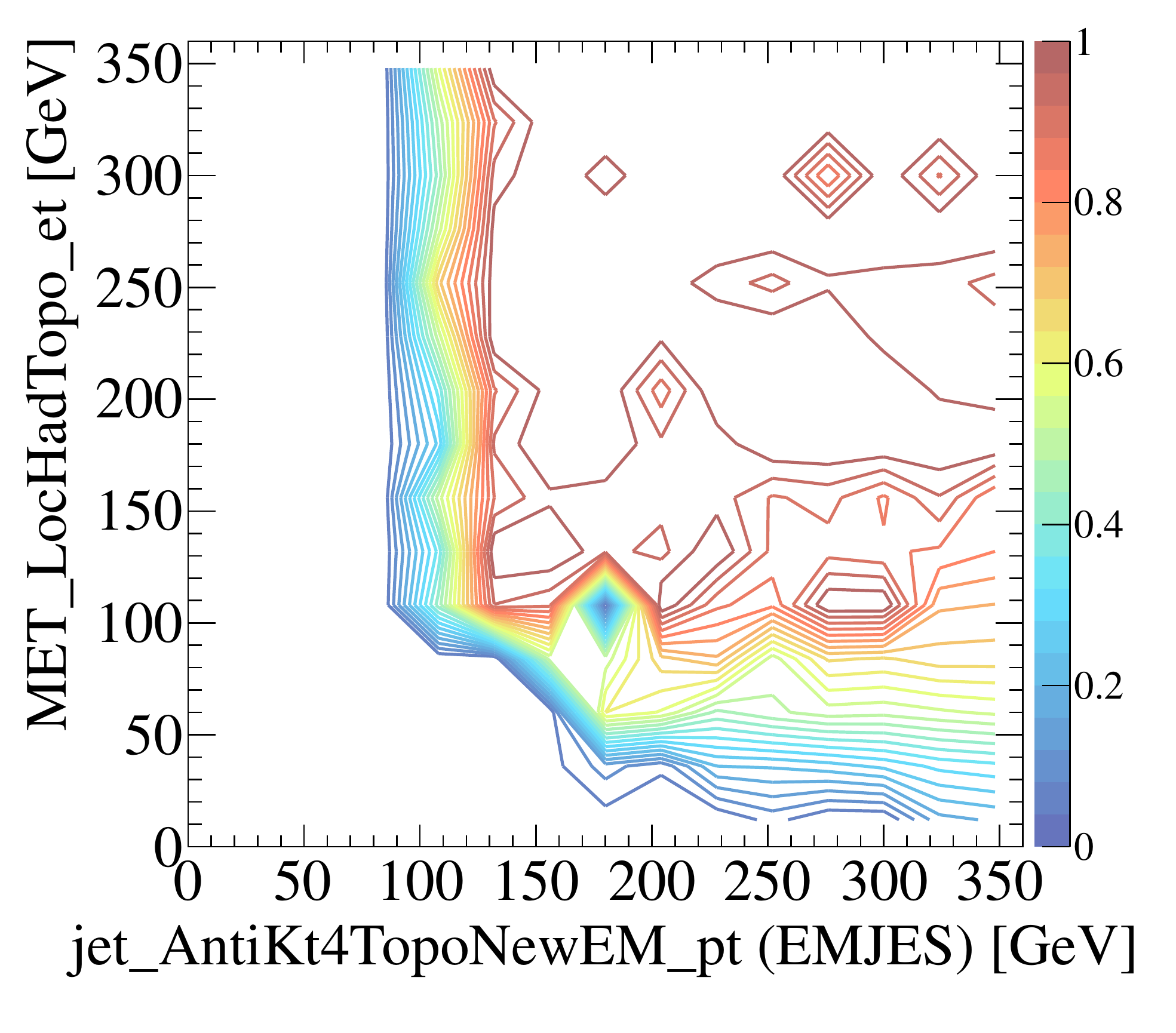}
  \hfill
  \incgraphics{width=\widthtwoplots}{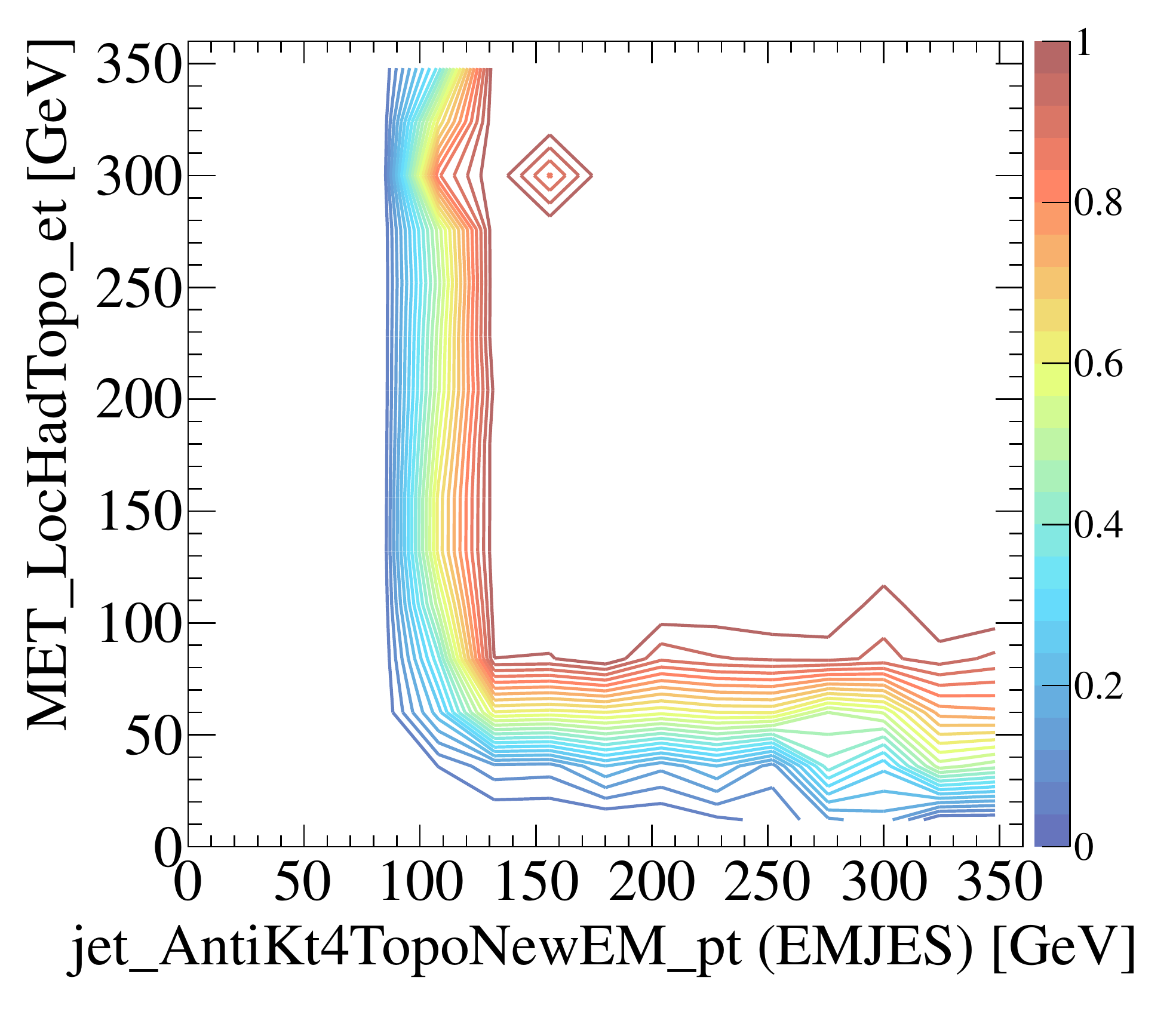}
  \\
  \incgraphics{width=\widthtwoplots}{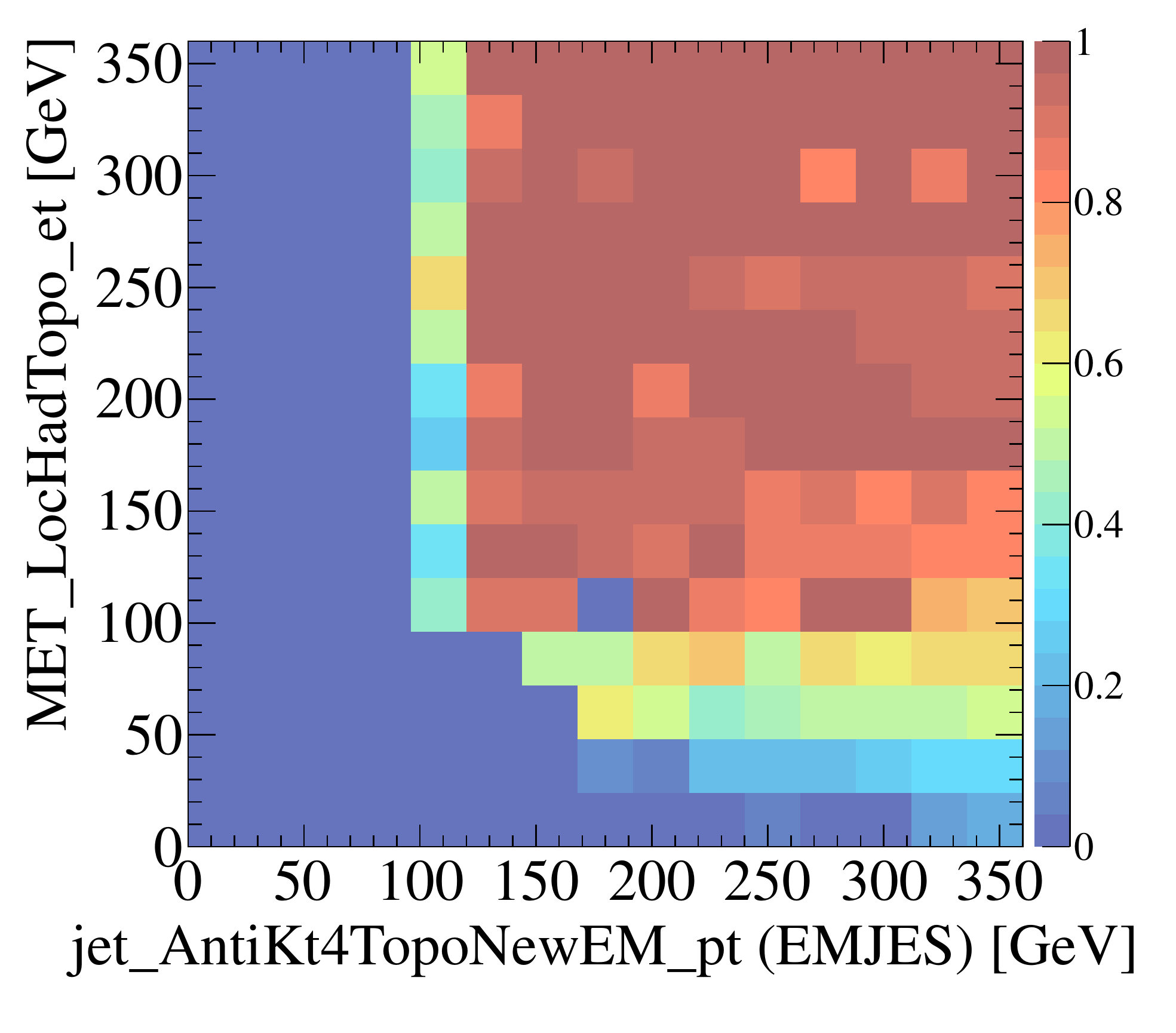}
  \hfill
  \incgraphics{width=\widthtwoplots}{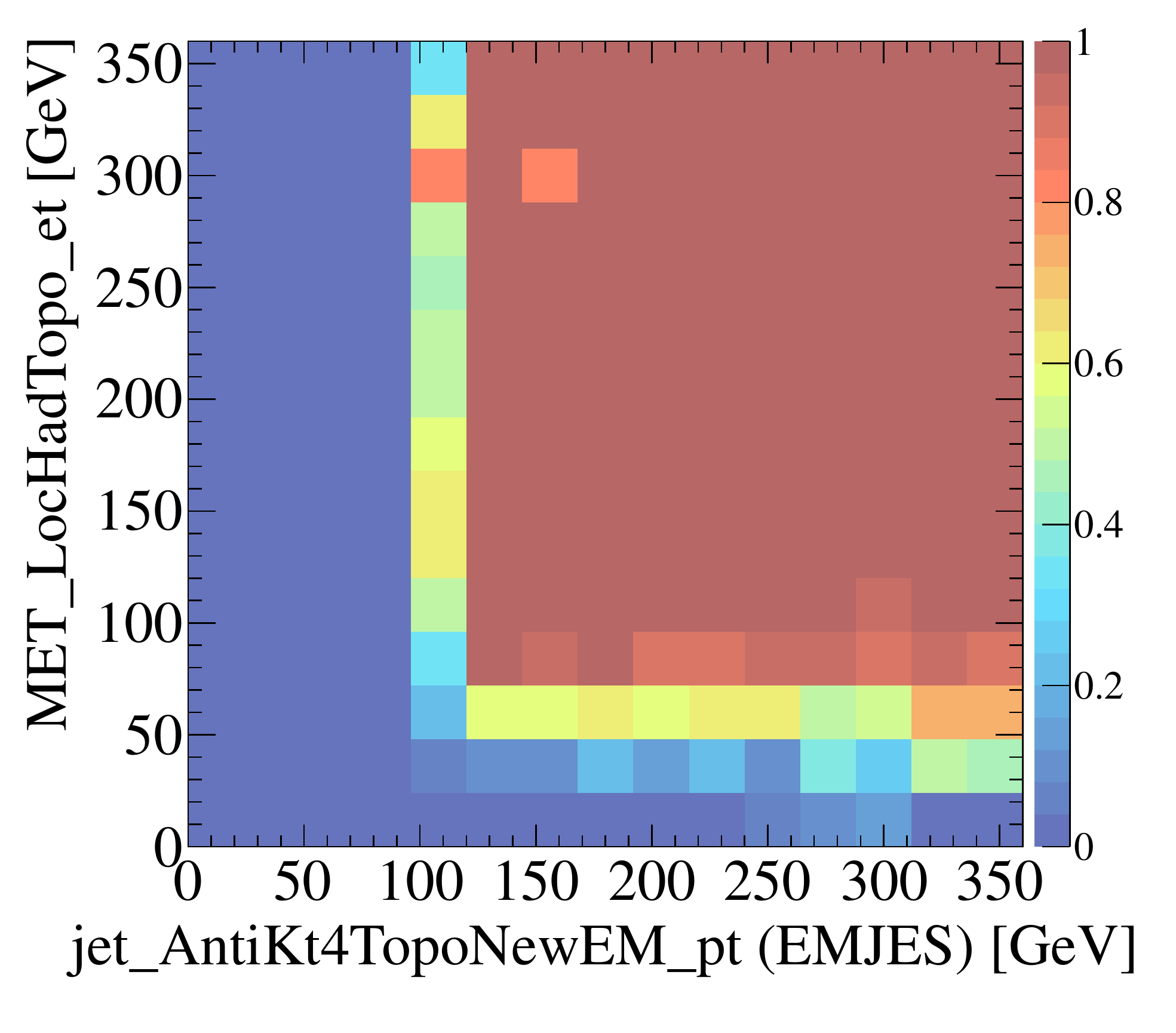}
  \caption{
    Two-dimensional efficiency plots of the combined \jetmet trigger \trigger{EF_j75_a4tc_EFFS_xe45_loose_noMu} in Monte Carlo events,
    for a QCD sample (left) and for a $Z\to\nu\nu$ + jets sample (right).
    The two upper plots show the efficiencies as contours,
    which highlights details in the regions where the efficiencies change,
    in particular in the turn-on regions.
    The two lower plots use a color-coding of the efficiencies,
    thereby making it easier to distinguish regions with low and high efficiencies.
  }
  \label{fig:results_trigger_performance_measurements_2d_efficiency_MC_LocHadTopo_2d}
\end{figure}

As outlined in the previous section on methods to compute trigger efficiencies,
in Monte Carlo the trigger efficiencies can be determined using simple event counting,
\ie relying on the results of the simulation of the trigger.
Both of the Monte Carlo samples from above are split into several subsamples with different \revised{production} cross sections
and therefore different weights.
The subsamples of the $Z\to\nu\nu$ sample differ in the number of additional partons,
\revised{while} the QCD sample is binned in the $\hat{p}_T$ exchanged in the hard \revised{scattering}
(cf. \Secs{sec:explain_qcd_samples} and \ref{sec:appendix_MC_datasets}), %
the subsamples being denoted by J0 through J6\footnote{
  J7 and J8 are not used, because J0 through J6 suffice to fill the phase space region covered by the plots,
  and J7 and J8 are not expected to contribute to the combination due to their low cross section.
}.
This means that the results from Monte Carlo counting on each individual sample
need to be combined using the method presented in \Sec{sec:results_trigger_performance_addition_weighted_samples}
in order to obtain a physically meaningful efficiency for the full sample.
This combination is shown in the two plots in \Fig{fig:results_trigger_performance_measurements_2d_efficiency_MC_LocHadTopo_2d},
where the trigger efficiency is presented using a two-dimensional, color-coded histogram,
with the offline \pt of the jets at \EMJES scale (cf. \Sec{sec:software_jet_energy_scale}) %
on the horizontal and the offline \met on the vertical axis.
Note that in this thesis, offline \met variants reconstructed by different algorithms are used.
The flavor of the \met reconstruction can be read off from the axis title and is explained in \Sec{sec:software_reconstruction_met}.
In general, the standard for the offline \met reconstruction used in the computation of trigger efficiencies
will be the \met from topological cluster cells calibrated using weights from the local hadronic calibration %
(LocHadTopo)
as used in \Fig{fig:results_trigger_performance_measurements_2d_efficiency_MC_LocHadTopo_2d}.
This \met comes very close to how the \met calculation is done by the trigger system,
and therefore gives the best offline reference.
The analog of the plots which are shown in this section
with the offline \met variant used for the SUSY zero-lepton analysis instead %
can be found in the Appendix in \Sec{sec:appendix_trigger_efficiency_plots_Monte_Carlo}. %
The binning is chosen \revised{such as} to give quadratic bins
with sizes that are a compromise between smaller statistical fluctuations for larger bins
and better sensitivity to features of the turn-on region for smaller bins.

The two plots to the left in \Fig{fig:results_trigger_performance_measurements_2d_efficiency_MC_LocHadTopo_2d}
show the trigger efficiency of the combined \jetmet trigger with the lowest thresholds running in 2011,
\trigger{EF_j75_a4tc_EFFS_xe45_loose_noMu}, for the QCD sample,
the two plots to the right the efficiency for the same trigger, but for $Z\to\nu\nu$.
This trigger applies a threshold of \GeV{75} for the jet part and \GeV{45} for the \met part.
The plots \revised{with $Z\to\nu\nu$ events and} real \met exhibit much sharper turn-on curves,
in particular as function of \met,
and less fluctuations in the plateau,
where the efficiency is almost everywhere close to one.
In these plots,
it can therefore be seen best that the \met turn-on of the \jetmet trigger
actually lies around the nominal value of \GeV{45},
whereas the jet turn-on only begins at values around \GeV{100},
much higher than the nominal \GeV{75}.
This is due to the fact that the offline and trigger jet calibration is different.
The trigger jets are calibrated at EM scale, %
whereas for the offline jets, the \EMJES calibrated \pt is used,
so that the same jet will be reconstructed with higher \pt offline than within the trigger.
This shifts the turn-on to higher values,
which can also be seen in the projections in \Fig{fig:results_trigger_performance_measurements_2d_efficiency_MC_LocHadTopo_projections_per_subsample}
discussed below\footnote{
  Note that there is no general problem in using offline jets at \ac{EM} scale instead.
  Indeed, the \pt at EM scale would be a more natural variable to plot the trigger efficiency as function of,
  but as most of the \jetmet trigger studies have been done for the official zero-lepton Supersymmetry analysis,
  here it has be conceded to use the same offline variable as in the analysis.
  As a different calibration of the jets does not induce
  such a fundamental difference in the offline variable with respect to the trigger variable
  as it is seen when using a different offline \met,
  the conclusions stay the same apart from the shift in the turn-on to higher values.
  The last fact makes it possible to directly read off from the turn-on curves
  the threshold for the offline jet \pt cut
  which is needed to only select events in the plateau region of the trigger.
}.
In the lower left plot for the QCD sample,
the blue region at small \met and jet \pt values cuts into the lower left corner of the plateau region.
This is due to the dominance of the J0 and J1 subsample in this region,
and is revealed by the two-dimensional \revised{efficiency plots} per subsample.
For the J3 and higher subsamples,
this region looks more like in the right plot for \revised{the} $Z\to\nu\nu$ sample.
This matches the observation that in the high $\hat{p}_T$ subsamples of the QCD sample,
the contribution of real \met to the total \met should be comparable to the real \met in $Z\to\nu\nu$. %

\begin{figure}
  \centering
  \incgraphics{width=\widthtwoplots}{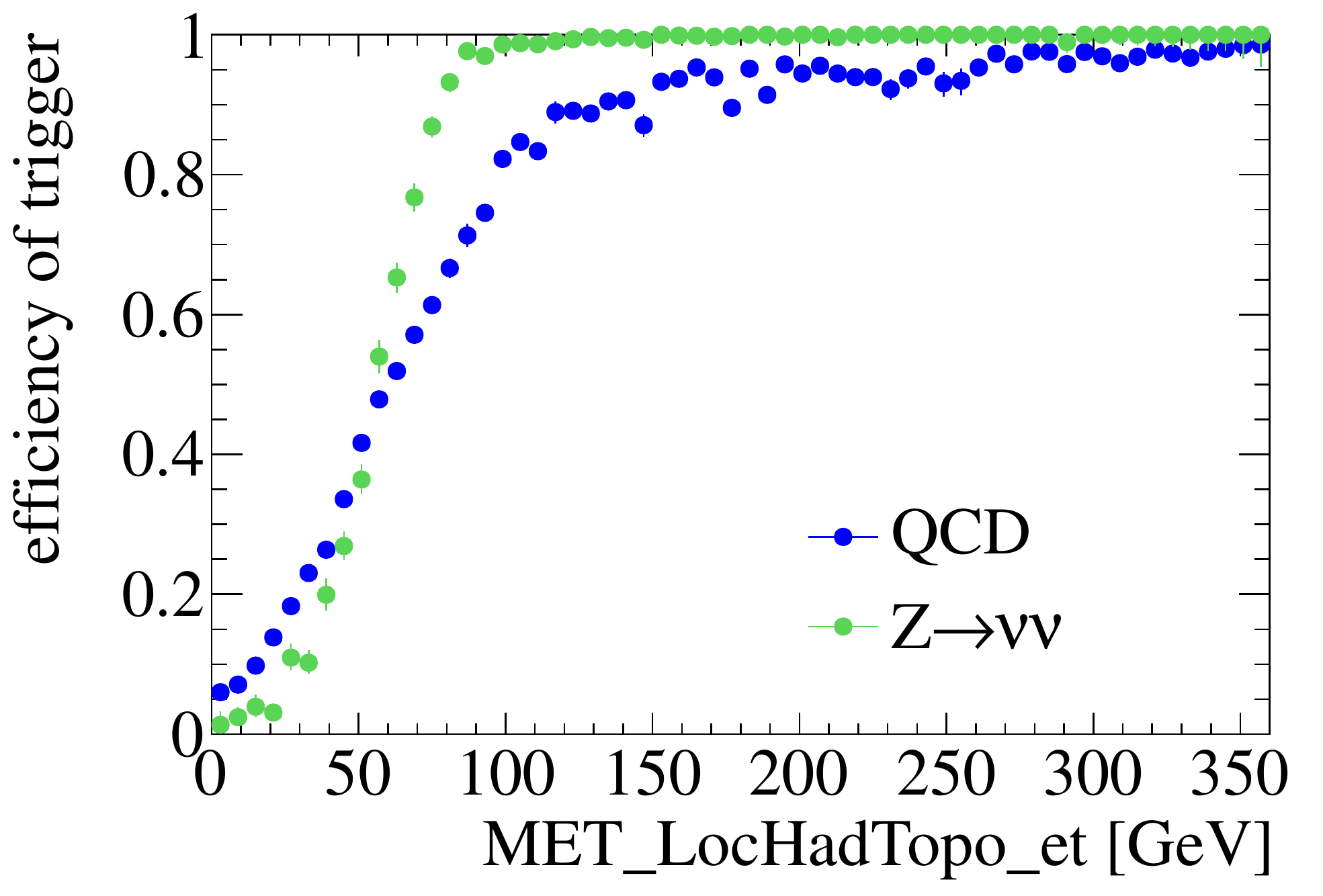}
  \hfill
  \incgraphics{width=\widthtwoplots}{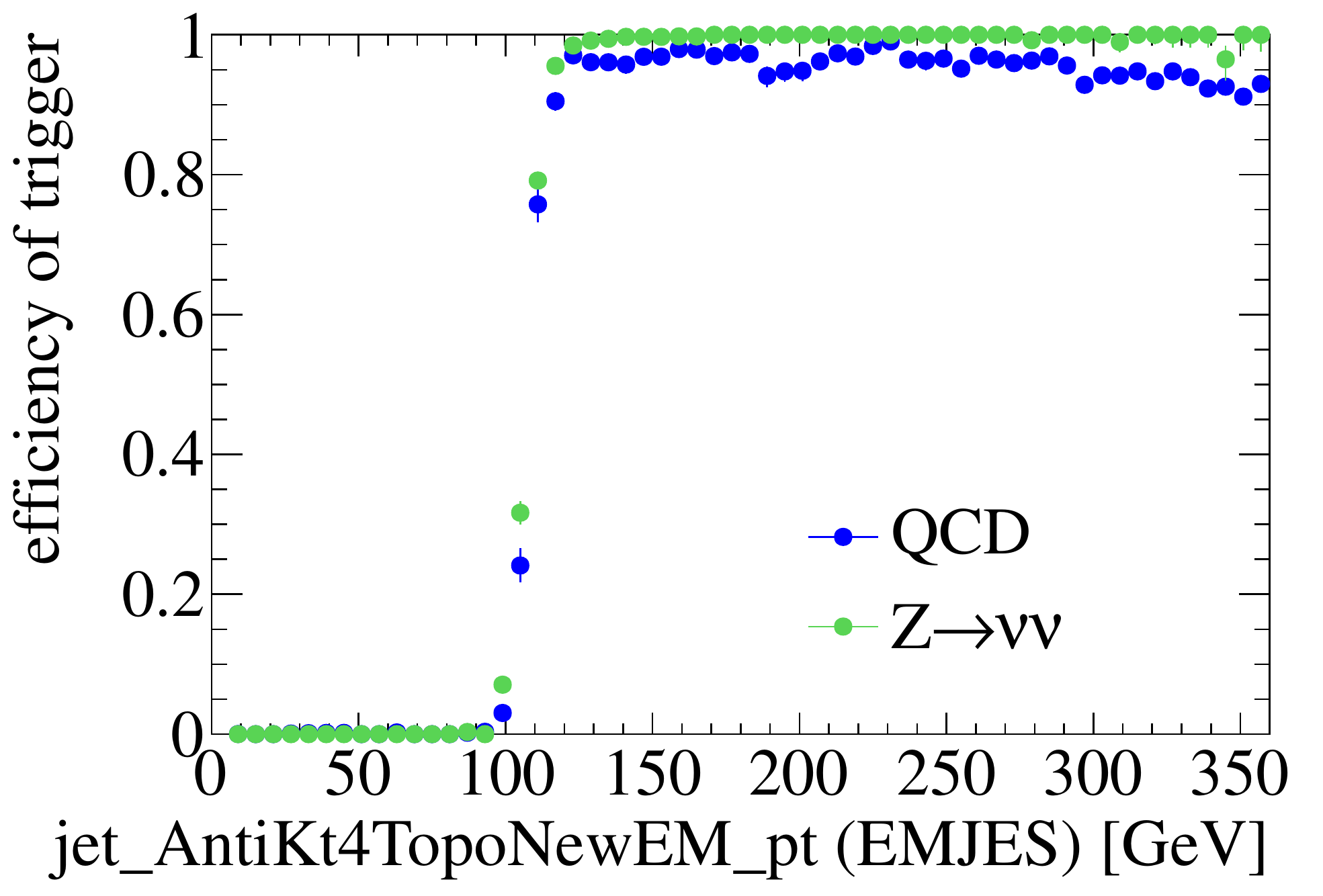}
  \caption{
    Projections of the efficiency of the combined \jetmet trigger \trigger{EF_j75_a4tc_EFFS_xe45_loose_noMu} in Monte Carlo events
    for the combined QCD sample and for the combined $Z\to\nu\nu$ + jets sample.
    Left: onto offline \met, right: onto jet \pt,
    after cuts on the respective orthogonal variable at \GeV{130}.
  }
  \label{fig:results_trigger_performance_measurements_2d_efficiency_MC_LocHadTopo_projections}
\end{figure}

\begin{figure}
  \centering
  \incgraphics{width=\widthtwoplots}{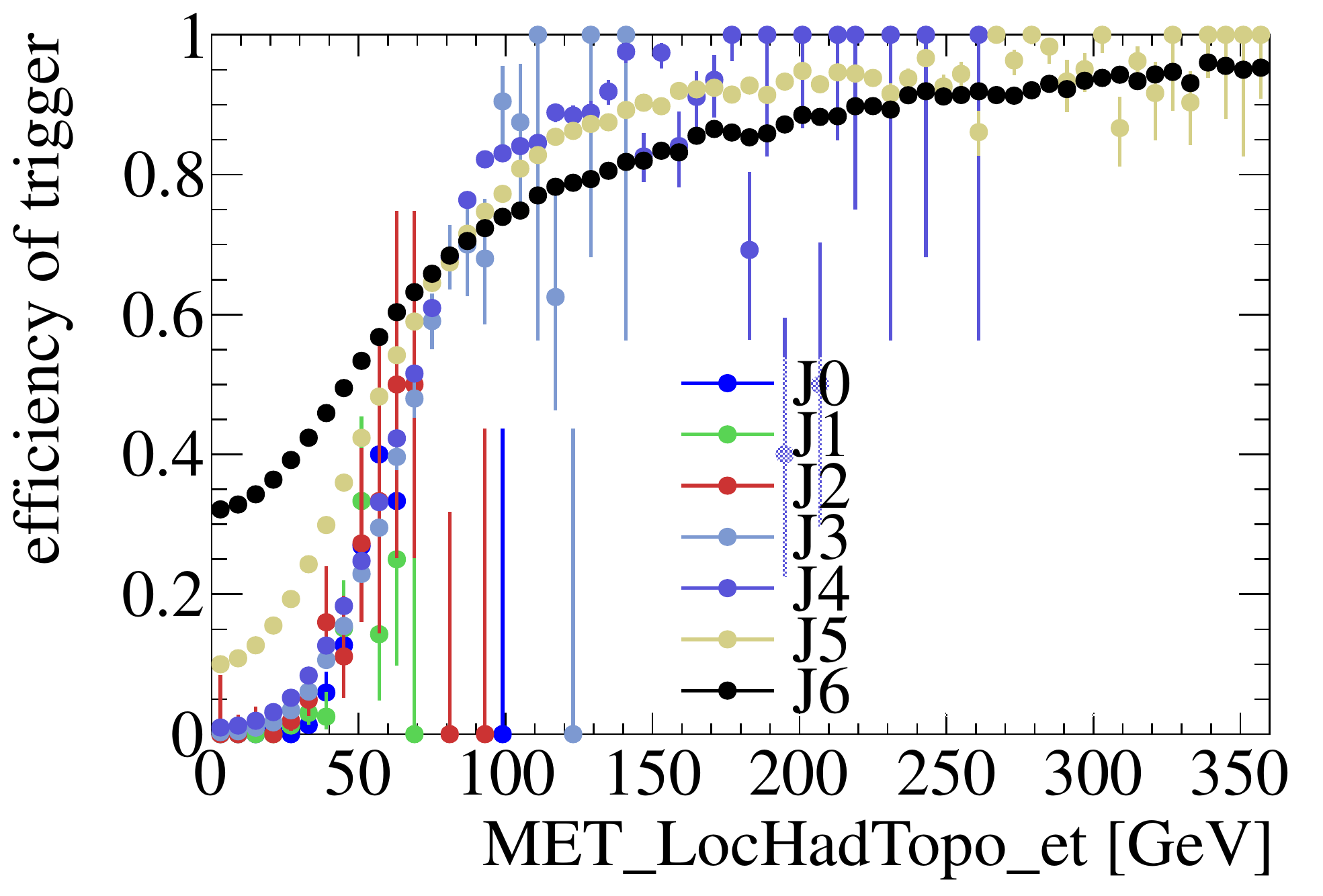}
  \hfill
  \incgraphics{width=\widthtwoplots}{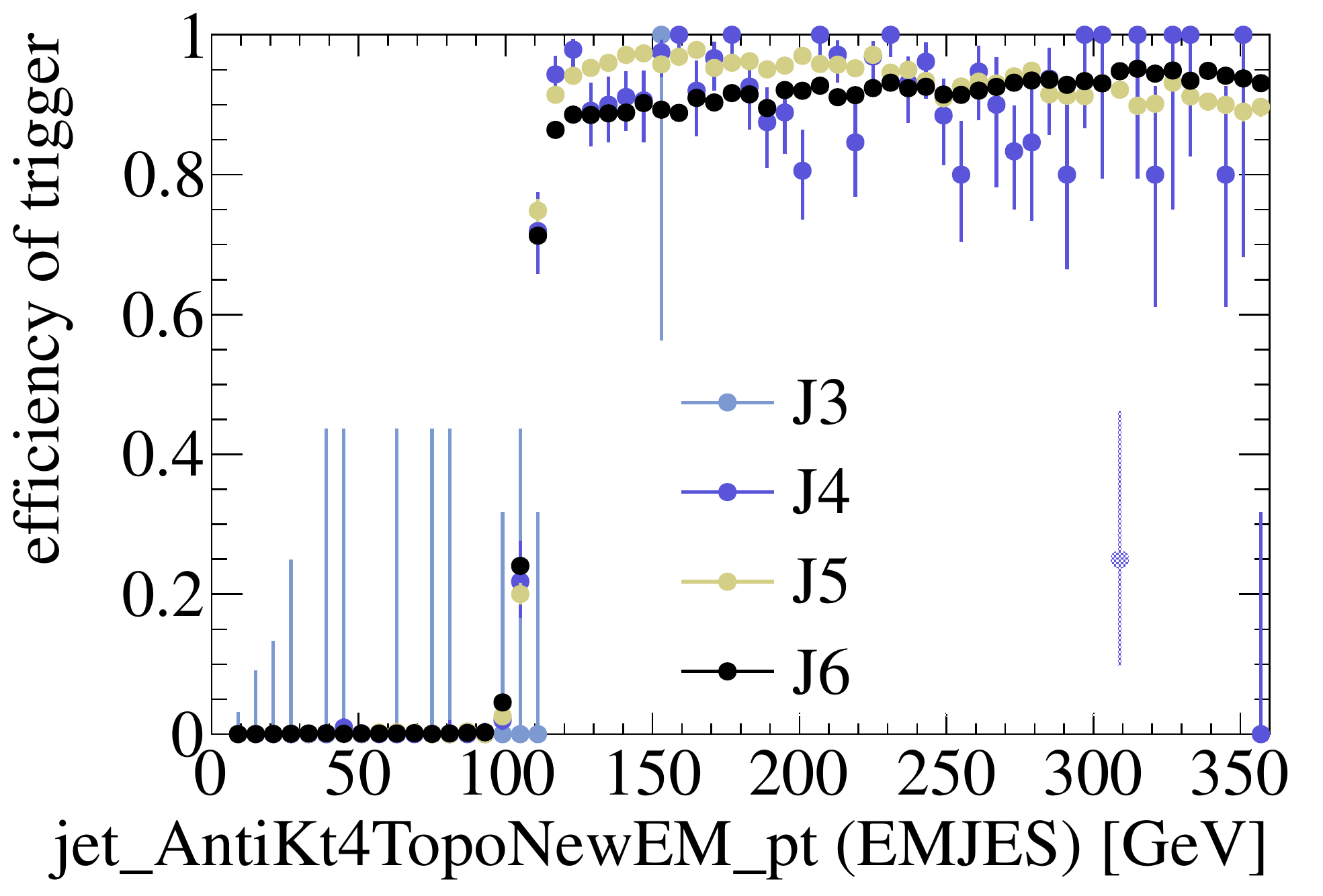}
  \\
  \incgraphics{width=\widthtwoplots}{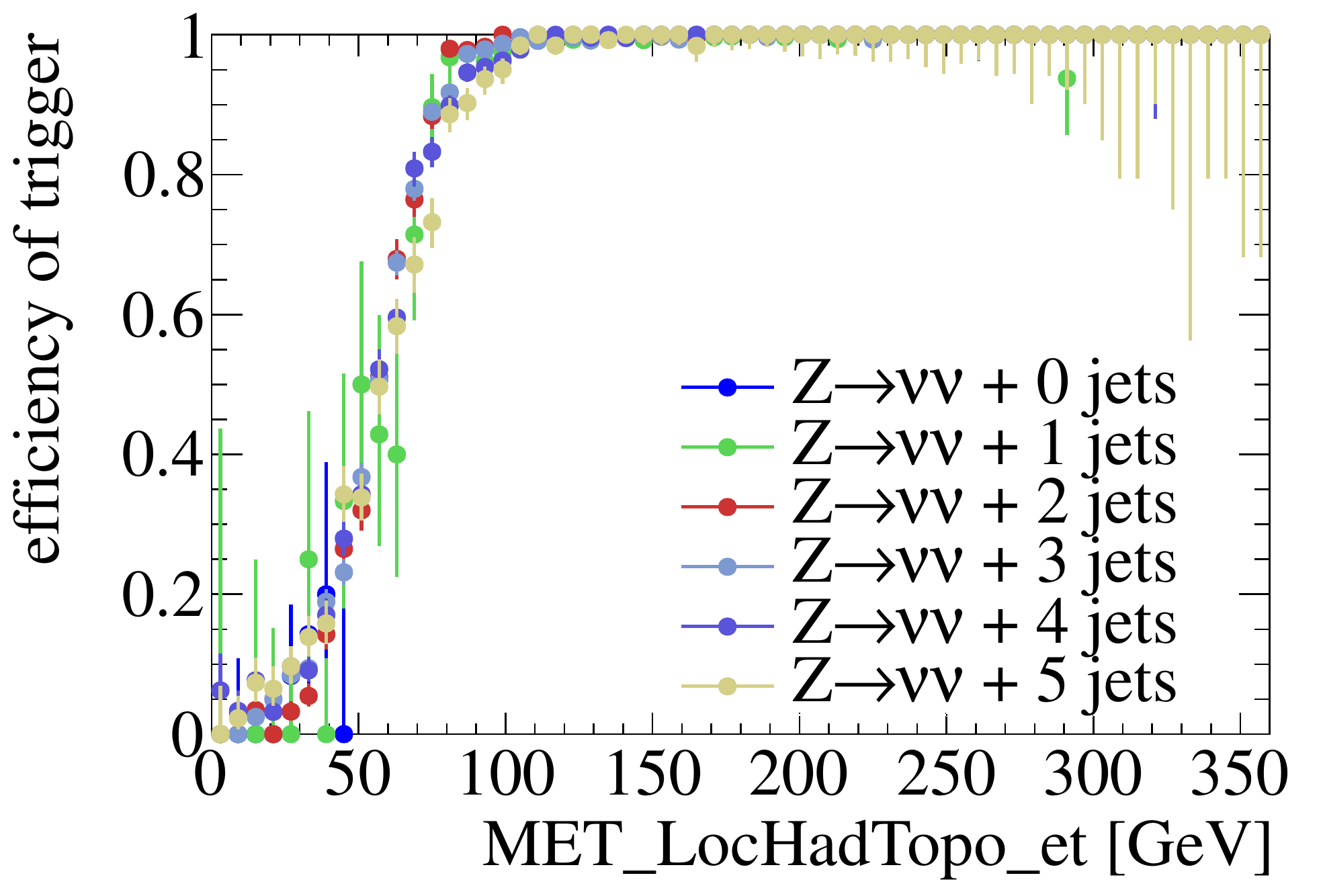}
  \hfill
  \incgraphics{width=\widthtwoplots}{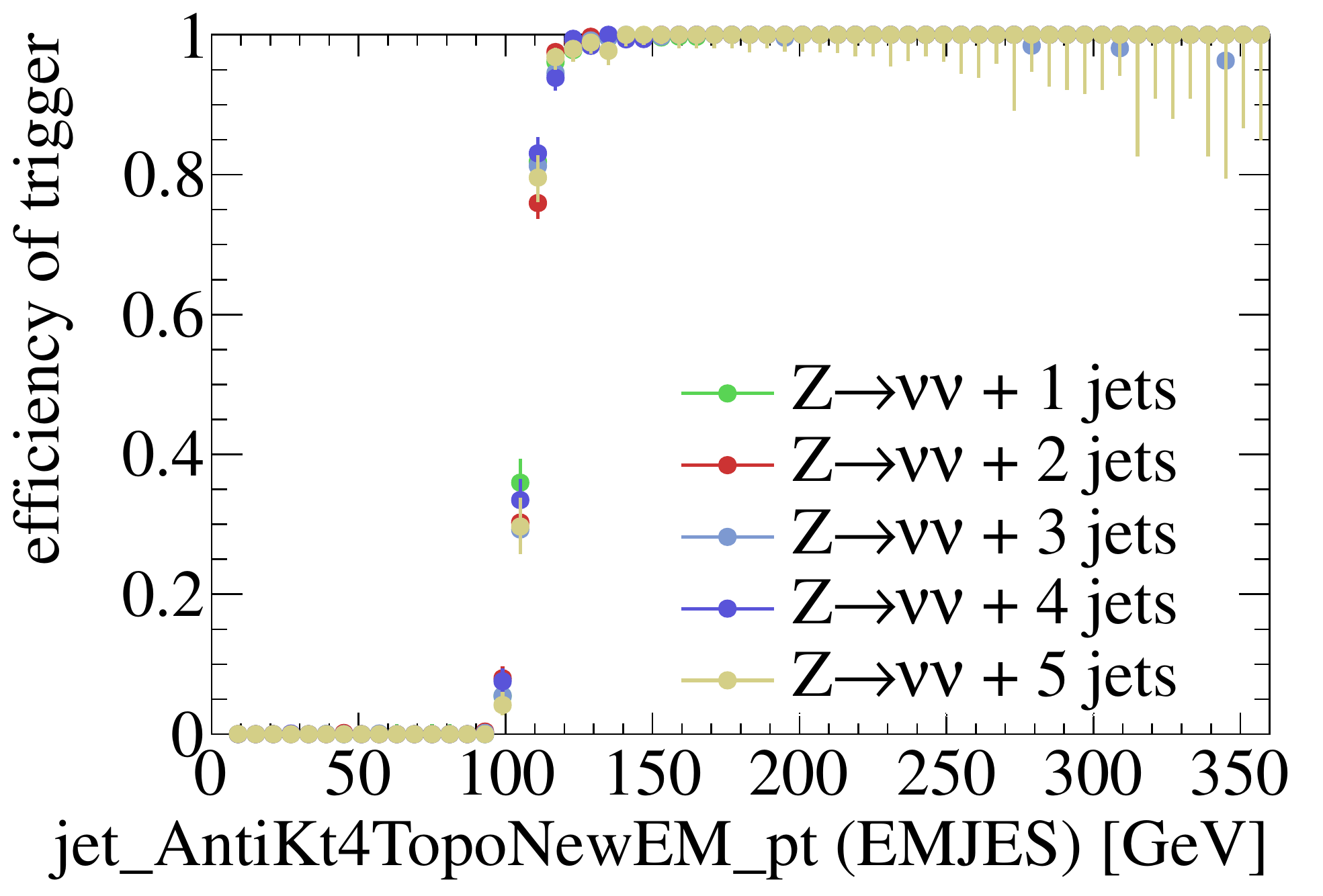}
  \caption{
    Projections of the efficiency of the combined \jetmet trigger \trigger{EF_j75_a4tc_EFFS_xe45_loose_noMu} in Monte Carlo events
    for each subsample of the QCD sample (top) and the $Z\to\nu\nu$ + jets sample (bottom) individually.
    Left: onto offline \met, right: onto jet \pt,
    after cuts on the respective orthogonal variable at \GeV{130}.
  }
  \label{fig:results_trigger_performance_measurements_2d_efficiency_MC_LocHadTopo_projections_per_subsample}
\end{figure}

As in the two-dimensional plots it is \revised{difficult to visualize} error bars,
often projections of the two-dimensional histograms onto one of their axes will be employed.
These projections are created from all bins above a minimum value on the respective orthogonal axis,
\ie when projecting onto the \met axis,
all bins above a given jet \pt threshold are used and vice versa.
The threshold is chosen such that the efficiency of the jet part of the combined trigger has reached its plateau value.
Otherwise, the projection would yield an average over the full range of jet \pt values
and, regarded as function of \met,
the efficiency would not reach an efficiency of one in the plateau,
because of the bins with an efficiency of zero for low jet \pt.
As can be seen in particular in the plots to the left in \Fig{fig:results_trigger_performance_measurements_2d_efficiency_MC_LocHadTopo_2d},
tracing the turn-on region around \met values of \GeV{100} along the horizontal axis,
the behavior of the turn-on may change as function of the orthogonal variable.
Here, as function of jet \pt,
the turn-on region of the \met part of the combined trigger gets broader.
Features like this of the two-dimensional structure of the trigger efficiency are naturally lost in projections,
so that it is always advisable to consider the two-dimensional plots, too.
\Fig{fig:results_trigger_performance_measurements_2d_efficiency_MC_LocHadTopo_projections} displays
the projections onto the \met and jet \pt axes for both samples,
created using the combination of all subsamples.
The efficiencies per subsample are shown in 
\Fig{fig:results_trigger_performance_measurements_2d_efficiency_MC_LocHadTopo_projections_per_subsample},
which demonstrates the large differences between the QCD subsamples,
affecting in particular the \met part, but much less so the jet part.
The $Z\to\nu\nu$ subsamples all behave alike with respect to the trigger turn-on.
The projections also show that the width of the turn-on region is very different,
the turn-on of the jet part being much sharper than the one of \met.
This is a typical difference between jet and \met triggers.
Note that some of the lower subsamples do not appear in the right plots in \Fig{fig:results_trigger_performance_measurements_2d_efficiency_MC_LocHadTopo_projections_per_subsample}
because no events remain after the cut on \met.

The results of the studies on Monte Carlo in this section
demonstrate a strong dependence of the outcome of the efficiency measurement on the event sample which is used as input.
They thereby underline the importance of finding a way to measure the efficiencies in data on an unbiased,
\revised{representative} sample of events.
It may be tempting to say
that in an analysis which is looking for Supersymmetry,
the selection is designed to accept only events which have real \met,
but that is not what is actually happening in the evaluation of the event selection,
where event counts from Monte Carlo and data are compared
without taking into account \eg the origin of the \met which makes an event pass the selection cuts.

\subsection{Measurements of Trigger Efficiencies in Data}
\label{sec:results_trigger_performance_measurements_data}
As was mentioned in the introduction,
the goal of the efficiency studies in this thesis
is the measurement of the efficiencies of the combined \jetmet triggers,
which are employed by the \ATLAS Supersymmetry group in the zero-lepton analysis
as well as in the analysis which is presented in \Sec{sec:analysis_susysearch}.
The trigger menus have undergone a fast-paced evolution in particular in 2010,
following the requirements of the increasing instantaneous luminosity.
In addition, they are different between 2010 and 2011 due to changes in the jet trigger algorithms.
Altogether, there have been a number of changes in the \ATLAS trigger between 2010 and 2011
which \revised{necessitate} changes in the measurement of the efficiencies
with respect to which triggers are of interest and which methods can be employed.
Therefore, the measurements of the trigger efficiencies are presented and discussed separately for 2010 and 2011.
The first part of this section deals with the trigger conditions and efficiency measurements in 2010,
which were coined by the \revised{rapid} increase in the instantaneous luminosity,
entailing constant changes in the structure of the trigger menu,
and the necessity to test which methods are suitable and feasible to determine the trigger efficiencies.
This part is also documented in an internal \ATLAS note \cite{ATL-COM-DAQ-2011-027}.
In the second part, the changes with respect to 2010 and efficiencies of newly introduced triggers are discussed,
the steadier beam conditions allowing for consistency checks between different periods.
The studies presented in this section %
have become part of the supplementary documentation of the official zero-lepton analyses
performed by the \ATLAS Supersymmetry group in 2010 and 2011
\cite{ATL-PHYS-INT-2011-009,ATL-PHYS-INT-2011-055,ATL-PHYS-INT-2011-085}.

\section{Measurements of Trigger Efficiencies in Data Taken in 2010}
\label{sec:results_trigger_performance_measurements_data_2010}

Before coming to the actual measurements,
a brief summary of the \jetmet triggers used by \ATLAS in 2010 will be presented.
Two different groups of \jetmet trigger chains have been defined in the trigger menus in 2010,
which not only differ in the value of their thresholds at the three trigger levels,
but also by which trigger objects cuts are applied to.
As at \acl{EF} level event rejection based on jet cuts has not been activated in 2010,
all jet triggers only cut at \acl{L1} and \acl{L2}.
The \ac{EF} hypothesis algorithms are set to forced accept.
This also holds for the combined \jetmet triggers.
Unless otherwise stated,
all of the rest of this section refers to 2010.

\subsection{\texorpdfstring{Combined \Jetmet Triggers in the Trigger Menu} {Combined JetMET Triggers in the Trigger Menu}}

\begin{table}
  \centering
  \begin{tabular}{lll}
    \toprule
    \centering Name of chain & Cut on jet $E_T$ [GeV] & Cut on \met [GeV]\\
    \midrule
    \trigger{EF_j50_jetNoEF_xe20, 30, 40_noMu}
    & L1: 30                    & L1: 10, 15, 25 \\
    & L2: 45                    & L2: 12, 20, 30 \\
    & EF: ---                   & EF: 20, 30, 40 \\
    \midrule
    \trigger{EF_j75_jetNoEF_EFxe20, 25, \dots, 40_noMu}
    & L1: 55           & L1: --- \\
    & L2: 70           & L2: --- \\
    & EF: ---          & EF: 20, 25, \dots, 40 \\
    \bottomrule
  \end{tabular}
  \caption{
    \Jetmet chains in the trigger menu \tagname{\trigger{Physics_pp_v1}},
    which was used at the end of $pp$ data taking in 2010.
    The table shows the cuts which are applied on the online measurement of \met and jet~$E_T$
    at Level~1, Level~2 and Event Filter for two different groups of \jetmet triggers.
  }
  \label{tab:results_trigger_performance_jetmet_chains_2010}
\end{table}

\Tab{tab:results_trigger_performance_jetmet_chains_2010} gives an overview of the two groups of \jetmet triggers
included in the trigger menu of 2010
in terms of the values of the thresholds applied to the transverse energy of the jets and the missing transverse energy at each trigger level.
The first group in the upper block of the table contains the ``traditional'' \jetmet trigger chains,
seeded by trigger items at \ac{L1} which combine cuts both on jets and \met.
There are three chains in this group,
which comprise the same set of jet cuts,
but differ by their thresholds on \met,
which are 20, 30 and \unit[40]{GeV} at \ac{EF} level
and correspondingly lower cuts on \met at \ac{L1} and \ac{L2}.
The other group of triggers is seeded by the L1 item \trigger{L1_J55}, \ie L1 jets with $E_T\geq \GeV{55}$, %
includes a \GeV{70} cut on jets at L2, but only applies cuts on \met at EF.
In this group, there are five triggers,
which again only differ in their cuts on \met,
whereas their cuts on jets are identical.
The motivation for having chains which are seeded by jets at L1 and cut at \met only at EF
is to profit from the higher resolution of the \met measurement at EF
\revised{owing to} to the higher latency affordable at this trigger level.
\trigger{L1_J55} could be kept unprescaled throughout the whole data-taking period of 2010, %
whereas the chains seeded by the lowest \jetmet trigger item at L1 with \revised{a L1~\met} cut of \unit[10]{GeV}
had to be deactivated towards the end of the data-taking period. %
Note that the (first) number in the name of the \jetmet chains refers to the \ac{EF} jet threshold
(\GeV{50} or \GeV{75}, respectively),
which for these chains is in general \GeV{5} higher than the \ac{L2} cut.
As the names of the \jetmet chains are quite long and easy to confuse,
in the following the first group of triggers in \Tab{tab:results_trigger_performance_jetmet_chains_2010}
will be referred to as \revised{``full-chain \jetmet triggers''}\inindex{full-chain JetMET trigger@Full-chain \jetmet trigger}
and the second group as \revised{``jet-seeded \jetmet triggers''}\inindex{Jet-seeded JetMET trigger@Jet-seeded \jetmet trigger}.

\subsection{\texorpdfstring{Sample Triggers for Measuring \Jetmet Trigger Efficiencies} {Sample Triggers for Measuring JetMET Trigger Efficiencies}}
\label{sec:results_trigger_performance_sample_triggers}

\begin{table}
  \centering
  \begin{tabular}{lll}
    \toprule
    Target trigger & \multicolumn{2}{c}{Potential sample triggers}\\
    \cmidrule{2-3}
     & Jet type & \met type  \\
    \midrule
    \trigger{EF_j50_jetNoEF_xe20,30,40_noMu} & \trigger{L1_J30} & \trigger{L1_XE10,15,25}  \\
     & \trigger{L2_j45} & \trigger{L2_xe12,20,30_noMu}   \\
     & \trigger{EF_j50_jetNoEF} & \trigger{EF_xe20,30,40_noMu}  \\
    \trigger{EF_j75_jetNoEF_EFxe20,25,\dots,40_noMu} & \trigger{L1_J55} & --- \\
     & \trigger{L2_j70} & --- \\
     & \trigger{EF_j75_jetNoEF} & (\trigger{EF_xe15_unbiased_noMu})  \\
    \bottomrule
  \end{tabular}
  \caption{
    Overview of the sample triggers that can be used for bootstrapping \jetmet triggers in 2010
    taking \Eq{eq:bootstrap_requirement} into account.
  }
  \label{tab:results_trigger_performace_potential_sample_triggers_2010}
\end{table}

For \jetmet triggers, two types of triggers fulfill \Eq{eq:bootstrap_requirement},
\revised{provided} the thresholds are low enough: single jet triggers and \met triggers.
This allows to bootstrap \jetmet triggers on two different samples,
one taken with a jet trigger, the other with an \met trigger,
and to cross-check the results for consistency.
Taking \Eq{eq:bootstrap_requirement} as guideline,
the set of sample triggers given in \Tab{tab:results_trigger_performace_potential_sample_triggers_2010} is identified,
which can be used to measure the efficiencies of the combined \jetmet triggers.
The table shows for both groups of triggers, jet-seeded and full-chain,
all possible jet and \met triggers which can be used as sample trigger,
together with their thresholds at Level~1 and Level~2 in terms of the items and chain names at these levels.
Consistency checks for the jet-seeded triggers using \met sample triggers in addition to jet sample triggers are not possible,
because the jet-seeded triggers only cut on \met at EF level,
and therefore a sample would be needed taken with an EF-only \met trigger.
There is only one such trigger, \trigger{EF_xe15_unbiased_noMu}, %
in the menu,
which cannot provide sufficient statistics. %

\begin{table}
  \centering
  \begin{tabular}{lll}
    \toprule
    Target trigger & \multicolumn{2}{c}{EF sample triggers with highest count}\\
    \cmidrule{2-3}
                   & \multicolumn{1}{c}{Jet type} & \multicolumn{1}{c}{\met type} \\
    \midrule
    \jetmetf{20} & \trigger{EF_j50_jetNoEF} (\numprint{616911}) & \trigger{EF_xe20_noMu} (\numprint{817636}) \\
    \jetmetf{30} & \trigger{EF_j50_jetNoEF} (\numprint{616911}) & \trigger{EF_xe20_noMu} (\numprint{817636}) \\
                 &                                              & \trigger{EF_xe30_noMu} (\numprint{892808}) \\
    \jetmetf{40} & \trigger{EF_j50_jetNoEF} (\numprint{616911}) & \trigger{EF_xe30_tight_noMu} \\
                 &                                              & \phantom{\trigger{EF_xe30_noMu}} (\numprint{1309130}) \\
    All jet-seeded triggers & \trigger{EF_j75_jetNoEF} (\numprint{2266903}) & \multicolumn{1}{c}{---}  \\
    \bottomrule
  \end{tabular}
  \caption{
    Overview of the triggers with highest counts
    which can be used as sample triggers for the respective target triggers
    in periods G~--~I.
    The number of events selected by each trigger is given in brackets
    to give a feeling for the number of events needed to generate turn-on curves such as shown in this section.
  }
  \label{tab:results_trigger_performance_trigger_counts_2010}
\end{table}

All event samples for the measurements of trigger efficiencies are selected
by requiring a positive EF decision of a single trigger chain.
In the following,
it is explained how the selection of this sample trigger is optimized to obtain the maximal possible sample sizes.
The fastest way to find out which trigger provides the largest statistics is
to read this information from the \ATLAS \propername{COOL} database,
where the number of events per trigger, run and luminosity block are stored.
This allows to compare the absolute number of events after prescales
in luminosity blocks marked as good in the corresponding \ac{GRL}
which were triggered by \eg the various \met triggers with different thresholds,
and to select the trigger which has the highest count of accepted events.
Note that \propername{COOL} only holds the trigger counts for periods~G and later.
As the prescales evolve over time, the trigger chains with highest counts vary between the different periods.
The trigger with the highest count summed over all three periods is chosen as sample trigger.

\Tab{tab:results_trigger_performance_trigger_counts_2010} shows the resulting optimal choices of the sample triggers
for the respective target triggers and the size of the event samples.
For \trigger{EF_j50_jetNoEF_xe30_noMu},
both \trigger{EF_xe20_noMu} and \trigger{EF_xe30_noMu} yield comparable sample sizes.
In this case,
using the trigger with the higher threshold gives better results,
because it has a better coverage of the plateau region of the target trigger for high \met values.
Note that \trigger{EF_xe30_tight_noMu} cannot be used for \jetmetf{30}
due to its higher thresholds at L1 and L2,
which would violate \Eq{eq:bootstrap_requirement}.
The cross-checks on the \tagname{Muons} stream require a muon trigger as sample trigger.
\trigger{EF_mu13} has the highest trigger count of all %
single muon triggers.

It turns out that the sample trigger with the highest event count is always the one
with the highest threshold that is allowed by condition \Eq{eq:bootstrap_requirement}.
The conclusion from this is that probably not much could be gained
by including triggers with lower thresholds in a logical OR with the optimal sample trigger
because events are scarce especially in regions with high \met or jet \pt,
and including chains with lower thresholds will not help there.
This, of course, only holds if the chosen sample triggers with the highest possible thresholds are unprescaled.

\subsection{Data Selection}
\label{sec:results_trigger_performance_dataselection_2010}

\revised{Unless stated otherwise, data from periods 2010~G, H, and~I will be used in the following},
as the trigger \revised{menu} has been relatively stable in these periods.
\revised{Moreover, they} comprise the latest and largest part (about \percent{92}) %
of the data collected in 2010 in terms of luminosity (about $\unit[41]{pb^{-1}}$ of recorded luminosity \cite{URL_DataLuminosity}).
All of this data has been reconstructed with \Athena in release 15. %
Mostly, the \tagname{NTUP\_SUSY} data format has been used,
which is produced from all relevant data streams apart from the \tagname{MinBias} stream.
The event selection relies on basic cuts that are common to most physics analyses in order to keep the results general
and to discard events which suffer from bad detector conditions, %
while retaining as many events as possible to have an event sample as large as possible.
The following event-level cuts have been applied:
Events must belong to luminosity blocks that have been marked as suitable
for physics analysis in the \ac{GRL} generated for the \ATLAS SUSY group.
Events are rejected if they contain a bad jet arising from hardware problems, cosmic-ray showers,
or general problems with the beam quality,
as defined by the \ATLAS Jet/Etmiss combined performance group in their updated recommendation
for loose cleaning of release 15 data after the PISA workshop in 2010 \cite{URL_JetCleaning, ATLAS-CONF-2010-038} (cf. \Sec{sec:results_0lanalysis_jet_cleaning}). %
A selection based on the number of reconstructed primary vertices,
as is \revised{made} in physics analyses to reject events with no collisions, is not applied.

As offline reference for jet triggers the jet collection used by the \ATLAS Supersymmetry group is chosen,
which contains offline jets reconstructed using an \antikt jet finding algorithm \cite{Cacciari2008} %
with a cone size of $0.4$ based on topological clusters (\tagname{AntiKt4H1Topo}, cf. also \Secs{sec:software_jet_reconstruction} and \ref{sec:analysis_object_identification_and_overlap_removal}). %
The energy is calibrated by applying a pseudorapidity and \pt dependent jet-energy-scale factor
to the energy measurement at the electromagnetic scale, denoted by \EMJES in the labels of the axes. %
For the turn-on curves, only offline jets within $|\eta| < 2.8$ \revised{are} used,
and no explicit minimum \pt requirement \revised{is} made.
Two different variants of offline \met are considered as offline reference:
\tagname{MET\_Topo} as primary reference,
and \tagname{MET\_EMJES\_RefFinal\_CellOutEM} as alternative (cf. \Sec{sec:software_reconstruction_met}),
which is the definition of \met the \ATLAS Supersymmetry group has agreed on to be used in publications analysing 2010 data,
and which will be used to facilitate conclusions from the analyser's point of view.
\revised{To} avoid the long name,
it will be referred to as the \revised{``\met definition of the SUSY group''} or \revised{``SUSY \met''} for short. %

\newcommand{\rsf}[2]{#2} %

\subsection{Jet Triggers}
\label{sec:results_trigger_performance_measurements_jet_triggers_2010}

\begin{figure}
  \centering
  \rsf{
    \incgraphicsclose{angle=0,width=\widthtwoplots}{postprocess_turnon_data10_F+G+H+I_Minbias_02-00516+517+518+519_EF_j30-50-75_jetNoEF_on_EF_mbMbts_1_eff_Note} %
  }{
    \incgraphics{angle=0,width=\widthtwoplots}{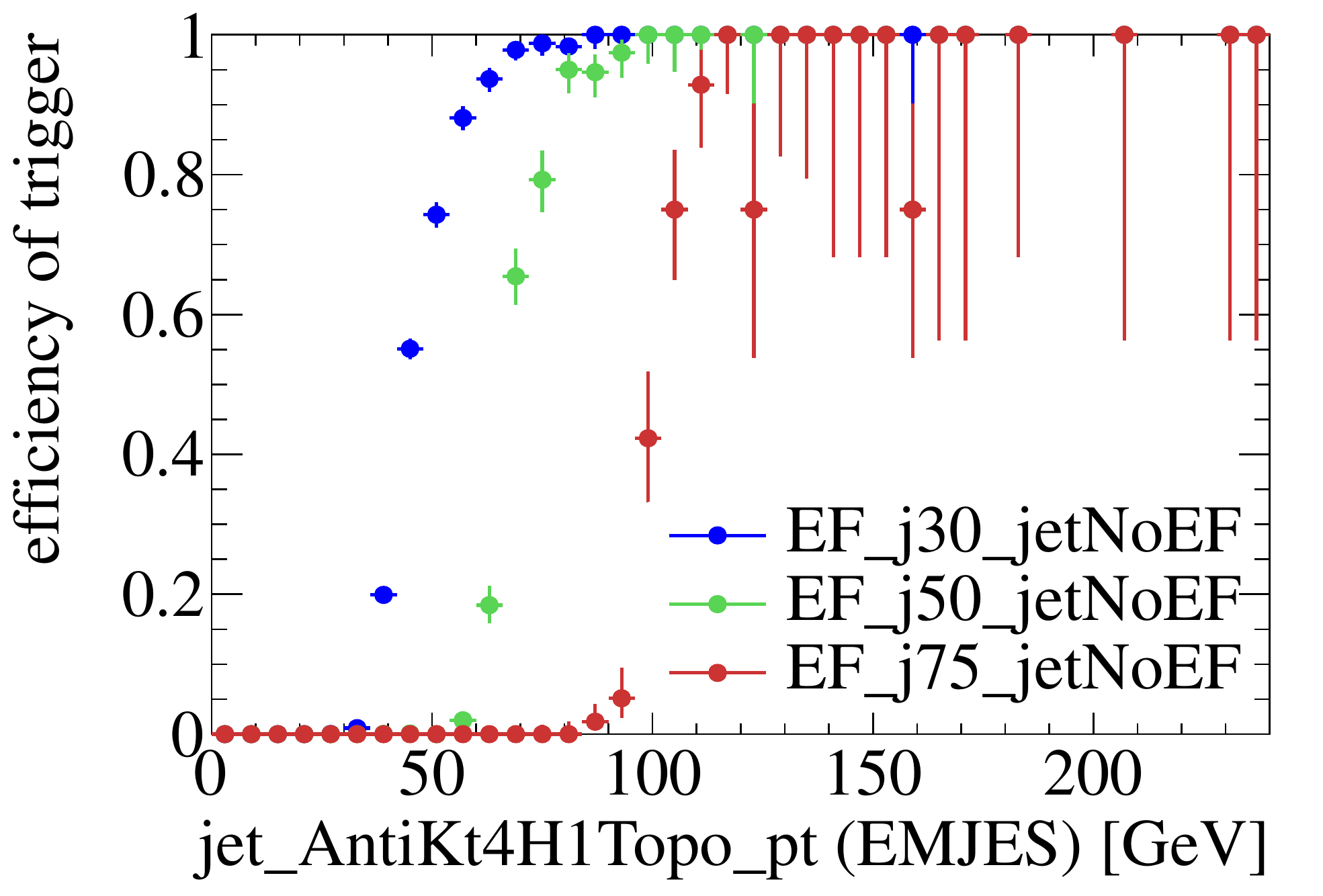} %
  }
    \hfill
  \rsf{
    \incgraphicsclose{width=\widthtwoplots}{postprocess_turnon_data10_G+H+I_JetTauEtmiss_02-00484_EF_j50-75_jetNoEF_on_EF_j30_jetNoEF_bs516+517+518+519_Note}
  }{
    \incgraphics{width=\widthtwoplots}{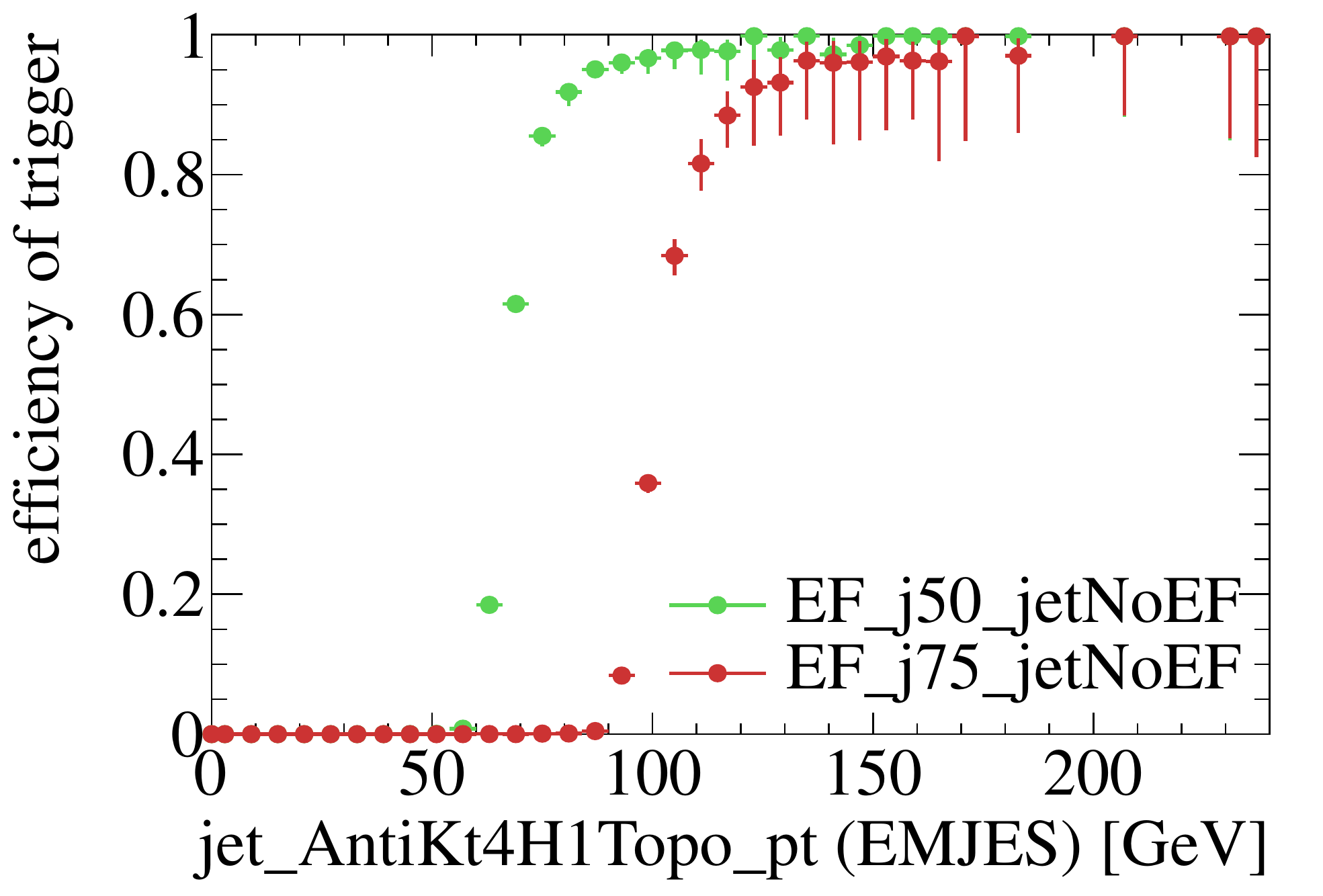}
  }
  \caption{
    Efficiency curves for a set of jet triggers,
    some of which will be used as sample triggers for the computation of \jetmet efficiencies.
    Left: Direct measurement on \tagname{MinBias} stream, periods F~--~I, %
      using \trigger{EF_mbMbts_1_eff} as sample trigger.
    Right: Bootstrapping on \tagname{JetTauEtmiss} stream with \jet{30} as sample trigger, periods G~--~I.
  }
  \label{fig:results_trigger_performance_measurements_turnons_jet_triggers}
\end{figure}

Bootstrapping the \jetmet turn-on curves from events collected with jet sample triggers requires to compute the jet-trigger efficiencies first.
There are several possibilities to measure the jet-trigger turn-on curves:
\begin{itemize}
  \item A direct measurement on events from the \tagname{MinBias} stream taken with a minimum-bias trigger.
  \item Bootstrapping, using a jet trigger with some even lower threshold as sample trigger on the \tagname{JetTauEtmiss} stream.
  \item Using a muon trigger as an orthogonal trigger on the \tagname{Muons} stream.
\end{itemize}
Note that the muon-triggered events do not \revised{yield} a completely unbiased sample
because requiring a muon enhances the admixture of jets from heavy flavor decays. %
\Fig{fig:results_trigger_performance_measurements_turnons_jet_triggers} shows the turn-on curves
of the jet triggers \jet{30}, \jet{50} and \jet{75}.
The left plot shows a direct measurement on the \tagname{MinBias} stream using \trigger{EF_mbMbts_1_eff} as sample trigger,
with data from periods F through I.
The turn-on curves from the four different periods agree among each other within the statistical uncertainties, %
\ie the trigger is stable enough to allow a combination of these periods.
It can be seen in \Fig{fig:results_trigger_performance_measurements_turnons_jet_triggers}
that the efficiency of \jet{50} has large statistical uncertainties when measured on the \tagname{MinBias} stream,
and the uncertainties for \jet{75} are even larger, of course.
As these triggers are to be used as sample triggers for bootstrapping the \jetmet trigger efficiencies,
the efficiencies of these two jet triggers need to be determined with higher statistics
on the \tagname{JetTauEtmiss} stream with the bootstrap method,
otherwise the very large uncertainties from the measurement on the \tagname{MinBias} stream
would deteriorate the turn-on curves obtained for the combined \jetmet triggers.
The resulting turn-on curves for \jet{50} and \jet{75} are shown in the right plot in \Fig{fig:results_trigger_performance_measurements_turnons_jet_triggers},
using bootstrapping and \jet{30} as sample trigger on the \tagname{JetTauEtmiss} stream.
The data is from periods G~--~I.
It has been checked that the efficiencies in these periods are consistent
so that it is justifiable to combine them. %
Note that comparing the left and right plot,
within the uncertainties, the efficiencies from the \tagname{MinBias} and the \tagname{JetTauEtmiss} stream agree.
This is an important cross-check
because the potential non-execution of the L2 jet reconstruction algorithms in events triggered by minimum-bias triggers
might bias the jet trigger efficiencies measured on the \tagname{MinBias} stream,
a problem that will be explained in more detail at the end of \Sec{sec:results_trigger_performance_measurements_missing_l2_jets}. %

All turn-on curves shown in the plots above and the following plots
for jet and combined \jetmet triggers are integrated over pseudorapidity $\eta$ and azimuthal angle $\phi$. %
It should be mentioned that there is a drop in efficiency
of the jet triggers as a function of $\eta$,
limited to the transition region around $|\eta| = 1.5$ between the barrel and the endcap calorimeters \cite{ATLAS-CONF-2010-094}. %
The efficiency is lower because in this region only a part of the calorimeter is used in the L1 trigger in 2010. %
The drop affects the \jetmet triggers in the same manner.
In order to collect sufficient statistics for the \jetmet trigger turn-on curves,
which require additional binning in \met,
it was decided to relinquish binning in $\eta$ here.

\subsection{Missing Transverse Energy Triggers}
\label{sec:results_trigger_performance_measurements_MET_triggers_2010}

It has \revised{been} explained above that for bootstrapping \jetmet triggers,
both parts of the combined trigger,
jets and \met, can be used as sample triggers.
Actually, this is only true for the full-chain triggers
because there is no \met trigger cutting only at EF (cf. \Sec{sec:results_trigger_performance_sample_triggers}).
For the employment of \met triggers as sample trigger however,
the problem lies in the measurement of the turn-on curves for the \met triggers themselves.
The main difficulty is to define a good sample of events to measure the \met trigger efficiency on.
Using an unbiased sample from the \tagname{MinBias} stream is not possible
because it runs out of statistics before reaching the plateau of even the lowest useful \met trigger which is \trigger{EF_xe20_noMu}.
Bootstrapping \trigger{EF_xe20_noMu} from a lower \met trigger is not feasible,
because there is only one lower trigger, \trigger{EF_xe15_noMu},
which is so close that it suffers from the problem of insufficient statistics,~too.
Hence, only orthogonal triggers remain as \revised{a} viable alternative for the sample selection.
Here, muon triggers will be used,
which can be considered as orthogonal triggers with respect to the \met and \jetmet triggers
because both jets and \met are quantities which in the trigger are calculated from calorimeter measurements only  (cf. \Sec{sec:tdaq_met}).
Note that turn-on curves for \met triggers can also be measured
on event samples with decays of $W$ bosons to electrons and muons \cite{TriggerPerf2010, Aracena:1303327}. %
\revised{Alas,} these are not unbiased in the general sense discussed above.

\begin{figure}
  \centering
  \rsf{
    \incgraphicsclose{width=\widthsingleplot}{postprocess_turnon_data10_F+G+H+I_Minbias_02-00555+556+557+558_EF_xe30_noMu_on_EF_mbMbts_1_eff_bsfilej484+485+486,bs559+560+563+562,cut3_Note}
  }{
    \incgraphics{width=\widthsingleplot}{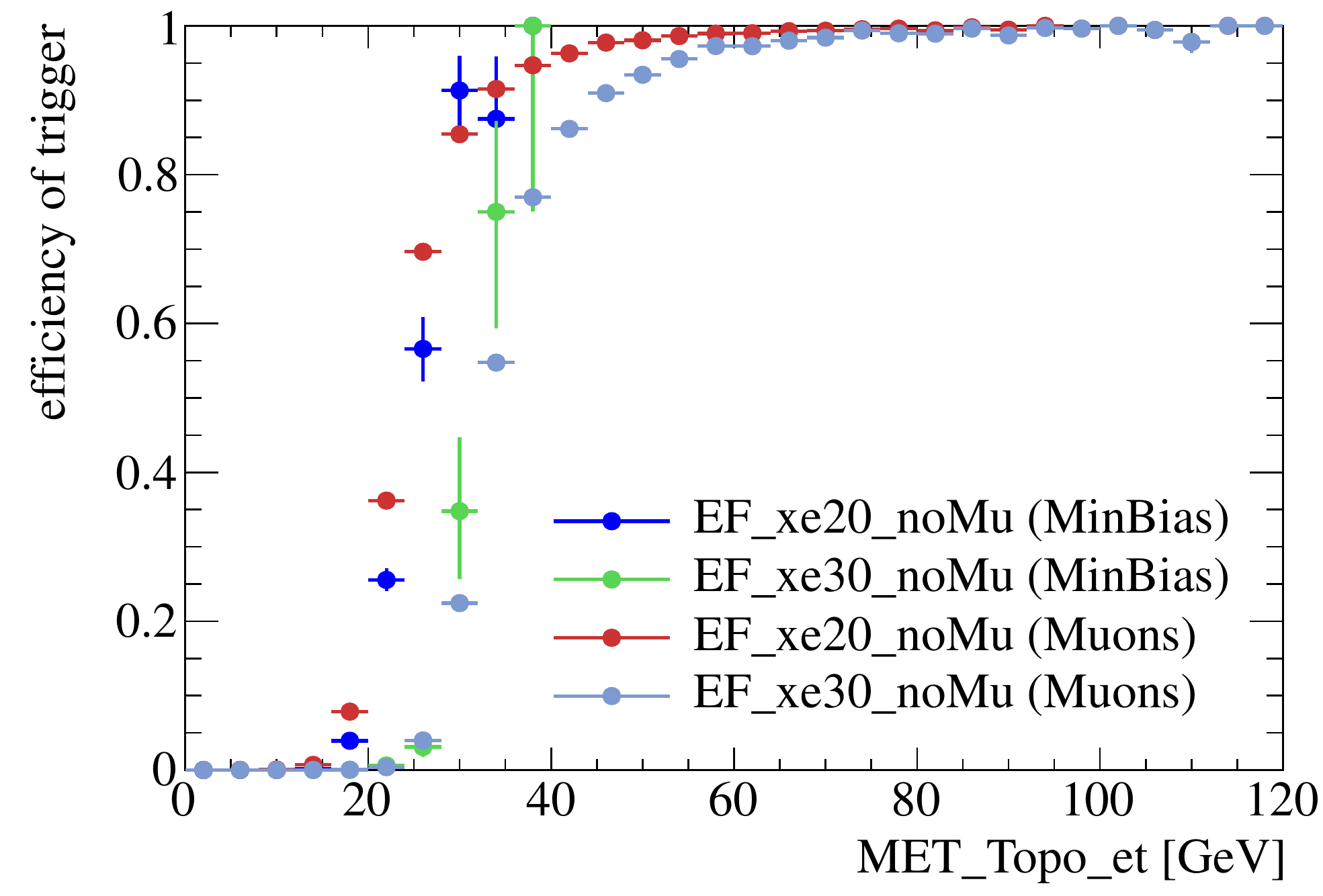}
  }
  \caption{
    Comparison of the turn-on curves for the \met triggers \trigger{EF_xe20_noMu} and \trigger{EF_xe30_noMu}, %
    using the minimum-bias trigger \trigger{EF_mbMbts_1_eff} (MinBias)
    and the single muon trigger \trigger{EF_mu13} (Muons) as sample triggers, respectively.
  }
  \label{fig:results_trigger_performance_measurements_turnonmet}
\end{figure}

In the following, muon triggers will be used as orthogonal triggers with respect to \met.
This allows to determine the \met turn-on, and use \met triggers for bootstrapping \jetmet triggers.
Later on, in \Sec{sec:results_trigger_performance_measurements_crosscheckmuons_2010}, %
muon triggers will again prove useful as an overall cross-check of the \jetmet efficiencies from bootstrapping.
\Fig{fig:results_trigger_performance_measurements_turnonmet} shows the turn-on curves of
two \met triggers with thresholds of $20$ and \unit[30]{GeV} to be used later on for bootstrapping of \jetmet trigger efficiencies.
It compares the efficiency estimates on events from the \tagname{MinBias} stream,
using the minimum-bias trigger \trigger{EF_mbMbts_1_eff} as sample trigger,
with those on events from the \tagname{Muons} stream, using the single muon trigger \trigger{EF_mu13} as sample trigger.
The agreement between the two samples is far from perfect,
but it is clear that only the \tagname{Muons} stream can provide enough statistics to measure the full turn-on curve.
A potential overestimation of the efficiencies in the turn-on region when using the \tagname{Muons} stream has to be kept in mind.
It should be noted that the efficiencies in the turn-on region also vary slightly with the threshold of the muon trigger which is chosen as sample trigger,
decreasing with increasing muon trigger threshold,
as can be seen in the left plot in \Fig{fig:results_trigger_performance_measurements_met_turnon_susymet_met_turnon_muonthreshold},
where the turn-on curves of \trigger{EF_xe30_noMu} on event samples taken with single muon triggers with different thresholds are compared.

\begin{figure}
  \centering
  \incgraphics{width=\widthtwoplots}{{{postprocess_turnon_data10_G+H+I_Muons_02-00438+439+440_EF_xe30_noMu_on_EF_mu20_sort}}} %
    \hfill
  \incgraphics{width=\widthtwoplots}{{{postprocess_turnon_data10_G+H+I_Muons_02-00521+522+523_EF_xe55_noMu_on_EF_mu13_SUSYMET}}} %
  \caption{
    Left: Comparison of the turn-on curve for \trigger{EF_xe30_noMu},
    on event samples taken with single muon triggers with different thresholds (given in the name in GeV),
    using \tagname{MET\_Topo} as offline reference.
    \newline %
    Right: Turn-on curves for several \met triggers using SUSY \met as offline reference,
    measured on events taken with \trigger{EF_mu13}.
    Note that for high offline \met values the points for all four triggers fall together.
  }
  \label{fig:results_trigger_performance_measurements_met_turnon_susymet_met_turnon_muonthreshold}
\end{figure}

When computing the turn-on curves of \met triggers with respect to the SUSY definition of offline \met,
using \revised{muon} triggers to obtain an orthogonal event selection has proven to exhibit undesired effects. %
The reason is that muons are included in the SUSY \met,
whereas in the trigger \met they are not.
Therefore, the \met turn-on as function of SUSY \met does not reach a plateau for high values of offline \met,
but instead drops again with increasing \met,
as can be seen in the right plot in \Fig{fig:results_trigger_performance_measurements_met_turnon_susymet_met_turnon_muonthreshold}
for four \met triggers with different thresholds.
This behavior is presumably due to events where muons and neutrinos are produced back-to-back.
In these events, there is real \met coming from the neutrino.
This real \met \revised{enters} the computation of SUSY \met,
giving high values for SUSY \met via the measurement of the transverse momentum of the muon.
But as the trigger \met does not include the muon,
it cannot measure the imbalance in $E_T$ due to the energy taken away by the neutrino
so that it measures only little or no \met,
and the probability of the \met trigger chains to fire is low.

\begin{figure}
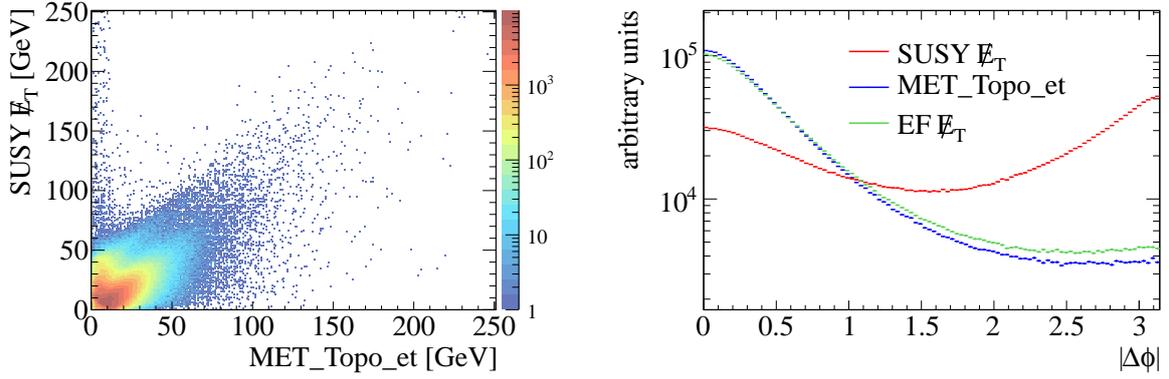

  \centering
  \incgraphics{width=\widthtwoplots}{{{plot.ntuple_METcorrelation_NTUP_SUSY_2_1_Note}}}
    \hfill
  \incgraphics{width=\widthtwoplots}{{{plot.ntuple_METcorrelation_NTUP_SUSY_1Note}}}
  \caption{
    Comparisons between \tagname{MET\_Topo} \met and the SUSY definition of \met on events sampled with \trigger{EF_mu13} from the \tagname{Muons} stream, period I.
    Left: Correlation of the two different \met definitions in events with exactly one reconstructed muon with \pt above \GeV{10}.
    Right: Distribution of the angle between the \met direction and the direction of the leading muon in the transverse plane,
    showing a completely different behavior for \tagname{MET\_Topo} and SUSY \met,
    whereas the distributions of \met for EF and \tagname{MET\_Topo} are very similar.
  }
  \label{fig:results_trigger_performance_measurements_compare_met_definitions}
\end{figure}

\Fig{fig:results_trigger_performance_measurements_compare_met_definitions} shows two comparisons between \tagname{MET\_Topo} and the SUSY definition of \met
on events sampled with \trigger{EF_mu13} from the \tagname{Muons} stream in period I.
In the left plot, showing the correlation between the two different \met definitions
in events with exactly one reconstructed muon with \pt above \GeV{10},
note the vertical structure which can be seen at the left border of the plot,
corresponding to events with vanishing values of \tagname{MET\_Topo} but high SUSY \met.
It can be seen that, taking horizontal slices of this plot,
the events from this vertical feature become predominant in number for high \met,
which explains the decrease of the \met trigger efficiency
when measured as function of the SUSY \met, but not when using \tagname{MET\_Topo} as offline reference.
The right plot in \Fig{fig:results_trigger_performance_measurements_compare_met_definitions} demonstrates the difference
between the two \met definitions arising from the muon correction.
The distribution of the angle between the \met direction and the direction of the leading muon in the transverse plane
peaks at zero for \tagname{MET\_Topo} and EF \met,
whereas for SUSY \met most events have \met and the leading muon going into anti-parallel directions.
This corroborates the argument that events with muons and neutrinos produced back-to-back
are responsible for the problems arising when measuring the \met trigger efficiency with SUSY \met as offline reference,
which is enhanced on the \tagname{Muons} stream due to the selection of only events in which a muon was identified by the trigger.

\subsection{\texorpdfstring{Combined \Jetmet Triggers} {Combined JetMET Triggers}}
\label{sec:results_trigger_performance_measurements_jetmet_triggers_2010}

The measurements of the efficiencies of the jet and \met triggers which have been discussed above
can now be used to determine the efficiencies of the combined \jetmet triggers with the bootstrapping method.
The two different types of \jetmet triggers are first discussed separately,
first the jet-seeded and then the full chain triggers,
and then compared against each other.
As offline reference for \met,
both \tagname{MET\_Topo} and, where possible and expedient, also the SUSY definition are used.
Often, the turn-on curves are shown as contour plots.
Note that this plotting style is not intended to allow to read off exact efficiency values.
Instead, it ought to provide an overview of the behavior of the \jetmet trigger at one glance,
in particular with regards to where the plateau region begins,
how broad the turn-on is and how it behaves as function of the respective orthogonal offline variable.
Projections will be used to compare triggers with different thresholds or of different type.

\subsubsection{Jet-seeded Triggers}

As discussed in \Sec{sec:results_trigger_performance_sample_triggers}, %
no \met trigger can be used to bootstrap the jet-seeded %
combined \jetmet triggers,
because all \met triggers include cuts at L1 and thus violate condition \eqref{eq:bootstrap_requirement}.
Therefore, only jet triggers will be considered for this type of \jetmet trigger.
According to \Tab{tab:results_trigger_performance_trigger_counts_2010},
\trigger{EF_j75_jetNoEF} is the best sample trigger here.
It was shown above that its efficiency cannot be measured on a minimum-bias sample due to lack of statistics,
but that it has to be taken from the \tagname{JetTauEtmiss} stream.
These events can be sampled using the jet trigger \trigger{EF_j30_jetNoEF},
which reaches its plateau for offline jet \pt values below the onset of the turn-on region of \trigger{EF_j75_jetNoEF}
so that in principle no bootstrap is needed,
but for consistency with the \jetmetfall turn-on curves that are shown later and for which using the bootstrap method is essential,
it is done here as well.
The complete sequence used to compute jet or \jetmet trigger efficiencies, respectively, thus is the following:

\vspace*{1mm}
  Jet\phantom{+ \met}: \trigger{EF_mbMbts_1_eff}
    $\xrightarrow[\tagname{MinBias}]{\text{counting}}$ \jet{30} %
    $\xrightarrow[\tagname{JetTauEtmiss}]{\text{bootstrapping}}$ \jet{75},

\vspace*{1mm}
  Jet + \met: \jet{75}
    $\xrightarrow[\tagname{JetTauEtmiss}]{\text{bootstrapping}}$ \jetmetj{*}.
\vspace*{1mm}

\noindent
The method used to compute the turn-on is given above the arrows and the name
of the stream from which the events are sampled is given below.
The sample trigger is written to the left and the target trigger to the right of the arrows.

\begin{figure}
  \centering
  \rsf{
    \incgraphicsclose{width=\widthtwoplots}{postprocess_turnon_data10_G+H+I_JetTauEtmiss_02-00509+510+511_EF_EFonly_xe25_noMuL1_J55L2_j70_on_EF_j75_jetNoEF_bsfilej484+485+486,bs516+517+518+519,cut3_Note}
  }{
    \incgraphics{width=\widthtwoplots}{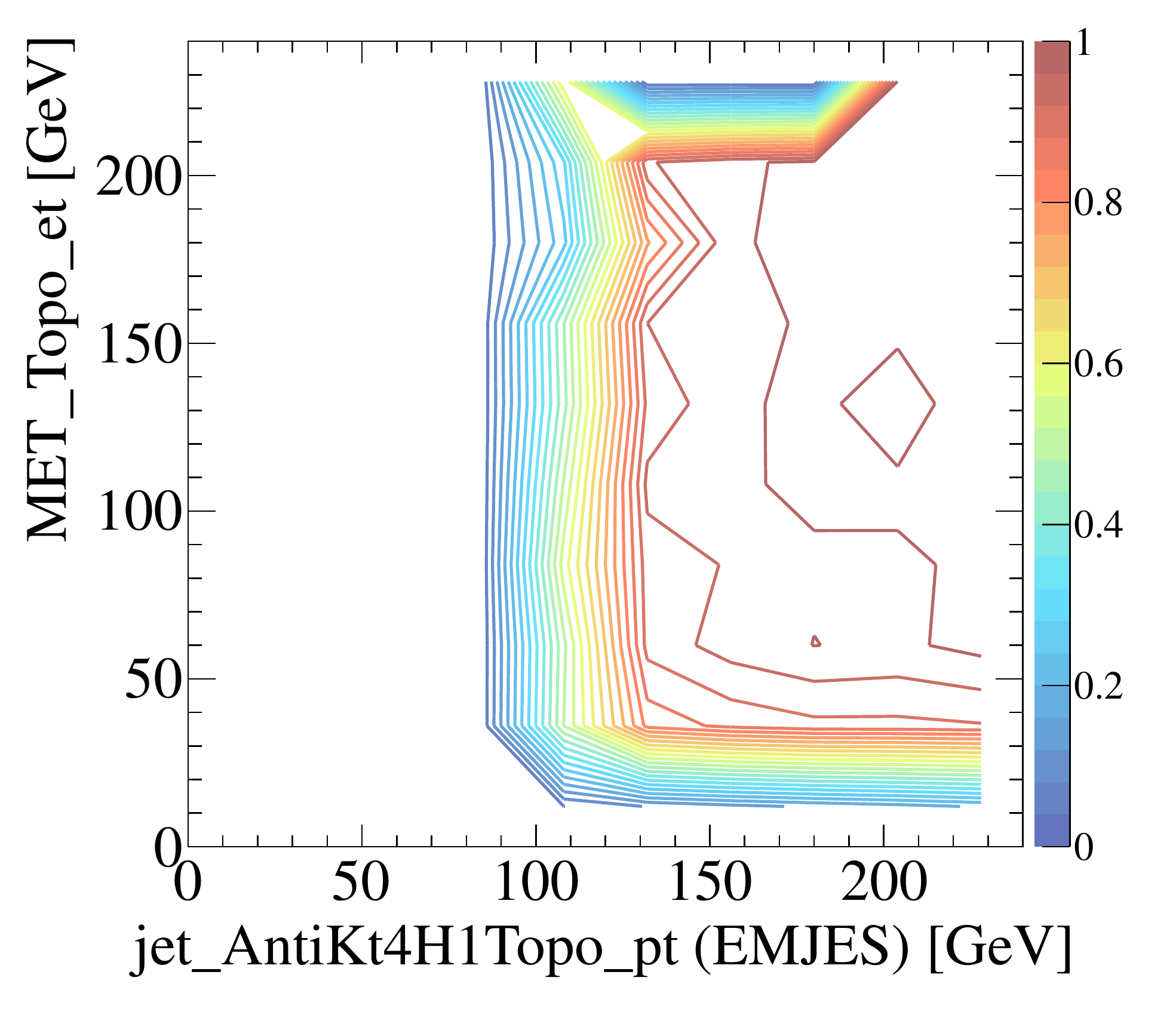}
  }
  \hfill
  \rsf{
    \incgraphicsclose{width=\widthtwoplots}{postprocess_turnon_data10_G+H+I_JetTauEtmiss_02-00509+510+511_EF_EFonly_xe25_noMuL1_J55L2_j70_on_EF_j75_jetNoEF_bsfilej484+485+486,bs516+517+518+519,cut3_text_Note}
  }{
    \incgraphics{width=\widthtwoplots}{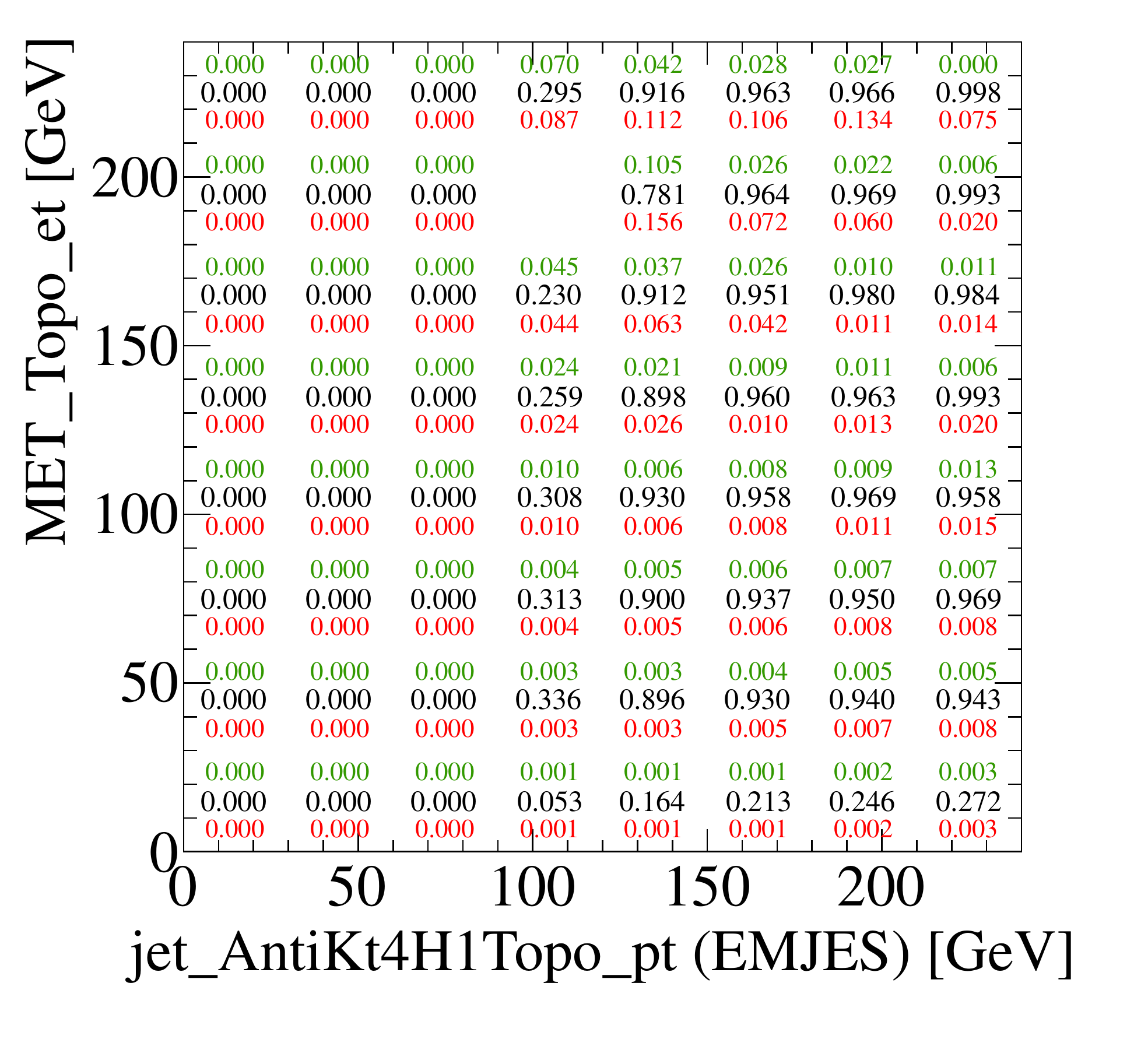}
  }
  \\
  \incgraphics{width=\widthtwoplots}{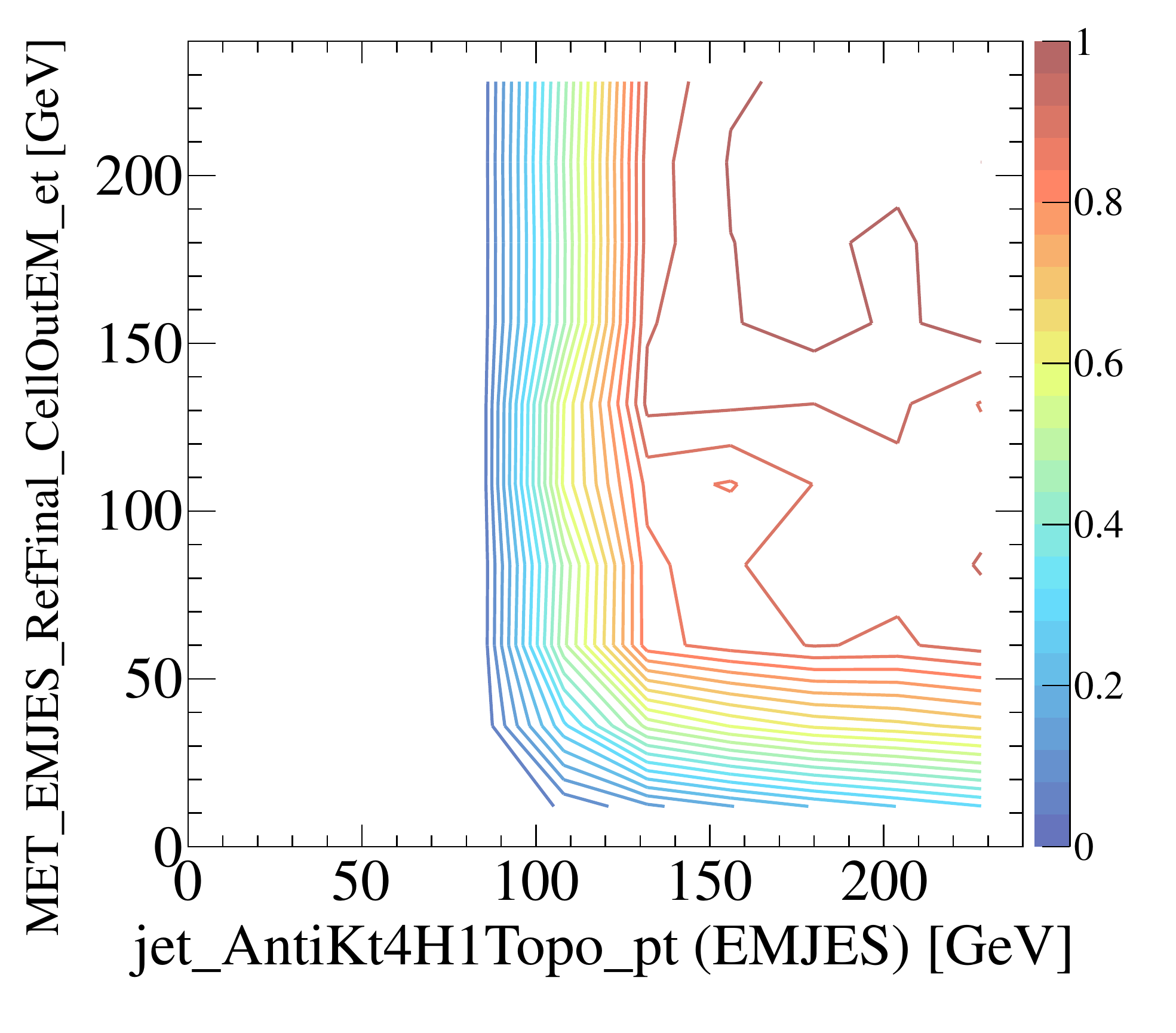}
    \hfill
  \incgraphics{width=\widthtwoplots}{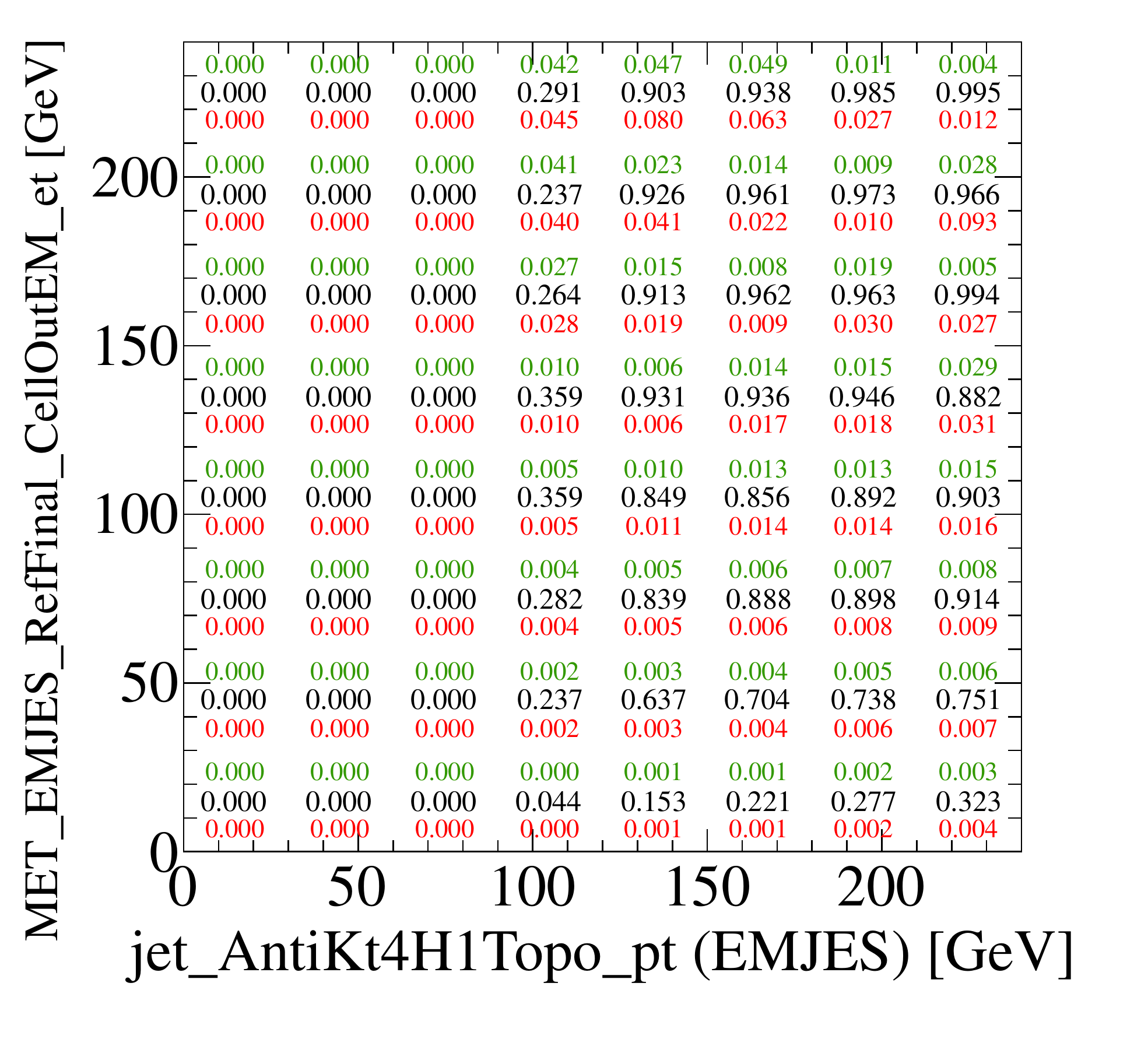}
  
  \caption{
    Efficiency of jet-seeded \jetmet trigger \jetmetj{25}
    from bootstrapping using \tagname{MET\_Topo} (top) and the SUSY \met (bottom) as offline reference
    and \trigger{EF_j75_jetNoEF} as sample trigger.
    Left: contour plot,
    right: efficiency estimates together with upper (green / above) and lower (red / below) errors for each bin.} %
  \label{fig:results_trigger_performance_measurements_turnons_jetmetj25}
\end{figure}

The upper plots in Figure \ref{fig:results_trigger_performance_measurements_turnons_jetmetj25} show the two-dimensional turn-on curve
of the jet-seeded combined \jetmet trigger \jetmetj{25} as function of \tagname{MET\_Topo} and jet \pt.
This trigger is the most important amongst the \jetmet triggers,
because it has been used as primary trigger to select events for the search for Supersymmetry in zero-lepton final states in 2010 data.
To the left, a contour plot of the efficiencies is shown,
the right plot gives the estimates of the efficiencies as numbers printed in black,
adding the upper and lower uncertainties above and below the efficiency values in green and red, respectively.
The empty spot in the right plot is one bin which happens to have no entries in the denominator. %
It can be seen that the plateau has a well-distinguished rectangular shape with an efficiency of about \percent{90},
going to values above \percent{95} for $\met > \GeV{90}$ and jet $\pt > \GeV{150}$.
A slight increase of the trigger efficiency with jet \pt can be seen outside the plateau region for events with little or no \met.
This can be explained by the fact that due to the lower resolution of the \met calculation in the trigger with respect to offline
in events which contain high-energetic jets,
it is more likely that a larger amount of (fake) \met is reconstructed online,
and these events are therefore more likely to also pass the \met threshold of the trigger.
The bump in the horizontally running turn-on region of the jet part of the combined trigger
as well as the seeming drop of efficiency for very high \met are due to a lack of statistics.

The lower plots in \Fig{fig:results_trigger_performance_measurements_turnons_jetmetj25} show the efficiencies of the same trigger,
but using the \met definition of the \ATLAS SUSY group instead of \tagname{MET\_Topo}.
The events from which all four plots were produced are the same,
they only differ in the way of computing the offline \met. %
The lower plots, in particular the left one, facilitate a comparison of the trigger efficiency
and the offline cuts used in the zero-lepton analysis of the \ATLAS SUSY group and in this thesis in \Sec{sec:analysis_susysearch},
where cuts on the missing transverse energy at \GeV{100}
and on the transverse momentum of the leading jet \pt at \GeV{120} were applied,
using exactly the same offline variables as plotted on the axes of the turn-on curve shown in these plots.
Because of the calibration factors,
the values of the SUSY \met are usually larger than those of \tagname{MET\_Topo} for the same event.
Neglecting the muon contributions,
the vertical axis is thus roughly a streched version of what is shown in the upper plots,
so that the vertically running turn-on region of the \met part gets broadened,
whereas the position and width of the turn-on of the jet part remain the same.
Note that in these turn-on curves no correlation between \met and jets can be seen,
whereas, when looking at the denominator and enumerator histograms separately,
the correlation is visible as a diagonal feature. %

\subsubsection{Full-Chain Triggers}

\begin{figure}
  \centering
  \rsf{
    \incgraphicsclose{width=\widthtwoplots}{postprocess_turnon_data10_G+H+I_JetTauEtmiss_02-00509+510+511_EF_xe30_noMuL1_J30L2_j45_on_EF_j50_jetNoEF_bsfilej484+485+486,bs516+517+518+519,cut3_Note} %
  }{
    \incgraphics{width=\widthtwoplots}{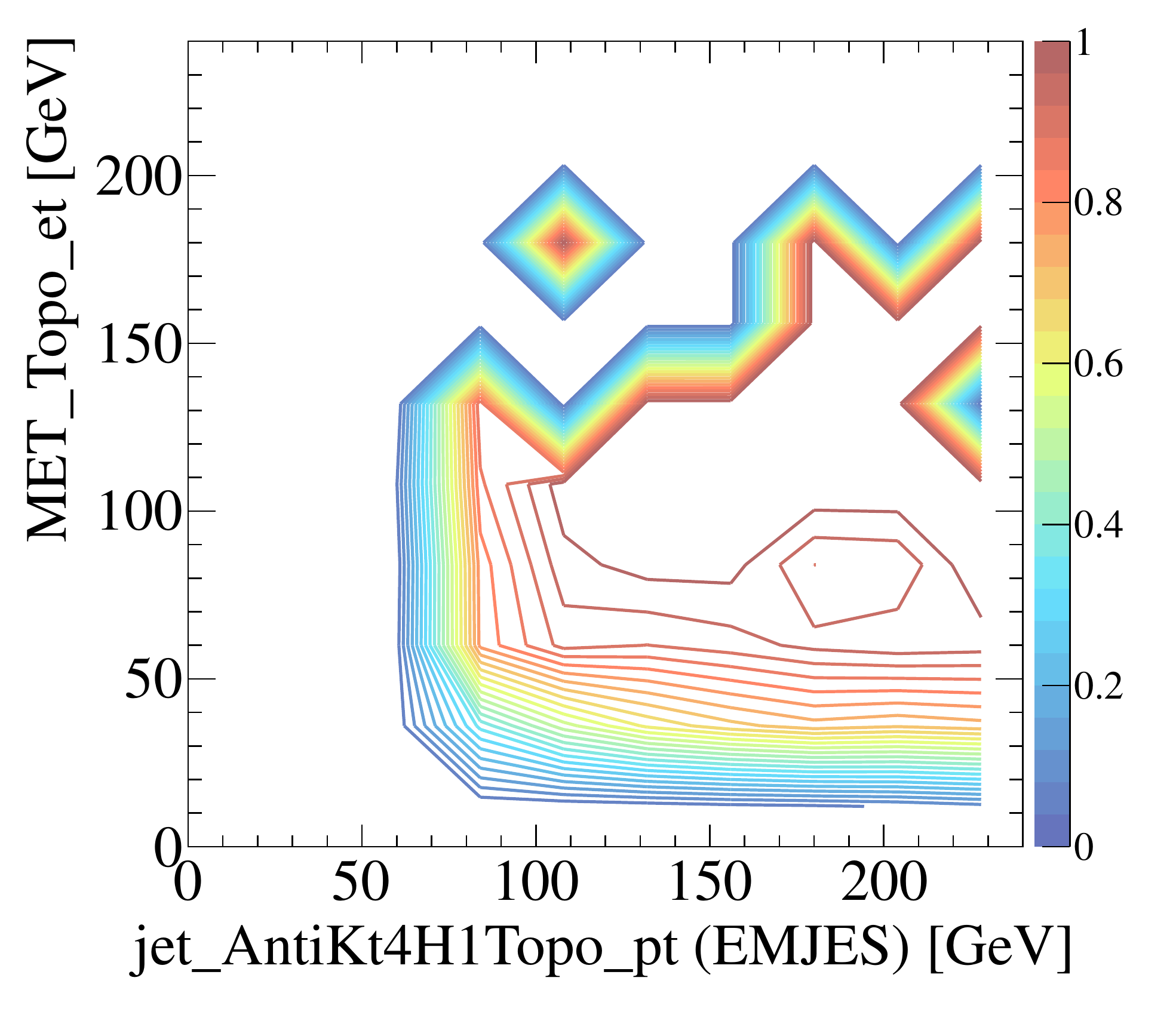}
  }
    \hfill
  \rsf{
    \incgraphicsclose{width=\widthtwoplots}{postprocess_turnon_data10_G+H+I_JetTauEtmiss_02-00547+548+549_EF_xe30_noMuL1_J30L2_j45_on_EF_j50_jetNoEF_bsfilej484+485+486,bs516+517+518+519,cut3_SUSYMET_Note}
  }{
    \incgraphics{width=\widthtwoplots}{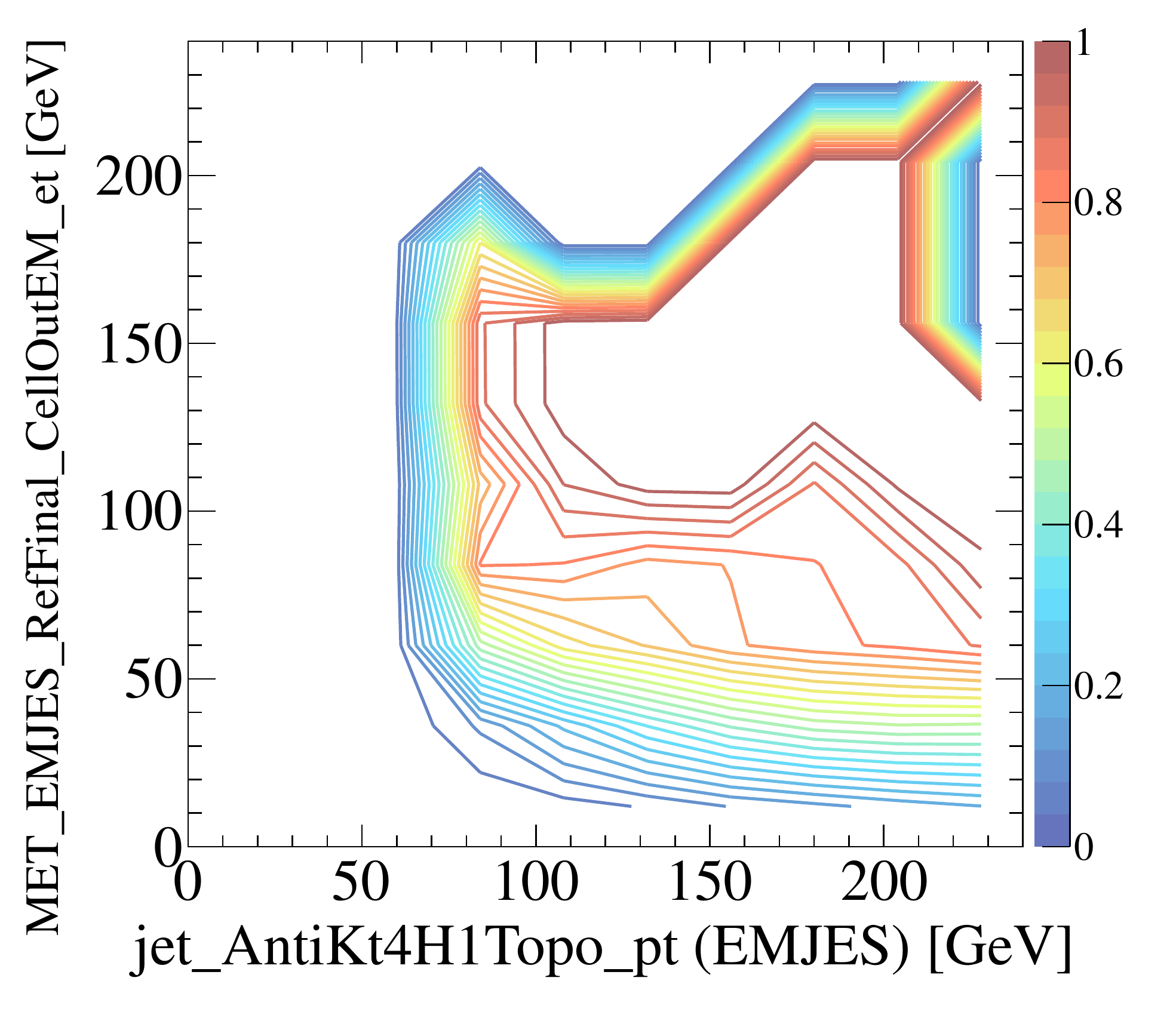}
  }
  \caption{
    Efficiency of the full-chain \jetmet trigger \jetmetf{30} from bootstrapping,
    using the jet trigger \jet{50} as sample trigger.
    Left: using \tagname{MET\_Topo} as offline reference,
    right: using the SUSY \met definition.
  }
  \label{fig:results_trigger_performance_measurements_turnons_jetmetf30_jetsample}
\end{figure}

\begin{figure}
  \centering
  \incgraphics{width=\widthtwoplots}{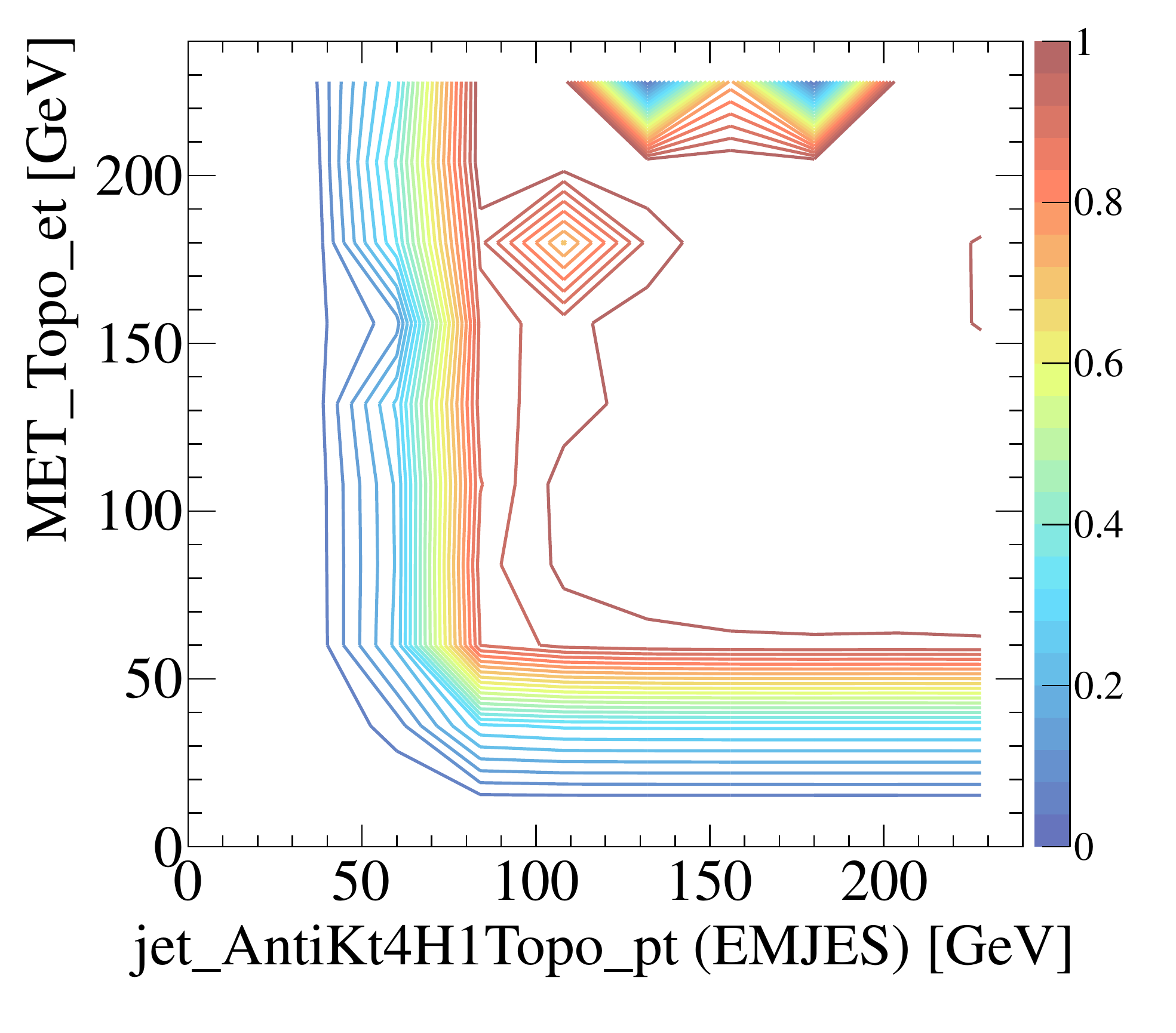} %
    \hfill
  \incgraphics{width=\widthtwoplots}{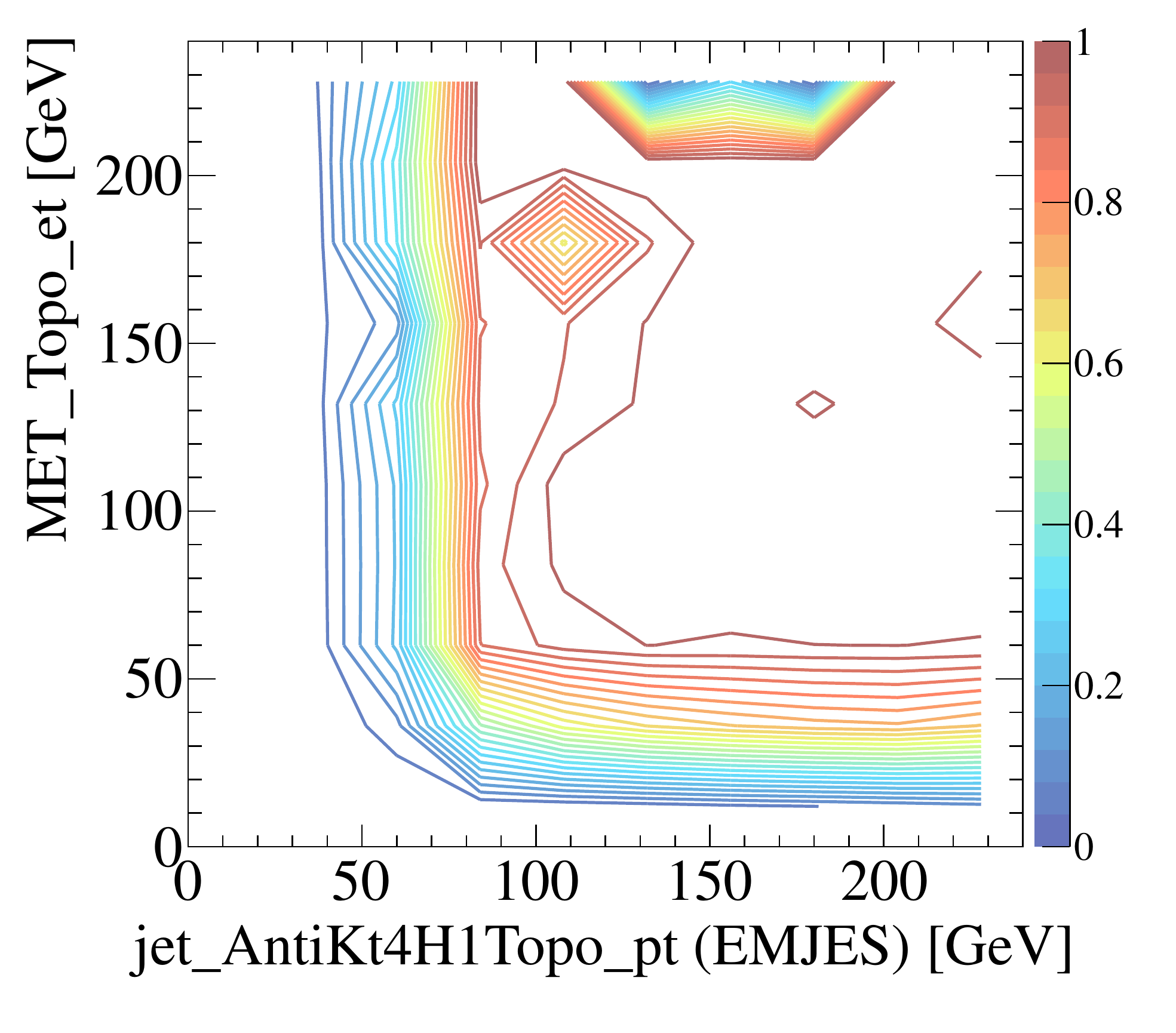} %
  \caption{
    Efficiency of the full-chain \jetmet trigger \jetmetf{30} from bootstrapping,
    using \trigger{EF_xe30_noMu} as sample trigger.
    Both plots use \tagname{MET\_Topo} as offline reference.
    Left: using a one-dimensional turn-on for bootstrapping,
    right: using a two-dimensional turn-on which is also binned in jet \pt.}
  \label{fig:results_trigger_performance_measurements_turnons_jetmetf30_topo}
\end{figure}

The \jetmet triggers including the full \met chain all have lower jet thresholds than the one studied above
(cf. \Tab{tab:results_trigger_performance_jetmet_chains_2010}),
which prohibits the use of \jet{75} as sample trigger.
The next lower jet trigger which can be used as sample trigger is \jet{50}.
The full chain of computations is then

\vspace*{1mm}
  Jet\phantom{+ \met}: \trigger{EF_mbMbts_1_eff}
    $\xrightarrow[\tagname{MinBias}]{\text{counting}}$ \jet{30}
    $\xrightarrow[\tagname{JetTauEtmiss}]{\text{bootstrapping}}$ \jet{50},

\vspace*{1mm}
  Jet + \met: \jet{50}
    $\xrightarrow[\tagname{JetTauEtmiss}]{\text{bootstrapping}}$ \jetmetf{*}.
\vspace*{1mm}

\Fig{fig:results_trigger_performance_measurements_turnons_jetmetf30_jetsample} shows the two-dimensional turn-on curves of the full-chain \jetmet trigger \jetmetf{30},
which has a threshold of \unit[30]{GeV} on \met.
The efficiencies are computed on an event sample taken with \jet{50} via bootstrapping.
In the left plot, \tagname{MET\_Topo} \met is used as offline variable,
in the right plot, the SUSY definition of \met is used.
It is obvious that in particular the high \met range needs more statistics.
While this was not possible for the jet-seeded triggers,
it is possible to determine the turn-on of the full-chain triggers also on event samples taken with \met triggers.
The result is shown in the left plot in \fig{fig:results_trigger_performance_measurements_turnons_jetmetf30_topo},
where \trigger{EF_xe30_noMu} has been used as sample trigger.
The turn-on of the sample trigger has been computed on the \tagname{Muons} stream,
as was discussed in \Sec{sec:results_trigger_performance_measurements_MET_triggers_2010}. %
The computation thus comprises two steps:

\vspace*{1mm}
  Jet\phantom{+ \met}: \trigger{EF_mu13}
    $\xrightarrow[\tagname{Muons}]{\text{counting}}$ \trigger{EF_xe30_noMu}

\vspace*{1mm}
  Jet + \met: \trigger{EF_xe30_noMu}
    $\xrightarrow[\tagname{JetTauEtmiss}]{\text{bootstrapping}}$ \jetmetf{30}.
\vspace*{1mm}

As can be seen in \Fig{fig:results_trigger_performance_measurements_turnons_jetmetf30_topo},
using the \met trigger yields higher statistics and a better coverage
of the plateau for high \met values than the jet sample trigger in \Fig{fig:results_trigger_performance_measurements_turnons_jetmetf30_jetsample}.
An interesting observation is that while the position of the plateau is unchanged,
the turn-on region of the \met part in the left plot of \Fig{fig:results_trigger_performance_measurements_turnons_jetmetf30_topo} does not show the variation
with jet \pt that is clearly visible in \Fig{fig:results_trigger_performance_measurements_turnons_jetmetf30_jetsample}.
The reason for this is that the \emph{relative} efficiency of the combined \jetmet trigger
on the sample taken with \trigger{EF_xe30_noMu} is close to one in this range
so that the shape visible in the plot is completely dominated by the turn-on behavior of the sample trigger \trigger{EF_xe30_noMu}.
When using a one-dimensional turn-on of \trigger{EF_xe30_noMu} for bootstrapping,
as was done in this plot,
the \pt dependence of the turn-on therefore is lost.
In a similar spirit, the slow turn-on of the jet part,
which can be seen in the left plot in \Fig{fig:results_trigger_performance_measurements_turnons_jetmetf30_topo}
in the vertical slice between $40$ and \GeV{60} of offline jet \pt,
is not reproduced in \Fig{fig:results_trigger_performance_measurements_turnons_jetmetf30_jetsample} due to the steep turn-on of the jet sample trigger. %

The jet-\pt dependence of the \met turn-on of the combined \jetmet trigger can be recovered
using a two-dimensional turn-on curve of the \met sample trigger,
where in addition to the binning in offline \met on the first axis, the second axis is binned in jet \pt.
This is possible thanks to the high statistics available from the \tagname{Muons} stream.
The result of bootstrapping using a two-dimensional turn-on of the sample trigger
is shown in the right plot of \Fig{fig:results_trigger_performance_measurements_turnons_jetmetf30_topo}.
In comparison to the left plot from the same figure and to the left plot in \Fig{fig:results_trigger_performance_measurements_turnons_jetmetf30_jetsample},
it becomes clear that the two-dimensional turn-on better reproduces the shape of the turn-on of the \met part.

Corresponding plots using an \met sample trigger with SUSY \met as offline reference are not possible,
because the \met trigger turn-ons cannot be measured properly on the \tagname{Muons} stream as discussed above.
Finally, it needs to be said that the caveats concerning missing L2 jet information
given in \Sec{sec:results_trigger_performance_measurements_crosscheckmuons_2010} %
also hold here when using \met triggers as sample triggers.

\subsubsection{Comparisons of Different \Jetmet Triggers}

In the remaining part of this section,
one-di\-men\-sional projections of the two-dimensional turn-on curves
of the combined \jetmet triggers will be presented,
making possible a direct comparison between different thresholds and trigger types in one plot.
For consistency among the different types of triggers,
only jet sample triggers will be used here,
because these are the only ones that can be used for bootstrapping the jet-seeded chains, too.
\begin{figure}
  \centering
  \rsf{
    \incgraphicsclose{width=\widthtwoplots}{postprocess_turnon_data10_G+H+I_JetTauEtmiss_02-00509+510+511_EF_xe40_noMuL1_J30L2_j45_on_EF_j50_jetNoEF_jetCut140_Note}
  }{
    \incgraphics{width=\widthtwoplots}{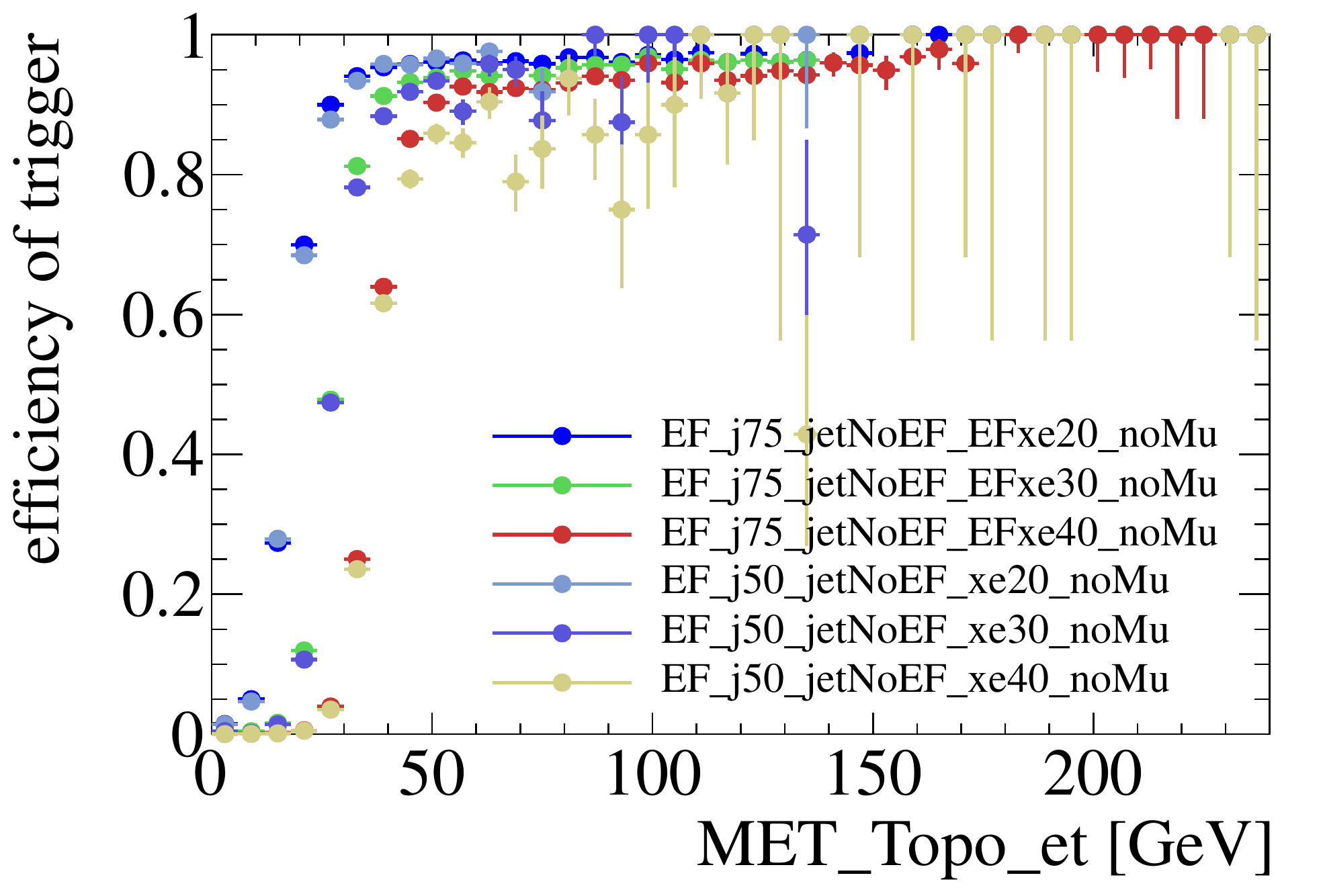}
  }
    \hfill
  \rsf{
    \incgraphicsclose{width=\widthtwoplots}{postprocess_turnon_data10_G+H+I_JetTauEtmiss_02-00487+488+489_EF_xe40_noMuL1_J30L2_j45_on_EF_j50_jetNoEF_cutJet140_SUSYMET_Note}
  }{
    \incgraphics{width=\widthtwoplots}{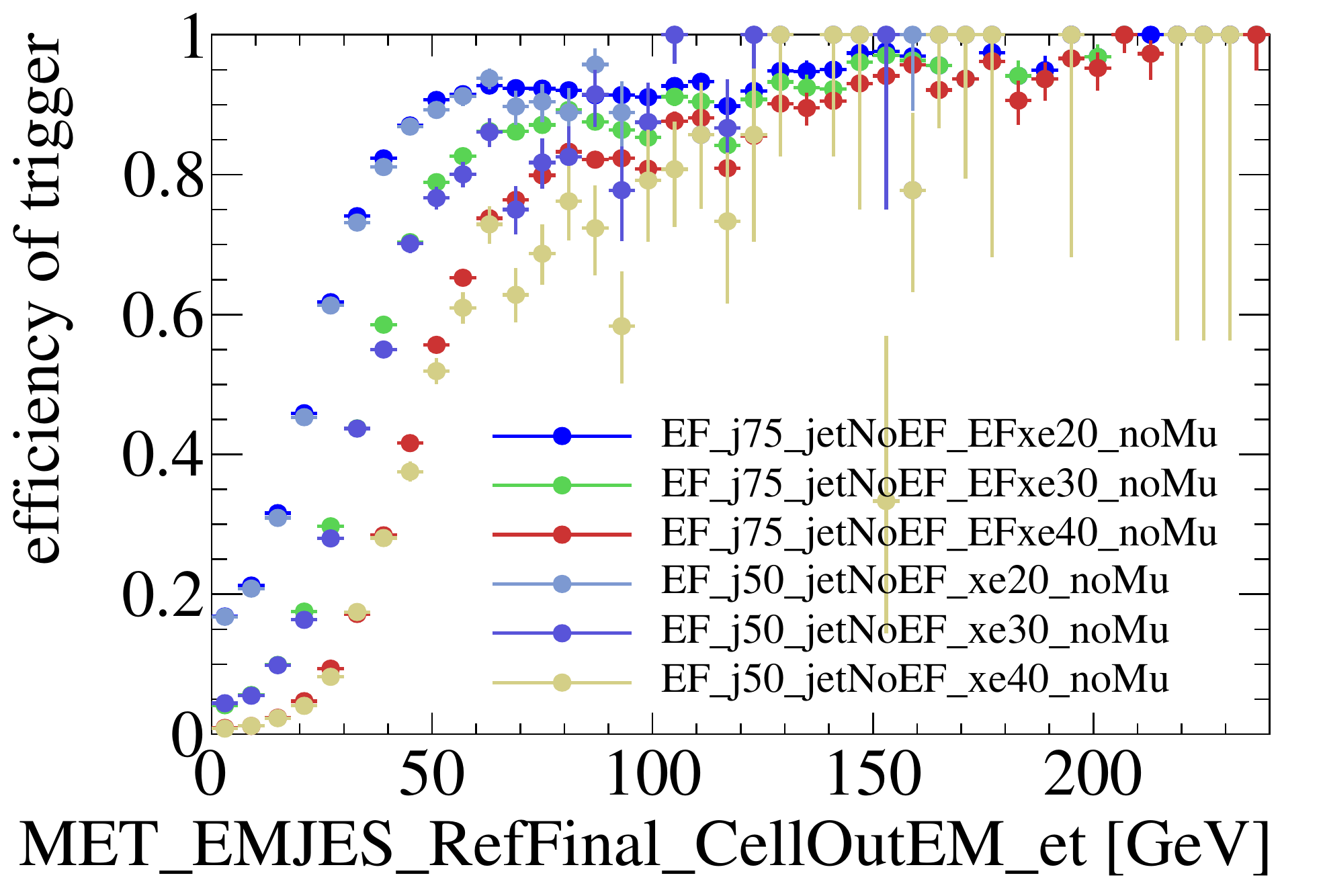}
  }
  \caption{
    Efficiency of the \met part of both types of \jetmet triggers with thresholds on \met at $20$, $30$ and \GeV{40},
    selecting events with offline jet \pt of at least \GeV{140}.
    Left plot: as function of \tagname{MET\_Topo}, right plot: as function of SUSY \met.
    The sample triggers are \trigger{EF_j75_jetNoEF} and \trigger{EF_j50_jetNoEF} for the jet-seeded and full-chain triggers, respectively.
  }
  \label{fig:results_trigger_performance_measurements_turnons_jetmet_projection_MET}
\end{figure}

\Fig{fig:results_trigger_performance_measurements_turnons_jetmet_projection_MET} shows the turn-on behavior of the \met part
of the two groups of jet-seeded and full-chain \jetmet triggers,
presenting from each group three triggers with thresholds on \met of $20$, $30$, and \GeV{40}
and using \jet{75} and \jet{50} as sample trigger for the jet-seeded (with a threshold of \GeV{75} at EF)
and full-chain triggers (with a threshold of \GeV{50} at EF), respectively.
The left plot uses \tagname{MET\_Topo} as offline \met, the right plot the SUSY \met definition.
In both cases, the projection is done after a cut on offline jet \pt at \GeV{140}.
The left plot shows a plateau efficiency around \percent{95}, whereas it appears to be little bit lower in the right plot.
In both plots, the efficiency of the jet-seeded triggers is slightly higher in the turn-on region with respect to the full-chain triggers,
whereas the plateau value is consistent within error bars.

\begin{figure}
  \centering
  \rsf{
    \incgraphicsdraft{width=\widthtwoplots}{postprocess_turnon_data10_G+H+I_JetTauEtmiss_02-00487+488+489_EF_EFonly_xe65_noMuL1_J55L2_j70_on_EF_j75_jetNoEF_cutJet140_SUSYMET_Note}
  }{
    \incgraphics{width=\widthtwoplots}{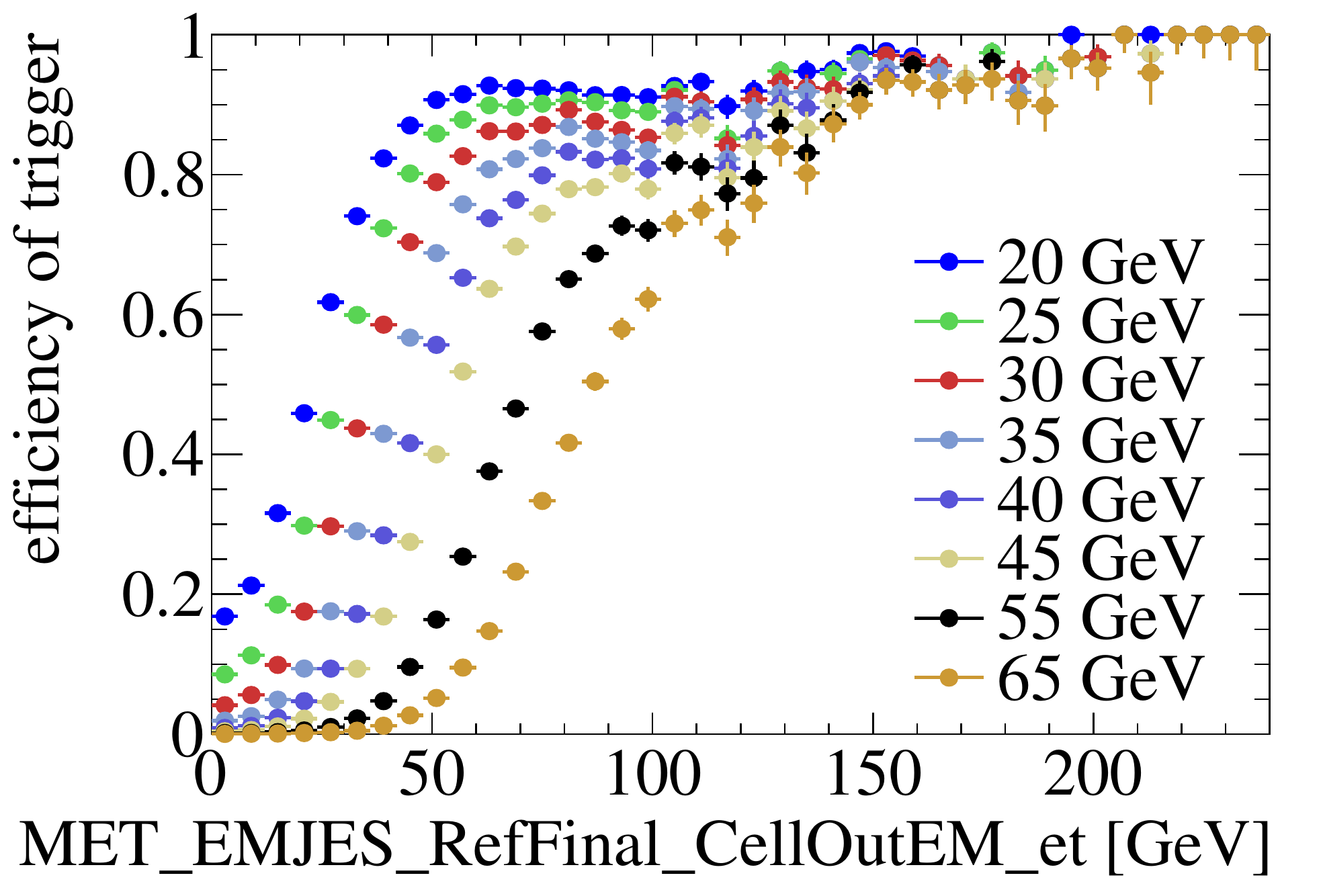}
  }
    \hfill
  \rsf{
    \incgraphicsdraft{width=\widthtwoplots}{postprocess_turnon_data10_G+H+I_JetTauEtmiss_02-00509+510+511_EF_EFonly_xe25-35-45_noMuL1_J55L2_j70+EF_xe20-30-40_on_EF_j75-50_jetNoEF_bsfilej484+485+486,bs516+517+518+519,cut3_Note}
  }{
    \incgraphics{width=\widthtwoplots}{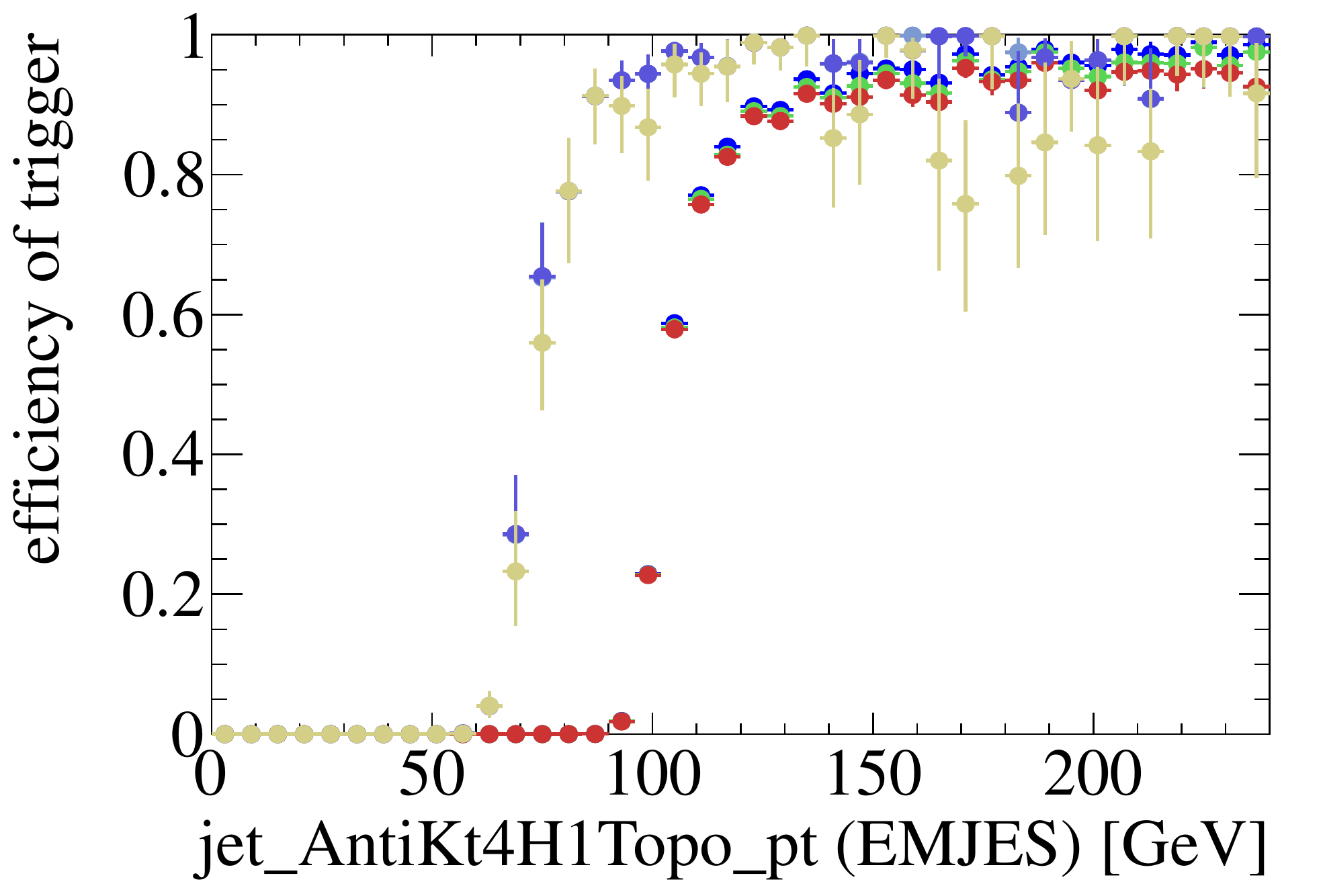}
  }
  \caption[fragile]{
    Left plot: Efficiency of the \met part of the jet-seeded triggers \jetmetj{*},
    after a cut on offline jet \pt at \GeV{140}
    as function of SUSY \met.
    The legend gives the cut on \met at EF.
    This plot includes three of the curves shown in the right plot of \Fig{fig:results_trigger_performance_measurements_turnons_jetmet_projection_MET}.

    Right plot: Efficiency of the jet part of both types of \jetmet triggers,
    after a cut on \tagname{MET\_Topo} at \GeV{70} as function of offline jet \pt.
    The two types of \jetmet triggers have different EF jet thresholds of $50$ and \GeV{75},
    therefore the curves fall into two groups with the same shape.
    The legend for the right plot %
    is the same as in the plots in \Fig{fig:results_trigger_performance_measurements_turnons_jetmet_projection_MET}.
  }
  \label{fig:results_trigger_performance_measurements_turnons_jetmet_projection_jet}
\end{figure}

A slight dependence of the plateau value on the \met threshold value is visible, too,
and becomes even more distinguished in the left plot in \Fig{fig:results_trigger_performance_measurements_turnons_jetmet_projection_jet},
where the turn-on behavior of the \met part of only the jet-seeded triggers is shown as function of SUSY \met.
Between $100$ and \GeV{200}, the plateau behavior of these triggers is not stable,
but all thresholds show a wavy structure as function of SUSY \met. %
Raising the offline cut on jet \pt seems to slightly dampen the variations,
but a lack of statistics impedes a final conclusion here. %
In the right plot of the same figure,
the turn-on behavior of the jet part of the \jetmet triggers is compared,
again showing efficiencies for both groups of \jetmet triggers.
As the two types of \jetmet triggers differ by their EF jet threshold
and the dependence on the \met threshold seems to be low after the offline \met cut on \tagname{MET\_Topo}
which is applied for the projection,
the curves fall into two groups,
each exhibiting approximately the same shape.
It can be seen that for \jetmetf{40} (in yellow),
the \met cut of \GeV{70} is too low
because this trigger has not yet reached its plateau efficiency at that value of \met.
The efficiencies therefore appear to be lower than those of the other two full-chain \jetmet triggers with thresholds at $20$ and \GeV{30},
which fall exactly together.

\subsubsection{Cross-checks Using Muon Triggers as Orthogonal Sample Triggers}
\label{sec:results_trigger_performance_measurements_crosscheckmuons_2010}

\begin{figure}
  \centering
  \incgraphics{width=\widthtwoplots}{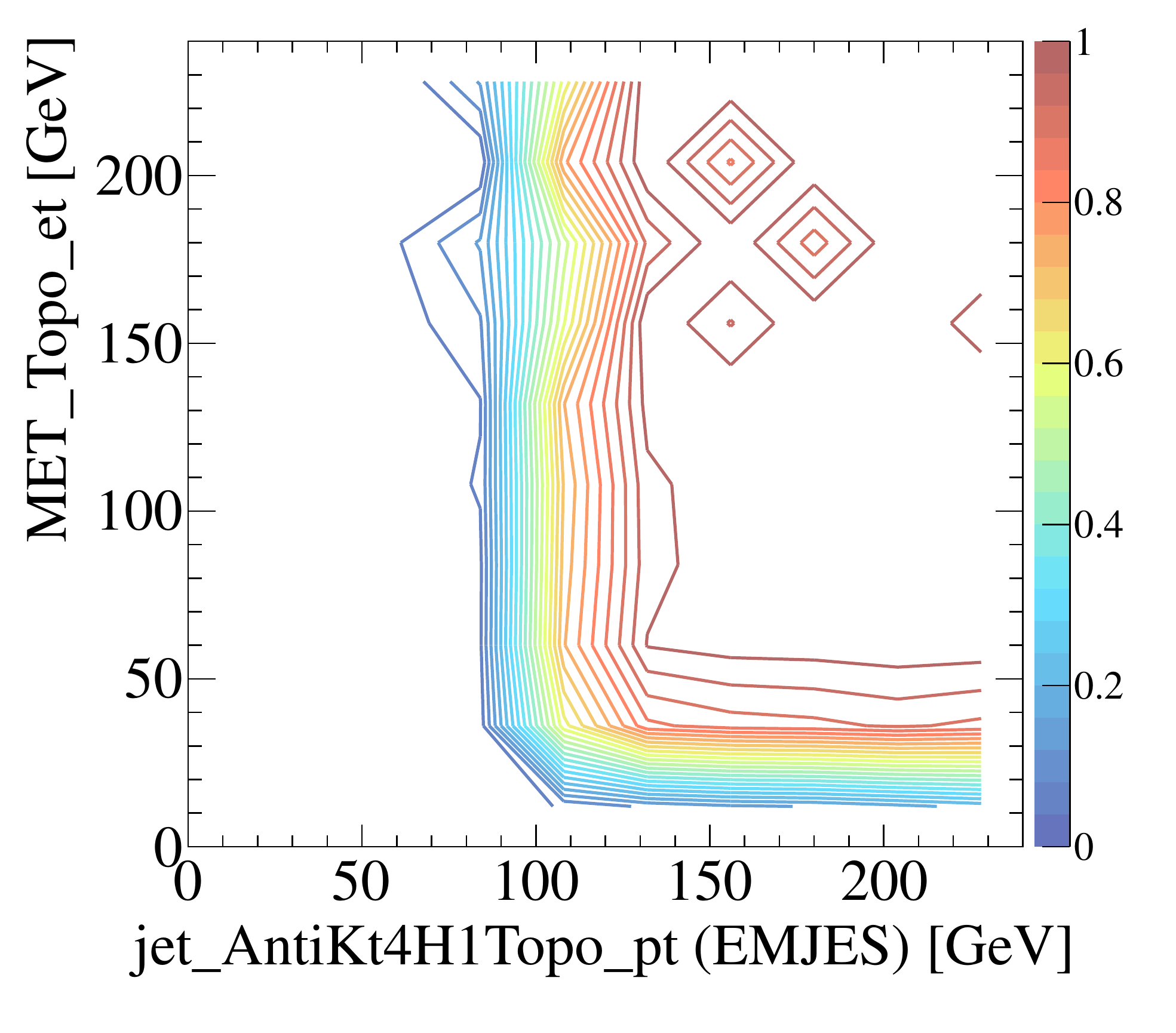} %
    \hfill
  \incgraphics{width=\widthtwoplots}{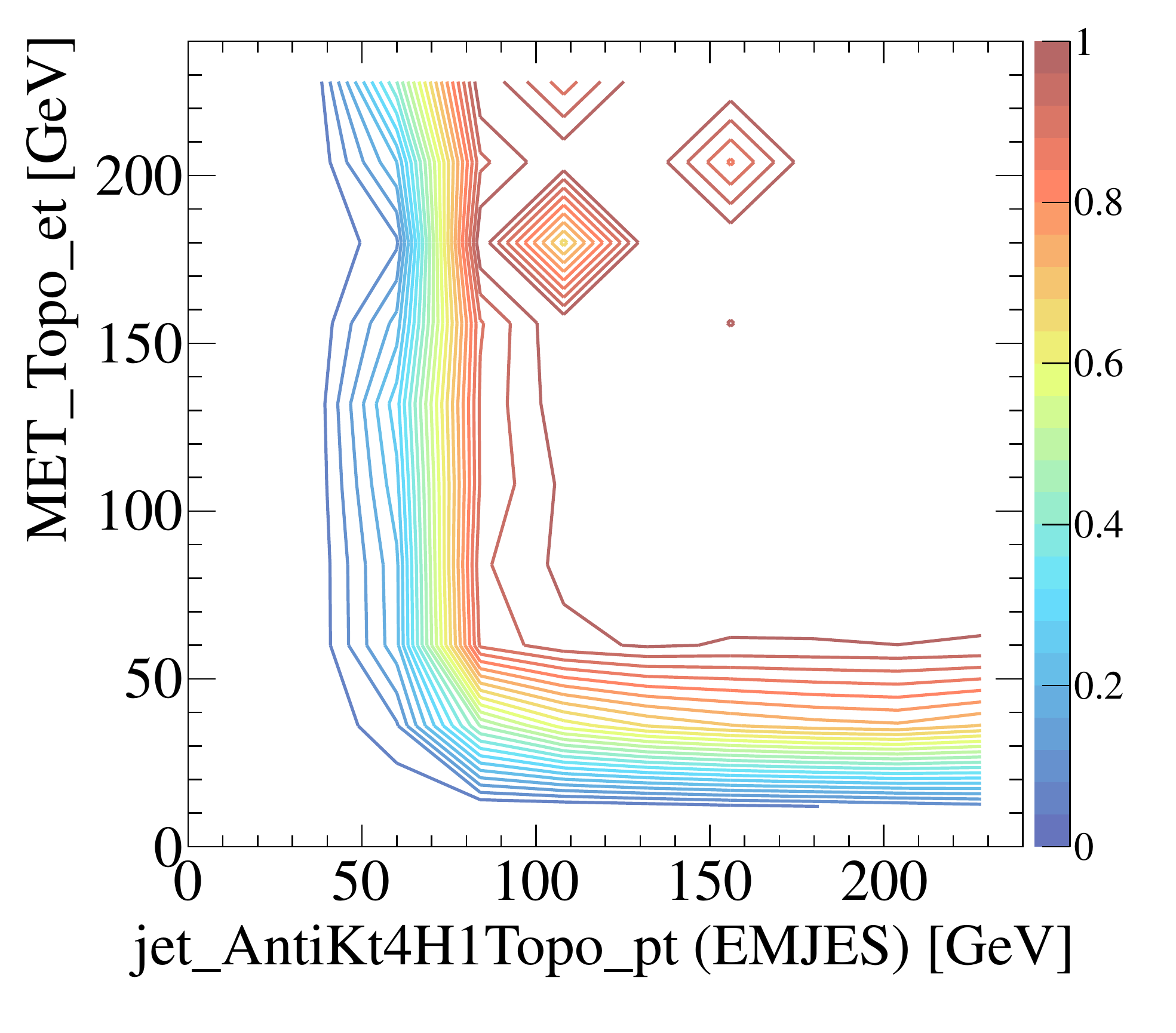}
  \caption{
    Efficiency of \jetmet triggers measured on the \tagname{Muons} stream
    under the assumption of orthogonality of the muon triggers with respect to jets and \met.
    Left: \jetmetj{25}, right: \jetmetf{30}.
    Both plots use \met from the \tagname{MET\_Topo} container as offline reference.
  }
  \label{fig:results_trigger_performance_measurements_turnons_jetmet_muons}
\end{figure}

The \tagname{Muons} stream has been assigned approximately the same bandwidth as the \tagname{Jet\-Tau\-Etmiss} stream %
and therefore offers a promising alternative to cross-check the bootstrap results.
Again, \trigger{EF_mu13} is used to select the event sample,
and \Fig{fig:results_trigger_performance_measurements_turnons_jetmet_muons} shows the turn-on curves of
the jet-seeded trigger \jetmetj{25} and the full-chain trigger \jetmetf{30},
in both plots using \tagname{MET\_Topo} as offline \met variable.
The two plots need to be compared to the upper plots in \Fig{fig:results_trigger_performance_measurements_turnons_jetmetj25}
and \Figs{fig:results_trigger_performance_measurements_turnons_jetmetf30_jetsample}
and \ref{fig:results_trigger_performance_measurements_turnons_jetmetf30_topo}, respectively,
and agree quite well with the results obtained from bootstrapping.
Note however that the turn-on curves of the \jetmet triggers computed from the \tagname{Muons} stream
are reliable only as long as the assumption of orthogonality of the muon triggers holds.
The corresponding plots with the SUSY definition of \met instead of \tagname{MET\_Topo} suffer from the same problem
which was explained for the \met turn-on curves in \Sec{sec:results_trigger_performance_measurements_MET_triggers_2010} %
and are therefore not suitable as cross-check.

\label{sec:results_trigger_performance_measurements_missing_l2_jets}
It must be stressed that in those events for which the feature-extraction algorithms
which reconstruct the \ac{L2} jet triggers objects have not been run,
because no \ac{L1} jet trigger has accepted the event due to prescales and thereby seeded the execution of the feature-extraction algorithms at \ac{L2},
the trigger efficiency cannot be correctly evaluated from the trigger objects as it is done here using the trigger emulation.
The trigger emulation would always reject the event
because it does not find any \ac{L2} jets,
even though the event possibly may have jets with sufficient energy which would have fired the emulated trigger
if the L1 accept signal had not been suppressed by prescales.
For the jet-seeded \jetmet triggers,
this is not a problem
because they require \trigger{L1_J55},
which has been running unprescaled throughout 2010 %
so that in all events where any of the jet-seeded \jetmet triggers could possibly have fired,
at least \trigger{L1_J55} must have fired, too, and seeded the reconstruction of trigger jets at L2.
For the full-chain \jetmet triggers,
the argumentation is not as straightforward,
because they require only \trigger{L1_J30},
which has become heavily prescaled since period 2010~F. %
This means that in events which have L1 jets between $30$ and \GeV{55},
the prescale of \trigger{L1_J30} will inhibit the creation of L2 jets in most events.
Events containing only L1 jets with smaller energy can be ignored
because those events would not fire the full-chain \jetmet triggers anyway.
Events containing L1 jets with larger energy also trigger the unprescaled \trigger{L1_J55} item,
and L2 jets become again available.
In any case,
the agreement of the turn-on curves using the \tagname{Muons} stream or \met sample triggers on the \tagname{JetTauEtmiss} stream,
in which case the L2 jets may also be missing,
and those using jet sample triggers on the \tagname{JetTauEtmiss} stream
(cf. in particular \Figs{fig:results_trigger_performance_measurements_turnons_jetmetf30_jetsample} and \ref{fig:results_trigger_performance_measurements_turnons_jetmetf30_topo}),
where L2 jets are always available due to the sample trigger requirement,
suggests that the impact of the potential lack of jet information at L2 can be safely neglected for 2010.
The same conclusion can be drawn from \Fig{fig:results_trigger_performance_measurements_turnons_jet_triggers} for the jet trigger efficiencies.

\section{Conclusions for 2010}
\label{sec:results_trigger_performance_measurements_conclusions_2010}

In the above, %
the application of the bootstrapping method to compute efficiency estimates of combined \jetmet triggers has been described,
and its application to the two different groups of \jetmet chains
that were defined in the 2010 data-taking menu for proton-proton collisions in the \ATLAS experiment was demonstrated.
It was shown how bootstrapping can be used to obtain unbiased turn-on efficiencies,
and the performance of the two different groups of \jetmet trigger chains was compared.
Making use of different types of sample triggers, based on jets, \met, and muons,
allowed to compare their coverage of the relevant part of phase space
and to do several cross-checks of the results yielding satisfying agreement.

The studies have been carried out using two different offline \met definitions,
\tagname{MET\_Topo}, which in terms of calibration and composition is similar to the computation of \met in the trigger,
and a refined calibration used by the \ATLAS Supersymmetry group for their physics analyses.
It has become clear that care has to be taken not only when choosing the thresholds of the physics triggers,
but also to define sample triggers that can \revised{be} used to collect event samples of sufficient size,
on which the efficiencies and the turn-on behavior of the primary triggers employed in physics analyses can be determined.
It should be stressed that,
while being a good cross-check for the \tagname{MET\_Topo} variant of \met,
the muon triggers cannot be used as sample triggers for turn-ons when using the SUSY \met definition
(or probably any \met which includes muon contributions) as offline reference, as was shown here.
For computing the turn-on curves of \jetmet triggers with respect to the offline SUSY \met definition
\met triggers could therefore not be used as sample trigger,
because no suitable turn-on of these to perform bootstrapping could be obtained.

\section{Measurements of Trigger Efficiencies in Data Taken in 2011}
\label{sec:results_trigger_performance_measurements_data_2011}

Combined \jetmet triggers have remained the primary physics triggers for the zero-lepton studies of the \ATLAS Supersymmetry group in 2011.
This section documents the continuation of the studies which were begun in 2010 and presented above.
The efficiency studies in 2011 profit a lot from the experience gained in 2010
in terms of which methods and sample triggers are suitable to measure the efficiencies of the combined \jetmet triggers.
Although the general type of the primary triggers has remained the same,
there were a number of changes in the trigger menu and in the trigger system itself,
which are explained in detail in the next section.

\subsection{\texorpdfstring{Overview of Changes and New \Jetmet Triggers} {Overview of Changes and New JetMET Triggers}}

\begin{table}
  \centering
  \begin{threeparttable}
  \begin{tabular}{lll}
    \toprule
    \centering Name of chain & Cut on jet $E_T$ [GeV] & Cut on \met [GeV]\\
    \midrule
    \trigger{EF_j75_a4tc_EFFS_xe45,55_loose_noMu}\tnote{1}
    & L1: 50                    & L1: 20, 25/30\tnote{3} \\
    & L2: 70                    & L2: 20, 25 \\
    & EF: 75                    & EF: 45, 55 \\
    \trigger{EF_j75_a4tc_EFFS_xe55_noMu}\tnote{2}
    & L1: 50                    & L1: 35 \\
    & L2: 70                    & L2: 35 \\
    & EF: 75                    & EF: 55 \\
    \trigger{EF_j80_a4tc_EFFS_xe60_noMu}\tnote{2}
    & L1: 50                    & L1: 35 \\
    & L2: 75                    & L2: 40 \\
    & EF: 80                    & EF: 60 \\
    \bottomrule
  \end{tabular}
  \begin{tablenotes}\footnotesize
    \item[1] There is also the chain \trigger{EF_j75_a4tc_EFFS_xe40_loose_noMu}, but it has never been running.
    \item[2] This chain has been introduced in 2011~I.
    \item[3] The definition of \trigger{EF_j75_a4tc_EFFS_xe55_loose_noMu} has changed starting with period 2011~I in that the L1 threshold on \met was raised to \GeV{30}.
  \end{tablenotes}
  \caption{
    Overview of the relevant %
    set of combined \jetmet trigger chains
    which were defined in the trigger menu \tagname{Physics\_pp\_v2} for data taking in 2011.
    The table shows the cuts which are applied to the values of \met and jet $E_T$
    measured online at Level~1, Level~2 and Event Filter.
  }
  \end{threeparttable}
  \label{tab:results_trigger_performance_jetmet_chains_2011}
\end{table}

The most important change between 2010 and 2011 with respect to the combined \jetmet triggers
concerns the jet trigger algorithm at Event Filter level,
which has been replaced by a full-scan algorithm (indicated by ``EFFS'' in the trigger name) %
and has been used for rejecting events for the first time in 2011 (cf. \Sec{sec:tdaq_jet}).
The efficiencies of triggers involving jets are therefore expected to potentially exhibit fundamental changes.

\Tab{tab:results_trigger_performance_jetmet_chains_2011} shows the relevant combined \jetmet triggers
which are defined in the trigger menu in 2011.
The thresholds lie in the same order of magnitude as the highest thresholds which were defined in 2010.
Jet-seeded chains are no longer included in the menu,
since the increased instantaneous luminosity prohibits keeping low threshold jet items at Level~1 unprescaled
because the input rate to Level~2 would be too high.
As a too-close spacing of the cuts at Level~1 and Event Filter on \met deteriorates the \met trigger performance
due to the very broad turn-on of the \met at Level~1,
the spacing of the thresholds at Level~1 and Level~2 and Event Filter is kept relatively large,
as can be seen from the table.
The combined trigger \trigger{EF_j75_a4tc_EFFS_xe45_loose_noMu} has been the primary trigger
for the first analyses of 2011 data by the \ATLAS Supersymmetry group \cite{ATL-PHYS-INT-2011-055,ATL-PHYS-INT-2011-085}.

Due to the difficulties with the measurements of the \met turn-on curves,
in particular when using \met with muon corrections as offline reference,
which were discussed above in the context of the trigger efficiencies in 2010,
and because of the larger statistics coming through jet triggers,
whereas the low threshold \met triggers had to be deactivated or heavily prescaled,
in 2011, \met triggers have not been employed for bootstrapping.
At the end of this section, a brief cross-check is given though.

The event selection is done in the same way as in 2010,
applying the SUSY \ac{GRL} and the jet cleaning in its updated version for \Athena release 16 \cite{URL_JetCleaning}. %
Which periods were used to generate the turn-on curves \revised{is} specified below for each plot \revised{individually}.
If not stated otherwise,
data from all available periods in 2011 is used, \ie up to period 2011~K,
except for the first three periods~A, B and~C.
Periods~A and B have been mostly for testing after the winter shutdown of the \ac{LHC} and do not contain much usable data,
and 2011~C has a reduced center-of-mass energy of \TeV{2.76}. %
The object definitions have been updated following the recommendations of the performance groups
and the changes in the zero-lepton analysis of the \ATLAS SUSY group.
Instead of \tagname{MET\_Topo},
\tagname{MET\_LocHadTopo} is used,
as recommended by the \ATLAS Jet/Etmiss combined performance group for 2011 \cite{URL_JetMETRecommendation2010}. %
The SUSY \met definition in 2011 is called \tagname{MET\_Simplified20\_RefFinal}\footnote{
  Note that the term ``SUSY \met'' in the following therefore denotes a different version of offline \met than in the plots for 2010,
  which in most plots is obvious from the axis title.
},
and the new jet collection is called \tagname{AntiKt4TopoNewEM},
but still contains jets calibrated at \EMJES scale (cf. \Secs{sec:software_reconstruction_met} and \ref{sec:software_jet_reconstruction}).

\subsection{Jet Triggers}
\label{sec:results_trigger_performance_measurements_jet_triggers_2011}

\begin{table}
  \centering
  \begin{tabular}{lll}
    \toprule
    EF chain name & L2 chain name & L1 item name \\
    \midrule
    \trigger{EF_j10_a4tc_EFFS}  & \trigger{L2_rd0_filled_NoAlg}  & \trigger{L1_RD0_FILLED} \\
    \trigger{EF_j15_a4tc_EFFS}  & \trigger{L2_rd0_filled_NoAlg}  & \trigger{L1_RD0_FILLED} \\
    \trigger{EF_j20_a4tc_EFFS}  & \trigger{L2_rd0_filled_NoAlg}  & \trigger{L1_RD0_FILLED} \\
    \trigger{EF_j30_a4tc_EFFS}  & \trigger{L2_j25}  & \trigger{L1_J10} \\
    \trigger{EF_j40_a4tc_EFFS}  & \trigger{L2_j35}  & \trigger{L1_J15} \\
    \trigger{EF_j55_a4tc_EFFS}  & \trigger{L2_j50}    & \trigger{L1_J30} \\
    \trigger{EF_j75_a4tc_EFFS}  & \trigger{L2_j70}    & \trigger{L1_J50} \\
    \bottomrule
  \end{tabular}
  \caption{
    Overview of the single jet trigger chains
    which were defined in the trigger menu \tagname{Physics\_pp\_v2} for data taking in 2011.
    The table shows the chain or item names at Level~1, Level~2 and Event Filter,
    the two-digit numbers specify the thresholds on trigger jets in GeV (cf. \Sec{sec:tdaq_trigger_nomenclature}). %
  }
  \label{tab:results_trigger_performance_jet_chains_2011}
\end{table}

The jet triggers will play an even more important role as sample triggers
for the measurement of the combined \jetmet trigger efficiencies in 2011 than they did in 2010.
In particular, the efficiencies of the jet trigger \trigger{EF_j75_a4tc_EFFS} will need to be studied
because this trigger will be the sample trigger for all efficiencies of \jetmet triggers
presented in \Sec{sec:results_trigger_performance_measurements_jetmet_triggers_2011}. %
\Tab{tab:results_trigger_performance_jet_chains_2011} gives an overview of the jet trigger chains
which were defined in the trigger menu \tagname{Physics\_pp\_v2} for data taking in 2011.
They fall into two groups:
The lowest chains with thresholds of 10, 15 and \GeV{20} at Event Filter level
are seeded by random triggers at Level~1 and Level~2 (\trigger{L1_RD0_FILLED} and \trigger{L2_rd0_filled_NoAlg}),
and only work with trigger jet objects at Event Filter level.
The chains with higher thresholds at Event Filter are full chains again
in the sense that they require trigger jets to have been reconstructed at Level~1 and Level~2, too.

The efficiency of \trigger{EF_j75_a4tc_EFFS} needs to be determined using a lower jet trigger as sample trigger,
in the same way as was done in 2010 for the jet triggers.
The lowest jet trigger fulfilling the bootstrapping condition \eqref{eq:bootstrap_requirement} is \trigger{EF_j30_a4tc_EFFS},
which is the lowest jet trigger chain not seeded by random triggers (cf. \Tab{tab:results_trigger_performance_jet_chains_2011}).
The first step, therefore, is to obtain the efficiency of \trigger{EF_j30_a4tc_EFFS}.
A direct measurement of the efficiency of \trigger{EF_j30_a4tc_EFFS} in terms of the full chain of Level~1, Level~2 and Event Filter,
\ie the product of the efficiencies at all three trigger levels in \ATLAS,
cannot be done on an event sample selected with the random trigger \trigger{EF_rd0_filled_NoAlg}, %
because of the high prescales of the respective jet triggers at Level~1 and Level~2,
leading to the already explained problem that trigger objects are missing.
This means that the trigger efficiencies would be strongly underestimated and biased by events which are actually only recorded,
because they have also fired a jet trigger with a higher threshold,
which led to the execution of the feature-extraction algorithms at the higher trigger levels.
However, it is possible to compute the efficiencies for the three random-seeded jet triggers \trigger{EF_j10,15,20_a4tc_EFFS} using such a sample,
because if the random trigger chain (\trigger{EF_rd0_filled_NoAlg} seeded by \trigger{L2_rd0_filled_NoAlg}) fires at \ac{L2},
the feature-extraction algorithms for jets at Event Filter level are executed according to \Tab{tab:results_trigger_performance_jet_chains_2011}.
The efficiency of \trigger{EF_j20_a4tc_EFFS} will therefore be measured like this and employed below.

\begin{figure}
  \centering
  \incgraphics{width=\widthtwoplots}{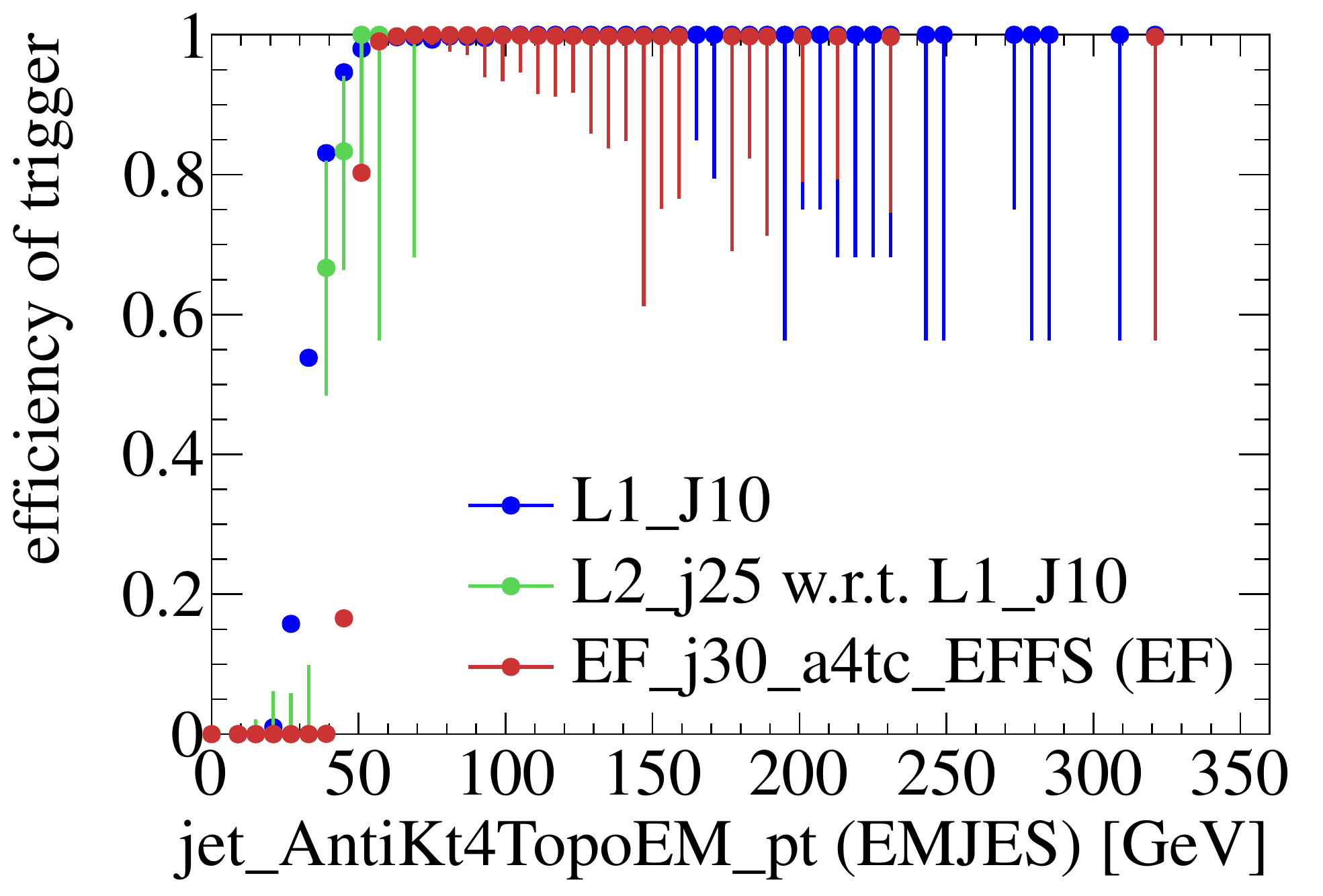}
  \hfill
  \incgraphics{width=\widthtwoplots}{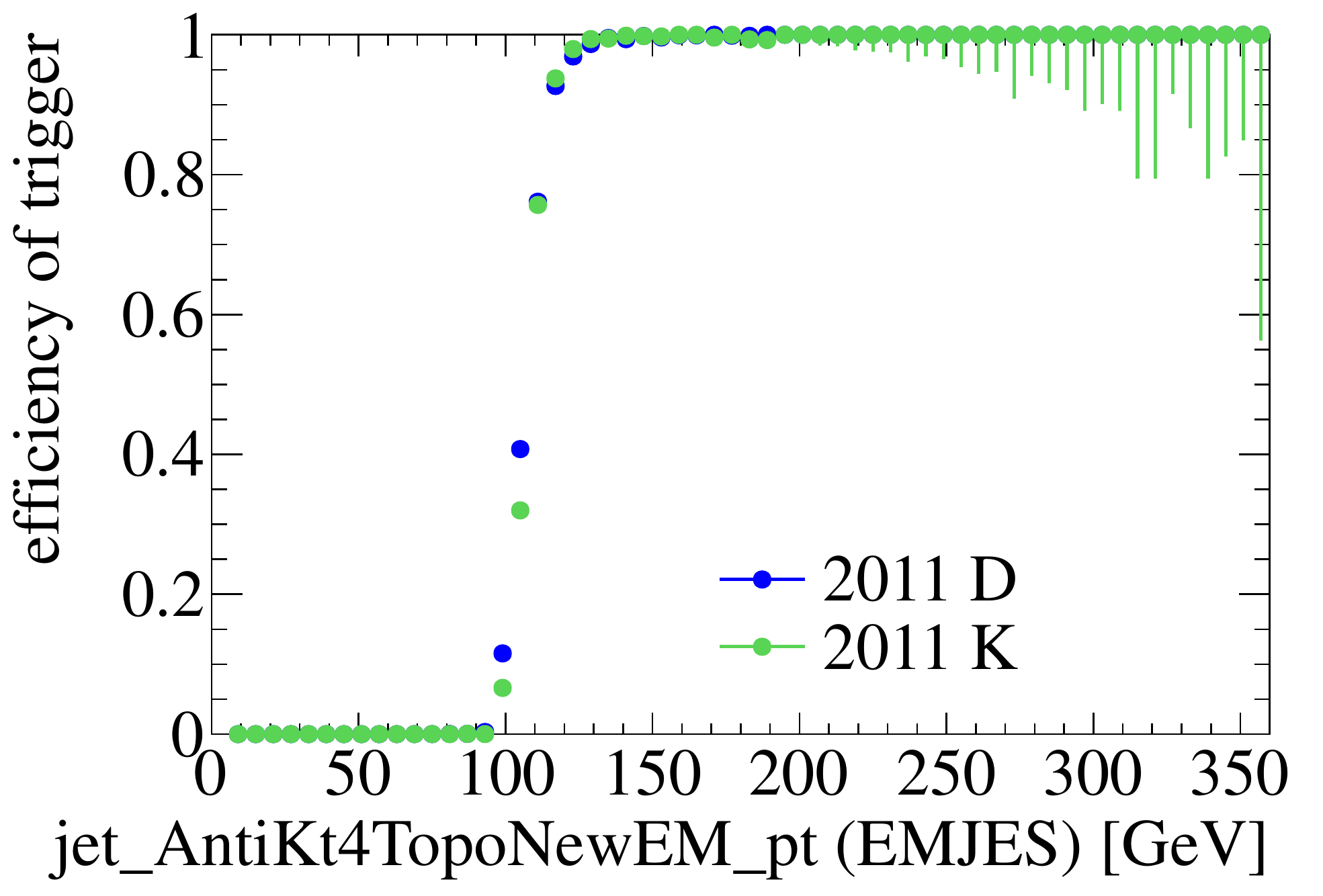}
  \caption{
    Left: Efficiencies of the three levels of the trigger chain \trigger{EF_j30_a4tc_EFFS},
    measuring L1 and L2 on a sample of events taken with the random trigger \trigger{EF_rd0_filled_NoAlg} from the \tagname{MinBias} stream,
    and EF with bootstrapping on \trigger{EF_j20_a4tc_EFFS} on the \tagname{JetTauEtmiss} stream.
    Note that for L2, this is the relative efficiency with respect to L1,
    and that the curve for \trigger{EF_j30_a4tc_EFFS} shows the efficiency of the EF level only
    and not the efficiency of the full chain.
    \newline
    Right: Comparison of the efficiency of the jet trigger \trigger{EF_j75_a4tc_EFFS} in periods 2011~D and~K,
    on a sample of events taken with \trigger{EF_j30_a4tc_EFFS}.
    No bootstrapping is applied.
  }
  \label{fig:results_trigger_performance_measurements_turnons_jet_crosscheck_K}
\end{figure}

As a consequence, in order to obtain the efficiency of \trigger{EF_j30_a4tc_EFFS},
the efficiencies of all three levels need to be measured separately:
\begin{itemize}
  \item At Level~1, the efficiency for \trigger{L1_J10} can be measured on an event sample taken with a random trigger.
  \item At Level~2, the efficiency of \trigger{L2_j25} relative to \trigger{L1_J10} can be measured
    by selecting a sample of events in which both \trigger{EF_rd0_filled_NoAlg} and \trigger{L1_J10} fire. %
    This will, of course, give very low statistics.
    Note that it is not enough to require \trigger{L1_J10} only
    because not all events passing \trigger{L1_J10} will be recorded due to prescales,
    and the sample obtained by requiring \trigger{L1_J10} is thus not an unbiased sample.
  \item The efficiency of \trigger{EF_j30_a4tc_EFFS} with respect to Level~2 can be measured either
    by again requiring both \trigger{EF_rd0_filled_NoAlg} and \trigger{L2_j25} to fire,
    or on a sample of events triggered by \trigger{L2_j25} and recorded due to the active pass-through factor at EF level for this chain, %
    or~--- as is done here~--- using bootstrapping on events taken with \trigger{EF_j20_a4tc_EFFS},
    the efficiency of which can be measured on random-triggered events.
\end{itemize}
This is obviously more complicated than a direct measurement of the turn-on curve for \trigger{EF_j30_a4tc_EFFS},
but for the following it will be sufficient to convince oneself
that this trigger reaches its plateau before the onset of the turn-on region of \trigger{EF_j75_a4tc_EFFS},
which is around \GeV{100}.
The correction of the sample trigger bias in the measurement of the efficiency of \trigger{EF_j75_a4tc_EFFS}
in a sample taken with \trigger{EF_j30_a4tc_EFFS} can then be relinquished.
The left plot in \Fig{fig:results_trigger_performance_measurements_turnons_jet_crosscheck_K}
shows the efficiencies of \trigger{L1_J10}, \trigger{L2_j25} and \trigger{EF_j30_{\allowbreak}a4tc_EFFS},
which comprise the full chain \trigger{EF_j30_a4tc_EFFS}\footnote{
  The convention by which the full chain bears the same name as the Event Filter part of the chain
  is a relic from the early commissioning of the trigger
  when event rejection was not activated for all trigger levels
  and the prefix was needed to make clear up to which level the performance of the chain is studied.
  Recently, it has become common to refer to the full chain by leaving out the prefix, \eg \trigger{j30_a4tc_EFFS}. %
}.
The efficiency for Level~2 is measured relative to its lower chain \trigger{L1_J10},
\ie the denominator of the efficiency of \trigger{L2_j25} is filled with the \pt of jets
from events which fire both \trigger{EF_rd0_filled_NoAlg} and \trigger{L1_J10}.
The efficiency for \trigger{EF_j30_a4tc_EFFS} is measured on events taken with \trigger{EF_j20_a4tc_EFFS},
by applying bootstrapping.
Note again that the efficiency here refers only to the Event Filter level of the trigger chain,
not to the combined efficiency of all trigger levels.
As expected, statistics is very low for \ac{L2} due to the conjunction of two trigger requirements.
Still, the plot allows to estimate the position of the turn-on
and the beginning of the plateau region
of the full chain \trigger{EF_j30_a4tc_EFFS} quite accurately.
It shows that Level~1 and Level~2 should not effect a large shift of the efficiency of the full chain to the right,
and it can be seen that the full turn-on,
\ie the product of the efficiencies of all three trigger levels,
should indeed reach \percent{100} safely below \GeV{100},
and even below \GeV{70},
which would be the condition for measuring also \trigger{EF_j55_a4tc_EFFS} on a sample taken by \trigger{EF_j30_a4tc_EFFS} directly.
\begin{figure}
  \centering
  \incgraphics{width=\widthsingleplot}{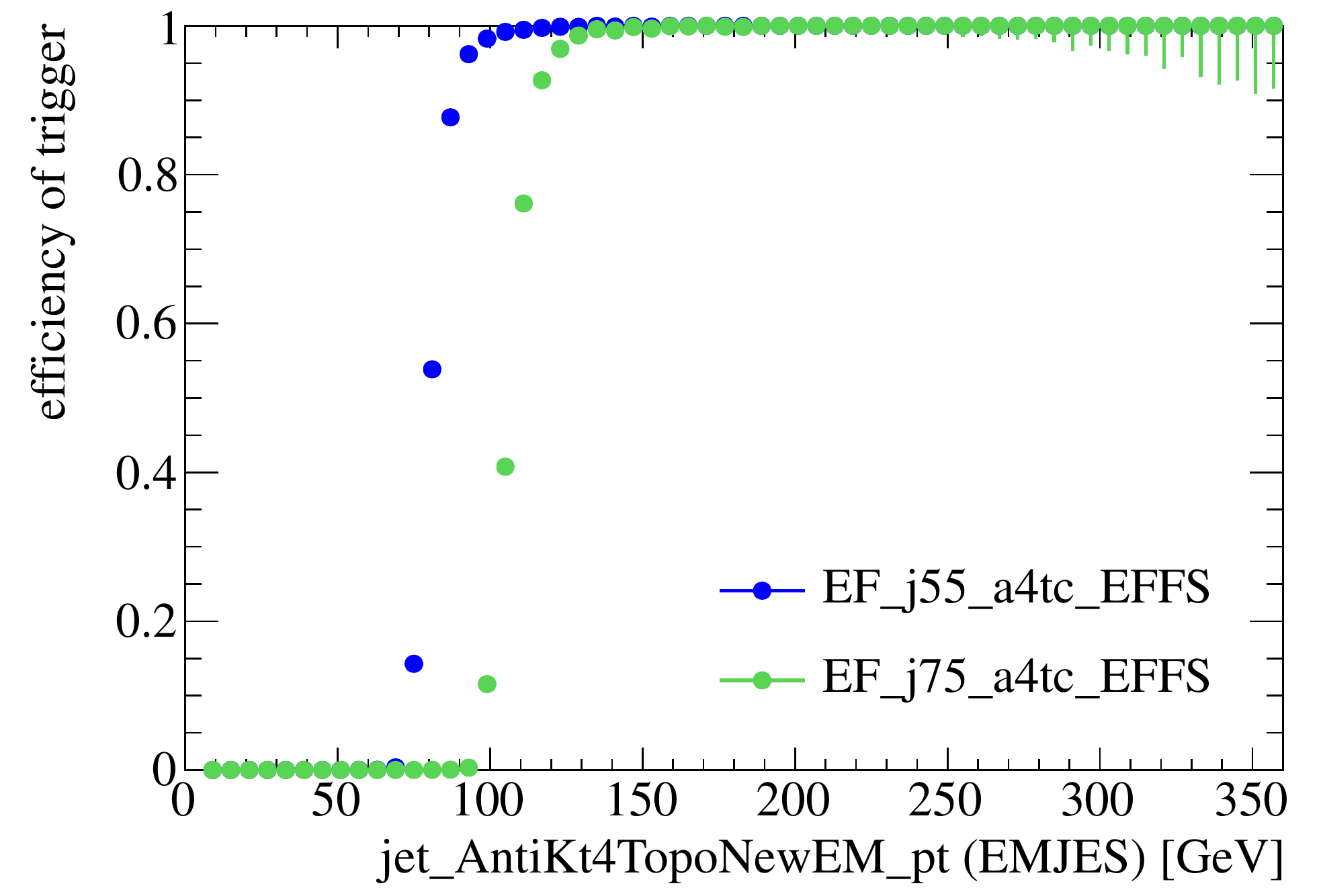}
  \caption{
    Efficiency of jet %
    triggers with EF thresholds of $55$ and \GeV{75}
    measured on the \tagname{JetTauEtmiss} stream on an event sample taken with \trigger{EF_j30_a4tc_EFFS}.
    No bootstrapping is applied.
  }
  \label{fig:results_trigger_performance_measurements_turnons_jet_55_75}
\end{figure}

The efficiency of \trigger{EF_j75_a4tc_EFFS} is plotted in \Fig{fig:results_trigger_performance_measurements_turnons_jet_55_75},
together with the efficiency of \trigger{EF_j55_a4tc_EFFS}.
No bootstrapping has been done to correct for a bias of \trigger{EF_j30_a4tc_EFFS} in the event sample.
Instead, the plot relies on the assumption that \trigger{EF_j30_a4tc_EFFS} reaches its efficiency plateau
well below the onset of the turn-on region of \trigger{EF_j75_a4tc_EFFS},
which has been demonstrated above.
Only data from period 2011~D has been used to produce this plot,
but this is sufficient to cover the full range up to \GeV{360} in offline jet \pt,
which will be the upper limit chosen for the measurement of the \jetmet trigger efficiencies in the following.
The increase in statistics which could be achieved by summing over periods E~--~G in addition to period~D is very small. %
To make sure that the efficiency of this sample trigger is stable and valid also for data taken in later periods,
a cross-check has been done by measuring the efficiency of \trigger{EF_j55_a4tc_EFFS} again in period 2011~K.
The result is shown in the right plot in \Fig{fig:results_trigger_performance_measurements_turnons_jet_crosscheck_K},
where no changes with respect to the horizontal position of the turn-on region can be observed.
The slight decrease of the efficiency in the turn-on region
may be attributed to the pile-up noise suppression which was introduced between periods~D and K. %

Recent trigger performance plots published by \ATLAS for 2011 data \cite{ATL-COM-DAQ-2011-063}
confirm that the efficiency of the jet trigger \trigger{EF_j40_a4tc_EFFS} with Event Filter threshold of \GeV{40}
reaches a plateau efficiency near \percent{100} at a calibrated offline jet $E_T$ of around $70$ to \GeV{75}. %
Using the jet trigger \trigger{EF_j30_a4tc_EFFS} with a lower Event Filter threshold of \GeV{30} as sample trigger
to determine the efficiency of the jet trigger \trigger{EF_j75_a4tc_EFFS} with an Event Filter threshold of \GeV{75}
therefore will give correct results even without the need to apply the bootstrap method.

\subsection{\texorpdfstring{Combined \Jetmet Triggers} {Combined JetMET Triggers}}
\label{sec:results_trigger_performance_measurements_jetmet_triggers_2011}

\begin{figure}
  \centering
  \incgraphics{width=\widthtwoplots}{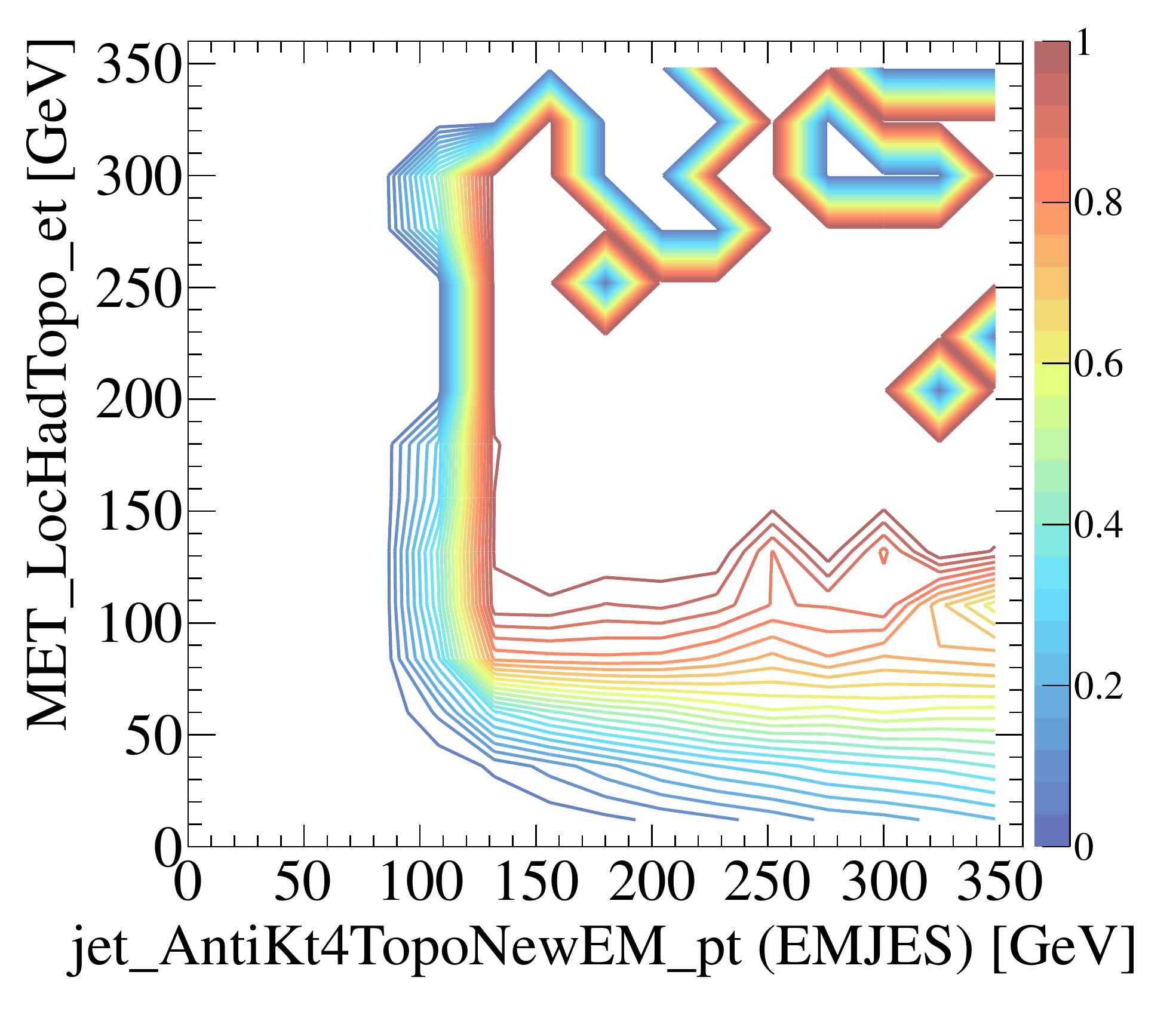}
  \hfill
  \incgraphics{width=\widthtwoplots}{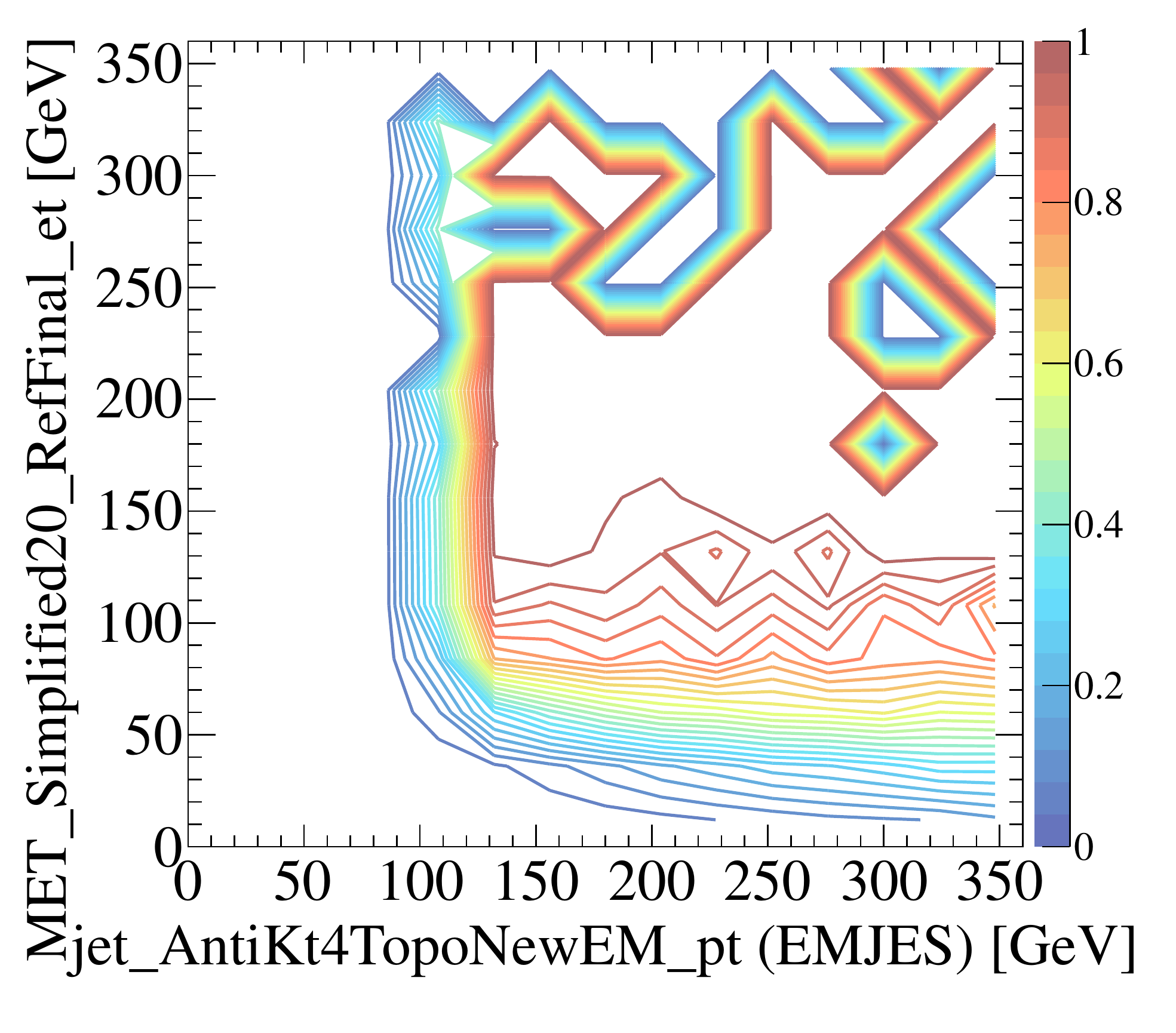}
  \\
  \incgraphics{width=\widthtwoplots}{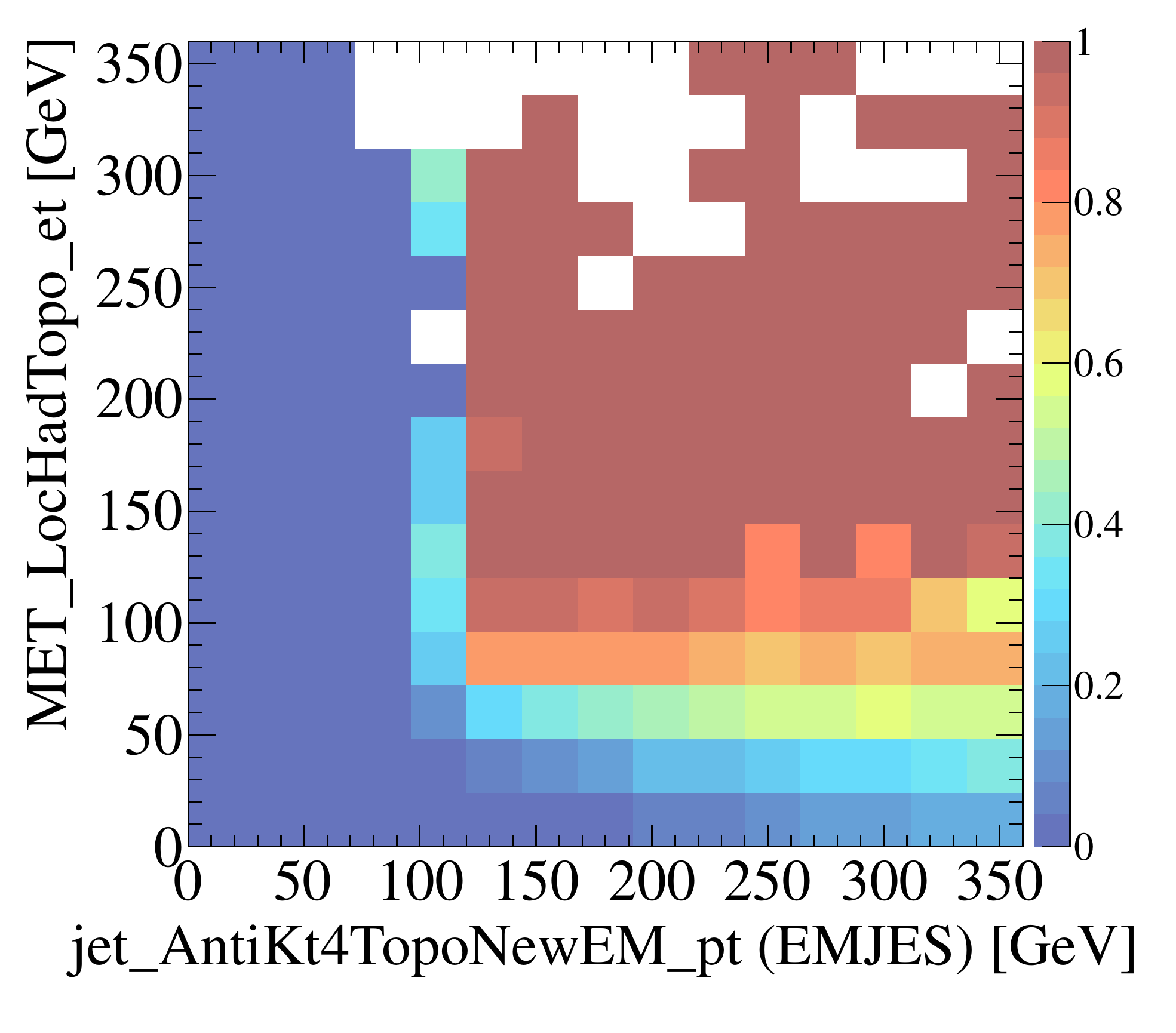}
  \hfill
  \incgraphics{width=\widthtwoplots}{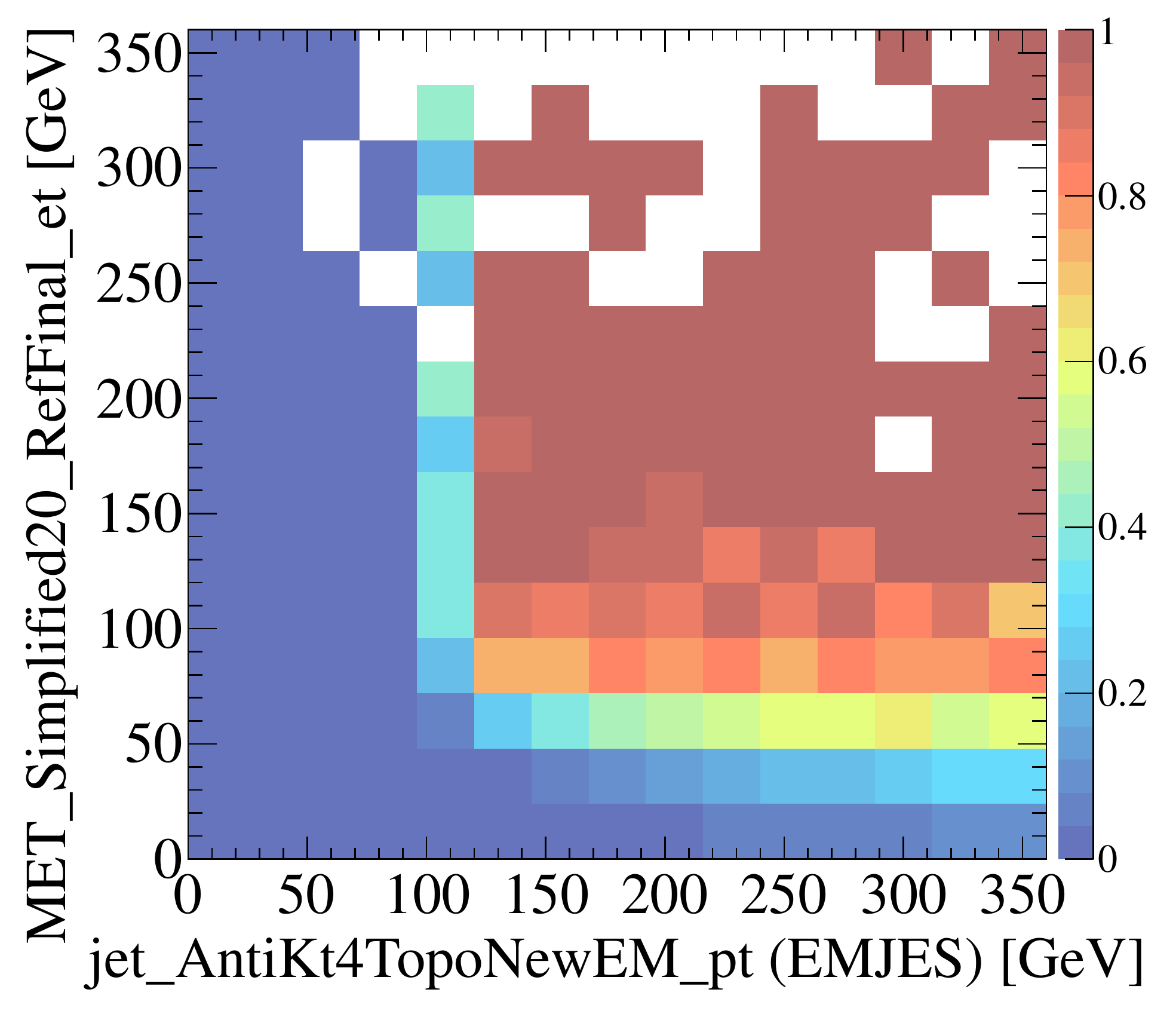}
  \caption{
    Efficiency of the \jetmet trigger \trigger{EF_j75_a4tc_EFFS_xe45_loose_noMu} from bootstrapping,
    using the jet trigger \trigger{EF_j75_a4tc_EFFS} as sample trigger and data from periods 2011 D~--~K.
    The efficiencies are shown as contour and as color-coded plots,
    using \tagname{MET\_LocHadTopo} as offline reference (left column)
    and using the SUSY \met definition (right column).
    White spots arise from empty bins in the denominator histograms due to insufficient statistics.
  }
  \label{fig:results_trigger_performance_measurements_turnons_jetmet_45_contour}
\end{figure}

\begin{figure}
  \centering
  \incgraphics{width=\widthtwoplots}{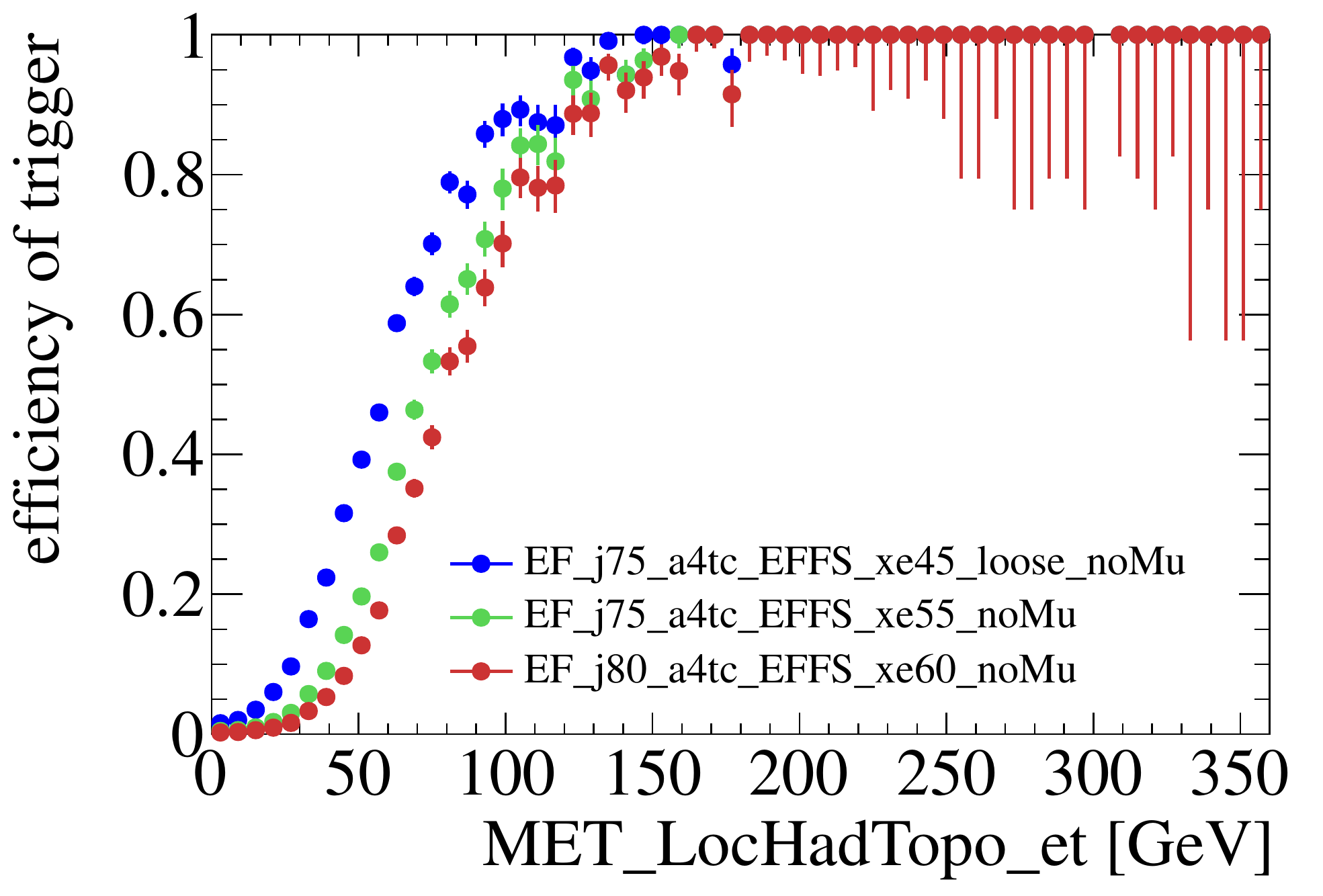}
  \hfill
  \incgraphics{width=\widthtwoplots}{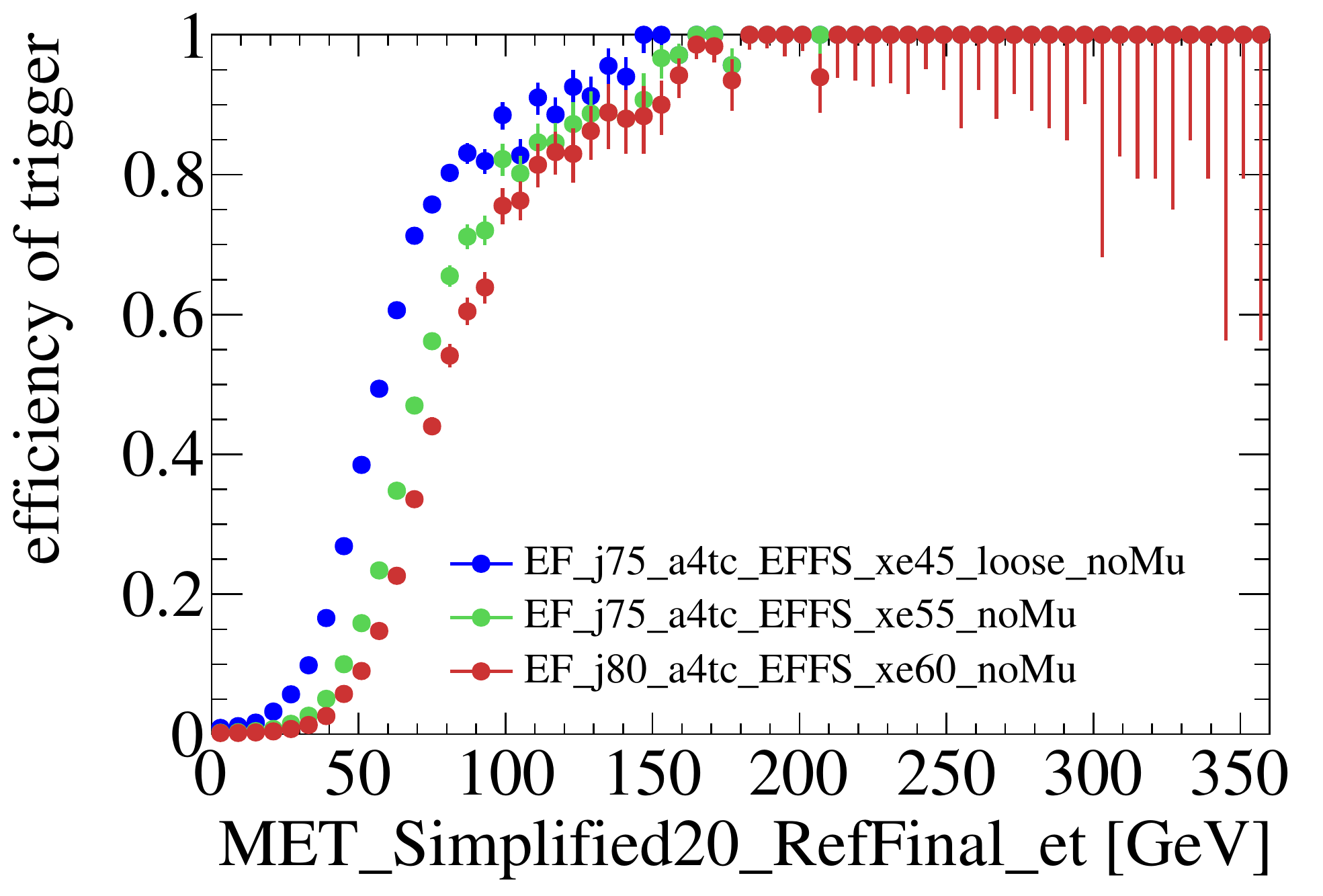}
  \\
  \incgraphics{width=\widthtwoplots}{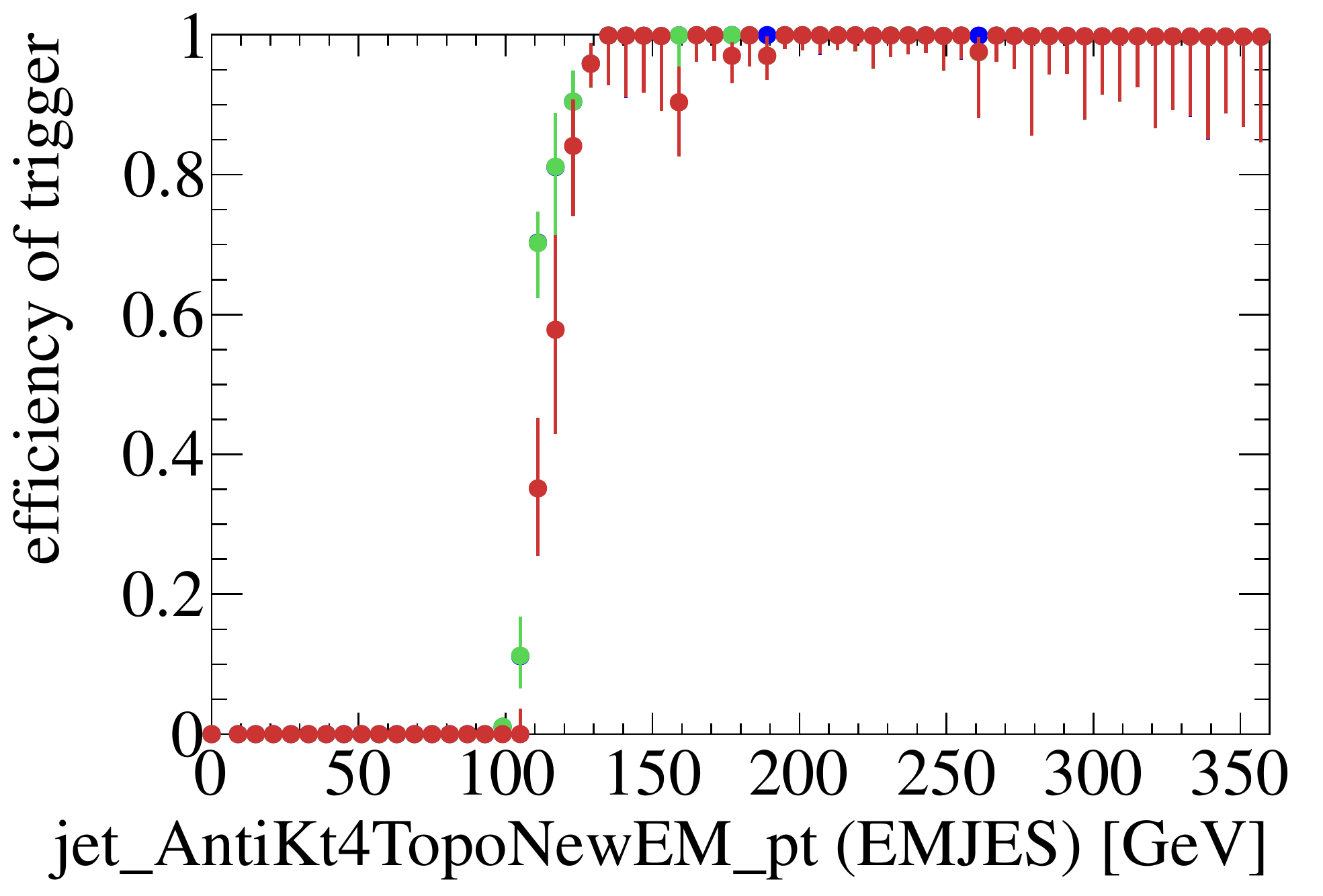}
  \hfill
  \incgraphics{width=\widthtwoplots}{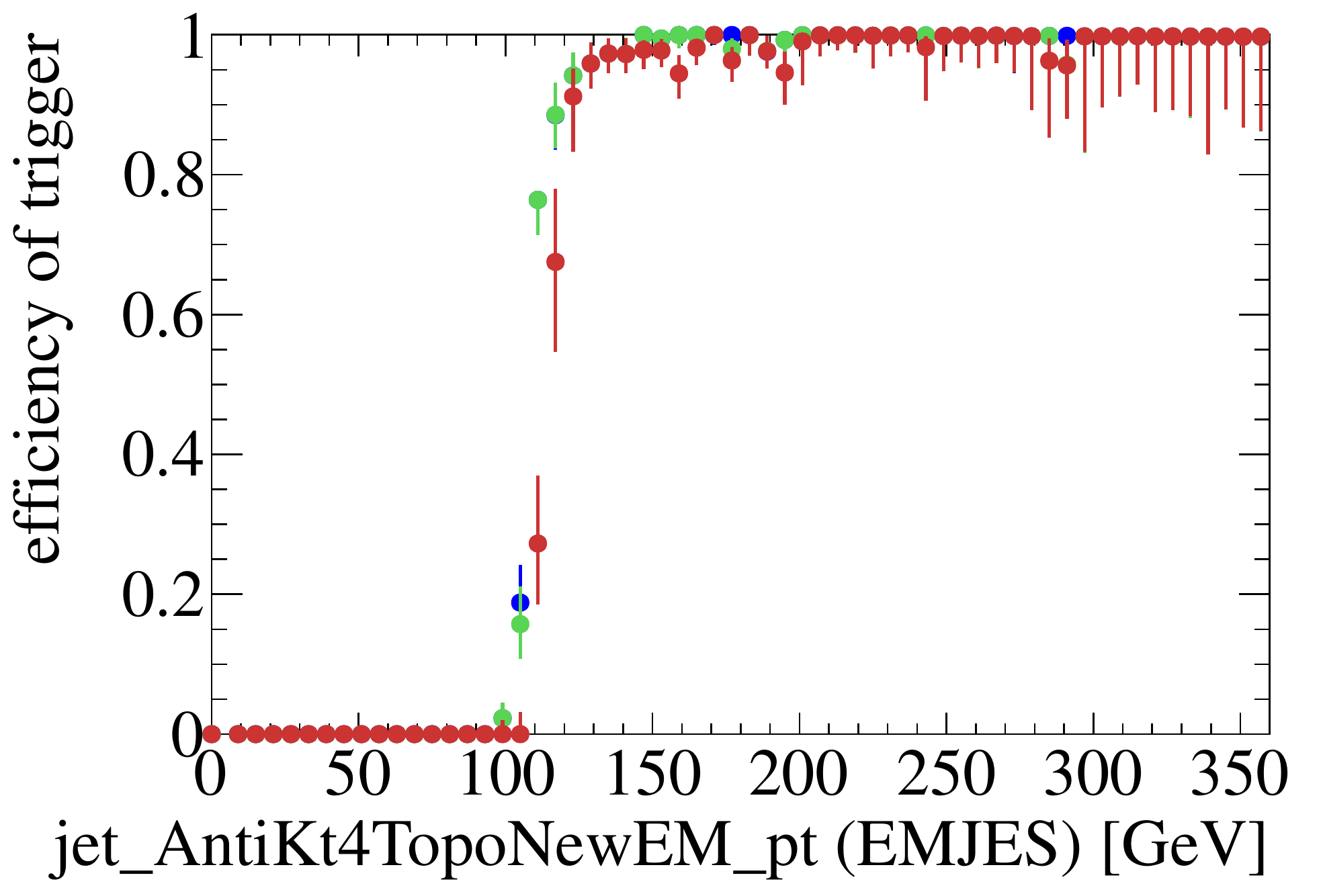}
  \caption{
    Efficiencies of three \jetmet triggers from bootstrapping,
    using the jet trigger \trigger{EF_j75_a4tc_EFFS} as sample trigger and data from periods 2011 D~--~K.
    Projections onto the \met~axis (top) and jet \pt axis (bottom) are shown,
    after cuts on the respective orthogonal variable at \GeV{150}.
    In the left column, \tagname{MET\_LocHadTopo} is the offline reference,
    in the right column the SUSY \met definition.
    (The legend in the upper plots also applies to the lower plots.)
  }
  \label{fig:results_trigger_performance_measurements_turnons_jetmet_projections}
\end{figure}

The efficiency of the combined \jetmet trigger \trigger{EF_j75_a4tc_EFFS_xe45_loose_noMu}
has been studied right from the beginning of data-taking in 2011,
updating the plots with new data whenever it became available.
Here, only the most recent plots including all so-far available data are shown,
which includes periods D~--~K.
The four plots in \Fig{fig:results_trigger_performance_measurements_turnons_jetmet_45_contour} show the efficiency
of this trigger with two different offline \met references,
\tagname{MET\_LocHadTopo} and the SUSY \met definition,
and two different plot styles, as contour plots and as color-coded plots.

\Fig{fig:results_trigger_performance_measurements_turnons_jetmet_projections} shows projections
of the efficiencies onto the \met and jet \pt axes,
after cuts on the respective orthogonal variable at \GeV{150}.
In these plots, it is possible to compare several triggers.
In addition to \trigger{EF_j75_a4tc_EFFS_xe45_loose_noMu}
the primary physics trigger at the beginning of 2011,
two other triggers are included which were introduced in period 2011~I,
when the former primary physics trigger with an \met threshold of \GeV{45} had to be prescaled
because its rate had become too high due to the increased instantaneous luminosity.

These three \jetmet triggers all differ in the thresholds on \met ($45$, $55$ and \GeV{60} at \ac{EF}),
but only one of them has a higher threshold also on jet \pt ($80$ instead of \GeV{75} at \ac{EF}).
Correspondingly, in the projection onto jet \pt
two of the three curves \revised{coincide},
and even the turn-on of the trigger with the higher jet \pt threshold is very close,
as the increase in the threshold is only \GeV{5}.
The spacing of the triggers with respect to their \met turn-ons also agrees with the expectations,
their thresholds at Event Filter being different by \GeV{10} and \GeV{5}, respectively.
In the upper right plot in \Fig{fig:results_trigger_performance_measurements_turnons_jetmet_projections},
it can be seen that the offline cut of \GeV{150} on SUSY \met is too low to bring the \jetmet trigger with the highest threshold of \GeV{60} on \met in its plateau,
with the consequence that in the lower right plot,
the efficiency of this trigger seems to be a little lower than \percent{100}.
For the projection onto \met, in particular the upper right plot, which uses the SUSY \met definition,
seems to hint at a slight decrease in efficiency in the region between roughly $100$ and \GeV{150} of offline \met.
It is much less pronounced for \tagname{MET\_LocHadTopo} (if present at all),
and may be connected to details of the definition of SUSY \met.
Note that although the new combined \jetmet triggers were only defined in periods 2011~I and later,
their performance can still be studied in earlier periods
thanks to the emulation of the trigger decision for periods where these \revised{triggers} were not yet defined
(cf. \Sec{sec:tdaq_explain_trigger_emulation}).
This gives a significant gain in statistics.

\subsubsection{Consistency of Periods}
\label{sec:results_trigger_performance_2011_consistency}
\begin{figure}
  \centering
  \incgraphics{width=\widthtwoplots}{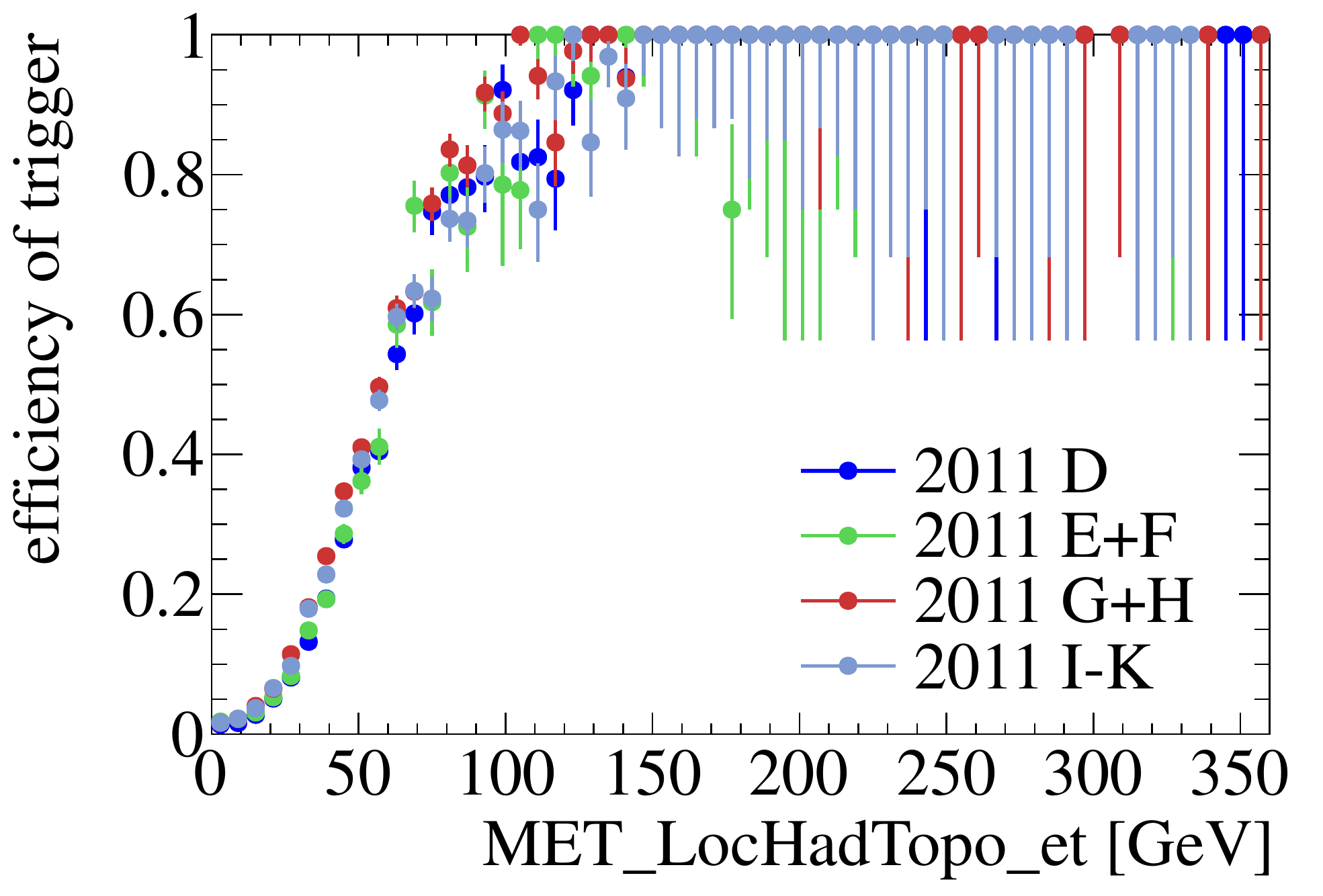}
  \hfill
  \incgraphics{width=\widthtwoplots}{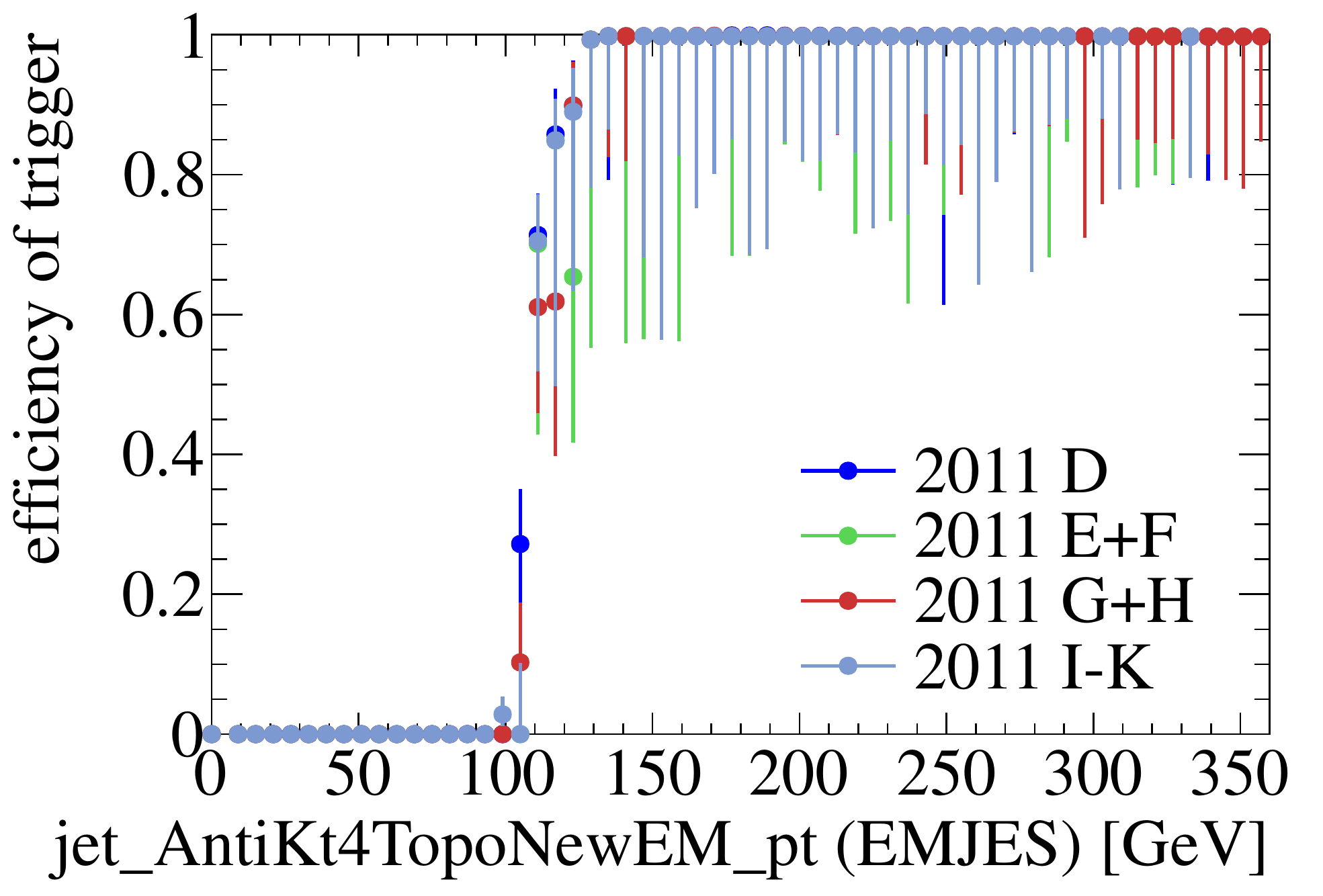}
  \caption{
    Efficiency of the \jetmet trigger \trigger{EF_j75_a4tc_EFFS_xe45_loose_noMu} from bootstrapping,
    using the jet trigger \trigger{EF_j75_a4tc_EFFS} as sample trigger.
    Here, the efficiencies are shown for small sets of periods separately,
    using the usual projections after cuts onto the respective orthogonal variable at \GeV{150}.
    The offline reference for \met is \tagname{MET\_LocHadTopo}.
  }
  \label{fig:results_trigger_performance_measurements_turnons_jetmet_per_period}
\end{figure}

As there have been a number of changes and incidents with potential impact on the \jetmet trigger efficiencies,
a cross-check has been done of the kind
that the efficiencies of the triggers are plotted separately for four small sets of data,
each combining only data from between one and three periods.
The result is shown in \Fig{fig:results_trigger_performance_measurements_turnons_jetmet_per_period},
again only for the primary trigger \trigger{EF_j75_a4tc_EFFS_xe45_loose_noMu},
because only this trigger has actually been used for taking data up to period 2011~I\footnote{
  The plots for the other two triggers are very similar to the ones shown in \Fig{fig:results_trigger_performance_measurements_turnons_jetmet_per_period}.
}.
The four sets contain data from periods~D, E~--~F, G~--~H and I~--~K.
This division has been chosen because period 2011~E is the first period
in which calorimeter measurements were affected by the LAr~hole (cf. \Sec{sec:experimental_setup_lar_hole}).
In period~G, the noise suppresion cuts in the trigger have been adjusted to account for pile-up contributions,
and beginning with period~I,
the read-out of $4$~of the $6$~front-end boards constituting the LAr~hole could be re-established.
In the two projection plots,
\tagname{MET\_LocHadTopo} is used as offline reference,
but the plots for SUSY \met basically show the same features.
The left plot, with the projection onto \met, indicates a small difference between periods D~--~F and G~--~K,
the trigger efficiency for small values of \met being higher in periods G~--~K.
This is, however, the opposite of what would be expected from the change (\ie increase) of the noise thresholds
which should lead to a decrease in the \met seen by the trigger and thus a lower efficiency.
Note that the projection on jet \pt is actually not expected to exhibit a strong dependency on the period,
because the turn-on behavior it shows is likely dominated by the turn-on that is used for bootstrapping,
\ie the one of the sample trigger \trigger{EF_j75_a4tc_EFFS},
because it has the same cuts on jet \pt as the combined trigger itself.
As the same turn-on from period~D is used for all periods,
the projection onto jet \pt is expected to agree for all periods.
The stability of the sample trigger and thus the jet part of the combined trigger
is demonstrated by the right plot in \Fig{fig:results_trigger_performance_measurements_turnons_jet_crosscheck_K} from above.

In general, the efficiencies from all four sets of data are in good agreement.
The onset of the plateau region suffers from low statistics and apparent fluctuations in the efficiency measurements as mentioned above.
From the contour plots in \Fig{fig:results_trigger_performance_measurements_turnons_jetmet_45_contour},
this behavior seems to be coming dominantly from the region of high jet \pt, %
and thus really be due to the limited statistics rather than being a genuine trigger issue.
In any case, it should be investigated further when more data becomes available.

\subsubsection{Cross-checks Using Muon Triggers as Orthogonal Sample Triggers}

\begin{figure}
  \centering
  \incgraphics{width=\widthtwoplots}{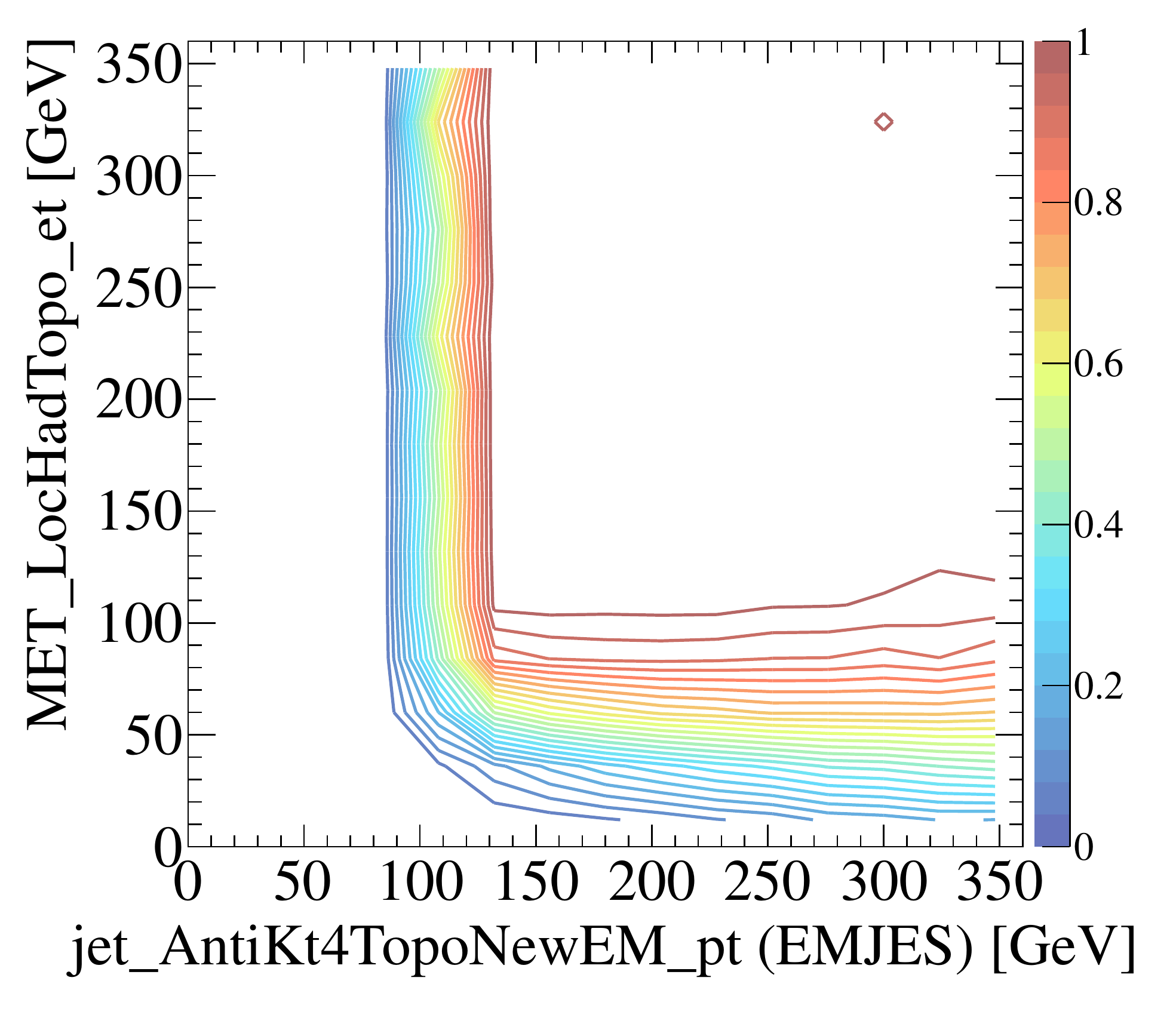}
  \hfill
  \incgraphics{width=\widthtwoplots}{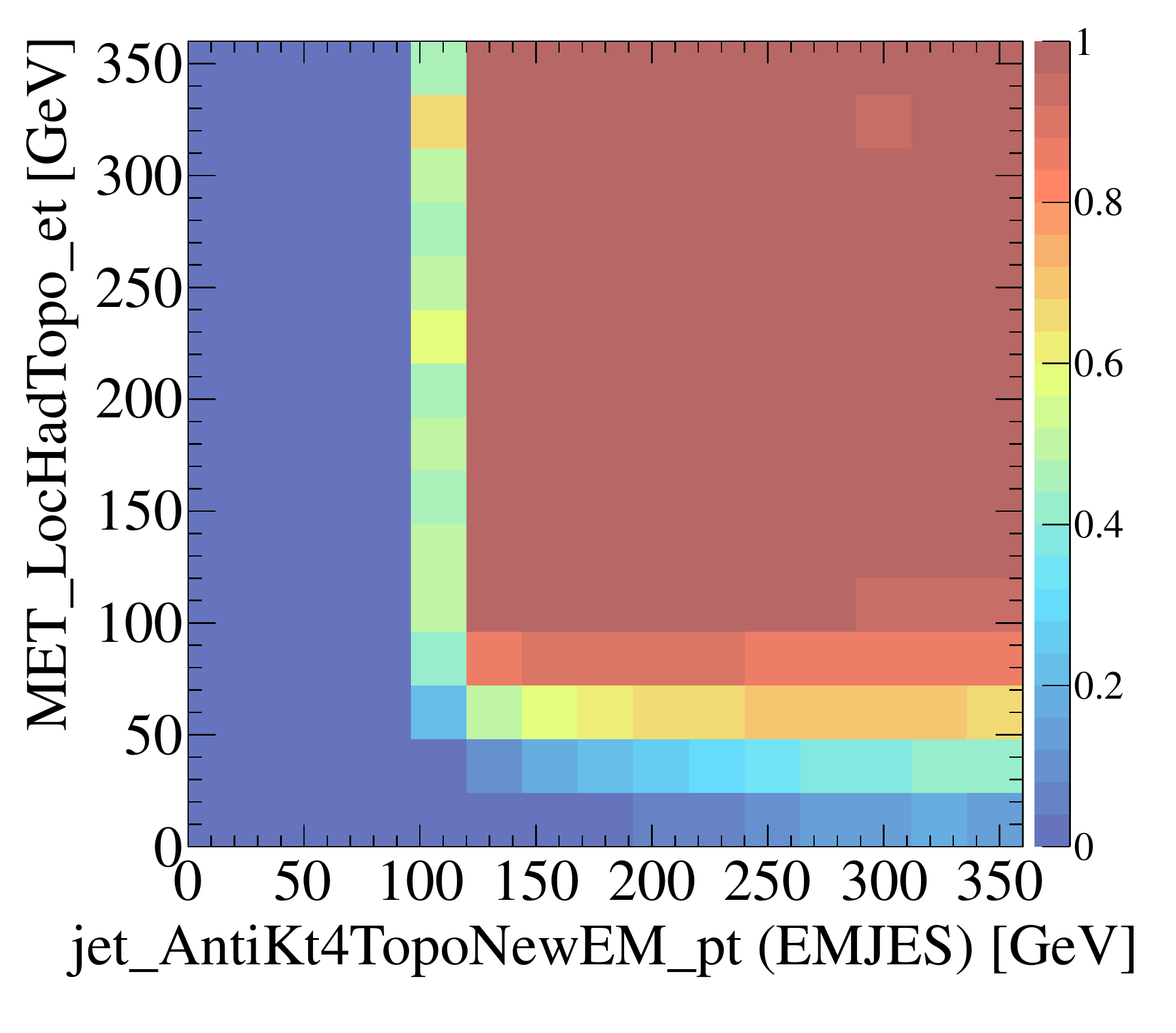}
  \caption{
    Efficiency of the \jetmet trigger \trigger{EF_j75_a4tc_EFFS_xe45_loose_noMu} on the \tagname{Muons} stream,
    using \trigger{EF_mu18} as sample trigger under the assumption of orthogonality.
    The offline reference for \met is \tagname{MET\_LocHadTopo}.
    The two plots are produced from the same data and only differ in style.
  }
  \label{fig:results_trigger_performance_measurements_turnons_jetmet_muons_2d}
\end{figure}

\begin{table}
  \centering
  \begin{tabular}{llll}
    \toprule
    \centering Name of chain & Sample trigger & \multicolumn{2}{l}{Lower chains}\\
    \midrule
    \trigger{EF_j75_a4tc_EFFS_xe45,55_loose_noMu}
    & \trigger{EF_xe30_noMu}
    & \trigger{L2_xe20_noMu}
    & \trigger{L1_XE20} \\
    \trigger{EF_j75_a4tc_EFFS_xe55_noMu}
    & \trigger{EF_xe40_noMu}
    & \trigger{L2_xe30_noMu}
    & \trigger{L1_XE30} \\
    \trigger{EF_j80_a4tc_EFFS_xe60_noMu}
    & \trigger{EF_xe50_noMu} 
    & \trigger{L2_xe35_noMu} 
    & \trigger{L1_XE35} \\
    \bottomrule
  \end{tabular}
  \caption{
    Overview of \met triggers with highest thresholds that respect the bootstrapping condition \eqref{eq:bootstrap_requirement}
    and can thus be used as sample triggers for the combined \jetmet triggers.
    The lower chain names make clear the thresholds applied at Level~1 and Level~2.
  }
  \label{tab:results_trigger_performance_jetmet_met_sample_triggers_2011}
\end{table}

\begin{figure}
  \centering
  \incgraphics{width=\widthtwoplots}{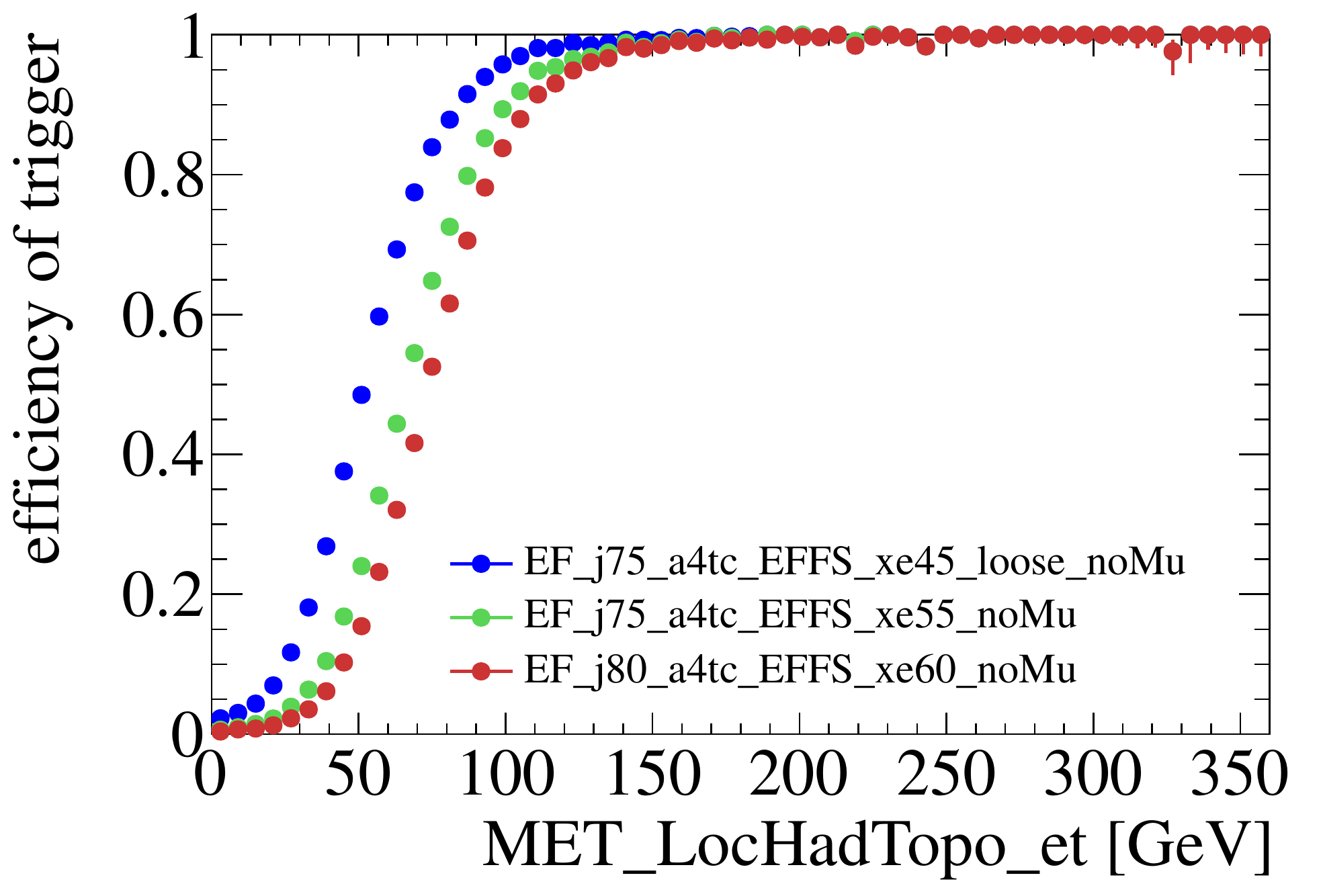}
  \hfill
  \incgraphics{width=\widthtwoplots}{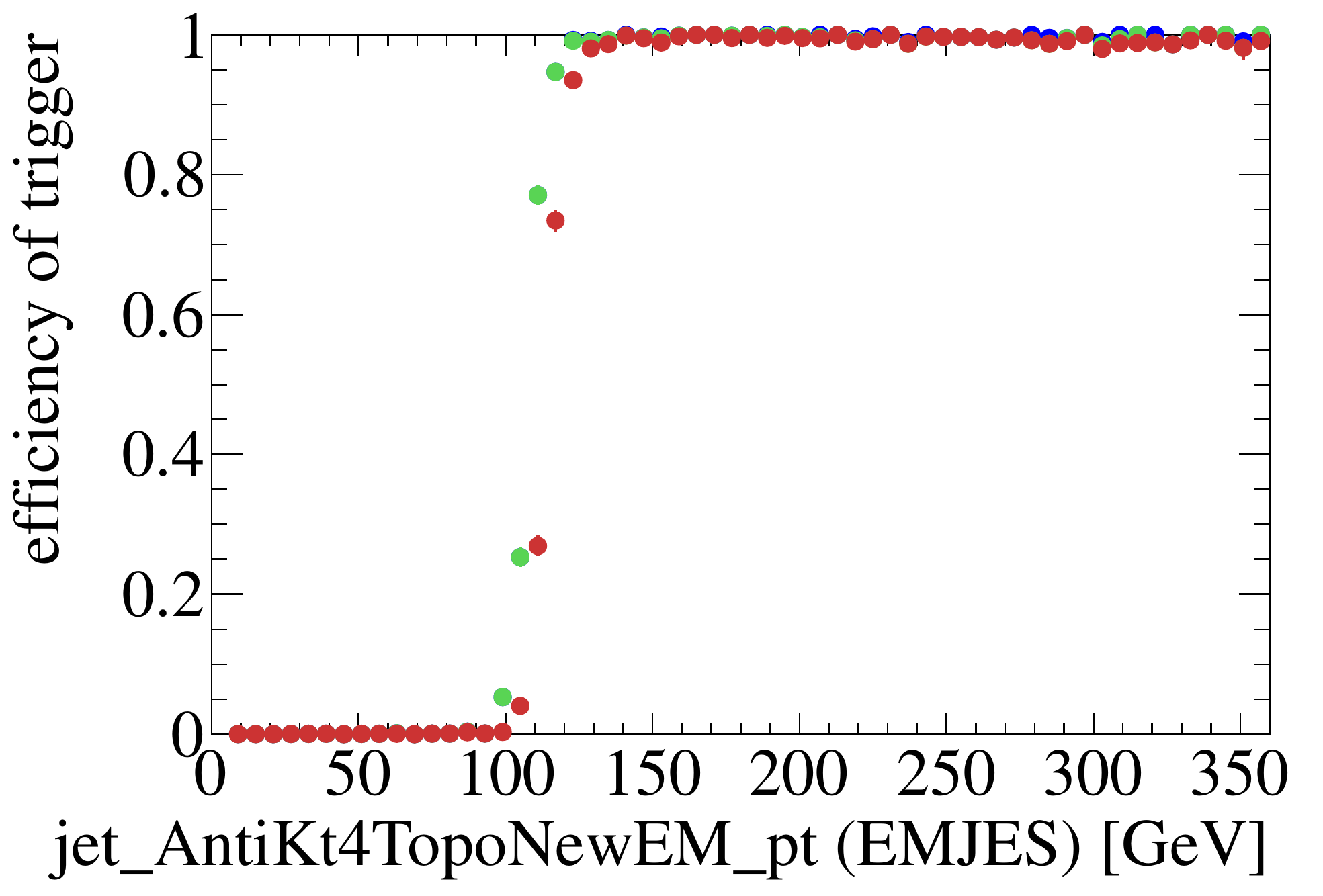}
  \caption{
    Efficiency of three \jetmet triggers on the \tagname{Muons} stream,
    using \trigger{EF_mu18} as sample trigger under the assumption of orthogonality
    and projecting onto the \met (left) and jet \pt (right) axis.
    (The legend in the left plot also applies for the right one.)
  }
  \label{fig:results_trigger_performance_measurements_turnons_jetmet_muons_projections}
\end{figure}

The idea of doing a cross-check of the efficiencies measured on the \tagname{JetTauEtmiss} stream
under the assumption of muon triggers being orthogonal to the \jetmet trigger definition,
thus allowing for the use of a muon-triggered event sample,
has been picked up for the efficiencies in 2011 again.
\Fig{fig:results_trigger_performance_measurements_turnons_jetmet_muons_2d} shows two-dimensional turn-on curves
for \trigger{EF_j75_a4tc_EFFS_{\allowbreak}xe45_loose_noMu} on the sample of events
selected with the single muon trigger \trigger{EF_mu18} from the \tagname{Muons} stream in periods 2011 I~--~K. %
A synopsis together with the two other \jetmet triggers introduced in 2011~I is shown in the projection plots
in \Fig{fig:results_trigger_performance_measurements_turnons_jetmet_muons_projections}.
All plots have very high statistics and therefore give a very good coverage of the plateau regions.
Noticably, the dip in the efficiency
which can be observed in \Fig{fig:results_trigger_performance_measurements_turnons_jetmet_projections}
is not reproduced in these plots at all.
This indicates that it is either only a problem of statistics,
or it is related to the composition of \met.
In the muon-triggered event sample, as explained above,
the \met composition is dominated by real \met
so that, if the dip is due to fake \met, \eg from jet mismeasurements,
it will not show up in the efficiency plots from the \tagname{Muons} stream.

\begin{figure}
  \centering
  \incgraphics{width=\widthtwoplots}{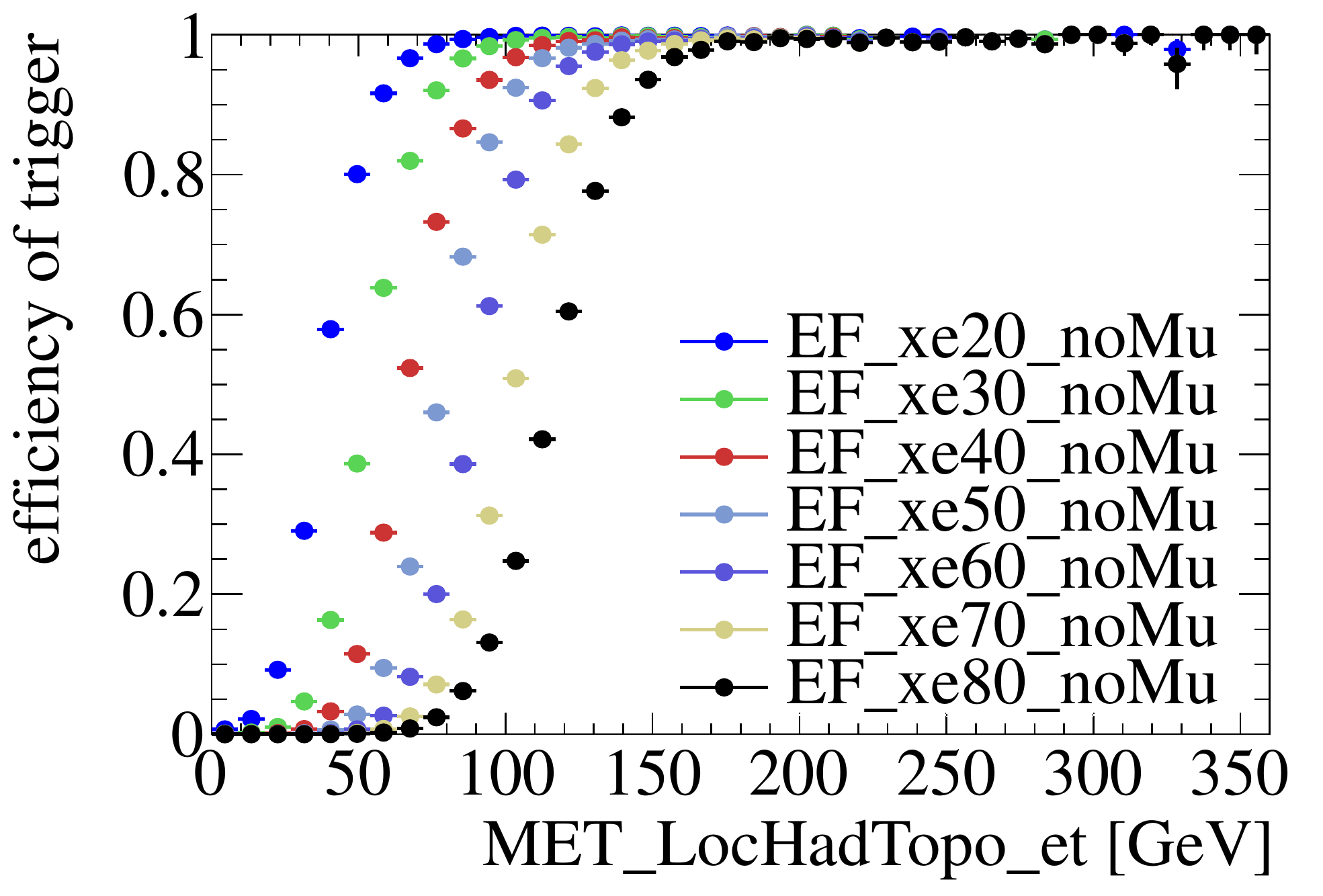} %
  \hfill
  \incgraphics{width=\widthtwoplots}{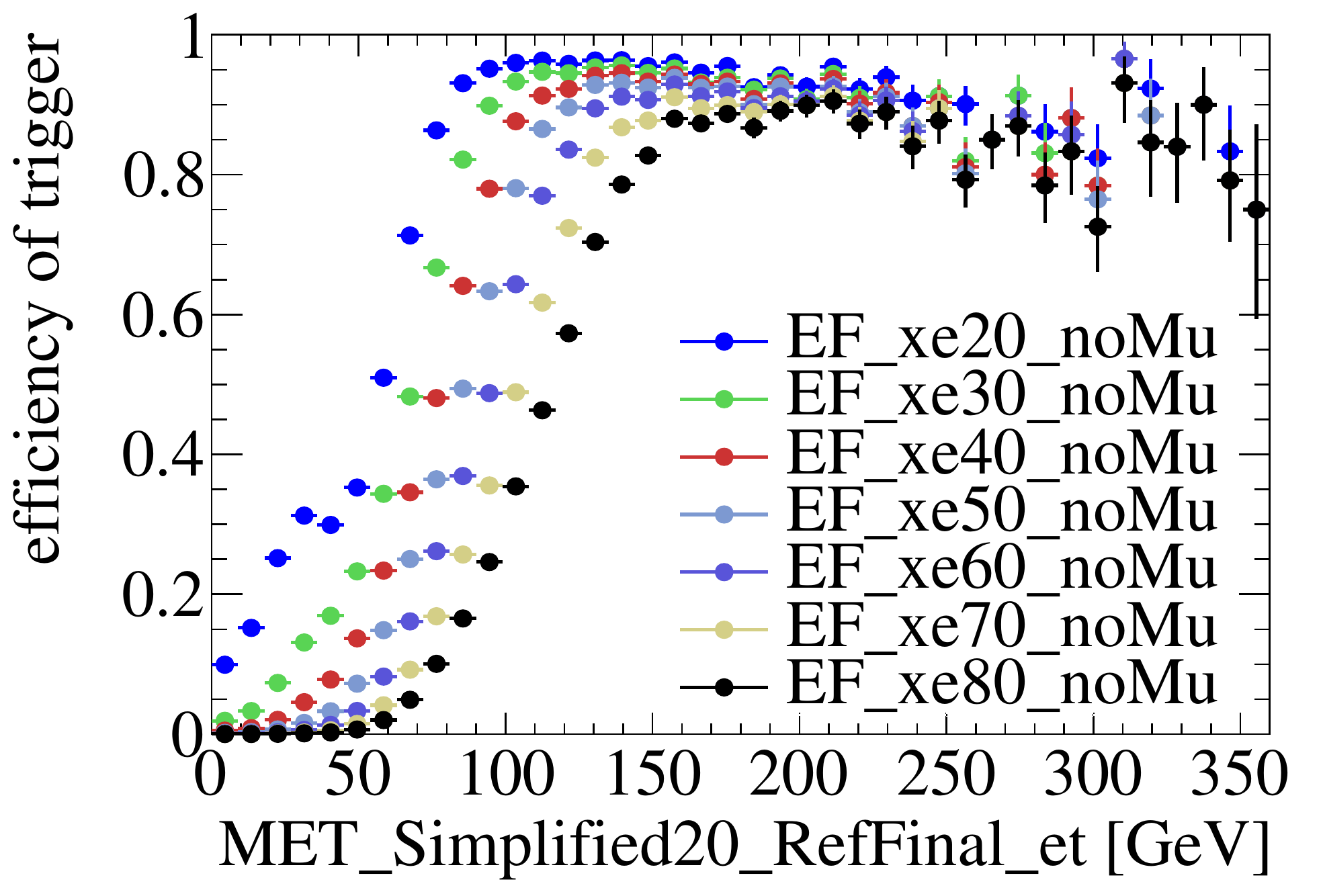} %
  \caption{
    Efficiency of \met triggers with thresholds between 20 and \GeV{80} on the \tagname{Muons} stream,
    using \trigger{EF_mu18} as sample trigger under the assumption of orthogonality,
    as function of \tagname{MET\_LocHadTopo} (left) and SUSY \met (right).
  }
  \label{fig:results_trigger_performance_measurements_turnons_met_2011}
\end{figure}

\Fig{fig:results_trigger_performance_measurements_turnons_met_2011} shows the turn-on curves for several triggers
with different thresholds on the missing transverse energy,
as they are computed on events taken with the muon trigger \trigger{EF_mu18} as sample trigger, under the assumption of orthogonality.
In the left plot, \tagname{MET\_LocHadTopo} is the offline reference for \met,
in the right plot the SUSY \met definition.
The comparison of the efficiencies using the SUSY \met definition in the right plot in \Fig{fig:results_trigger_performance_measurements_turnons_met_2011}
to the right plot in \Fig{fig:results_trigger_performance_measurements_met_turnon_susymet_met_turnon_muonthreshold} shows that,
although the instability of the plateau when using SUSY \met as offline reference
appears to be less pronounced than it is in 2010 data,
it definitely still can be seen.
Note also the hump in the efficiency curve clearly visible for \trigger{EF_xe20_noMu}
at the typical position of the peak of the transverse energy spectrum
of muons produced in $W$ decays. %
The improvement over 2010 may be explained by the updated \met definition
as well as by the different composition of \met in muon-triggered events,
which in 2011 will have a higher level of fake \met due to the increased number of overlaid in-time pile-up events.

\begin{figure}
  \centering
  \incgraphics{width=\widthtwoplots}{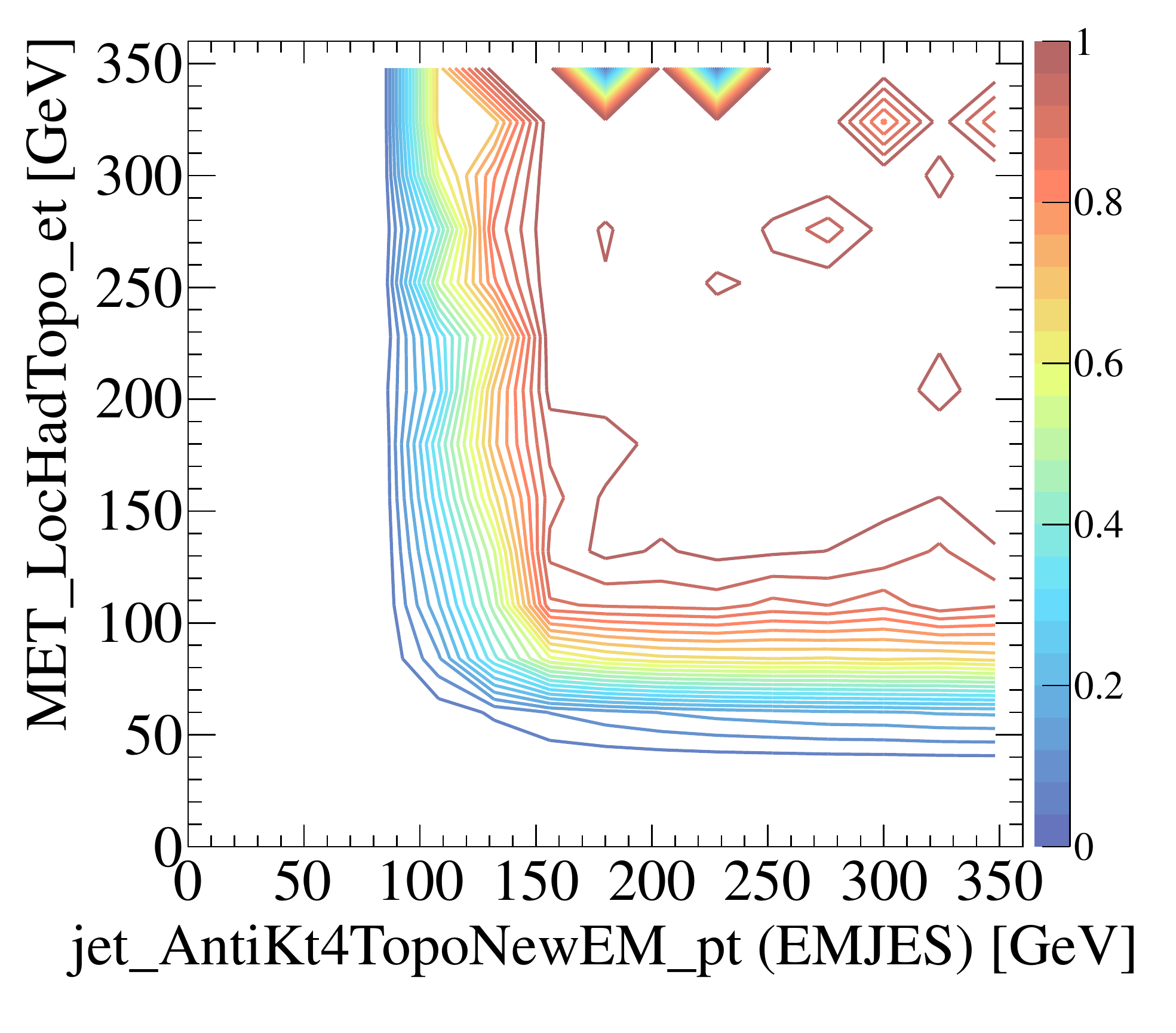}
  \hfill
  \incgraphics{width=\widthtwoplots}{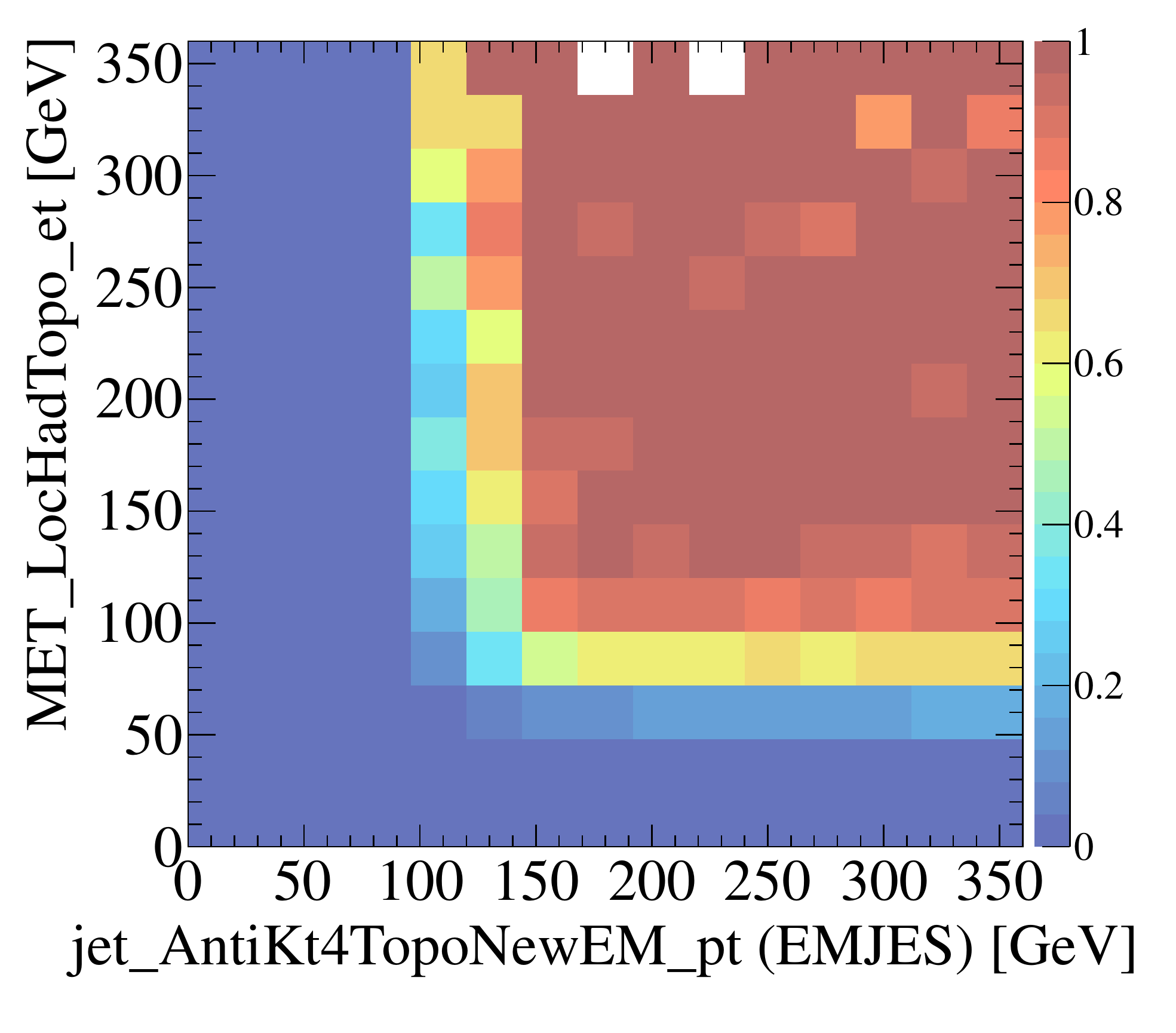}
  \caption{
    Efficiency of the combined \jetmet triggers \trigger{EF_j75_a4tc_EFFS_xe55_noMu}
    bootstrapped from an event sample taken with \trigger{EF_xe40_noMu}.
  }
  \label{fig:results_trigger_performance_measurements_turnons_jetmet_bootstrap_met}
\end{figure}

In terms of event counts,
the \met trigger \trigger{EF_xe20_noMu} even gives slightly better statistics than the jet trigger \trigger{EF_j75_a4tc_EFFS},
but it cannot compete with this jet trigger, which has been used as sample trigger above.
Due to the large spacing of its \met threshold at Event Filter level with respect to the combined triggers,
there are too few events left at high \met to cover the plateau region.
For the combined \jetmet triggers with higher thresholds on \met,
which already have or will supersede the current primary triggers,
\met triggers as sample triggers may become an appealing option again
if the problem of how to obtain the sample trigger turn-on can be solved.
For \tagname{MET\_LocHadTopo} as offline reference,
the turn-on curves for the \met trigger from the \tagname{Muons} stream can be used for bootstrapping.
An example is given in \Fig{fig:results_trigger_performance_measurements_turnons_jetmet_bootstrap_met},
where the efficiency of the combined \jetmet trigger \trigger{EF_j75_a4tc_EFFS_xe55_noMu},
which was introduced in 2011~I,
is shown on an event sample taken with \trigger{EF_xe40_noMu} after bootstrapping.
The plot demonstrates that this \met trigger gives a very good coverage of the plateau region,
better in particular at high \met values in comparison to \Fig{fig:results_trigger_performance_measurements_turnons_jetmet_45_contour}.
Noticeably, the vertical jet turn-on region in this plot is much broader and extends to higher offline values
than it does in \Fig{fig:results_trigger_performance_measurements_turnons_jetmet_45_contour} using a jet sample trigger.
This suggests that in the \met triggered events used to produce the plot,
the problem of missing trigger jet objects again deteriorates the measurement of the jet turn-on.
The \met turn-on on the other hand is much steeper.
This can be explained by the fact that the sample trigger turn-on is measured on an event sample collected with a muon trigger,
which is dominated by real \met rather than the fake \met found in most events in the \tagname{JetTauEtmiss} stream,
and thus has a sharper turn-on,
as could be seen \eg in \Fig{fig:results_trigger_performance_measurements_2d_efficiency_MC_LocHadTopo_projections}.
Through the bootstrapping which is applied,
the turn-on of the \met trigger will shape the turn-on of the \met part of the combined trigger.
This is not a problem with the bootstrapping procedure itself,
but highlights again the fact that the sample trigger turn-on used for bootstrapping needs to be unbiased, too.

\section{Conclusions for 2011}

In \Sec{sec:results_trigger_performance_measurements_data_2011}, the efficiencies of the combined \jetmet trigger in the 2011 trigger menu have been studied.
The focus has been on the physics triggers which are used by the \ATLAS Supersymmetry group.
The computation of the turn-on curves largely relies on the jet trigger \trigger{EF_j75_a4tc_EFFS} as sample trigger,
and employs the bootstrap method to correct for its bias in the event sample.
The results built upon experience from the studies in 2010 and constitute a continuation of these studies.
A number of additional cross-checks have been presented,
showing an overall good agreement and demonstrating the consistency of the efficiencies in different periods,
despite many changes in the trigger.

With respect to the official Supersymmetry analysis,
the offline cuts follow the trigger requirements to be in the plateau of the trigger efficiency:
It has been decided to use an offline cut of \GeV{130} on both offline \met and jet \pt
for the analysis using \trigger{EF_j75_a4tc_EFFS_{\allowbreak}xe45_loose_noMu} in the zero-lepton channel \cite{ATL-PHYS-INT-2011-085}. %
There is an apparent tension between this cut and the results on the latest periods I~--~K %
because the beginning of the plateau seems to be affected by a drop in efficiency,
which could not be seen previously due to the smaller statistics.
There are several possible explanations.
It is conceivable that the change in the turn-on is caused
by the increasing level of in-time pile-up and changing composition of \met in the events.
It could be due to a crossing of the \met turn-on curves of the different trigger levels constituting the \jetmet chain,
but it also cannot be ruled out that it is an artefact that will vanish with increasing statistics.
In any case, this drop cannot be seen in events collected with muon triggers
so that \revised{it} can be assumed that in events dominated by real \met
the trigger is not afflicted by this decrease of efficiency.
This is also consistent with the argument of crossing turn-ons,
because if the turn-ons of the different trigger levels are steeper in this sample due to the different composition of \met,
they would possibly not cross for this type of events.
The efficiencies of the two new \jetmet triggers introduced in period~I
could be measured on the full dataset thanks to trigger emulation.
For the next update of the Supersymmetry analysis\footnote{
  Moriond 2012
},
the trigger \trigger{EF_j75_a4tc_EFFS_xe55_noMu} will be become the primary physics trigger,
which necessitates to increase the offline \met cut to presumably \GeV{150}.
Using \met triggers as sample triggers might become interesting due to the higher statistics,
because the currently used jet sample trigger is already heavily prescaled.
Nevertheless, there are two challenges that would have to be met:
First, the \met turn-on needs to be measured,
which is hindered by finding an unbiased event sample and by the difference between SUSY \met and trigger \met.
Second, there is the problem of missing trigger jets,
which by using jet sample triggers can be circumvented elegantly.

An approach that would be interesting to try and adopt in future analyses relying on this type of trigger
would be to find the value of the offline variable
at which the trigger efficiency exceeds a given threshold
and an uncertainty on this value.
This could then be converted into a systematic uncertainty in the analysis.
One way to find this value is to fit the turn-on curve with \eg a linear combination or a product of two Gauss error functions,
which would be able to account for a crossing of two turn-on curves. %
Such an approach could also fully exploit the computation of the uncertainties on the trigger efficiencies,
which would then enter the analysis in terms of an additional systematic uncertainty.
So far, systematic uncertainties coming from the trigger efficiency have been neglected in the zero-lepton analyses,
because the offline cuts were chosen such that the efficiency of the primary trigger is close to \percent{100},
and the systematic uncertainties arising from the trigger efficiency expected to be small compared to other sources of systematic errors.
With a better control of the other uncertainties,
which are dominating the total uncertainty at the moment,
and an on-going optimization of the analysis for sensitivity,
a refined treatment of the trigger uncertainties will become more important.
Besides, the progression of the instantaneous luminosity might render it unfavorable
to adapt the offline cuts to the trigger requirements to be in the plateau.
In particular, when fitting of kinematic distributions instead of event counting will be introduced,
the impact of the variation of the trigger efficiency between different bins needs to be evaluated.
Furthermore, it would be interesting to study the correlation of jet \pt and \met,
which is found in particular in events where jet mismeasurements make up for a large part of the \met,
and to estimate its influence on the trigger efficiencies.

\newpage
\setcounter{footnote}{1} %

\chapter{\texorpdfstring{A Model for \met Trigger Rates} {A Model for MET Trigger Rates}}
\label{sec:results_met_model}

Missing transverse energy (\met) is one of the most involved measurements at a particle detector,
because it is basically a sum over all particles that are produced in the proton-proton collisions.
Its resolution will therefore suffer from all possible sources of mismeasurements of all subdetectors combined in the sum.
What concerns the trigger, the measurement of \met is additionally complicated by the fact
that due to time constraints it is prohibitive, at least at Level~1, to do a sum over actually all detector components.
In practice this means that measurements from the muon spectrometers cannot be included in the \met sum at Level~1 (cf. \Sec{sec:tdaq_met}).
Currently, also at Level~2 and Event Filter no muon correction is employed to make the trigger decision,
although here the muon contribution is computed and could in principle be used.
Furthermore, the \met measurements in general (and even more \sumet measurements) will directly receive contributions from pile-up,
which leads to a strong dependence of the trigger rates on the activity in the event and the level of pile-up.

At the \acl{LHC}, the trigger rates are dominated by \acs{QCD} processes (cf. \Fig{fig:lhc_cross_sections}). %
In these processes, events with a large real \met are very rare
because no invisible particles like neutrinos are produced in the hard \revised{scattering} process.
Neutrinos may still be produced in jets originating from heavy quarks,
but these neutrinos mostly carry away only comparably small energies.
The largest contribution to the \met measured by the detector will thus be fake \met
due to resolution effects in the calorimeter (cf. \Sec{sec:general_missing_transverse_energy}).
As \met, being the magnitude of a vector quantity, is by definition a positive quantity,
fluctuations due to noise do not cancel,
and \met will attain a positive mean even if there is no real \met.

\section{Motivation}
\label{sec:results_met_model_motivation}

As was said in the introduction,
the \met measurement depends on a lot of different inputs.
It will in particular also be directly dependent on the activity in the event.
The higher the activity and the more collisions take place at the same time and the more particles are produced,
the larger the sum of the energy deposited in the calorimeter systems of the detector will be.
The impact of this dependence can be seen by comparing the rate of different types of triggers
as function of the instantaneous luminosity \Linst and the average number of interactions\footnote{
  Interactions means proton-proton collisions here,
  happening when two bunches of protons are brought to collision.
}
per bunch-crossing, which is denoted as \avgmu (cf. \Sec{sec:experimental_setup_general_define_avgmu}).
The trigger rate plots in this section which show rate measurements from the \ATLAS trigger system
have all been produced from the data stored in the \COOL database,
as described in \Sec{sec:results_trigger_rates_describe_COOL_plots}.

\begin{figure}
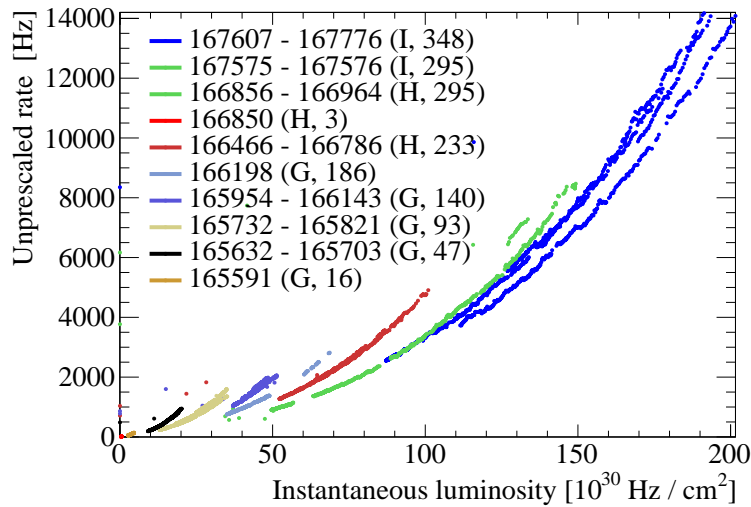

  \centering
  \incgraphics{width=\widthsingleplot}{{{COOL.plot_rates_L1_XE20_2010I,2010H,2010G_withGRL_Xoffllumi_cbnbc}}}
  \caption{
    Rate of the \met trigger at Level~1 with a threshold of \GeV{20}, \trigger{L1_XE20}, in runs from periods G~--~I of 2010,
    drawn as function of the instantaneous luminosity.
    The colors correspond to different numbers of colliding bunches $n_\text{coll}$,
    which are given in brackets together with the period.
    The intervals in the legend are the numbers of the runs,
    which have been grouped according to the number of colliding bunches.
  }
  \label{fig:results_met_model_rate_L1_XE20_lumi}
\end{figure}

\Fig{fig:results_met_model_rate_L1_XE20_lumi} shows the rates of one of the Level~1 \met triggers in \ATLAS,
from periods G~--~I of 2010 data taking, as a function of the instantaneous luminosity.
The plot exhibits two important features:
\begin{itemize}
  \item 
    Unlike the rate of other triggers, for example the jet trigger for which the rates are shown in \Fig{fig:trigger_rate_example_L1_J30},
    the rate of this trigger is not a linear function of the instantaneous luminosity,
    but depends on the luminosity with an exponent larger than one.
    This means that the rate of this trigger grows stronger than the instantaneous luminosity,
    and it therefore needs to be watched carefully in order not to exceed the bandwidth capacities of the \acs{TDAQ} system.
    This is a major motivation for trying to achieve a better understanding of what causes this behavior.
  \item
    The rate shows a discontinuous behavior,
    as if parts of the curve had not been properly scaled to match the others.
    This does not come from prescales.
    These have been taken into account and corrected for when drawing the rates for the different runs.
    (Which is confirmed by the fact that this trigger has been prescaled for the first time in run 167607 from period~I.) %
    The steps which can be observed in this plot do not only occur at the boundaries of periods,
    where trigger rates might be expected to change due to changing detector conditions, but also within periods.
    Their origin can be explained by looking at the bunch structure of the \LHC fills in the different runs.
\end{itemize}

\renewcommand{\arraystretch}{1} %
\begin{table}
  \centering
  \begin{tabular}{crrlcrrlcrr}
    \toprule
    Run no. & $n_\text{coll}$ & $n_\text{train}$ &  & Run no.& $n_\text{coll}$ & $n_\text{train}$ &  & Run no.& $n_\text{coll}$ & $n_\text{train}$\\
    \cmidrule{1-3}\cmidrule{5-7}\cmidrule{9-11}
    165591 & 16 & 2 &  & 166097 & 140 & 18 &  & 166925 & 295 & 38\\
    165632 & 47 & 6 &  & 166142 & 140 & 18 &  & 166927 & 295 & 38\\
    165703 & 47 & 6 &  & 166143 & 140 & 18 &  & 166964 & 295 & 38\\
    165732 & 93 & 12 &  & 166198 & 186 & 24 &  & 167575 & 295 & 38\\
    165767 & 93 & 12 &  & 166305 & 186 & 24 &  & 167576 & 295 & 38\\
    165815 & 93 & 12 &  & 166383 & 186 & 24 &  & 167607 & 348 & 46\\
    165817 & 93 & 12 &  & 166466 & 233 & 30 &  & 167661 & 348 & 46\\
    165818 & 93 & 12 &  & 166658 & 233 & 30 &  & 167680 & 348 & 46\\
    165821 & 93 & 12 &  & 166786 & 233 & 30 &  & 167776 & 348 & 46\\
    165954 & 140 & 18 &  & 166850 & 3 & 3 &  & 167844 & 348 & 46\\
    165956 & 140 & 18 &  & 166856 & 295 & 38 &  &  &  & \\
    166094 & 140 & 18 &  & 166924 & 295 & 38 &  &  &  & \\
    \bottomrule
  \end{tabular}
  \caption{
    Number and structure of the proton bunches colliding in \ATLAS in the runs from period G~--~I in 2010.
    The table lists the run number together with the number of colliding bunches $n_\text{coll}$
    and the number of bunch trains $n_\text{train}$ in which the colliding bunches are organized.
  }
  \label{tab:results_met_model_bunch_structure_2010}
\end{table}
\arraystretchdefault

\Tab{tab:results_met_model_bunch_structure_2010} shows the number and structure of the proton bunches
which are brought to collision %
at the interaction point in \ATLAS.
Given are the number of colliding bunches $n_\text{coll}$ and the number of trains $n_\text{train}$
in which these colliding bunches are organized (cf. \Sec{sec:experimental_setup_LHC}).
Note that these numbers only refer to the period of time in which the LHC has declared stable beams.
In particular before that, in the phase of bunch injection into the collider ring, the number of bunches is lower,
but these luminosity blocks are cut away by the \ac{GRL}.
Moreover, the trigger system is not activated
(\ie not issuing trigger signals) before stable beams have been declared.
The spacing of bunches within the bunch trains is always (at least) five empty buckets 
between two colliding proton bunches,
corresponding to a spacing of (at least) \unit[150]{ns}.

Comparing the positions of the steps in the trigger rate plot in \Fig{fig:results_met_model_rate_L1_XE20_lumi}
to the changes in the bunch structure from \Tab{tab:results_met_model_bunch_structure_2010} shows a correlation between the two:
whenever the number of bunches is increased for a fixed instantaneous luminosity,
the rate of the \met trigger becomes smaller.
This is made obvious by the coloring that has been used in \Fig{fig:results_met_model_rate_L1_XE20_lumi}.
The runs have been grouped by the number of colliding bunches in \ATLAS,
and the color changes whenever the number of colliding bunches changes.
All the different, apparently disconnected pieces of the rate curves have different colors,
\ie correspond to different numbers of colliding bunches.

The origin of the effect is a combination of two factors:
the sensitivity of the \met rates to the total activity in the event
and the relation between the instantaneous luminosity and the number of bunches,
which is simply
\begin{equation}
  \Linst = \sum^{n_\text{bunch}}_{i=1} \Linst^1,
\end{equation}
assuming in the following for simplicity that the instantaneous luminosity per bunch $\Linst^1$ is the same for all colliding bunches.
From this equation, it follows that, for a larger $n_\text{bunch}$ at a fixed instantaneous luminosity,
the luminosity per bunch becomes smaller,
and therefore the activity within this bunch crossing becomes smaller.

It should be stressed again that this behavior is typical for the \met and also for the \sumet triggers, %
but different for triggers like the aforementioned jet triggers,
which also may have a dependence on the event activity,
but this dependence is small enough to be neglected
so that the jet trigger rate only depends on the total instantaneous luminosity,
independent of whether this total instantaneous luminosity is generated by a few or by many colliding bunches (cf. \Fig{fig:trigger_rate_example_L1_J30}). %

\revised{
\begin{figure}
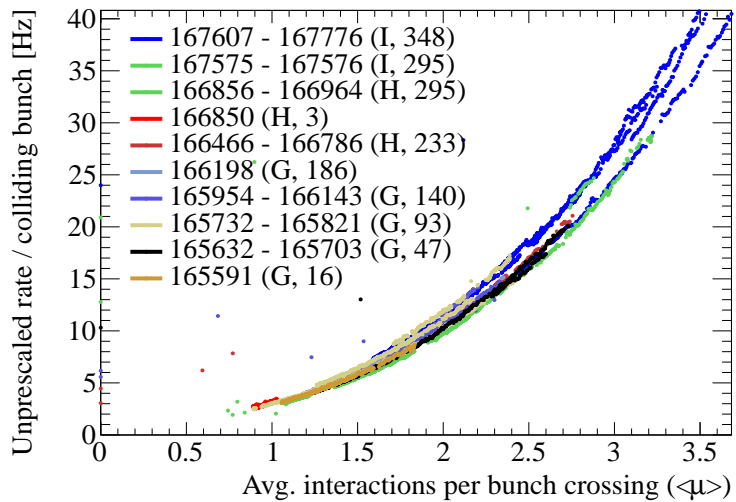

  \centering
  \incgraphics{width=\widthsingleplot}{{{COOL.plot_rates_L1_XE20_2010I,2010H,2010G_withGRL_Xofflmu_cbnbc}}} %
  \caption{
    Rate of the \met trigger \trigger{L1_XE20} in runs from periods G~--~I of 2010,
    drawn as function of the average number of concurrent interactions \avgmu.
    Again the colors correspond to different numbers of colliding bunches $n_\text{coll}$ which are given in brackets together with the period.
    The intervals in the legend are the numbers of the runs
    which have been grouped according to the number of colliding bunches.
  }
  \label{fig:results_met_model_rate_L1_XE20_avgmu}
\end{figure}
}

\Fig{fig:results_met_model_rate_L1_XE20_avgmu} shows the rate of the \trigger{L1_XE20} \met trigger
for the same periods as in \Fig{fig:results_met_model_rate_L1_XE20_lumi},
but plotted as function of the average number of interactions per bunch crossing,
which is related to the instantaneous luminosity by
$\avgmu \sim {{\cal L}_\text{inst} } / {n_\text{coll}}$ (cf. \Sec{sec:results_trigger_rates_describe_COOL_plots}).
Indeed, after rescaling with the number of colliding bunches,
the curves for all runs show a consistent behavior.
The conclusion from this for the model building is that,
instead of regarding rates as function of instantaneous luminosity,
for the \met triggers the event activity, \ie the average number of interactions per bunch crossing \avgmu, is a more suitable choice
because it does not depend on the number of bunches,
which would otherwise have to be included as an additional parameter in the model.

\section{Overview}

In this section, a model will be presented which is aimed at helping to understand and explain
the behavior of the missing transverse energy (\met) and sum of transverse energy (\sumet) triggers.
Although the model has been developed using data from the \ATLAS experiment, it is deliberately kept very general.
This allows to transfer the conclusions made here to trigger systems at other collider experiments.
Besides, another strong motivation for keeping the model as simple as possible
is that this simplicity allows to understand the essential features which drive the trigger behavior.
The model does not seek to replace a full Monte Carlo detector simulation
and cannot possibly describe the \met distribution found in data in all detail.
Instead, it is designed to be sufficiently precise to allow predictions of the pile-up dependence of the trigger rates,
while still exhibiting the connections between the underlying ideas and the resulting impact on the rate,
and to thereby complement existing Monte Carlo simulations.

The goal of the model is to describe the fake \met contribution coming from the limited detector resolution
and to reproduce the \sumet and \met spectra,
from which the trigger rates can then be computed.
A key aspect is to account for in-time pile-up to be able to predict changes
as function of the average number of interactions per bunch-crossing \avgmu.
Another important ingredient is that the \met resolution is independent of \avgmu
and only depends on \sumet, as will be shown below.

The model presented here has been developed in collaboration with the \ATLAS \met trigger group.
The documentation of the model is also published in an \ATLAS note \cite{ATL-COM-DAQ-2010-218}.

\section{Model Building}

\subsection{\texorpdfstring{\met Shape} {MET Shape}}

\begin{figure}
  \centering
  \incgraphics{width=\widthsingleplot}{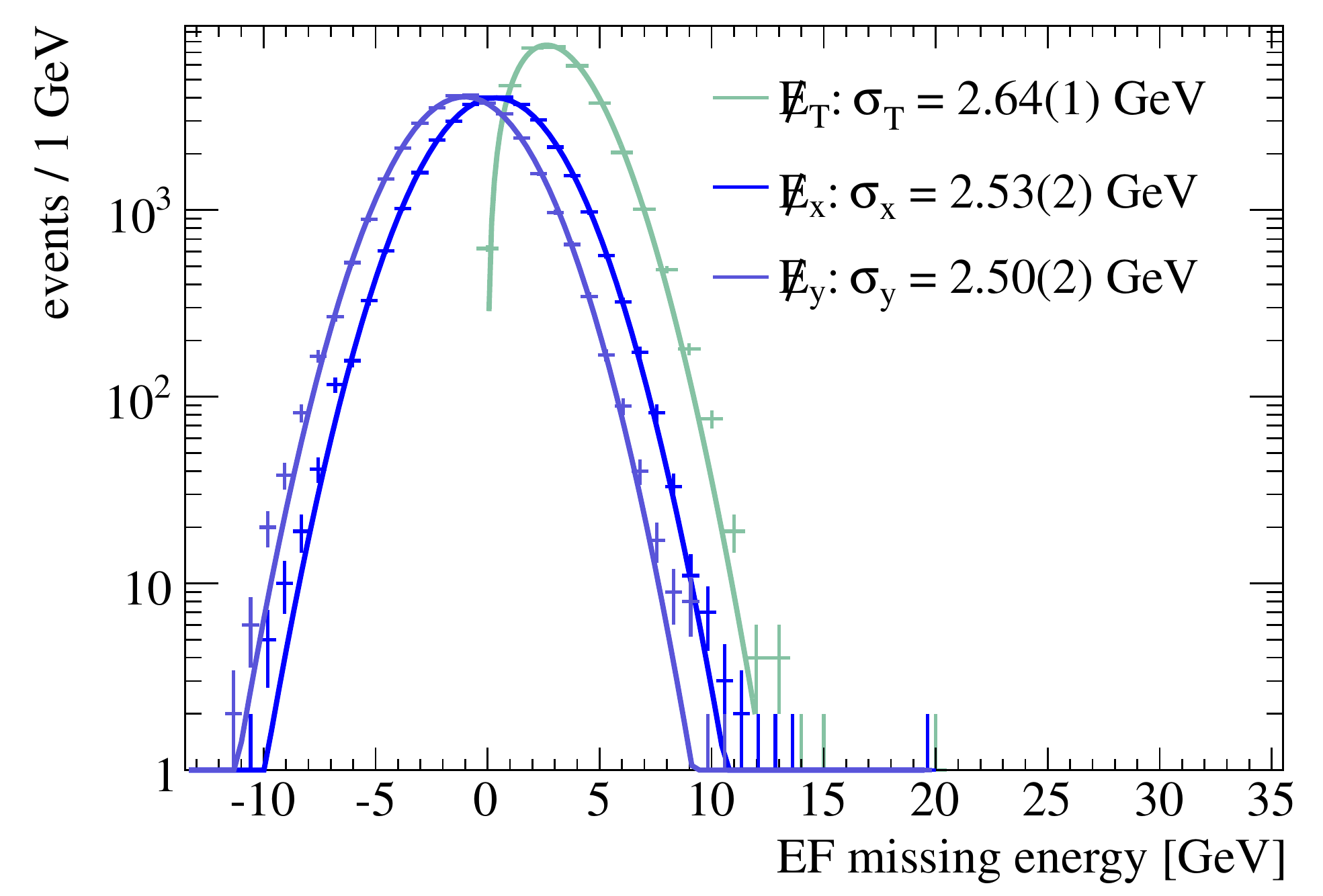} %
  \caption{
    Distribution of EF \mex, \mey and \met for minimum-bias events from run 167776 (period 2010~I)
    which have $\GeV{53} < \sumet < \GeV{75}$.
    The solid lines are fits to the data points with a Gaussian (for \mex and \mey) or \met distribution,
    for which the relevant parameters are given in the legend.
  }
  \label{fig:results_met_model_mexy}
\end{figure}

\Fig{fig:results_met_model_mexy} shows the distributions of \mex, \mey and \met measured at Event Filter,
for a sample of events  with EF \sumet in a given range, $\GeV{53} < \sumet < \GeV{75}$.
The events are taken from run 167776 in period 2010~I after applying the \ac{GRL}.
They are selected by the minimum-bias trigger \trigger{EF_mbMbts_1_eff}
and can therefore be assumed to consist mostly of events
which do not contain real \met from particles invisible to the detector,
carrying away part of the energy unnoticed.
All of the \met is therefore fake \met coming from the limited detector resolution.
The measurement of \met or its components in the transverse plane, \mex and \mey,
is thus, in fact, a measurement of the detector resolution.
As can be seen from the fit in \Fig{fig:results_met_model_mexy},
the bulk of the distribution of \mex and \mey is well described by a Gaussian\footnote{
  In the following, a lot of probability density functions for quantities like \met and \sumet will be given,
  often decorated with additional Poisson distributions.
  A compromise between an abstract and an explicit notation has been found
  in that it was decided to abandon a lean notation in favor of a more explicit notation,
  which will hopefully allow for a fluent reading.
  The expression $P(Q = x|\vec\alpha)$ should be understood to denote
  the density of probabilities of the values $x$ which can be attained by a physical quantity $Q$ (\eg \sumet or \met) %
  for a given set of parameters $\vec\alpha$.
},
\begin{equation}
  N_0 \, P(\mex = x|\sigma_x,\,x_0) = \frac{N_0}{\sqrt{2\pi\sigma_x^2}} \exp\left({ -\frac{(x-x_0)^2}{2\sigma_x^2} }\right),
  \label{eq:results_met_model_gauss_mex}
\end{equation}
and correspondingly for $\mey$.
The values for the width parameter $\sigma_{x,y}$ of the Gaussians,
together with their uncertainties as they are obtained from the fit,
are given in the legend.
The normalization $N_0$ is not of importance here,
and also the small shifts $x_0$ and $y_0$ will be neglected in the following and in the model in general
because this allows to derive an analytical form of the distribution of \met.
The distribution of \met evidently does not follow a Gaussian,
but is described by
\begin{equation}
  N_0 \, P(\met = x|\sigma_T) = \frac{N_0}{\sigma_T^2} \, x \exp\left({ -\frac{x^2}{2\sigma_T^2} }\right).
  \label{eq:appendix_met_tobeproven_repeated}
\end{equation}
The derivation of this form is given in \Sec{sec:appendix_analytic_form_met} in the Appendix, %
and follows from the geometrical relation $\met = \sqrt{(\mex)^2 + (\mey)^2}$
under the assumption that \mex and \mey can be described by Gaussian distributions with zero mean.
In particular, if $\sigma_x = \sigma_y \asdefined \sigma$, then also $\sigma_T = \sigma$.
A comparison of \Eqs{eq:results_met_model_gauss_mex} and \eqref{eq:appendix_met_tobeproven_repeated}
shows that even if the mean of both \mex and \mey is zero,
the measurement of \met will still yield a non-vanishing value leading to fake \met directly correlated to the resolution.

The value of $\sigma_T$ obtained from the fit in \Fig{fig:results_met_model_mexy} is also given in the legend.
It can be seen that $\sigma_x$ and $\sigma_y$ do not match perfectly, but agree within less than two standard deviations.
In the same spirit, also $\sigma_T$, which ought to be equal to $\sigma_x$ and $\sigma_y$,
does not match perfectly, but is slightly larger due to the small shift of \mey.
Still, the overall agreement shows that the \met and its components can be reasonably well described by the given forms,
which shows that the underlying assumption of the dominance of fake \met due to the limited detector resolution is correct.

The reason that the event sample used to produce the plot in \Fig{fig:results_met_model_mexy}
has been selected from a narrow slice of \sumet
is that the resolution of the $x$ and $y$ components of the missing energy does depend on \sumet.
But apart from that, the resolution is a detector parameter that is independent from pile-up,
\ie from the average number of interactions per bunch crossing. %
One important step for the model building is therefore to find a parametrization of the
resolution of \met as function of \sumet
and to measure this on data.
How this can be done is demonstrated below,
and the result can be seen in \Fig{fig:results_met_model_resolution_MET_167776},
which also shows that the parametrization of the resolution is independent from the activity in the event due to pile-up.
On the other hand, the dependence of the \met resolution on \sumet means
that in order to model the \met distribution, a model for the underlying \sumet distribution is needed as well.
The next step is therefore to analyse the distribution of \sumet.

\subsection{\texorpdfstring{\sumet Shape} {SumET Shape}}

\begin{figure}
  \centering
  \incgraphics{width=\widthtwoplots}{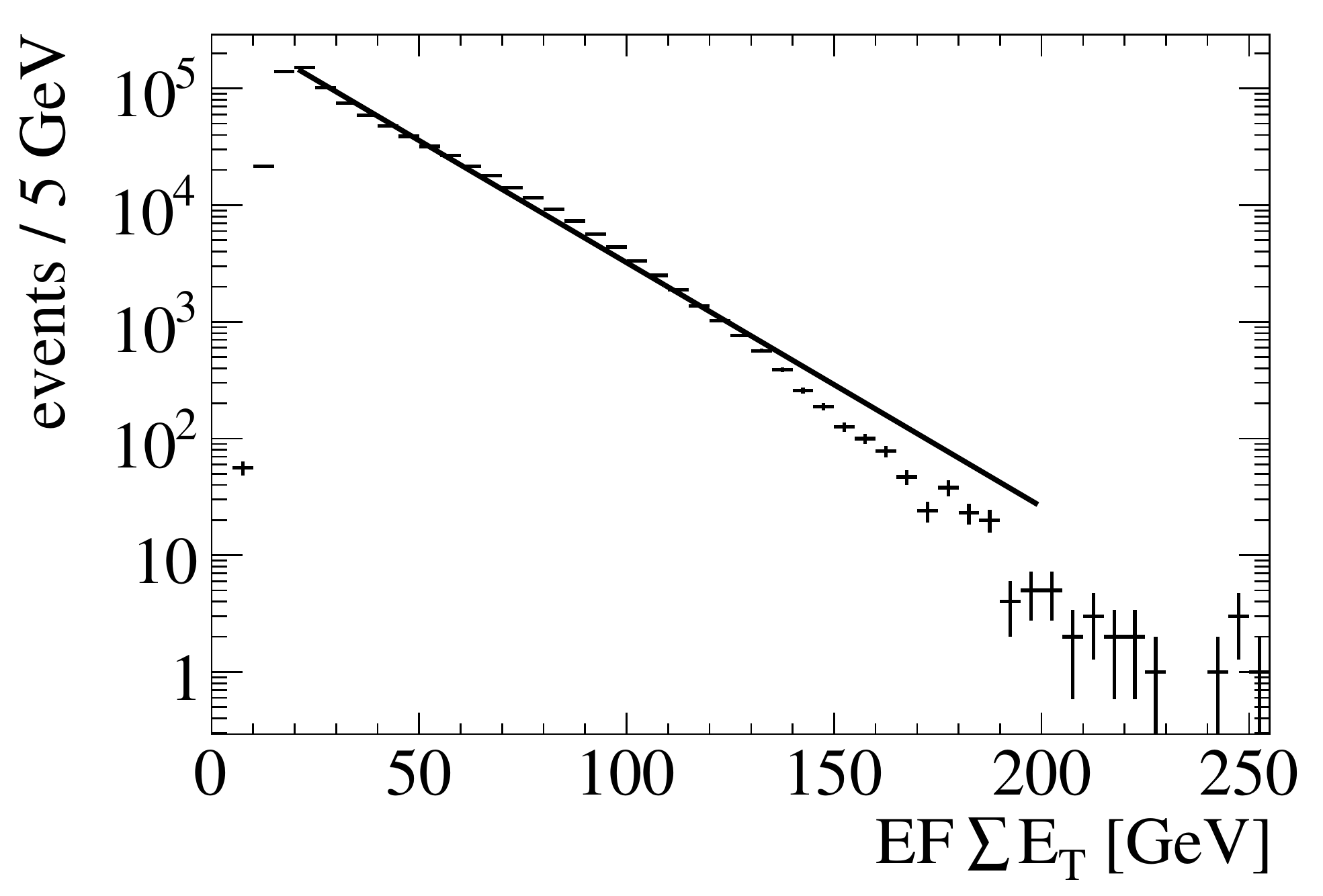}
  \incgraphics{width=\widthtwoplots}{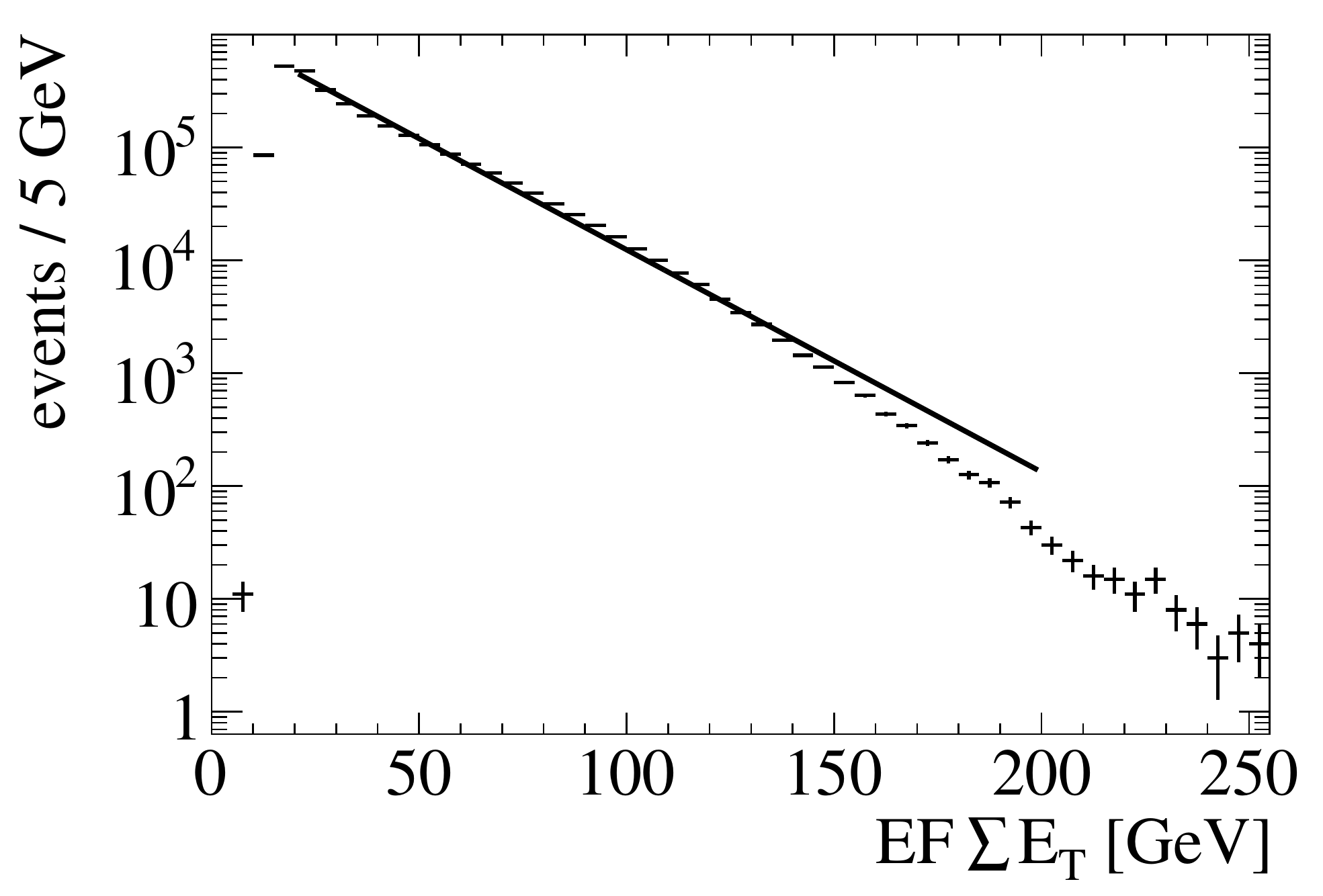}
  \caption{
    Distribution of EF \sumet in minimum-bias Monte Carlo (left) and data taken with a minimum-bias trigger (right) from an early run with low luminosity.
    The solid line in both plots is a fit with an exponentially falling function $f(x) = N \exp(-\lambda x)$,
    yielding $\lambda_\text{MC} = 0.0482(1)$ in the left and $\lambda_\text{data} = 0.0454(1)$ in the right plot.
  }
  \label{fig:results_met_model_sumet1}
\end{figure}

The baseline of the model of the \sumet shape is an exponential distribution
describing the energy measured in the calorimeter in events with exactly one interaction
and without significant contributions of real \met.
The plots in \Fig{fig:results_met_model_sumet1} motivate this assumption by showing the distribution of \sumet measured in the Event Filter
in minimum-bias Monte Carlo (left) and in data (right).
The minimum-bias Monte Carlo generated with \propername{Pythia} in the left plot is without pile-up.
Only events with exactly one interaction are simulated, and no event selection is done.
The data used in the right plot is triggered by a minimum-bias trigger \trigger{EF_mbMbts_1} %
in data from early 2010, %
where the instantaneous luminosity was so low that only rarely more than one interaction would take place at the same time,
and the contamination by pile-up events should be low.
The average instantaneous luminosity in this run was $\instlumi{2\ten{28}}$. %
The average number of interactions per bunch crossing averaged over the whole run was about $0.1$.
In addition, in the data sample only events with exactly one reconstructed primary vertex were included.
Fits with exponential functions $f(x) = N \exp(-\lambda x)$ have been done in both plots with $x = \sumet/\unit[]{GeV}$,
yielding a slope of $\lambda_\text{MC} = 0.0480(1)$ for Monte Carlo 
and $\lambda_\text{data} = 0.0454(1)$ for data.
(Only the range between $20$ and \GeV{200} has been used for the fit as indicated by the range over which the solid line is drawn.)
The fitted slopes in data and Monte Carlo thus do not agree within their uncertainties,
but the modelling of \sumet in minimum-bias events is known not to be ideal,
and besides there may still be a small contamination of the data sample by events with more than one concurrent interaction,
leading to a slightly harder spectrum and thus a slope which is a bit smaller. %

Evidently, the assumption of \sumet following an exponential distribution
is not perfect for the full range of \sumet,
but it has the important advantage that this distribution can be treated analytically
and still adequately reflects the overall behavior of the spectrum.
Two deviations can be seen:
For values below approximately \GeV{20}, the spectrum of \sumet drops very steeply.
This is not an effect of the trigger being a minimum-bias trigger,
but also found using a random trigger. %
It is a result of the one-sided cut applied to the measurement of cell energies in the Event Filter (cf. \Sec{sec:tdaq_met})
and discussed below.
For high values, in the tail of the distribution, the slope gets steeper.
As this effect can be seen both in data and in Monte Carlo which has exactly one simulated minimum-bias interaction per event,
it is unlikely to be caused by a mixture of events with different slopes due to different numbers of concurrent interactions.

\begin{figure}
  \centering
  \incgraphics{width=\widthtwoplots}{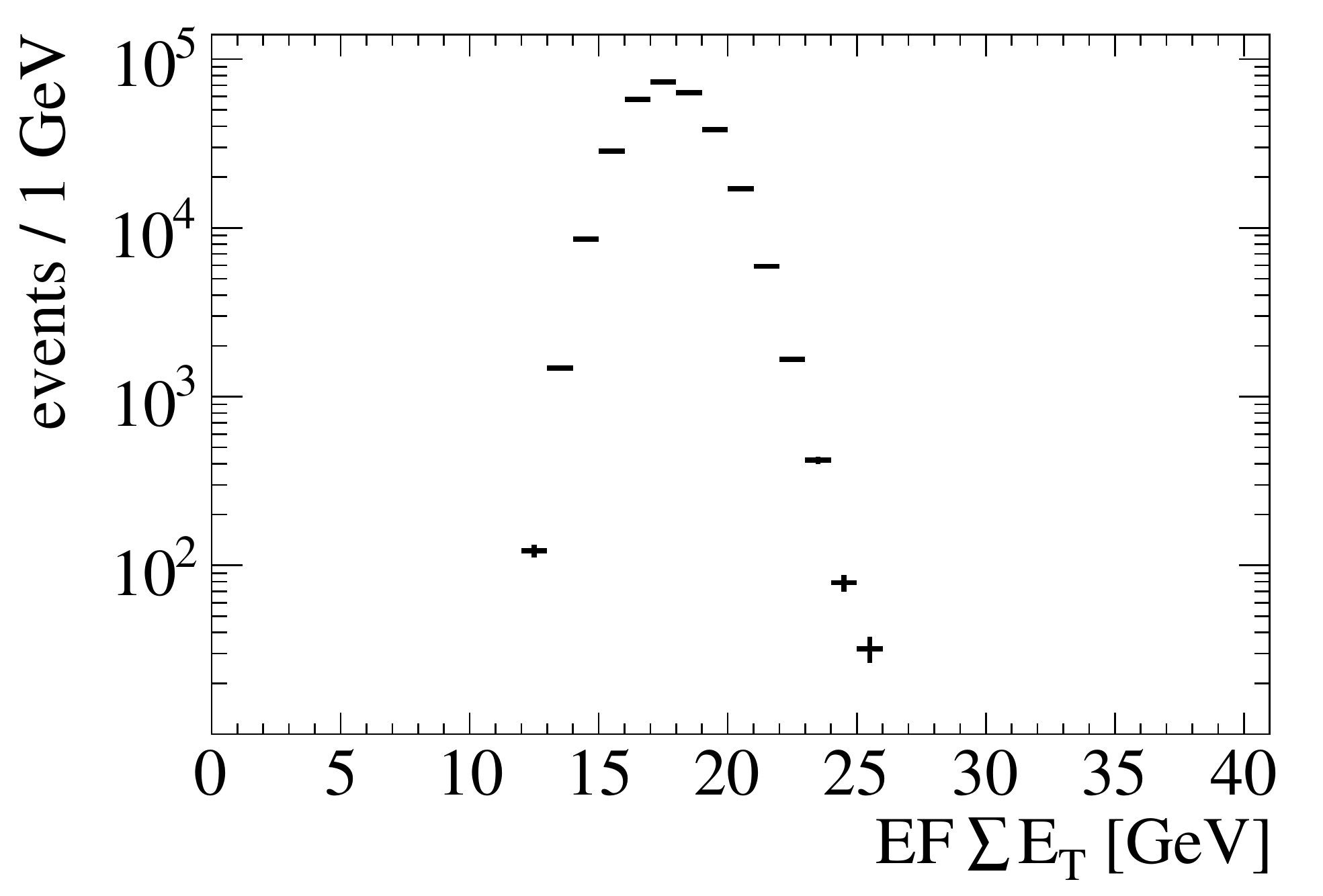}
  \incgraphics{width=\widthtwoplots}{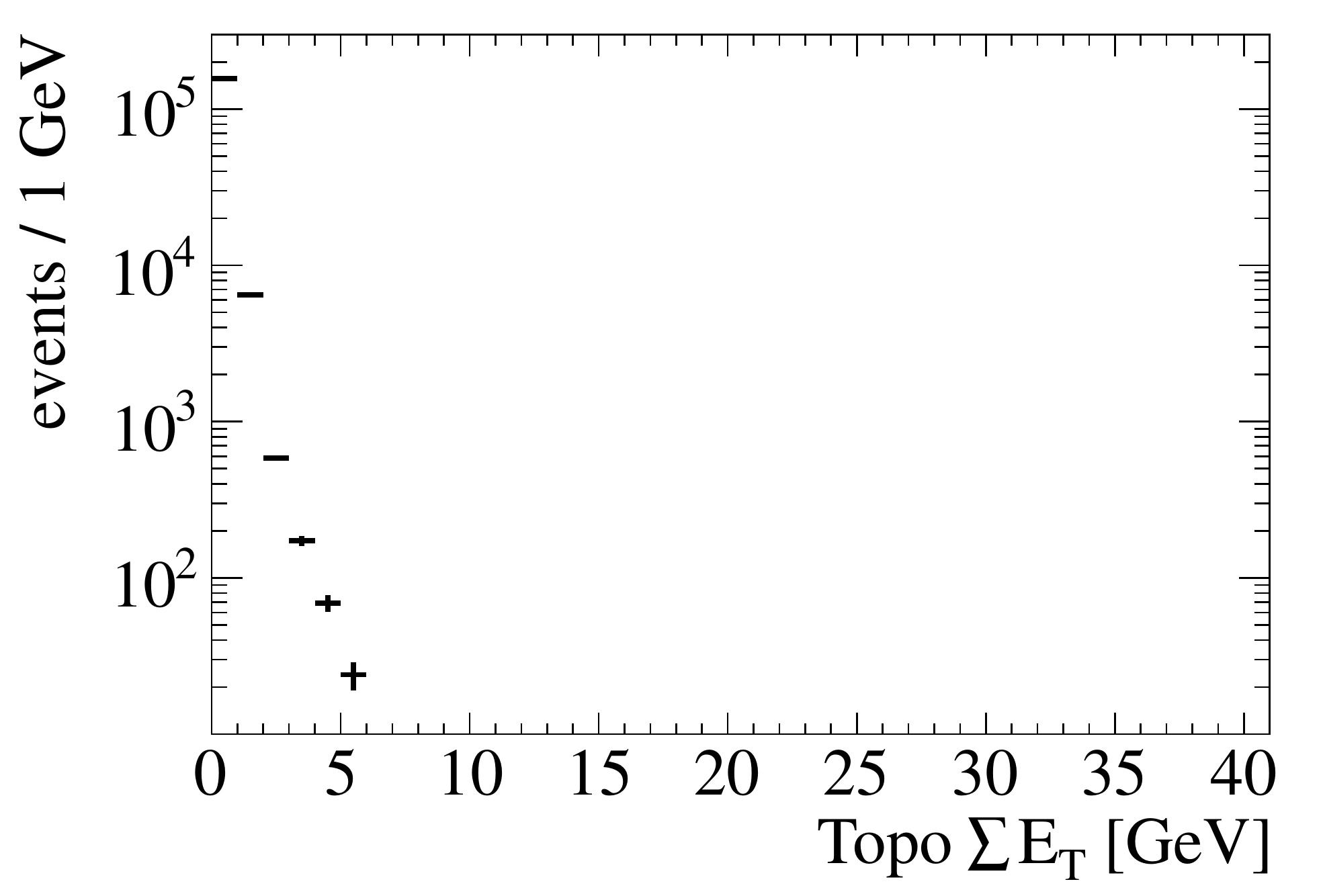}
  \caption{
    Distribution of EF \sumet (left) and offline \sumet (right, Topo calibration) in 2011 data (period 2011~D)
    taken with a random trigger on empty bunch crossings with no proton-proton collisions.
  }
  \label{fig:results_met_model_sumet0}
\end{figure}

\Fig{fig:results_met_model_sumet0} shows in the left plot the \sumet spectrum computed online in the trigger at Event Filter level
and in the right plot the offline \met obtained from the Topo calibration (cf. \Sec{sec:software_reconstruction_met}).
Both plots use data from period 2011~D, %
where the minimum-bias triggers have been replaced by random triggers. %
The events have been selected using the random trigger \trigger{EF_rd0_empty_NoAlg},
which triggers on empty bunch crossings\footnote{
  The term empty bunch crossing refers to LHC time buckets (cf. \Sec{sec:experimental_setup_LHC})
  into which no proton bunch has been injected in this particular LHC fill.
  They therefore contain no (or at least only negligibly few) protons
  so that no proton-proton interactions are expected when these ``empty bunches'' cross.
}
and the probability is thus very high that no interaction has taken place in these events.
The comparison between online and offline \sumet shows the impact of the one-sided noise cuts
which are implemented in the trigger at Event Filter level.
They lead to a bias
which appears as a shift of the mean of the Gaussian distribution\footnote{
  It turns out that using a log-normal distribution gives an even better fit.
  This can be motivated by applying the central limit theorem
  not directly to the sum of energy measurements over a large number of calorimeter cells,
  taking the measurements as independent and identically distributed random variables (i.i.d.),
  but taking the order of magnitude of the measurements as i.i.d. random variables,
  such that the logarithm of the measurements is i.i.d., \ie
  $P_0(\sumet = x|x_0,\,\sigma) = \frac{1}{x \sigma \sqrt{2 \pi}}\, \exp({-\frac{(\ln x - x_0)^2}{2\sigma^2}})$, $x>0$. %
}
of the online \sumet measurements from zero to about \GeV{18}.
This one-sided cut is unique to the trigger and cannot be seen in the offline \sumet spectrum,
which peaks at zero as would be expected.

The shift arising from this measurement bias is included in the model
by subtracting it from the \sumet values
which \revised{enter} the fits when extracting the parametrization.
The argument of the analytic form of the \sumet distribution, cf. \Eq{eq:appendix_sumet_tobeproven_repeated}, is replaced by $x-x_0$,
where $x_0 = \GeV{11.5}$ for 2010 and $x_0 = \GeV{17.6}$ for 2011 is the mean of a fit to the peak of \sumet in empty events.
That the \sumet in empty events in 2011 is higher than in 2010
can be attributed to the higher instantaneous luminosity,
leading to an overall higher activity in the detector.
This higher activity can be transferred to empty bunch crossings either by instrumental effects,
\eg due to the lag in the response to energy depositions in the calorimeter,
\ie the time the detector needs to reach its zero state again,
or by physical effects, \eg higher remaining activity from previous collisions and, in general, a higher background level in the cavern.
Note that subtracting the full shift probably tends to be an overestimation of the pedestal,
as part of the sources of \sumet contributing to the shift in empty events
will also contribute to \sumet in collision events that the model seeks to describe.
This would be interesting to study in more detail.
However, it would require to develop a method to discriminate between the different contributions of fake \met.

As said above, the benefit of using an exponential distribution to model the shape of \sumet 
is that the resulting shape of \sumet for $\mu$ concurrent interactions under these assumptions can be derived analytically.
This derivation is shown in \Sec{sec:appendix_analytic_form_sumet} and yields (cf. \Eq{eq:appendix_sumet_tobeproven})
\begin{equation}
  P\left(\sumet = x | \mu\right) = 
    \frac{ \lambda^\mu }{(\mu-1)!} \, x^{\mu-1} \exp(-\lambda x).
  \label{eq:appendix_sumet_tobeproven_repeated}
\end{equation}
The domain of this probability distribution is restricted to non-negative values.
In the end, a model for \sumet as function of the average number of interactions per bunch crossing \avgmu is required.
The distribution of \sumet for a given \avgmu is simply the Poisson-weighted  
sum of the distributions of \sumet for a fixed number of interactions per bunch crossing given by \Eq{eq:appendix_sumet_tobeproven_repeated},
\begin{equation}
  P\left(\sumet = x | \avgmu\right) = {\cal N}_0 \sum_{\mu=1}^\infty P\left(\sumet = x | \mu\right) \, \Poisson{\mu}{\avgmu}.
  \label{eq:results_met_model_full_sumet}
\end{equation}
Note that in the sum the term $\mu=0$ has been left out,
although it has non-zero weight in the Poisson distribution.
The distribution therefore needs an additional normalization factor ${\cal N}_0$.
This is done because the model does not include events without any interaction,
which will contribute only a negligible amount of energy in the calorimeters anyway.
This can be seen in the left plot in \Fig{fig:results_met_model_sumet0},
where it should be remembered that the actual sum of the energy depositions needs to be corrected for the shift from the measurement bias.
Concerning the \sumet triggers, even the lowest \sumet thresholds are of the order of a few hundred GeV at Event Filter %
and therefore lie safely above the peak stemming from the events with no interactions.
The normalization factor ${\cal N}_0$ is simply given by the Poisson weight of the $\mu=0$ bin 
for the given value of the model parameter \avgmu, %
\begin{equation}
  {\cal N}_0 = \frac{1}{\sum_{\mu=1}^\infty\Poisson{\mu}{\avgmu}} = \frac1{1-\Poisson{0}{\avgmu}}.
\end{equation}

\Eq{eq:results_met_model_full_sumet}, giving the shape of \sumet with \avgmu as parameter, completes the model.
Using the distribution of \met as described by \Eq{eq:appendix_met_tobeproven_repeated}
and the knowledge that the resolution $\sigma_T$ is dependent on \sumet,
the complete analytic form for the \met distribution can now be written down:
\begin{align}
  P(\met = x|\avgmu)
    &= \frac{1}{\sigma_T(\sumet)^2} \,x \exp\left({ -\frac{x^2}{2\sigma_T(\sumet)^2} }\right) \\
    &= \int_0^\infty \frac{1}{\sigma_T(y)^2} \,x \exp\left({ -\frac{x^2}{2\sigma_T(y)^2} }\right) \, P\left(\sumet = y | \avgmu\right) \intd y \label{eq:results_met_model_full_met_but_one} \\
    &= {\cal N}_0 \sum_{\mu=1}^\infty   \Poisson{\mu}{\avgmu}   \int_0^\infty
       \frac{ P\left(\sumet = y | \mu\right) }{ \sigma_T(y)^2 }
       \, x \exp\left({ -\frac{x^2}{2\sigma_T(y)^2} }\right) \intd y,
  \label{eq:results_met_model_full_met}
\end{align}
with $P\left(\sumet = y | \mu\right)$ given by \Eq{eq:appendix_sumet_tobeproven_repeated}.

\subsection{Trigger Rates}

For any modelled trigger quantity $\Omega$ the average probability $P_\text{ev}$ of an event to issue a trigger
is the probability of the trigger measurement to exceed the given trigger threshold $\Omega_i$,
\begin{equation}
  P_\text{ev}(\Omega_i) = \int_{\Omega_i}^\infty P(\Omega = x) \intd x,
\end{equation}
not writing dependencies on additional parameters explicitly.
Note that this only holds for one trigger level and does not take correlations of the measurements at different levels into account.
To compute the absolute trigger rate $R(\Omega_i)$, this probability needs to be multiplied with the
total cross section $\sigma_\text{tot}$ and the instantaneous luminosity $\Linst$,
\begin{equation}
  R(\Omega_i) = P_\text{ev}(\Omega_i) \, \sigma_\text{tot} \, \Linst.
  \label{eq:results_met_model_absolute_rate}
\end{equation}
This normalization is difficult to obtain, but when computing relative rates, assuming without loss of generality $\Omega_i < \Omega_j$,
\begin{equation}
  \frac{R(\Omega_i)}{R(\Omega_j)} = 
    \frac{ P_\text{ev}(\Omega_i) }{ P_\text{ev}(\Omega_j) } =
    \frac{ \int_{\Omega_i}^\infty P(\Omega = x) \intd x }{ \int_{\Omega_j}^\infty P(\Omega = x) \intd x } =
    1 + \frac{ \int_{\Omega_i}^{\Omega_j} P(\Omega = x) \intd x }{ \int_{\Omega_j}^\infty P(\Omega = x) \intd x },
  \label{eq:results_met_model_relative_rate}
\end{equation}
it cancels out.
In this context it should be stressed that the low tail of the trigger measurements, below the lowest trigger threshold,
neither affects the relative nor the absolute trigger rates,
and thus discrepancies between the model and the real spectrum in this range
do not deteriorate the quality of the model predictions.

\subsection{Summary}

The full model for the prediction of the \met trigger rates
for a given value of the average number of interactions per bunch crossing \avgmu
is thus comprised of three steps:
\begin{enumerate}
  \item Compute the \sumet spectrum.
    This step uses \avgmu and the slope of the exponential distribution of the energy
    found for events with one proton-proton interaction and the \sumet offset as input,
    and employs \Eq{eq:results_met_model_full_sumet} to obtain the distribution of \sumet.
    The assumptions used in this step are that the \sumet for one interaction per bunch crossing, $\mu=1$,
    can be described by an exponential distribution,
    and that \sumet from $\mu$ concurrent interactions
    can be described as an overlay of independent \sumet contributions from $\mu$ individual interactions%
    \footnote{
      That this part of the computation can be done in a first independent step
      is somewhat concealed in \Eq{eq:results_met_model_full_met},
      but clear from the fact that the model \sumet shape does not depend on \met.
    }.
  \item Compute the \met spectrum.
    This step uses the parametrization of the detector resolution
    as a function of \sumet
    and the \sumet distribution obtained in the previous step as input.
    It employs \Eq{eq:results_met_model_full_met_but_one} to obtain the distribution of \met.
    The assumptions used in this step are that the detector resolution is independent of \avgmu, %
    which will be shown for \ATLAS in \Sec{sec:results_met_model_met_resolution}, %
    and that all \met is fake \met coming from the limited detector resolution.
    Note that the model assumptions also imply that the resolution parameter for \mex, \mey and \met is the same.
  \item Compute the trigger rates.
    This step uses the \met and the \sumet spectrum from the previous steps as inputs,
    together with a set of trigger thresholds as parameters.
    The trigger rate is then given by the integral over the spectrum above the given threshold,
    as in \Eqs{eq:results_met_model_absolute_rate} and \eqref{eq:results_met_model_relative_rate}.
    Here, only relative trigger rates are computed.
\end{enumerate}
The number of input parameters of the model is, following the guiding principle of simplicity, very small:
There are the slope $\lambda$ of \sumet for events with a single interaction,
the offset of \sumet estimated from the \sumet distribution in empty bunch crossings
and two parameters describing the resolution of \met as function of \sumet.
The target average number of concurrent interactions \avgmu is an external parameter.
In the following, the extraction of the input parameters from data is described.
Afterwards, the model is used to produce trigger rate predictions.

\subsection{Implementation}
The actual implementation of the model closely follows the summary given above.
Some care needs to be taken when choosing the cut-off for the integrals and summations which have upper limits sent to infinity.
The function evaluating \Eq{eq:results_met_model_full_sumet} for given \sumet includes a loop over $\mu$.
This loop runs from $1$ to $\min(\{20, \avgmu+5\sqrt{\avgmu}\})$, %
where the last term comes from an approximation of the Poisson distribution by a Gaussian,
which is cut off at 5 standard deviations to the positive side of its mean.
The \sumet distribution is obtained by filling a histogram,
calling this function once for every bin.
The \met distribution is then obtained by looping over the previously filled histogram holding the \sumet distribution,
and for each bin computing the corresponding resolution of \met.
A histogram with the \met distribution is filled by summing up the contributions from all \sumet bins with the appropriate weight,
which is given by the content of the respective \sumet bin.
It is thus important to choose the range of the \sumet histogram suitably
because in particular the tail of the \met distribution will receive significant contributions from relatively high \sumet values.
A fixed value of \TeV{2} has been used here.
This could be improved by dynamically adjusting the upper bounds automatically during the computation as needed.

\section{Determination of Input Parameters}
One of the two important inputs to the model which need to be measured on data
is the slope of the exponential distribution of the sum of the transverse energy
in events which have one proton-proton interaction.
As the actual number of interactions which have taken place %
in a data event is unknown,
this number has to be estimated from some suitable quantity which can be reconstructed.
The average number of interactions per bunch-crossing \avgmu
is the expectation value of a Poisson distribution and as such only meaningful for a sample of events.
Currently, the best per-event estimate for the
number of concurrent proton-proton interactions is the number of reconstructed primary vertices.
The reconstruction of primary vertices has an intrinsic efficiency $\eff{vx,reco}\leq1$,
so that the number of reconstructed primary vertices $n_\text{vx,reco}$ is expected to be smaller than \avgmu.
However,
assuming that the reconstruction efficiency is independent of the number of primary vertices,
it is easy to show %
that the distribution of $n_\text{vx,reco}$ for a sample of many events also follows a Poisson distribution
and that its expectation value $\langle n_\text{vx,reco} \rangle$ is proportional to \avgmu,
with a coefficient given by $\eff{vx,reco}$.
The quality of this estimation will be discussed in the first part of this section.

\subsection{Number of Concurrent Collisions}
The primary vertices are reconstructed by finding intersections of tracks reconstructed in the Inner Detector
and applying a suitable set of cuts (cf. \Sec{sec:software_vertex_reconstruction}).
In physics analyses, often a cut is applied on the number of tracks associated to a primary vertex.
For instance, in the analysis presented in \Sec{sec:analysis_susysearch} at least 5 tracks are required
to select only events which have a reliably reconstructed primary vertex.
The algorithm which creates the collection of reconstructed primary vertices is written 
in such a way that one \index{dummy vertex} with no associated tracks is always included in addition to the reconstructed primary vertices.
This has to be taken into account when no cut on the number of associated tracks is made and can be seen in the following figure.

\begin{figure}
  \centering
  \incgraphics{width=\widthtwoplots}{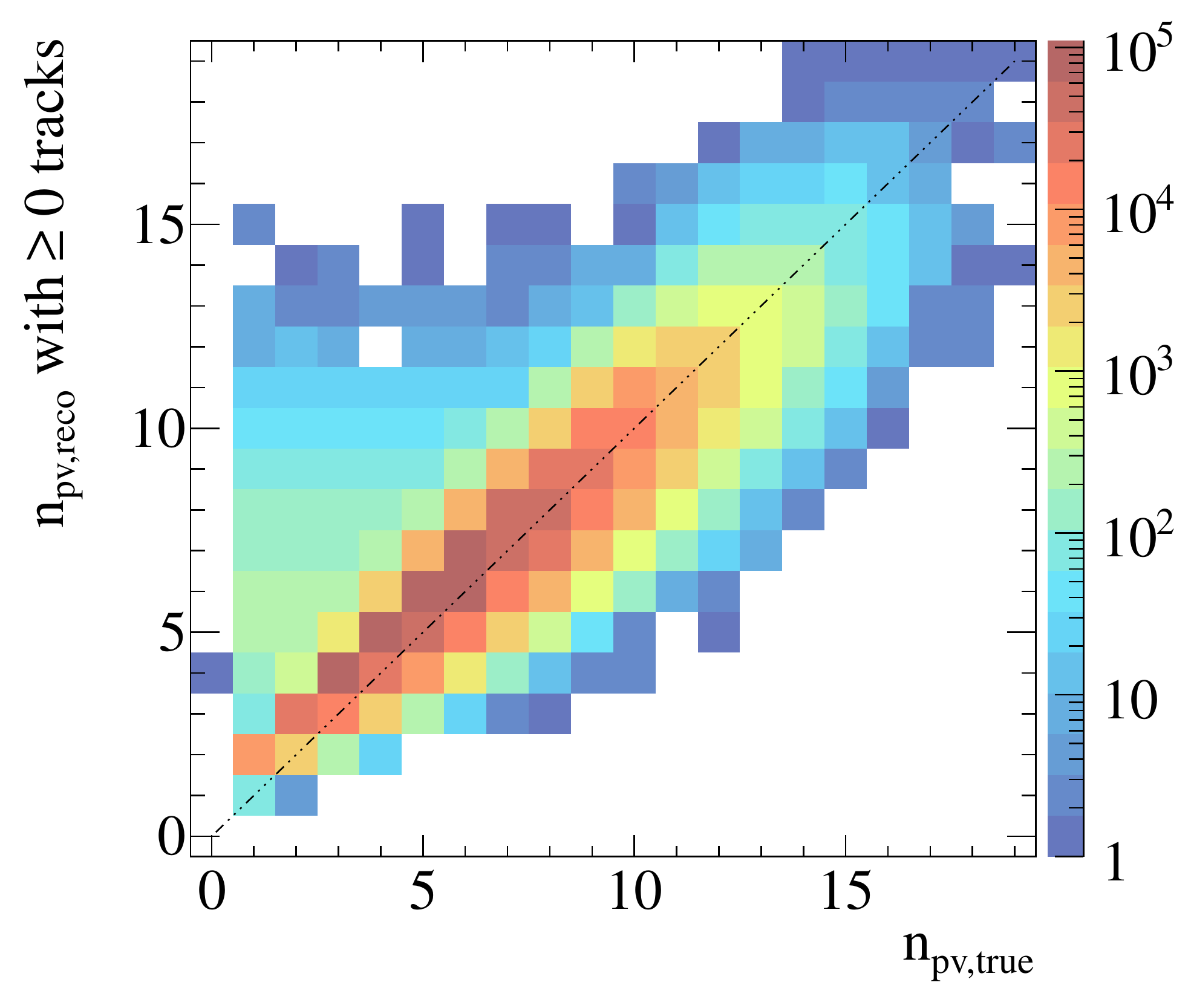}
  \incgraphics{width=\widthtwoplots}{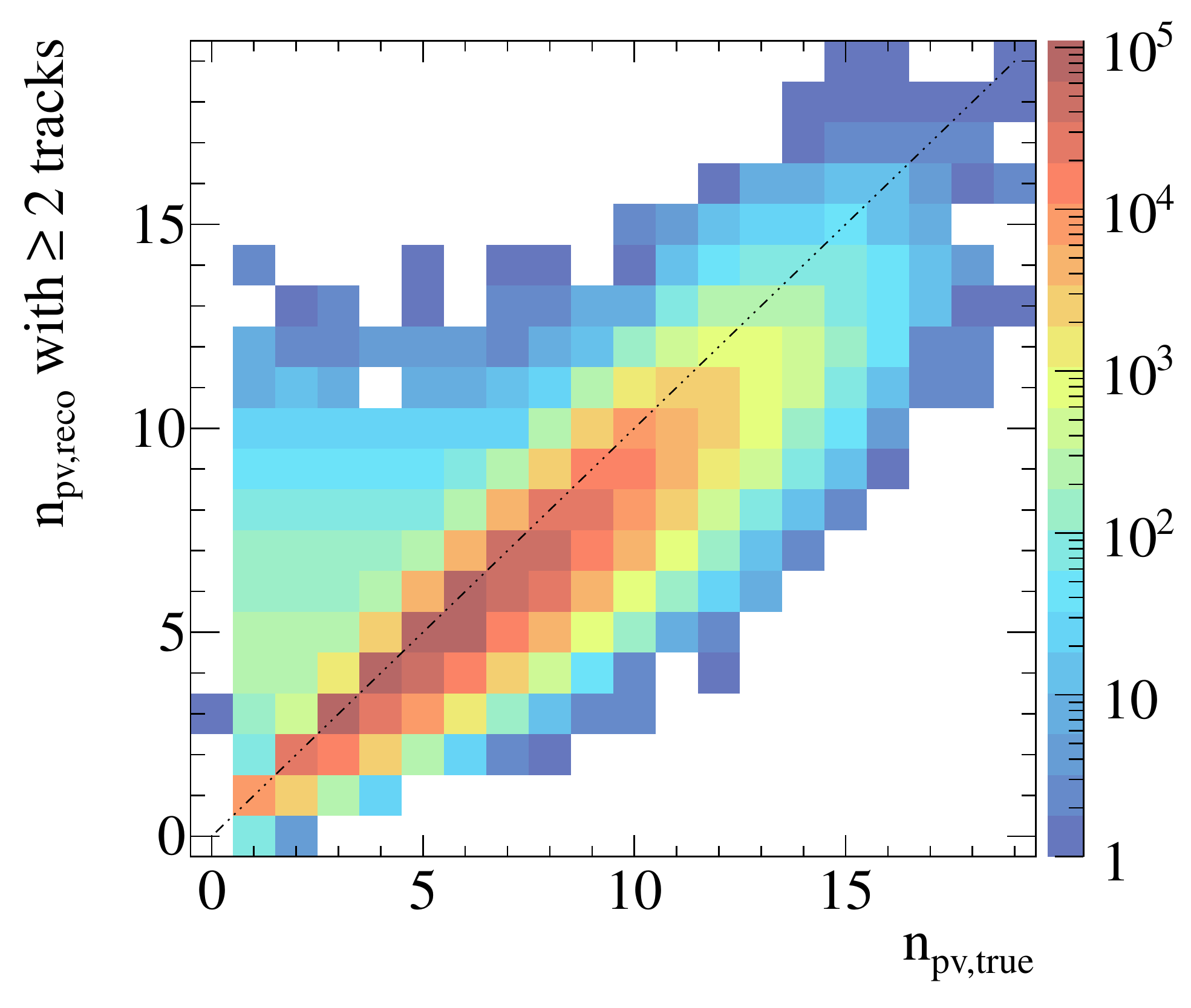}
  \incgraphics{width=\widthtwoplots}{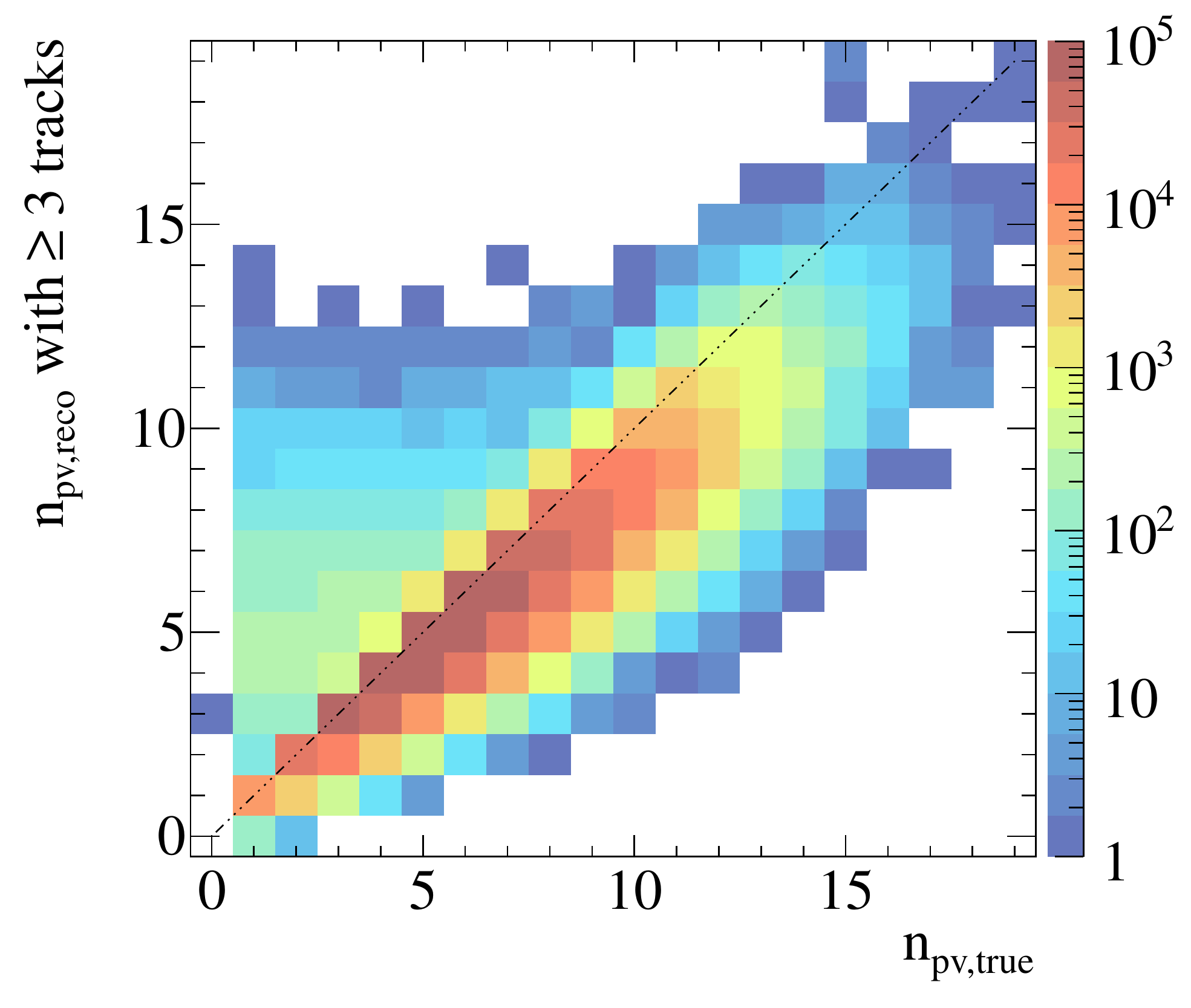}
  \incgraphics{width=\widthtwoplots}{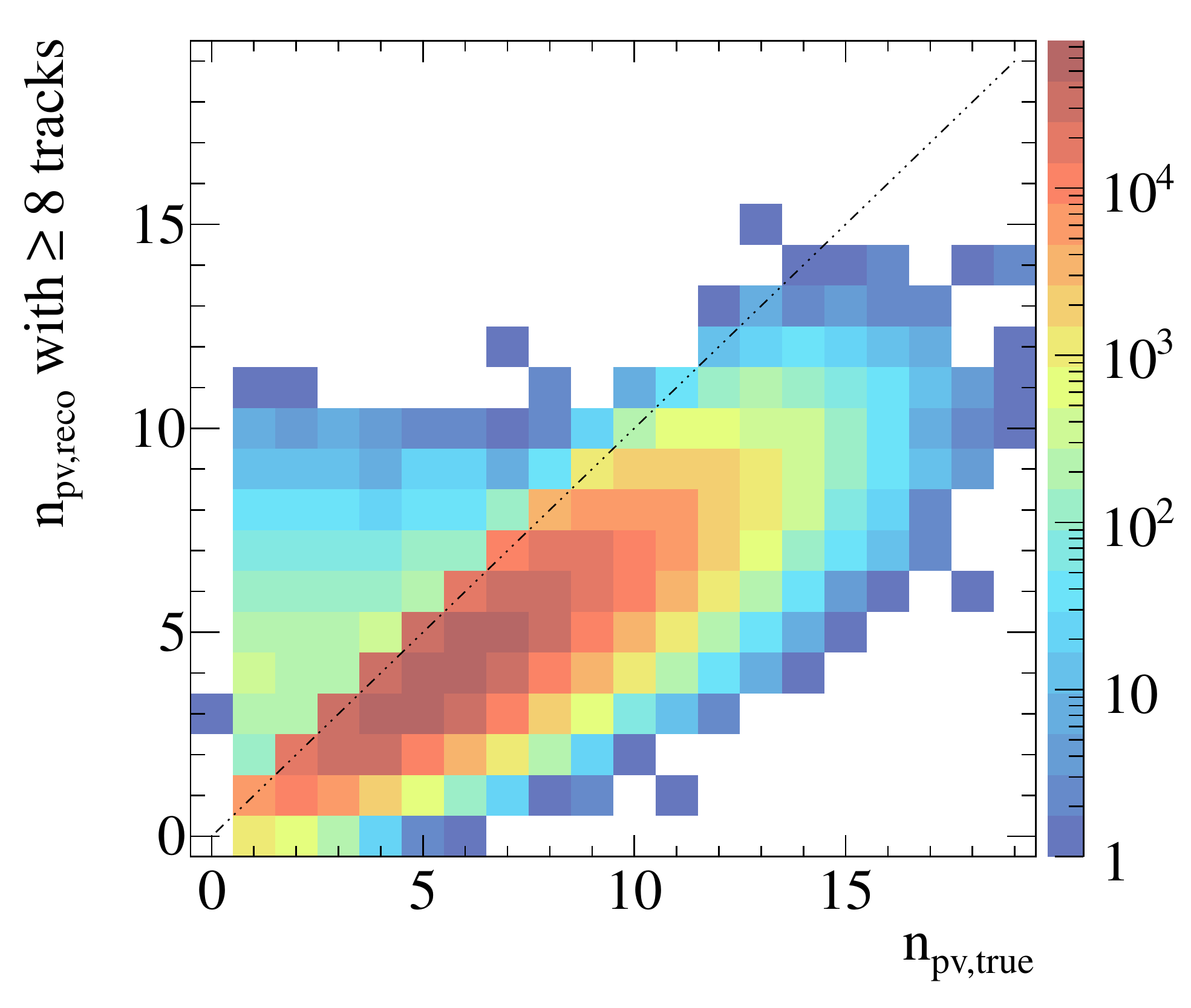}
  \caption{
    Two-dimensional correlation plots of the number of reconstructed primary vertices (\nreco, vertical axis),
    counting only vertices with a given minimum number of associated tracks,
    and the true number of interactions ($n_\text{pv,true}$, horizontal axis) for Monte Carlo minimum-bias events.
    Four criteria have been used to count the number of reconstructed primary vertices,
    requiring at least $0$, $2$, $3$ or $8$ associated tracks.
    The resulting two plots in the top row are identical,
    the left one being shifted up by $1$.
  }
  \label{fig:results_met_model_correlation_PVs_MC}
\end{figure}

The plots in \Fig{fig:results_met_model_correlation_PVs_MC} shows the correlation of the
number of reconstructed primary vertices and the true number of interactions for \tagname{MC09} Monte Carlo samples
which contain minimum-bias events generated with \propername{Pythia}
and have an average number of interactions per bunch crossing  of $\avgmu = 5+1$\footnote{
  Writing the average number of interactions per bunch crossing as $\avgmu = n+1$ shall indicate
  that every event contains one interaction according to the physics process which defines the Monte Carlo sample,
  and on top several minimum-bias interactions from pile-up.
  The number of additional interactions is Poisson distributed with an expectation value which here is 5.
  Note that the resulting total number of interactions per bunch crossing
  is different from a Poisson distribution with an expectation value of 6.
}.
The four plots differ in the way the number of reconstructed primary vertices is counted:
Only reconstructed primary vertices which have at least 0~(upper left), 2~(upper right), 3~(lower left) or 8~(lower right) associated tracks ($n_\text{track})$ are counted.
Comparing the two plots in the upper row makes obvious
that requiring two tracks only removes the dummy vertex with no associated tracks.
Apart from that, the distribution stays the same.
A large fraction (\percent{54} for $n_\text{track} \geq 2$) %
of events lies on the diagonal marked by the black line.
Here, the reconstructed number of primary vertices corresponds to the true number of interactions.
For most of the rest of the events, the number of primary vertices is smaller than the true number of interactions, mostly by one,
which comes from the inefficiency of the reconstruction algorithm for primary vertices.
In constrast to this, an overestimation of the true number of interactions is very rare
because this could only be due to the accidental crossing of two or more tracks.
For higher numbers of true primary vertices, the distribution around the diagonal gets broader as can be expected.
Comparing the plots with $n_\text{track} \geq 2$ and $n_\text{track} \geq 3$ shows that already here
requiring more tracks starts to make the agreement between the reconstructed number of primary vertices and the true number of interactions worse.
For $n_\text{track} \geq 3$, the number of events on the diagonal is about \percent{49}.
The number of interactions is underestimated from the number of reconstructed primary vertices.
Going to $n_\text{track}\geq 8$ makes this even worse, of course.
The conclusion is that in the following, in order to obtain an estimate of the number of interactions per event,
either no selection based on the number of tracks is done,
keeping in mind to subtract one from the number of reconstructed primary vertices to account for the dummy vertex,
or, equivalently as shown above, only reconstructed primary vertices with at least two associated tracks are counted.

\begin{figure}
  \centering
  \incgraphics{width=\widthsingleplot}{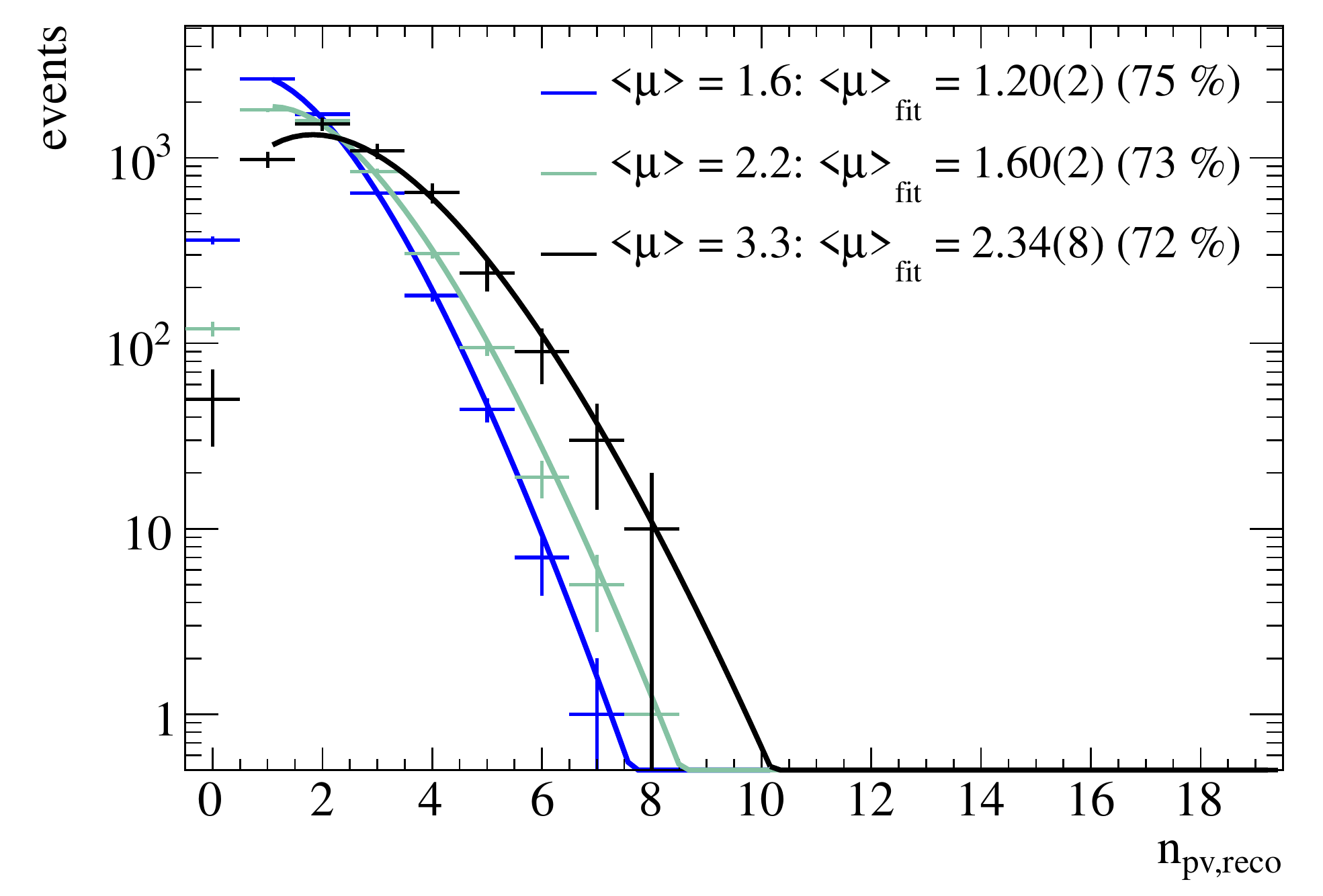} %
  \caption{
    Distribution of the number of reconstructed primary vertices \nreco in data from 2010 %
    for three small event samples with a narrow distribution of \avgmu.
    The data points are fitted with a Poisson distribution (solid lines) with free rate parameter $\mu$
    and the fit result $\avgmu_\text{fit}$ is given in the legend
    together with the ratio of $\avgmu_\text{fit}$ over the nominal \avgmu expressed as a percentage value in brackets.
    The black points have been scaled up by a factor 10. %
    (A corresponding plot for 2011 is shown in \Fig{fig:results_met_model_data_reco_PV_fit_2011} in the Appendix.) %
  }
  \label{fig:results_met_model_data_reco_PV_fit_2010}
\end{figure}

From the plots on Monte Carlo in \Fig{fig:results_met_model_correlation_PVs_MC},
it can be seen that the number of reconstructed vertices tends to underestimate the number of interactions.
It is therefore natural to expect this also to be the case for data,
and indeed this effect can also be made visible on data as will be discussed now.
The number of interactions per event in data is unknown
and therefore cannot be compared directly to the number of reconstructed vertices,
but for ensembles of events the average number of interactions can be estimated from the instantaneous luminosity,
which is available per luminosity block.
Again, this number is taken from the \COOL database,
as was done to produce the trigger rate plots.
For events from a small set of luminosity blocks over which the average number of interactions \avgmu does not vary too much,
the distribution of the number of reconstructed primary vertices is plotted in \Fig{fig:results_met_model_data_reco_PV_fit_2010}.
Three different sets of luminosity blocks from run 167776 in period 2010~I
have been used, corresponding to an \avgmu of approximately $1.6$ (luminosity blocks 541--546), $2.2$ (320--326) and $3.3$ (120--125), respectively.
The relative spread of the nominal value of \avgmu over the selected luminosity blocks is smaller than half a percent.
The events have been selected using the minimum-bias trigger \trigger{EF_mbMbts_1_eff},
and the SUSY \ac{GRL} has been applied.
A Poisson distribution has been fitted to the three distributions,
and the fit results for the expectation value of the Poisson distribution, $\avgmu_\text{fit}$, are listed in the legend.
For all three sets, $\avgmu_\text{fit}$ is about one quarter smaller than the expected average number of interactions \avgmu.
The number of interactions is thus again underestimated by the reconstructed number of primary vertices,
due to the inefficiency in the vertex reconstruction.
The underestimation tends to become larger when going to higher numbers of concurrent interactions.
This can be explained as a result of \index{vertex merging},
which occurs when in busier events with a higher number of tracks
the probability increases that two primary vertices are merged and mistaken to be the same vertex.
\Fig{fig:results_met_model_data_reco_PV_fit_2010} suggests that correction factors can be found
to improve the estimator of the number of concurrent interactions.
However, this only allows to correct the average number of interactions per bunch crossing \avgmu for event samples,
but does not help with the selection of events with a fixed number value of $\mu$,
for which a reliable event-by-event estimate of $\mu$ rather than \avgmu is needed.

\subsection{\texorpdfstring{\sumet Slope} {SumET Slope}}
Coming back now to the original question, the extraction of the slope of \sumet for events with one proton-proton interaction,
the conclusion from the above is that it is not easy
to reliably determine the number of interactions that have taken place in a given event.
When selecting a sample of events with exactly one interaction,
a major problem will be the contamination of this sample
by events in which actually more than one interaction has taken place,
but one or more of the primary vertices have been missed by the reconstruction algorithm.
This %
will lead to an underestimation of the slope
because events with a higher number of interactions have a harder \sumet spectrum
and thus enhance the tail of the distribution, leading to a seemingly smaller slope. %

\begin{figure}
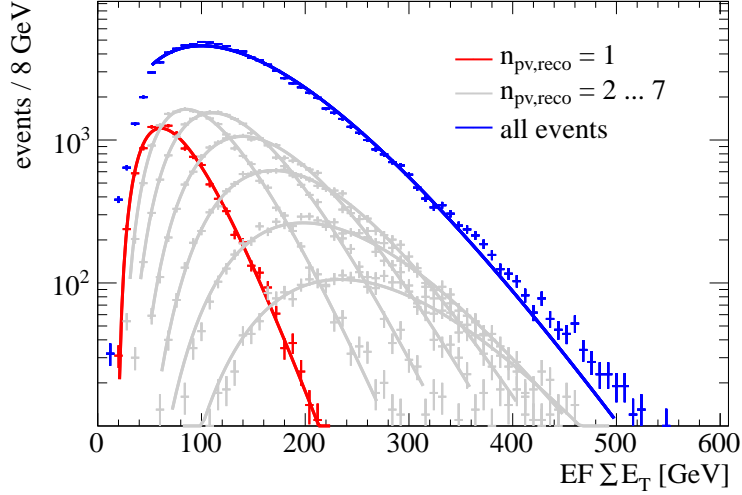

  \centering
  \incgraphics{width=\widthsingleplot}{{{plot_MET_model_11b_2011_Rd0_mu5.0_Arbeit}}}
  \caption{
    Comparison of fits to the Event Filter \sumet spectrum in events from 2011 data with $\avgmu \approx 5.0$.
    Shown are the distributions from selecting subsamples 
    by the number of reconstructed vertices \nreco being $1$ (red) and 2 through 7 (gray)
    as well as the full sample (blue).
    The data points represent the distributions in data,
    the curves are fits which are explained in the text.
    The intervals over which the lines are drawn indicate the fit ranges.
  }
  \label{fig:results_met_model_sumet_slope_fit_to_mu5}
\end{figure}
The solution to this is not to try and fit the \sumet distribution for exactly one interaction,
but instead to use the full model for \sumet, as given in \Eq{eq:results_met_model_full_sumet},
and to fit \sumet in a sample of events in which the average number of interactions follows, as usual, a Poisson distribution.
The data for this fit has been taken with the random trigger \trigger{EF_rd0_filled_NoAlg} in data from period 2011~D.
For each of the runs used to produce the plot,
only data from luminosity blocks was used which have an average number of interactions per bunch crossing \avgmu around $5.0$
(selecting events with $4.9 < \avgmu < 5.1$),
where \avgmu is taken from the \COOL database. %
This selection is needed to ensure that the distribution of \avgmu in the event sample closely follows a Poisson distribution as is assumed in the fit.

The plot in \Fig{fig:results_met_model_sumet_slope_fit_to_mu5} shows a comparison of several distributions of Event Filter \sumet on different subsamples.
The data points in red and gray correspond to subsamples in which the number of reconstructed vertices \nreco is $1$ or $2$ up to $7$, respectively.
The red points with $\nreco=1$ make obvious that selecting one reconstructed vertex
no longer gives a pure sample of events with one concurrent interaction,
as was the case in \Fig{fig:results_met_model_sumet1},
because it is no longer an exponential distribution,
which would follow a straight line in logarithmic scaling.
The red and gray data points are fitted using the simple form of \sumet in \Eq{eq:appendix_sumet_tobeproven_repeated},
where the slope $\lambda$ and the parameter $\mu$ are left free in the fit.
This allows the fitted curves to closely follow the data points,
but the good agreement is misleading
because the fit yields values for $\lambda$ which vary strongly with \nreco.
This contradicts the model assumption of \sumet being the result of an overlay of $\mu$ independent spectra
with the same fixed slope $\lambda$ for all values of $\mu$.
This is, of course, again a result of the contamination of the samples by events with $\mu$ larger than the selected \nreco,
which is also reflected in the fact
that the value of $\avgmu_\text{fit}$ obtained from the fit
is larger than \nreco for small \nreco, but less so for high \nreco.

All this confirms that it is not possible to extract the slope $\lambda$ from a subsample with given \nreco.
In constrast, the blue data points in \Fig{fig:results_met_model_sumet_slope_fit_to_mu5}
show the distribution of \sumet in the full sample,
with no selection on \nreco and fitted with the full \sumet model from \Eq{eq:results_met_model_full_sumet}.
In the fit, both \avgmu and the slope $\lambda$ of \sumet for $\mu=1$ are left free.
The fit yields $\avgmu_\text{fit} = 4.96(5)$ and a slope of $\lambda_\text{fit} = 0.0410(4)$, %
which is compatible with the selected \avgmu
and reasonably close to the slope obtained above for data from the fit in \Fig{fig:results_met_model_sumet1}.
The same procedure has been carried out on data from 2010 taken with a minimum-bias trigger.
The result is shown in the Appendix in \Fig{fig:results_met_model_sumet_slope_fit_to_MB_2010} %
and is in very good agreement with the value from 2011 data.
This approach is therefore found suitable to determine the slope used as input to the model for the shape and rate predictions.

\subsection{Missing Energy Resolution}
\label{sec:results_met_model_met_resolution}

\begin{figure}
  \centering
  \incgraphics{width=\widthtwoplots}{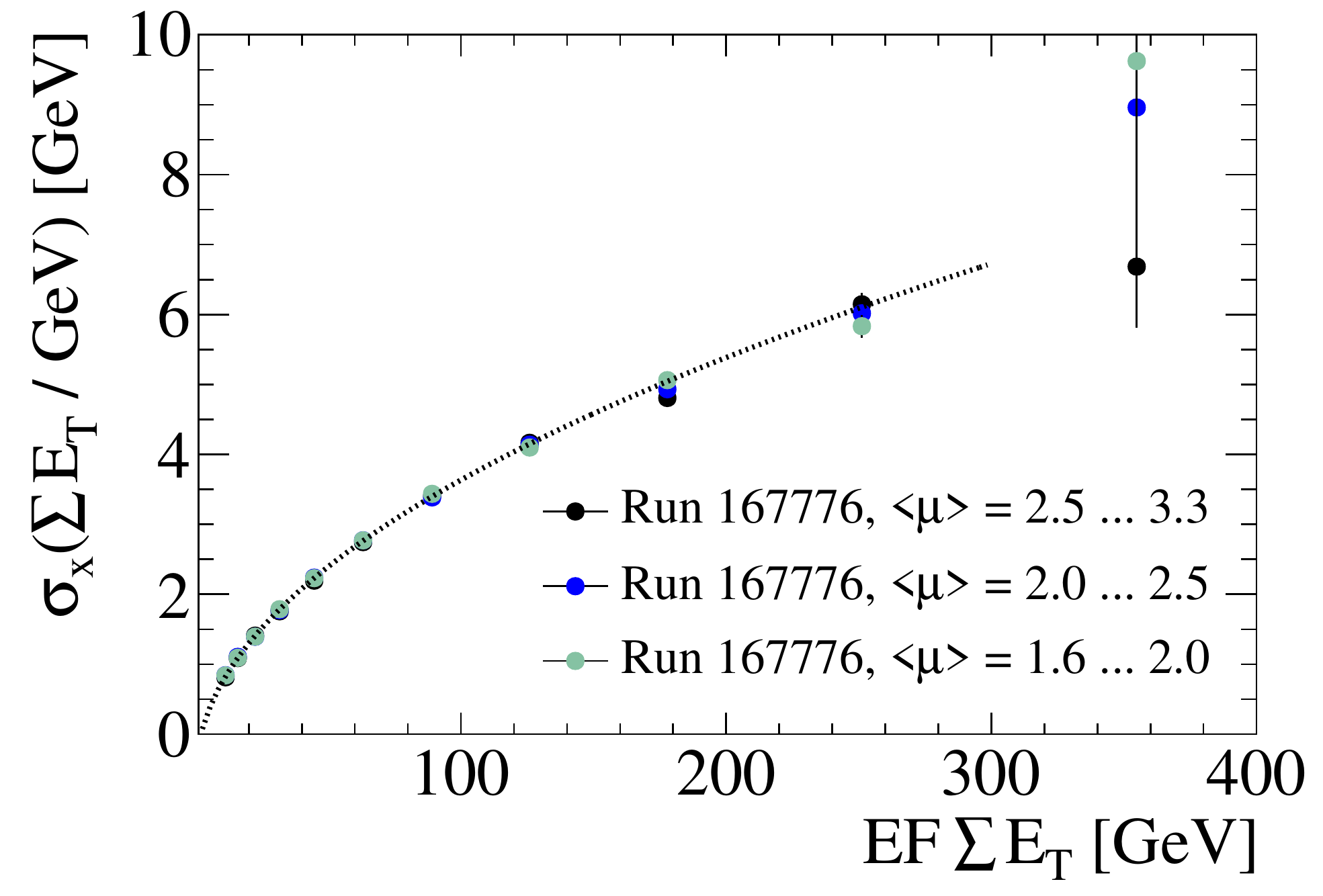}
  \incgraphics{width=\widthtwoplots}{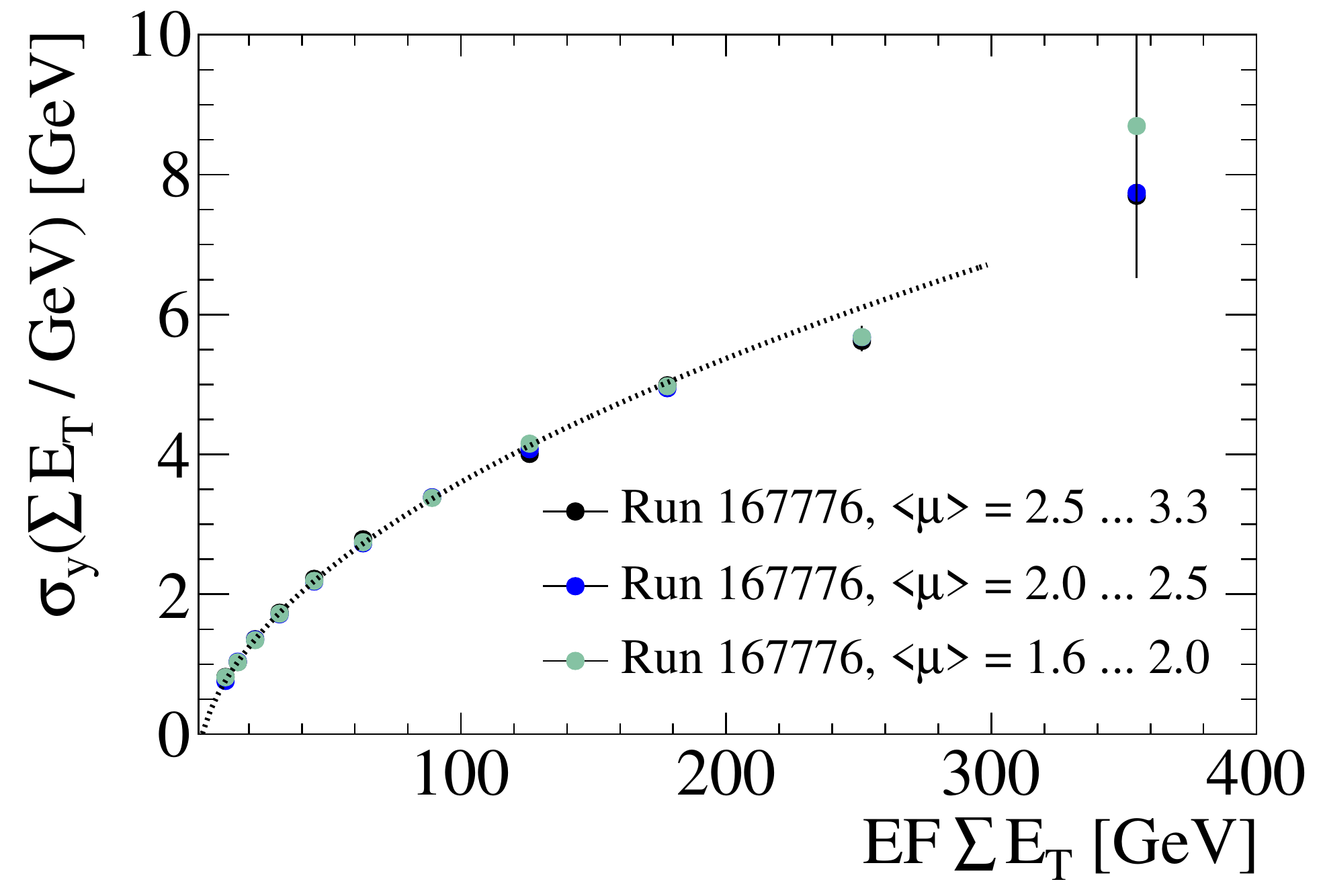}
  \caption{
    Resolution of \ac{EF} \mex (left) and \mey (right) plotted as function of EF \sumet for run 167776 from period 2010~I.
    Three event samples with different average numbers of concurrent interactions \avgmu are compared.
    The black dotted line is a fit with a square root function.
  }
  \label{fig:results_met_model_resolution_MET_167776}
\end{figure}

The other important input to the model is the resolution of the \met measurement in the detector.
\Fig{fig:results_met_model_resolution_MET_167776} shows the resolution of \mex and \mey measured by the trigger at Event Filter level
as function of Event Filter \sumet for run 167776 from period 2010~I.
The resolution is computed by doing Gaussian fits of the EF \mex and \mey distribution
for small slices of \sumet as shown in \Fig{fig:results_met_model_mexy}.
The fit result for $\sigma$
is used as resolution parameter.
Dots in three different colors mark three event samples,
which differ in the average numbers of concurrent interactions \avgmu,
as shown in the legend of the plot.
The error bars on the dots are the uncertainties on the resolution parameter obtained from the Gaussian fit.
The comparison between the three samples shows that the resolution in very good approximation is independent from \avgmu.
It can therefore be assumed to be a detector parameter that is independent of the pile-up level.

To derive a parametrization which can later be used as input to the model,
a square root function $f(x) = a+b\sqrt{x}$ is fitted to the combined set of dots from all three event samples.
The functional form of the fit is inspired by the general parametrization
of the energy resolution of the calorimeter in \Eq{eq:experimental_setup_general_calorimeter_resolution_parametrization}.
The parameters obtained from the fit
marked by the black dotted line in the figure
are given in \Tab{tab:results_met_model_resolution_parametrization}.
Concerning the seemingly large value of the constant term $a$,
it should be noted that the shape of the curve is dominated by the square root term with prefactor $b$,
as is obvious from the plot.
The fitted function gives negative values for the resolution for small values of $\sumet$ below about \GeV{2}.
These negative values for the resolution are unphysical, of course,
but will not affect the model as such small \sumet values never occur in the model
due to the inclusion of the pedestal of \sumet for empty events as described above.
What is furthermore apparent from the values for $b$ is that $\sigma_y$ is slightly larger than $\sigma_x$ for large \sumet.
This is a result of the shift of the mean of the $y$-component of the missing energy.
In the model, the mean of the fit parameters obtained for $\sigma_x$ and $\sigma_y$ is used
for the parametrization of the \met resolution because both \mex and \mey are assumed to have the same resolution.

\begin{table}
  \centering
  \begin{tabular}{lrr}
    \toprule
    Resolution [GeV] & \multicolumn{1}{c}{$a$} & \multicolumn{1}{c}{$b$} \\
    \midrule 
    $\sigma_x(\sumet / \unit[]{GeV})$ & $-0.584 \pm 0.008$ & $0.421 \pm 0.002$ \\
    $\sigma_y(\sumet / \unit[]{GeV})$ & $-0.658 \pm 0.008$ & $0.425 \pm 0.002$ \\ %
    \bottomrule
  \end{tabular}
  \caption{%
    Parametrization of the resolution of \mex and \mey, denoted $\sigma_x$ and $\sigma_y$, as function of \sumet.
    The parameter values are obtained from the square root fit $f(x) = a+b\sqrt{x}$ shown as the black dotted line in the resolution plot
    in \Fig{fig:results_met_model_resolution_MET_167776} as function of $x = \sumet / \unit[]{GeV}$.
  }
  \label{tab:results_met_model_resolution_parametrization}
\end{table}

\begin{figure}
  \centering
  \incgraphics{width=\widthsingleplot}{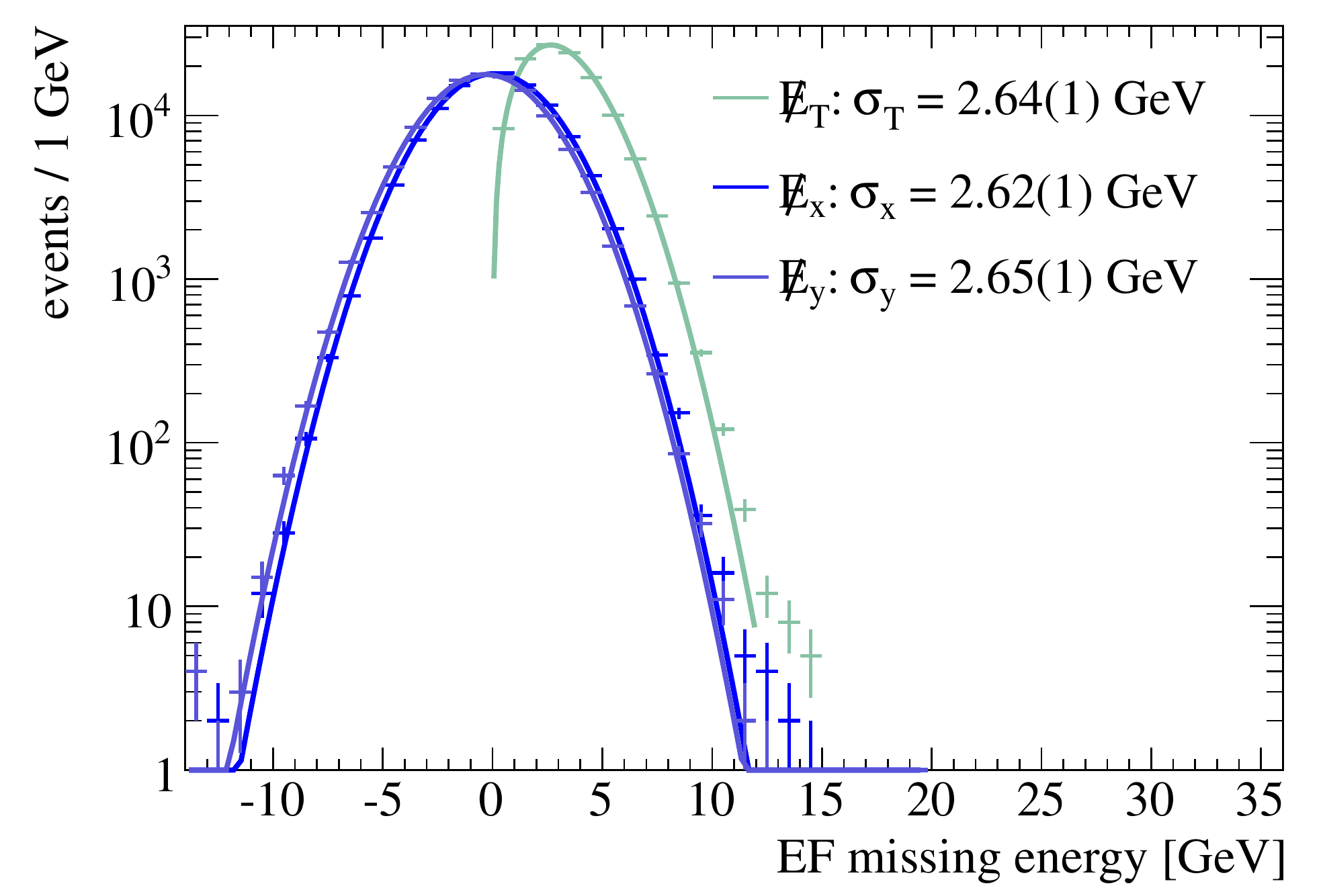}
  \caption{
    Distribution of EF \mex, \mey and \met for events
    from period 2011~D
    which have $\GeV{53} < \sumet < \GeV{75}$,
    taken with a random trigger.
    The solid lines are fits to the data points with a Gaussian (for \mex and \mey) or \met distribution,
    for which the relevant parameters are given in the legend.
  }
  \label{fig:results_met_model_mexy_2011}
\end{figure}
\Fig{fig:results_met_model_mexy_2011} shows
the distributions of the components of the missing energy along the $x$- and $y$-direction
as well as its transverse component,
for the same slice of \sumet as in \Fig{fig:results_met_model_mexy},
but for data taken with a random trigger in period~D from 2011 instead of 2010.
All three distributions are fitted with the appropriate functions.
The fact that the same resolution parameter $\sigma_T$ is found as in the fit from 2010
confirms the assumption that the resolution parameters from 2010 can also be applied in 2011.
\section{Model Predictions}
\label{sec:results_met_model_predictions}

In this section, the model will be applied and compared to data taken in 2010 and 2011.
The data taken in each of these two years will be treated separately
in the following two sections,
due to considerable changes in the trigger and the instantaneous luminosity.
In addition, within 2011 there have been further changes in the trigger,
so that for 2011 only data up to and including period 2011~F will be used.

\subsection{Performance in 2010}
\label{sec:results_met_model_predictions_2010}

\subsubsection{Shape Predictions}

\begin{figure}
  \centering
  \incgraphics{width=\widthsingleplot}{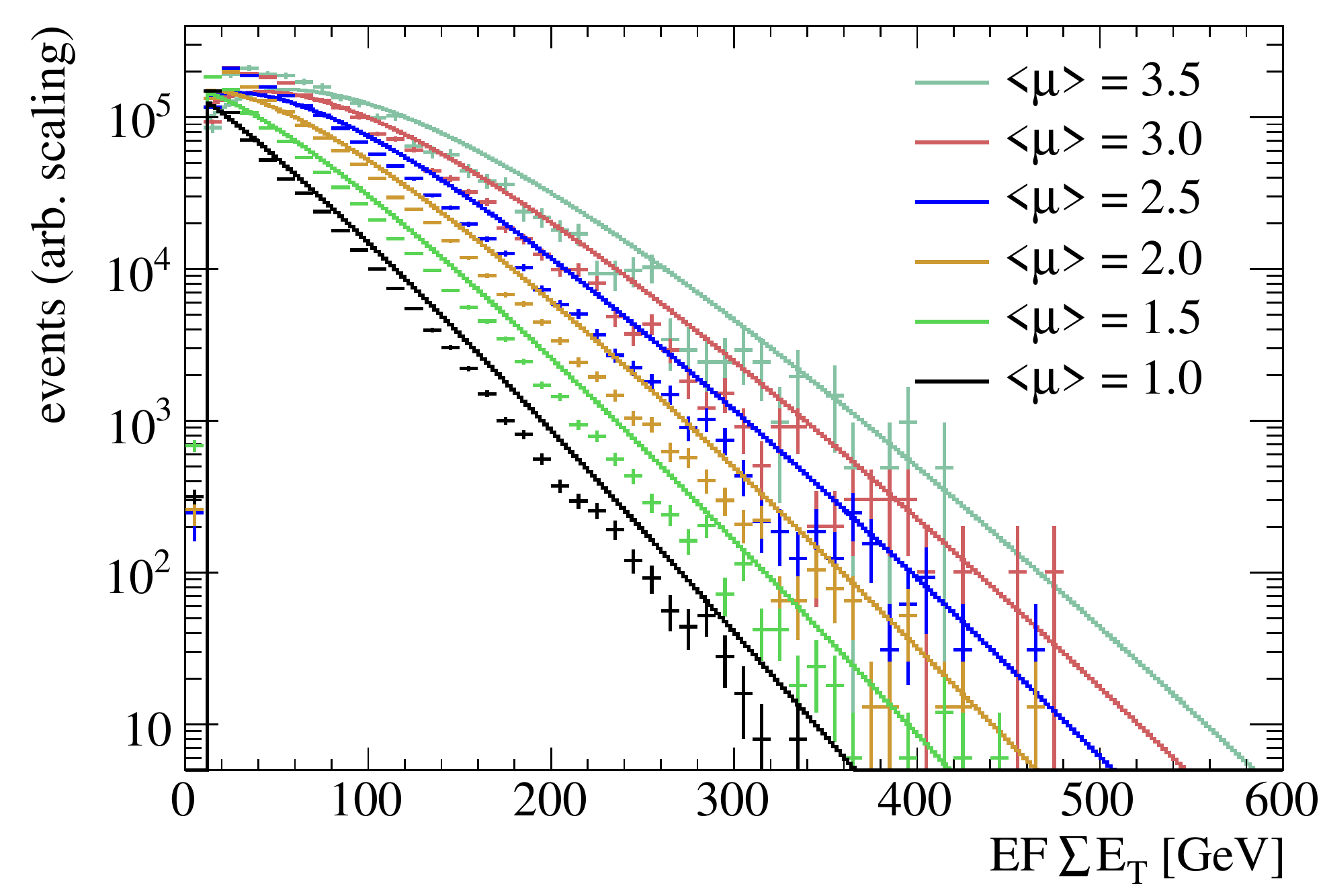}
  \incgraphics{width=\widthsingleplot}{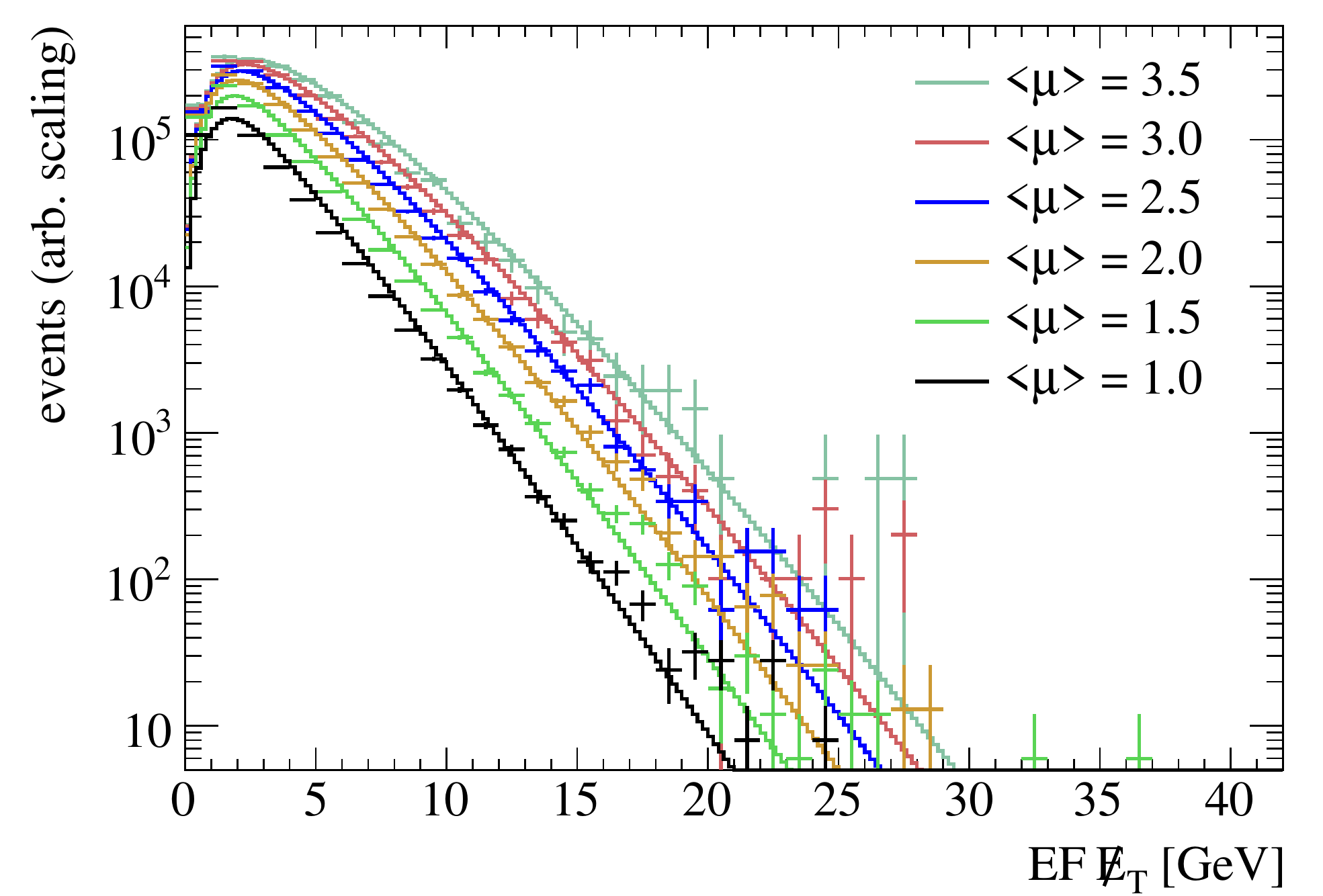}
  \caption{
    Comparison of the distribution of Event Filter \sumet (top) and \met (bottom),
    from the model prediction (lines) and 2010 data (points)
    for several values of \avgmu.
    The normalization of the data points is chosen such that the tails are clearly visible.
  }
  \label{fig:results_met_model_predictions_shapes_2010}
\end{figure}

\Fig{fig:results_met_model_predictions_shapes_2010} compares the predictions of the model
for Event Filter \sumet and \met to the respective spectrum measured in data from 2010.
The points represent data which has been collected
from several runs in periods F~--~I,
using the minimum-bias trigger \trigger{EF_mbMbts_1_eff}\footnote{
  In 2010, the random triggers were prescaled so heavily that they cannot be used for these studies.
  On the other hand, in 2011 the minimum-bias triggers are prescaled heavily
  because due to the high activity they would probably trigger on any bunch crossing anyway,
  and thus they have been replaced by purely random triggers.
  Therefore, in 2010 a minimum-bias trigger is used,
  whereas in 2011 a random trigger is used.
}
after applying the \ac{GRL} defined by the \ATLAS Supersymmetry group.
The range of luminosity blocks contributing to the plot has been chosen for each run such that
$\avgmu_\text{LB}$, computed from the instantaneous luminosity for a given luminosity block,
lies within a range
$|\avgmu_\text{LB} - \avgmu| < 0.1$,
with $\avgmu$ ranging between $1$ and $3.5$ in steps of $0.5$
for the six sets of data points in the plot.
This covers most of the range of \avgmu values that is available is 2010 data, the peak value being $3.8$. %

The curves in the plot give the model predictions for the respective value of \avgmu.
They have been generated using as input the detector resolution measured in 2010 data
and the slope of \sumet for one concurrent interaction obtained as described above\footnote{
  Note that for the fractional values of $\mu$,
  the same analytic form from \Eq{eq:results_met_model_full_sumet} is used for \sumet,
  extending (without proof) its definition from integer $\mu$ to fractional $\mu$
  by using the Gamma function $\Gamma(n) = (n-1)!$, %
  which generalizes the factorial function to real (and complex) arguments $n$.
}.
The normalization of the data points and curves has been chosen such
that the tails of the distributions are clearly visible,
by shifting the distributions so that they do not cross.
Comparing the Event Filter \sumet between data and model,
the distributions in data fit much better the predictions of the model
for values of \avgmu which are smaller by about $0.5$,
in particular for $\avgmu \geq 2.0$.
Indeed, this effect can also be captured by fitting the model \sumet function to the measured \sumet distribution in data
(cf. \Fig{fig:results_met_model_data_reco_PV_fit_2011} in the Appendix)
and seems to affect primarily data from 2010,
where \avgmu is in general smaller than in 2011.
Normalizing data and the model prediction in the \sumet range above a given threshold of \eg~\GeV{50} %
gives a better agreement,
confirming that it is especially difficult to model the low part of the \sumet distribution.
The \met distribution is stable against this effect
and the agreement between data and model prediction is satisfactory for all \avgmu,
as can be seen in the lower plot in \Fig{fig:results_met_model_predictions_shapes_2010}.

\subsubsection{Rate Predictions}

\begin{figure}
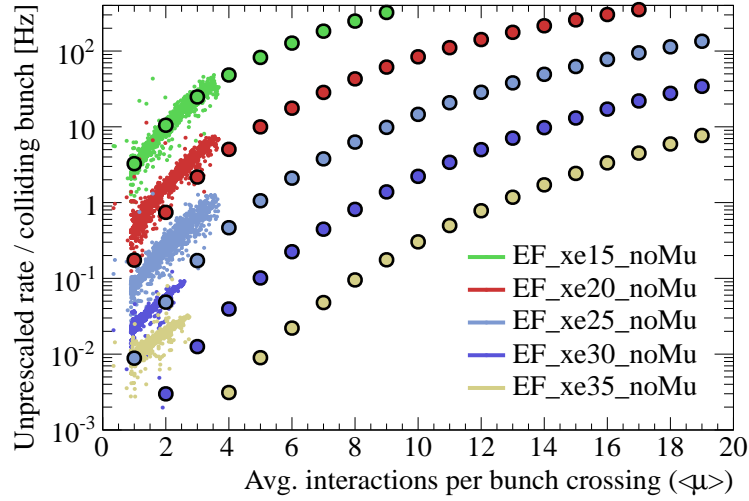

  \centering
  \incgraphics{width=\widthsingleplot}{{{COOL.combine_rates_own2010_rate2010IHG_avg2_EF_xe15_noMu.pdf-rate2010IHG_avg2_EF_xe35_noMu.pdf_182726-189425}}}
  \caption{
    Comparison of predicted and observed rates for missing transverse energy triggers
    with different thresholds ($15$, $20$, \dots, \GeV{35}) in 2010, %
    as function of the average number of interactions per bunch crossing \avgmu.
    The filled circles are the model predictions,
    the clouds of small dots represent the trigger rate\revised{ } measurements.
  }
  \label{fig:results_met_model_predictions_rates_2010}
\end{figure}

The spectra of Event Filter \met in \Fig{fig:results_met_model_predictions_shapes_2010} for different values of \avgmu
show good agreement between prediction and observation.
From the predicted spectra, the \met trigger rates can now be computed by integration according to \Eq{eq:results_met_model_relative_rate}.
As the model cannot predict absolute rates, but only relative rates,
the rate predictions are normalized to the measured trigger rate of the lowest \met trigger.
The comparison of predicted and measured rates is shown in \Fig{fig:results_met_model_predictions_rates_2010}.
The clouds of points represent the measured trigger rates,
which are obtained from the \ac{COOL} database
and normalized to the number of colliding bunches,
as described in detail in \Sec{sec:results_met_model_motivation}.
To reduce the spread, each point is an average over two luminosity blocks in this plot.
The predicted rates are computed for integer values of \avgmu and marked by filled circles
in the color matching the corresponding measured rates.
The green points and markers correspond to the lowest threshold of \GeV{15},
and the measured rates for $\avgmu = 2$ for this threshold are used to normalize the model predictions. %

Note that the prediction of the rate from the \met spectrum measured by the Event Filter
only encompasses one level of the three-level trigger system of \ATLAS.
A simulation of more than one level is not possible with the current approach
because correlations between the measurements at the different levels are not included in the model.
However, the prediction works as long as the overlap between the different levels is small enough.
To put it the other way round,
if the spacing between Level~2 and Event Filter is large enough,
the Event Filter sees an undistorted spectrum of \met,
and therefore its trigger rate is consistent with the (properly normalized) integral
of the \met spectrum above the Event Filter \met threshold.

It is no surprise that the rates for the lowest threshold of \GeV{15} agree well with the prediction,
as this is the rate which the prediction is normalized to.
Note, however, that this normalization only uses the point with $\avgmu = 2$.
The agreement of the predictions for $\avgmu = 1$ and $3$ with the measured rates is not automatic,
but shows that the dependence of the rates on $\avgmu$ is correctly described by the model,
at least over the small range where rate measurements are available in 2010.
On the other hand, the plot also shows that already for the next higher threshold of \GeV{20},
the prediction underestimates the measured rate by a factor of about $2$,
and for higher thresholds the discrepancy gets larger.
Thus, even the spacing of the thresholds in terms of rates is apparently not well described by the model.
The reason for the large discrepancies for higher thresholds lies in the composition of \met,
which arises from several sources.
Only one of these, namely fake \met from the limited detector resolution, is included in the model.
This source of \met is no longer dominant for high trigger thresholds,
which leads to an underestimation of the measured trigger rates by the model prediction.
This is explained in more detail below, but first the results for 2011 are presented.

\subsection{Performance in 2011}
\label{sec:results_met_model_predictions_2011}

One of the noticable differences between data taken in 2010 and 2011 is that,
as part of the progression in the instantaneous luminosity,
in 2011 also the bunch spacing was reduced,
going from \unit[150]{ns} in 2010~I down to \unit[50]{ns} is 2011~D, %
so that out-of-time pile-up possibly can no longer be neglected\footnote{
  The \ATLAS \ac{TDR} states a peaking time of the bipolar shapers
  in the read-out of the liquid-argon Calorimeters of approximately \unit[35]{ns} \cite{ATLASTDR1999}.
  In fact, a (small) destructive interference due to out-of-time pile-up 
  can be expected even for large bunch spacings up to \unit[500]{ns} \cite{ATLASDetector2008}. %
}.
It is not \latein{a priori} clear that the model would still give sensible results.
As will become clear from the plots in this section, however,
the agreement for the shapes in 2011 is even better than in 2010.

\subsubsection{Shape Predictions}

\begin{figure}
  \centering
  \incgraphics{width=\widthsingleplot}{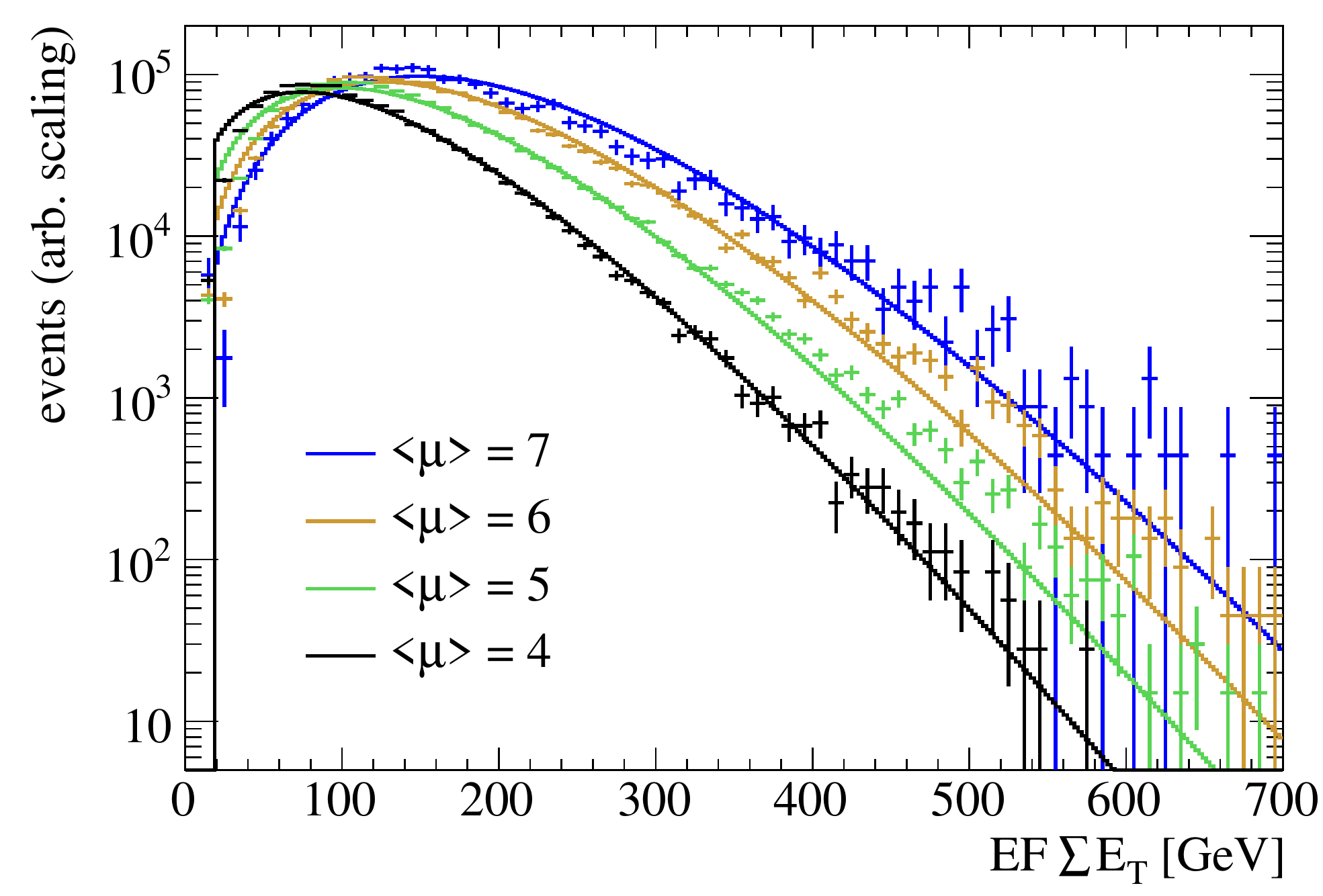}
  \incgraphics{width=\widthsingleplot}{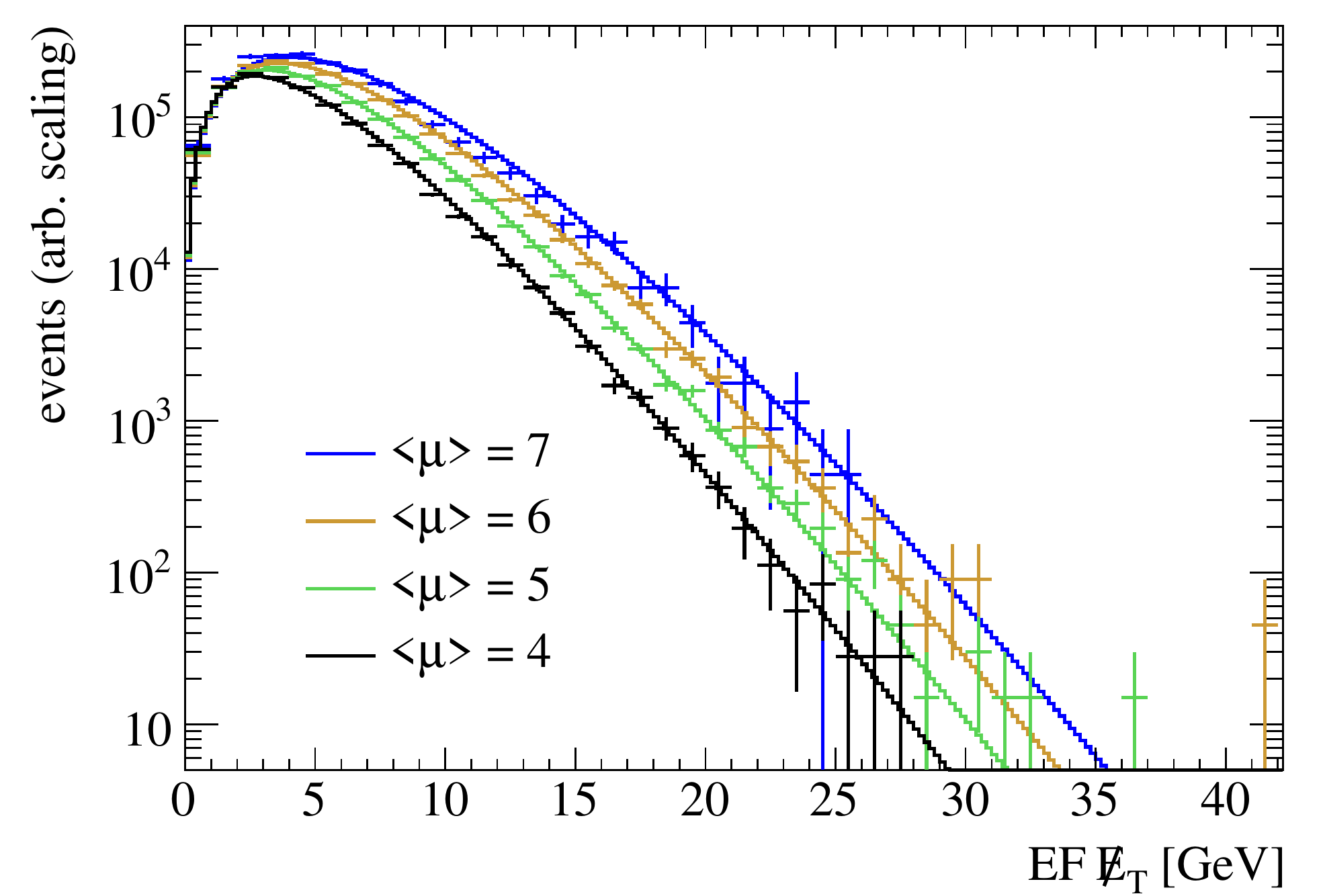}
  \caption{
    Comparison of the distribution of Event Filter \sumet (top) and \met (bottom),
    from the model prediction (lines) and 2011 data (points)
    for several values of \avgmu.
    The normalization of the data points is chosen such that the tails are clearly visible.
  }
  \label{fig:results_met_model_predictions_shapes_2011}
\end{figure}
In \Fig{fig:results_met_model_predictions_shapes_2011},
the predictions of the model and the spectra measured in data from period 2011~D
for the Event Filter \sumet and \met are shown.
The data has been collected from several runs,
this time using a random trigger, \trigger{EF_rd0_filled_NoAlg},
and applying the SUSY \ac{GRL}.
Again, different subsamples are shown,
which are selected by requiring $\avgmu_\text{LB}$ to lie within a range %
$|\avgmu_\text{LB} - \avgmu| < 0.1$,
with $\avgmu \in \{4,\,5,\,6,\,7\}$ for the four different sets of data points.
The four curves compare the model predictions for the respective value of \avgmu
to the observed spectra from data,
which are plotted as data points.
The model predictions use the detector resolution from 2010 as input,
and the slope of \sumet for one concurrent interaction,
which has been extracted from the fit of the full model to the \sumet spectrum for $\avgmu \sim 5$.
The normalization of the data points and curves has been chosen such that the tails of the distributions are clearly visible.

It can be seen from these plots that the model gives a very good description for both \met and \sumet.
The good description of the \sumet spectrum for $\avgmu=5$ may be attributed to the fact
that the slope $\lambda$ which is used as input to the model was extracted from this subsample of events,
but still the value of \avgmu was left free in the fit from which $\lambda$ was taken,
and is set to $5$ here,
so the good agreement is not completely artificial.
The other three \sumet distributions and all distributions for \met
follow from \Eqs{eq:results_met_model_full_sumet} and \eqref{eq:results_met_model_full_met} by adjusting only \avgmu.

\subsubsection{Rate Predictions}

\begin{figure}
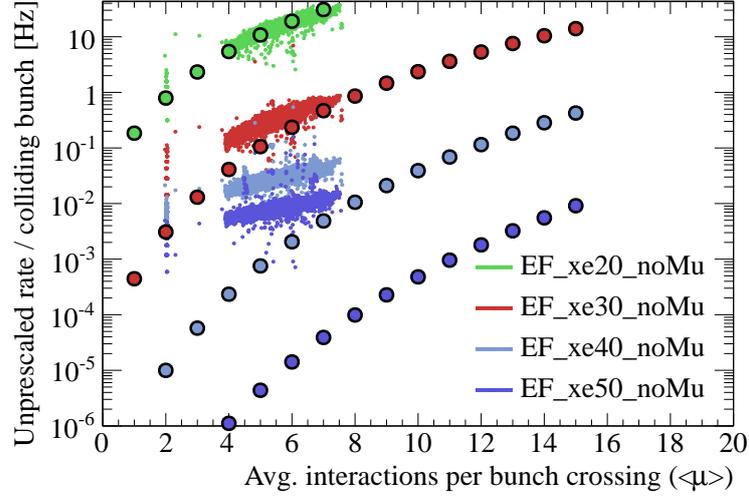

  \centering
  \incgraphics{width=\widthsingleplot}{{{COOL.combine_rates_own2011_rate2011K-D_avg2_EF_xe20_noMu.pdf-rate2011K-D_avg2_EF_xe50_noMu.pdf_179710-182519}}}
  \caption{
    Comparison of predicted and observed rates for missing transverse energy triggers with different thresholds ($20$, $30$, $40$, \GeV{50}) in 2011,
    as function of the average number of interactions per bunch crossing \avgmu.
    The filled circles are the model predictions,
    the clouds of small dots represent the trigger rates measurements.
  }
  \label{fig:results_met_model_predictions_rates_2011}
\end{figure}

\Fig{fig:results_met_model_predictions_rates_2011} shows the predictions
of the missing transverse energy trigger rates compared to the rates measured in 2011, periods 2011~D --~F. %
Runs from later periods have higher \avgmu (almost up to $10$, cf. \eg \Fig{fig:results_met_model_rate_high_EF_triggers}), %
but cannot be compared to previous periods nor to the model predictions
due to changes in the trigger which are discussed below.
Like for 2010, also here the predicted rates are normalized to the lowest threshold,
which is \GeV{20} for 2011.
Again, the dependence on \avgmu of the rate of this trigger threshold seems to be well described by the model,
but unfortunately less so for the next higher threshold,
which exhibits a slower increase of its rate with increasing \avgmu than is predicted by the model.
Note that here, the spacing of the two lowest thresholds is twice as large as in 2010,
but the disagreement for the second-lowest threshold is not much worse than in 2010.
For higher thresholds, the predicted rates stay behind the observed rates
because again for the higher thresholds other components of \met become important
which are not included in the model.

\section{Discussion and Outlook}

In this section, a model has been presented to predict the \met trigger rates as function of the average number of interactions per bunch-crossing \avgmu.
The basic idea is that the rates are dominated by fake \met from the limited detector resolution.
That this assumption is indeed a good description is confirmed 
by the fact that the functional forms which are implied by this assumption
yields consistent resolution parameters for \mex, \mey and \met when fitted to data.
The resolution depends strongly on \sumet and increases (\ie becomes worse) with higher \sumet.
The baseline for the explanation of the stronger than linear increase of the rates is therefore
that higher pile-up means more activity in the event,
which implies higher calorimeter occupancy and higher \sumet,
and thus an increased resolution from larger fluctuations.
This in turn leads to higher trigger rates.
It has been shown that the \met resolution is, however, independent of \avgmu,
\ie its pile-up dependence is swallowed by the parametrization as function of \sumet.

Using only three input parameters, the model already provides a good description of both the \met and \sumet shapes as function of \avgmu.
The predicted rates are only in good agreement with the actual trigger rates for the lowest thresholds.
Their dependence on \avgmu seems to well described, underlining the predictive power of the model for high \avgmu.
The good agreement of the shapes, together with the deviations in the rates,
means that there must be a change in the \met and \sumet spectra
outside the range over which the model and the observed spectra are compared.
This is indeed found when plotting \met and \sumet over an extended range,
which is only possible by using a combination of \met triggers in addition to the minimum-bias and random triggers employed so far.

\subsection{\texorpdfstring{Extended \met and \sumet Spectra} {Extended MET and SumET Spectra}}

\begin{figure}
  \centering
  \incgraphics{width=\widthtwoplots}{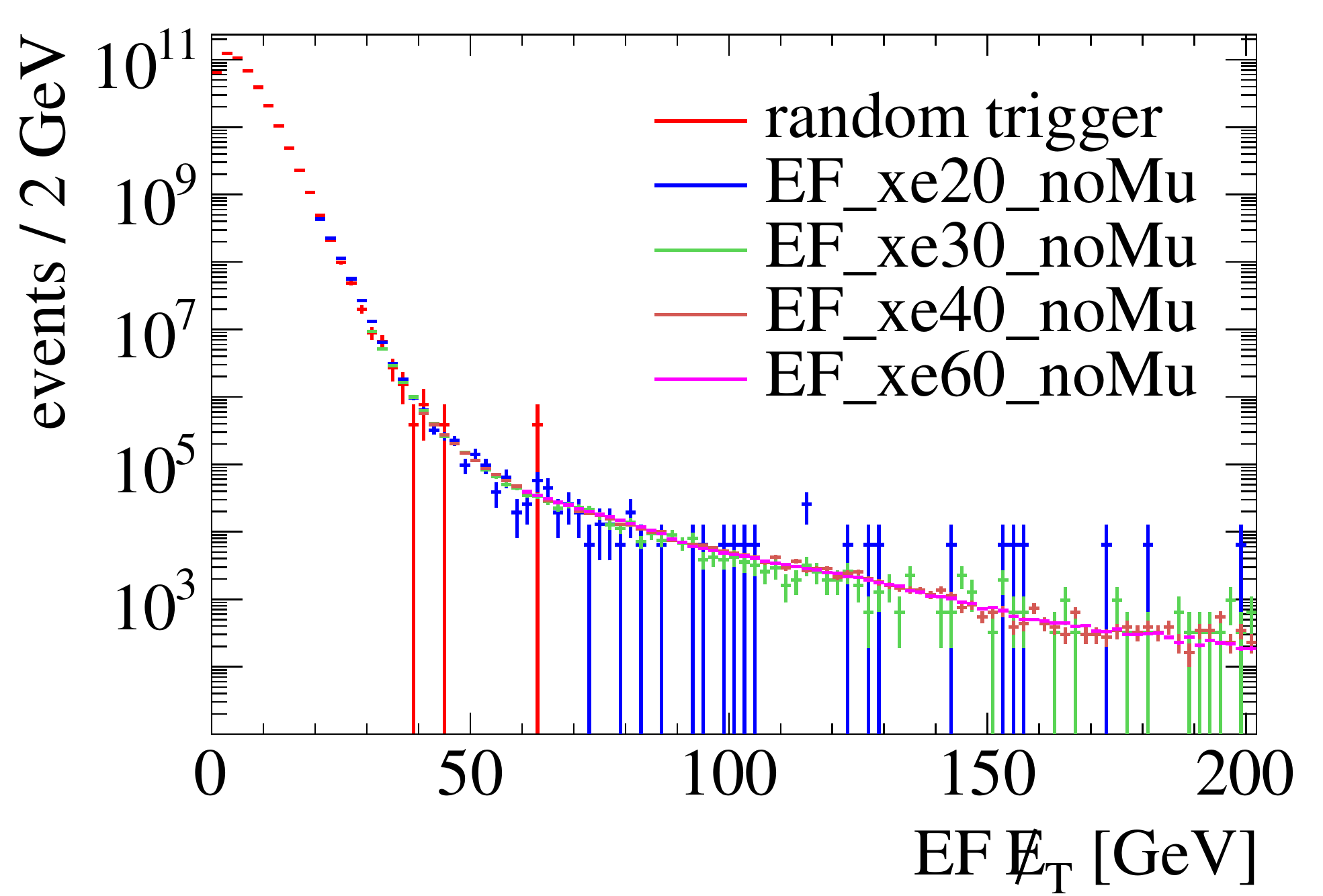}
  \incgraphics{width=\widthtwoplots}{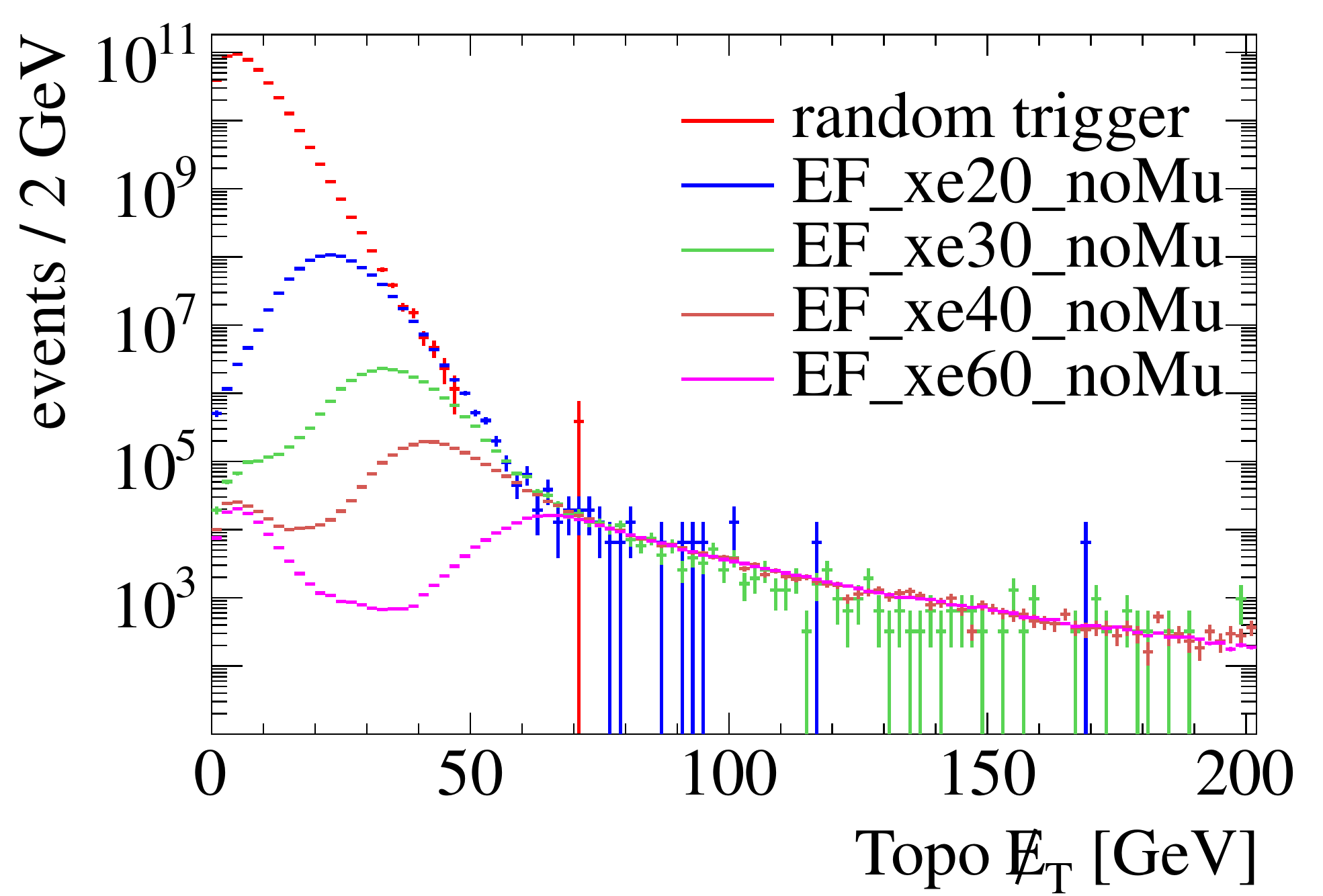}
  \caption{
    Spectrum of \met constructed by overlaying the spectra
    from several samples of events taken with different triggers in period 2011~D.
    Left: \met at Event Filter, right: offline \met (Topo calibration).
    The normalization of the different sets was done in the left plot, as described in the text,
    and then applied to the right plot without changes.
  }
  \label{fig:results_met_model_extended_ef}
  \label{fig:results_met_model_extended_topo}
\end{figure}

To make the change in the slope of \met visible, a large range of \met needs to be covered.
Due to the steeply falling nature of \met, this range cannot be covered by a single trigger alone.
In an example \met spectrum taken with a random trigger in 2011,
$3\ten{6}$ raw events ($1.1\ten{6}$ after applying the \ac{GRL} and trigger selection)
cover the range up to roughly \GeV{35}.
(This is the distribution in red in \Fig{fig:results_met_model_extended_ef}.)
The slope indicates that a factor of 60 times more events is needed to reach higher in \met by \GeV{10}.
Extrapolating this and assuming no change in the slope\footnote{
  The assumption that the slope is constant up to much larger value of \met is only correct for the contribution of fake \met,
  as will become clear below.
},
\eg $1.4\ten{9}$ raw events would be needed to cover the \met spectrum up to \GeV{50}, %
which would in turn correspond to a raw integrated luminosity of \ifb{35} with the prescales of the existing random trigger dataset, %
many times more than the LHC has delivered in 2011.

What is done, therefore, in the plots in \Fig{fig:results_met_model_extended_ef}, %
is to use spectra taken with several different triggers,
which cover different ranges of \met,
and to normalize the event samples by matching them in the overlapping regions. %
A first justification that this normalization indeed is correct can be drawn from the fact
that the shapes in the overlapping regions fit well.
Below, another and more objective justification will be given.
The triggers used to collect the different data samples are
a random trigger on filled bunch crossings (with colliding bunches)
for the lowest part of the spectrum (\trigger{EF_rd0_filled_NoAlg})
and four \met triggers with different thresholds (\trigger{EF_xe20,30,40,60_noMu}).
Every \met trigger covers the EF \met range above its threshold.
For higher \met, the statistics gets lower, until at the next threshold the next higher \met trigger sets in.
For the offline \met, the coverage is smeared out, as can be seen in the plot.
Both plots cover a range between \GeV{0} and \GeV{200} of \met,
in which the effective number of events per bin varies between about $10^2$ and $10^{11}$.
Note that this, of course, is not the actual event count per bin,
which is far below $10^{11}$,
but the number of events multiplied by the corresponding normalization factors.

\begin{table}
  \setlength{\tabcolsep}{4.9pt} %
  \centering
    \begin{tabular}{lllrd{4.3}SS}
      \toprule
      Stream & Period & Trigger & Events & \multicolumn{1}{l}{\Lint [$\unit[]{nb^{-1}}$]} & \multicolumn{1}{c}{$f_\text{pre}$} & \multicolumn{1}{c}{$f_\text{match}$} \\
      \midrule
      \tagname{MinBias} & D (8 runs) & \trigger{EF_rd0_filled_NoAlg} & \numprint{1140870} & 0.556 & $8\cdot 16.5$ & 60 \\
      \tagname{JetTauEtmiss} &
      D (4 runs) & 
           \trigger{EF_xe20_noMu} & \numprint{139793} & \numprint{    9}.15  & 22.2 & 20 \\
        && \trigger{EF_xe30_noMu} & \numprint{ 72313} & \numprint{  203}.0   & 15.3 & 14 \\
        && \trigger{EF_xe40_noMu} & \numprint{105871} & \numprint{ 3098}     & 23.5 & 23 \\
        && \trigger{EF_xe60_noMu} & \numprint{408883} & \numprint{72650}     &  & \\
      \bottomrule
    \end{tabular}
  \caption{
    Details of the normalization of the \met spectrum and the data used in \Fig{fig:results_met_model_extended_ef}.
    The left-most five columns hold information about which data was used,
    including the number of events and the respective integrated luminosity \Lint.
    The last two columns display the relative normalization factors,
    computed from prescales ($f_\text{pre}$)
    or from a matching in the overlapping regions ($f_\text{match}$),
    where \eg $f_\text{match} = 60$
    is the relative normalization of the data samples from the \trigger{EF_rd0_filled_NoAlg} and the \trigger{EF_xe20_noMu} triggers.
  }
  \label{tab:results_met_model_extended_spectrum_details}
  \resettabcolsep  %
\end{table}

\Tab{tab:results_met_model_extended_spectrum_details} shows
details of the normalization of the \met spectrum and the data used in \Fig{fig:results_met_model_extended_ef}.
For all spectra, data from period 2011~D is used, selecting a number of runs from this period arbitrarily.
The total number of events after applying the \ac{GRL} from all selected runs
and the integrated luminosity \Lint are also shown in the table.
The integrated luminosity can be used 
to compute the relative normalization factors $f_\text{pre}$,
which are needed to match the spectra taken with the different triggers.
They are just the relative prescale factors
and these can be computed from the ratio of the integrated luminosity of the data samples
taken with the different triggers.
The values of the integrated luminosity, which are stated in the table,
have been computed using \propername{iLumiCalc} (see \Sec{sec:luminosity_measurement}).
The factors $f_\text{pre}$ in the penultimate column can be compared to the factors $f_\text{match}$,
which are the normalization factors obtained from the overlap of the different distributions.

For the \met triggers, the values of $f_\text{pre}$ and $f_\text{match}$ agree very well,
the values for $f_\text{pre}$ being slightly larger,
which is probably due to some bias when matching the distributions in the overlap regions.
(No errors are given on these numbers as these are not vital for the discussion.)
The first value for $f_\text{pre}$ in the table needs some explanation.
In addition to the prescales set at the three trigger levels,
this trigger has an intristic prescale-like selection,
which is configured to pick on average one out of eight bunch crossings at random.
The raw rate of this trigger therefore is $f_\text{LHC} / 8 \approx \unit[5]{MHz}$,
where $f_\text{LHC}$ is the nominal rate of bunch collisions in \ATLAS.
Therefore, an additional prescale of eight on top of the relative prescale
to the lowest \met trigger needs to be included in the normalization.
As can be seen from the table though, this does not give the correct result,
but is too large by a factor of about two compared to the overlap normalization.
What causes this discrepancy is not clear.
Explaining it by the ratio of filled bunches over the total number of buckets,
on which the nominal rate of bunch collisions is based, comes to nothing
because this ratio is too low (up to $598/3564$) %
for the set of runs that have been used,
and it should affect the rates of all triggers in the same way, including the \met triggers.

\begin{figure}
  \centering
  \incgraphics{width=\widthtwoplots}{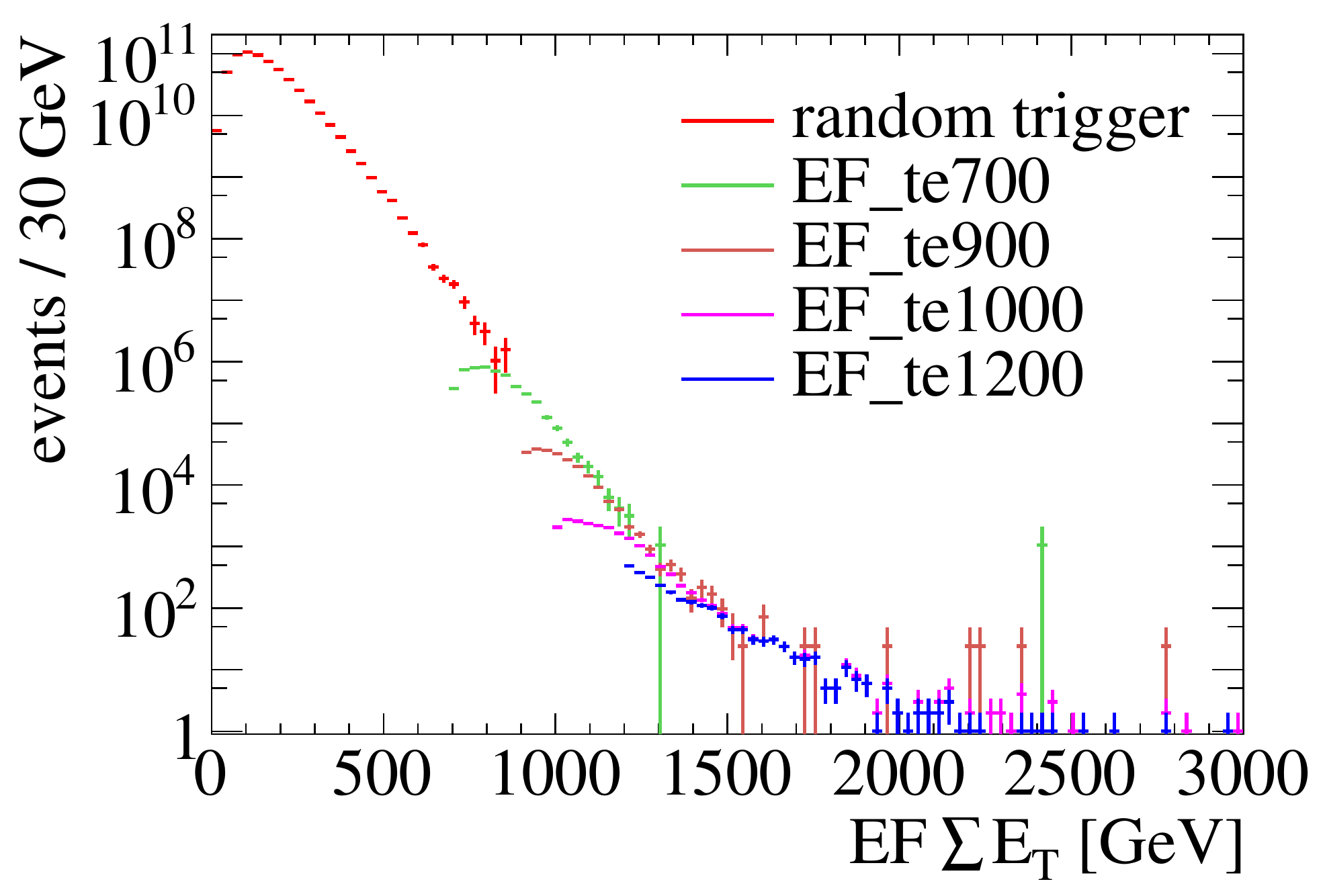}
  \incgraphics{width=\widthtwoplots}{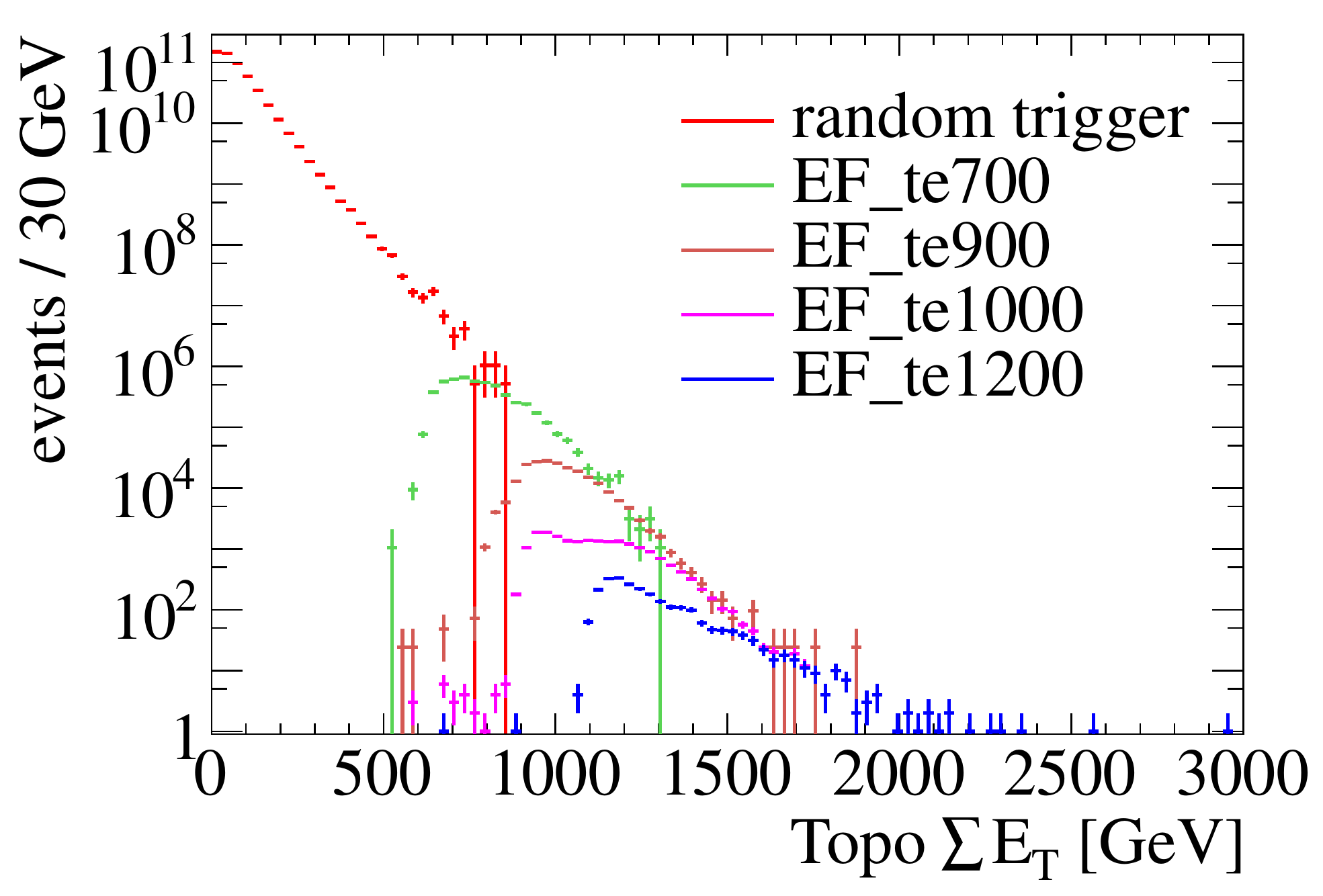}
  \caption{
    Spectrum of \sumet constructed by overlaying the spectra
    from several samples of events taken with different triggers in period 2011~D.
    Left: \sumet at Event Filter, right: offline \sumet (Topo calibration).
    The normalization of the different sets was done based on prescales for both plots, as described in the text.
    The nominal trigger thresholds at Event Filter level are as usual given in the chain name in GeV.
  }
  \label{fig:results_met_model_extended_sumet_ef}
  \label{fig:results_met_model_extended_sumet_topo}
\end{figure}

\Fig{fig:results_met_model_extended_sumet_ef} shows the spectrum of \sumet
and was produced in a similar way as \Fig{fig:results_met_model_extended_ef},
but this time using \sumet triggers instead of the \met triggers.
The lowest spectrum is again taken with the same random trigger as before and spans quite a substantial range of \sumet.
To be consistent with the \met spectrum in \Fig{fig:results_met_model_extended_ef},
the spectrum taken with the random trigger has been scaled by an additional factor of four instead of eight.
In constrast to the \met triggers, no manual normalization was possible for \sumet
because the overlap of the spectra taken with \sumet triggers is not sufficient.
Instead, only the prescale-based scaling was applied.
A striking difference between the spectra for EF \met and EF \sumet in \Figs{fig:results_met_model_extended_ef} and \ref{fig:results_met_model_extended_sumet_ef}
is that the individual \sumet spectra taken with \sumet triggers are not monotonically falling functions,
but reach a maximum which lies above the nominal trigger threshold.
This is an indication that the respective lower triggers (\trigger{L1_TE} at Level~1 and \trigger{L2_te} at Level~2)
have not reached their plateau at the threshold value of the EF cut on \sumet.
In fact, the L1 and L2 \sumet triggers have a very slow turn-on, %
so that,
in order not to have to go to too high values at EF and still be able to stay within the rate limits at L2,
this kind of selection overlap between the trigger levels is unavoidable. %
Another difference to the \met spectrum is that the \sumet spectrum seems to be rather well described
by a single exponentially falling distribution over the range from \GeV{200} up to \TeV{1.5},
whereas the \met spectrum shows a clear change in its slope as discussed above.
Note that no selection on the number of interactions has been done,
so that this \sumet spectrum arises from a mixture of events with different numbers of interactions
and the slope is not the same as the slope for events with exactly one interaction used as input parameter to the model.

\subsection{\texorpdfstring{Other Sources of \met} {Other Sources of MET}}
\begin{figure}
  \centering
  \incgraphics{width=\widthsingleplot}{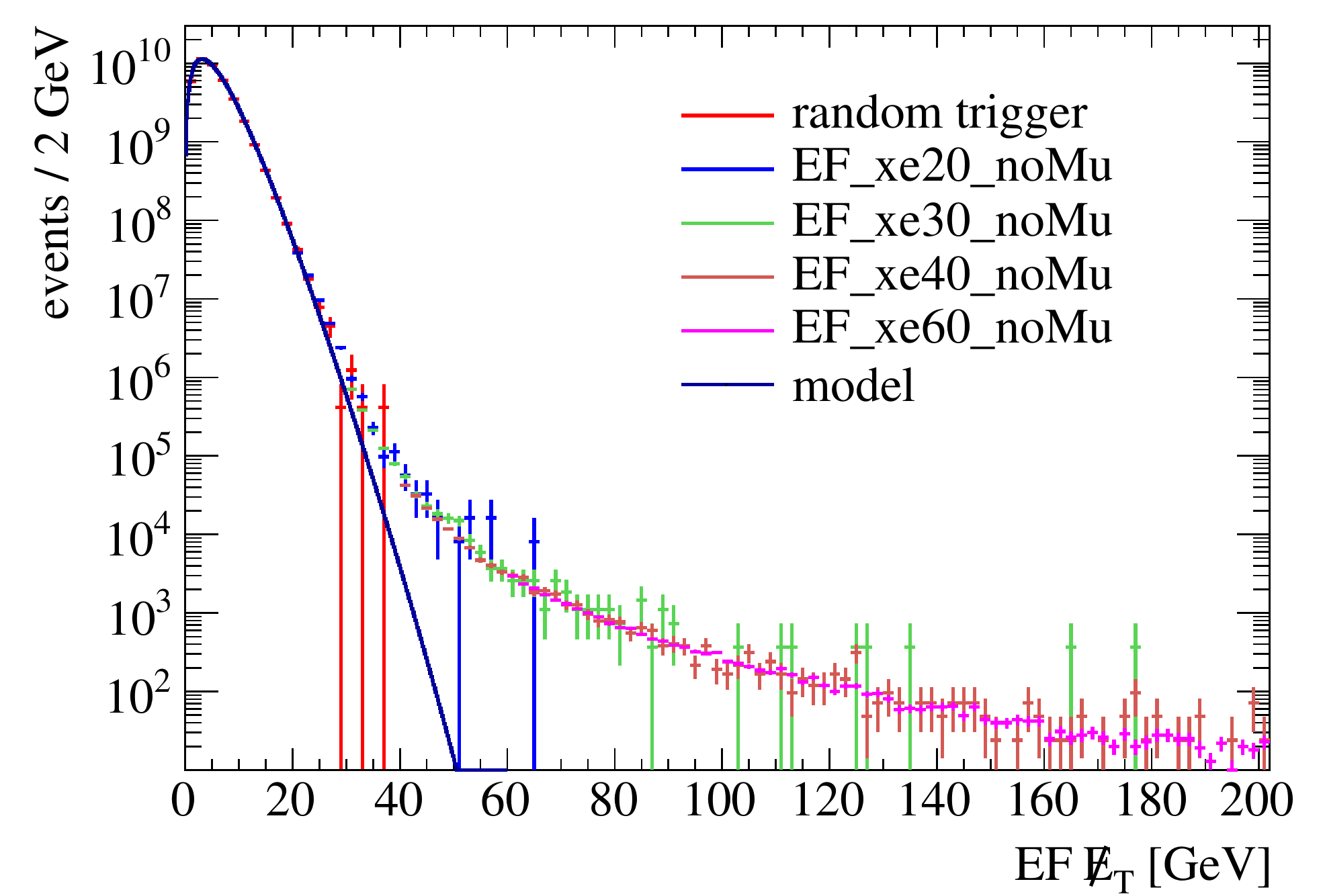}
  \caption{
    Spectrum of \met constructed by overlaying the spectra
    from several samples of events taken with different triggers and restricted to luminosity blocks with $\avgmu \approx 5$.
    In addition the model prediction is plotted (solid line).
  }
  \label{fig:results_met_model_extended_met_ef_mu5}
\end{figure}
To allow a direct comparison of the model prediction and the extended \met spectrum,
the plot in \Fig{fig:results_met_model_extended_met_ef_mu5} repeats the \met spectrum from \Fig{fig:results_met_model_extended_ef},
but this time again with an additional selection,
which requires events to be from luminosity blocks for which $\avgmu \approx 5$.
The black curve in the plot is the model prediction for this value of \avgmu.
The change in the slope of the \met spectrum happens around \GeV{40},
but the deviation from the prediction by the model already starts below this value:
for \met values above this value, the observed \met spectrum lies above the prediction.
It is clear that the \met rates for larger thresholds cannot be expected to be well described by the model,
and indeed, the deviation from the plot is reflected in the underestimation of the rate for the trigger
with an \met threshold of \GeV{30} in \Fig{fig:results_met_model_predictions_rates_2011}.
The underlying reason for this is that the model includes only one source of \met,
the fake \met from resolution effects,
which is dominant at low \met,
but the \met measurement receives additional contributions from other sources at higher values:
\begin{itemize}
  \item \met from mismeasured jets leading to an imbalance of the total energy sum
    is expected to be proportional to the number of jets which are produced per second
    and thus scales linearly with the instantaneous luminosity.
    This \met contribution shall also be called fake \met, as opposed to real \met from invisible particles.
    Yet it is of different origin than the resolution-based fake \met and not to be confused with it.
  \item The trigger rate due to real \met from physics processes
    is proportional to the number of invisible objects 
    produced per second and thus also scales linearly with instantaneous luminosity.
  \item Detector problems like ``hot'' cells constantly giving large contributions to the energy measurement
    are independent of \avgmu and instantaneous luminosity.
\end{itemize}
The rates attributed to all of these sources are expected to grow slower with \avgmu
than the contribution to the trigger rate from fake \met due to the detector resolution.
Indeed, the dependence on \avgmu for higher \met triggers in the range of \met
where the slope in the extended \met spectrum deviates from the model prediction is linear,
confirming that here the above sources drive the trigger rate.
This linear dependence can be seen in \Fig{fig:results_met_model_rate_high_EF_triggers},
where the rates of \met triggers with thresholds of $60$, $70$, $80$ and \GeV{90} are plotted as function of \avgmu
in periods I~--~K of 2011.
Note that in this plot,
the vertical axis is not on logarithmic scale.
The spread of the data points is relatively large,
although the values are averaged over 8 luminosity blocks,
due to the low rates which have larger statistical fluctuations.

\begin{figure}
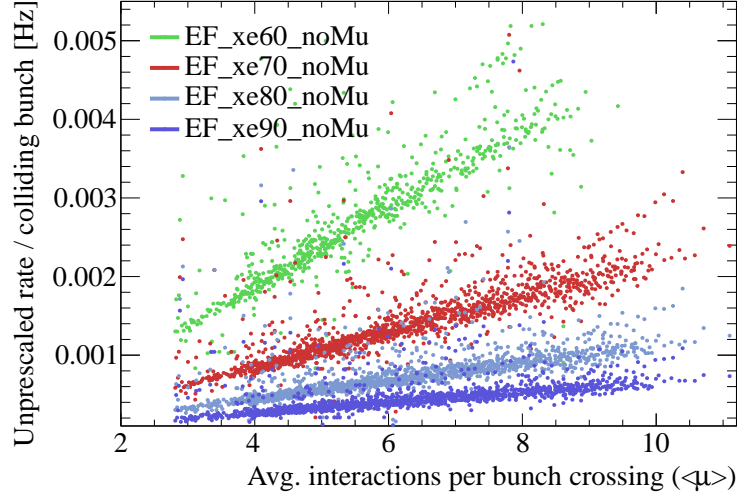

  \centering
  \incgraphics{width=\widthsingleplot}{{{COOL.combine_rates_own2011_rate2011K-D_avg8_EF_xe60_noMu.pdf-rate2011K-D_avg8_EF_xe90_noMu.pdf_185353-189425}}}
  \caption{
    Rates of several high-threshold \met triggers in runs from period I~--~K of 2011. %
    A linear behavior as function of the average number of interactions per bunch crossing \avgmu can be seen.
  }
  \label{fig:results_met_model_rate_high_EF_triggers}
\end{figure}

Thus, even though the model only accounts for one source of \met,
it covers the most problematic source of fake \met,
the one which gives rise to non-linear trigger rates,
and its predictions can be understood as a lower bound for the trigger rates.
With increasing \avgmu, fake \met from resolution effects will become the dominant contribution for higher and higher thresholds,
and the model is important for estimating the contribution of this type of fake \met when going to very high \avgmu.
The estimation can continuously be adapted and improved as new data for higher \avgmu becomes available.

\subsection{Pile-up Noise Suppression}
One example of a change in the \ATLAS trigger system in 2011 that necessitates a recalibration of the model
is the introduction of the pile-up noise suppression in the online system with the start of period 2011~G.
The noise suppression uses an energy-dependent \ac{RMS} value
to do a one-sided cut for each cell,
requiring the energy to lie above $3$ RMS.
The set of RMS values on which the cuts are based has been replaced
by a set which is tuned to a calorimeter activity corresponding to eight pile-up interactions, %
whereas before it has only been accounting for electronic noise.

\begin{figure}
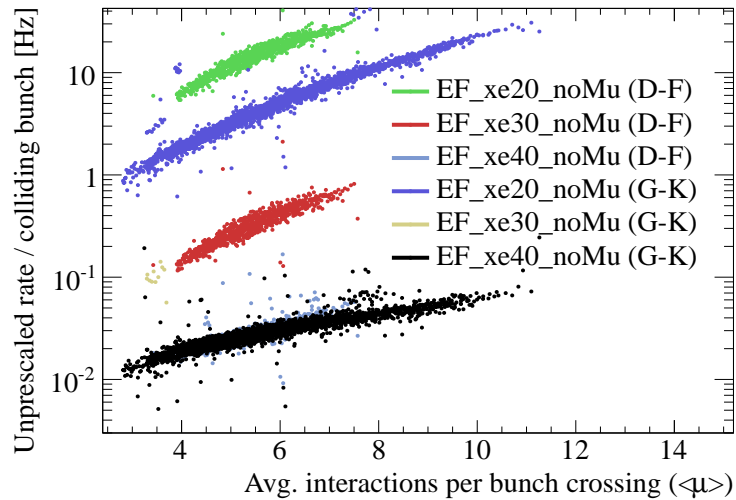

  \centering
  \incgraphics{width=\widthsingleplot}{{{COOL.combine_rates_own2011_rate2011K-D_avg8_EF_xe20_noMu.pdf-rate2011K-D_avg8_EF_xe40_noMu.pdf_182726-189425}}}
  \caption{
    Rates of the three \met triggers \trigger{EF_xe20,30,40_noMu}
    as function of the average number of interactions per bunch-crossing
    compared between periods 2011 D~--~F (without)
    and G~--~K (with additional pile-up noise suppression in the trigger).
    For \trigger{EF_xe40_noMu} the points for periods D~--~F are mostly hidden behind G~--~K and of comparable spread.
  }
  \label{fig:results_met_model_predictions_low_met_rates}
\end{figure}
The rates of the three lowest \met triggers in 2011 are plotted in \Fig{fig:results_met_model_predictions_low_met_rates} for periods D~--~F,
before the introduction of the pile-up suppression in the noise cuts,
and for periods G~--~K,
which have the pile-up suppression.
The impact on the rates is clearly visible for \trigger{EF_xe20_noMu}.
\trigger{EF_xe40_noMu} has already such a high threshold that its rate is not strongly affected by the pile-up noise suppression.
The intermediate threshold \trigger{EF_xe30_noMu} has been deactivated is period~G and later periods,
except for one run (185353) with very low instantaneous luminosity (peak value: \instlumi{1.2\ten{30}}), %
so that almost no data for this trigger with the pile-up noise suppression in place is available.

This change at the beginning of period~G is the reason why the above plots,
in which the trigger rates in 2011 between prediction and measurement are compared,
use periods D~--~F only.
(Periods A and B have been mostly for testing after the winter shutdown of the \ac{LHC} and do not contain much usable data,
2011~C is left out because it has a reduced center-of-mass energy of \TeV{2.76}.) %

The aim of including the updated noise suppression was to reduce the rates of the \met triggers.
As increasing the thresholds directly reduces the \met seen by the trigger, this has naturally been achieved.
Still, the strong dependence of the rates on \avgmu appears to be unchanged,
considering the slope of the rates of \trigger{EF_xe20_noMu} in \Fig{fig:results_met_model_predictions_low_met_rates}.
With respect to the model, the impact of the new noise suppression scheme
on the resolution of the \met measurement in the trigger and possibly the \sumet spectrum needs to be studied,
and the input parameters of the model need to be updated accordingly to obtain meaningful comparisons for the new periods.
This will be the next step when pursuing the development of the model further.
Other possibilities for future developments of the model are discussed in the remainder of this section.

So far, no uncertainties are available for the rates predicted by the model.
The extraction of the input parameters yields uncertainties on these parameters
which have been stated above and which can be propagated to the spectra and trigger rates.
This can help to assess the reliability of the extrapolation to higher pile-up levels
and the sensitivity of the model predictions to the input parameters.
Furthermore, there will also be systematic errors arising from the modeling itself.
To quantize these, careful studies are needed.
It has been stressed several times that the model is deliberately kept simple.
Nevertheless, now that the underlying mechanisms driving the trigger rates seem to be understood quite well,
the above discussion invites to do more checks and extend the existing model
to incorporate some more of the features found in the observed \met spectrum,
which it currently does not cover.
With respect to the implementation,
the numerical stability could be checked.
In the computation of the shapes, often very small values are used,
and it would be good to confirm that the numerical precision at which the computations are carried out does not have a significant impact on the predictions.
Predictions of the rates of the \sumet triggers have not been attempted so far,
as they are not widely used in \ATLAS,
and the prediction of \sumet seems to be less stable than that of \met.
Yet it would be interesting to do some studies of how the predictions of \sumet trigger rates compare against observed rates.
It may also be interesting to check the influence of varying bunch sizes on the trigger rates:
Not all bunches which are injected into the LHC contain the same number of protons.
Depending on the magnitude of this variation,
there may be a dependence of the trigger rate on the bunch crossing,
which at least for triggers with low \met thresholds will not average out,
due to the non-linear dependence of the trigger rate on the average number of interactions within a bunch crossing.
As said before, the model cannot predict the behavior of the full trigger chain.
Whether it makes sense to try and include correlations between different trigger levels in the model is not obvious.
It would, in any case, require a complete change of how the model works,
and as long as the spacing of the thresholds at the different trigger levels is sufficiently large,
there does not seem to be an urgent need to do so.
Another possible extension
which would probably promise the largest benefit with respect to the rate predictions
is to include the fake \met coming from jet mismeasurements in the model.
This would require to find an analytic description
which at least phenomenologically can describe the \met contribution from jet mismeasurements.
Its dependence on \avgmu should simply be linear as explained above.
In any case, the model has shown that mismeasurements of the energy of jets make up an important part of the trigger rate,
so that it might be worthwhile to consider how the \met trigger itself can be improved to be more stable against this type of fake \met,
\eg by vetoing events in which the direction of the missing transverse energy and the leading jet or a jet with significant energy are anti-aligned.
Another offspring of the \met trigger studies is the development of \met significance triggers,
which are briefly presented in the last part of this chapter.
\subsection{\texorpdfstring{\met Significance Triggers} {MET Significance Triggers}}
\label{sec:results_met_model_predictions_xs_triggers}
\inindex{XS trigger}

\begin{figure}
  \centering
  \incgraphics{width=\widthtwoplots}{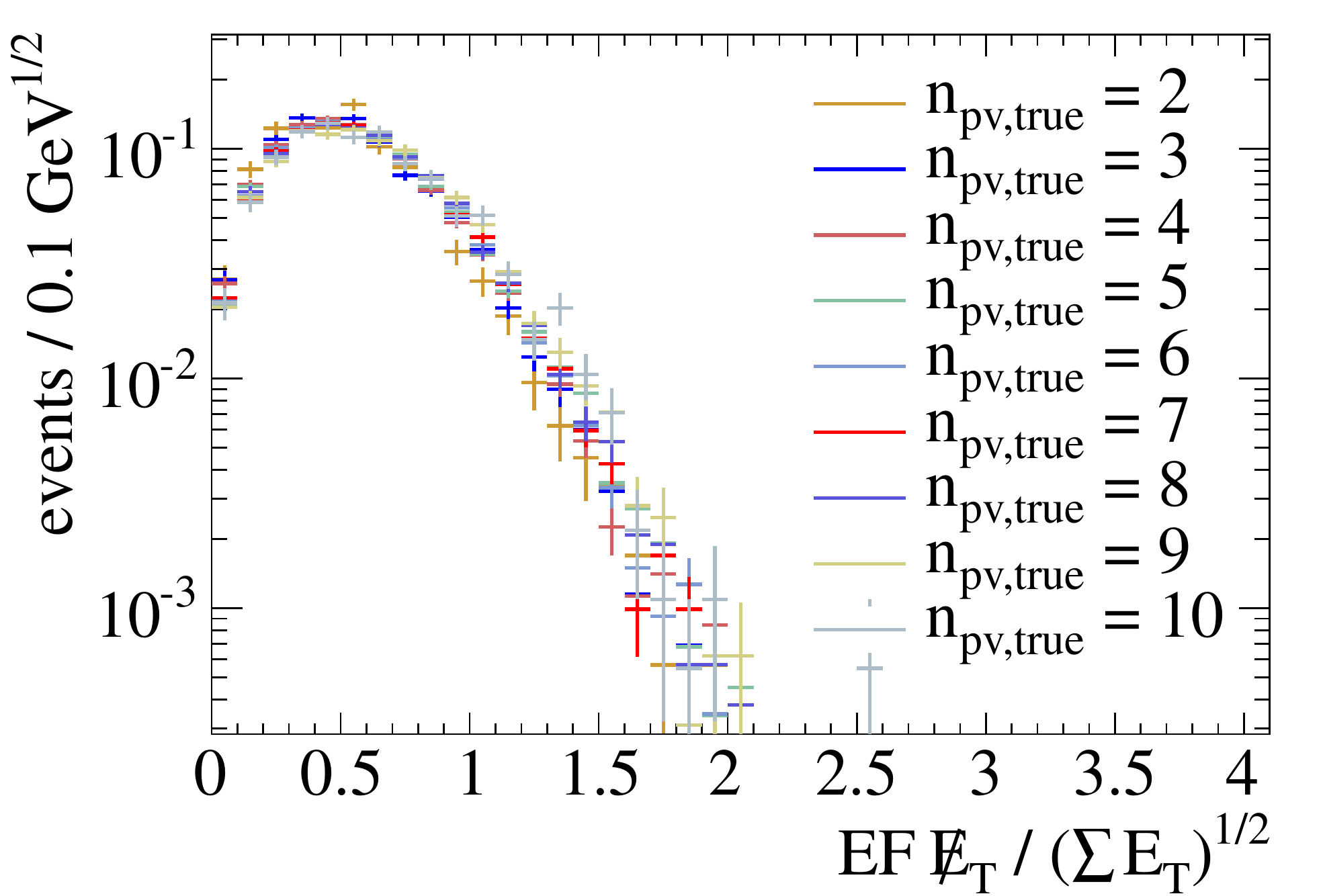}
  \incgraphics{width=\widthtwoplots}{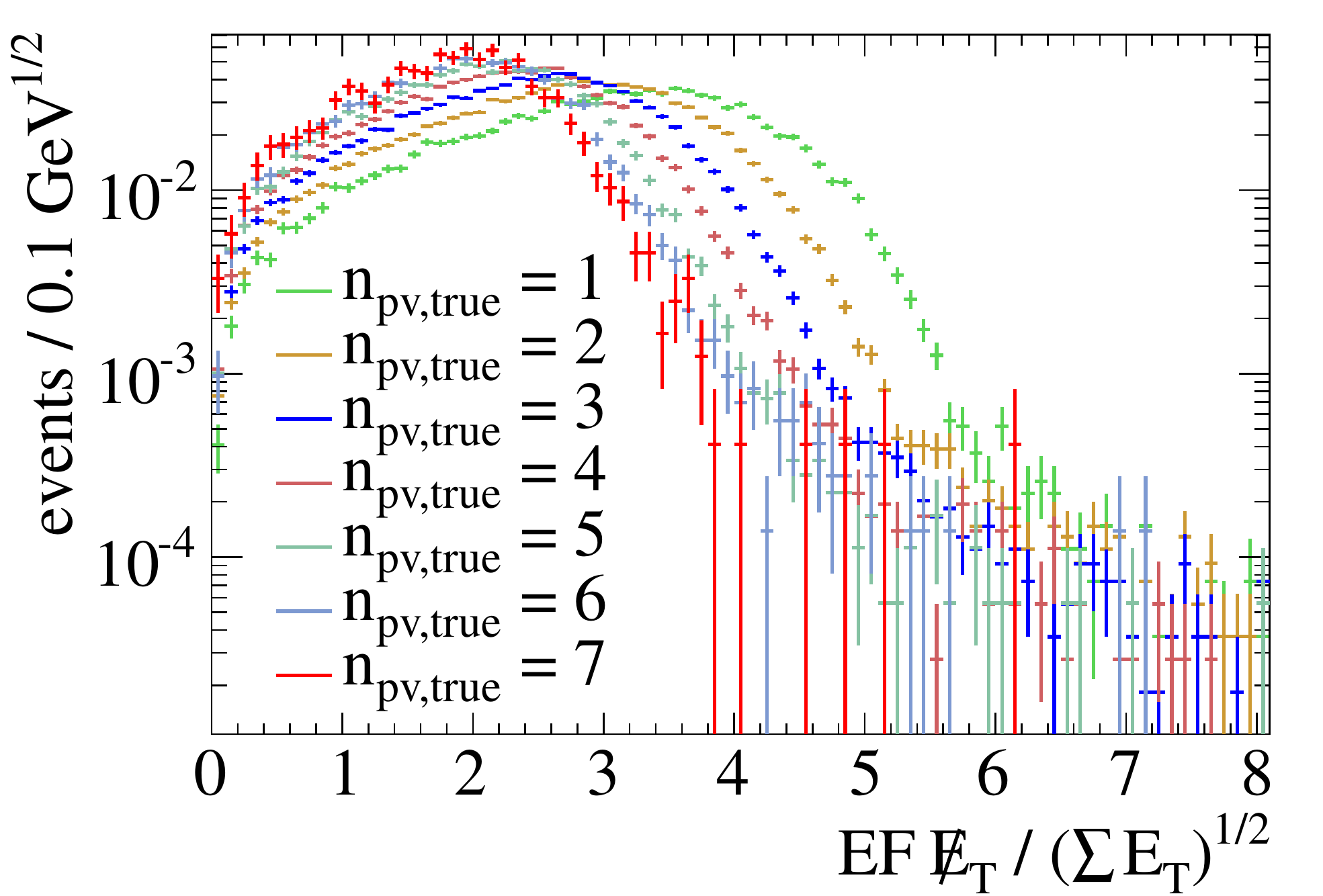}
  \caption{
    Distribution of the ratio $\met/\sqrt{\sumet}$,
    as measured by the trigger at Event Filter level
    in simulated Monte Carlo (\tagname{mc09}) events with pile-up,
    with both \met and \sumet in \unit[GeV]{}.
    The events are split up by the number of true primary vertices $n_\text{pv,true}$.
    Left: Minimum-bias events, right: $W\to e\nu_e$ decays.
    All distributions are normalized to unity to allow to compare their shapes.
  }
  \label{fig:results_met_model_predictions_xs_examples}
\end{figure}

One of the conclusions drawn from the dominance of fake \met with respect to the trigger rates
is to design a new type of triggers,
which use the \met significance instead of the measurement of \met alone.
The \met significance XS is given by the ratio of \met and $\sqrt{\sumet}$.
The actual implementation in the trigger is
\begin{equation}
  \text{XS} \definedas \frac{ \met / \unit[]{GeV} }{ A + B \sqrt{\sumet / \unit[]{GeV}} + C (\sumet / \unit[]{GeV}) },
  \label{eq:results_met_model_predictions_XS_definition}
\end{equation}
where $A$, $B$ and $C$ are constants which allow to account for the offset due to noise suppression
and to tune the online implementation of the trigger\footnote{
  The description here focuses on the main concept of the online implementation,
  leaving out additional accept and reject conditions which are introduced to ensure robustness in edge cases.
}.
Roughly speaking, resolution effects are proportional to the square root of the sum of the energy
which is deposited in the calorimeter.
Thus, the denominator normalizes the measured \met to the total activity in the event
and levels out the impact of in-time pile-up on the \met measurement,
which otherwise gives increasing contributions to \sumet as the instantaneous luminosity per bunch increases.
Following this argument, the rates of the \met significance triggers are expected
to be much more stable against in-time pile-up than the those of the \met or \sumet triggers.
The general behavior of this type of XS triggers is studied here on two types of Monte Carlo events.
Both plots in \Fig{fig:results_met_model_predictions_xs_examples} show the distribution of the ratio $\met/\sqrt{\sumet}$ measured at Event Filter level.
The different distributions within each plot stand for different selections
based on the true number of interactions per event, $n_\text{pv,true}$.
The left plot is produced from Monte Carlo with minimum-bias events
overlaid with an average number of five additional interactions from pile-up per event.
It clearly shows that in this sample, $\met/\sqrt{\sumet}$ does not depend on the number of interactions
because indeed all distributions have the same shape,
independent of the number of interactions,
and fall together when normalized to one,
as is done in this plot.
For this type of background rate, coming from minimum-bias events with negligible real \met,
the XS triggers are therefore expected to be stable against pile-up contributions.
The right plot shows the distribution of XS on a very different Monte Carlo sample
containing $W\to e\nu_e$ events,
again overlaid with on average five additional interactions from pile-up per event.
In this type of events, a significant contribution to \met is real \met from the neutrino.
With increasing pile-up, the normalization by $\sqrt{\sumet}$ shifts the peak in the spectrum to the left.
(Note that the range of the horizontal axis is twice the range in the left plot
so that the peak still lies at higher values on the horizontal axis than the peak from the minimum-bias background.)
A direct consequence is that the contribution from events with real \met to the trigger rate
will decrease with increasing pile-up,
which may be regarded an unusual if not undesired behavior.
In any case, this is important to remember
when considering to use this type of triggers for selecting events for physics analyses
or doing studies of trigger and signal efficiencies on these events.
It would also be interesting to study how big the gain of the XS triggers over \met triggers 
from the improved stability with respect to pile-up is,
\eg whether the efficiency for certain (signal) types of events is improved considerably.

\begin{figure}
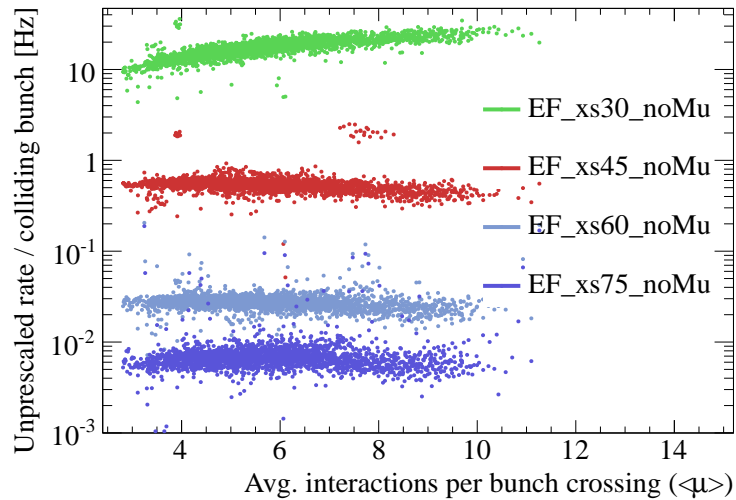

  \centering
  \incgraphics{width=\widthsingleplot}{{{COOL.combine_rates_own2011_rate2011K-G_avg8_EF_xs30_noMu.pdf-rate2011K-G_avg8_EF_xs75_noMu.pdf_182726-189425}}}
  \caption{
    Rates of several \met significance triggers with different thresholds
    as function of the average number of interactions per bunch-crossing \avgmu
    from 2011 data, periods 2011 G~--~K.
    The meaning of the threshold values is explained in the text.
  }
  \label{fig:results_met_model_predictions_xs_rates}
\end{figure}
The XS-based triggers have been running online since the middle of 2011 in \ATLAS.
In the plot in \Fig{fig:results_met_model_predictions_xs_rates},
the rates of four XS triggers from periods G -- K in 2011 are shown,
in which the \met significance triggers were running for the first time with the final calibration.
The nomenclature for these triggers is such that the number in the trigger name is
$10$ times the value of XS as defined in \Eq{eq:results_met_model_predictions_XS_definition}. %
Indeed, the measured trigger rates show no or only a weak dependence of the average number of interactions per bunch-crossing \avgmu.
\trigger{EF_xs30_noMu} is the lowest XS threshold,
which had to be heavily prescaled already in 2011~G, %
as can be deduced from the rate plot in \Fig{fig:results_met_model_predictions_xs_rates}
(note that the rate needs to be multiplied by the number of colliding bunches).
Nonetheless, from the right plot in \Fig{fig:results_met_model_predictions_xs_examples},
it already rejects most of this type of events for high levels of pile-up,
\ie has a low signal efficiency for $W\to e\nu_e$ events,
which shows that there are limitations for this new type of trigger.

\newpage
\setcounter{footnote}{1} %

\chapter{Search for Supersymmetry}
\label{sec:analysis_susysearch}

In this chapter,
a search 
in data taken with the \ATLAS detector in the year 2010
for signatures from decays of supersymmetric particles
is presented.
First, the search channel is introduced,
and the outline of the analysis is given.
Afterwards, the data sample and event selection are described,
including the object definitions on which the event selection is based
and the definition of the signal regions. %
This is followed by the description of the Monte Carlo samples for the Standard Model backgrounds
and the signal grids which implement the supersymmetric model scenarios.
The uncertainties on the measurements in data and Monte Carlo are discussed,
and the result of the analysis is presented in terms of the obtained event counts.
From the event counts and the systematic uncertainties
limits are computed on the mass parameters of two different supersymmetric models.

\section{Introduction}

\subsection{The Zero-Lepton Channel}

Supersymmetry can manifest itself through many different signatures in a collider experiment.
Based on the type of signature that is searched for,
a rough characterization of different analyses can be done,
for example by specifying the multiplicity of particles in the final state.
In this thesis, the zero-lepton channel is studied,
meaning that in the selected events no leptons appear in the final state.
Instead, the targeted signature consists of several hard (\ie high-energetic) jets
and a large amount of missing transverse energy,
which is carried away by a heavy particle that evades detection
because it only weakly interacts with matter.

\begin{figure}[t]
  \centering
  \incgraphics{width=\widthtwoplots}{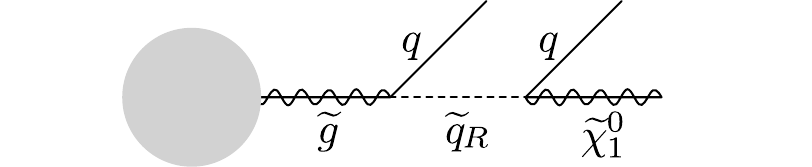}
  \incgraphics{width=\widthtwoplots}{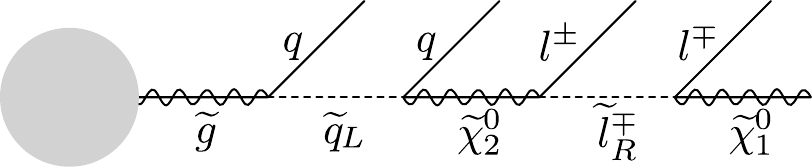}
  \hfill
  \caption{
    Left: A supersymmetric decay chain which produces the detector signature being searched for in the zero-lepton channel:
    jets, no leptons and missing transverse energy.
    \newline
    Right: For comparison, a longer supersymmetric decay chain in which an opposite-sign lepton pair is produced.
  }
  \label{fig:analysis_susy_decay_chain}
\end{figure}

The left diagram in \Fig{fig:analysis_susy_decay_chain} shows an example of a supersymmetric decay chain
which yields the detector signature of the signal events searched for in the zero-lepton channel.
Due to $R$-parity conservation,
the decay chain ends in the \ac{LSP},
and every event contains an even number of these decay chains.
The missing transverse energy is due to the \ac{LSP},
the hard jets are produced in the decays of the heavy supersymmetric particles.
For comparison, in the right diagram a longer supersymmetric cascade decay is shown,
in which an opposite-sign lepton pair is produced.
Which decay modes are preferred depends on the mixings of the neutralinos \cite{SUSYPrimerMartin1997}.
In the analysis at hand, where leptons are vetoed,
the term leptons is understood to only include electrons and muons, but not tau leptons.
Tau leptons are excluded because of their decay properties,
giving a very different detector signature.
They have very short lifetimes
and in contrast to electrons and muons have enough mass to decay into hadrons, giving rise to hadronic jets.
They are thus harder to identify than electrons and in particular muons,
which give clean signatures with high efficiencies and low fake rates.
Jets from hadronic tau decays are always interpreted as ordinary jets, no tagging is used.
To trigger on the signature of the supersymmetric signal events,
jet triggers, missing transverse energy triggers
or combinations of both can be used.
The benefits of using combined triggers have been demonstrated in \Sec{sec:results_trigger_rates_benefits_combined_triggers}.
Consequently, the primary triggers for the zero-lepton channel analysis are combined \jetmet triggers.
The efficiencies of these triggers have been discussed in \Sec{sec:triggerefficiencies},
for 2010 in particular in \Sec{sec:results_trigger_performance_measurements_data_2010}.
The offline cuts of the analysis ensure
that for events passing all cuts, %
the trigger can be assumed to be fully efficient.

\subsection{Outline of the Analysis}

In the general approach for a search for new physics that is persued here,
observed data is compared to theoretical expectations
which are predicted by the Standard Model of particle physics.
As it is not possible to describe the expected detector response in detail analytically,
Monte Carlo simulations are employed.
They encode the expectations of detector signals,
both from Standard Model processes and from new types of physics processes that are being searched for,
in terms of simulated data.
The simulated data is made to resemble the actual detector measurements as closely as possible.
If deviations are found in event counts, distributions of kinematic variables or other quantities,
which cannot be explained by uncertainties in the measurement itself,
these discrepancies may be interpreted in terms of physics beyond the Standard model. %
If the inclusion of the new physics processes can explain the data,
it has to be checked
whether this model is superior
to the wealth of competing models
which may also be used to explain the data.
If the model is better than all other (of equal simplicity),
then it has good chances of superseding parts of previous models,
but this is a process at timescales of years or decades.
If no deviations are found,
the results can still be interpreted in the framework of some concrete model including new physics,
by setting limits on the probability for this model to be realized in nature,
or at least on its parameters which have impact on the probability to find traces of this model at the detector.
Usually, new physics processes are supposed to lead to deviations in terms of an excess of events
rather than finding fewer events than expected.
An important counterexample are neutrino oscillations (cf. \Sec{sec:theory_neutrino_masses_sm}), %
which are the solution to the solar neutrino problem,
denoting the fact that when the first measurements of the flow of solar neutrinos to the earth were done,
much less neutrinos (of a given flavor) were found than expected.
This was resolved by %
giving up the assumption of massless neutrinos.
Assuming instead that neutrinos have a small but non-zero mass,
they can oscillate between different flavors.
Summing up neutrinos of all flavors, the flux was found to agree with the predictions again \cite{SNO2005}.
Thus in principle, new physics processes can lead to a smaller number of observed events,
but in the following, it will be assumed that Supersymmetry only yields a positive contribution to the event count
on top of the Standard Model background.

In counting analyses, the question whether physics beyond the Standard Model exists,
is answered by looking at the number or events or the distribution of kinematic quantities
after applying a set of cuts which define a \index{signal region}.
The term region refers to a certain subset of the phase space of events,
or, better, the physics objects the events are composed of.
The event count in the signal region is then compared to the expectation from Monte Carlo.
The significance of the result depends
on how well the contamination of the signal region with background events can be predicted.
As the Monte Carlo may suffer from uncertainties and systematic bias,
many analyses try to be as little dependent on Monte Carlo as possible,
and to constrain the prediction of the background contamination in the signal region
by additional measurements in \index{control region}s.
The control regions are designed to be enriched with background events,
in order to allow a better determination of the background level,
which then is extrapolated to the signal region.
The extrapolation is done using transfer functions,
which in the simplest case are linear factors,
giving the ratio of the selection efficiencies
\ie the number of events expected to be observed in the control and signal region.
In case of several types of backgrounds and several control regions,
the transfer functions become a set of linear equations. %
Transfer functions can be measured on Monte Carlo.
This still relies on Monte Carlo,
but no longer on the absolute normalization,
which is often most affected by uncertainties.
For the present analysis %
the official definitions of the signal regions
which are used within the \ATLAS SUSY group for the zero-lepton analysis of 2010 data are adopted.
This allows to validate the results of the official results,
and on top to take additional aspects into account.
An updated calibration for the luminosity is used,
giving a slightly smaller value for the integrated luminosity (see \Sec{sec:analysis_luminosity_calculation}).
The impact of pile-up,
which was neglected in the official analysis,
is studied,
and two different methods for the interpretation of the results in terms of limits on the mass parameters are compared.

The studies of the \ATLAS SUSY group show good agreement
between the Monte Carlo predictions for the Standard Model backgrounds
and cross-checks from measurements in control regions on real data.
Therefore,
the Monte Carlo predictions of the backgrounds are used as central values here,
and discrepancies taken into account in terms of systematic uncertainties.
The background studies using control regions which were done by the \ATLAS SUSY group are not repeated here.
Instead, for these aspects,
the following analysis relies on the results of the background studies given in the internal note \cite{ATL-PHYS-INT-2011-009},
the relevant parts of which will be summarized
when discussing the normalization of the background Monte Carlo samples in \Sec{sec:analysis_background_normalization}. %

\section{Integrated Luminosity, Data Samples and Trigger Selection}
\label{sec:analysis_luminosity_calculation}
The integrated luminosity is a measure for the amount of collision data that has been collected (cf. \Sec{sec:definition_luminosity}).
It is needed for almost every kind of analysis of collision data.
In analyses that compare data to Monte Carlo predictions, as is done here,
it enters as a normalization factor to scale the Monte Carlo event count to the number of events in the dataset.

For the calculation of the integrated luminosity of the 2010 dataset,
which is evaluated in the analysis presented in this thesis,
the official \propername{iLumiCalc} tool described in \Sec{sec:luminosity_measurement} is used.
As live-fraction trigger, needed to estimate the dead time of the trigger system, %
the minimum-bias trigger \trigger{L1_MBTS_2} is used,
and the database tag for the luminosity calibration is \texttt{OflLumi-7TeV-003},
following the recommendations for 2010 $pp$ collision data \cite{URL_iLumiCalc}. %
\afterpage{%
\renewcommand{\abovecaptionskip}{9pt}
\begin{landscape}
  \centering
  \vspace*{5mm} %
  \begin{threeparttable}
    \begin{tabular}{lllp{28mm}d{5.2}}
      \toprule
      Period & Run numbers & Trigger requirement for event selection & Online~trigger & \multicolumn{1}{p{20mm}}{Integrated luminosity in nb$^{-1}$} \\
      \midrule
      A & 152166 -- 153200 & \trigger{L1_J55} + \trigger{L2_j70} + \trigger{EF_EFonly_xe25_noMu} & \trigger{L1_J55} & 0.31\\
      B & 153565 -- 155160 & \trigger{L1_J55} + \trigger{L2_j70} + \trigger{EF_EFonly_xe25_noMu} & \trigger{L1_J55} & 7.69\\
      C & 155228 -- 156682 & \trigger{L1_J55} + \trigger{L2_j70} + \trigger{EF_EFonly_xe25_noMu} & \trigger{L1_J55} & 8.14\\
      D & 158045 -- 159224 & \trigger{L1_J55} + \trigger{L2_j70} + \trigger{EF_EFonly_xe25_noMu} & \trigger{L1_J55} & 278\\
      E & 160387 -- 161948 & \trigger{L1_J55} + \trigger{L2_j70} + \trigger{EF_EFonly_xe25_noMu} & \trigger{EF_j75_jetNoCut} & 906\\
      F & 162347 -- 162882 & \trigger{L1_J55} + \trigger{L2_j70} + \trigger{EF_EFonly_xe25_noMu} & \trigger{EF_j75_jetNoCut} & \numprint{1652}\\
      G1 & 165591 -- 165632 & \trigger{EF_j75_jetNoEF_EFxe25_noMu} &  & \\
      G2 & 165703 -- 165732 & \trigger{EF_j75_jetNoEF_EFxe25_noMu} &  & \\
      G3a & 165767 & \trigger{EF_j75_jetNoEF_EFxe25_noMu} &  & \\
      G3b & 165815 & \trigger{EF_j75_jetNoEF} + \trigger{EF_EFonly_xe25_noMu} & \trigger{EF_j75_jetNoEF}\rdelim\}{3}{1mm} & \multirow{3}{*}{536}\\
      G4 & 165817 -- 165818 & \trigger{EF_j75_jetNoEF} + \trigger{EF_EFonly_xe25_noMu} & \trigger{EF_j75_jetNoEF} & \\
      G5a & 165821 & \trigger{EF_j75_jetNoEF} + \trigger{EF_EFonly_xe25_noMu} & \trigger{EF_j75_jetNoEF} & \\
      G5b & 165954 -- 166143 & \trigger{EF_j75_jetNoEF_EFxe25_noMu} &  & \\
      G6 & 166198 -- 166383 & \trigger{EF_j75_jetNoEF_EFxe25_noMu} &  & \numprint{4915}\tnote{$\dagger$}\\
      H & 166466 -- 166964 & \trigger{EF_j75_jetNoEF_EFxe25_noMu} &  &  \numprint{6803}         \\
      I & 167575 -- 167844 & \trigger{EF_j75_jetNoEF_EFxe25_noMu} &  & \numprint{18319}         \\
      \midrule
      \multicolumn{4}{l}{Sum over all periods} & \numprint{33425}\\
      \bottomrule
    \end{tabular}
    \begin{tablenotes}\footnotesize
      \item[$\dagger$] $\unit[4915]{nb^{-1}}$ is the integrated luminosity for period G excluding G3b, G4, and G5a.
    \end{tablenotes}
    \caption{
      Integrated luminosities and triggers per period for $pp$ collisions taken in 2010.
      The online trigger used for data collection
      is given if it is different from the trigger requirement imposed offline to select events.
    }
    \label{tab:analysis_luminosities2010}
  \end{threeparttable}
\end{landscape}
\abovecaptionskipdefault
}

\label{sec:analysis_explain_triggers}
\Tab{tab:analysis_luminosities2010} gives an overview of the luminosity per data-taking period in 2010,
calculated using the \propername{iLumiCalc} tool.
Note that the resulting value for the total integrated luminosity in 2010 is about \percent{4} lower than the value
which was officially used by the \ATLAS SUSY group for the first analyses of 2010 data \cite{SUSYAnalyse0lepton2010}
because at that time the luminosity calibration implemented in \texttt{OflLumi-7TeV-003} was not yet available  (cf. \Sec{sec:luminosity_measurement}).
The run numbers are not consecutive because the pool of run numbers is shared between different \ATLAS projects,
and not all run numbers are given to runs in which data for physics analysis is taken.
In total, an integrated luminosity of
\begin{equation}
  \Lint^\text{2010} \definedas \int_{2010} \!\! \Linst(t) \intd t = \unit[33.4 \pm 1.2]{pb^{-1}} %
  \label{eq:analysis_integrated_luminosity_full_dataset_2010}
\end{equation}
is computed for the dataset taken in 2010,
which is used for the search for Supersymmetry in this thesis.
The data was taken by the \ATLAS collaboration between 30th March 2010 and 27th October 2010.

\Tab{tab:analysis_luminosities2010} also contains details
about the trigger requirements for the analysis per period.
Due to the constant changes in the trigger menu in the start-up phase of the LHC,
the trigger selection is not homogeneous,
which is reflected in the table.
The trigger used for the event selection in the analysis is a combined \jetmet trigger.
As this trigger was not available in the trigger menu in all periods,
a different, looser trigger is used to select events and the \jetmet trigger decision is emulated offline.
Therefore, in the table there is, for some periods, a distinction being made
between the trigger which was used online to record selected events
and which therefore is relevant to the luminosity calculation in terms of prescales,
and the trigger requirement that is later imposed to select events in the analysis.
The latter must, of course, select a subset of the previous.
Otherwise the event sample would lack events and probably be systematically biased in a very subtle way,
because the composition of events would be determined by the configuration of the trigger menu,
in particular which triggers were running in parallel,
rather than physics properties and the trigger efficiency.

In periods A~--~D, the \ac{HLT} was still being commissioned: it was running, but not rejecting events.
Every event that passed L1 was recorded,
and in an analysis it was enough to require a positive L1 trigger decision to select a valid event sample.
From period E onwards,
the HLT has commenced normal operation.
The \jetmet trigger was not yet implemented and events therefore selected
with the jet trigger \trigger{EF_j75_jetNoCut},
the suffix jetNoCut implying that still no cut was done in the HLT.
From period~G onwards,
the combined \jetmet trigger, the primary trigger for the analysis presented in this thesis,
became available and the online trigger decision could therefore be used directly to select events.
At the same time, event rejection by the HLT was also activated for the jet slice,
although the jet algorithms at EF were still in pass-through mode and have never been rejecting events in 2010
(therefore the suffix jetNoEF),
so that effectively only the L2 jet algorithm was activated beginning with period~G.
In a subset of runs (165815~-- 165821) from this period,
the \jetmet triggers were disabled,
but all events can be recovered by using \trigger{EF_j75_jetNoEF} which was running unprescaled in these runs.
For periods~H and~I, the \jetmet trigger could be used again.

The Monte Carlo version used in this thesis is \tagname{mc09\_7TeV}.
The data format used as input for the analysis scripts is the \tagname{NTUP\_SUSY} ntuple format (see \Sec{sec:definition_dataformats}).
This format can be read by \ROOT directly and is produced from data and Monte Carlo \ac{AOD} files
in the central production with the SUSY D3PD maker,
which is a derivate of the official D3PD maker and maintained by the \ATLAS SUSY group.
To ensure homogeneity of the data with regard to the code used for the reconstruction and the calibration,
all ntuples used in this thesis have been produced with version \tagname{06-20-00} of SUSY D3PD maker
in \Athena release \tagname{AtlasProduction 16.0.2.2}\footnote{%
  Note that although this is release 16 of \Athena,
  the reconstruction of the data has been done in release 15,
  \eg in \tagname{AtlasProduction 15.6.13.7} for datasets with tag \tagname{f299} such as
  \tagname{data10\_7TeV.00167776.physics\_JetTauEtmiss.merge.AOD.f299\_m639},
  from which \tagname{data10\_7TeV.00167776.physics\_JetTauEtmiss.merge.NTUP\_SUSY.f299\_m639\_p305} is produced.
}.
These datasets bear the processing tag \tagname{p305}
and have been produced without skimming
and thus contain all events from the respective datasets.
The same holds for the data samples.
The \tagname{NTUP\_SUSY} ntuples with production tag \tagname{p305} are used,
their project name is \tagname{data10}. %

\section{Event Selection}
\label{sec:analyis_description_cutflow}

For the analysis of \ATLAS 2010 data
implementing the search for Supersymmetry in the channel with jets, missing transverse energy and no identified leptons in the final state,
the official cuts developed within the \ATLAS SUSY group have been implemented.
It has been checked that the results on data
in terms of event counts after cuts
are consistent with the official analysis.
In particular, the number of data events reaching each of the signal regions
agrees exactly with the official analysis published in \cite{SUSYAnalyse0lepton2010}.
Details about the definition of the signal regions are given below.

The cuts used in the analysis are described in detail in this section.
To give a first idea of how the analysis proceeds,
the cuts can be grouped and summed up in the following scheme:
\begin{enumerate}
  \item Selection of collision candidates and event clean-up.
    This step filters out bad data from runs with known detector problems, but also implements the trigger selection.
  \item Reject events with reconstructed leptons to define the zero-lepton search channel.
    As mentioned in the introduction, only electrons and muons are considered in the lepton veto.
  \item Cuts on reconstructed jets to select events of a given jet multiplicity corresponding to the jet multiplicity of the signal regions.
    This defines the sub-channel.
  \item Cuts on \met which is also part of the baseline selection.
  \item Cuts on derived event variables that further refine the signal regions.
    These are \meff and \mttwo (cf. \Sec{sec:reconstruction_effective_contransverse_mass}).
\end{enumerate}

\begin{table}
  \centering
  \begin{tabular}{rlrd{5.2}c}
    \toprule
    Cutflow step & Cut description   & Event count & \multicolumn{1}{c}{$\epsilon\sub{cut}$ [\%]} & SR\\
    \midrule 
    1 & Raw events   & \numprint{268706385} &  & \\ %
    2 & GRL   & \numprint{229004362} & 85.2 & \\
    3 & Main trigger   & \numprint{3889200} & 1.7 & \\
    4 & Jet cleaning   & \numprint{3848197} & 98.9 & \\
    5 & Vertex requirement   & \numprint{3838202} & 99.7 & \\
    6 & Veto on crack electrons   & \numprint{3833805} & 99.9 & \\
          & \multicolumn{3}{l}{(Object identification and overlap removal)} &  \\
    7 & Veto on leptons   & \numprint{3787202} & 98.8 & \\
    8 & Leading jet $\pt > \GeV{120}$   & \numprint{1910736} & 50.5 & \\
    9 & \qquad Monojet %
          & \numprint{49839} & 2.6 & \\
    10 & \qquad Charge fraction  & \numprint{45237} & 90.8 & \\
    11 & Second jet $\pt >\GeV{40}$   & \numprint{1789007} & 93.6 & \\
    12 & \qquad Charge fraction  & \numprint{1763609} & 98.6 & \\
    13 & \qquad $\met > \GeV{100}$  & \numprint{9121} & 0.5 & \\
    14 & \qquad $\Delta\phi(\text{jet}, \met) > 0.4$  & 934 & 10.2 & \\
    15 & \qquad \qquad $\met/\meff > 0.3$ & 621 & 66.5 & \\
    16 & \qquad \qquad $\meff > \GeV{500}$ & 87 & 14.0 & A\\
    17 & \qquad $\mttwo > \GeV{300}$  & 11 & 1.2 & B\\
    18 & Third jet $\pt > \GeV{40}$   & \numprint{680058} & 38.0 & \\
    19 & \qquad $\met > \GeV{100}$  & \numprint{4510} & 0.7 & \\
    20 & \qquad $\Delta\phi(\text{jet}, \met) > 0.4$  & 397 & 8.8 & \\
    21 & \qquad $\met/\meff > 0.25$  & 198 & 49.9 & \\
    22 & \qquad $\meff > \GeV{500}$  & 66 & 33.3 & C\\
    23 & \qquad $\meff > \TeV{1}$  & 2 & 3.0 & D\\
    24 & Fourth jet $\pt > \GeV{40}$   & \numprint{152405} & 22.4 & \\
    \bottomrule
  \end{tabular}
  \caption{
    Overview of the complete cutflow of the analysis.
    A short description of the cuts is given
    together with the number of events in data left after each cut
    and the efficiency of the cut $\epsilon\sub{cut}$.
    In the last column, the four signal regions are denoted with the letters~A through~D (cf. \Tab{tab:analysis_overview_signal_regions}). %
    The indention of the cut description hints at the hierarchy of the cuts and is explained in the text.
  }
  \label{tab:analysis_cutflow_data}
\end{table}

\Tab{tab:analysis_cutflow_data} shows the sequence of cuts, collectively called \index{cutflow},
together with the number of events in data left after each cut.
The efficiency of the cuts $\epsilon\sub{cut}$ on data is also listed\footnote{
  Assuming Poisson statistics and using Gaussian error propagation,
  rarely more than two digits are significant for the efficiencies,
  even for large event counts.
}.
Note that not all cuts are part of the definition of a signal region.
A few steps in the cutflow are not evaluated when setting limits.
These are the cuts selecting events with a monojet (step 9 and 10) or $4$- (or more) jet topology (step 24).
They are included for completeness and in order to make sure that the numbering of the cuts is consistent everywhere.

About 270 millions of raw events from the \tagname{JetTauEtmiss} stream are processed,
leaving about 230 millions events in luminosity blocks which are marked as suitable for physics analysis
according to the \ac{GRL} defined by the SUSY group (cf. \Sec{sec:analysis_explain_grl}).
The triggers used to select a well-defined sample of events from the stream are listed in \Tab{tab:analysis_luminosities2010} as explained above.
For the Monte Carlo samples,
a combination of \trigger{L2_j70} and \trigger{EF_EFonly_xe25_noMu}
is used as trigger.
\trigger{EF_EFonly_xe25_noMu} refers to an emulated trigger (cf. \Sec{sec:tdaq_explain_trigger_emulation}),
which cuts at \GeV{25} on the measurement of missing transverse energy obtained by the trigger at Event Filter level.

\subsection{Jet Cleaning}
\label{sec:results_0lanalysis_jet_cleaning}
The \index{jet cleaning} procedure implies a veto on events
in which the reconstruction of one or more jets is considered to be unreliable
due to suspicious values in variables quantifying the reconstruction quality \cite{ATLAS-CONF-2010-038}. %
To this aim, every jet is classified as either good or bad\footnote{
  The third category containing ``ugly jets'' is not used in this analysis.
},
where different sets of criteria can be used to define bad jets.
Bad jets in this sense are energy deposits in the calorimeters from various sources
other than the true jets from the proton-proton interaction.
These include hardware problems, LHC beam conditions and cosmic-ray showers.

\begin{table}
  \centering
  \begin{tabular}{lcccc}
    \toprule
    Target && Loose & Tight \\
    \midrule
    EM coherent noise && $f_\text{EM} > 0.95$ and $|Q| > 0.8$ & $f_\text{EM} > 0.90$ and $|Q| > 0.6$ \\
    \multirow{2}{*}{HEC spikes} & \rdelim\{{2}{0mm} & $f_\text{HEC} > 0.8$ and $N_{90} \leq 5$ & $f_\text{HEC} > 0.3$ and $|Q| > 0.3$ \\
    && $f_\text{HEC} > 0.5$ and $|Q| > 0.5$ & $f_\text{HEC} > 1-|Q|$ \\
    \multirow{3}{7em}{Cosmic rays and beam background} & \rdelim\{{3}{0mm} & $|t\sub{jet}| > \unit[25]{ns}$  \\
    && $f_\text{EM} < 0.05$ & $f_\text{EM} < 0.10$ \\
    && $f_\text{max} > 0.99$ and $|\eta_\text{EM}| < 2$ & $f_\text{max} > 0.95$ and $|\eta_\text{EM}| < 2$ \\
    \bottomrule
  \end{tabular}
  \caption{
    Defining conditions for a jet to be flagged as bad
    according to the loose and tight jet cleaning recommendations \cite{ATLAS-CONF-2010-038,JetCleaningPresentation2010}.
    Only one of the conditions listed in the two rows %
    has to be fulfilled.
    The variables appearing in the table are explained in the text.
  }
  \label{tab:analysis_definition_bad_jets}
\end{table}
Both the loose and the tight definition for bad jets recommended by the Jet/Etmiss combined performance group
for detector data reconstructed with \Athena in release 15 are used \cite{URL_JetCleaning}.
A jet is marked as bad under the loose or tight selection, respectively,
if any of the conditions listed in the corresponding column in \Tab{tab:analysis_definition_bad_jets} is met.
Note that jets fulfilling any of the loose conditions also are considered bad 
when applying a selection with the stricter tight conditions.
Here, stricter selection means that more jets are considered bad and thus more events rejected,
\ie the set of bad jets under the tight selection is the union of
all bad jets from the loose selection
plus all jets fulfilling any of the conditions listed under the tight selection criteria.
The cuts listed in the table target different sources of potential mismeasurements or misidentifications of jets
and make use of the following variables:
\begin{itemize}
\item $f_\text{EM}$ and $f_\text{HEC}$ are the fraction of the energy of the jet deposited in the electromagnetic and the hadronic end-cap calorimeter, respectively.
\item $Q$ is the fraction of jet energy from LAr cells with a quality factor larger than 4000. %
  The cell quality factor is given by the quadratic difference $\sum_i (a_i-a_i^0)^{2}$ between the samplings of the
  measured pulse $a_{i}$ and the reference pulse shape $a_{i}^0$ that is used to reconstruct the cell energy.
  It can be used to detect energy accounted for by coherent noise.
\item $N_{90}$ is the minimum number of cells containing at least \percent{90} of the jet energy.
  An unusually small value for $N_{90}$ is a good indication for single-cell jets
  which are due to spikes affecting only one or a few cells due to cross-talk.
\item The timing variable $t\sub{jet}$ is the energy-squared-weighted cell time,
  defined with respect to the event time
  and used to find jets reconstructed from large out-of-time energy depositions,
  \eg due to photons produced by cosmic ray muons.
\item $f_\text{max}$ gives the maximum energy fraction deposited in only one calorimeter layer.
  In the central region,
  jets should spread out over more than one layer.
\item The $\eta_\text{EM}$ value used in this definition is the pseudorapidity of the jet at EM scale\footnote{
  Although a rescaling of the energy does not change the jet direction \eta,
  small discrepancies between the two \eta variables can be observed
  which may occur when calculating the direction from reweighted cell contributions.
  }. %
\end{itemize}
Events containing a bad jet according to the loose definition with $\pt > \GeV{20}$ at \EMJES scale are discarded.
The inefficiency of this cut is small,
as can be seen from \Tab{tab:analysis_luminosities2010}.
The tight definition is applied later on in the 2-jet sub-channel.
Events passing the jet cleaning are in the next step required to have at least
one reconstructed primary vertex with $\geq 5$ tracks associated to it,
to filter out empty events in which no hard interaction has taken place.
Afterwards, a veto is applied,
rejecting events if they contain a reconstructed electron
which fulfills the object identification criteria described below
and goes into the region $1.37 < |\eta_\text{cluster}| < 1.52$.
This transition region between the barrel and the end-cap electromagnetic calorimeters %
is expected to have poorer performance, %
because of the large amount of passive material in front of the first active layer of the electromagnetic calorimeter. %

\subsection{Object Identification}
\label{sec:analysis_object_identification_and_overlap_removal}
The next step is the \index{object identification} and overlap removal.
Here, no filtering of events is done,
but a promotion of candidate objects to (selected) reconstructed objects is made,
rejecting objects which do not fulfill the selection criteria.
This step is called object identification, object definition or object selection.
The criteria used for the selection of physics objects are summarized in 
\Tabs{tab:analysis_object_definitions_ej} and \ref{tab:analysis_object_definitions_m} and explained in detail below.
This step also includes an overlap removal in which objects are deleted
which overlap geometrically, are of different physics type (electron, muon or jet),
but are probably reconstructed from different interpretations of detector signals
caused by the same real physics object.

\begin{table}
  \setlength{\tabcolsep}{5pt} %
  \centering
  \begin{tabular}{lll}
    \toprule
    Cut & Electron & Jet\\
    \midrule
    Reconstruction algorithm & \code{AuthorElectron} or \code{AuthorSofte} & \revised{\Antikt}, \EMJES scale\\ %
    Acceptance & $\pt > \GeV{10}$ and $|\eta| < 2.47$ & $\pt > \GeV{20}$ and $|\eta| < 2.5$\\
    Isolation & --- & ---\\
    Overlap & removed if $0.2 < \Delta R_\text{min}^\text{jet} < 0.4$ & removed if $\Delta R_\text{min}^e < 0.2$\\
    Quality cuts & RobustMedium & ---\\
     & \tagname{OQMaps} & \\
    \bottomrule
  \end{tabular}
  \caption{
    Object identification criteria for electrons and jets.
    $\Delta R_\text{min}^\text{a}$ is the smallest distance in \eta-$\phi$ space to an object of type $a$.
  }
  \label{tab:analysis_object_definitions_ej}
  \resettabcolsep %
\end{table}

\subsubsection{Electrons}
\label{sec:analysis_object_definitions_electrons}
The selection of electron candidates follows the recommendations of the ElectronGamma combined performance group
for the identification of electrons within the region $|\eta|<2.5$ covered by the Inner Detector \cite{ElectronPerformance2010}.
It is based on a number of rectangular cuts (as opposed to multivariate techniques),
defined in three reference sets which allow to trade efficiency for purity of the electron sample obtained.
In the analysis, electrons reconstructed by either only the cluster-based algorithm,
or by both the cluster-based and track-based algorithm are used (cf. \Sec{sec:software_electron_reconstruction}).
The quality and purity cuts defined in the \tagname{RobustMedium} set are applied.
This requires the electron to satisfy both the loose \cite{ATLAS-CONF-2010-005}
and the updated (robuster) version of the medium cuts \cite{ATL-PHYS-INT-2010-126},
which, among other changes, takes into account differences in the shower shape between data and Monte Carlo. %
The acceptance cuts for electrons are $\pt > \GeV{10}$ and $|\eta| < 2.47$.
In addition, electrons must not fall into a region which suffers from detector problems.
This is ensured by using the Object Quality maps (\tagname{OQMaps}) provided by the ElectronGamma group. %
Selected electrons that fall into the transition region $1.37 < |\eta_\text{cluster}| < 1.52$
trigger the rejection of the event as a whole.

\subsubsection{Jets}
For the jets there are only few requirements on top of the jet cleaning as described above. %
All jets reconstructed by the \antikt algorithm with a radius parameter $\Delta R = 0.4$ from topological clusters are used, %
if they have $\pt > \GeV{20}$ and $|\eta| < 2.5$.
Another common value for the radius parameter in the jet reconstruction algorithms in \ATLAS is $\Delta R = 0.6$,
but as many jets are expected in typical signal events,
the smaller cone radius is used.
The energy of jets is calibrated at \EMJES scale,
as described in \Sec{sec:software_jet_energy_scale}.
This follows the recommendations of the \ATLAS Jet and \met Group for physics analyses on 2010 data \cite{URL_JetMETRecommendation2010}.

\begin{table}
  \centering
  \begin{tabular}{lll}
    \toprule
    Cut & Muon & \\
    \midrule
    Reconstruction algorithm & \tagname{Staco} (Combined or segment tagged) & \\
    Acceptance & $\pt > \GeV{10}$ and $|\eta| < 2.4$ & \\
    Isolation & $\sum\limits_{\Delta R < 0.2}\pt^\text{track} < \GeV{1.8}$ & \\
    Overlap & removed if $\Delta R_\text{min}^\text{jet} < 0.4$ & \\
    Quality cuts & $n_\text{pixel} \geq 1$, $n_\text{SCT} \geq 6$ & \\
     & for $|\eta| < 1.9$: $n > 5$ and $n_\text{TRT}^\text{outliers} < 0.9n$ & \\
     & for $|\eta| \geq 1.9$ and $n > 5$: $n_\text{TRT}^\text{outliers} < 0.9n$ & \\
     & for combined muons: $\chi^2_\text{match} \leq 150$ & \\
     & for combined muons with $\pt < \GeV{50}$: \\
     & \quad $p_\text{MS}^\text{extrapol.} - p_\text{ID} > -0.4 p_\text{ID}$ & \\
    \bottomrule
  \end{tabular}
  \caption{
    Object identification criteria for muons.
    The quality cuts correspond to the recommendation from the Muon combined performance group for release 15
    and are explained in the text,
    together with the variables they are based on.
  }
  \label{tab:analysis_object_definitions_m}
\end{table}

\subsubsection{Muons}
Muons are taken from the container filled by the \tagname{Staco} reconstruction algorithms.
Following the recommendations from the Muon combined performance group for release 15, %
combined muons and segment-tagged muons are used (cf. \Sec{sec:software_muon_reconstruction}),
and the respective cuts applied \cite{URL_MuonRecoRel15}:
To make sure that identified muons have traversed the Inner Detector
with hits associated to the fitted track from all sub-detectors,
the number of pixel hits $n_\text{pixel}$ is required to be at least one
and in addition six or more hits in the \ac{SCT} on the muon track are required.
Within the acceptance of the \ac{TRT},
a minimum number of associated hits is required,
plus a limit on the number of TRT outliers on the muon track $n_\text{TRT}^\text{outliers}$ is set,
where $n = n_\text{TRT}^\text{hits} + n_\text{TRT}^\text{outliers}$,
and $n_\text{TRT}^\text{hits}$ is the number of TRT hits on the muon track.
For combined muons, in addition to the above cuts,
the quality of the combination is tested
by requiring that the $\chi^2_\text{match}$ of the matching of the tracks
from the muon spectrometer and the Inner Detector at the perigee must be smaller than $150$,
and for combined muons with $\pt < \GeV{50}$,
a cut on the difference
between the extrapolated momentum measurement from the muon spectrometer $p_\text{MS}^\text{extrapol.}$
and the Inner Detector $p_\text{ID}$ is also applied.
The acceptance cuts for all muons are $\pt > \GeV{10}$ and $|\eta| < 2.4$.
To select only isolated muons,
a relatively tight cut is applied on the summed \pt of tracks,
which must be smaller than \GeV{1.8} within a cone of $\Delta R = 0.2$ around the muon.
\subsubsection{Missing Transverse Energy}
The missing transverse energy that is used in the analysis is called \tagname{MET\_EMJES\_Ref\-Final\_{\allowbreak}CellOutEM}
and calculated with an object-based refined \met algorithm,
as described at the end of \Sec{sec:software_reconstruction_met}. %
Two fixes need to be applied.
As the tool for the computation of the \met does not allow for the muon selection criteria from above,
the muon term is subtracted from the refined \met and replaced 
by adding the negative $x$ and $y$ components of the muon momentum to the \met components.
Similarly, the \met calculation is corrected
by subtracting $p_{x,y}$ from the missing energy
for all electrons that pass the RobustMedium,
but not the medium purity criteria,
and lie within a distance of $\Delta R<0.2$ to a jet.
Then, $p_{x,y}$ of the corresponding jet is added
to the respective components of the missing energy.
This corrects for the fact that the wrong electron identification is used in the \met computation in data. %
In addition to these fixes,
changes in \met arising from the calculation of the jet-energy scale and jet-energy resolution uncertainties described below are taken into account.

\subsubsection{Overlap Removal and Lepton Veto}
\inindex{Overlap removal}
Having thus selected the physics objects for the following analysis,
any occurring overlaps need to be resolved.
This is done according to the following strategy:
\begin{enumerate}
  \item Selected jets are removed if they fall within a cone of $\Delta R < 0.2$ around a selected electron.
    The idea behind this is that if both a jet and an electron are reconstructed so close geometrically,
    the real physics object is more likely to be an electron rather than a jet.
  \item Selected electrons are removed if they fall within a cone of $\Delta R < 0.4$ around one of the remaining selected jets,
    because they are then believed to have been produced in a (hadronic) jet.
  \item Along the same lines, selected muons are removed if they fall within a cone of $\Delta R < 0.4$ around one of the remaining selected jets.
\end{enumerate}
(The $\Delta R$ distance is computed using uncalibrated (EM scale) coordinates of the jet here.)

\inindex{Lepton veto}
Events that contain identified leptons (electrons or muons passing the above selection) after the overlap removal are vetoed.
This veto prevents events from entering the zero-lepton channel analysis
that are also selected in other SUSY search channels like the one-, two- or multi-lepton analyses
which include leptons in their selection.
It thereby avoids double counting of events,
so that it is easier to compute combined limits from the results of search channels with different lepton multiplicities.
Note that in constrast to this,
the jet selection within the zero-lepton channel is inclusive,
so that events passing the three-jet selection may also appear in the signal regions with two jets.
This needs to be remembered when setting limits and is discussed in \Sec{sec:analysis_interpretation}.
It is important for the limit-setting procedure,
because the signal regions cannot be taken as independent search channels.

\subsection{Jet Multiplicity and Sub-Channel Selection}
The rest of the cuts after the lepton veto no longer involve leptons,
but only jets, \met and derived event quantities, namely \meff and \mttwo.
The cuts are straightforward and lead to four different signal regions described below. %
The indentations in \Tab{tab:analysis_cutflow_data} mark branching points in the cutflow.
Cuts that are indented are not relevant for following cuts on a lower indentation level.
The cut in step 11, for example, is applied to all events left after step 8,
regardless of the two intermediate steps.

\subsubsection{Monojet Channel}
All sub-channels require a leading jet with $\pt > \GeV{120}$.
Steps 9 and 10 define the monojet channel.
In step 9, events are rejected if they contain any further %
jets with $\pt > \GeV{30}$
or if they contain jets (the leading or further jets with $\GeV{20} < \pt < \GeV{30}$)
that are marked as bad according to the tight selection criteria from \Tab{tab:analysis_definition_bad_jets}.
Step 10 reject events which have a jet with $\pt \geq \GeV{120}$, $|\eta| \leq 2$ 
and a charge fraction $f_\text{charge} \leq 0.02$,
where the charge fraction is defined as the summed \pt of all tracks associated with the jet,
divided by the \pt of the jet at \EMJES scale.
The monojet channel does not define any signal region, but is kept for completeness.

\subsubsection{2-Jet Channel} %
Steps 11 and 12 reject events if they do not have a second jet with at least $\pt > \GeV{40}$,
if they have a bad jet according to the tight definition
or a jet with $\pt \geq \GeV{120}$, $|\eta| \leq 2$
and a charge fraction $f_\text{charge} \leq 0.02$.
The last step of the baseline selection in the 2-jet channel then requires $\met > \GeV{100}$
in order to obtain an event sample for which the trigger can be assumed to be fully efficient.
Steps 14~-- 17 define the first two signal regions~A and~B (see below).

\subsubsection{3-Jet Channel}
Step 18 requires events to have a third jet with at least $\pt > \GeV{40}$.
In contrast to the 2-jet channel, in the 3-jet channel the tight jet cleaning is not enforced,
only a simplified additional jet cleaning is done:
Events are removed if any of the leading three jets falls within $|\eta| \leq 2$,
has a charge fraction $f_\text{charge} \leq 0.02$
and the fraction of its energy deposited in the electromagnetic calorimeter is $f_\text{EM} \geq 0.98$.
Step 19 completes the baseline selection in the 3-jet channel, requiring $\met > \GeV{100}$.
The following steps define the signal regions~C and~D,
with~D being a subset of~C.

\subsubsection{4-Jet Channel}
The 4-jet channel requires a fourth jet with $\pt > \GeV{40}$ and 
the same simplified jet cleaning as in the 3-jet channel, but extended to all four leading jets.
It is kept for completeness and is not used for limit setting.

\subsection{Signal Regions}
\label{sec:analysis_signal_regions} %

\begin{table}
  \centering
  \begin{tabular}{lcccc}
    \toprule
    Signal region & A & B & C & D\\ 
    \midrule 
    Cutflow step      & 16 & 17 & 22 & 23 \\ %
    Target final state & squark--squark & squark--squark & gluino--gluino & squark--gluino \\
    Jet multiplicity & $\geq2$ & $\geq2$ & $\geq3$ & $\geq3$ \\
      $\met / \meff$ & $>0.3$ & --- & $>0.25$ & $>0.25$ \\ 
      \meff  [GeV] & $>500$ & --- & $>500$ & $>1000$ \\
      \mttwo [GeV] & ---  & $>300$ & ---  & --- \\
    \bottomrule 
  \end{tabular}
  \caption{
    Overview of the signal regions defined for the analysis \cite{SUSYAnalyse0lepton2010}.
    A cut at $\met > \GeV{100}$ and $\Delta \phi(\text{jet}, \met) > 0.4$ is common to all four signal regions
    as well as other cuts from the baseline selection which is explained in the text
    and summarized in \Tab{tab:analysis_cutflow_data}.
  }
  \label{tab:analysis_overview_signal_regions}
\end{table}

Four \index{signal region}s are used in the analysis,
optimized for different supersymmetric final states.
The definitions of the signal regions are given in \Tab{tab:analysis_overview_signal_regions},
following \cite{SUSYAnalyse0lepton2010}\footnote{
  In some places in the internal note \cite{ATL-PHYS-INT-2011-009}, %
  the letters assigned to the signal regions are swapped, A$\leftrightarrow$C and B$\leftrightarrow$D.
}, together with the final state they target.
The cutflow for the baseline selection has been explained in the previous section.
On top of the baseline selection, all four signal regions have a cut 
on the minimum angle between the vector of the direction of the missing energy and
the direction of the three leading jets with $\pt > \GeV{40}$ at \EMJES scale
projected onto the transverse plain.
The minimum angle is denoted as $\Delta \phi(\text{jet}, \met)$ and required to be larger than $0.4$.
The signal regions are then defined by cuts on
the effective mass \meff, the stransverse mass $\mttwo$
or the ratio of missing transverse energy and effective mass $\met/\meff$
as summarized in \Tab{tab:analysis_overview_signal_regions} or the cutflow in \Tab{tab:analysis_cutflow_data}.
The cut on the ratio of missing transverse energy and effective mass $\met/\meff$
aims at quantifying the significance of the measured \met. %
As said above, the four signal regions are not orthogonal,
which is obvious for~C and~D,
where every event from~D is also contained in~C,
but also the other signal regions may overlap.

Signal regions~A and~B are optimized for squark-squark pair production
which, under the assumption that $\squark \to q\gluino$ is kinematically forbidden,
will give two jets from each of the squarks decaying to a quark plus a neutralino or chargino.
If squarks may decay into gluinos, even more jets are produced,
but to include all events only two jets are required.
Signal regions~C and~D require three or more jets
as they arise in the decay of pair-produced gluinos, $\gluino \gluino$,
or from the associated production of a squark together with a gluino, $\squark \gluino$.
The gluino can only decay through a squark, $\gluino \to q\squark$,
which increases the number of jets with respect to the direct production of squark pairs.
Cutting on a higher number of jets, as is done in the three-jet channel,
then gives better sensitivity to the SUSY signal over the background.
In signal region~D, tighter cuts than in signal region~C can be used
due to the higher total cross section of the associated production compared to gluino pair production. %

\section{Standard Model Backgrounds}
\label{sec:results_0lanalysis_backgrounds}
This section will discuss the Standard Model backgrounds
which are relevant for the search for Supersymmetry in the zero-lepton channel,
before in \Sec{sec:analysis_mc_sm_background} %
the Monte Carlo samples
that are used to model the Standard Model background will be presented in detail.

\inindex{Signal-to-background ratio}
In a search for new physics,
where ``new'' refers to any type of physics that goes beyond the current version of the Standard Model,
all known processes from the Standard Model comprise the background to the search.
Usually background processes dominate over the theorized new physics processes in terms of cross section and hence event counts;
back\-ground-free searches are the exception.
Using suitable cuts on kinematic variables, event properties or derived quantities,
the signal is separated from the background,
with the goal that in the signal regions,
the ratio of signal events over background events should be as high as possible.
The ratio of the number of signal events $s$ and background events $b$
is an important figure of merit when optimizing the cuts in the analysis.
Searches usually optimize the ratio \ssqrtb,
motivated by the fact that this ratio gives the approximate number of standard deviations $\sigma \sim \sqrt{b}$
that the background needs to fluctuate upwards to fake a potential signal.
For measurements, typically the ratio \soverb is considered instead. %
Note that \soverb is constant as function of the luminosity,
whereas \ssqrtb grows as the square-root of the luminosity.

The optimal cuts are a compromise between purity and efficiency in the signal region.
The purity measures the fraction of signal events in all events that reach the signal region, %
and the efficiency denotes the ratio of signal events that remain after the cuts defining the signal region
to the original number of signal events.
Monte Carlo serves as a test bed to optimize cuts with respect to purity and efficiency.
Depending on their similarity to the signal topology,
some types of backgrounds can be suppressed efficiently, others cannot.
Backgrounds that are indistinguishable from signal events, %
because they give the same detector signature,
are called irreducible.

The number of background events reaching the signal region needs to be estimated both for irreducible and reducible backgrounds.
It may prove more difficult to estimate the amount of background contamination 
for backgrounds which are easy to cut away, but have very large cross sections,
than for backgrounds which are more signal-like, but also have relatively small cross sections.
At the \ac{LHC}, in contrast to lepton colliders, the event rate is completely dominated by \ac{QCD} processes,
giving rise to events with lots of hadronic jets.
The cross section for this QCD background is many orders of magnitude larger
than for all potential signal processes (cf. \Fig{fig:lhc_cross_sections}).
QCD events are expected to contain only little real \met,
which can arise from neutrinos produced in jets,
but most of the \met in QCD events will be fake \met from mismeasurement of jets. %
The QCD background can therefore be strongly reduced by cutting on \met,
the ratio of \met and \meff,
or the angle between \met and the leading jet,
so that in the analysis, it will only be one of the minor backgrounds.
Still, the main challenge with respect to the QCD background is
that events with high jet multiplicities are difficult to model in Monte Carlo (see \eg \cite{WjetsD02011}), %
so that the predictions of the total cross section and the number of QCD events entering the signal region 
from Monte Carlo suffer from large uncertainties.

Standard Model processes that yield real \met and jets are very similar to the signal topologies and make up the main background.
Three types of events contribute to this background:
\begin{itemize}
  \item \Wjets.
    $W$ bosons that decay leptonically are an important background due to the \met from the neutrino created in the decay.
    If the lepton is misidentified, out of acceptance or is a tau lepton, the event will pass the lepton veto,
    and additional jets will make it pass the jet selection.
    The dominating contribution in the signal regions here comes from the decays of $W$ bosons including a tau lepton.
    In addition to leptonic $W$ decays, $W$ decays to bottom-antibottom quark pairs are considered explicitly.
  \item \Zjets. %
    \Zjets events are the second important background to the analysis.
    They can in particular give rise to an irreducible background of the zero-lepton search,
    if the $Z$ boson decays into a neutrino-antineutrino pair.
    Together with the additional jets, this gives a signature that exactly matches the topology
    of events which are looked for in the jet plus \met based zero-lepton search.
    Less likely to pass the selection are $Z$ boson decays to electron-positron and muon-antimuon pairs,
    because it is unlikely that both leptons fail to be identified,
    and therefore the lepton veto leads to a significant reduction of these two decay channels.
    For $Z\to\tau^+\tau^-$, the lepton veto is not effective,
    as tau leptons are not regarded as leptons here,
    still electrons and muons from tau decays may trigger the lepton veto.
  \item Top quarks.
    As the top quark is the heaviest of the Standard Model particles,
    it is not surprising that its production cross section depends strongly on the center-of-mass energy.
    At the moment, top quarks contribute one of the subleading backgrounds,
    in between QCD processes and electroweak vector boson production.
    Two different types of contributions are considered in the analysis,
    the production of top quark pairs and the less likely single top production.
    Top quarks can be assumed to always decay via $t \to W b$,
    so that every top quark produces at least one heavy $b$-jet,
    plus possibly \met from the subsequent decay of the $W$ boson.
    Top quark decays are therefore similar in their signature to \Wjets events which also produce jets and \met,
    so that most of the above said holds.
    Due the topology of their decays,
    there are a number of kinematic constraints which allow to identify top quark events,
    but this is not exploited in the present analysis.
\end{itemize}

\section{Monte Carlo for Standard Model Backgrounds}
\label{sec:analysis_mc_sm_background}

\subsection{Generators}

In summary, the following classes of background events are considered:
Decays of $Z$ and $W$ bosons,
decays of pairwise or singly produced top quarks,
and inclusive jet events.
The latter are modeled via $2\to2$ matrix elements
for the hard \revised{scattering} at leading order,
giving two hard jets \cite{Combridge1977234}, %
which make up the main component of the \ac{QCD} background with a very large cross section.
Top quark production is not included in the QCD background sample.

The $W$ and $Z$ samples are generated with \propername{Alpgen} \cite{Alpgen2003},
with \propername{Herwig} \cite{Herwig2001,Herwig2002} for the parton showering and fragmentation
and \propername{JIMMY} \cite{Jimmy1996} for the underlying event.
The top quark pair samples were generated with \propername{MC@NLO} \cite{MCatNLO2002,MCatNLO2003},
as were the samples with single top quarks \cite{MCatNLOSingleTop2006,MCatNLOWtop2008},
both in combination again with \propername{Herwig} and \propername{JIMMY}.
The inclusive jet samples were generated with \propername{Pythia} \cite{Pythia6.4}.

\inindex{Monte Carlo tune}
Different parton distribution functions and tunes of the minimum-bias and the un\-der\-ly\-ing-event description
are used in \ATLAS in combination with the different Monte Carlo generators \cite{ATL-PHYS-PUB-2010-002}.
The tunes are modifications of phenomenological model parameters
so that the MC predictions describe existing data (from other collider experiments such as those at the Tevatron) as well as possible.
For the generation of the inclusive jet samples in the \propername{Pythia} generator at leading order,
the modified leading order parton distribution functions \texttt{MRST 2007 LO$^*$} are used \cite{MRSTLO2008}.
They are an optimized version of the \ac{LO} parton distributions,
introduced to amend the flaws observed for both \ac{LO} and \ac{NLO} parton distributions.
They also aim at an improved consistency of cross sections and differential distributions with NLO calculations
when using this optimized parton distribution with \ac{LO} generators.
For the generation of the events with top quark decays with \propername{MC@NLO},
the \texttt{CTEQ6.6} \ac{PDF} set \cite{CTEQ6.6,URL_CTEQ} is used
for the matrix element, parton shower and underlying event.
The $W$ and \Zjets samples were generated using \texttt{CTEQ6L1} \cite{CTEQ6}. %

\subsection{Cross Sections and Background Normalization}
\label{sec:analysis_background_normalization}
\inindex{Background normalization}
A detailed list of the cross sections for the Monte Carlo samples that have been used
is given in the Appendix in \Tabs{tab:analysis_MC_datasets_details_WZ} and \ref{tab:analysis_MC_datasets_details_top_dijet}.
The cross sections $\sigma$ which are given in these tables
in most cases are the official cross sections used within the \ATLAS Supersymmetry group \cite{URL_SUSYxsec}.
They include \kfactors and generator efficiencies.
In general, whenever possible,
a data-driven normalization of the Monte Carlo samples,
using additional measurements in dedicated control regions,
is preferred over the normalization using the bare cross section as predicted from theory,
because this cross section, being an independent prediction, cannot account for efficiencies of the detector.
Furthermore, distributions of kinematic variables
may not be correctly reflected in the Monte Carlo simulations.
Elaborate studies of the background estimates from Monte Carlo have been done within the \ATLAS SUSY group,
using different approaches for each of the backgrounds listed above \cite{ATL-PHYS-INT-2011-009}.
In general, very good agreement with the Monte Carlo prediction is found.
This justifies to use the Monte Carlo predictions as a first approximation
and not to repeat similar studies in the scope of this thesis.
For completeness, the main results from \cite{ATL-PHYS-INT-2011-009} are briefly summarized in the following, %
before coming to the description of the normalization of the QCD background in \Sec{sec:analysis_normalization_qcd_background}.

\subsubsection{\texorpdfstring{\WJets and \ZJets} {W+Jets and Z+Jets}}

Cross sections at \ac{NNLO} for the $W$ and $Z$ samples have been computed with the \propername{FEWZ} code. %
For the $Z\to l^\pm l^\mp$ processes, $l \in \{e,\mu,\tau\}$,
the cross section at \ac{LO} given by the generator \propername{Alpgen}
has been scaled with a \kfactor common to all subprocesses,
such as to give a total cross section of \unit[1.069]{nb} for each of the three lepton flavors.
In the same way, the cross sections for $W^+\to l^+\nu_l$ and $W^-\to l^-\bar{\nu}_l$ %
have been scaled with a common \kfactor,
so that the total cross section amounts to \unit[10.46]{nb} per flavor.
The same cross section is assumed for all three flavors to enforce lepton universality,
which holds in weak interactions up to minor corrections from the mass differences of the produced particles.
Finally, for $Z\to\nu\bar{\nu}$, a NNLO cross section of \unit[5.817]{nb} is used as total cross section summing over the three lepton flavors.
Note that the sum of the cross sections for all $Z\to\nu\bar{\nu}$ processes in \Tab{tab:analysis_MC_datasets_details_WZ}
does not give \unit[5.817]{nb},
because the generator efficiency in this case is significantly smaller than $1.0$,
and as stated above the cross sections listed in this table are the final cross sections used for normalization,
which also include the generator efficiencies.
The production of bottom quark pairs in $W$ boson decays,
$W\to b\bar{b}$, is included via a separate MC sample.
This sample has been found to have a reasonably small overlap
with the other $W$~+ (light) jets samples %
of around $4$ \cite{ATL-PHYS-INT-2010-132} to \percent{10} \cite{URL_TOPxsec}, %
so that it can be used in combination with these.
The difference between the $W$~+ light jets samples and the $W\to b\bar{b}$ sample is
that the $W$~+ light jets samples include only hard jets from the matrix element
produced by gluons or quarks from the first two generations,
which are assumed to be massless for this purpose.
Jets arising from $b$ quarks can also be found in this sample,
then having been produced in the hadron shower,
but those jets will predominantly have $\pt < \GeV{15}$ \cite{ATL-PHYS-INT-2010-132}. %
For the cross sections of the samples describing this decay channel,
the cross section computed by the generators is used %
and normalized with a \kfactor of $1.22$ as given in \cite{ATL-PHYS-INT-2010-132}. %

\label{sec:analysis_explain_SUSY_WZ_uncertainties}
For the $W$~and \Zjets background, the number of events in the data sample is too small
to allow for a purely data-driven estimate of the background normalization. %
Hence, the prediction from \propername{Alpgen} Monte Carlo is used as central value,
which can be justified by the good agreement with data.
The systematic uncertainties on the central value,
such as the jet-energy scale, the jet-energy resolution and luminosity,
are derived from Monte Carlo,
as explained in \Sec{sec:analysis_uncertainties}. %
Other uncertainties arise from control measurements,
where the control region statistics,
being the dominant uncertainty in the comparison between data and Monte Carlo,
is taken as a systematic uncertainty of the Monte Carlo predictions.
For \Zjets, the control measurement is done on a sample of events with $Z\to l^+l^-$ decays and additional jets.
Here, the uncertainties are the control region statistics and the acceptance,
for which a conservative \percent{100} for the unmeasured phase space region at high \eta is assumed.
The uncertainty from the control region statistics is taken as a replacement for the theoretical uncertainties,
which are expected to be smaller.
It includes, for the high mass signal regions,
for which the $\meff$ or $m_\text{T2}$ cuts on the control regions had to be relaxed to have any events at all,
an uncertainty from the necessary extrapolation from the control region to the signal region.
Both acceptance and control region statistics give contributions to the uncertainty
which are of the size of the jet-energy scale uncertainty which is the dominating uncertainty otherwise (cf. \Tab{tab:analysis_uncertainties_values}). %
For \Wjets, the control measurement uses $W\to l\nu$ + jets. %
In addition to the control region statistics,
the lepton reconstruction efficiency is taken into account as a systematic uncertainty.
For both $W$ and \Zjets, alternative methods have been tested
and have been found to give consistent results among each other and with the \propername{Alpgen} predictions.
The uncertainty from the control region statistics is always the dominating uncertainty,
again together with the jet-energy scale uncertainty.

\subsubsection{Top Quarks}

For the Monte Carlo sample with decays of pair-produced top quarks,
the \ac{NNLO} cross section of \unit[160.79]{pb} declared in the \ac{AMI} is used,
which is consistent with the value recommended by the SUSY group.
This cross section is shared between two subsamples,
one of which includes only events where both top quarks decay hadronically, $t\bar{t} \to WWbb \to bbqqqq$,
and the other the rest of events where one or both top quarks decay leptonically, $t\bar{t} \to WWbb \to bbqql\nu \text{ or } bbl\nu l\nu$.
Top quark decays give a very high jet multiplicity and, in case of leptonic decays, also \met,
and taking into account the comparably large cross section at the center-of-mass energy of the LHC,
they are an important background.
For the production of single top quarks in the $s$ and $t$ channel,
a \ac{NLO} %
cross section of \unit[7.15]{pb} and \unit[0.47]{pb}, respectively,
and \unit[14.6]{pb} for the associated production of a top quark and a $W$ boson is assumed. %
This is similar to the values recommended by the SUSY group and follows \cite{ATL-PHYS-INT-2010-132}. %
A recent computation including higher-order corrections at next-to-next-to-leading-logarithm accuracy %
gives slightly higher values for $s$-channel and $Wt$ production \cite{Kidonakis2010}.

Top events are a sub-leading background in the signal regions.
A partially data-driven cross-check has been done,
using a control region measurement in the same way as for the other backgrounds.
The control region has the same jet requirements as the signal regions,
but with an additional muon, a $b$-tagged jet and $\GeV{40} < \met < \GeV{100}$.
No additional uncertainties on top of the uncertainties
which are discussed in detail in \Sec{sec:analysis_uncertainties} %
are assumed in \cite{ATL-PHYS-INT-2011-009}.
Also, no theoretical uncertainty is given,
because it is found to be small compared to statistical and jet-energy scale uncertainties
from several other cross-checks.

\subsubsection{Inclusive Jets (QCD)}

\begin{table}
  \centering
  \begin{tabular}{clrr}
    \toprule
    Subsample & MC ID & $\hat p_{T,\text{min}}$ [GeV] & $\hat p_{T,\text{max}}$ [GeV]\\
    \midrule
    J0 & 105009 & 8 & 17\\
    J1 & 105010 & 17 & 35\\
    J2 & 105011 & 35 & 70\\
    J3 & 105012 & 70 & 140\\
    J4 & 105013 & 140 & 280\\
    J5 & 105014 & 280 & 560\\
    J6 & 105015 & 560 & 1120\\
    J7 & 105016 & 1120 & 2240\\
    J8 & 105017 & 2240 & $\infty$\\
    \bottomrule
  \end{tabular}
  \caption{The $\hat {p}_T$ intervals covered by the subsamples of the \propername{Pythia} Monte Carlo inclusive jet samples used in \ATLAS.}
  \label{tab:analyis_Pythia_dijet_ptslices}
\end{table}

\label{sec:explain_qcd_samples}
The inclusive jet sample generated with \propername{Pythia}, modeling the \ac{QCD} background, is split into several subsamples.
The splitting is implemented by specifying a \pt range,
setting the \code{CKIN(3)} and \code{CKIN(4)} variables of the \propername{Pythia} generator
which determine the range of allowed $\hat{p}_T$ values for the hard $2\to2$ process.
$\hat{p}_T$ is the transverse momentum defined in the rest frame of the hard interaction \cite{Pythia6.4}. %
This slicing is advisable because of the steeply falling cross section as function of \pt,
which would otherwise make it impossible to cover all of the large \pt range with reasonable statistics.
The large differences in the cross sections between the slices
that can be found in \Tab{tab:analysis_MC_datasets_details_top_dijet}
make obvious the need for slicing to get sufficient statistics.
How the slicing translates into $\hat{p}_T$ intervals is shown in \Tab{tab:analyis_Pythia_dijet_ptslices}.
In general, the higher the number in J$n$, the higher the $\hat{p}_T$ values and consequently the lower the cross section.
Samples J0 through J7 are exclusive samples, J8 is the highest inclusive bin without an upper bound on $\hat{p}_T$.
J8 is %
not available as \tagname{NTUP\_SUSY} and therefore not included in the analysis.
Its cross section including \kfactor is $\unit[8\ten{-6}]{pb}$ %
and thus negligibly small.
The cross sections given in \Tab{tab:analysis_MC_datasets_details_top_dijet} for the \ac{QCD} samples
are the generator cross sections provided in the \ac{AMI},
multiplied by a scale factor of $1.28$ that is discussed below.
The values used by the \ATLAS SUSY group differ from these by about \percent{1}.
QCD creates an overwhelming background before cuts,
which can be efficiently reduced to give a small background after cuts in the signal regions.
It is nevertheless difficult to model in Monte Carlo, and therefore needs to be well controlled.
QCD events that enter the signal regions must have high jet multiplicity,
and are likely not well modeled in $2\to2$ \propername{Pythia} Monte Carlo.
The baseline method in \cite{ATL-PHYS-INT-2011-009} is a control region measurement based on an inverted $\Delta\phi$ cut.
From the control region measurements,
an effective scale factor is derived
\begin{equation}
  k_R = \frac{N_\text{data}^\text{CR} - N_\text{MC(W+Z+top)}^\text{CR}}{N_\text{MC(QCD)}^\text{CR}}
  \qquad \text{ with } \qquad
  N_\text{exp}^\text{SR} = k_R N_\text{MC(QCD)}^\text{SR}.
  \label{eq:analysis_QCD_kfactor}
\end{equation}
An inverted $\met / \meff$ cut has also been checked
instead of the inverted $\Delta\phi$ cut,
and gives results which are well compatible within the uncertainties.
The reason to prefer the inverted $\Delta\phi$ cut is that it contains events
with large missing momentum due to jet mismeasurements so that more systematics will cancel out.
The central value of the QCD background normalization is not the bare Monte Carlo prediction,
but its value multiplied by the $k_R$-factor.
The uncertainties on this include generator systematics, heavy flavor systematics
and non-QCD background uncertainties. %
The combined uncertainties on the nominal value for the QCD prediction
lie above \percent{100} for all four control regions,
because a conservative \percent{100} systematic uncertainty
from varying detector conditions is assigned to the ratio. %

The scale factor used for normalization of the QCD Monte Carlo to data
that has been adopted in this thesis is $k_R = 1.28$.
It is motivated below and consistent with all values for $k_R$ from \cite{ATL-PHYS-INT-2011-009}. %
It can be regarded as a simple normalization,
making the rate of data and Monte Carlo agree after the \met cut.

\subsection{Normalization of the QCD Background} %
\label{sec:analysis_normalization_qcd_background}
\begin{figure}
  \centering
  \incgraphics{width=\widthsingleplot}{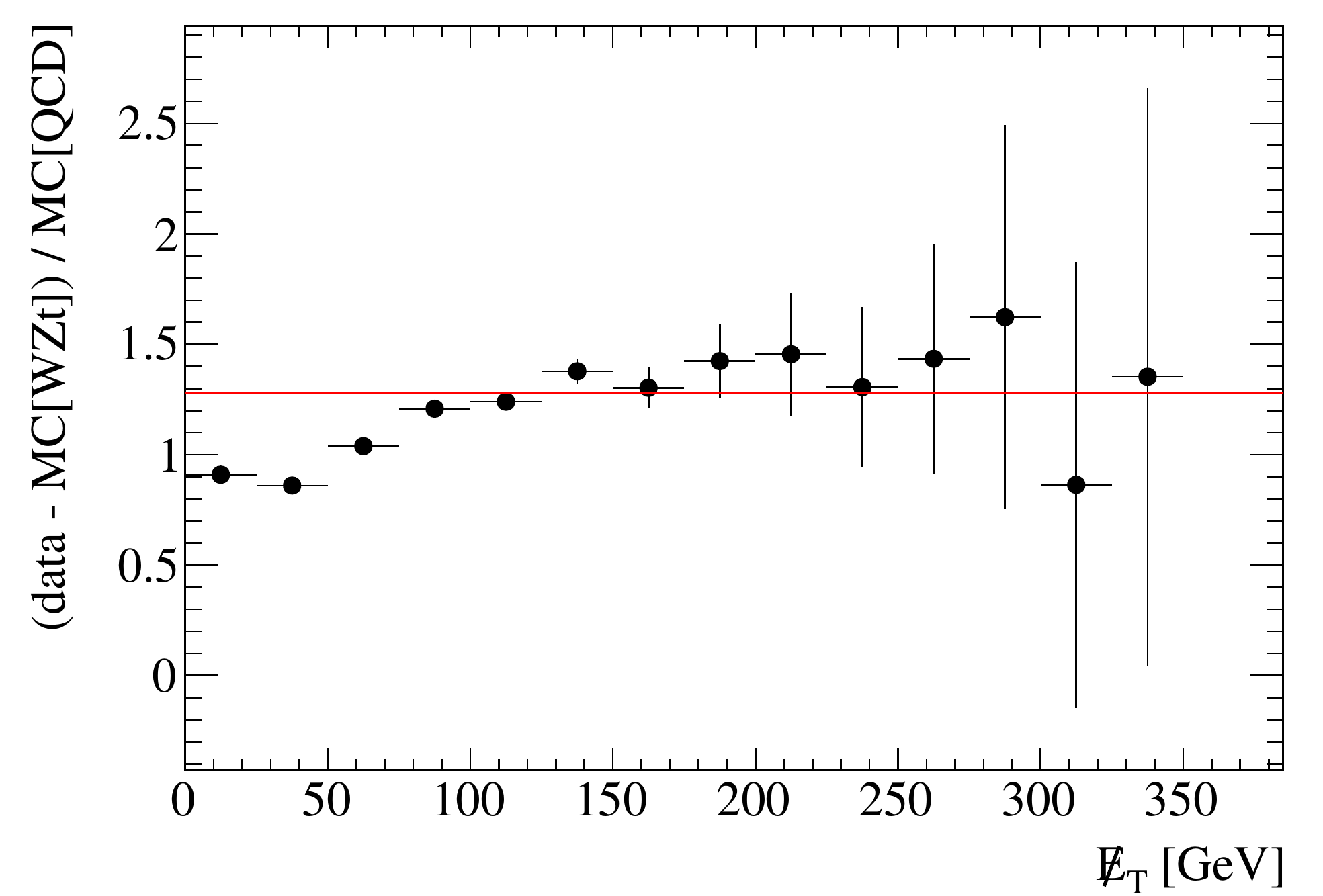}
  \caption{
    Ratio of the estimation of the number of QCD events in data and in Monte Carlo,
    using the missing transverse energy \met for the binning of the horizontal axis.
    The red line marks the value $1.28$ of the scale factor that is applied to the cross section
    of the QCD Monte Carlo in the analysis.
    ``MC[WZt]'' denotes the sum of events in weak boson and top quark Monte Carlo samples.
  }
  \label{fig:analysis_qcd_kfactor_met}
\end{figure}

\inindex{QCD normalization}
\Fig{fig:analysis_qcd_kfactor_met} shows the factor for the normalization of the \ac{QCD} background
that would be needed to make data to Monte Carlo agree, binned in \met.
It is computed from a direct comparison of the Monte Carlo and data
after the cuts up to and including step 12 from the cutflow in \Tab{tab:analysis_cutflow_data},
\ie before the \met cut in the 2-jet channel.
This scale factor is computed for each bin according to \Eq{eq:analysis_QCD_kfactor}
as the ratio of the
difference of the event counts in data and the sum from all background Monte Carlo samples,
apart from the \propername{Pythia} QCD samples,
over the number of QCD events in Monte Carlo.
Obviously, the scale factor is not independent of \met.
For small \met it is below one,
for large values it settles around a value
that is consistent with $1.28$, the value derived by the \ATLAS SUSY group from different data-driven methods as said above.
The good agreement between these values
can be taken as a sanity check and motivation to use the value of $1.28$
as scale factor for the normalization of the \propername{Pythia} QCD samples.

\begin{figure}
  \centering
  \incgraphics{width=\widthtwoplots}{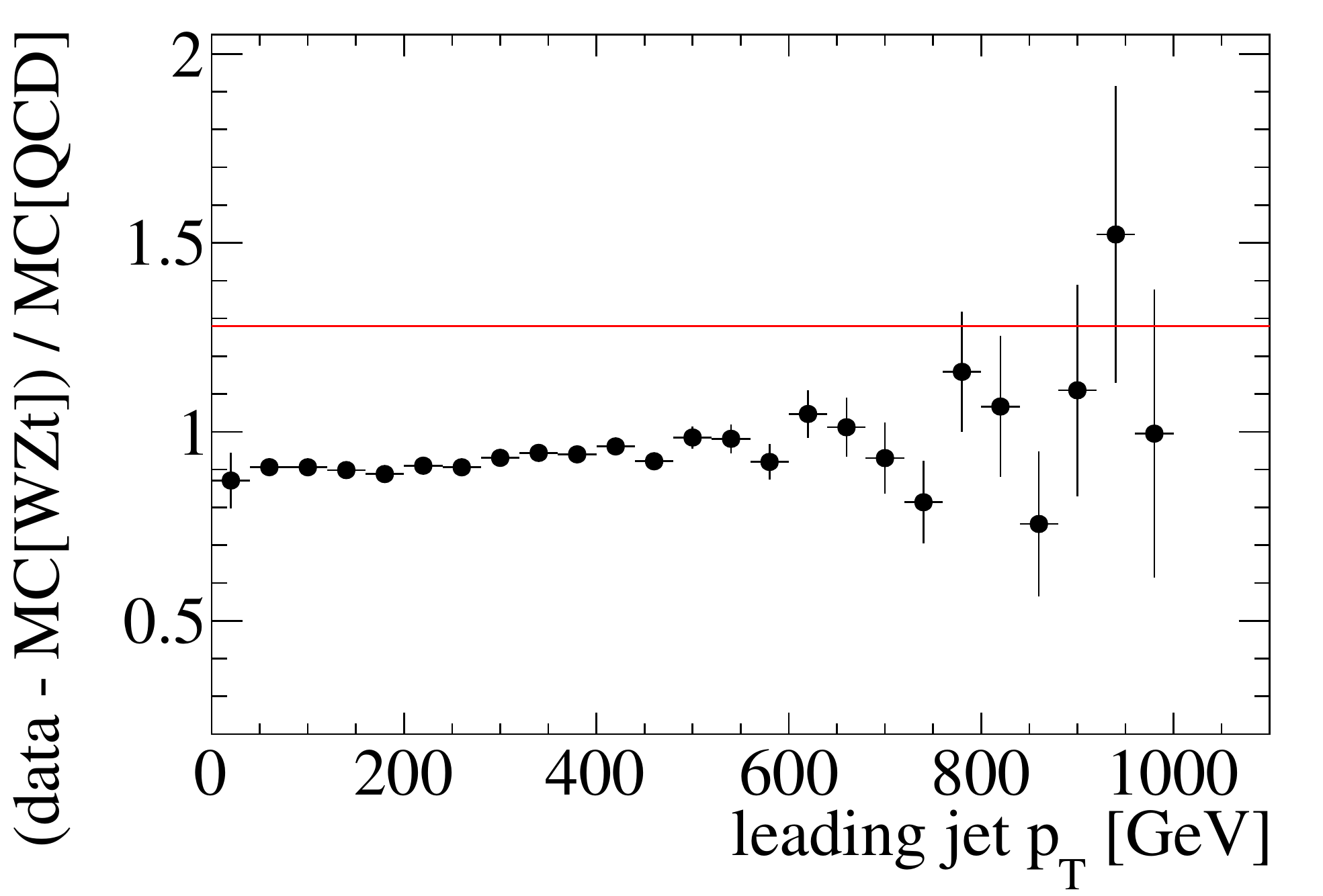}
  \hfill
  \incgraphics{width=\widthtwoplots}{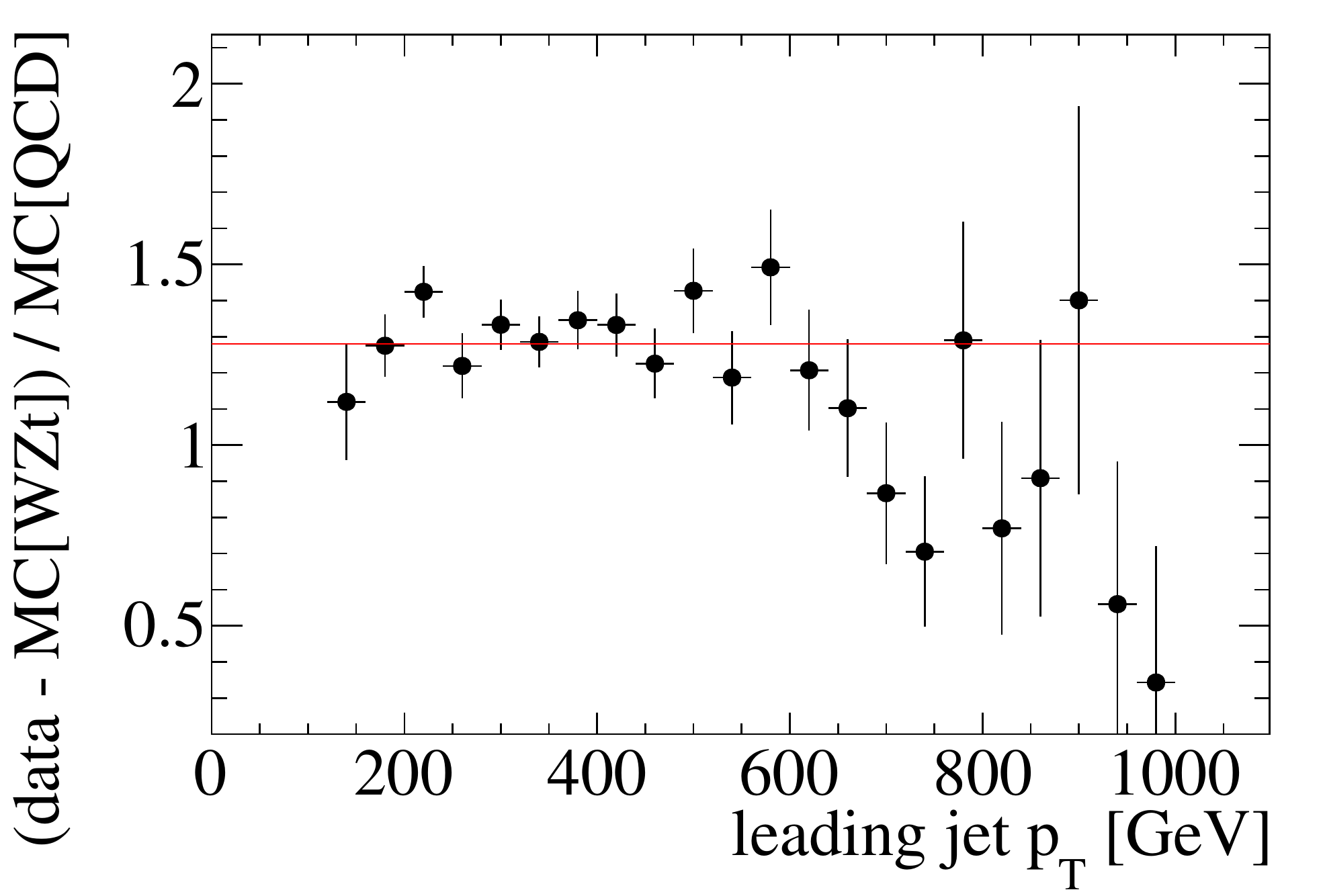}
  \caption{
    Ratio of the estimation of the number of QCD events in data and in Monte Carlo,
    using the \pt of the leading jet at \EMJES scale for the binning of the horizontal axis.
    The red line again marks the value $1.28$.
    The left plot is before, the right plot after cuts on \met and leading jet \pt.
  }
  \label{fig:analysis_qcd_kfactor_jets}
\end{figure}

The dependence of the scale factor on \met indicates
that the distribution of \met in Monte Carlo
does not describe the shape of \met in data well,
at least for small values
and especially in the \propername{Pythia} QCD sample
which makes up the dominating background at this stage (cf. \Fig{fig:analysis_distribution_cut11MET}). %
However, this part of phase space is cut away by the baseline selection cuts which cut on $\met > \GeV{100}$.
After the cut on \met at \GeV{100},
a normalization factor of $1.28$ can be expected to give a good agreement of data and Monte Carlo.
\Fig{fig:analysis_qcd_kfactor_jets} shows scale factors which are computed in the same way,
but binned in another important kinematic variable, the \pt of the leading jet.
The two plots show the distributions at different steps in the cutflow (cf. \Tab{tab:analysis_cutflow_data}).
The left plot in this figure is after step 7,
\ie directly after the lepton veto and without any cuts on the leading jet.
The dependence of the scale factor on this variable is much less pronounced.
The scale factor is smaller than $1.28$ over the whole range,
but this is due to the dominance of events with small \met,
which have a smaller scale factor as shown in the previous plot in \Fig{fig:analysis_qcd_kfactor_met}.
After applying the cuts up to step 13,
thereby including the cut $\met > \GeV{100}$,
the scale factors again are consistent with $1.28$ over the whole range,
as can be seen in the right plot in the same figure.

\subsection{Pile-up Effects in Monte Carlo}

The policy for producing Monte Carlo with pile-up contributions has changed from 2010 to 2011,
following the evolution of the beam conditions.
In 2010, in addition to the samples without pile-up,
samples have been produced in which the events
are overlaid with minimum-bias events before the reconstruction step.
The average number of additional minimum-bias events was fixed to two or five,
with a large bunch spacing of \unit[900]{ns} and accordingly only in-time pile-up. %
In 2011, instead of using a fixed average number of overlaid events,
this number was varied,
with an equal number of events being produced for Poisson expectation values of six to ten,
and less for lower and higher values (from zero up to 18). %
Moreover, the simulated bunch train structure was modified,
using now three bunch trains with \unit[225]{ns} separation
and a bunch spacing of \unit[50]{ns} within the trains,
consistent with the online conditions and leading to both in-time and out-of-time pile-up.

The Monte Carlo samples which are used to model the background as described above do not contain pile-up contributions.
However, in the analysis which is presented in this chapter,
the impact of pile-up is taken into account in terms of a systematic uncertainty,
using a different Monte Carlo reconstruction that includes pile-up contributions.
The Monte Carlo for all background types described above,
except for the QCD samples,
are available without pile-up,
as well as with pile-up from two overlaid minimum-bias events.
The baseline analysis is done using the Monte Carlo samples without pile-up contributions.
The difference to this central value when using the pile-up samples instead,
is taken to be a systematic uncertainty from pile-up contributions.
This approach is chosen,
because a reweighting of the Monte Carlo events to match the distribution of pile-up in data is not possible.
Such a reweighting would require to have Monte Carlo events which are overlaid with a range of different numbers of additional pile-up interactions.
As for the \propername{Pythia} QCD samples no pile-up version is available,
an estimation of the size of the pile-up uncertainty has been tried on \propername{Alpgen} QCD samples instead.
It has been found that the size of these samples is too small to have sufficient statistics in the signal regions.
Therefore, for the QCD samples,
the pile-up uncertainty is set to \percent{100}. %

For the signal samples described in \Sec{sec:analysis_SUSY_GRIDs},
no Monte Carlo version with overlaid pile-up is available.
Supersymmetry events in general are assumed to give hard (high energy) objects,
whereas the overlaid pile-up events are rather soft
and contribute little to the variables that are cut on to select events.
Therefore, for the SUSY signal,
neglecting pile-up ought to be an acceptable approximation for 2010, %
where the average number of interactions per bunch crossing in most runs lay below $3$. %
For 2011, this approximation should be revisited,
the pile-up level being several times as high as in 2010.

\section{Signal Grids}
\label{sec:analysis_SUSY_GRIDs}

\begin{figure}
  \centering
  \incgraphics{width=\widthtwoplots}{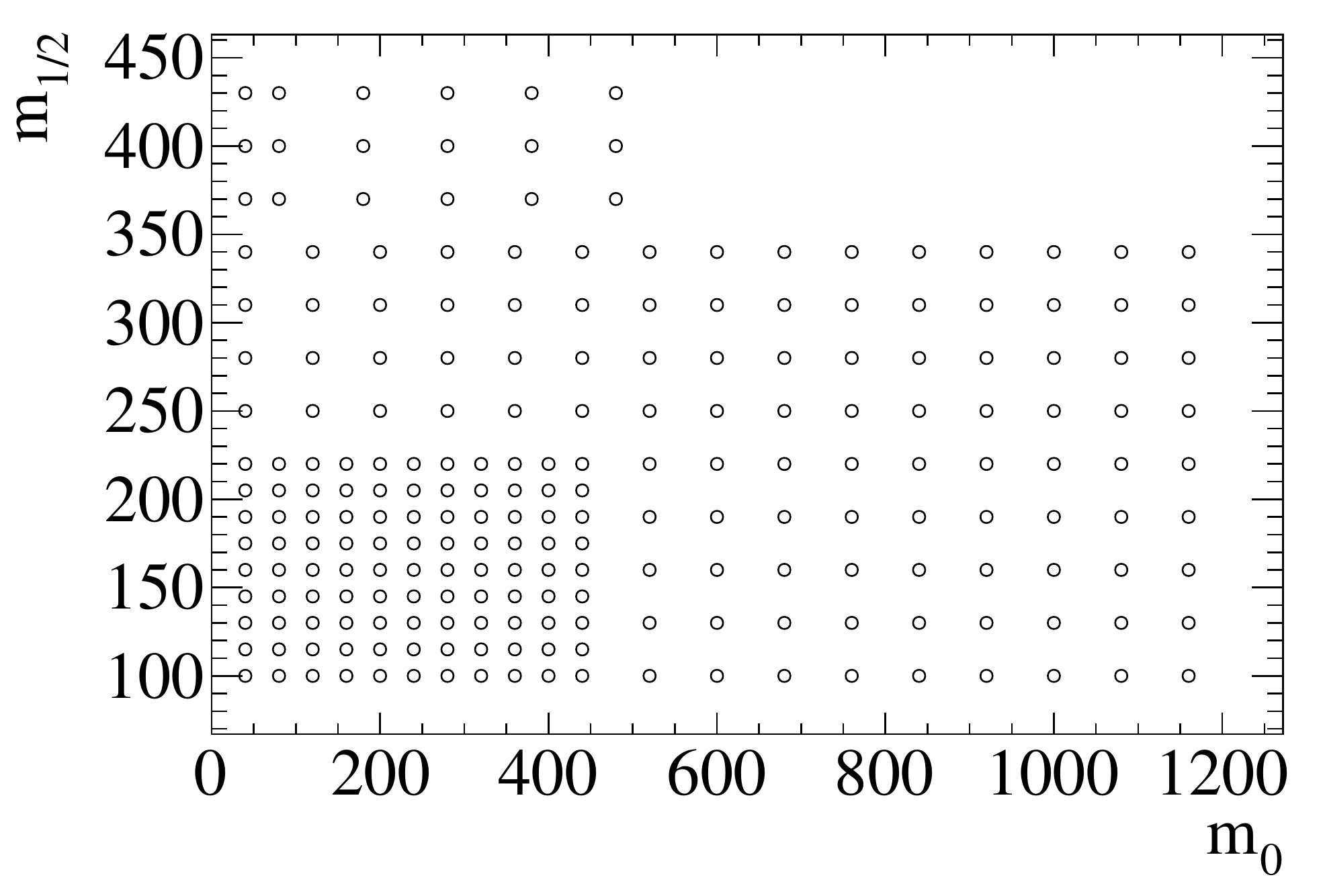}
  \hfill
  \incgraphics{width=\widthtwoplots}{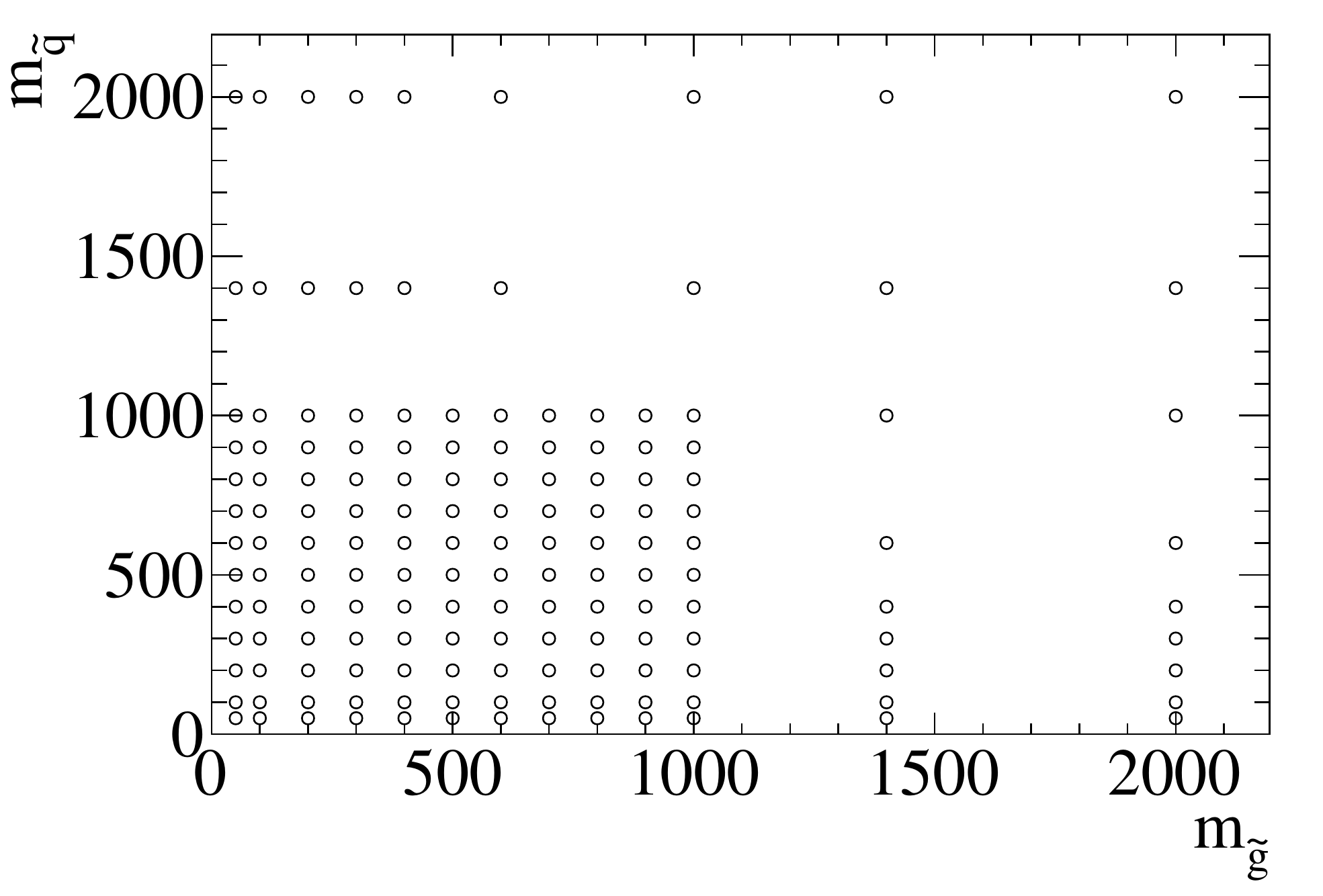}
  \caption{
    Overview of the points defined in the two grids used in this thesis.
    Left:~mSUGRA grid, defined in the \mzero-\moh plane.
    Right: MSSM grid, defined in the gluino- ($m_\gluino$) and squark-mass ($m_\squark$) plane.
  }
  \label{fig:analysis_grids_points}
\end{figure}

As explained in \Sec{sec:theory_supersymmetry},
the parameter space of Supersymmetry models has a very large dimensionality.
A restriction to a specific mechanism of Supersymmetry breaking like the mininal supergravity framework (cf. \Sec{sec:theory_bsm_model_minimal_supergravity})
reduces the number of free parameters considerably.
Still, this number often is too large to scan the complete phase space,
mainly due to limitations in computing time and storage space
needed to generate Monte Carlo for the necessary number of grid points.
To ease the presentation of results, for example when visualizing exclusion limits,
often two-dimensional subspaces of the parameter space are chosen,
which are then scanned using a fixed grid of parameter sets for which Monte Carlo is generated.
In this thesis, two such grids of points in parameter space are used.
Each point in parameter space corresponds to a specific choice of SUSY scenario,
and for each point a sample of Monte Carlo events is generated.

\subsection{mSUGRA Grid}

\afterpage{%
  \begin{landscape}
  \begin{figure}
    \centering
    \begin{overpic}[width=\widthsingleplot]{{{susy_SU_120_100_0_3.slha}}}
      \put(14,58){$\mzero = \GeV{120}$, $\moh = \GeV{100}$}
    \end{overpic}
    \hspace{1cm}
    \begin{overpic}[width=\widthsingleplot]{{{susy_SU_1000_100_0_3.slha}}}
      \put(14,58){$\mzero = \GeV{1000}$, $\moh = \GeV{100}$}
    \end{overpic}
    \begin{overpic}[width=\widthsingleplot]{{{susy_SU_120_340_0_3.slha}}}
      \put(14,58){$\mzero = \GeV{120}$, $\moh = \GeV{340}$}
    \end{overpic}
    \hspace{1cm}
    \begin{overpic}[width=\widthsingleplot]{{{susy_SU_1000_340_0_3.slha}}}
      \put(14,58){$\mzero = \GeV{1000}$, $\moh = \GeV{340}$}
    \end{overpic}
    \caption{
      Mass spectra of the supersymmetric particles
      at four example points of the \ac{mSUGRA} grid,
      with low $\moh = \GeV{100}$ (upper plots) and high $\moh = \GeV{340}$ (lower plots),
      and with low $\mzero = \GeV{120}$ (left plots) and high $\mzero = \GeV{1000}$ (right plots).
      In all plots $\tan\beta = 3$, $\signmu = +1$ and $A_0 = \GeV{0}$.
      The arrows indicate decays with a branching ratio of at least \percent{1}.
      Labels for almost degenerate mass states have been merged.
      Squarks ($\squark_{L,R}$), sleptons ($\tilde l_{L,R}$) and sneutrinos ($\tilde\nu_L$) include the first two generations.
    }
    \label{fig:analysis_mass_spectrum_msugra}
  \end{figure}
  \end{landscape}
}

The first grid is defined in the plane spanned by the \mzero and \moh parameters of \ac{mSUGRA}
for a relatively low value of $\tan \beta = 3$.
The trilinear coupling is set to $A_0 = \GeV{0}$,  %
and the sign of the Higgsino mass parameter is positive, $\mu > 0$. %
The Monte Carlo event generation has been done with \propername{Herwig++} \cite{Herwig++,Herwig++23}
and \propername{ISAJET 7.79} \cite{Isajet769,URL_ISAJET},
using ISASUGRA from this program suite as spectrum calculator
to calculate the weak scale parameters from those at the high scale,
and the \tagname{MRST 2007 LO$^*$} parton distributions \cite{CTEQ6}. %

The mass range that is covered by this grid is 
$\GeV{40} < \mzero < \GeV{1160}$ and $\GeV{100} < \moh < \GeV{430}$,
with a varying density of grid points that for low \mzero-\moh values is four times as high as in the rest of the plane
to allow for a finer resolution of the exclusion contour in this region.
An overview of the points defined in this grid is shown in the left plot in \Fig{fig:analysis_grids_points}
in which each marker corresponds to one \ac{MC} dataset.
Note that the position of the dots in \Fig{fig:analysis_grids_points} is exact,
whereas in the two-dimensional color-coded histograms in the following
the grid points may appear to be slightly shifted,
but this is only an artefact from the binning of the histograms.
The number of events within the MC datasets varies between approximately $5\ten{3}$ and $1.5\ten{5}$.
This grid consists of 222 points and will be referred to as the mSUGRA grid\inindex{mSUGRA grid} in the following. %

\Fig{fig:analysis_mass_spectrum_msugra} shows four examples for mass spectra from the mSUGRA grid with low and high \mzero and \moh,
where low and high refer to the mass range covered by the mSUGRA Monte Carlo grid used for the analysis.
The two upper plots show the spectrum for $\moh = \GeV{100}$, the lower plots for $\moh = \GeV{340}$;
the two plots to the left have \mzero set to \GeV{120}, the two plots to the right have $\mzero = \GeV{1000}$.
All other parameters, $\tan\beta$, \signmu and $A_0$ are set to the values used for the production of the mSUGRA grid.
The dashed lines show the most important decays with a branching ratio of at least \percent{1},
the thickness and opacity of the line indicating at the branching ratio.

\subsection{MSSM Grid}
\begin{figure}
  \centering
  \incgraphics{width=\widthsingleplot}{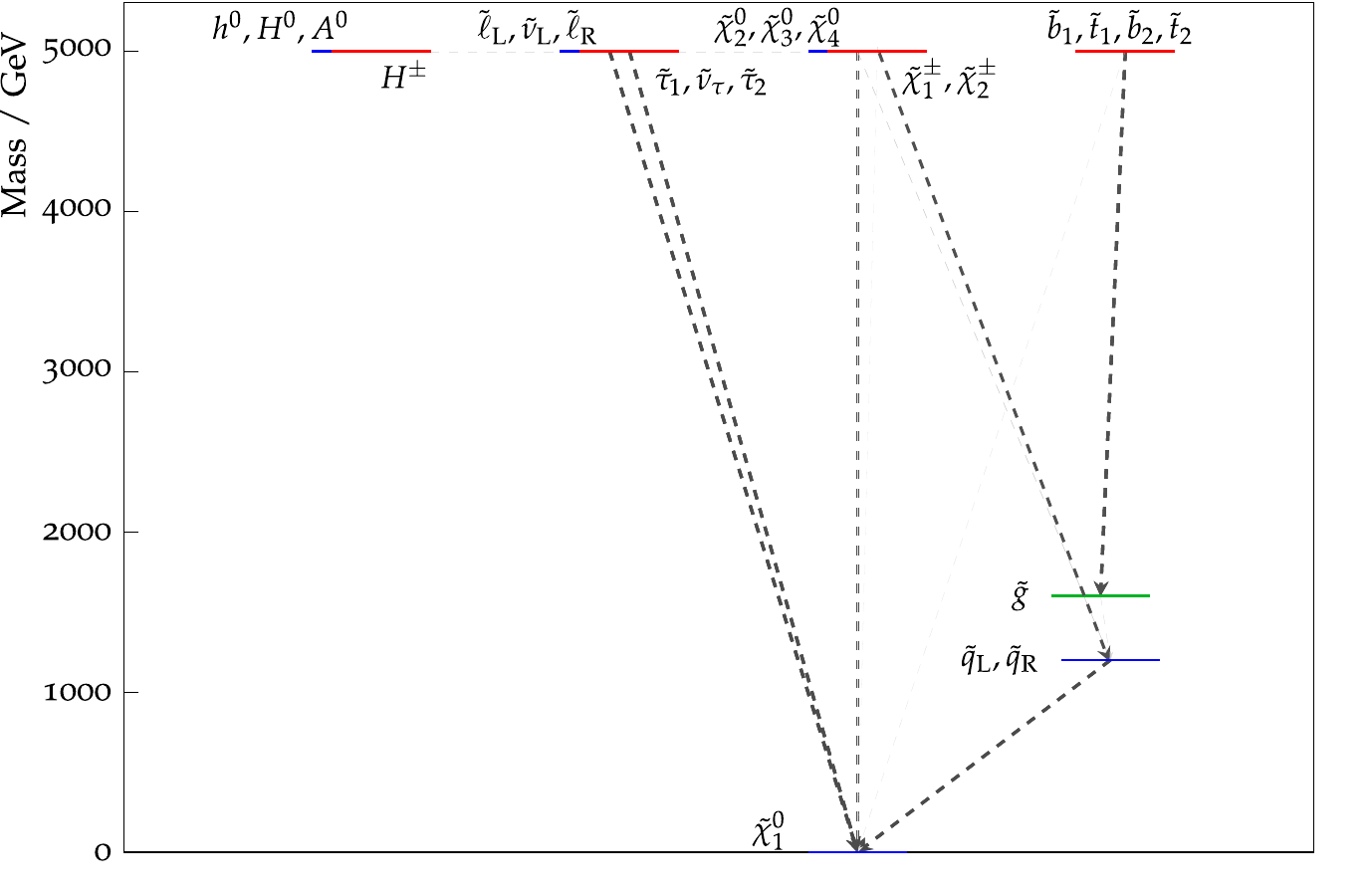}
  \caption{
    Mass spectrum of the supersymmetric particles at one example point of the MSSM grid,
    with $m_\squark = \GeV{1200}$, $m_\gluino = \GeV{1600}$ and a massless lightest neutralino \neutralinon{1}.
    The masses of all other supersymmetric particles including $\tilde b$ and $\tilde t$ are set to \TeV{5}.
    Arrows indicate decays with a branching ratio of at least \percent{1}.
  }
  \label{fig:analysis_mass_spectrum_mssm_example}
\end{figure}

The second grid is a non-universal supergravity model with minimal particle content, %
in which the grid points have a fixed, low value for the \ac{LSP}
and vary in the masses of the squarks and gluinos,
both covering the same range of $\GeV{50} < m_\gluino < \GeV{2000}$ and $\GeV{50} < m_\squark < \GeV{2000}$.
This type of grid is available for three different masses of the \ac{LSP}, $0$, $50$, and \GeV{95},
where for the massive \acp{LSP} only points have been simulated
for which the lighter of $m_\gluino$ and $m_\squark$ is less than $3 \cdot m_\text{LSP}$. %
The masses of all other supersymmetric particles,
including the masses of the third generation of scalar quarks, are set to \TeV{5} to decouple them.
The mass spectra at the different grid points therefore all are very simple,
and identical except for the squark and gluino masses.
An example is given in \Fig{fig:analysis_mass_spectrum_mssm_example}.
The arrows indicate decays with a branching ratio larger than \percent{1}.
Particles without outgoing arrows are not stable, but dominantly decay
to non-supersymmetric particles which are not shown in this plot.
The $H^\pm$ bosons \eg decay almost always to $t\bar b$ or $\bar t b$ pairs, %
and the gluino decays to all light squark flavors with the same branching ratio of $1/16$. %
The mass spectrum for this grid is generated using ISASUGRA in a non-universal supergravity model,
using a set of soft-symmetry breaking parameters that should approximately give the desired masses.
The masses returned by ISASUGRA slightly differ from those specified, and are therefore edited by hand,
then ISASUSY is used to calculate the branching ratios and decay widths \cite{URL_MSSMSquarkGluinoGRID}.

In the following, only the grid with a massless \ac{LSP} is considered,
because it covers the full plane within the limits given above.
This grid consists of 165 points and will be referred to as the \index{MSSM grid} in the following.
An overview of the points defined in this grid is shown in the right plot in \Fig{fig:analysis_grids_points}.
Note that this plot only shows 153 points, because the remaining points lie at much larger values of the masses, at around \TeV{5},
and are not relevant in the context of the following analysis.
The number of events within the MC datasets is approximately $10^4$ for all grid points,
apart for very few exceptions for which probably some of the reconstruction jobs failed. %

\begin{figure}
  \centering
  \incgraphics{width=\widthtwoplots}{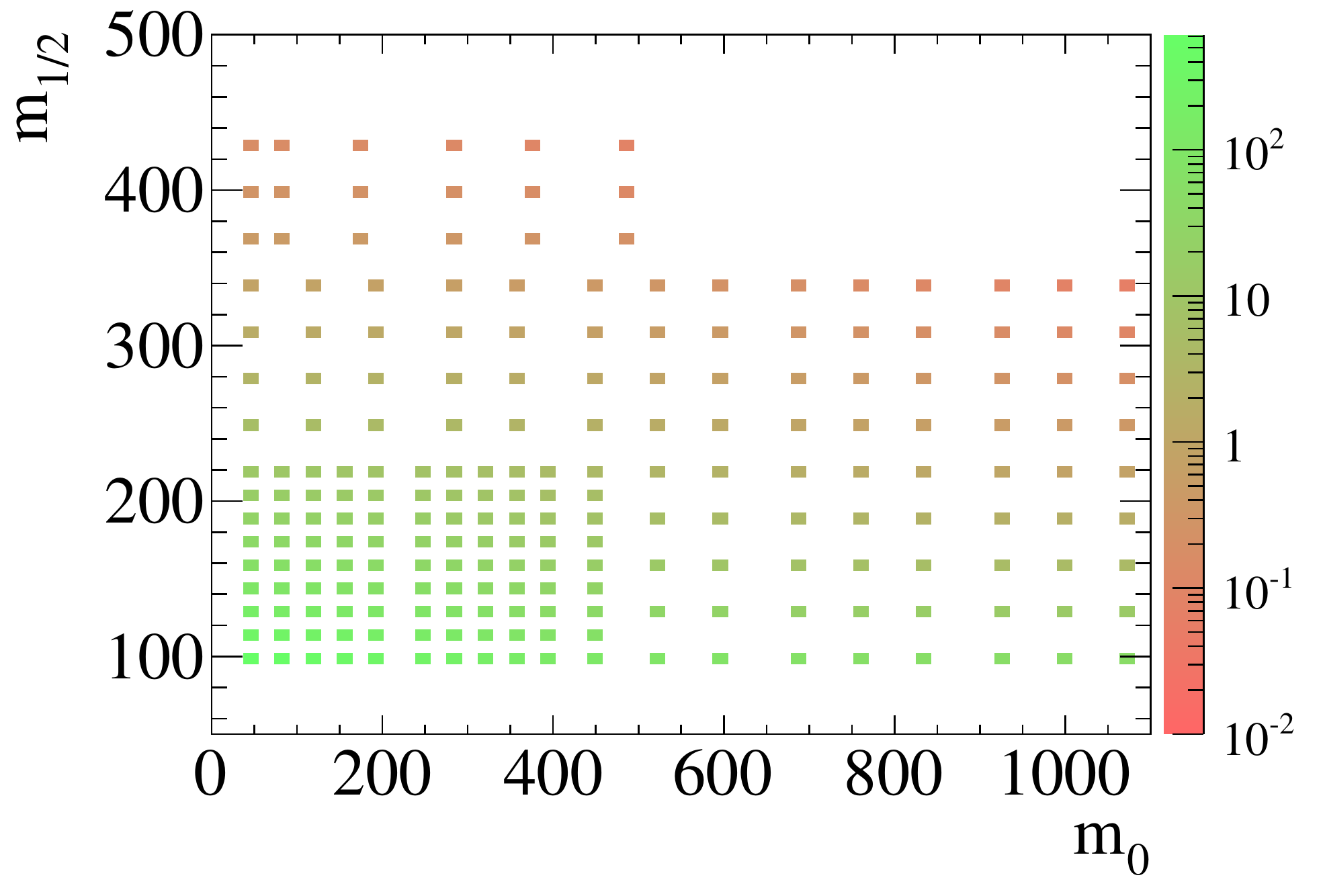}
  \hfill
  \incgraphics{width=\widthtwoplots}{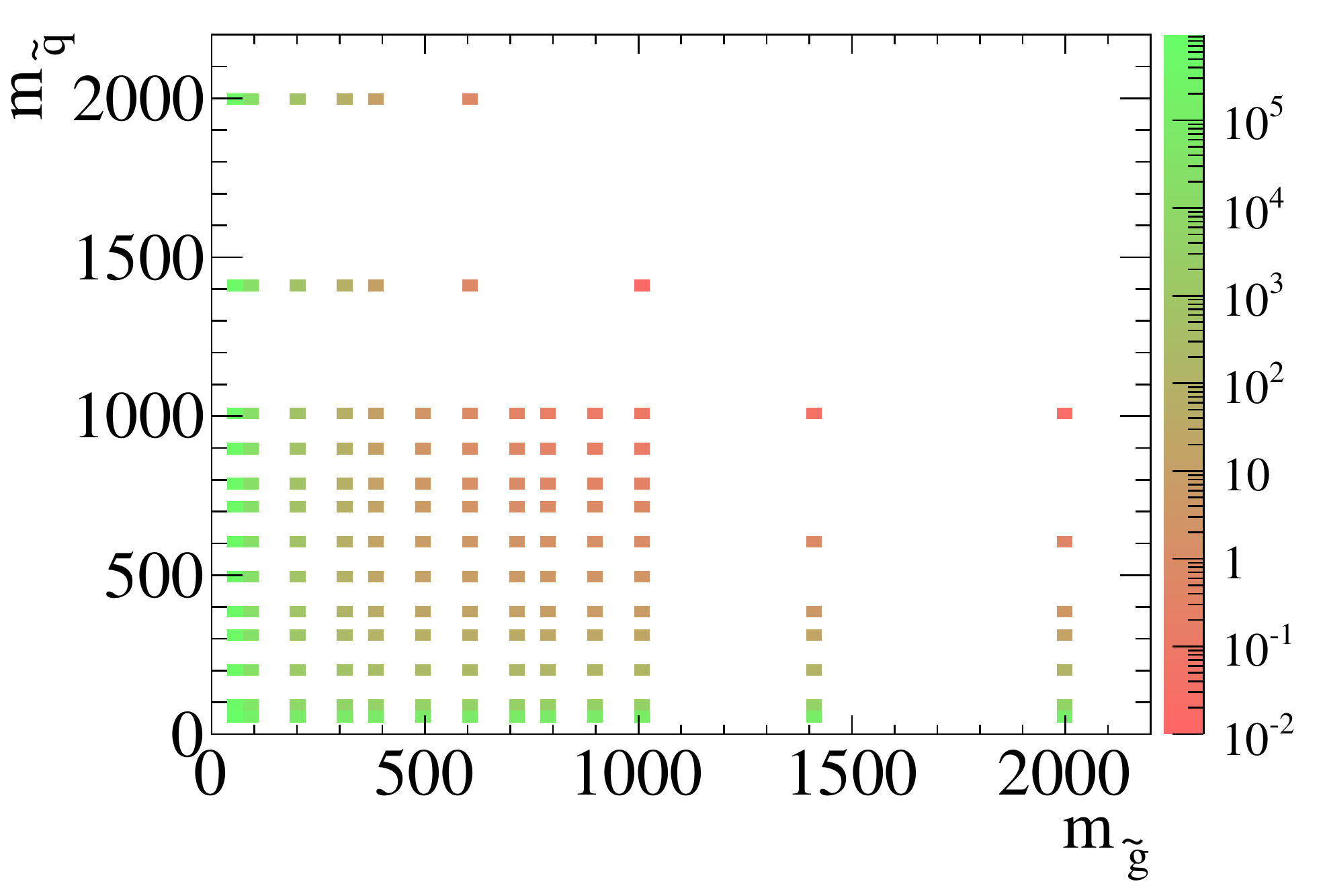}
  \caption{
    Cross sections of the SUSY grid Monte Carlo samples in picobarn, %
    left: mSUGRA grid, right: MSSM grid.
  }
  \label{fig:analysis_overview_grid_cross_sections}
\end{figure}
In \Fig{fig:analysis_overview_grid_cross_sections},
the cross sections of the SUSY grid Monte Carlo samples
for the \ac{mSUGRA} grid (left plot) and the \ac{MSSM} grid (right plot) are shown.
In both cases, the highest cross sections are found for low masses, of course.
In the mSUGRA grid, the cross section depends strongly on \moh and less so on \mzero.
For the MSSM grid, the dependence on the squark and gluino masses is comparable.
Note that the MSSM grid covers a range up to much larger cross sections than the mSUGRA grid
which are attained at grid points with very low squark and gluino masses.

\subsection{Final-State Dependent Cross Sections}
\label{sec:results_0lanalysis_final_state_dependent_xsec}

There is a large variety of different Monte Carlo samples for Standard Model processes,
which together make up the background in a search for new physics.
Instead of having one big Monte Carlo sample containing all possible types of events,
for several reasons, as outlined in \Sec{sec:software_MC_split_MC}, %
it makes more sense
to simulate different classes of physics processes individually and combine them again later.
In the combination, an appropriate weight is applied to each event,
which is computed from the cross section of the respective physics process.

For the SUSY signal samples, there is only one sample of Monte Carlo events for each point in parameter space.
The distinction between different processes,
all of which are of supersymmetric nature,
is done afterwards in the analysis.
Using the Monte Carlo truth information, every event is identified as belonging to a certain class.
The classes are defined by the particles in the final state of the hard \revised{scattering} process,
which contains two supersymmetric, pair-produced particles, \eg two squarks or a squark and a gluino.
(Although these two particles stand at the beginning of the decay cascade, they are the final state 
from the point-of-view of the generator of the hard \revised{scattering} process.)
Instead of using one overall cross section for the full \ac{MC} sample to compute an average weight for each event,
this procedure allows to use individual weights for each subprocess that produces SUSY particles.
These subprocesses do not only have different cross sections,
but also the acceptance of the event selection that is done in the analysis
will depend on the particles that were produced,
and therefore will vary significantly between different final states.
In a simplified example, a selection requiring two jets would mostly select squark pairs,
because (assuming $\squark \to q\gluino$ is kinematically forbidden)
the two final state squarks will each decay to a quark giving a jet plus a neutralino or chargino. %
Requiring four jets would prefer gluino pairs,
each of which can only decay through a squark, $\gluino \to q\squark$,
and therefore likely give two quarks, \latein{ergo} two jets, in the decay chain (see \eg \cite{SUSYPrimerMartin1997}). %
Using the overall cross section is not wrong, but yields an averaged result.
Different final states may have different \kfactors,
and for SUSY processes the \kfactors tend to be large \cite{Beenakker1997}, %
so that going from the generator cross section to the \ac{NLO} cross section will enhance different final states by different factors.

\begin{table}
  \centering
  \begin{tabular}{llp{1em}ll}
    \toprule
    ID & Process && ID & Process\\
    \cmidrule{1-2}\cmidrule{4-5}
    1 & squark -- gluino     && 71, 72, 73, 74 & $\neutralinon{1,2,3,4}$ -- gluino\\
    2 & gluino -- gluino     && 75, 76 & $\charginopn{1, 2}$ -- gluino\\
    3 & squark -- squark     && 77, 78 & $\charginomn{1, 2}$ -- gluino\\
    4 & squark -- antisquark && 81, 82, 83, 84 & $\neutralinon{1,2,3,4}$ -- squark\\
    51 & sbottom 1 pair      && 85, 86 & $\charginopn{1, 2}$ -- squark\\
    52 & sbottom 2 pair      && 87, 88 & $\charginomn{1, 2}$ -- squark\\
    61 & stop -- antistop 1  &&  & \\
    62 & stop -- antistop 2  &&  & \\
    \bottomrule
  \end{tabular}
  \caption{ID numbers of the final states used to assign events to classes with given cross sections.}
  \label{tab:analysis_subprocesses_ids}
\end{table}

Computing the cross sections takes quite a lot of \ac{CPU} time,
because there are many grid points, many final states for each,
and the estimation of the systematic uncertainties (cf. \Sec{sec:analysis_signal_xsec_uncertainties})
again increases the computational effort by a significant factor.
The cross sections are therefore computed centrally by the \ATLAS SUSY group using \propername{Prospino}
and provided as a ROOT file for each of the SUSY grids.
Not all possible final states are considered in this computation,
but only those that are expected to be dominant. %
\Tab{tab:analysis_subprocesses_ids} lists~24 classes of events that are defined by the final state,
together with the integer numbers which are used as their ID in the official cross section tables.

For the MSSM grid, cross sections are available only for final states with IDs 1~--~4.
For the mSUGRA grid, also the other~20 final states from \Tab{tab:analysis_subprocesses_ids} are taken into account.
Of course, these~24 final states do not cover all possibilities,
and events which in the analysis are identified as not belonging
to one of the~24 classes are ignored, because their cross section in unknown.
Ignoring these unidentified events
is equivalent to assuming a cross section of zero for the unidentified processes
and means that the expected number of signal events is artificically reduced,
leading to exclusion limits that are more conservative than those that would be obtained when including all events.
\Tabs{tab:susygrid_missing_states_mssm} and \ref{tab:susygrid_missing_states_tanbeta3} in the Appendix %
give an overview of all final states that are encountered in the Monte Carlo signal grid samples.
From \Tab{tab:susygrid_missing_states_mssm}, it is clear that in the MSSM samples
using final states with IDs 1~--~4 is mostly sufficient,
because only these types of final states exist in the Monte Carlo,
apart from final states where a neutralino is produced in association with a gluino or squark
and some more, negligibly small exceptions.
In particular, final states never include scalar bottom nor scalar top quarks.
Looking at \Fig{fig:analysis_mass_spectrum_mssm_example} again,
this makes clear that the heavy supersymmetric particles are indeed decoupled,
because only the much lighter squarks and gluinos are produced,
and only their decays are relevant.
For the mSUGRA grid, the variety of final states is larger,
including scalar bottom and scalar top quarks and
also associated production  %
with charginos and neutralinos other than the lightest neutralino. %

According to the documentation of \propername{Prospino},
for final states including squarks (ID~1, 3 and 4 in \Tab{tab:analysis_subprocesses_ids}), %
a summation over both squark chiralities
and over all possible squark flavors is understood, excluding stops \cite{Beenakker1996}.
The sum over the squark flavors is due to the computation
being done in a supergravity-inspired model in which all squarks are assumed to have a common mass.
This is not correct for stops, and these are therefore excluded from the final states.
The official statement from the \ATLAS SUSY group is that scalar bottoms are included,
but in the updated version of \propername{Prospino},
which was used to compute the cross sections used by the SUSY group,
scalar bottom quarks can be treated separately and are not included in squark--gluino and squark--squark by default,
whereas they are included by default in squark--antisquark\footnote{
  This can be seen in the output of the official version of the \propername{Prospino} code used by the SUSY group, %
  where an integer value \code{isquark*} is running from $-4$ through $4$ for \code{final\_state\_in = 'ss'},
  which corresponds to the different squark flavors and does not include scalar bottoms.
  Scalar bottoms would be $-5$ and $5$ \cite{PCPlehn1}, values that only appear
  when computing the cross section for squark--antisquark final states (\code{final\_state\_in = 'sb'}).
}.
The final states with ID~1 and 3,
listed in the table as squark--squark and squark--gluino,
also include a summation over charge-conjugate final states \cite{Beenakker1996,Beenakker1997}. %
The final states with ID~51 and 52 are taken to refer to sbottom--antisbottom production
corresponding to the respective final state option in \propername{Prospino}\footnote{
  \code{final\_state\_in = 'bb'} %
}.
Final states with ID~61 and 62 refer to $\tilde{t}_1\bar{\tilde{t}}_1$ and $\tilde{t}_2\bar{\tilde{t}}_2$.
Mixed pairs ($\tilde{t}_1\bar{\tilde{t}}_2$ and $\tilde{t}_2\bar{\tilde{t}}_1$) 
can only be produced at higher orders (${\cal O}(\as^4)$) at strongly suppressed rates \cite{Beenakker1998}, %
which can also be seen from \Tab{tab:susygrid_missing_states_tanbeta3},
where $\tilde{t}_1\bar{\tilde{t}}_1$ appears about $6000$ times as often as $\tilde{t}_1\bar{\tilde{t}}_2$. %

From the above considerations, the following mapping is derived.
Final states that consist of a squark--antisquark pair,
one of which or both may be scalar bottom squarks, are all included in ID~4,
independently of which of the two mixed mass states %
the sbottom and~/ or antisbottom is.
The cross sections computed for ID~51 and 52 are therefore not used in the analysis,
because they are a subset of all squark--antisquark final states %
and overlap with ID~4.
Events with final states including scalar bottom quarks that do not fall into the class with ID~4,
\eg sbottom--sbottom or sbottom--gluino, are ignored,
because they are also not included in the cross section computation by \propername{Prospino} for ID~1, nor for ID~3.

With respect to the implementation,
the classification of events according to the final state of the hard process,
as given in \Tab{tab:analysis_subprocesses_ids},
is done by evaluating the Monte Carlo truth information,
searching the Monte Carlo tree for the two SUSY particles that have a Standard Model particle as ancestor.
From the identity of these particles,
it can be determined to which of the defined classes the event belongs.
If it does not match any of the classes, the event is discarded.

\begin{figure}
  \centering
  \incgraphics{width=\widthtwoplots}{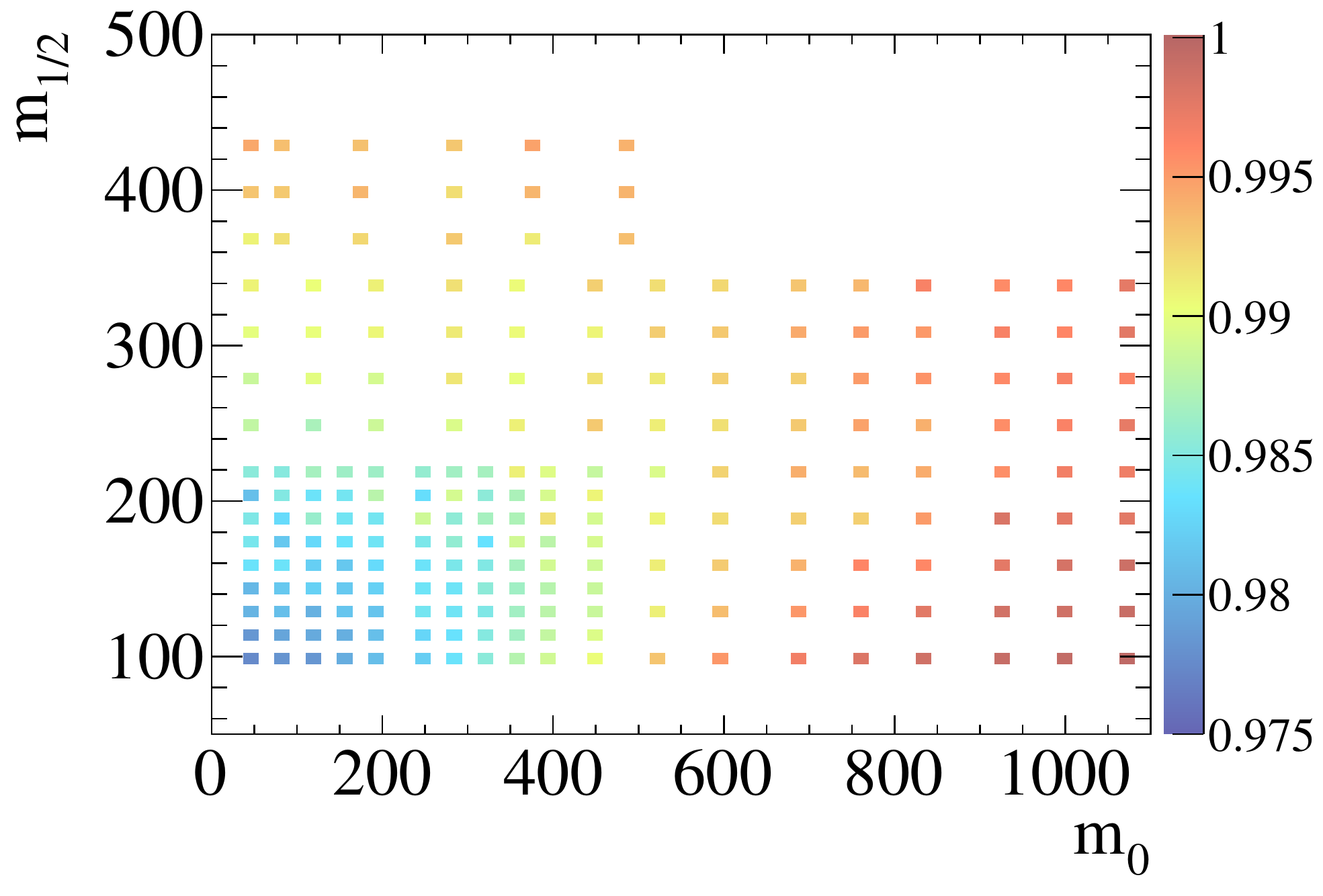}
  \hfill
  \incgraphics{width=\widthtwoplots}{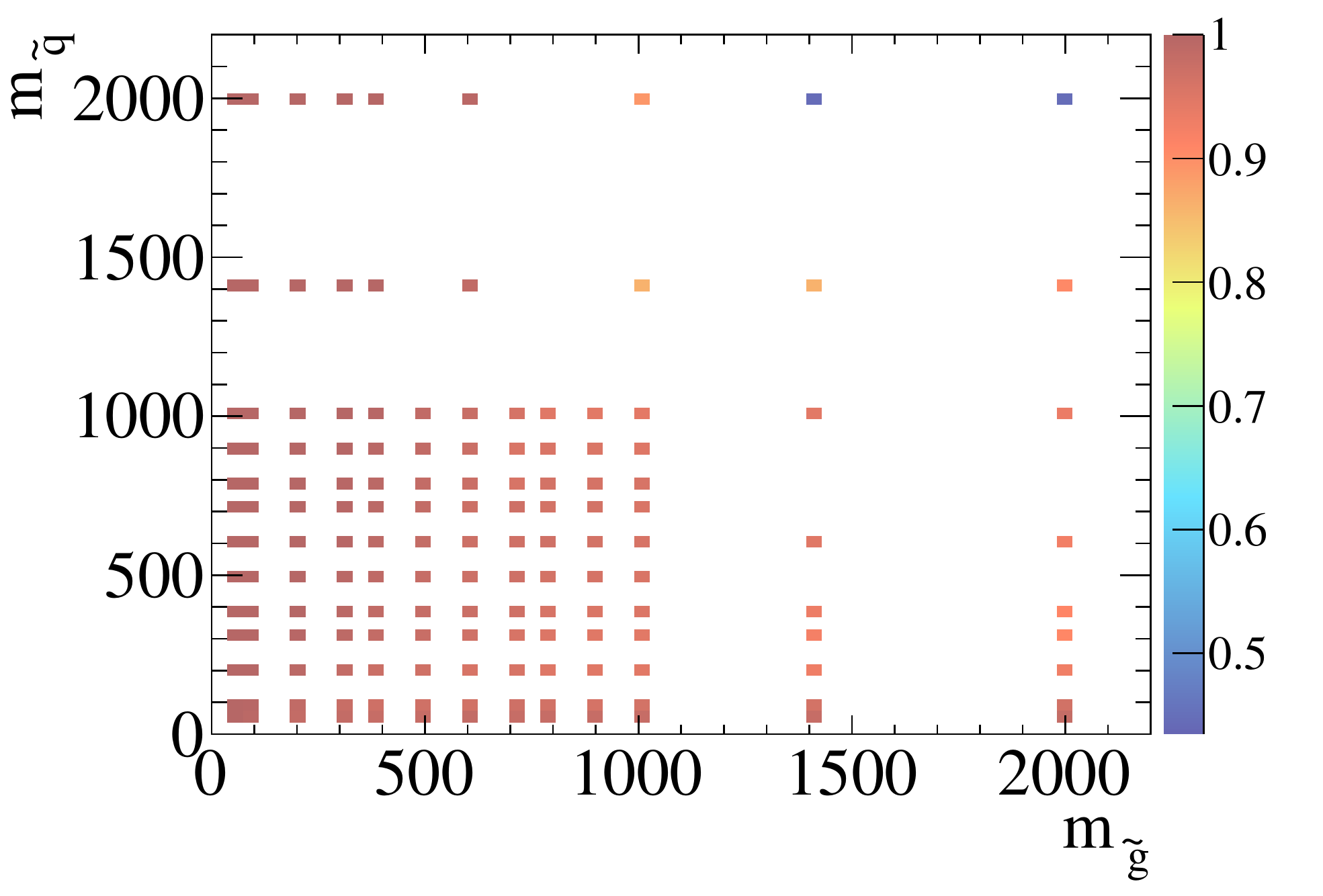}
  \caption{
    Fraction of events in the Monte Carlo samples from the signal grids
    for which the cross section is known.
    Left: mSUGRA grid, right: MSSM grid.
  }
  \label{fig:analysis_impact_neglected_events}
\end{figure}
The impact of neglecting certain classes of events is discussed briefly now.
It can be estimated from the fraction of events that do not fall into one of the classes
for which the cross section is known\footnote{
  The overall magnitude of this fraction of this can also be taken from the counts in \Tab{tab:susygrid_missing_states_tanbeta3}.
}.
If this fraction is small, this is an indication that the impact should be negligible.
\Fig{fig:analysis_impact_neglected_events} shows two plots,
with the fraction of events in the grid samples for which the cross section is known,
\ie which fall into one of the classes as defined above,
for which the \propername{Prospino} \ac{NLO} cross section is available.
The left plot shows the fractions for the \ac{mSUGRA} grid, the right one for the \ac{MSSM} grid.
For the \ac{mSUGRA} grid, the fractions increase from about \percent{98} in the lower left corner
to nearly a \percent{100} close to the right and top border.
In the \ac{MSSM} grid, for low $m_\gluino$,
the fraction of known events is close to \percent{100},
and falls off to the right, going down to \percent{94} in the dense grid region ($m_\gluino$ and $m_\squark$ below \TeV{1}).
In the upper right corner, the fraction is only \percent{45}.
Because of the large masses of the squarks and gluinos,
the associated production of squarks and gluinos with neutralinos \neutralino becomes dominating here.
Note that the yield of the different final states with squarks and gluinos depends in different ways
on the squark and gluino masses \cite{Beenakker1997}. %
For $m_\squark \approx m_\gluino$, squark--gluino production dominates at low masses,
whereas at higher masses of about \TeV{1} squark-pair production has the highest yield.
From the plots in \Fig{fig:analysis_impact_neglected_events}, it can be concluded
that the impact of the neglected events should not deteriorate the power of the analysis too much.
A complementary cross-check is presented in \Sec{sec:appendix_impact_unidentified_events}.
In any case, as said above, neglecting events
which are contained in the Monte Carlo will make the limits more conservative,
but it is safe in that in cannot lead to false exclusions.

\section{Statistical and Systematic Uncertainties}
\label{sec:analysis_uncertainties}

\begin{table}
  \setlength{\tabcolsep}{5pt} %
  \centering
  \begin{threeparttable}
    \begin{tabular}{lcccc}
      \toprule
      Source & Measured & Signal / Background & Symmetric & Correlated\\
      \midrule
      Statistics & on MC & S / B & yes & no\\
      Integrated luminosity & fixed & S / B & yes & yes\\
      Jet resolution & on MC & S / B & yes & yes\\
      Jet-energy scale & on MC & S / B & no & yes\\
      Pile-up contribution & on MC & -- / B & yes & yes \\ %
      Background normalization & fixed & -- / B & yes & no\\
      Signal normalization & fixed & S / -- & yes & no\\
      \bottomrule
    \end{tabular}
    \caption{
      Overview of systematic uncertainties. %
      The second column describes how the impact of the uncertainty was estimated,
      the third states which Monte Carlo the uncertainty applies to.
      The last two columns state whether this uncertainty is symmetric
      and whether it is correlated between the signal and the different background contributions.
      Further explanations can be found in the text.
    }
    \label{tab:analysis_overview_systematic_uncertainties}
  \end{threeparttable}
  \resettabcolsep %
\end{table}

\inindex{Systematic uncertainties}
The statistical evaluation of the observed data requires a careful treatment of its systematic and statistical uncertainties,
because only then it becomes possible to arrive at a conclusion which can be said to hold at a certain probability.
\Tab{tab:analysis_overview_systematic_uncertainties} gives an overview of the systematic uncertainties that were considered.
Statistical uncertainties are included in this list, too,
because they enter the limit setting prodedure (see \Sec{sec:analysis_interpretation})
in a fashion very similar to the systematic uncertainties.
It is common to assume the number of observed events in the data to be a fixed number without uncertainties
and to assign the statistical and systematic uncertainties to the Monte Carlo prediction.
The uncertainties are discussed in detail below.
A table with the values of the uncertainties is included in \Sec{sec:analysis_summary_limit_input}.

All uncertainties are modeled as Gaussian distributions with one exception.
The jet-energy scale is modeled as a Gaussian distribution,
where the two halves,
corresponding to upwards and downwards fluctuations,
may have different widths to take into account the asymmetric nature of the jet-energy scale uncertainty.
The distribution may thus be discontinuous at zero.
The pile-up contribution is also modeled as a Gaussian distribution,
allowing for upwards as well as downwards fluctuations.
One might expect pile-up contributions to only increase the number of events after cuts,
because pile-up always adds energy to the event,
but there are \eg cases in which objects which lead to an event veto
profit from additional energy due to overlaid pile-up events,
and pass a selection which they would not have without pile-up.
Indeed, although increases in event numbers dominate,
negative changes in the event numbers are found, too, when including pile-up. %
The fluctuations covered by systematic uncertainties can either be independent
between different background types and the signal,
such as the purely stochastic fluctuations due to limited Monte Carlo statistics,
or they can be correlated such as the uncertainty on the luminosity,
which will lead to an up- or downscaling of all backgrounds and the signal Monte Carlo alike.
In general, there are more possible levels of correlations between systematic uncertainties, %
between different physics processes, between different regions of phase space,
or, on a larger scale, between different analyses or experiments.
It is important to take correlations between uncertainties into account,
because fluctuations due to uncertainties will lead to larger variations in the outcome
if the uncertainties are correlated.
(Uncorrelated uncertainties are added in quadrature, whereas absolutely correlated uncertainties need to be added linearly.)
The last column in \Tab{tab:analysis_overview_systematic_uncertainties} 
shows which systematic uncertainties are taken to be correlated across the signal and background Monte Carlo samples.
The uncertainty on the signal normalization affects only the signal Monte Carlo sample,
so there is no possible correlation.

\subsection{Statistical Uncertainty}

The \index{statistical uncertainty} is determined by the number of events available within the Monte Carlo samples.
These numbers are given in \Tabs{tab:analysis_MC_datasets_details_WZ} and \ref{tab:analysis_MC_datasets_details_top_dijet},
together with the integrated luminosity the complete Monte Carlo dataset corresponds to.
As for almost all samples (all but QCD) this luminosity is comfortably above the integrated luminosity of the data,
the statistical uncertainty is not expected to be a dominant contribution,
at least unless very harsh cuts are applied as in some of the signal regions.
It can be directly evaluated from Monte Carlo,
assuming Poisson distributions for the bin contents or the number of events at a certain step of the cutflow, respectively.
If $N_i$ is the number of events from Monte Carlo sample $i$ surviving all cuts up to a certain stage,
the uncertainty $\sigma_i$ on the contribution from this sample 
is taken to be the square root of the number of events $\sqrt{N_i}$,
which is just the standard deviation of a Poisson distribution with expectation value $N_i$.
The total uncertainty is the weighted sum over all Monte Carlo samples.
Denoting the weight of sample $i$ with $w_i$, the total uncertainty $\sigma_N$ is therefore given by
\begin{equation}
  \sigma_N = \sqrt{\sum_i w_i^2 \sigma_i^2} = \sqrt{\sum_i w_i^2 N_i}.
\end{equation}
This follows from the standard rule for the propagation of (uncorrelated) errors, %
which assumes Gaussian distributions for the individual uncertainties.
This assumption is justified if $N_i$ is sufficiently large.

\subsection{Uncertainty on the Jet-Energy Scale}

\inindex{Jet-energy scale uncertainty}
The Jet/Etmiss combined performance group provides an official tool called \propername{JESUncertaintyProvider},
that can be used to assess the uncertainties on the \acl{JES} \cite{ATLAS-CONF-2010-056,ATLAS-CONF-2011-007}.
For data reconstructed with \Athena release 15, as is used here,
the values on which the uncertainties from this tool are based
have been computed from Monte Carlo samples and in-situ measurements for the calorimeter response uncertainty. %
The updated \ac{JES} uncertainty for release 16 would additionally include an individual treatment
of isolated jets and close-by jets and take differences in the quark and gluon response into account \cite{ATLAS-CONF-2011-032}.

In this analysis, version \tagname{JetUncertainties-{\allowbreak}00-02-00} of the \propername{JESUncertaintyProvider} tool is used.
The impact of the \ac{JES} uncertainty is propagated to the event counts in the signal regions
by rerunning the Monte Carlo twice, once with the \ac{JES} scaled down by the given uncertainty and once with \ac{JES} scaled up.
Implementation-wise, the \propername{JESUncertaintyProvider} tool is called once for each jet
with the \eta and \pt of the jet and the number of vertices reconstructed for this event.
The number of vertices is used to account for the energy contribution to calorimeter jets
from multiple proton-proton interactions in pile-up events
which is not included in the current JES calibration.
Instead, it is taken to give a separate contribution to the systematic uncertainty on the jet-energy scale. %
As the uncertainties have only been determined for a \pt range from \GeV{15} to \TeV{1}
and for a \eta range within $\pm4.5$,
the tool only accepts these ranges for the values of the input parameters.
For jets which fall outside these ranges,
the closest values within the allowed ranges are therefore used to query the uncertainty.
The tool returns an uncertainty $\Delta$ for the given parameters,
and the \pt and energy of the jet are multiplied with a factor $1\pm\Delta$ for the computation of the upwards and downwards fluctuation
due to the \ac{JES} uncertainty, respectively.
The \met is also updated to take the change in jet \pt into account.

\subsection{Uncertainty on the Jet-Energy Resolution}

To compute the uncertainty on the jet-energy resolution\inindex{Jet-energy resolution},
another tool called \propername{JetEnergyResolutionProvider} is provided,
which returns two numbers, the relative jet transverse momentum resolution $\sigma_{\pt}$ and its uncertainty $\sigma_r$,
both as function of jet \pt and rapidity $y$.
The recipe to convert these numbers into a systematic uncertainty on the event count
and reproduce the slightly worse resolution in data (see \Sec{sec:jet_resolution})
is to calculate a smearing factor $\sigma_s$ as
\begin{equation}
  \sigma_s = \sqrt{\left(\sigma_{\pt} + \sigma_r\right)^2 - \left(\sigma_{\pt}\right)^2}.
  \label{eq:analysis_JER_smearing_factor}
\end{equation}
The energy and \pt of the jet is then smeared with a Gaussian of width $\sigma_s$ centered around zero,
\begin{equation}
  \pt \mapsto \pt \cdot \left( 1 + \operatorname{Gauss}(0, \sigma_s) \right),
\end{equation}
and equivalently for the energy.
Here, $\sigma_{\pt}$ is assumed to correspond to the resolution in Monte Carlo $\sigma_\text{MC}$,
and $\left(\sigma_{\pt} + \sigma_r\right)$ to the resolution in data $\sigma_\text{data}$. 
\Eq{eq:analysis_JER_smearing_factor} follows from the fact that 
in the smearing the standard deviations add quadratically,
$\sigma_\text{data}^2 = \sigma_\text{MC}^2 + \sigma_s^2$.

The Monte Carlo is rerun with the smeared \pt and an
updated \met taking the change in the jet \pt into account,
and the difference in the event count is used as the absolute systematic uncertainty.
As the values provided by the \propername{JetEnergyResolutionProvider}
have only been determined and validated for a \pt range from \GeV{20} to \GeV{500}
and for a $|\eta|$ range within $2.8$,
the tool only accepts these ranges for the input parameters.
For jets which fall outside these ranges,
the closest values within the allowed ranges are therefore used to query the tool.
Note that even though only jets that lie within the allowed parameter range of the tools
will pass the cuts imposed in the event selection (see \Sec{sec:analyis_description_cutflow}),
in both cases, \ac{JES} and \ac{JER}, jets outside this region
need to be considered, too,
because of the propagation of the change in jet \pt to \met.
To do so, limiting the parameters as described above is necessary.
The \ac{JER} uncertainty is a correlated uncertainty,
although it is computed event-wise with a smearing based on a random distribution
and therefore not as obviously correlated as the \ac{JES} uncertainty.
But the change in event numbers induced by a deviation of the true \ac{JER} from its assumed value within the uncertainty
will affect all Monte Carlo samples
in the same way as a deviation of the true \ac{JES} from its assumed value
does affect all samples in a correlated manner.

\subsection{Uncertainty on the Integrated Luminosity}

The computation of the integrated luminosity has been explained in \Sec{sec:analysis_luminosity_calculation}.
The luminosity uncertainty is taken from \cite{ATLAS-CONF-2011-011},
which gives the official value of \percent{$\pm 3.4$} for the average relative uncertainty
on the central value of the updated luminosity measurement in 2010 (cf. \Sec{sec:luminosity_measurement}).
This uncertainty has a direct impact on the normalization,
so that no rerunning of the Monte Carlo is necessary to estimate the variation in event numbers caused by this uncertainty.
It affects signal and background in the same way,
and is therefore taken to be a correlated uncertainty for signal and background.

\subsection{Uncertainty from Pile-Up}

\inindex{Pile-up uncertainty}
For the top quark and vector boson samples,
the uncertainty due to in-time pile-up effects
is estimated by rerunning the selection on Monte Carlo samples including two additional overlaid pile-up interactions per event.
From the difference in event numbers with respect to the nominal results without pile-up
a relative uncertainty is computed.
For the \ac{QCD} background, this is not possible
because no pile-up version of the \propername{Pythia} QCD samples is available.
There are \propername{Alpgen} QCD samples which are available both with and without pile-up, %
but these samples are rather small in terms of event numbers or corresponding sample luminosity,
so that in three out of the four signal regions no events are left. %
A conservative estimate of the uncertainty from pile-up of \percent{100} has therefore been assigned to the \propername{Pythia} QCD background. %

\subsection{Theoretical Uncertainties on the Background Normalization}

\begin{table}
  \centering
  \begin{tabular}{llcc}
    \toprule
    && \multicolumn{2}{c}{Uncertainty [\%]} \\
    \multicolumn{2}{l}{Signal region} & \Zjets & \Wjets\\
    \midrule
    A & 2 jets + $\meff  >  \GeV{500}$ & 40 & 21\\
    B & 2 jets + $\mttwo >  \GeV{300}$ & 68 & 71\\
    C & 3 jets + $\meff  >  \GeV{500}$ & 44 & 23\\
    D & 3 jets + $\meff  > \GeV{1000}$ & 73 & 55\\
    \bottomrule
  \end{tabular}
  \caption{
    Relative uncertainty in percent on the \Wjets and \Zjets background in the four signal regions as given in \cite{ATL-PHYS-INT-2011-009}.
  }
  \label{tab:analysis_uncertainties_susy}
\end{table}

Theoretical uncertainties\inindex{Theoretical uncertainty}
include uncertainties on the distributions of kinematic variables within the sample
and uncertainties on the predicted cross section of a Monte Carlo sample as a whole,
which then give an uncertainty on the normalization of the events in this sample.
For the analysis at hand, details about the uncertainties on the cross sections of the signal samples
are presented in below. %
The uncertainties on the normalization of the Standard Model backgrounds can be expected to be under better control due to existing measurements
and are assumed to be covered by the uncertainties taken from the official SUSY group studies summarized in \Sec{sec:analysis_mc_sm_background}.
As described in \Sec{sec:analysis_explain_SUSY_WZ_uncertainties},
further studies within the \ATLAS Supersymmetry group
on the uncertainties of the Monte Carlo for the \Wjets and \Zjets background have been done.
As the uncertainties from the control region measurements are found to give a large contribution
to the total uncertainty for these samples,
these results are used here in addition to the uncertainties described above.
The values in \Tab{tab:analysis_uncertainties_susy} are taken from \cite{ATL-PHYS-INT-2011-009},
and state the relative uncertainty in percent on the \Wjets and \Zjets background in the four signal regions
from acceptance and control region statistics (for \Zjets) and from lepton efficiency and control region statistics (for \Wjets).

\subsection{Theoretical Uncertainties on the Signal Normalization}
\label{sec:analysis_signal_xsec_uncertainties}
The cross sections for the pair-production of SUSY particles in the scenarios
specified by the grid points have been computed by the \ATLAS SUSY group at \ac{NLO} order
using \propername{Prospino 2.1} \cite{Beenakker1996,URL_Prospino}. %
In addition to the cross sections themselves,
theoretical uncertainties on these coming from several sources are provided \cite{VanDerLeeuw2010}:
\begin{itemize}
  \item \texttt{relUncPDF}:
    The relative uncertainty from the \acf{PDF} that is used.
    In the computation of the cross sections,
    the PDF set \texttt{CTEQ6.6} \cite{CTEQ6.6,URL_CTEQ} from the CTEQ group has been used,
    which has 22 free parameters and is extrapolated from measurements and therefore entailed with its own systematic uncertainties
    that need to be propagated.
    The central fit PDF gives the nominal cross section and
    the uncertainty on this value is estimated by recomputing the cross sections
    for each of the 44 eigenvector PDF sets %
    and combining the results as
    $1.645 \Delta_\text{PDF} = \frac12 \sqrt{ \sum_i (\sigma_+^i - \sigma_-^i )^2 }$,
    where the sum runs over the 22 upwards and downwards variations which give $\sigma_\pm$.
    (This is the Hessian formalism for the uncertainty analysis \cite{Pumplin2001}, %
    where the parton parameter space is spanned by a set of orthonormal eigenvectors. %
    The factor $1.645$ converts the \percent{90} \CL into a \percent{68} \CL) %
    To obtain the relative uncertainty,
    $\Delta_\text{PDF}$ is divided by the central value of the cross section.
  \item \texttt{relScaleUncHalfQ, relScaleUnc2Q}:
    At \ac{NLO}, the combined re\-nor\-malization and factorization scale $Q$ must be fixed by an external (hadronic) scale.
    \propername{Prospino} uses the average mass of the produced massive particles 
    or the transverse mass of the detected final-state particle as value of $Q$ \cite{Beenakker1996}.
    The relative uncertainty from the choice of this scale
    is determined by setting the scale once to twice and once to half the value
    used for the computation of the central value of the cross section,
    and recomputing the cross section.
    The differences of the resulting cross section with respect to the central value give the uncertainty.
  \item \texttt{relProspinoError}:
    The relative internal inaccurary of the calculation,
    as it is returned by \propername{Pros\-pi\-no}.
    This gives only a small contribution to the total uncertainty (cf. \Tab{tab:analysis_theory_uncertainties_values}).
\end{itemize}
For those points of the 2010 MSSM grid which overlap with the 2011 MSSM grid,
also the relative uncertainty $\Delta(\as)$ on the cross section
from the uncertainty in the strong coupling constant $\alpha_s$ is available:
\begin{itemize}
  \item \texttt{relUncAlphaS}: This uncertainty, arising from the uncertainty of the strong coupling constant,
    is defined as half the absolute difference between the cross sections obtained 
    using the two extreme $\alpha_S$ variations $AS_{-2}$ and $AS_{+2}$ from CTEQ6.6AS \cite{CTEQ6.6AS}.
\end{itemize}

\begin{figure}
  \centering
  \incgraphics{width=\widthsingleplot}{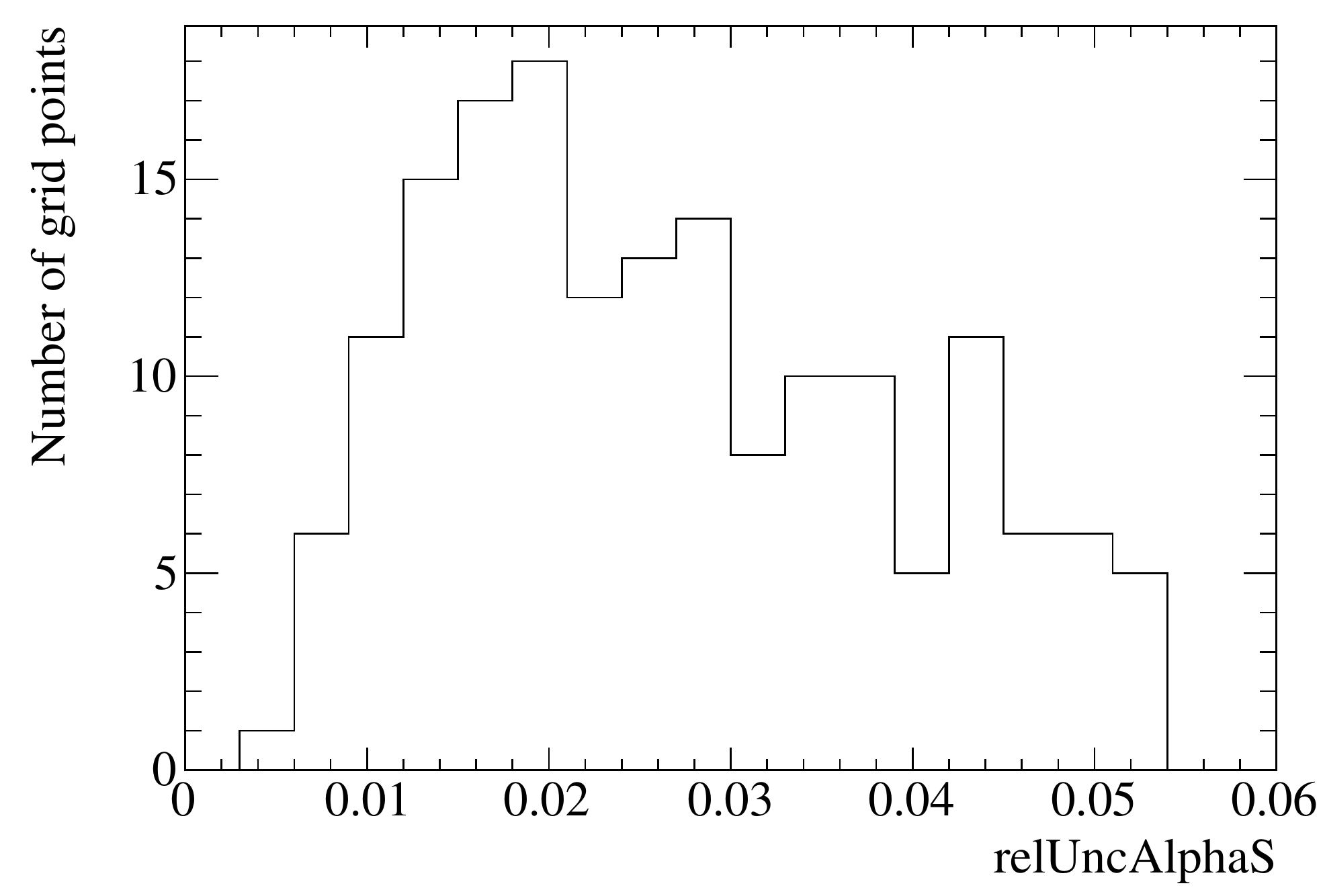}
  \caption{
    Distribution of the relative uncertainty from the variation of $\alpha_S$
    for the subset of points from the $\tan \beta = 10$ grid
    that fall into the \mzero-\moh range covered by the $\tan \beta = 3$ grid.
  }
  \label{fig:tanbeta10_alphas_distribution}
\end{figure}

\revised{%
For the 2010 mSUGRA grid with \tanbeta{3},
the uncertainty arising from \as has not been computed centrally.
This uncertainty therefore needs to be estimated from the \tanbeta{10} grid,
for which the distribution of the $\alpha_S$ uncertainties is shown in \Fig{fig:tanbeta10_alphas_distribution}.
As the \tanbeta{10} grid extends further out into the \mzero-\moh plane than the \tanbeta{3} grid does,
and the $\alpha_S$ uncertainties increase with \mzero and \moh,
only the subset of grid points from the \tanbeta{10} grid
which fall into the range covered by the \tanbeta{3} grid ($\mzero < \GeV{1160}$, $\GeV{90} < \moh < \GeV{430}$)
has been included in \Fig{fig:tanbeta10_alphas_distribution}. %
For each grid point,
the $\as$ uncertainty is an average over all defined event classes (as defined in \Sec{sec:results_0lanalysis_final_state_dependent_xsec}),
weighted with the corresponding cross section,
so that each grid point corresponds to one entry in the histogram in \Fig{fig:tanbeta10_alphas_distribution}.
}

\begin{table}
  \centering
  \begin{threeparttable}
    \begin{tabular}{crrd{4.4}crrd{4.4}l}
      \toprule
       & \multicolumn{4}{c}{$\tan \beta = 3$ grid} & \multicolumn{4}{c}{$\tan \beta = 10$ grid} \\
      \cmidrule(lr){2-5}  \cmidrule(lr){6-9}
      dominant final state & $m_0$ & \moh & \multicolumn{1}{c}{$\sigma$} & $\Delta(\as)$ & $m_0$ & \moh & \multicolumn{1}{c}{$\sigma$} & $\Delta(\as)$\\
      \midrule
      squark -- squark & 80 & 100 & 260 & 0.0013 & 100 & 90 & 381 & 0.0005\\
      squark -- gluino & 200 & 190 & 8.66 & 0.0122 & 180 & 180 & 12.1 & 0.0112\\
      squark -- squark & 180 & 400 & 0.102 & 0.0225 & 180 & 390 & 0.120 & 0.0225\\
      squark -- gluino & 1000 & 190 & 0.387 & 0.0304 & 980 & 180 & 0.499 & 0.0288\\
      squark -- gluino & 1000 & 340 & 0.0326 & 0.0479 & 980 & 330 & 0.0409 & 0.0460\\
      \bottomrule
    \end{tabular}
  \end{threeparttable}
  \caption{
    Relative uncertainty $\Delta(\as)$ on the cross section $\sigma$ arising
    from the uncertainty in the strong coupling constant \as
    for five grid points of the $\tan \beta = 3$ mSUGRA grid,
    as calculated with \propername{Prospino}.
    The uncertainties are compared to those of grid points with $\tan \beta = 10$ and similar values of $m_0$ and \moh.
    All cross sections and uncertainties are given for the dominant subprocess stated in the first column.
    Masses are given in GeV, cross sections in pb.
  }
  \label{tab:analysis_alphas_crosscheck}
\end{table}

To check how the \as uncertainty depends on $\tan \beta$ in general,
for a random sample of five grid points from the mSUGRA grid for $\tan \beta = 3$,
the \as uncertainty has been calculated here for the two dominant subprocesses using \propername{Prospino}.
\Tab{tab:analysis_alphas_crosscheck} shows the results for the respective dominant subprocess.
It is found that the \as uncertainty $\Delta(\as)$ for $\tan \beta = 3$ is larger than for $\tan \beta = 10$,
but only slightly so.
Only for the point in this sample with very low \mzero and \moh,
the deviation is larger, the uncertainty for $\tan \beta = 3$ being more than twice as big as for $\tan \beta = 10$,
but the value of the relative uncertainty itself for this point is very small.
A conservative value of \percent{3} is therefore assumed for the \alphas uncertainty in the following.

With respect to the final input to the limit setting,
a total uncertainty on the normalization of each of the signal Monte Carlo samples is computed
as the quadratic sum of the individual uncertainties described above.
For the scale uncertainty (\texttt{relScaleUncHalfQ} and \texttt{relScaleUnc2Q}),
the larger of the two values obtained when doubling or halving the scale is used.
The uncertainty on the signal normalization
is computed as a total uncertainty for the whole sample,
where the uncertainties are weighted with the cross sections of the individual final states.
\Tab{tab:analysis_theory_uncertainties_values} gives an overview of the size of the individual uncertainties in percent,
by listing the minimum, maximum and average value of their distributions over all points of the two signal grids as well as the standard deviation.
The scale and the \ac{PDF} uncertainties clearly dominate among the systematic uncertainties that are considered.
The uncertainties from the computation in \propername{Prospino} itself
and the uncertainty from the variation of the strong coupling constant are much smaller.
As described above for the \ac{mSUGRA} grid,
a fixed value is assumed for \texttt{relUncAlphaS} for all grid points.

\begin{table}
  \centering
  \begin{tabular}{ll*{4}{d{2.2}}}
    \toprule
    Signal grid & Uncertainty & \multicolumn{1}{l}{Min} [\%] & \multicolumn{1}{l}{Max} [\%] & \multicolumn{1}{l}{Mean} [\%] & \multicolumn{1}{l}{Std. dev.} [\%]\\
    \midrule
    mSUGRA & \texttt{relUncPDF}  & 3.0 & 24.3 & 10.0 & 5.4\\
     & \texttt{relScale}  & 6.0 & 15.5 & 8.6 & 2.6\\
     & \texttt{relProspinoError}  & 0.05 & 0.15 & 0.07 & 0.02\\
     & \texttt{relUncAlphaS} &  & & 3.0  & \\
    MSSM & \texttt{relUncPDF}  & 1.5 & 49.9 & 8.1 & 7.1\\
     & \texttt{relScale}  & 6.1 & 26.0 & 11.7 & 3.9\\
     & \texttt{relProspinoError}  & 0.06 & 0.30 & 0.11 & 0.04\\
     & \texttt{relUncAlphaS}  & 0.4 & 10.2 & 2.2 & 1.7\\
    \bottomrule
  \end{tabular}
  \caption{
    Statistical properties of the distributions of the theoretical uncertainties on the cross sections in the two signal grids.
    Listed are the minimum (Min), maximum (Max) and average value (Mean) of the distributions of the uncertainties from all grid points,
    together with the standard deviation (Std. dev.).
  }
  \label{tab:analysis_theory_uncertainties_values}
\end{table}

\newpage

\section{Analysis Results in Numbers}
\label{sec:analysis_summary_limit_input}

This section summarizes the event counts in data and in Monte Carlo,
following from the selection according to the cutflow presented in \Sec{sec:analyis_description_cutflow},
as well as the statistical and systematic uncertainties on these numbers\footnote{
  Note that the distinction between statistical and systematic errors is not always sharp.
  In general, statistical errors are understood to be those that are affected by the dataset size,
  whereas systematic errors are independent of the dataset.
  In this sense,
  there is a statistical uncertainty due to the finite size of the Monte Carlo samples,
  which cannot be fixed by some auxiliary information, but it can be reduced by increasing the sample size,
  and there is a systematic uncertainty due to theory, \eg when comparing different Monte Carlo generators,
  the value of which will not go to zero for infinite sample sizes.
  On the other hand, though there are uncertainties which look at least intuitively like a systematic uncertainty,
  for example the lepton energy scale, %
  which is the largest uncertainty in the $W$ mass measurement at the Tevatron \cite{WmassTevatronRev2008,WmassTevatron2009}.
  But this systematic uncertainty gets smaller the more luminosity is collected.
}.
From these, the inputs are distilled which are finally fed into the limit-setting calculation.

\renewcommand{\arraystretch}{1.} %
\newcommand{\noc}{\multicolumn{1}{c}{}} %
\newcommand{\jescell}{\multirow{2}{*}{JES \parbox{1cm}{ {\delimitershortfall=5pt\rdelim\{{2}{0mm}} \hspace*{1.5mm} up \\ \hspace*{2.5mm} down}}} %

\afterpage{%
  \begin{landscape}
    \renewcommand{\arraystretch}{1.} %
    \begin{table}
      \setlength{\tabcolsep}{5.2pt} %
      \small
      \hspace*{-2mm} %
      \begin{tabular}{ll@{}*{5}{rrl}}
      \toprule
      && \multicolumn{3}{l}{$Z$+jets} & \multicolumn{3}{l}{$W$+jets} & \multicolumn{3}{l}{$t\bar t$} & \multicolumn{3}{l}{\propername{Pythia} dijets} & \multicolumn{3}{l}{Single top} \\
      \cmidrule(r){3-5} \cmidrule(r){6-8} \cmidrule(r){9-11} \cmidrule(r){12-14} \cmidrule(r){15-17}
      \multicolumn{2}{l}{cutflow step} & \multicolumn{1}{c}{$n_\text{exp}$} & $\Delta_\text{stat}$ & $\epsilon_\text{cut}$ & \multicolumn{1}{c}{$n_\text{exp}$} & $\Delta_\text{stat}$ & $\epsilon_\text{cut}$ & \multicolumn{1}{c}{$n_\text{exp}$} & $\Delta_\text{stat}$ & $\epsilon_\text{cut}$ & \multicolumn{1}{c}{$n_\text{exp}$} & $n_\text{stat}$ & $\epsilon_\text{cut}$ & \multicolumn{1}{c}{$n_\text{exp}$} & $\Delta_\text{stat}$ & $\epsilon_\text{cut}$\\
      \midrule
      1 & Raw events & \numprint{137867.6} & 141.5 &  & $1.05\ten{6}$ & 450.7 &  & \numprint{5370.4} & 7.8 &  & $4.52\ten{11}$ & $3.60\ten{8}$ &  & \numprint{1250.4} & 6.8 & \\
      2 & GRL & \numprint{137867.6} & 141.5 & 1.00 & $1.05\ten{6}$ & 450.7 & 1.00 & \numprint{5370.4} & 7.8 & 1.00 & $4.52\ten{11}$ & $3.60\ten{8}$ & 1.00 & \numprint{1250.4} & 6.8 & 1.00\\
      3 & Main trigger & \numprint{3418.9} & 25.3 & 0.02 & \numprint{11025.2} & 36.8 & 0.01 & \numprint{1541.3} & 3.6 & 0.29 & \numprint{5439501.5} & \numprint{32584.3} & 0.00 & 231.4 & 2.9 & 0.19\\
      4 & Jet cleaning & \numprint{3401.6} & 25.3 & 0.99 & \numprint{10955.4} & 36.7 & 0.99 & \numprint{1530.6} & 3.6 & 0.99 & \numprint{5427600.0} & \numprint{32550.9} & 1.00 & 229.9 & 2.9 & 0.99\\
      5 & Primary vertex & \numprint{3399.9} & 25.3 & 1.00 & \numprint{10951.2} & 36.7 & 1.00 & \numprint{1530.5} & 3.6 & 1.00 & \numprint{5427490.5} & \numprint{32550.8} & 1.00 & 229.7 & 2.9 & 1.00\\
      6 & Crack electrons & \numprint{3353.8} & 25.2 & 0.99 & \numprint{10795.6} & 36.5 & 0.99 & \numprint{1509.7} & 3.6 & 0.99 & \numprint{5415772.5} & \numprint{32471.9} & 1.00 & 226.5 & 2.9 & 0.99\\
      7 & Veto on leptons & \numprint{2306.3} & 23.3 & 0.69 & \numprint{4454.6} & 23.0 & 0.41 & 814.0 & 3.1 & 0.54 & \numprint{5353897.0} & \numprint{32326.6} & 0.99 & 104.5 & 2.0 & 0.46\\
      8 & Leading jet & \numprint{1369.3} & 17.7 & 0.59 & \numprint{2353.9} & 16.1 & 0.53 & 573.5 & 2.7 & 0.70 & \numprint{2715003.3} & \numprint{14496.2} & 0.51 & 66.0 & 1.6 & 0.63\\
      9 & Monojet& 552.8 & 13.3 & 0.40 & 541.5 & 9.1 & 0.23 & 3.1 & 0.1 & 0.01 & \numprint{72049.7} & \numprint{3463.0} & 0.03 & 3.4 & 0.4 & 0.05\\
      10 & Charge fraction & 547.7 & 13.3 & 0.99 & 534.4 & 9.0 & 0.99 & 3.1 & 0.1 & 1.00 & \numprint{71096.6} & \numprint{3454.5} & 0.99 & 3.4 & 0.4 & 1.00\\
      11 & Second jet & 707.2 & 10.7 & 0.52 & \numprint{1594.7} & 12.4 & 0.68 & 556.6 & 2.7 & 0.97 & \numprint{2544498.3} & \numprint{13033.8} & 0.94 & 59.3 & 1.5 & 0.90\\
      12 & Charge fraction & 699.6 & 10.6 & 0.99 & \numprint{1560.1} & 12.2 & 0.98 & 551.5 & 2.7 & 0.99 & \numprint{2515436.5} & \numprint{12856.8} & 0.99 & 58.6 & 1.5 & 0.99\\
      13 & $\met > \GeV{100}$ & 339.8 & 7.7 & 0.49 & 516.9 & 7.0 & 0.33 & 102.5 & 0.7 & 0.19 & \numprint{8165.4} & 212.8 & 0.00 & 14.6 & 0.7 & 0.25\\
      14 & $\Delta\phi(\text{jet}, \met) > 0.4$ & 286.8 & 7.2 & 0.84 & 377.6 & 5.9 & 0.73 & 75.1 & 0.5 & 0.73 & 295.0 & 70.4 & 0.04 & 10.7 & 0.6 & 0.74\\
      15 & $\met/\meff > 0.3$ & 244.8 & 6.7 & 0.85 & 308.3 & 5.4 & 0.82 & 59.0 & 0.5 & 0.79 & 126.1 & 68.2 & 0.43 & 8.6 & 0.6 & 0.80\\
      \rowcolor{lightblue} %
      16 & $\meff > \GeV{500}$ & 50.3 & 3.0 & 0.21 & 47.9 & 2.1 & 0.16 & 8.5 & 0.2 & 0.14 & 7.2 & 3.8 & 0.06 & 1.0 & 0.2 & 0.12\\
      \rowcolor{lightblue}
      17 & $\mttwo > \GeV{300}$ & 4.0 & 0.8 & 0.01 & 4.2 & 0.6 & 0.01 & 0.8 & 0.1 & 0.01 & 0.3 & 0.1 & 0.00 & 0.1 & 0.1 & 0.01\\
      18 & Third jet & 229.5 & 5.4 & 0.17 & 552.6 & 6.6 & 0.23 & 494.8 & 2.6 & 0.86 & \numprint{1001039.9} & \numprint{6694.7} & 0.37 & 37.6 & 1.2 & 0.57\\
      19 & $\met > \GeV{100}$ & 98.3 & 3.7 & 0.43 & 172.2 & 3.7 & 0.31 & 82.7 & 0.6 & 0.17 & \numprint{4125.6} & 170.2 & 0.00 & 7.0 & 0.5 & 0.19\\
      20 & $\Delta\phi(\text{jet}, \met) > 0.4$ & 71.2 & 3.2 & 0.72 & 111.1 & 3.0 & 0.65 & 57.1 & 0.5 & 0.69 & 144.1 & 15.3 & 0.03 & 4.3 & 0.4 & 0.61\\
      21 & $\met/\meff > 0.25$ & 57.3 & 2.9 & 0.81 & 82.1 & 2.6 & 0.74 & 43.4 & 0.4 & 0.76 & 18.1 & 6.6 & 0.13 & 3.3 & 0.3 & 0.78\\
      \rowcolor{lightblue}
      22 & $\meff > \GeV{500}$ & 26.4 & 2.0 & 0.46 & 33.5 & 1.6 & 0.41 & 15.3 & 0.2 & 0.35 & 7.2 & 3.8 & 0.40 & 1.1 & 0.2 & 0.32\\
      \rowcolor{lightblue}
      23 & $\meff > \TeV{1}$ & 0.8 & 0.3 & 0.01 & 1.1 & 0.3 & 0.01 & 0.2 & 0.0 & 0.00 & 0.1 & 0.1 & 0.00 & 0.1 & 0.1 & 0.03\\
      24 & Fourth jet & 52.0 & 2.2 & 0.04 & 136.2 & 3.2 & 0.06 & 349.8 & 2.3 & 0.61 & \numprint{232505.0} & \numprint{2536.3} & 0.09 & 19.7 & 0.8 & 0.30\\
      \bottomrule
      \end{tabular}
      \caption{
        Event counts as they follow from the analysis cutflow,
        showing the expected event numbers $n_\text{exp}$ from Monte Carlo,
        normalized to an integrated luminosity of %
        $\Lint^\text{2010} = \unit[33.4]{pb^{-1}}$.
        $\Delta_\text{stat}$ is the absolute statistical uncertainty on this number,
        and $\epsilon_\text{cut}$ is the efficiency of the cut on this sample with respect to the preceding cut (cf. \Tab{tab:analysis_cutflow_data}).
        The four rows defining the signal regions are highlighted.
      }
      \label{tab:analysis_cutflow_numbers_MC}
      \resettabcolsep %
    \end{table}
    \arraystretchdefault
  \end{landscape}

  \begin{table}

    \setlength{\tabcolsep}{6.4pt} %
    \begin{tabular}{ll*{4}{d{2.2}>{\,(}d{2.2}<{\,\%)}}}

      \toprule
      && \multicolumn{8}{c}{Signal region} \\
      \cmidrule{3-10}
      Background & Uncertainty & \multicolumn{1}{c}{A} & \noc & \multicolumn{1}{c}{B} & \noc & \multicolumn{1}{c}{C} & \noc & \multicolumn{1}{c}{D} & \noc  \\
      \midrule

      $W$ + jets & \textit{Event count} & 47.90 & \noc & 4.23 & \noc & 33.52 & \noc & 1.06 & \noc\\
      & Statistics & 2.06 & 4.3 & 0.60 & 14.1 & 1.63 & 4.9 & 0.28 & 26.8\\
      & Luminosity & 1.63 & 3.4 & 0.14 & 3.4 & 1.14 & 3.4 & 0.04 & 3.4\\
      & JER & 0.76 & 1.6 & 0.08 & 1.9 & 0.03 & 0.1 & 0.00 & 0.0\\
      & \jescell & 9.47 & 19.8 & 1.20 & 28.3 & 6.97 & 20.8 & 0.15 & 13.9\\
      &  & 7.92 & 16.5 & 0.70 & 16.6 & 6.16 & 18.4 & 0.30 & 28.0\\
      & Pile-up & 2.41 & 5.0 & 0.42 & 9.8 & 1.72 & 5.1 & 0.24 & 23.0\\
      & SUSY & 19.16 & 40.0 & 2.88 & 68.0 & 14.75 & 44.0 & 0.77 & 73.0\\
      $Z$ + jets & \textit{Event count} & 50.29 & \noc & 3.97 & \noc & 26.41 & \noc & 0.81 & \noc\\
      & Statistics & 3.02 & 6.0 & 0.81 & 20.5 & 1.95 & 7.4 & 0.34 & 42.3\\
      & Luminosity & 1.71 & 3.4 & 0.13 & 3.4 & 0.90 & 3.4 & 0.03 & 3.4\\
      & JER & 0.44 & 0.9 & 0.08 & 1.9 & 0.64 & 2.4 & 0.00 & 0.0\\
      & \jescell & 11.27 & 22.4 & 1.54 & 38.8 & 6.84 & 25.9 & 0.31 & 38.7\\
      &  & 9.22 & 18.3 & 0.73 & 18.5 & 4.88 & 18.5 & 0.00 & 0.0\\
      & Pile-up & 1.51 & 3.0 & 0.33 & 8.3 & 2.49 & 9.4 & 0.29 & 35.6\\
      & SUSY & 10.56 & 21.0 & 2.82 & 71.0 & 6.07 & 23.0 & 0.45 & 55.0\\
      $t\bar t$ & \textit{Event count} & 8.50 & \noc & 0.79 & \noc & 15.27 & \noc & 0.18 & \noc\\
      & Statistics & 0.18 & 2.1 & 0.06 & 7.0 & 0.24 & 1.6 & 0.03 & 14.6\\
      & Luminosity & 0.29 & 3.4 & 0.03 & 3.4 & 0.52 & 3.4 & 0.01 & 3.4\\
      & JER & 0.07 & 0.8 & 0.05 & 5.9 & 0.45 & 3.0 & 0.00 & 0.0\\
      & \jescell & 1.78 & 20.9 & 0.22 & 28.4 & 3.83 & 25.1 & 0.07 & 38.3\\
      &  & 1.63 & 19.2 & 0.18 & 22.5 & 2.94 & 19.2 & 0.05 & 25.5\\
      & Pile-up & 0.09 & 1.0 & 0.02 & 2.4 & 0.21 & 1.4 & 0.07 & 39.9\\
      Single top & \textit{Event count} & 1.02 & \noc & 0.12 & \noc & 1.07 & \noc & 0.11 & \noc\\
      & Statistics & 0.19 & 18.7 & 0.07 & 54.3 & 0.19 & 18.1 & 0.06 & 55.9\\
      & Luminosity & 0.03 & 3.4 & 0.00 & 3.4 & 0.04 & 3.4 & 0.00 & 3.4\\
      & JER & 0.05 & 4.8 & 0.04 & 30.3 & 0.00 & 0.3 & 0.00 & 0.0\\
      & \jescell & 0.29 & 28.1 & 0.04 & 30.3 & 0.21 & 19.5 & 0.04 & 32.3\\
      &  & 0.20 & 19.2 & 0.04 & 31.9 & 0.28 & 25.8 & 0.07 & 64.5\\
      & Pile-up & 0.05 & 4.9 & 0.00 & 1.0 & 0.02 & 1.6 & 0.02 & 20.9\\
      \multirow{2}{4.5em}{\propername{Pythia} dijets} & \textit{Event count} & 7.18 & \noc & 0.28 & \noc & 7.19 & \noc & 0.09 & \noc\\
      & Statistics & 3.81 & 53.1 & 0.13 & 44.2 & 3.81 & 53.0 & 0.07 & 83.0\\
      & Luminosity & 0.24 & 3.4 & 0.01 & 3.4 & 0.24 & 3.4 & 0.00 & 3.4\\
      & JER & 0.14 & 2.0 & 0.00 & 0.4 & 0.36 & 5.0 & 0.00 & 0.0\\
      & \jescell & 2.83 & 39.4 & 0.09 & 31.5 & 0.07 & 1.0 & 0.07 & 82.8\\
      &  & 2.83 & 39.4 & 0.01 & 4.2 & 2.76 & 38.3 & 0.00 & 0.0\\
      & Pile-up & 7.18 & 100.0 & 0.28 & 100.0 & 7.19 & 100.0 & 0.09 & 100.0\\
      \bottomrule
    \end{tabular}
    \caption{
      Systematic uncertainties as absolute event counts and relative numbers (in brackets)
      in the four signal regions,
      broken up by background type and source of uncertainty.
      The rows labelled ``Event count'' do not state an uncertainty
      but the number of Monte Carlo events from this sample reaching the respective signal region.
      ``SUSY'' stands for uncertainties adopted from the official analysis (see \Sec{sec:analysis_explain_SUSY_WZ_uncertainties}). %
      \revised{%
      Rounding makes small uncertainties appear as $0.00$,
      which should be read as $<0.01$.}
    }
    \label{tab:analysis_uncertainties_values}
  \resettabcolsep %
  \end{table}
  
  \arraystretchdefault
}

\Tab{tab:analysis_cutflow_numbers_MC} lists the event counts from the cutflow on Monte Carlo
for the different types of background samples,
sorted from the left to the right by the event counts in the signal regions.
The number of expected events $n_\text{exp}$ is shown
together with the absolute statistical uncertainty $\Delta_\text{stat}$
and the cut efficiency $\epsilon_\text{cut}$ on the respective samples with respect to the preceding cut\footnote{
  Note that not always the previous step is the preceding step in the cutflow, cf. \Tab{tab:analysis_cutflow_data}.
}.
The rows which correspond to the four signal regions are highlighted.
The table shows which of the backgrounds gives the largest contribution
and allows to compare the efficiency of the cuts with respect to the suppression of the backgrounds.
It demonstrates that a strong suppression of the otherwise dominating QCD background is achieved by cutting on \met.
The fraction of events with pairwise produced top quarks reaching the signal regions
is the largest among the background samples which are considered.
It is only small in terms of absolute event numbers due to the small production cross section,
but when the LHC will run at the nominal center-of-mass energy,
the background from top quark events will be of the same order as \Zjets and \Wjets. %

\Tab{tab:analysis_uncertainties_values} lists the systematic uncertainties in the four signal regions,
broken up by the different background types and sources of uncertainty,
which have been introduced above (cf. \Tab{tab:analysis_overview_systematic_uncertainties}).
The luminosity uncertainty has a fixed value for all background types and the signal.
The uncertainties arising from pile-up, the jet-energy scale and resolution are computed in the way described in \Sec{sec:analysis_uncertainties}.
The uncertainty on the background normalization is denoted as SUSY in this table
because the values from the official Supersymmetry analysis are used (cf. \Tab{tab:analysis_uncertainties_susy}).
Not listed is the theory uncertainty on the signal normalization
because this table contains only the numbers for the background Monte Carlo.

In \Tab{tab:analysis_signal_regions_event_count}
the final collection of numbers that make up the result of the analysis is given.
It contains the number of events counted in data $n_\text{data}$ in the signal regions,
according to the selection due to the cutflow from \Tab{tab:analysis_cutflow_data},
and the expected event counts $n_\text{exp}$ in the signal regions from Monte Carlo from \Tab{tab:analysis_cutflow_numbers_MC},
together with the total uncertainties from all the different sources.
The total uncertainties are derived from the sample uncertainties given in \Tab{tab:analysis_uncertainties_values}.
Depending on the type of uncertainty and its correlation among different backgrounds
as given in \Tab{tab:analysis_overview_systematic_uncertainties},
the total uncertainty is the arithmetic or quadratic sum of the respective absolute uncertainties from all background types.
For convenience, the numbers are again given as absolute and relative uncertainties.
For signal regions A and C,
which have the highest event counts,
the dominating uncertainty comes from the uncertainty of the jet-energy scale.
In signal regions B and D with much lower event counts,
the uncertainty on the background normalization %
becomes the dominant source of uncertainty.
In signal region D with the smallest event count,
also the limited Monte Carlo statistics gives an important contribution.
Another important uncertainty here is due to pile-up.

The table allows to compare the expected event counts in the signal regions from Standard Model backgrounds
modeled in the Monte Carlo samples to the event counts that were actually measured in data.
In signal regions~A, C and~D,
the Standard Model background expectation from Monte Carlo
exceeds the number of events measured in data.
In signal region~B,
the number of events measured in data is slightly above
the Standard Model background expectation from Monte Carlo,
but also well consistent within uncertainties.
This can be seen in \Tab{tab:analysis_signal_regions_event_count_summary},
which is a summary of \Tab{tab:analysis_signal_regions_event_count},
summing over all sources of uncertainties.

Due to this finding, showing no significant excess of the number of events beyond the Standard Model background expectation,
in the rest of this section the results will be interpreted in terms of exclusion limits.
The two signal grids presented in \Sec{sec:analysis_SUSY_GRIDs}
will be used to obtain the expected number of signal events
for different realizations of parameters within the simplied supergravity and mSUGRA model.

\begin{table}
  \centering
  \begin{tabular}{ll*{4}{d{3.1}>{(}d{2.4}<{\,\%)}}l}
    \toprule
     &  & \multicolumn{8}{c}{Signal region} &  \\
    \cmidrule{3-11}
     &  & \multicolumn{2}{c}{A} & \multicolumn{2}{c}{B} & \multicolumn{2}{c}{C} & \multicolumn{2}{c}{D} \\
    \midrule

     & $n_\text{data}$ & 87 & \noc & 11 & \noc & 66 & \noc & 2 & \noc\\
     & $n_\text{exp}$ & 114.9 & \noc & 9.4 & \noc & 83.5 & \noc & 2.2 & \noc\\

    \midrule

    \multirow{7}{*}{\Vert{Uncertainty}} & Statistics  & 5.3 & 4.6 & 1.0 & 10.9 & 4.6 & 5.5 & 0.5 & 20.3\\
     & Luminosity  & 3.9 & 3.4 & 0.3 & 3.4 & 2.8 & 3.4 & 0.1 & 3.4\\
     & JER  & 1.5 & 1.3 & 0.0 & 0.1 & 1.4 & 1.7 & 0.0 & 0.0\\ %
     & \jescell & 25.6 & 22.3 & 3.1 & 32.9 & 17.9 & 21.5 & 0.6 & 28.4\\
     &  & 21.8 & 19.0 & 1.7 & 17.7 & 17.0 & 20.4 & 0.4 & 18.5\\
     & Pile-up  & 11.2 & 9.8 & 1.1 & 11.2 & 11.6 & 13.9 & 0.7 & 31.8\\
     & SUSY  & 21.9 & 19.0 & 4.0 & 42.9 & 16.0 & 19.1 & 0.9 & 39.6\\

    \bottomrule
  \end{tabular}
  \caption{
    Summary of the numbers that constitute the result of the analysis.
    Given in the table are the event counts on data ($n_\text{data}$)
    and the expected counts on Monte Carlo ($n_\text{exp}$),
    together with the uncertainties on the Monte Carlo expectations
    in terms of absolute event counts and (in brackets) as percentage values.
    \revised{%
    Rounding to one decimal place makes small uncertainties appear as $0.0$,
    which should be read as $<0.1$.} %
  }
  \label{tab:analysis_signal_regions_event_count}
\end{table}

\renewcommand{\arraystretch}{1.3}
\begin{table}
  \centering
  \begin{tabular}{llrr@{}>{$\pm$}l@{}rr}
    \toprule
    && \multicolumn{2}{c}{Event counts} \\
    \cmidrule{3-7}
    \multicolumn{2}{l}{Signal region} & \multicolumn{1}{r}{$n_\text{data}$} & \multicolumn{4}{c}{$n_\text{exp} \pm \Delta_\text{stat} \pm \Delta_\text{syst} $} \\
    \midrule
        A  &  2 jets + $\meff  >  \GeV{500}$  & 87 & 114.9 & 5.3 & $^{+35.8}_{-33.1}$\\
    B  &  2 jets + $\mttwo >  \GeV{300}$  & 11 & 9.4 & 1.0 & $^{+5.2}_{-4.5}$\\
    C  &  3 jets + $\meff  >  \GeV{500}$  & 66 & 83.5 & 4.6 & $^{+26.9}_{-26.2}$\\
    D  &  3 jets + $\meff  > \GeV{1000}$  & 2 & 2.2 & 0.5 & $^{+1.3}_{-1.2}$\\

    \bottomrule
  \end{tabular}
  \caption{
    Summary of the analysis result in \Tab{tab:analysis_signal_regions_event_count},
    giving the event counts observed in data ($n_\text{data}$)
    and expected from Monte Carlo ($n_\text{exp}$),
    together with the statistical ($\Delta_\text{stat}$)  and total systematic uncertainties ($\Delta_\text{syst}$).
    The systematic uncertainties are asymmetric due to the asymmetric contribution from the uncertainty
    on the jet-energy scale.
  }
  \label{tab:analysis_signal_regions_event_count_summary}
\end{table}
\arraystretchdefault

\section{Kinematic Distributions}

A number of important distributions, on which the cuts in the analysis cutflow are based,
are collected in \Figs{fig:analysis_distribution_cut11MET} through \ref{fig:analysis_distribution_cut20Meff3}.
All of them show the distribution of the respective variable in data and in Monte Carlo using the same scheme:
The upper part of each plot shows several distributions in terms of event counts,
the data being represented by the black dots, for which the error bars give statistical uncertainties\footnote{
  Note that this is by convention different from the numbers given in the tables,
  where statistical uncertainties are given explicitly only for the event counts expected from Monte Carlo.
}.
The background expectations from Monte Carlo are represented by the shaded histograms,
stacking the \ac{QCD} (\propername{Pythia} dijets), $W$ + jets, $Z$ + jets, top pair and single top background expectations on top of each other in different colors.
They are normalized to the integrated luminosity of the 2010 dataset (cf. \Eq{eq:analysis_integrated_luminosity_full_dataset_2010}) and sorted,
so that the largest contribution is on top.
Note that the relative contributions from the different Standard Model backgrounds vary,
depending on which cuts have already been applied,
and the order of the stacking changes accordingly.
\ac{QCD} is the dominating background in the first plots,
but is replaced in later steps of the cutflow by the other backgrounds.
To give an impression of the expected additional event counts that would arise from a potential Supersymmetry signal,
the red dashed line on top of the background expectations
gives the Monte Carlo expectation for the supersymmetric benchmark scenario SU4,
a scenario with comparably low masses for the supersymmetric particles and a large cross section.
The vertical line with the arrow indicates the cut which is done on the variable shown in the histogram,
the tip of the arrow pointing into the direction of the events that are being kept.
In the lower part of each plot, observed data and Monte Carlo expectation are compared bin-by-bin
to provide an overview of where the Monte Carlo under- or overestimates the number of events observed in data at one glance.
The black markers in the lower part give the ratio of the number of data events and the expected number of events from the Monte Carlo expectation,
here without including a potential signal, \ie from Standard Model background processes only.
The error bars above and below the markers show the uncertainties on the ratio,
which stem from the statistical uncertainty on the number of data events.
The shaded regions above and below the markers depict the total uncertainties on the Standard Model background expectation,
summing all sources of uncertainties that have been discussed,
including the statistical uncertainty from the limited Monte Carlo sample size. %
The distributions shown in the plots are the distributions before the cut indicated by the arrow is applied.
The cuts in \Figs{fig:analysis_distribution_cut11MET} through \ref{fig:analysis_distribution_cut15mT2}
and \Figs{fig:analysis_distribution_cut17MET} through \ref{fig:analysis_distribution_cut20Meff3}
correspond to the steps~13 through~17 and~19 through~22 in \Tab{tab:analysis_cutflow_data}, respectively.

Specifically, the plots show the following distributions and cuts:
\Fig{fig:analysis_distribution_cut11MET} shows the distribution of \met before the cut at \GeV{100} in the 2-jet channel,
which reduces the \ac{QCD} background significantly.
\Fig{fig:analysis_distribution_cut12dPhi} shows the distribution of the angles $\Delta \phi(\text{jet}, \met)$ between the three leading jets %
with $\pt > \GeV{40}$ and the direction of \met in the transverse plane in the 2-jet channel,
before the cut at $0.4$ to get rid of events in which a hard jet is aligned with~\met,
and the \met potentially is due to a mismeasurement of the energy of this jet.
\Fig{fig:analysis_distribution_cut13METMeff2} shows the distribution of the ratio of the missing transverse energy and the effective mass in the 2-jet channel,
before the cut at $0.3$.
\Fig{fig:analysis_distribution_cut14Meff2} shows the distribution of the effective mass in the 2-jet channel,
before the cut at \GeV{500},
which defines signal region A.
\Fig{fig:analysis_distribution_cut15mT2} shows the distribution of the stransverse mass \mttwo,
the generalization of the transverse mass to pair decays,
before the cut at \GeV{300} in the 2-jet channel,
which defines signal region B.
\Fig{fig:analysis_distribution_cut17MET} shows the distribution of \met,
again before a cut at \GeV{100},
but this time for events which have three jets with $\pt > \GeV{120}$ for the leading and $\pt > \GeV{40}$ for the second and third jet.
\Fig{fig:analysis_distribution_cut18dPhi} shows the distribution of the angles $\Delta \phi(\text{jet}, \met)$,
before the cut at $0.4$,
this time in the three-jet channel.
\Fig{fig:analysis_distribution_cut19METMeff3} shows the distribution of the ratio of the missing transverse energy and the effective mass in the 3-jet channel,
before the cut at $0.25$.
\Fig{fig:analysis_distribution_cut20Meff3} shows the distribution of the effective mass in the 3-jet channel before the cut at \GeV{500},
which defines signal region~C.
The cut which defines signal region~D in the 3-jet channel is applied to the same distribution
but at \TeV{1}, so that only two events are left in data after this cut,
as can be seen in the same plot.

Note that the cuts which are indicated in the plots of $\Delta \phi(\text{jet}, \met)$ are slightly misleading
because an event passes this cut only if all of the three leading jets above \GeV{40} fulfill $\Delta \phi(\text{jet}, \met) > 0.4$,
which cannot be represented in the one-dimensional plot.
This explains why in \Fig{fig:analysis_distribution_cut12dPhi} it looks like \ac{QCD} is still the dominating background after the cut on $\Delta \phi(\text{jet}, \met)$,
but according to \Fig{fig:analysis_distribution_cut13METMeff2} it no longer is.
This is also confirmed by  \Tab{tab:analysis_cutflow_numbers_MC}. %

\begin{figure}
  \centering
    \incgraphics{width=\widthwideplot}{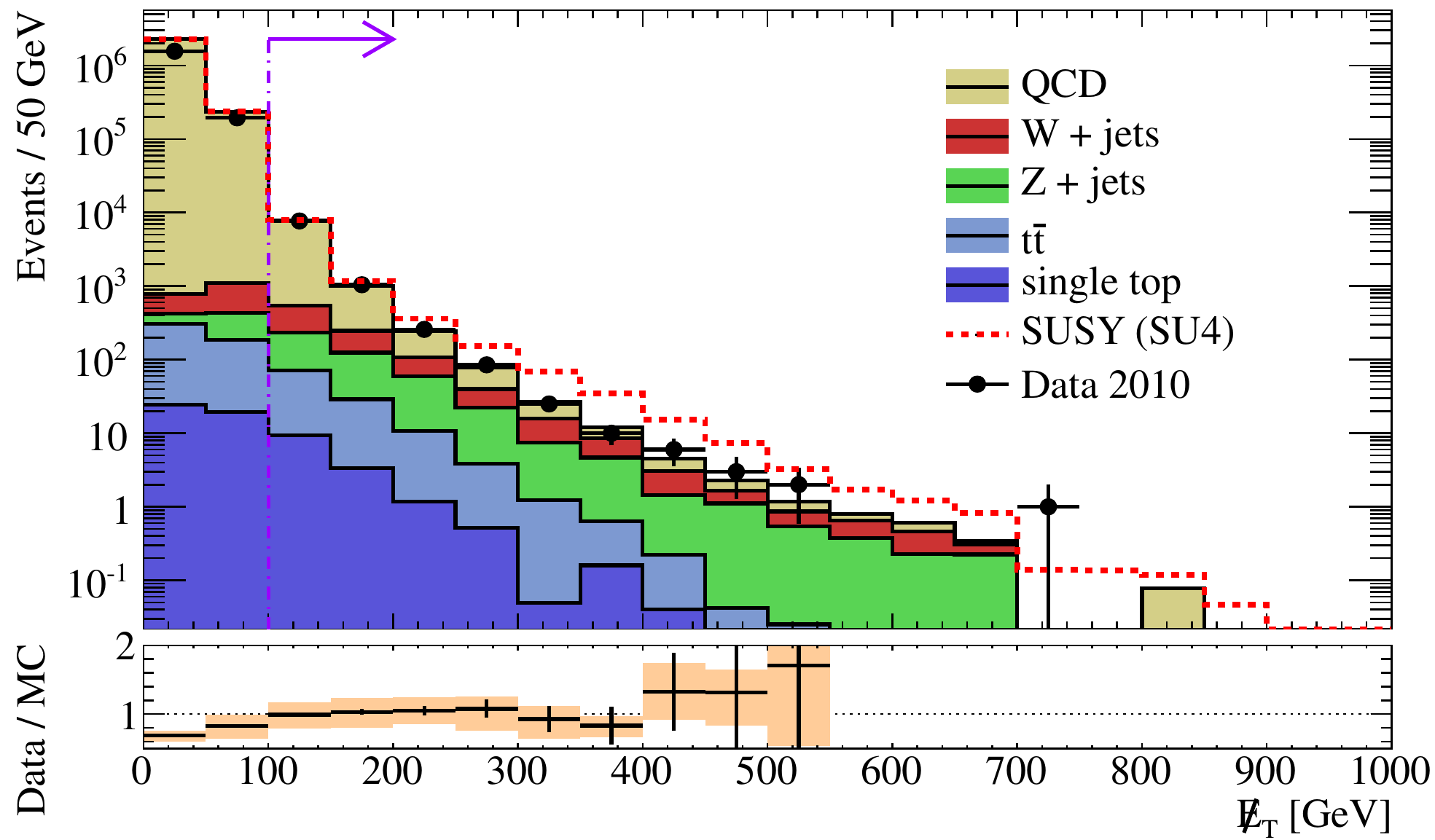}
  \caption{
    Distribution of \met before the cut at \GeV{100} in the 2-jet channel.
    The upper plot shows the distribution observed in data as dots,
    with error bars indicating the statistical error,
    and the Monte Carlo expectations for the Standard Model backgrounds as stacked shaded histograms,
    with the expected additional counts for an exemplary supersymmetric signal shown as the dashed red line on top.
    The arrow depicts the cut that is applied to the variable on the horizontal axis.
    The lower part shows the ratio of data and Monte Carlo,
    with error bars indicating the uncertainty on the ratio
    arising from the statistical uncertainty on the observed counts in data.
    The shaded regions represent the systematic uncertainties.
  }
  \label{fig:analysis_distribution_cut11MET}
\end{figure}

\begin{figure}
  \centering
    \incgraphics{width=\widthwideplot}{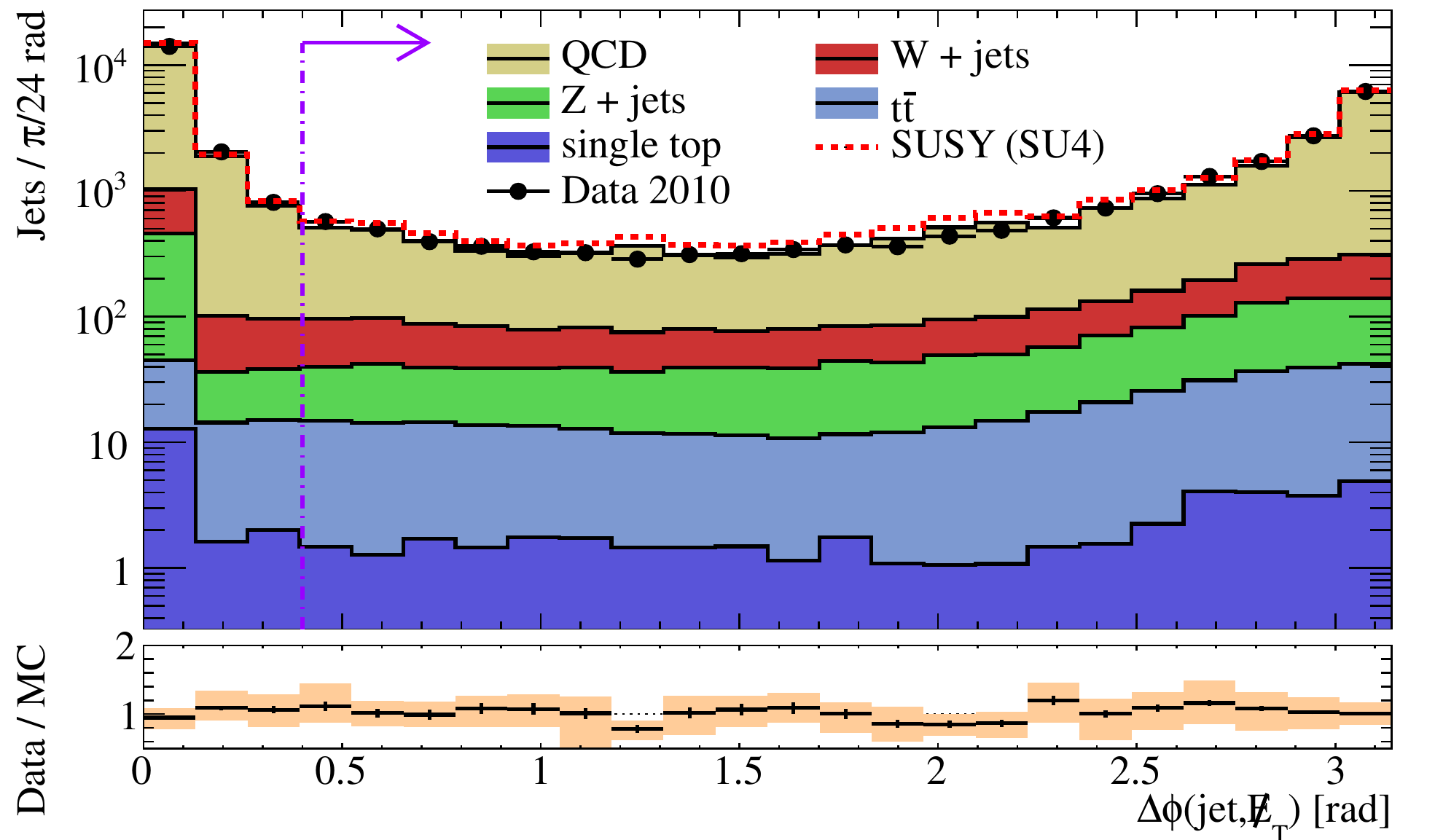}
  \caption{
    Distribution of the angles $\Delta \phi(\text{jet}, \met)$ between the three leading jets with $\pt > \GeV{40}$
    and the direction of \met in the transverse plane in the 2-jet channel,
    before the cut at~$0.4$.
    (Note that the cut at~$0.4$ in this histogram is slightly shifted with respect to the lower edge of the respective bin,
    but the discrepancy is negligible.)
  }
  \label{fig:analysis_distribution_cut12dPhi}
\end{figure}

\begin{figure}
  \centering
    \incgraphics{width=\widthwideplot}{{{postprocess_local_plots_cut13_METM_eff2_styled}}}
  \caption{
    Distribution of the ratio of the missing transverse energy and the effective mass in the 2-jet channel before the cut at~$0.3$.
  }
  \label{fig:analysis_distribution_cut13METMeff2} %
\end{figure}

\begin{figure}
  \centering
    \incgraphics{width=\widthwideplot}{{{postprocess_local_plots_cut14_M_eff_2_MeV_styled}}}
  \caption{
    Distribution of the effective mass in the 2-jet channel before the cut at \GeV{500},
    which defines signal region~A.
  }
  \label{fig:analysis_distribution_cut14Meff2} %
\end{figure}

\begin{figure}
  \centering
    \incgraphics{width=\widthwideplot}{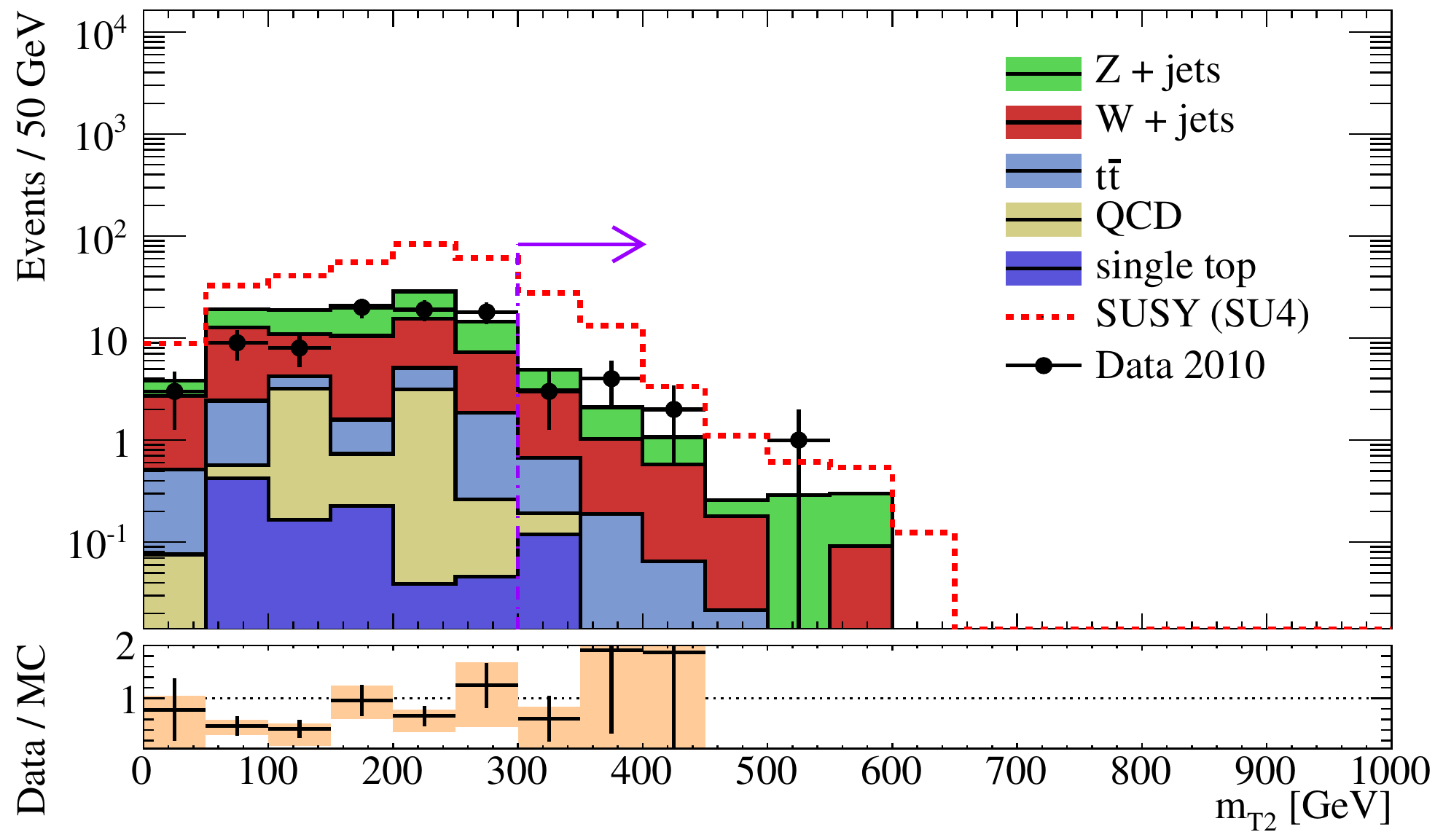}
  \caption{
    Distribution of the stransverse mass \mttwo before the cut at \GeV{300} in the 2-jet channel,
    which defines signal region~B.
  }
  \label{fig:analysis_distribution_cut15mT2}
\end{figure}

\begin{figure}
  \centering
    \incgraphics{width=\widthwideplot}{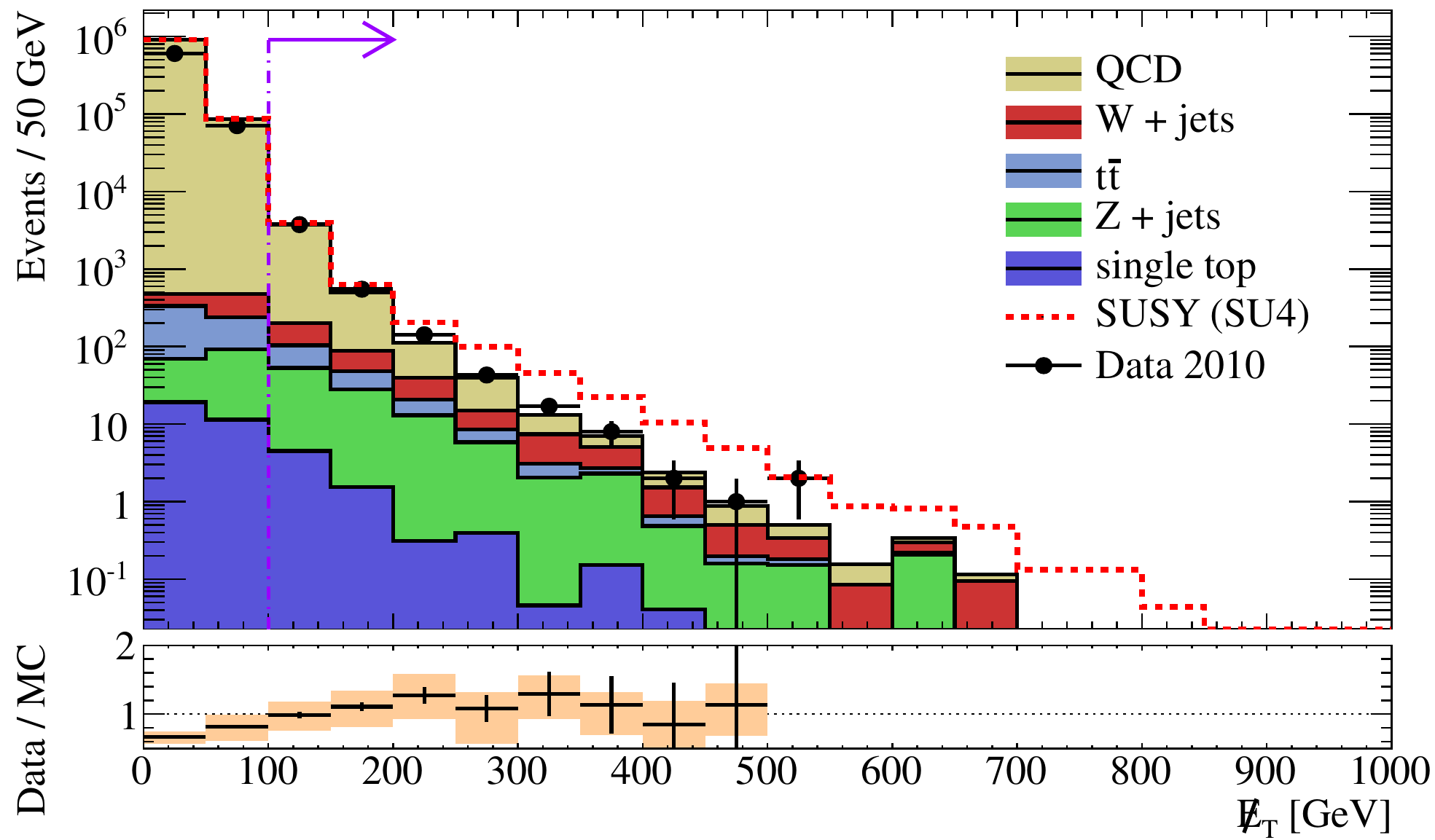}
  \caption{
    Distribution of \met before the cut at \GeV{100} in the 3-jet channel.
  }
  \label{fig:analysis_distribution_cut17MET}
\end{figure}

\begin{figure}
  \centering
    \incgraphics{width=\widthwideplot}{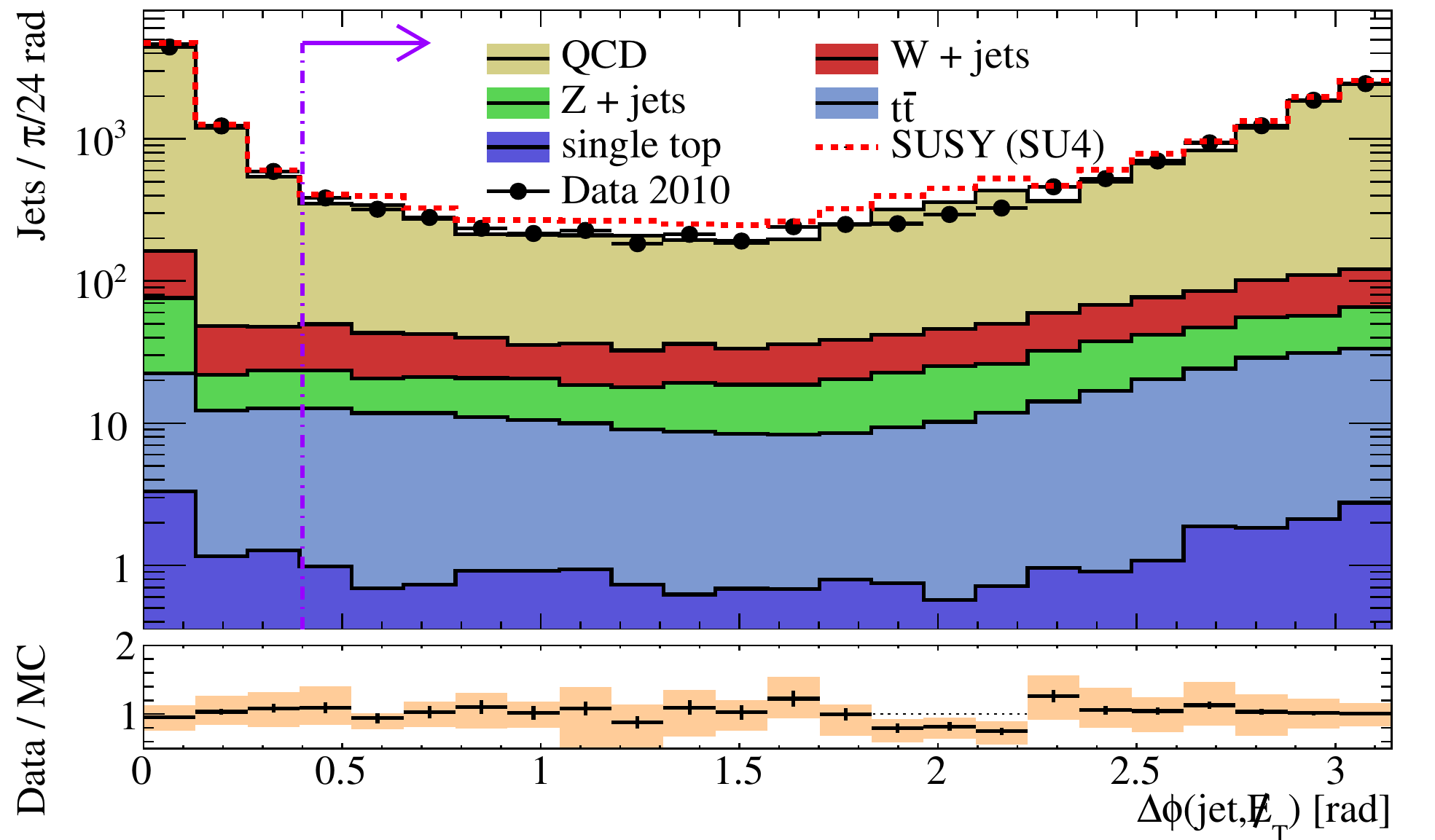}
  \caption{
    Distribution of the angles $\Delta \phi(\text{jet}, \met)$ between the three leading jets with $\pt > \GeV{40}$
    and the direction of \met in the transverse plane in the 3-jet channel
    before the cut at~$0.4$.
  }
  \label{fig:analysis_distribution_cut18dPhi}
\end{figure}

\begin{figure}
  \centering
    \incgraphics{width=\widthwideplot}{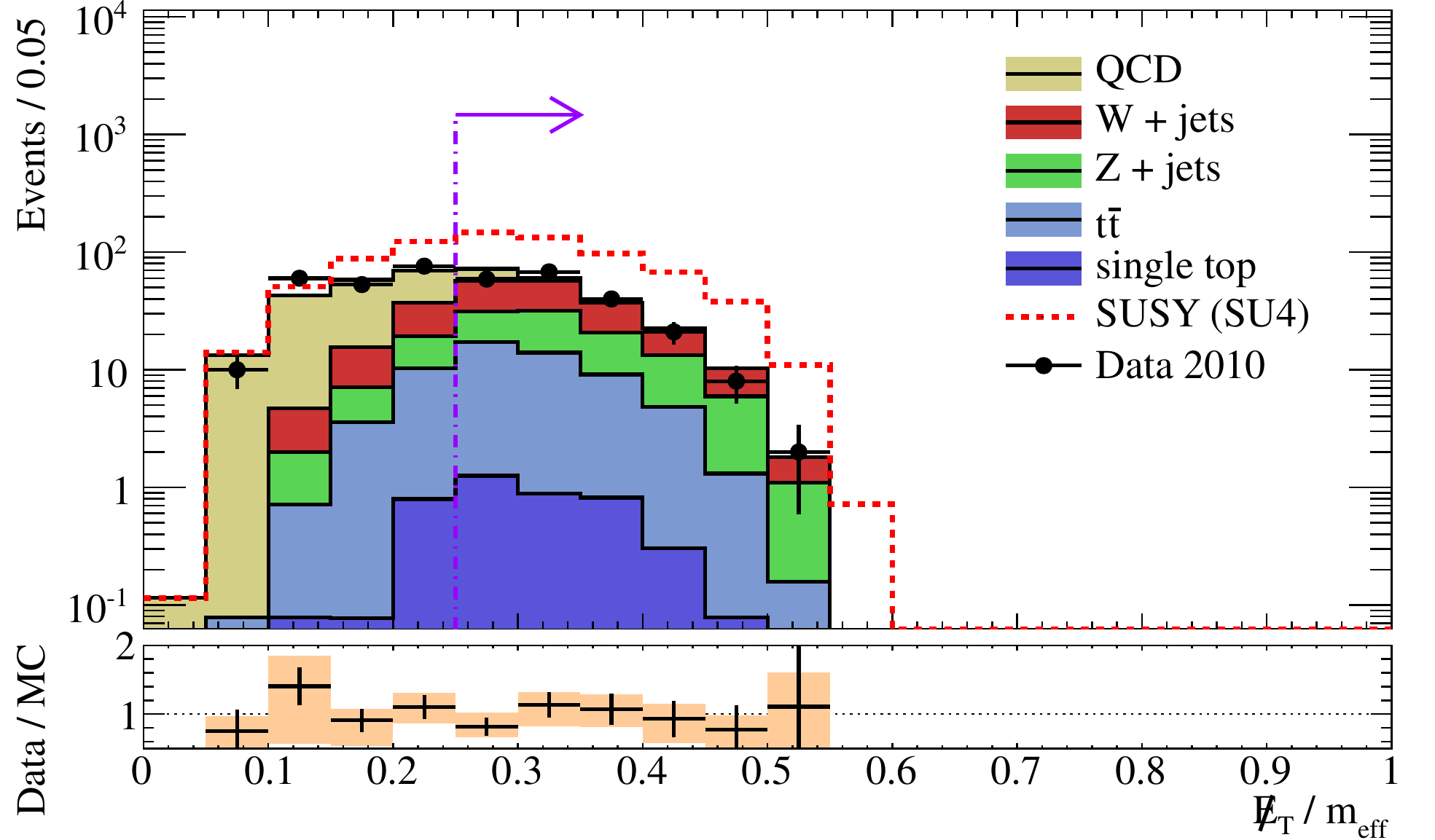}
  \caption{
    Distribution of the ratio of the missing transverse energy and the effective mass in the 3-jet channel before the cut at~$0.25$.
  }
  \label{fig:analysis_distribution_cut19METMeff3}
\end{figure}

\begin{figure}
  \centering
    \incgraphics{width=\widthwideplot}{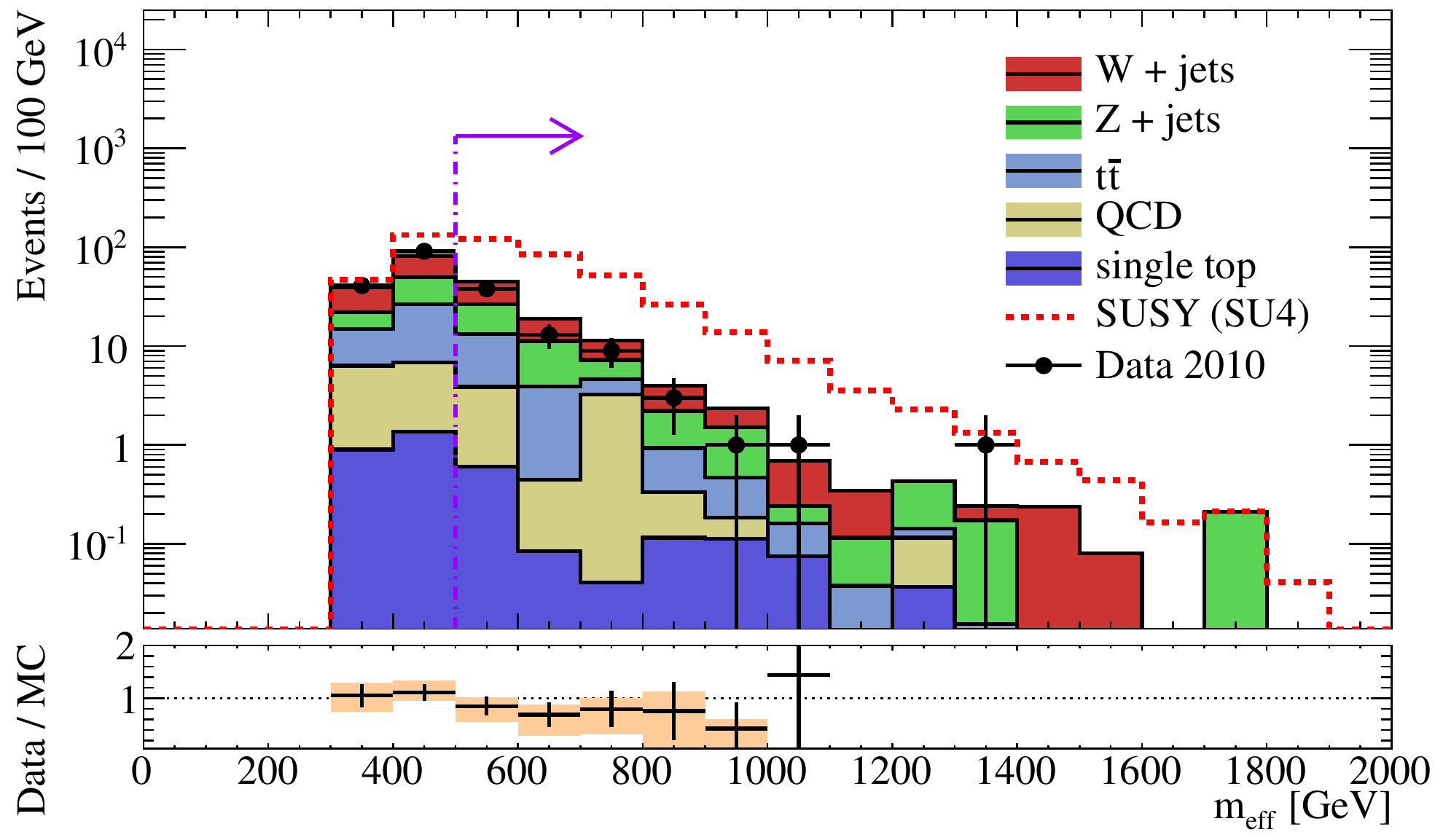}
  \caption{
    Distribution of the effective mass in the 3-jet channel before the cut at \GeV{500},
    which defines signal region~C.
    The cut which defines signal region D in the 3-jet channel is applied to the same distribution but at \TeV{1},
    reducing the number of observed events in data to only~two.
  }
  \label{fig:analysis_distribution_cut20Meff3}
\end{figure}

\section{Overview of Statistical Methods}
\label{sec:results_0lanalysis_statistical_methods}
Before continuing the evaluation of the numbers which constitute the outcome of the analysis in this thesis,
in this section, the statistical methods will be presented
that will be used to quantify the agreement between the data which has been taken
and the predictions from the Standard Model,
with or without the inclusion of Supersymmetry giving rise to physics processes beyond the Standard Model prediction.
Some cross-checks will also be presented,
which compare the performance of the two methods employed for the limit setting.

The important second step after having conducted an experiment is the interpretation of the results that were obtained.
Especially in high-energy physics,
where data-mining makes up a large fraction of the work, statistics plays a vital role.
Depending on the nature and maturity of the experiment,
potential methods for evaluating the outcome
range from straightforward approaches like counting experiments with rectangular cuts %
to sophisticated techniques like neural networks, boosted decision trees or support vector machines,
which aim at extracting the maximum possible information from the measurements
or are needed if no simple discriminating variable exists or the signal is much smaller than the background.
In general, the better the understanding of the detector and its environment, the more advanced methods can be employed.
For the first \ATLAS data which has been collected so far,
the focus is on techniques that are simple enough to still be reasonably intuitive.
On the one hand, the understanding of the detector naturally has not yet reached a state that can be called complete,
and most of the advanced techniques carry the risk of cloaking the causal relations of input and output due to their complexity.
On the other hand, most of the advanced techniques are tailored towards a very specific analysis model,
whereas at the moment more general approaches should be used, in order not to miss something new or bias the analysis towards a specific direction.
As soon as the instantaneous luminosity levels out and stable detector operation is fully established,
so that collecting data becomes day-to-day business,
and when the first hints emerge %
where to look for what,
the analysis will certainly be refined further.

Details about the limit-setting procedure based on the profile likelihood ratio,
which has been employed by the \ATLAS Supersymmetry group for the first analyses of 2010 and 2011 data,
can be found in \cite{ATL-PHYS-INT-2011-032}. %
To implement this procedure, the \propername{RooStats} framework is used.
A brief description of the main methods available in the \propername{RooStats} framework is given in \cite{Moneta2010}.
One of the standard references for statistical methods in high-energy physics is \cite{Barlow1989}. %
Other references are cited in the discussion below.

\subsection{Nomenclature}
The following section will introduce a number of terms relevant to the discussion
of the methods which are used to compute the limits.

One of the most important distributions for the following is the Poisson distribution.
Its probability mass function is given by
\begin{equation}
  \label{eq:def_poisson_distribution}
  \operatorname{Pois}(k | \lambda) = \exp(-\lambda)\frac{\lambda^k}{k!},
\end{equation}
with $k \in \mathbf{N}$.
It has one parameter $\lambda \in (0,\infty)$
and can be generalized to a continuous support by replacing the normalization $k!$ with $\Gamma(k+1)$.
The Poisson distribution is an approximation
of the Binomial distribution given in \Eq{eq:definition_binomial_1}
for rare events, %
\ie for $n\to\infty$ and $np\to\lambda$.
For large values of $\lambda$, due to the central limit theorem and up to a continuity correction, %
the Poisson distribution in turn can be approximated by a Gaussian distribution.
Both expectation value and variance of the Poisson distribution are equal to $\lambda$, the standard deviation is thus $\sqrt{\lambda}$.
The importance of this distribution stems from the fact that the number of entries in the bins of a histogram,
or outcomes of counting experiments in general,
are well modeled by a Poisson distribution for sufficiently large statistics. %
It also describes, for example, the distribution of radioactive decays per time of a source with constant activity.

\subsubsection{Statistical Inference}

The analysis of search results using statistical inference is done by hypothesis testing.
In hypothesis testing, to show the existence of some effect,
the opposite effect is hypothesized in the \index{null hypothesis} \hypo{0},
and it is then to be shown that \hypo{0} is not compatible with the experimental outcome.
Thus, the discovery of new physics or the exclusion of such processes within certain bounds 
is done by rejecting the opposite hypothesis at a given level of certainty.
One can either use the hypothesis of the existence of new physics processes in addition to the known Standard Model processes,
quantify the effect of such new processes\footnote{
  This implies assuming a certain signal strength.
  The signal strength is the ratio of the cross section used in %
  the hypothesis
  over the expected cross section from the underlying model (Standard Model or new physics).
}
and compute the probability that the data agree with this hypothesis.
If the data agree well already with the Standard Model predictions,
this probability will be low
and one can reject the hypothesis of new physics at a high level of certainty,
which will lead to an exclusion.
Or, one can test the hypothesis that there is no physics beyond the known (background) processes described within the Standard Model.
If the data do not agree with the background-only hypothesis, it will be excluded, effectively leading to the discovery of a new signal.
This does not say whether the model for the new physics is correct,
it only rejects the hypothesis that there is only background,
but it leaves open the question,
which of probably many possible explanations for the observed excess of events is correct.

Thus, for limit setting, the null hypothesis \hypo{0} states that both signal and background exist. %
This is called the signal-plus-background hypothesis \hyposb in the following.
The alternate hypothesis is that the hypothesized signal is absent or too small to be seen
and that the data can be explained by previously established theories.
This will be the background-only hypothesis \hypob.

Two types of errors are distinguished in hypothesis testing.
A \index{type I error} occurs when a true null hypothesis is rejected on basis of the data.
Taking the null hypothesis to be the signal-plus-background hypothesis,
this leads to a false exclusion of a true signal (false exclusion rate). %
\index{Type II error}s mean %
failing to reject a false null hypothesis,
corresponding then to a false discovery of an absent signal (false discovery rate). %
A test is called significant at a given level $\alpha$,
\eg significant at the \percent{5} level, %
if the probability for type I errors is less than or equal to $\alpha$.
The probability to not make a type II error,
\ie to correctly reject a false hypothesis, is called the power of a test. %
Taking again the null hypothesis to be \hyposb, this is the \index{exclusion potential}.
For making discoveries, the null hypothesis \hypo{0} would be \hypob,
and the power of the test would be the \index{discovery potential}.

The level of agreement between the observed data and the hypothesis is quantified by the $p$-value\inindex{p-value@$p$-value}.
The $p$-value is the probability to obtain data
which is at least as incompatible with the hypothesis to be tested as is the data at hand,
under the assumption of the validity of the hypothesis.
The hypothesis is reported to be excluded if the $p$-value is below a given threshold,
\ie if it is too unlikely to obtain data which is so incompatible with the hypothesis.
\removesection{ %
By the usual convention that
$ %
  Z = \Phi^{-1}(1-p),
$ %
with $\Phi$ being the inverse of the cumulative Gaussian distribution,
the $p$-value can also be converted into an equivalent significance value $Z$,
stating the number of standard deviations a Gaussian distributed variable %
would deviate from its expectation value if it had this $p$-value.
} %

\subsection{Frequentist and Bayesian Statistics}
In the discussion of experimental results, limit setting and confidence levels,
two different points of view concerning the interpretation of probability can be adopted,
which are called Frequentist and Bayesian.
The difference can be summarized like this \cite{Barlow1989}:

According to the Frequentist definition \inindex{Frequentist statistics},
the probability is the long-run expected frequency of occurrence,
  $p(A) = n_A / N$,
where $n_A$ is the number of times the event $A$ has been found in $N$ observations.
In the Frequentist view, the population mean is fixed, but unknown, and can only be estimated from the data. %
A confidence interval, centered at the sample mean, is inferred from the distribution of the sample mean, %
using the Neyman or confidence belt construction \cite{PDB2010},
such that with a given coverage probability $p_c$, {\eg} \percent{95}, %
the true, unknown mean is in this interval,
and if the experiment were repeated many times,
and more confidence intervals constructed in the same way,
a fraction $p_c$ of these would contain the true mean.
This should not be understood to mean
that the true mean has a probability of $1-p_c$ to lie outside the constructed interval,
but that,
if the true mean lies outside the constructed interval,
the probability to obtain the measured result is $1-p_c$.
Saying that the true value lies within the interval with probability $p_c$
is thus not a statement about the true value but about the interval limits.
One problem of the Frequentist approach is that it relies on the repeatability of the experiment.
Another is that it cannot incorporate limits on the measured parameter
that are physically impossible to violate, such as a mass having to be non-negative.
This is only possible in the Bayesian approach.
The Bayesian definition of probability \inindex{Bayesian statistics}
is a subjective probability.
Experimental evidence can and will modify an initial or \latein{a priori} degree of belief.
To do so, Bayes' theorem is invoked,
which requires specifying the \latein{a priori} distribution of the parameter values that are to be measured.
Using Bayes' theorem, a credible interval is constructed,
which then is reported to contain the true mean with a given probability.
(Note the subtle difference with respect to the Frequentist view,
where no concrete statement about the mean is being made.) %
If a proper prior is chosen, the outcome will be driven by the experimental observation,
and the influence of using different priors needs to be checked,
but will not lead to an unacceptable bias in the constructed limits.
It should be noted that in general priors are not invariant under a change of parameter,
although there are exceptions,
using \eg Jeffreys' rule \cite{PDB2010}. %
For both the Frequentist and the Bayesian approach,
it is clear that the larger the interval
(and the less the information content of or equivalently the less restrictive the statement),
the higher the confidence at which this statement can be made.
The \index{likelihood function} is an important object in the context of statistical inference.
For a joint probability density function (PDF) $f(\vec x|\vec\theta)$
which encodes the probability of finding the measurement $\vec x$
given the parameters of the model $\vec\theta$ for the data,
the likelihood function $L$ is constructed by evaluating the probability density function $f$
with the observed data $\vec x_\text{obs}$
and regarding it as a function of the parameters,
\ie $L = L(\vec \theta) = f(\vec \theta|\vec x_\text{obs})$.
The likelihood is not a probability density function for the parameters.
From a frequentist point-of-view this is not defined.
In Bayesian statistics, a posterior probability density function for the parameters
can be obtained from $L$ by multiplication with a prior probability density function $\pi(\vec\theta)$,
using Bayes' theorem:
\begin{equation}
  p(\vec\theta|\vec x) = \frac{f(\vec x|\vec\theta)\pi(\vec\theta)} {\int f(\vec x|\vec\theta')\pi(\vec\theta')\intd\vec\theta'}. %
\end{equation}
In case of one parameter $\theta$,
this allows to compute a credible interval $[\theta_\text{low}, \theta_\text{high}]$
with a given probability of containing the true value of the parameter by solving
\begin{equation}
  1 - \alpha = \int_{\theta_\text{low}}^{\theta_\text{high}} p(\theta|\vec x) \intd\theta %
  \label{eq:compute_credible_interval}
\end{equation}
for $\theta_\text{low}$ and $\theta_\text{high}$, where the threshold $\alpha$ is called the size of a test.
For $\theta_\text{low}\to-\infty$ this gives an upper limit only.
Note that in general the solution of \Eq{eq:compute_credible_interval} is not unique.
Often the smallest such interval is used.

The Bayesian approach allows to account for systematic uncertainties 
by including prior distributions on additional parameters in the likelihood function 
and marginalizing (integrating) over these.
The additional parameters are called nuisance parameters.
Limits on the measured parameter %
can also be included in the Bayesian limits by setting the prior probability to zero in the physically excluded parameter ranges.
For example, for hypothesized signals that can only contribute positively to the expected event count,
the prior for the signal strength could be chosen flat and non-negative.
Within this framework, also control measurements,
giving \eg an estimate of the background,
can readily be taken into account.

\subsection{Methods for Setting Exclusion Limits}
\label{sec:analysis_methods_description}
Two different methods are used in this thesis to set exclusion limits.
The first is the so-called \cls method,
introduced by the LEP experiments for the Higgs search \cite{Junk1999,Read2002}.
This method is currently (2011) recommended by the \ATLAS Statistics Forum. %
The second method is the \acl{PLR} method
which is used by the \ATLAS SUSY group in the first papers on searches for Supersymmetry in \ATLAS data \cite{ATL-PHYS-INT-2011-032}. %
Both methods can be used to obtain exclusion limits at a given confidence level,
\ie a upper (or lower) limit on some quantity 
such that all values above (or below) this limiting value are excluded at the given confidence level.
In the following, the mathematics behind the two methods is described.
The technical details of the actual implementation can be found in \Sec{sec:analysis_implementation}.

Unlike in experiments looking, for example, for the neutrino oscillations already mentioned in the introduction, %
where the signal hypothesis may predict an increased as well as a decreased event rate compared to the null hypothesis,
it will be assumed here that new physics processes can only increase the number of events observed in the signal region.
Downward fluctuations are thus interpreted to not indicate the presence of a signal,
and the signal strength is bound by zero from below. %

\subsubsection{\texorpdfstring{\cls Method} {CLs Method}}
The \cls method\inindex{CLs method@\cls method} is a method for the calculation of confidence intervals based on two alternative hypotheses,
the signal-plus-background hypothesis (\hyposb),
stating the existence of both signal and background,
and the background-only hypothesis (\hypob),
stating the existence of only background.
It uses Poisson statistics and allows for an easy combination of results of independent searches.
The \cls method and an iterative approximate method to compute combined exclusion confidence levels based on \cls 
have been described by Junk \cite{Junk1999}.
Read gives a generalized form, including probability density functions for the discriminating variables \cite{Read2002}
and a discussion of the differences between \cls and frequentist confidence intervals or Bayesian credible intervals 
in the context of Higgs searches at the \acl{LEP} \cite{Read2002,Read2002JoP}.

Binning the results of the analyses to be combined in their discriminant variables
allows to treat each bin as a statistically independent search channel of a counting experiment,
which can then be combined with the \cls method.
The likelihood for finding find $d_i$ events in channel $i$,
assuming the existence of both signal and background, \hyposb,
is given by a product of Poisson probabilities,
\begin{equation}
  \label{eq:poisson_likelihood_sb}
  L(d|s+b) = \prod_{i=1}^{n} \frac{\exp(-(s_i+b_i)) \, (s_i+b_i)^{d_i}}{d_i!},
\end{equation}
and in the same way,
assuming \hypob instead,
\begin{equation}
  \label{eq:poisson_likelihood_b}
  L(d|b) = \prod_{i=1}^{n} \frac{\exp(-b_i) \, {b_i}^{d_i}}{d_i!},
\end{equation}
where $n$ is the number of channels or bins.

A test statistic $X$ is now defined,
which discriminates signal-like outcomes from back\-ground-like ones using a likelihood ratio.
In searches for new physics, an appropriate likelihood ratio is given
by the ratio of the probability densities for the signal-plus-background hypothesis and the background-only hypothesis \cite{Junk1999},
\begin{align}
  X &= \frac{L(d|s+b)}{L(d|b)}. %
  \label{eq:first_version_cls_test_statistic}
\end{align}
$X$ increases monotonically with the number of observed events 
and induces an ordering on the space of possible outcomes. %
It ranks the possible experimental outcomes from the least to the most signal-like or, equivalently, from the most to least background-like.
This incorporates at this stage an implicit restriction to positive signals $s_i \geq 0$ as mentioned above.
Its distribution can then be used to compute the probability of obtaining a certain outcome.
Note that for vanishing backgrounds %
this test statistic cannot be used because it diverges for $\sum_i b_i = 0$.
Using \Eqs{eq:poisson_likelihood_sb} and \eqref{eq:poisson_likelihood_b},
\Eq{eq:first_version_cls_test_statistic} goes over into
\begin{equation}
  \label{eq:def_test_statistic}
  X = \exp\left(\sum_{i=1}^n s_i\right) \prod_{i=1}^n\left(1 + \frac{s_i}{b_i} \right)^{d_i}.
\end{equation}
Without destroying the ordering property, this can be written in logarithmic form,
\begin{equation}
  \label{eq:def_log_test_statistic}
  Q \definedas -2 \ln X = -2 \sum_{i=1}^n s_i - 2 \sum_{i=1}^n d_i \ln\left(1 + \frac{s_i}{b_i} \right), %
\end{equation}
which will also be used later in the implementation as it is numerically more stable.
The additional factor $-2$ may be motivated from Wilks' theorem (see below),
but here is purely conventional as Wilks' theorem is not used in this context.
A more general form which takes into account the probability density functions
of the discriminating variables is given in \cite{Read2002}. %

The probability to obtain a value of the test statistic which is less or equal to the value actually observed assuming the existence of signal and background \hyposb is
\begin{equation}
  \label{eq:p_clsb}
  P_{s+b}(X \leq X_\text{obs}) = \sum_{X(\{d_i'\}) \leq X_\text{obs}} \prod_{i=1}^n \frac{\exp\left(-(s_i+b_i)\right) \, (s_i+b_i)^{d_i'}}{d_i'!}.
\end{equation}
The product is a weighting term, given by the likelihood in \Eq{eq:poisson_likelihood_sb},
and $X_\text{obs} = X(\{d_i\})$ are the observed numbers of events.
In the same way,
$P_b$ is defined %
to give the probability that the test statistic is
less or equal to the value that is observed in the data,
assuming the existence of only background:
\begin{equation}
  \label{eq:p_clb}
  P_{b}(X \leq X_\text{obs}) = \sum_{X(\{d_i'\}) \leq X_\text{obs}} \prod_{i=1}^n \frac{\exp(-b_i) (b_i)^{d_i'}}{d_i'!}.
\end{equation}
Note that $X$ is written as a function of $\{d_i'\}$ only
because the signal and background expectations, $\{s_i\}$ and $\{b_i\}$,
which also go into the computation of $X$ as parameters,
stay the same in both cases,
when computing \clsb and when computing \clb.
In the continuous limit, this is sometimes written as %
\begin{equation}
  P_{s+b}(X \leq X_\text{obs}) = \int_{-\infty}^{X_\text{obs}} \frac{\intd P_{s+b}}{\intd X} \intd X, %
\end{equation}
with $\intd P_{s+b} / \intd X$ being the probability density of $X$ under \hyposb, %
and in the same way for $P_{b}(X \leq X_\text{obs})$, \latein{mutatis mutandis}. %

The confidence value for the background-only hypothesis, denoted \clb, is %
\begin{equation}
  \clb = P_b(X \leq X_\text{obs}).
\end{equation}
Note that calling this the confidence in the background hypothesis
(as is done in the first publications on \cls) can be misleading
because high values of \clb close to one %
show lack of agreement of the data with the background-only hypothesis,
and will be used to reject the background-only hypothesis and to quote a potential discovery:
If $1 - \clb < \alpha$, one would exclude the background-only hypothesis at a confidence level of $1-\alpha$,
interpreting $1 - \clb$ as the probability that the background fluctuates to give a distribution of observed events
at least as signal-like as the one observed in data.
Note that the background in the search result is accounted for in two ways. %
It goes into the test statistic, and it is used in the computation of the confidences. %

\clsb is the confidence level for the possibility of simultaneous presence of new particle production and background,
\begin{equation}
  \clsb = P_{s+b}(X \leq X_\text{obs}).
\end{equation}
Small values of \clsb disfavor the signal-plus-background hypothesis \hyposb,
so to exclude \hyposb %
at a given confidence level $1-\alpha$,
the condition $\clsb < \alpha$ needs to be fulfilled.

The criticism about \clsb is that when \eg using $\alpha = 0.05$,
the exclusion potential goes down only to (and stays at) \percent{5},
even for expected signal rates which are very small compared to the expected background \cite{Read2002}. %
This means that if the data has a downward fluctuation
so that the number of observed events is smaller than the number of expected background events,
any signal may be excluded at a high confidence level.
This is correct from the Frequentist point of view because the result of the interpretation is known
to be wrong in \percent{5} of the cases, and Frequentist intervals are statements about the data, not about the signal.
But this also means that by increasing the background expectation, an experiment can be improved \latein{a priori} %
so that it would turn out to be better to have an inferior experiment with a higher expected background level,
and even worse, the exclusion result can be improved a posteriori by increasing the background expectation \cite{Read2002JoP}.

The solution to avoid these unintuitive properties is to require 
that experiments without sensitivity to a particular model should not be able to exclude this model.
One therefore defines the ``Modified Frequentist confidence level'' \cls as
\begin{equation}
  \cls \definedas \clsb / \clb,
  \label{eq:definition_cls}
\end{equation}
which can be thought of as an approximate confidence in the signal-only hypothesis \cite{Read2002JoP},
and in terms of $p$-values can be formulated as
\begin{equation}
  \cls = \frac{ \text{$p$-value of signal-plus-background hypothesis \hyposb} }{ \text{$1-p$-value of background-only hypothesis \hypob} }.
  \label{eq:cls_pvalue_definition}
\end{equation}
Just like with \clsb, the signal-plus-background hypothesis \hyposb
is excluded at a confidence level of $1-\alpha$ if $\cls < \alpha$. %
As $\clb$ is smaller than or equal to one, \cls is always greater than or equal to \clsb,
\ie the models excluded by using \cls are a subset of those excluded using \clsb
so that the upper limit using \cls is higher and therefore weaker and more conservative.
Being a ratio of probabilities, \cls may become larger than one and is not a proper probability itself.

An important figure of merit of the average expected performance of an experiment
are the expected \cls and \clsb values under assumption of the background-only hypothesis\footnote{
  They are computed as the fraction of the distribution of the test statistic
  which lies above the median of the test statistic, assuming the validity of \hypob. %
  The expected \clb therefore is $0.5$, the quantile value of the median.
}.
If the expected \cls value (or \clsb, depending on the choice of which criterion to use)
lies above the threshold for the exclusion of a model,
then the experiment is not expected to be able
to exclude this model even in case the signal is truly absent.
It still may be able to exclude this model due to underfluctuations in the number of observed events,
if the sensitivity is not too small such that the renormalization in the \cls method prohibits the exclusion.
In general, it is therefore interesting to compare the expected and observed exclusion limits
to judge whether the experiment potentially should be able to extend its currently observed exclusion limits.
In order to get an impression of the typical variations in the exclusion potential,
in addition to the expected \cls values
the change in the expected \cls when the background fluctuates up or down by one standard deviation is of interest.
These fluctuations are often indicated as \index{one-sigma band}s above and below the expected \cls limits,
\eg when plotting limits on production cross sections as function of the particle mass
(usually the ratio to the theoretically expected cross section is shown).

\begin{figure}
  \centering
  \incgraphics{width=\widthtwoplots}{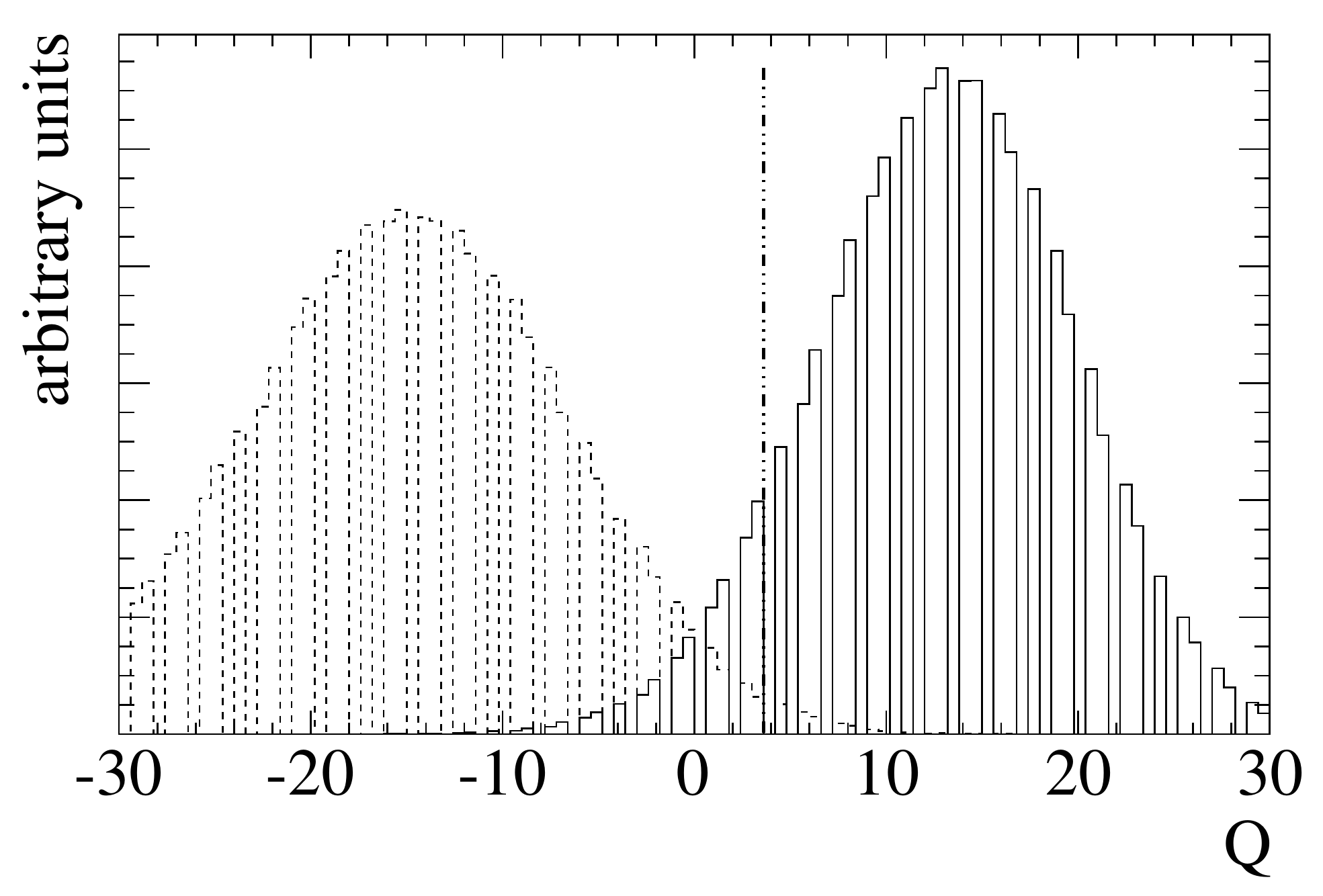}
  \hfill
  \incgraphics{width=\widthtwoplots}{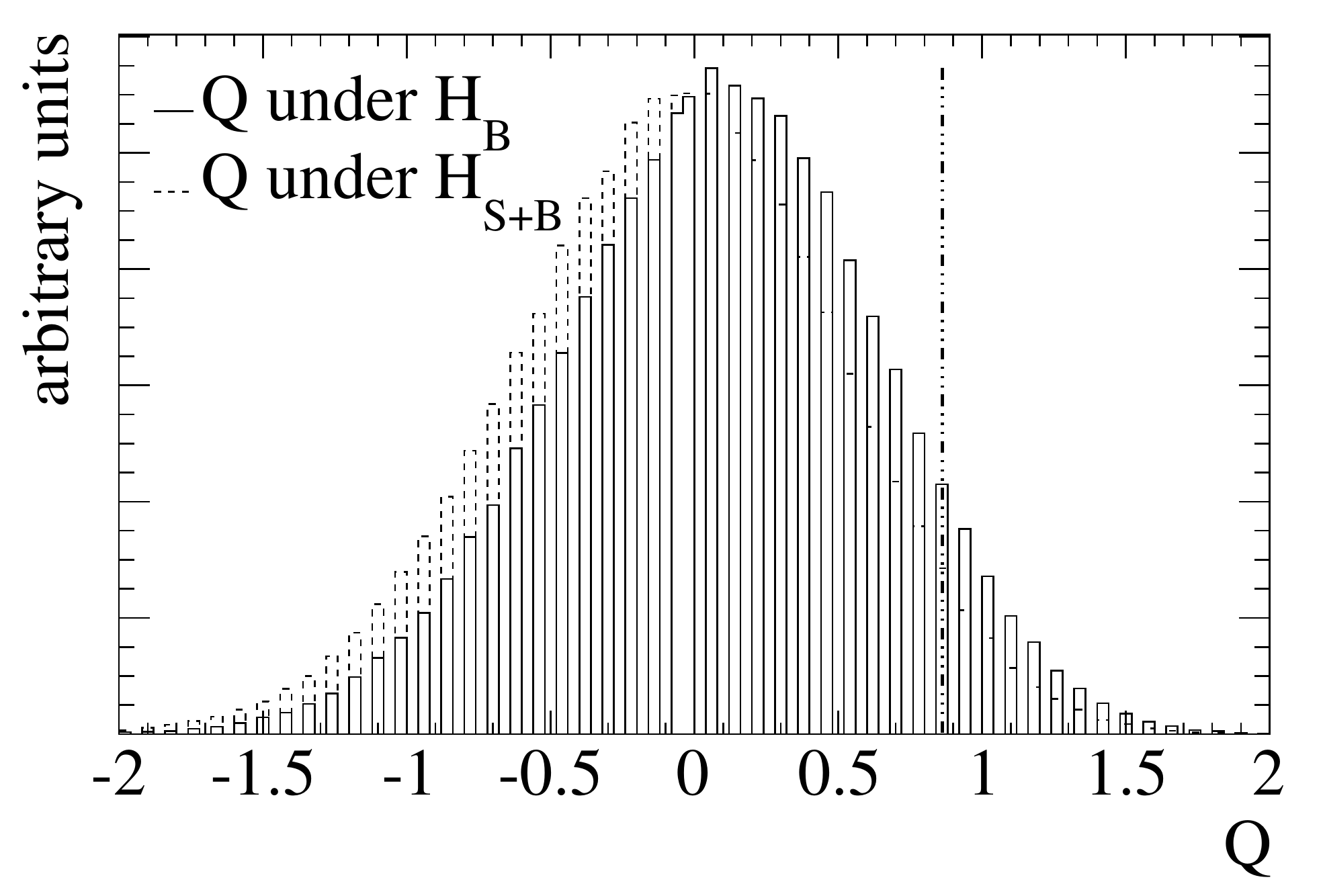}
  \caption{
    Example distributions of the test statistic $Q = -2\ln(X)$ under the two hypotheses \hyposb (dashed line) and \hypob (solid line), with 
    $d = 60$, $b = 50$ and $s = 30$ (left) and
    $d = 40$, $b = 50$ and $s = 2$ (right).
    The vertical line marks the observed value of the test statistic.
  }
  \label{fig:cls_teststatistic_example}
\end{figure}
\Fig{fig:cls_teststatistic_example} shows examples of the distribution of the
test statistic $Q = -2\ln(X)$ for the two hypotheses \hyposb and \hypob,
where $X$ is calculated according to \Eq{eq:def_test_statistic},
and its cumulative distribution is governed by \Eqs{eq:p_clsb} and \eqref{eq:p_clb}, respectively. %
In this type of plot, the distribution of $Q$ for \hyposb is always shifted left with respect to \hypob,
as can be seen from \Eq{eq:def_log_test_statistic} because $d_i$ on average is larger when evaluating the distribution for \hyposb. %
The histograms have been filled performing $2\ten5$ toy experiments,
varying the observed number of events randomly under the two hypotheses and computing each time $Q$.
As only one channel is used here for simplicity,
strong binning effects are visible due to the discreteness of the event numbers that go into the test statistic.
In the left plot, $d = 60$ observed, $b = 50$ expected background and $s = 30$ expected signal events have been used,
resulting in a good separation of the distributions for the two alternative hypotheses.
Both \cls and \clsb are around $0.01$ (\clb is $0.9$),
so having chosen $\alpha = 0.05$,
the signal hypothesis is excluded at a \percent{95} confidence level (\CL).
In general, for large significances the curves for \hyposb and \hypob are well separated so that
in case of the absence of signal,
\clb is close to one and the exclusion potential of \cls converges to that of \clsb.
For low sensitivity, the denominator in \Eq{eq:definition_cls} makes \cls larger.
An example of this is shown in the right plot in \Fig{fig:cls_teststatistic_example},
where the signal expectation $s = 2$ is very small compared to the background expectation $b = 50$ 
and the number of observed events is taken to be $d = 40$.
For this example, the signal would still be excluded at \percent{95} \CL using \clsb ($\clsb = 0.037$),
but not using \cls ($\cls = 0.56$), because \clb is also very small ($\clb = 0.066$).

So far, no uncertainties on the signal and background expectations have been taken into account.
Usually no uncertainties on the confidence level are quoted,
but instead the confidence limits are modified to allow for the experimental uncertainties,
meaning in the case of \cls that the higher the uncertainties, the broader the two distributions of the test statistic,
the longer the tails, the larger the overlap and the weaker the exclusion limits will be.
This can be implemented by doing toy experiments like they were used to fill the histograms in \Fig{fig:cls_teststatistic_example}
and smearing the expected number of signal and background events
with the corresponding uncertainties which will yield broader distributions for the test statistic.
Correlations between uncertainties can be taken into account
by fluctuating these numbers in a correlated manner in each pseudoexperiment.
In the end, these toy experiments are an approximation to averaging
over all possible outcomes for signal and background
given by the probability distributions of their systematic uncertainties.

The \cls method only yields a binary decision whether or not a given scenario (in terms of signal and background expectations)
is excluded at a given certainty level.
From this, upper limits on the signal strength can be found by starting in the excluded region
and reducing the number of expected signal events gradually,
which will give a monotonically increasing \cls value.
This is continued until \cls is no longer smaller than the given threshold
to find the upper limit on the number of signal events at a given certainty level.
Note that the confidence intervals constructed in this way do
neither have the same interpretation as Frequentist,
nor as Bayesian intervals \cite{Read2002}.

\begin{figure}
  \centering
  \incgraphics{width=\widthsingleplot}{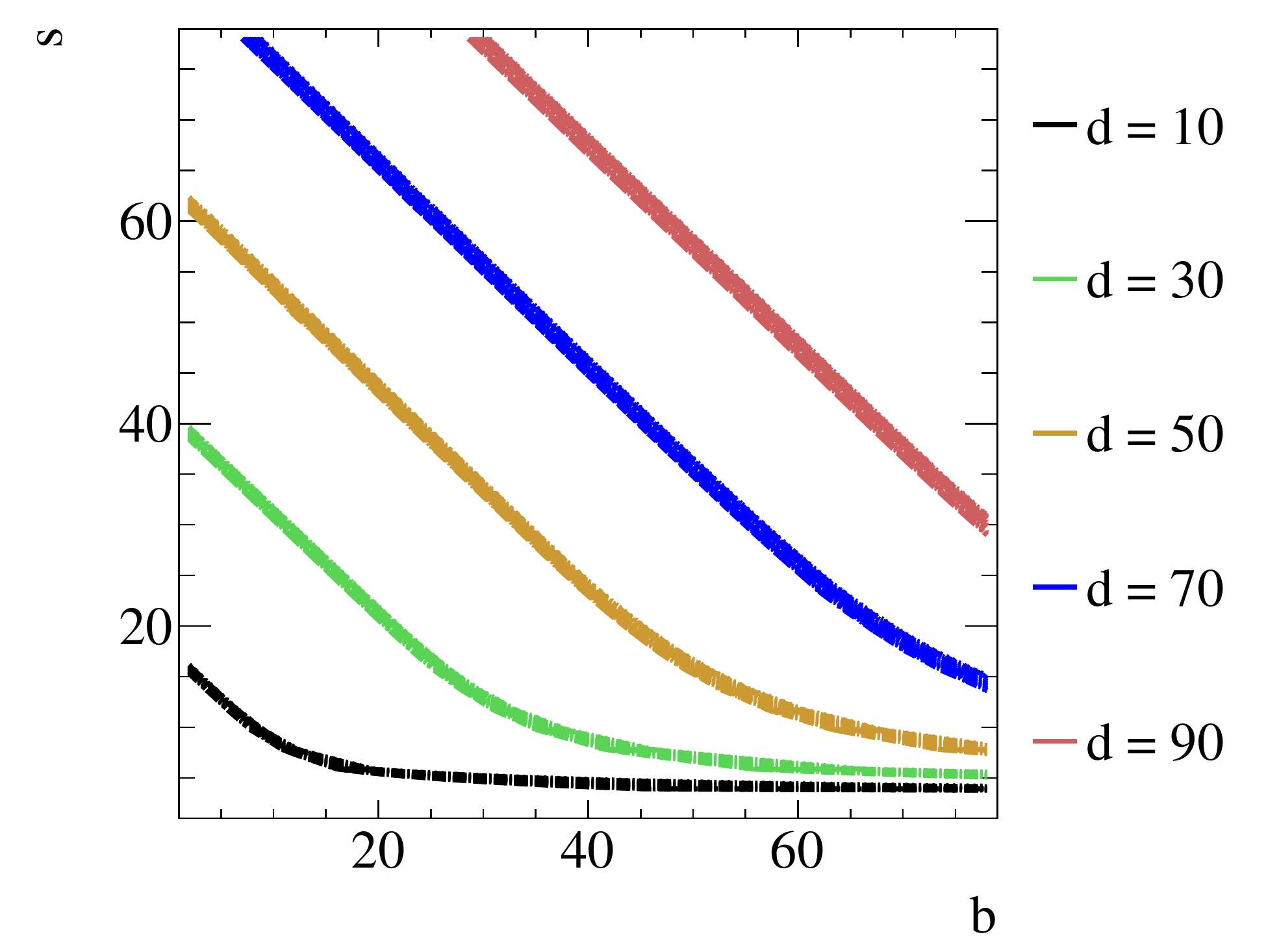}
  \caption{
    \cls from an analytic calculation for the simple case of only one search channel and without taking any uncertainties into account.
    The horizontal and the vertical axis are the number of background events $b$ and signal events $s$, respectively.
    The lines are the \percent{95} \CL contours for different values of the observed number of events $d$.
  }
  \label{fig:analysis_cls_approximation}
\end{figure}
For the very simple case of only one search channel ($n = 1$) and no uncertainties,
\cls can be calculated analytically without effort.
The $p$-values from \Eq{eq:cls_pvalue_definition} can then be computed directly
from the cumulative distribution function of the Poisson distribution.
In the enumerator of \Eq{eq:cls_pvalue_definition},
the \clsb values are symmetric under exchange of $s$ and $b$,
and so is \cls,
as long as \clb in the denominator is close to one.
For small $s$ though, the symmetry of \cls is broken.
This may seem counterintuitive because \clb only depends on $b$ and not on $s$,
but due to the symmetry of \clsb,
small $s$ correspond to large $b$,
for which the $p$-value of the background-only hypothesis increases and \clb thus goes down.
This can be seen in \Fig{fig:analysis_cls_approximation},
which shows exclusion limits obtained from the analytic calculation
in the plane spanned by the number of expected signal events $s$ and expected background events $b$. %
Five different hypothetical numbers of observed events $d$ are compared,
giving five contours in the plane where the computed value of \cls is $0.05$,
corresponding to an exclusion at a confidence level of \percent{95}.
In the plot in this figure, everything to the upper right
of the respective contour lines is excluded at the given confidence level.

\removesection{ %
\subsubsection{Power-Constrained Limits}
\inindex{Power-constrained limits}
An alternative to the \cls method is \acf{PCL} \cite{Cowan2011PCL}.
It uses $\clsb$ for setting exclusion limits
and addresses the issue of spurious exclusions by clipping the exclusion limits in regions where there is no or vanishing sensitivity:
If the resulting limit is more than one standard deviation more stringent than the expected limit, %
the expected limit plus one standard deviation is taken as the exclusion limit instead.
This can give more powerful limits than using \cls,
because as detailed above, \cls is always at least as conservative as \clsb,
but does not exclude parameter values where there is no sensitivity.
Of course, using one standard deviation as bound is arbitrary and therefore adds subjectivity to the resulting limit.
} %

\subsubsection{\acl{PLR} Method}
The second method which will be used later to evaluate and interpret the outcome of the analysis of \ATLAS data %
is the \acf{PLR}\inindex{PLR method} method.
It is described \eg in \cite{Cowan2011}.
Like \cls, which uses a likelihood ratio in the test statistic $X$,
the PLR method is also based on a likelihood ratio.
But unlike \cls,
the central step in the \ac{PLR} method involves fitting the data using a model that is not prescribed by the method itself.
This model has two classes of parameters.
The first class are the parameters of interest\inindex{Parameter of interest},
which are parameters that the evaluation of the measurements is aiming at,
for example particle masses or cross sections.
The parameters in the second class are called \index{nuisance parameter}s,
for which values are determined by the fit to the data, too,
but these values are not considered relevant and only introduced to allow the model
to give a better description of the data by including additional degrees of freedom.
Systematic uncertainties like signal efficiencies or the jet-energy scale are
typical examples of parameters in this second class.

The model that is used in the PLR method to describe the data is encoded in terms of a likelihood function
that assigns a likelihood value to every possible set of parameters and measurements:
\begin{equation}
  L(\vec\mu, \vec\nu) = \prod\limits_{\mathclap{i\in \text{events}}} P(\vec x_i|\vec\mu,\vec\nu).
\end{equation}
$P(\vec x|\vec\mu,\vec\nu)$ is the probability to obtain a measurement $\vec x$,
given the values $\vec \mu$ for the parameters of interest
and $\vec\nu$ for the nuisance parameters,
which characterize the shape of the individual PDFs that together constitute $P$.
This is the most general form,
where the product runs over all events.
In a binned analysis, it would run over all measurement channels or bins of the relevant histograms instead.
The typical form of $P$ is a product of Poisson probabilities for the observation of $d_j$ events in channel $j$,
with an expected number of $s_j$ signal and $b_j$ background events,
multiplied by additional PDFs describing subsidiary measurements
and the probabilities to observe a set of given values for the nuisance parameters.
Restricting %
to the signal strength parameter as the only parameter of interest $\mu$\footnote{
  The following can also be generalized to more than one parameter of interest \cite{Cowan2011}. %
},
the binned form of the likelihood can be written as
\begin{equation}
  \label{eq:PLR_likelihood_Poisson}
  L(\mu, \vec\nu) = \prod_{j=1}^n \Poisson{d_j}{\mu f_s(s_j, \vec\nu) + f_b(b_j, \vec\nu)} \cdot \prod_{k} p(\nu_k),
\end{equation}
where $n$ is the number of channels as before,
and $k$ runs over the nuisance parameters.
The \index{signal strength} $\mu$ is a simple scaling factor for the number of expected signal events.
$\mu = 0$ corresponds to the background-only hypothesis, $\mu = 1$ is the nominal signal hypothesis.
Since the nuisance parameters can have an influence on the number of expected events,
the Poisson probabilities have not been written in a way that they depend directly on $s$ and $b$,
but additional wrapper functions $f_s$ and $f_b$ have been inserted,
which incorporate this dependency and will be written out below.
The second product contains the probabilities $p$ to observe given values of the nuisance parameters, %
which here have been assumed to factor. %
These probability distributions are prior distributions in the sense of Bayesian statistics
and are often taken to be uniform or Gaussian.

Using the likelihood function $L$, the \index{profile likelihood ratio} $\lambda$ is defined as:
\begin{equation}
  \lambda(\mu) = \frac{L(\mu,\hat{\hat{\vec\nu}})}{L(\hat{\mu},\hat{\vec\nu})}.
  \label{eq:PLR_likelihood_ratio}
\end{equation}
In the denominator, $\hat{\mu}$ and $\hat{\vec\nu}$ are the maximum-likelihood estimators for $\mu$ and $\vec\nu$,
which are obtained by maximizing the likelihood $L$ over all possible values of $\mu$ and $\vec\nu$.
The enumerator is called the profile likelihood function. %
$\hat{\hat{\vec\nu}}$ in the enumerator is the conditional maximum-likelihood estimator,
which maximizes $L$ for a given value of $\mu$.
(The computation of the maximum is often called a maximum-likelihood fit.)
Note that the likelihood ratio $\lambda$ in the end only depends on the parameter of interest $\mu$.
As the likelihood is a product of probabilities, from the above definition it is clear that $0\leq\lambda(\mu)\leq1$.
$\lambda(\mu)$ approaches one if the data agrees well with the hypothesized value of $\mu$.
In many analyses, the number of signal events is assumed to be non-negative,
and therefore negative values of $\mu$ are excluded
because they would correspond to negative signal contributions.
As probabilities usually involve products of small numbers,
it is convenient to use the logarithm of the likelihood ratio 
\begin{equation}
  t(\mu) = -2 \ln \lambda(\mu)
  \label{eq:PLR_t_statistic}
\end{equation}
as test statistic,
which, as in the case of the \cls method, by convention includes a factor $-2$.
$t$~attains its minimum value of zero at $\mu = \hat\mu$,
and larger values of $t$ correspond to an increasing incompatibility 
between the data and the hypothesized value of the signal strength~$\mu$.

Knowing the probability distribution $f\!\left(t(\mu)|\mu\right)$ of the \acf{LLR} $t(\mu)$ for a fixed value $\mu$,
\begin{equation}
  p_\mu = \int_{t_\text{obs}}^\infty f\!\left(t(\mu)|\mu\right) \intd t(\mu)
\end{equation}
gives the probability to obtain a value for $t(\mu)$
equally or less compatible with the data
than the value $t_\text{obs}$ which has been observed under the assumption of the signal strength $\mu$.
If the $p$-value for a given $\mu$, $p_\mu$, is found to lie below a critical value $\alpha$,
then this value of the signal strength $\mu$ is said to be excluded at a confidence level $1 - \alpha$. %
From this, an upper limit on $\mu$ as the largest value for $\mu$ so that $p_\mu \leq \alpha$ can be derived.

It is therefore important to know $f\!\left(t(\mu)|\mu\right)$.
A number of approximations for the distributions of different variants of the \ac{LLR}
and analytic expressions for the resulting upper limits are presented in \cite{Cowan2011}. %
These approximations are based on results by Wald \cite{Wald1943} and Wilks \cite{Wilks1938},
stating that under certain optimality conditions and for sufficiently large statistics,
$t(\mu) = -2\ln \lambda(\mu)$ follows a $\chi^2$~distribution with $h-m$ degrees of freedom,
$h$ being the total number of parameters and $m$ being the number of nuisance parameters, \ie here $h-m = 1$.
This approximation is also used in the implementation of the \ac{PLR} method that is described below.

\subsection{\texorpdfstring{Comparison of the \cls and PLR Methods} {Comparison of the CLs and PLR Methods}}

Two different methods to compute limits have been presented above.
As the two methods are optimized with regard to slightly different aspects,
no exact agreement is expected, but still the limits that are obtained should be consistent.
This will be studied and discussed in this section.
First, a summary of important differences between \cls and PLR is given.

\begin{itemize}
  \item The main purpose of \cls is dealing with underfluctuations in case of small sensitivity,
    and to avoid exclusion of signals if the exclusion is not justified due to a lack of sensitivity.
    In PLR, no such mechanism is implemented\footnote{
      In principle, the ratio of two hypotheses
      could be used with PLR in a similar manner as done in \cls, too,
      instead of using the $p$-value of the hypothesis \hyposb alone.
    }
    and this will later become apparent through the fact that the limits obtained from \cls
    are slightly more conservative than those from the \ac{PLR} method.
  \item The procedure to compute limits using the \cls method
    only gives an upper limit on the number of signal events,
    which is fully acceptable when setting exclusion limits.
    For sufficiently high signal-to-background ratios,
    PLR may also yield a lower limit on the signal,
    depending on the number of observed events.
    This is visible in \Fig{fig:analysis_methods_comparison_cls_plr},
    which is discussed below.
  \item With respect to the implementation,
    the \ac{PDF} for the nuisance parameters in the \propername{Roo\-Stats} framework
    can be modeled to any level of complexity using custom \acp{PDF}.
    This allows to use an asymmetric distribution \eg to better take into account the nature of the asymmetric \ac{JES} uncertainties.
    In the \cls implementation employed here (\propername{TLimit}),
    this is not possible because all uncertainties are modeled as Gaussian distributions.
    Of course, this is not a general shortcome of the \cls method,
    which could as well be implemented to use other distributions for the uncertainties when doing the pseudoexperiments.
\end{itemize}

\begin{figure}
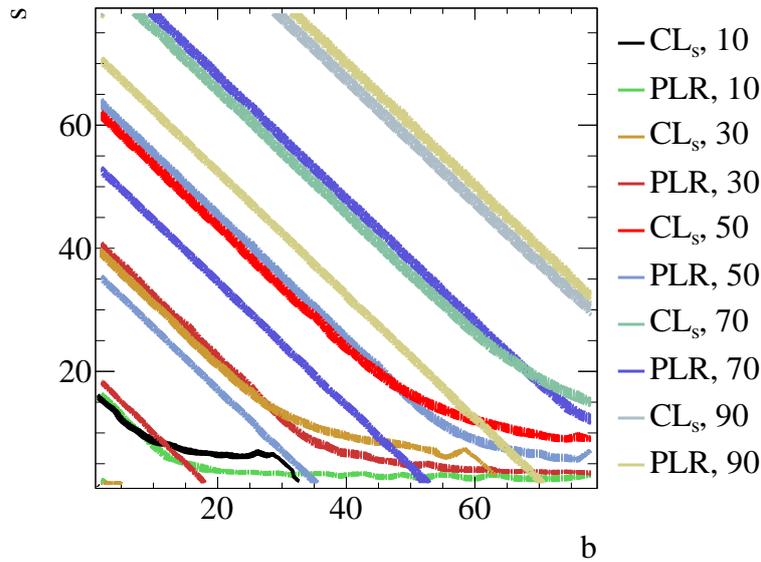

  \centering
  \incgraphics{width=\widthsingleplot}{{{LimitCalculators_test_exclusion_nobsall_uncert0.0_MCsteps1500000_sMC.compare}}}
  \caption{
    Comparison of exclusion limits from the \cls and the PLR method
    for different number of observed events (given in the legend)
    as function of the number of background events $b$ and signal events $s$.
  }
  \label{fig:analysis_methods_comparison_cls_plr}
\end{figure}
In \Fig{fig:analysis_methods_comparison_cls_plr},
the exclusion limits set by the \cls and PLR method are compared.
On the horizontal axis the number of background events $b$ and
on the vertical axis the number of signal events $s$ is varied between $0$ and $80$.
The contours shown in the plot correspond to five different numbers of observed events.
They mark where the \cls or $p$-value from PLR, respectively, are equal to $0.05$.
Everything to the top right of each line is excluded at a confidence level of \percent{95}.
The contour lines are symmetric under $s\leftrightarrow b$,
except for very small signal numbers.
The plot shows that the limits from the two different methods give the expected agreement,
the exclusion from \cls being more conservative for low expected signal strengths than the exclusion from \ac{PLR}.
It also shows, as mentioned above, that \ac{PLR}, due to the parabolic shape of the \ac{LLR},
may give an additional second contour at lower $s+b$ values.
In these cases, the outmost line is the exclusion limit of interest here.

Note that for the \cls method all uncertainties,
and for the \ac{PLR} method all but one, namely the \ac{JES} uncertainty,
are modeled using symmetric Gaussian distributions.
This would in principle allow to combine all uncertainties into one,
adding them up in the appropriate way given by their correlations.
The main advantage of keeping the uncertainties as distinct numbers is an easier handling of many channels
with different combinations of correlations between uncertainties,
which does not apply in the following,
where only one channel is used.

\section{Interpretation of Analysis Results}
\label{sec:analysis_interpretation}

In \Sec{sec:results_0lanalysis_statistical_methods}, the methods have been presented
which will be employed in the following to perform the final step of the search for Supersymmetry in \ATLAS 2010 data in the zero-lepton channel.
\cls and PLR will now be applied to the results of the cutflow and the uncertainties summarized in \Sec{sec:analysis_summary_limit_input}.
As no statistically significant excess over the Standard Model background expectations is observed in the data from 2010,
the event counts are interpreted by setting exclusion limits on processes that go beyond the Standard Model.
Here, the observable that comprises the search result simply is the number of events
that satisfy all cuts that define one of the four signal regions.
The limits will first be set in a model-independent way,
and then using the Monte Carlo grids presented in \Sec{sec:analysis_SUSY_GRIDs} to model the supersymmetric signal events.

\subsection{Model-Agnostic Limits}

\begin{table}
  \centering
  \begin{tabular}{llcccccccc}
    \toprule
     &  & \multicolumn{8}{l}{Signal Region} \\
    \cmidrule{3-10}
     &  & \multicolumn{2}{l}{A} & \multicolumn{2}{l}{B} & \multicolumn{2}{l}{C} & \multicolumn{2}{l}{D} \\
    \cmidrule(r){3-4} \cmidrule(r){5-6} \cmidrule(r){7-8} \cmidrule(r){9-10}
    Method & $\Delta_\text{tot}$ & $n_\text{signal}$ & $\sigma$ [pb] & $n_\text{signal}$ & $\sigma$ [pb] & $n_\text{signal}$ & $\sigma$ [pb] & $n_\text{signal}$ & $\sigma$ [pb]  \\
    \midrule
    PLR & 0.00 & 53.4 & 1.6 & 14.1 & 0.4 & 39.9 & 1.2 & 4.6 & 0.1\\
     & 0.20 & 56.8 & 1.7 & 15.4 & 0.5 & 42.6 & 1.3 & 4.9 & 0.1\\
     & 0.40 & 74.8 & 2.2 & 24.3 & 0.7 & 56.6 & 1.7 & 6.7 & 0.2\\
    \cls & 0.00 & 58.2 & 1.7 & 12.8 & 0.4 & 44.0 & 1.3 & 5.7 & 0.2\\
     & 0.20 & 62.7 & 1.9 & 14.2 & 0.4 & 47.5 & 1.4 & 6.1 & 0.2\\
     & 0.40 & 85.8 & 2.6 & 20.3 & 0.6 & 64.5 & 1.9 & 9.0 & 0.3\\
    \bottomrule
  \end{tabular}
  \caption{
    Model-agnostic upper limits on the number of signal events $n_\text{signal}$ and the effective cross section $\sigma$
    in the four signal regions, from PLR and \cls,
    comparing three different values of the total systematic uncertainty $\Delta_\text{tot}$.
  }
  \label{tab:analysis_model_agnostic_limits}.
\end{table}

Without choosing a specific set of parameters for Supersymmetry or even without specifying the type of \acf{BSM} physics,
a model-independent limit on the cross section of \ac{BSM} physics processes can be derived\footnote{
  This means apart from the assumption that they lead to an increase in event counts,
  not to a destructive interference with known physics processes.
}.
Of course, the value of this limit, being the result of a particular analysis, 
depends on the cuts which are applied to suppress Standard Model backgrounds,
and these are optimized for, and therefore prefer, particular signal event topologies.
Thus, the limit is not fully model-independent,
but at least it is model-agnostic.

Although in this approach the number of signal events reaching the signal regions defined by the cutflow is left free,
still an estimate of the relative uncertainties on the signal is needed.
Two different values are used and compared in this section, \percent{20} and \percent{40}, %
which approximately reflect the spread of the quadratic sum of all signal uncertainties at typical points within the mSUGRA signal grid. %
These total uncertainties are used in an uncorrelated way with respect to the background uncertainties,
and therefore have conservatively been rounded up.
On top of this uncorrelated uncertainty estimate,
the uncertainty on the luminosity is included when computing the limits,
and in addition to the two values given above,
the results assuming no uncertainty but the one from the luminosity are also shown.

\begin{figure}
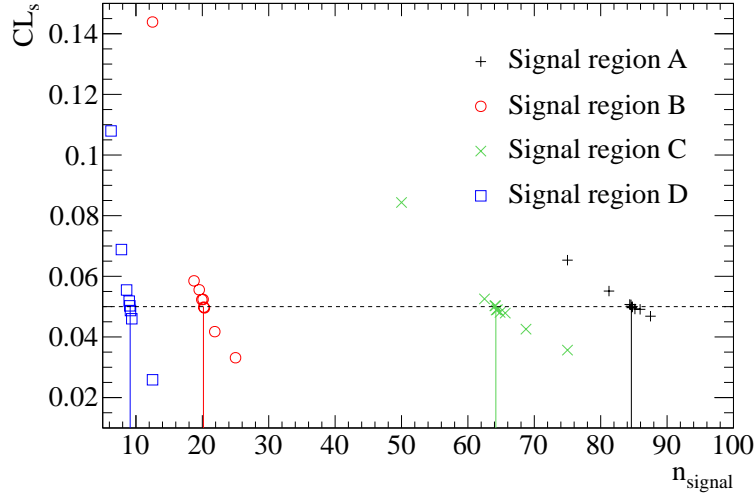

  \centering
  \incgraphics{width=\widthsingleplot}{{{postprocess_local_harvester_SUSYGRID_ModelAgnostic_CLs_Unc0.40}}}
  \caption{
    Evolution of \cls in the determination of the upper limit on the number of signal events using a bisection method.
    Each point corresponds to the \cls value computed for the midpoint of the current interval.
    The colors correspond to the four different signal regions.
    The horizontal line marks $\cls = 0.05$, the vertical line the final upper limit.
    In this plot, a total uncertainty of \percent{40} (plus the luminosity uncertainty) is assumed.
  }
  \label{fig:analysis_bisection_cls}
\end{figure}

The \ac{PLR} method directly returns an upper limit on the number of signal events $n_\text{signal}$,
fixing the signal strength to $\mu = 1$ and leaving $n_\text{signal}$ free as the parameter of interest,
then using the approximation for the distribution of the test statistic as described above.
To find an upper limit on $n_\text{signal}$ following the \cls prescription,
the set of possible values of $n_\text{signal}$ needs to be scanned,
which is done here using a simple bisection method with a fixed number of $10$ steps.
This is considerably faster than a scan of equidistant points. %
As the \cls implementation is based on Monte Carlo,
the resulting \cls values will exhibit random fluctuations
which may cause the bisectioning to exclude the wrong half of the interval,
but this will only happen when the interval sizes are already of the size of the Monte Carlo fluctuations.
The sequence of intervals has been checked with respect to this effect,
and the effect has been found to be negligible,
as can be seen in \Fig{fig:analysis_bisection_cls}.
This plot shows, for a total uncertainty of \percent{40} plus the luminosity uncertainty,
the development of \cls as it is computed when running the bisection method
for the four different signal regions.
Each dot corresponds to the \cls value computed for the midpoint of the current interval.
The horizontal line marks $\cls = 0.05$,
and the vertical line gives the final upper limit
as the mid of the interval obtained from the last iteration.
The fluctuations from the Monte Carlo evaluation are visible in the plot,
but in the relevant region the dots do not fluctuate by so much
that a significant impact on the limits is expected.
From the plot, a systematic uncertainty arising from the bisection method seems to be no larger than $\pm 1$ event.
(The slope of the dependence of \cls on $n_\text{signal}$ for the other two total uncertainties,
\percent{0} and \percent{20},
is larger
and the uncertainty on the limit from the method therefore is smaller.)

The upper limits on the number of signal events $n_\text{signal}$ from the two methods
are summarized in \Tab{tab:analysis_model_agnostic_limits} for the three different total uncertainties $\Delta_\text{tot}$,
which do not include the correlated luminosity uncertainty.
The table also lists the corresponding upper limit on the cross section calculated via $\sigma = n_\text{signal}/\Lint$,
where $\Lint$ is the total integrated luminosity from \Sec{sec:analysis_luminosity_calculation}.
Note that these limits naturally cannot take into account the signal efficiency
because no particular form of the signal or topology of signal events is assumed.
The limit is therefore understood to be on an effective cross section including signal efficiency and other factors %
as appropriate.
The confidence level is \percent{95} for both PLR and \cls.
The higher the systematic uncertainties, the larger, \ie weaker, are the limits on the number of signal events and the cross section.
Note that \cls always gives slightly weaker limits than \ac{PLR} except for signal region~B.
The harshest event selection, corresponding to signal region~D, gives the best limits among the four signal regions,
excluding \ac{BSM} physics processes with effective cross sections above $0.1$ to \unit[0.3]{pb}.
The dependence of these limits on the total uncertainty $\Delta_\text{tot}$ is strong,
as can be seen in the table.
No additional uncertainty on these limits has been evaluated.

\subsection{Exclusion of Specific SUSY Scenarios}

In this section, the results summarized in \Sec{sec:analysis_summary_limit_input}
are used as inputs to set limits and thereby exclude certain ranges of values
for the parameters of specific models of supersymmetric physics processes.
As the parameter space even in minimal Supersymmetry models with additional constraints has too high a number of dimensions,
only scans of low-dimensional subspaces of the full parameter space are possible.
This is done here using the signal grids from \Sec{sec:analysis_SUSY_GRIDs},
which implement two constrained models.
For the \ac{mSUGRA} model, the grid points span the \mzero-\moh plane,
the \ac{MSSM} grid consists of points in the gluino mass-squark mass plane.

\subsubsection{Selecting the Signal Region}
As the four signal regions are not independent,
it needs to be decided how to deal with their correlations.
The approach chosen here is to select the best signal region for each grid point individually.
This must be done without looking at the data,
\ie from the expected exclusion potential alone,
to avoid a bias similar to the look-elsewhere effect.
Thus, for each of the four signal regions,
the expected and observed limits are computed independently,
and the one with the best expected limit is adopted as the nominal result,
as has been done \eg in \cite{PRL102CDFLimit}. %
Several different measures can be used to determine the best exclusion potential:
the largest expected ratio of signal over background events,
the smallest expected \cls value
and the smallest expected $p$-value from the \ac{PLR} method.

\renewcommand{\abovecaptionskip}{-0pt}

\begin{figure}
  \centering
    \incgraphics{width=\widthtwoplots}{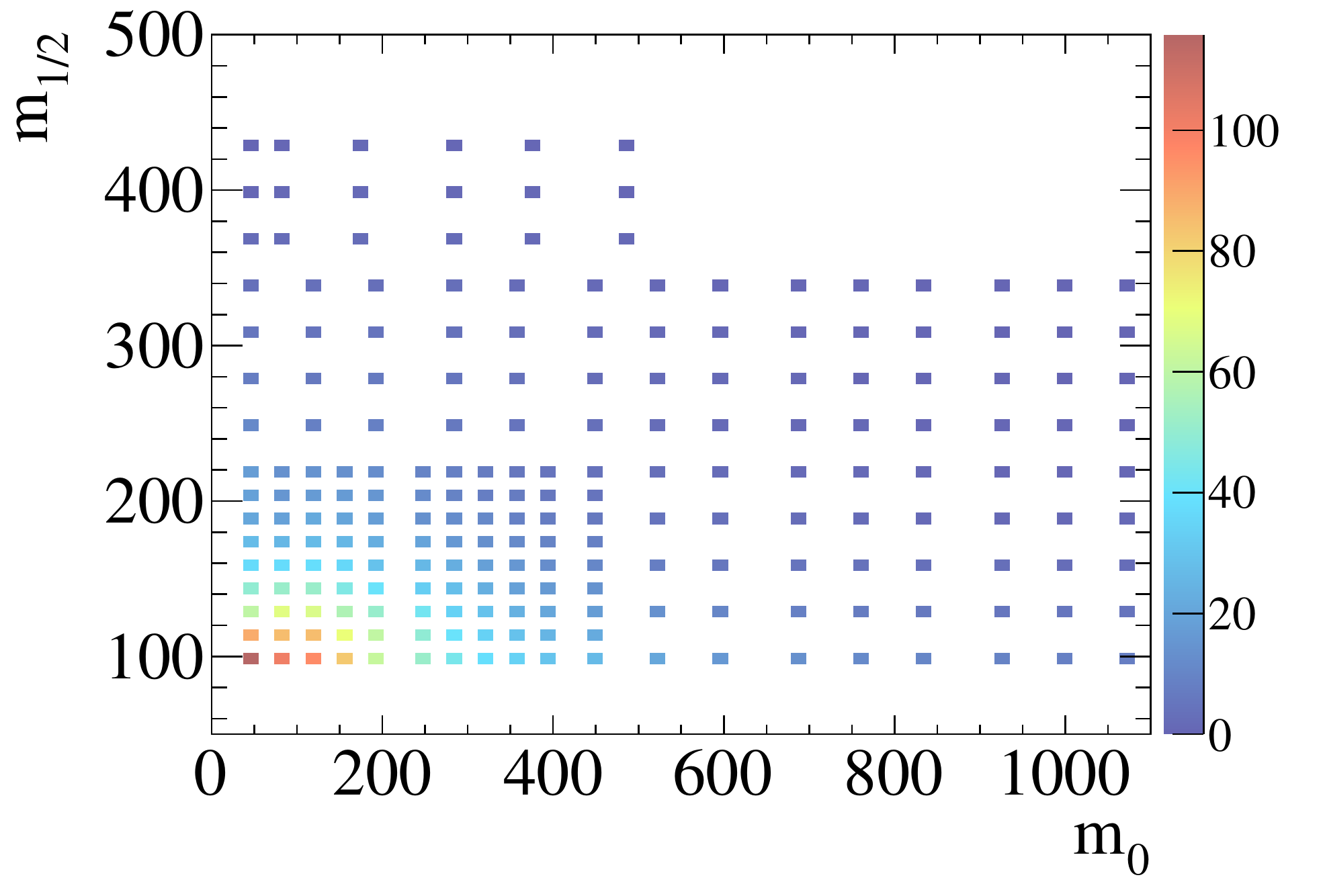}
    \hfill
    \incgraphics{width=\widthtwoplots}{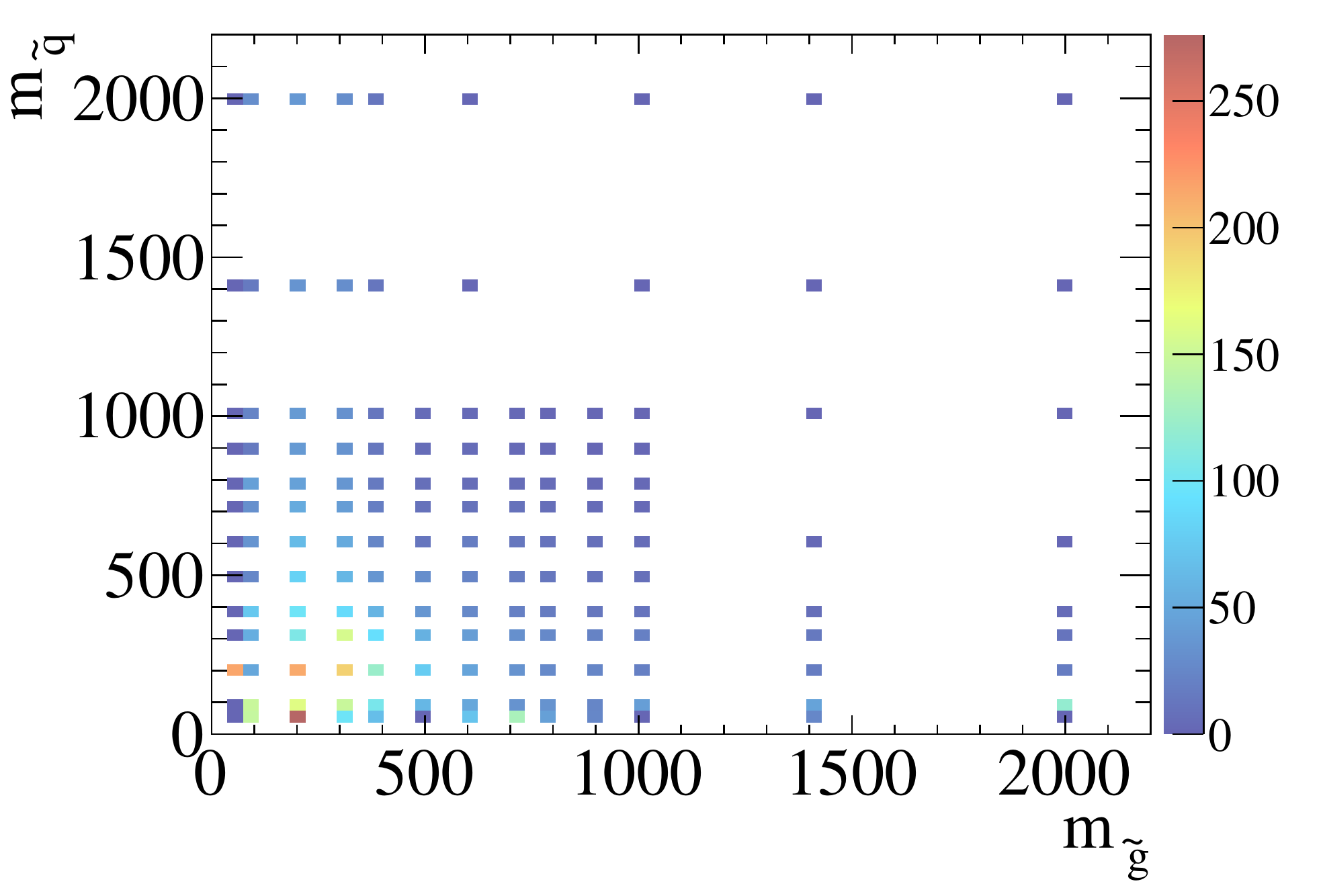}
  \caption{
    Best \ssqrtb ratio (\ie largest among the four signal regions) for the Monte Carlo samples in the signal grids.
    Left: mSUGRA grid, right: MSSM grid.
  }
  \label{fig:analysis_interpretation_bestssqrtb}
\end{figure}

\begin{figure}
  \centering
    \incgraphics{width=\widthtwoplots}{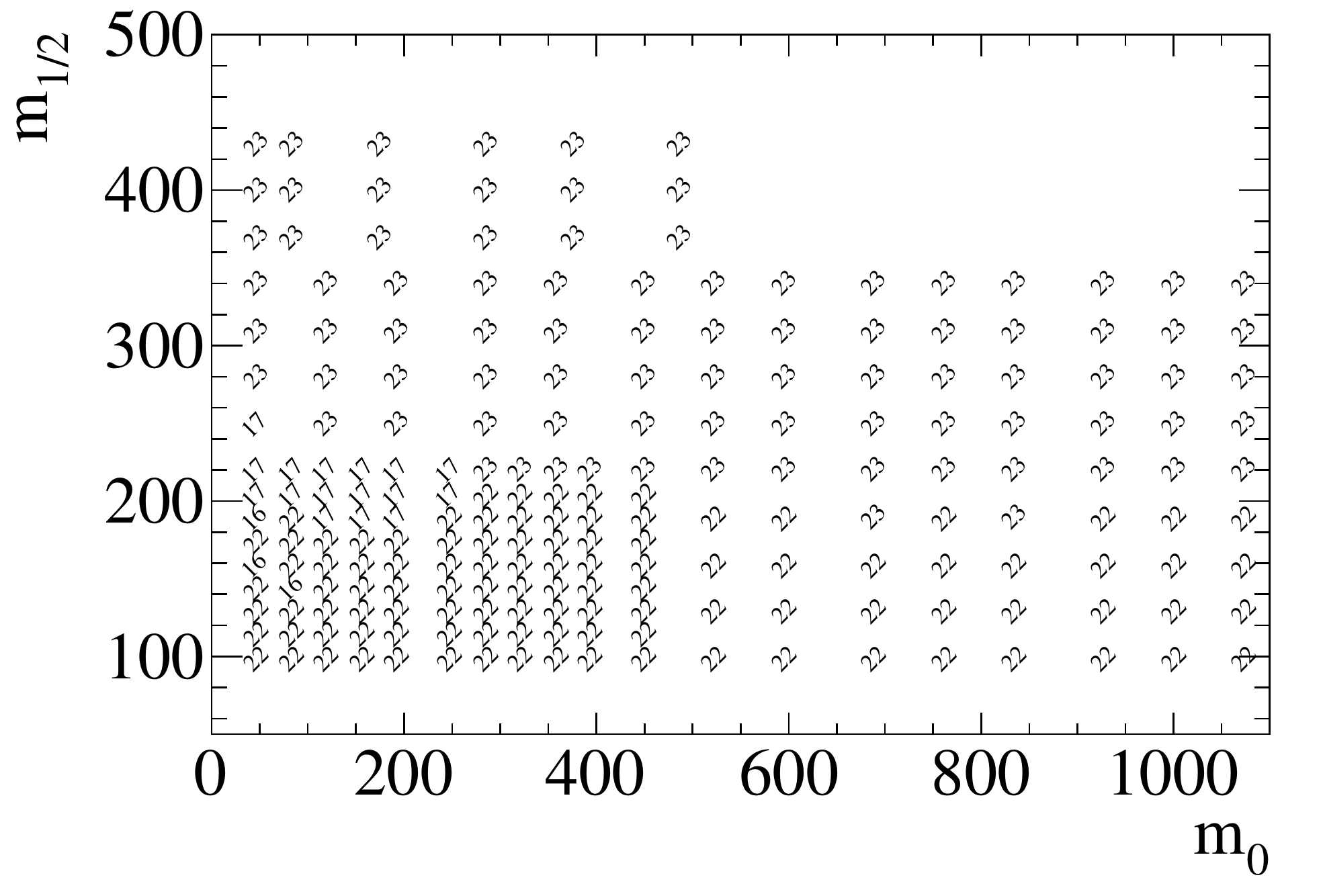}
    \hfill
    \incgraphics{width=\widthtwoplots}{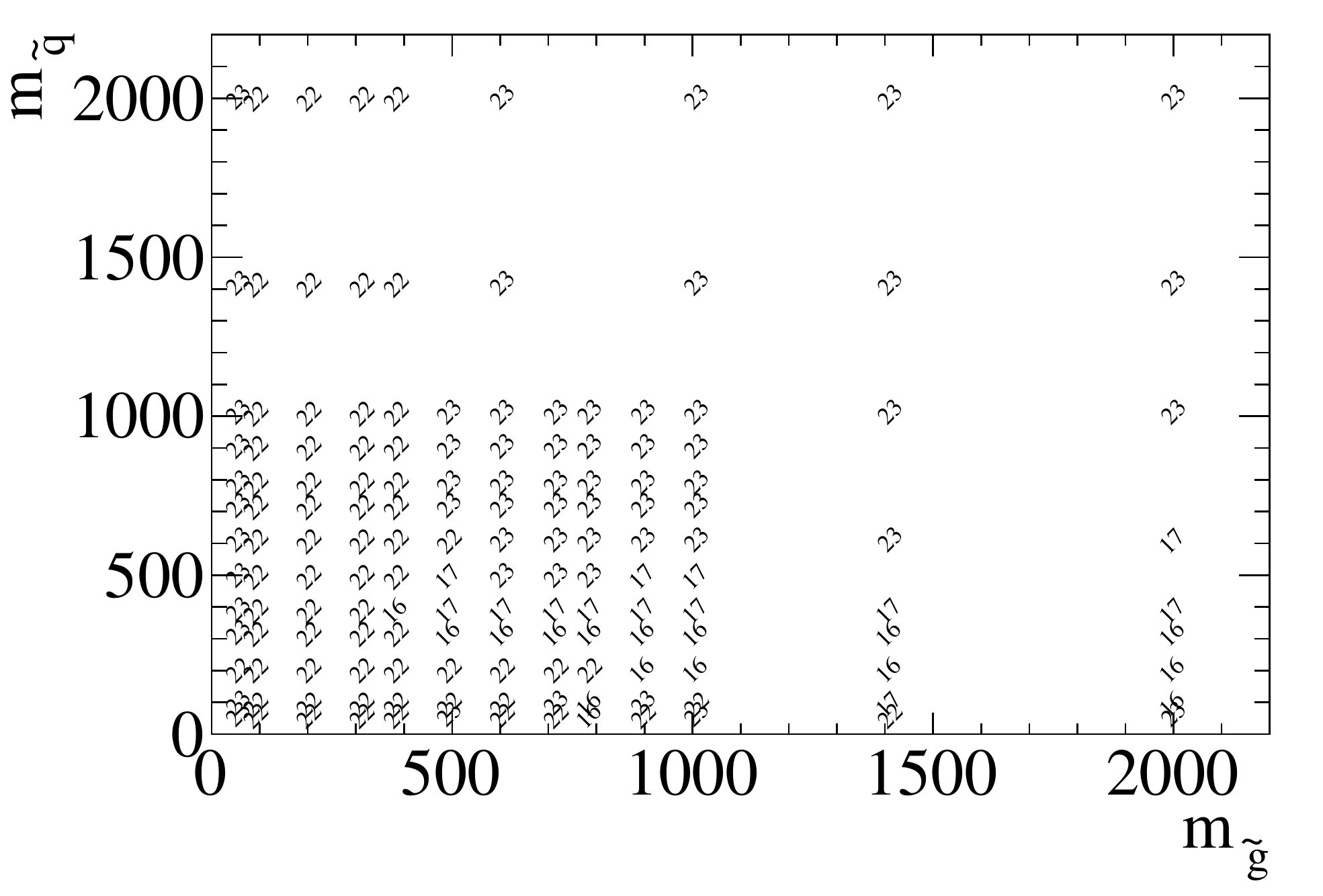}
  \caption{
    Signal region with the best \ssqrtb ratio,
    denoted by the number of its last defining step in the cutflow.
    Left: mSUGRA grid, right: MSSM grid.
  }
  \label{fig:analysis_interpretation_bestssqrtbi}
\end{figure}

\begin{figure}
  \centering
    \incgraphics{width=\widthtwoplots}{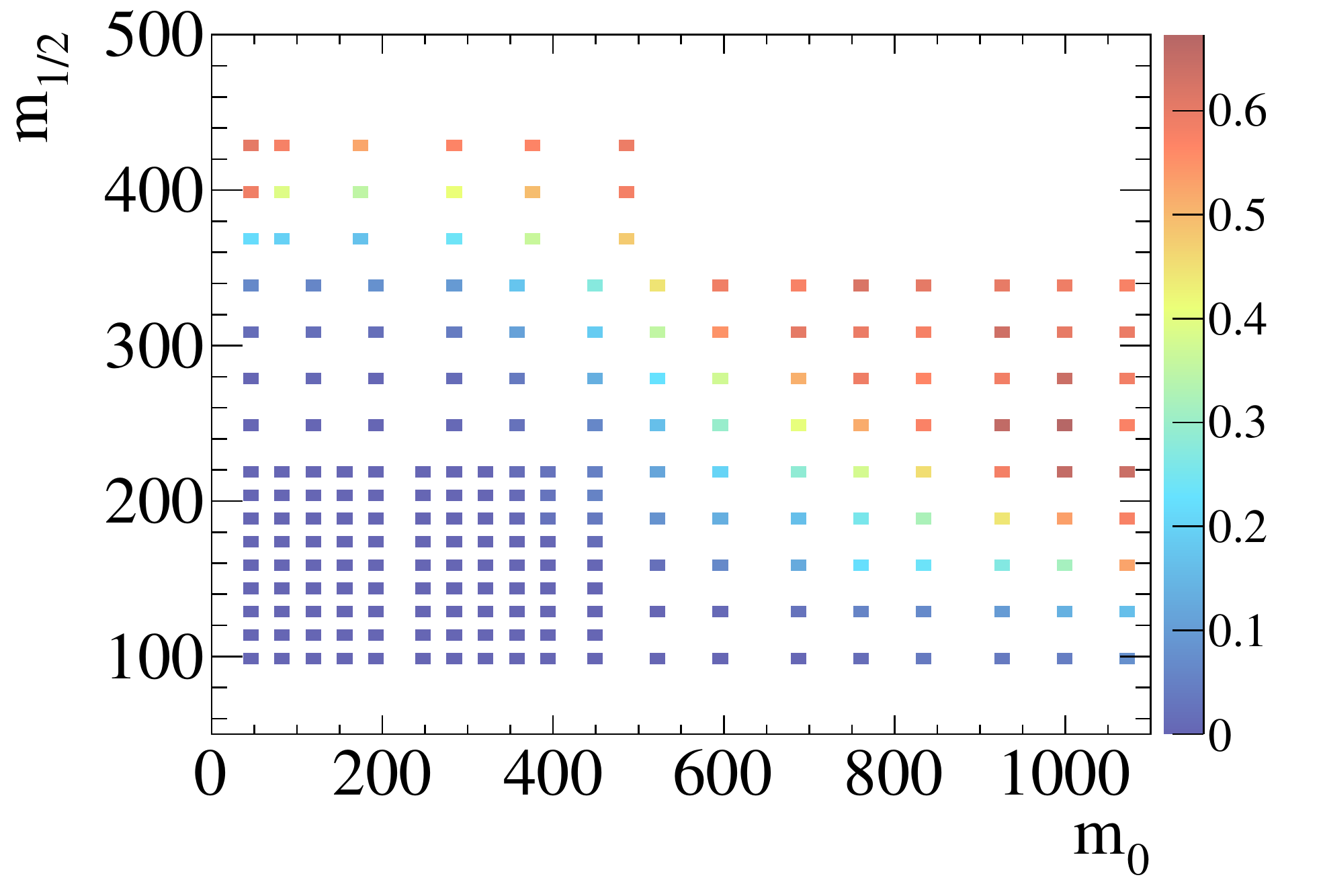}
    \hfill
    \incgraphics{width=\widthtwoplots}{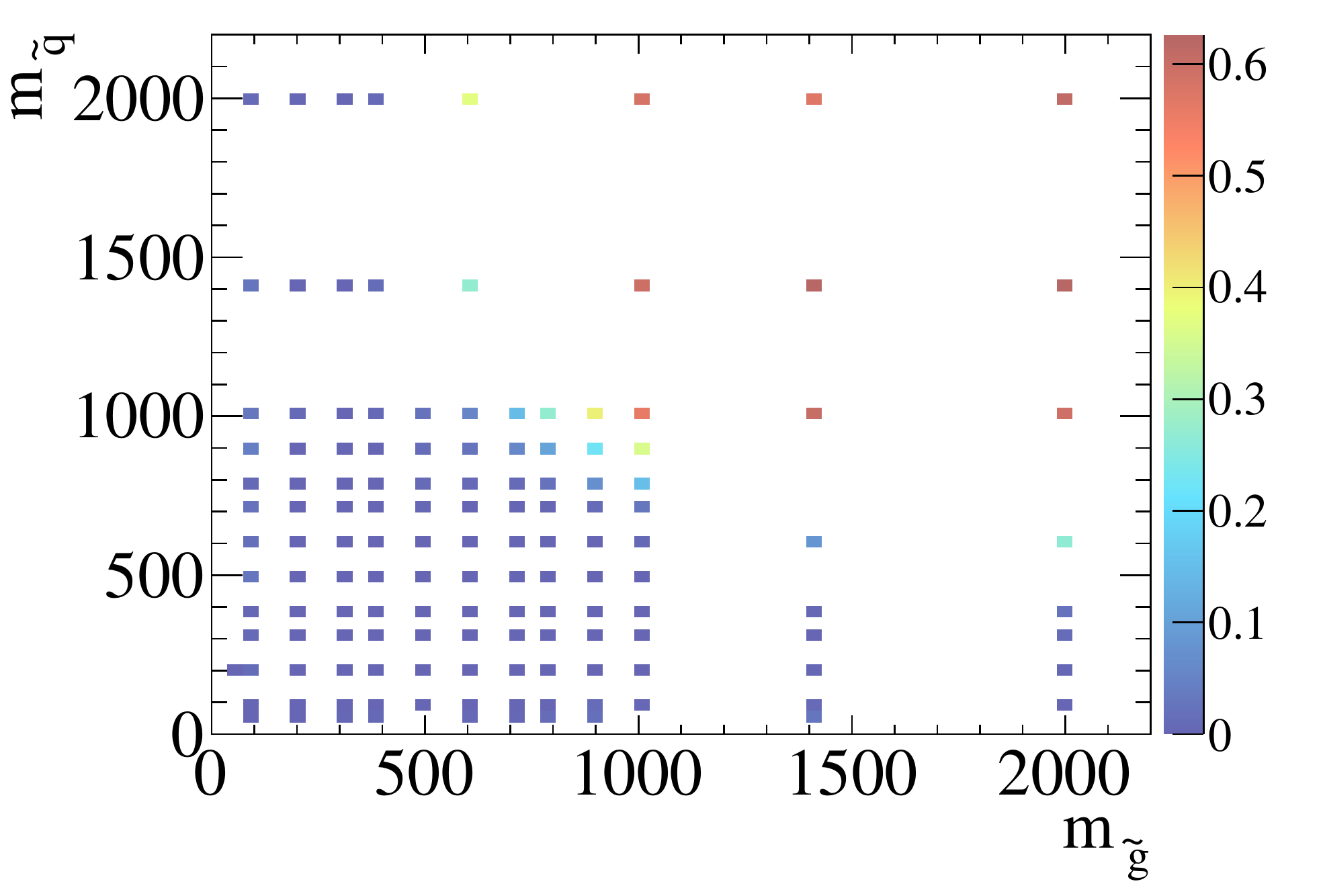}
  \caption{
    Best \cls value (\ie smallest amongst all four signal regions) expected under the assumption of background only for the Monte Carlo samples in the signal grids.
    Left: mSUGRA grid, right: MSSM grid.
  }
  \label{fig:analysis_interpretation_bestexpcls}
\end{figure}

\begin{figure}
  \centering
    \incgraphics{width=\widthtwoplots}{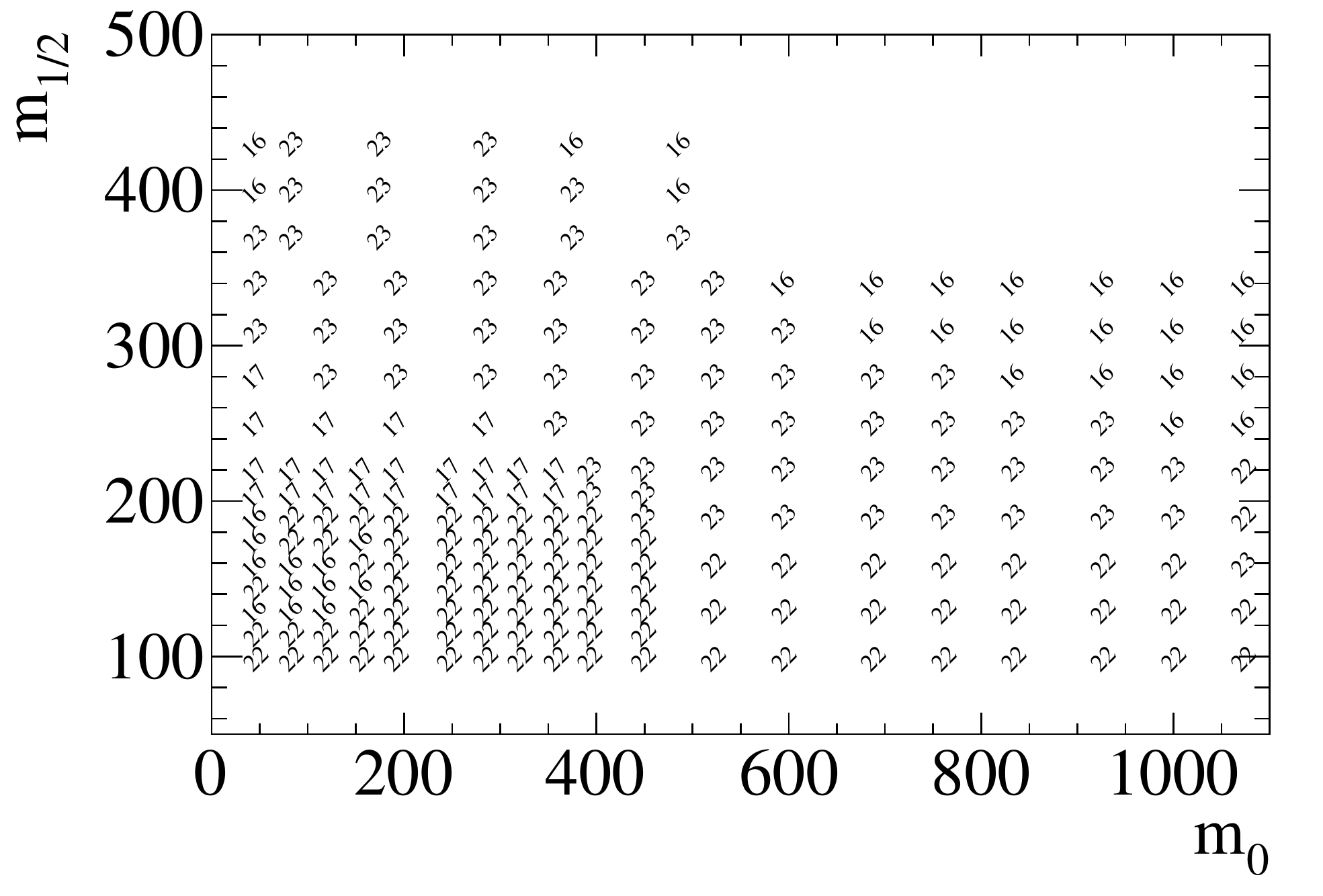}
    \hfill
    \incgraphics{width=\widthtwoplots}{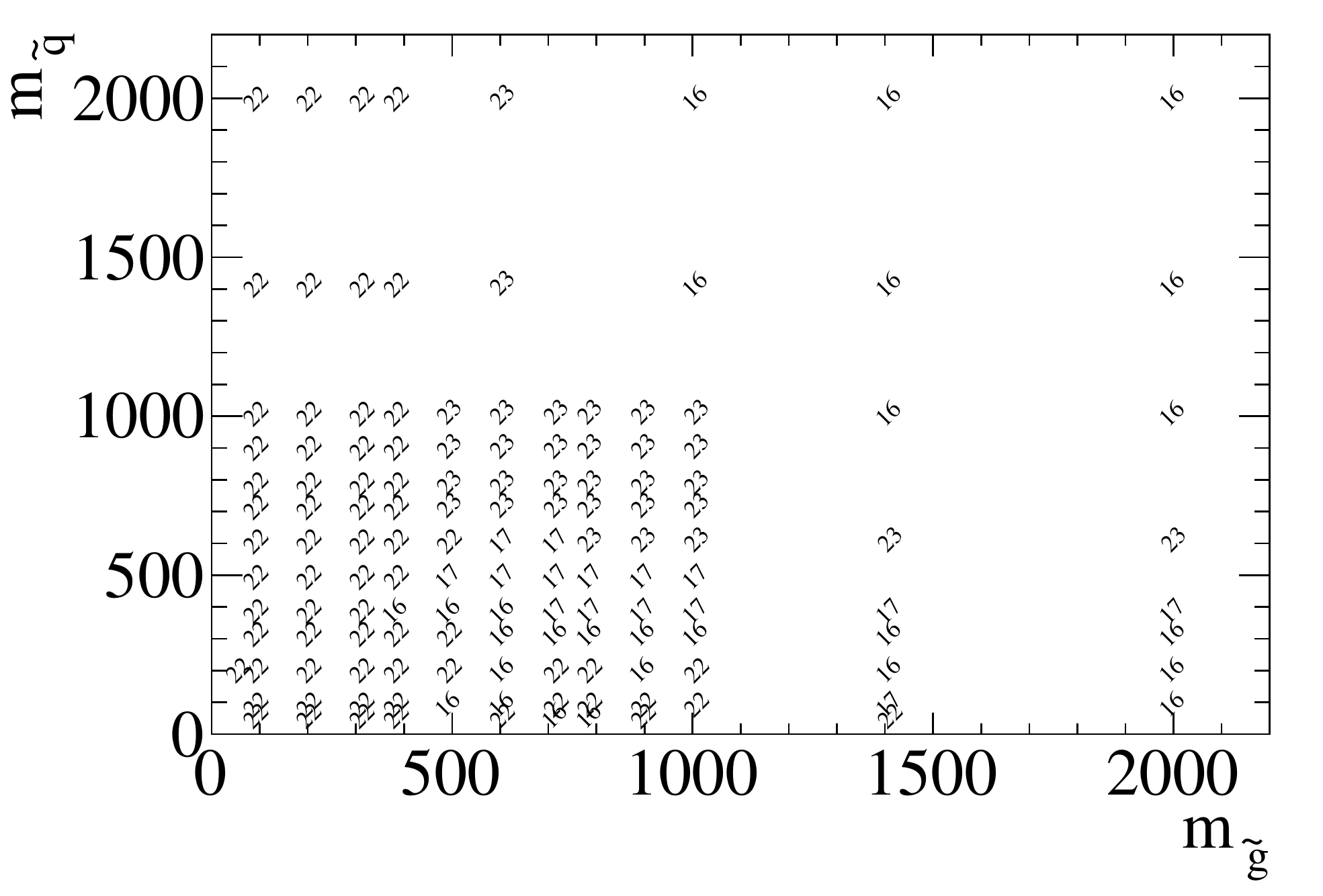}
  \caption{
    Signal region with the best \cls value expected under the assumption of background only,
    denoted by the number of its last defining step in the cutflow.
    Left: mSUGRA grid, right: MSSM grid.
  }
  \label{fig:analysis_interpretation_bestexpclsi}
\end{figure}

\begin{figure}
  \centering
    \incgraphics{width=\widthtwoplots}{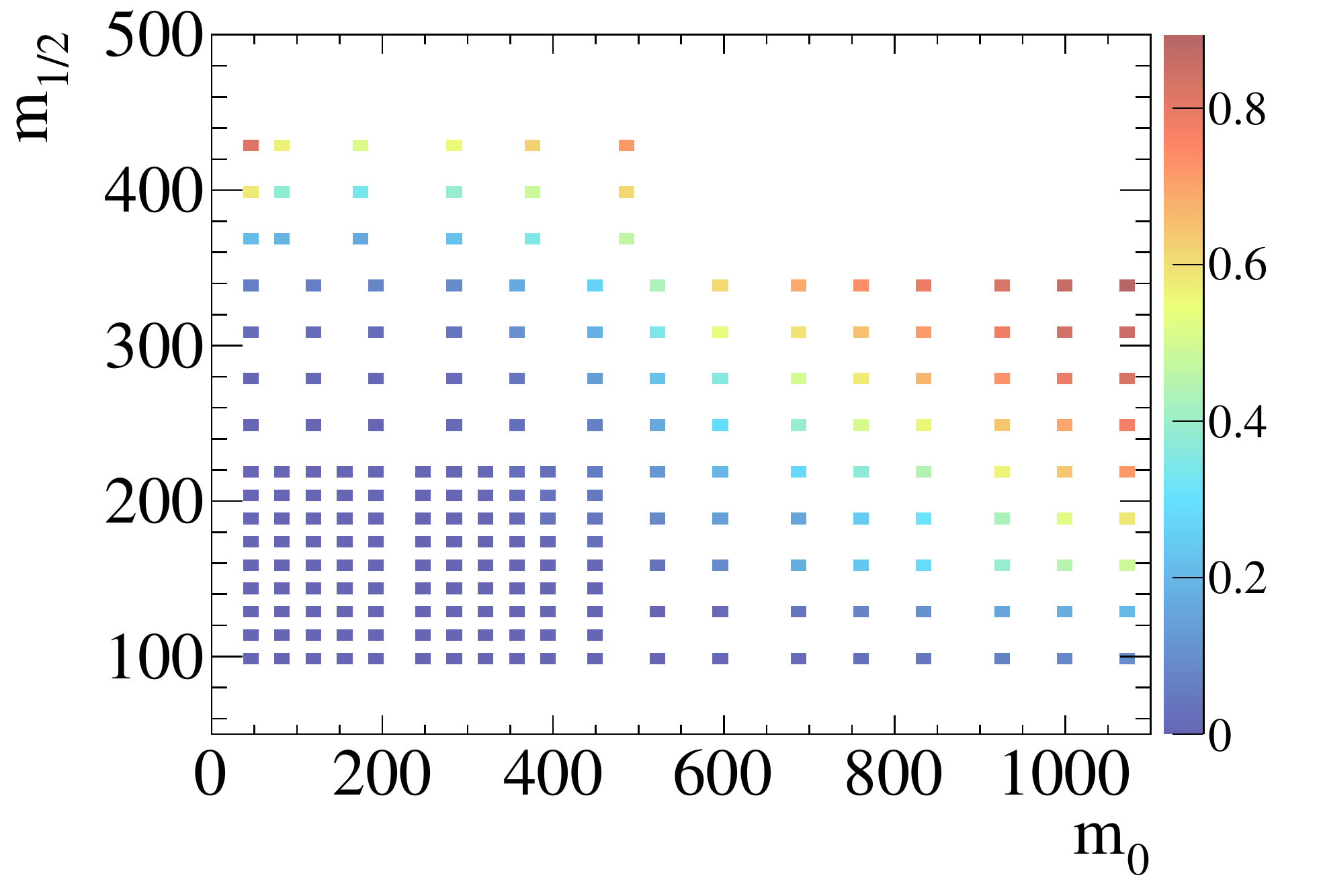}
    \hfill
    \incgraphics{width=\widthtwoplots}{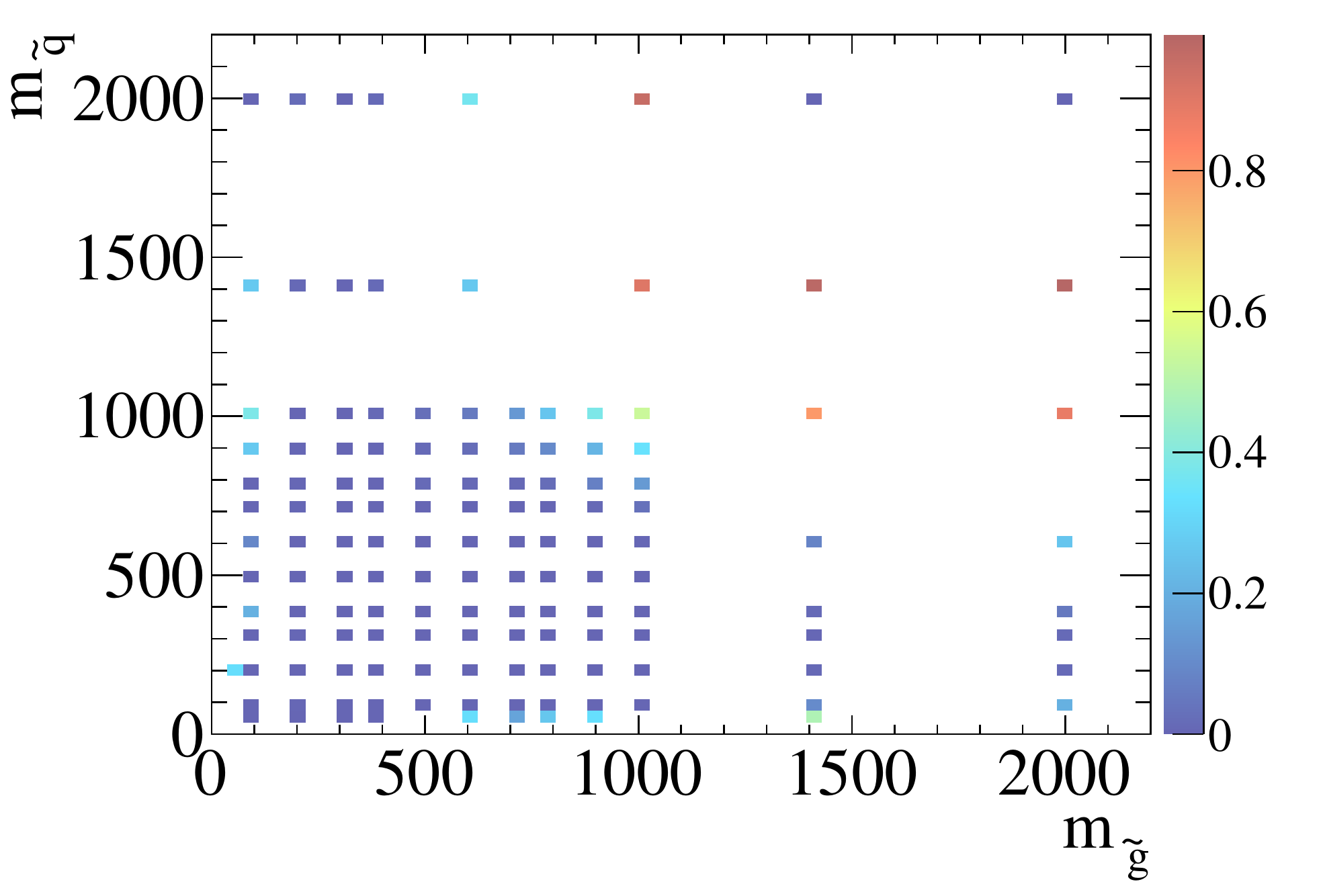}
  \caption{
    Best PLR value (\ie smallest amongst all four signal regions) expected under the assumption of background only for the Monte Carlo samples in the signal grids.
    Left: mSUGRA grid, right: MSSM grid.
  }
  \label{fig:analysis_interpretation_bestexpplr}
\end{figure}

\begin{figure}
  \centering
    \incgraphics{width=\widthtwoplots}{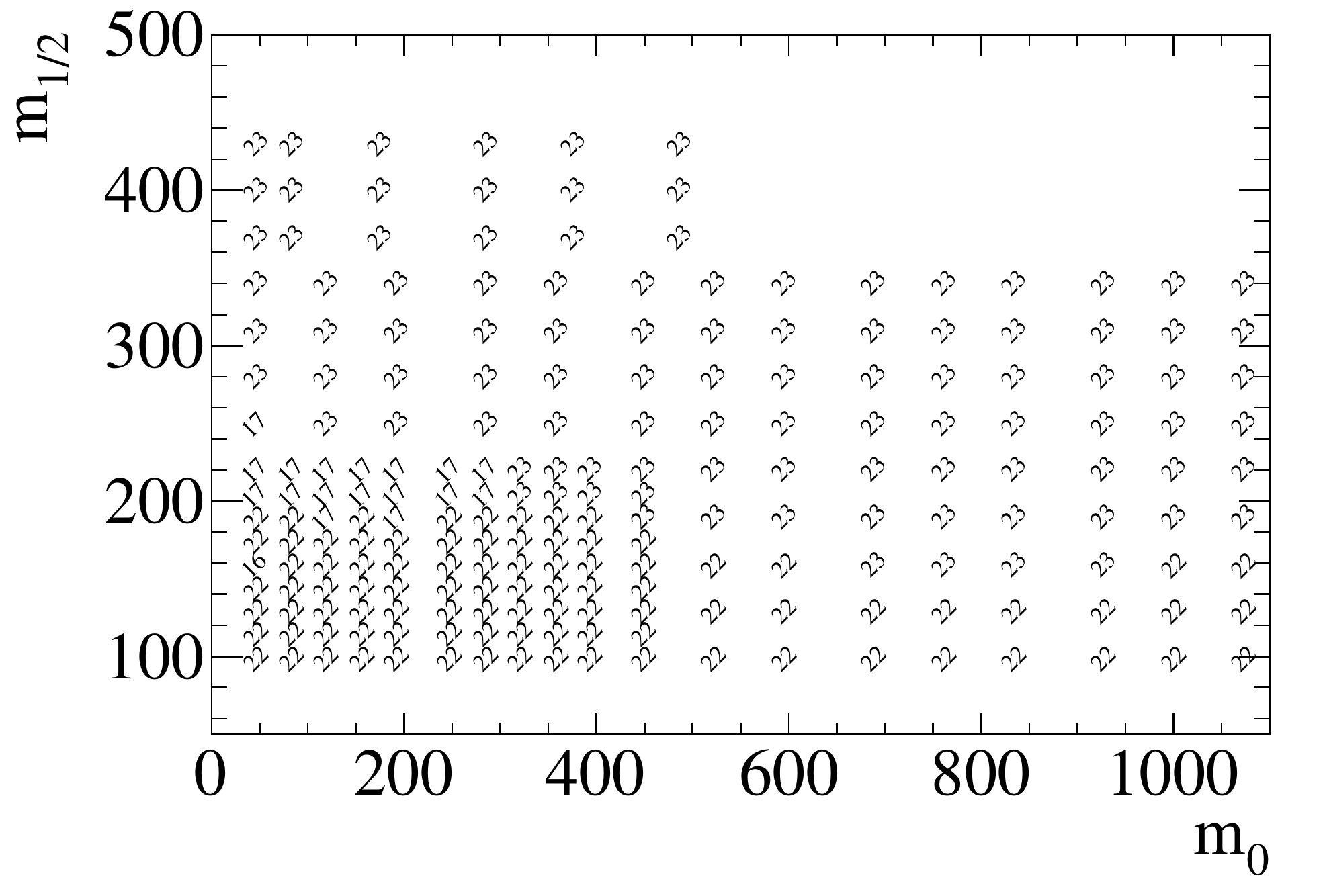}
    \hfill
    \incgraphics{width=\widthtwoplots}{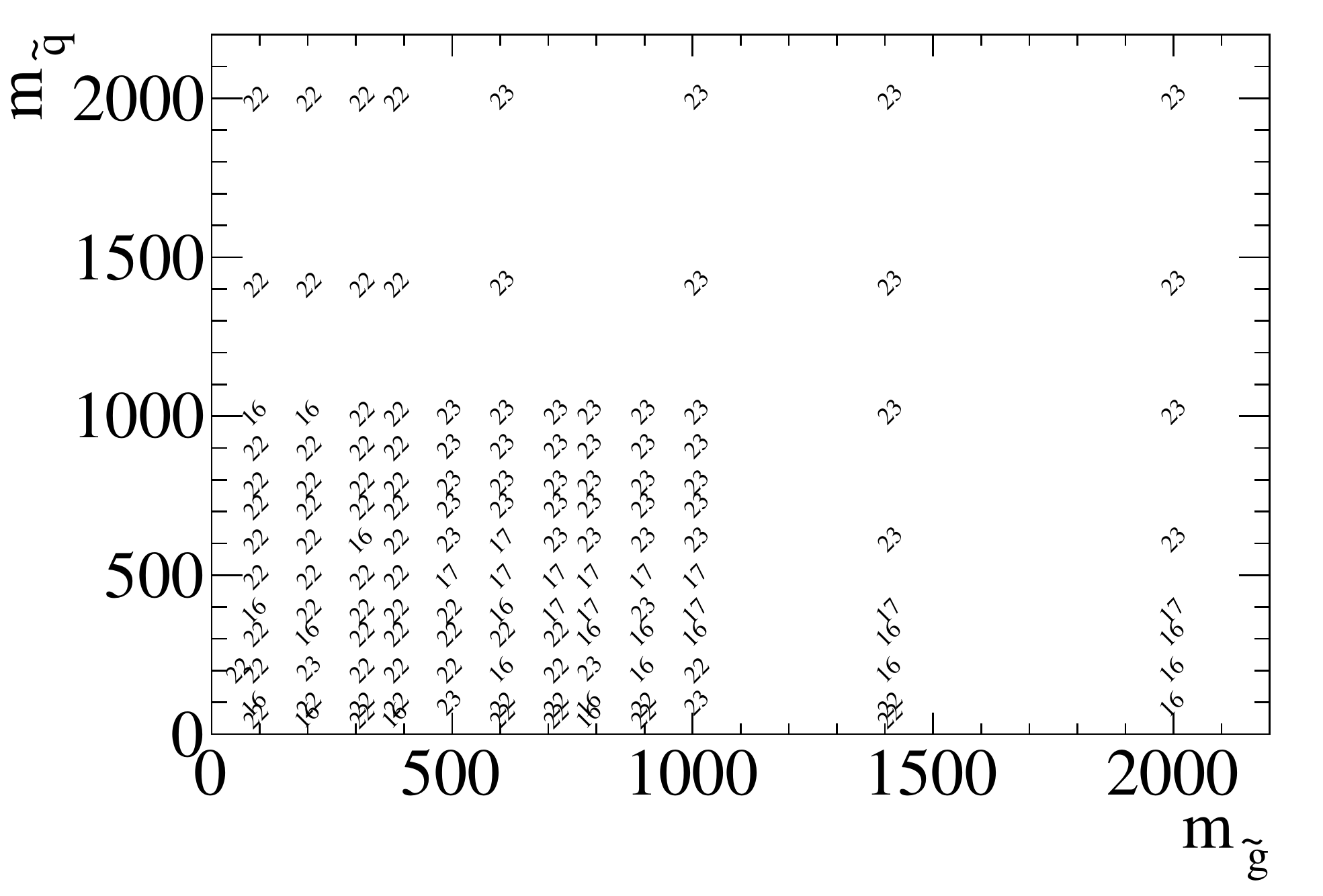}
  \caption{
    Signal region with the best PLR value expected under the assumption of background only,
    denoted by the number of its last defining step in the cutflow.
    Left: mSUGRA grid, right: MSSM grid.
  }
  \label{fig:analysis_interpretation_bestexpplri}
\end{figure}

\abovecaptionskipdefault

\Fig{fig:analysis_interpretation_bestssqrtb} shows,
for each of the Monte Carlo samples comprising the mSUGRA grid (left plot) and MSSM grid (right plot),
the best, \ie the highest, of the four \ssqrtb values %
obtained for each of the four signal regions at every point of the grid.
The plots in \Fig{fig:analysis_interpretation_bestssqrtbi} show
which of the four signal region provides this \ssqrtb ratio.
In these two plots, the number of the last step in the cutflow from \Tab{tab:analysis_cutflow_data},
defining the respective signal region is printed,
so that the numbers~16, 17, 22 and~23 denote signal regions~A, B, C and~D, respectively.

From the right plot in \Fig{fig:analysis_interpretation_bestssqrtb} for the MSSM grid,
one can see that the ratio \ssqrtb is largest in a vertical and a horizontal strip
with gluino or squark masses around \GeV{200}.
For higher masses, as well as for the grid points where one of the masses is very low, \ssqrtb becomes smaller.
As the background is the same for all grid points,
this means that the signal yield in these regions is lower.
Taking into account that the cross section falls off monotonically with increasing masses (cf. \Fig{fig:analysis_overview_grid_cross_sections}),
this in particular means that the much larger cross section for low masses %
cannot compensate for the loss in the signal efficiency for low squark and gluino masses,
due to the softer spectrum of the decay products,
so that the product of larger cross section and smaller signal efficiency still yields a smaller number of events expected in the signal regions.
For the mSUGRA grid in the left plot,
the behavior is different and no such peak at intermediate masses can be seen.
Here, the ratio \ssqrtb decreases in general radially,
with the highest values in the region where both \mzero and \moh are small.

In \Figs{fig:analysis_interpretation_bestexpcls} and \ref{fig:analysis_interpretation_bestexpclsi},
the same type of plots is shown, this time choosing the signal region according to the best \cls value,
where the best \cls value is the smallest \cls value expected under the assumption of the background-only hypothesis
and computed from the expected number of signal and background events without considering the observed data.
Again, \Fig{fig:analysis_interpretation_bestexpcls} shows the \cls values obtained for the mSUGRA (left) and MSSM grid (right),
and \Fig{fig:analysis_interpretation_bestexpclsi} shows which signal region yields this \cls value
and will thus be used in the limit-setting procedure to determine the \cls value for this particular grid point.
Finally, in \Figs{fig:analysis_interpretation_bestexpplr} and \ref{fig:analysis_interpretation_bestexpplri}
the same plots are shown, this time choosing the best signal region
in terms of the smallest expected PLR value under the background-only hypothesis.

The main difference between a selection based on \ssqrtb, \cls or PLR is that \cls and PLR take the uncertainties into account,
whereas \ssqrtb is purely based on the expected number of signal and background events in the signal regions.
For most grid points,
the signal region chosen according to the three measures is the same,
but there are differences,
with signal regions~A and~B chosen only rarely by \ssqrtb,
and more often by \cls in both grids,
and by \ac{PLR} in the \ac{MSSM} grid.
Note that for some of the \ac{MSSM} grid points,
all of the signal regions have a expected number of signal events of zero.
For these points $\ssqrtb = 0$, and neither the \cls nor the PLR value can be computed.
They are therefore left out in the \cls and PLR plots.

\subsubsection{Exclusion Plots for the mSUGRA Grid}

\begin{figure}
  \centering
    \incgraphics{width=\widthwideplot}{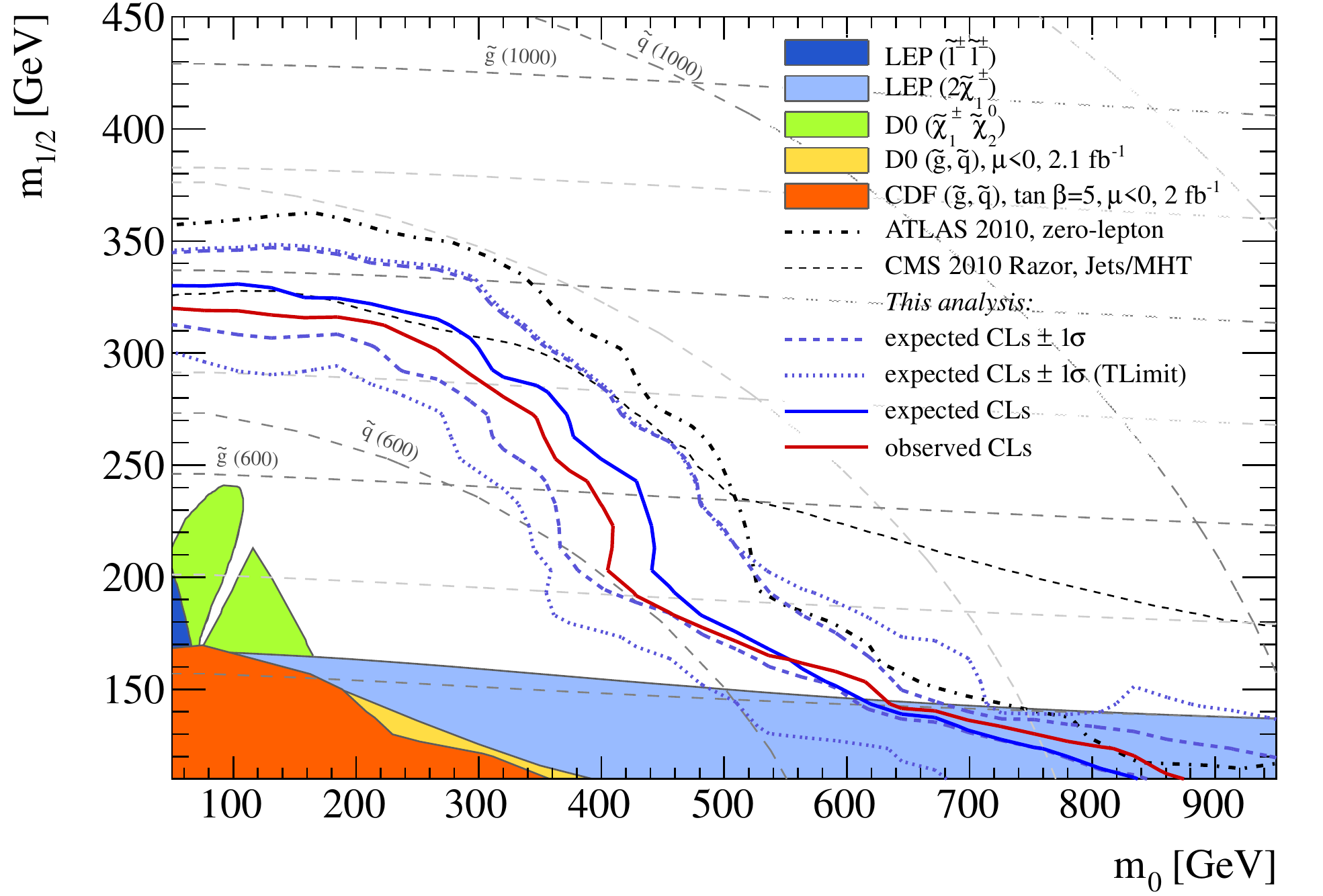}
  \caption{
    \percent{95} \CL exclusion limits in the mSUGRA plane spanned by \mzero and \moh.
    The red line marks the observed exclusion limit obtained from the \cls method in this analysis,
    the solid blue line gives the expected exclusion limit and
    the dashed and dotted blue lines mark the $\pm1\sigma$ bands from two different methods as explained in the text.
    Shaded areas indicate exclusion results from other experiments.
  }
  \label{fig:analysis_exclusion_limit_mSUGRA_CLs}
\end{figure}

\begin{figure}
  \centering
    \incgraphics{width=\widthwideplot}{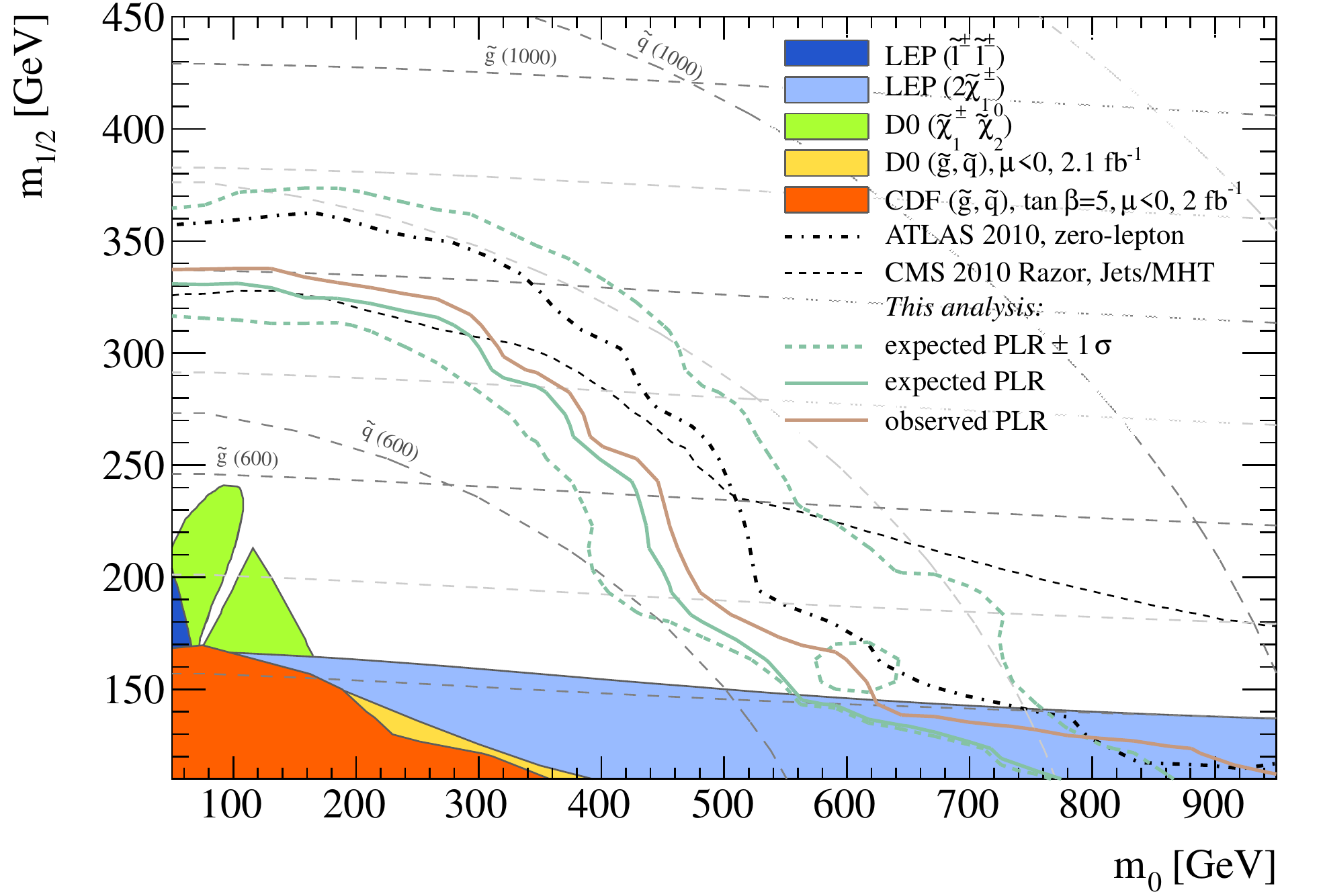}
  \caption{
    \percent{95} \CL exclusion limits in the mSUGRA plane spanned by \mzero and \moh.
    The brown line marks the observed exclusion limit obtained from the PLR method in this analysis,
    the solid green line gives the expected exclusion limit and
    the dashed green lines mark the $\pm1\sigma$ band.
    Shaded areas indicate exclusion results from other experiments.
  }
  \label{fig:analysis_exclusion_limit_mSUGRA_PLR}
\end{figure}

\begin{figure}
  \centering
    \incgraphics{width=\widthwideplot}{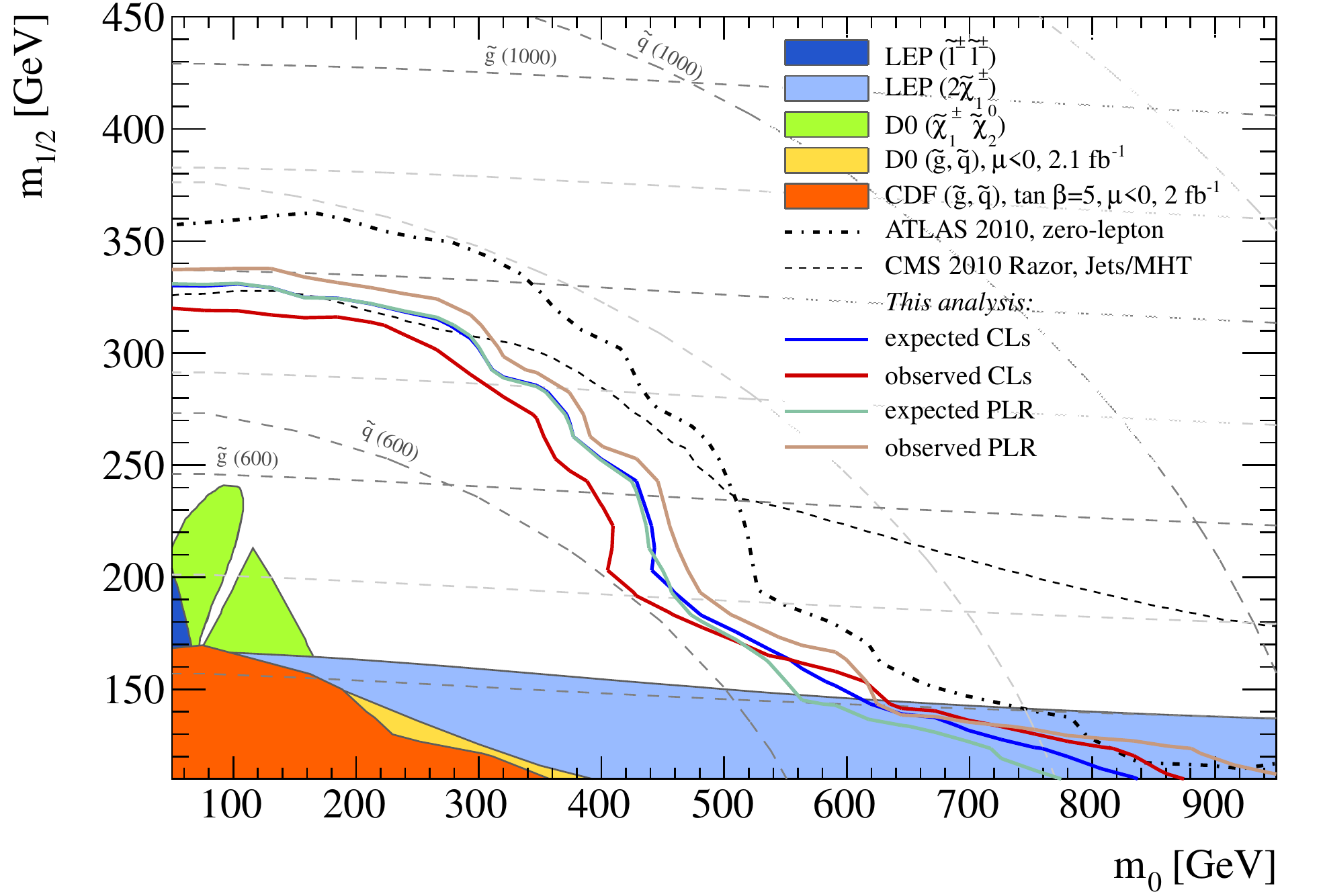}
  \caption{
    \percent{95} \CL exclusion limits in the mSUGRA plane spanned by \mzero and \moh,
    comparing in particular the results obtained in this analysis from the \cls method with those from PLR.
    The line styles and color coding are the same as in
    \Figs{fig:analysis_exclusion_limit_mSUGRA_CLs} and \ref{fig:analysis_exclusion_limit_mSUGRA_PLR}.
  }
  \label{fig:analysis_exclusion_limit_mSUGRA_Compare}
\end{figure}

\Fig{fig:analysis_exclusion_limit_mSUGRA_CLs} shows the \percent{95} \CL exclusion limits in the mSUGRA plane
in terms of the expected and observed exclusion limits,
as they are obtained from the \cls method in this analysis.
The expected limits mark the range of parameter space which the analysis is expected to be able to exclude at the given confidence level,
the observed limits mark the range that is excluded based on the observed data.
Differences between these two quantities may either arise from statistical fluctuations in the observed data,
from sources of systematic bias which have not been taken into account properly, %
or, in case the observed limit stays below the expected limits, from signal processes. %
\Figs{fig:analysis_exclusion_limit_mSUGRA_PLR}
shows the limits which have been computed using the PLR method,
and \Fig{fig:analysis_exclusion_limit_mSUGRA_Compare} shows
a comparison of the expected and observed limits from \cls and PLR in one plot.
In \Fig{fig:analysis_exclusion_limit_mSUGRA_Compare},
it can be seen that the expected limits for both \cls and PLR are the same,
but the observed limit from \cls is more conservative than the one from PLR as expected.

To give a feeling for the size of the impact of statistical fluctuations,
the $\pm1\sigma$ bands (colored dashed lines) demonstrate the variation of the expected limits
arising from up- and downwards fluctuations of the event counts by one standard deviation.
These variations are computed in two different ways.
For the \cls limit from \propername{TLimit},
they can be computed using the built-in methods from \code{TConfidenceLevel} (cf. \Sec{sec:analysis_implementation_cls}),
which varies the expected background up and down by one standard deviation
according to the distribution of the test statistic for the assumption of only background.
The second approach used here to compute the bands is to vary the background expectation %
up and down manually by a value that corresponds to a fluctuation of the background
by one standard deviation according to the total uncertainties on the background,
and to re-run the limit calculation.
This approach is possible for both \cls and PLR.
The two alternatives are different conceptually and therefore not \latein{a priori} expected to give the same results.
They are compared in \Fig{fig:analysis_exclusion_limit_mSUGRA_CLs}
because for PLR only the latter approach can be used.
As can be seen in this plot,
for the downwards fluctuations of the background, %
allowing to exclude a larger range,
the two methods give similar results ,
for the upwards fluctuations they do not.

Results from searches for Supersymmetry have been published by past and present experiments.
These publications set limits on many different quantities,
such as masses of supersymmetric particles and cross sections,
and use various different supersymmetric models and parameter sets.
Care therefore needs to be taken to gather a set of results which is consistent
in the choice of model and the values for its parameters.
For comparison, results from other collider experiments are shown as shaded areas 
in the exclusion plots in \Fig{fig:analysis_exclusion_limit_mSUGRA_CLs} through \Fig{fig:analysis_exclusion_limit_mSUGRA_Compare}.
The macros which draw these regions are taken from the official code of the \ATLAS Supersymmetry group.
In the following, it is described where these results can be found in literature and what are the respective model settings.
Note that parts of the plane are also excluded by theory constraints.
A small region with $\mzero \lesssim \GeV{60}$ and $\moh \lesssim \GeV{80}$ is excluded
because there is no electroweak symmetry breaking, %
but this region is smaller than the region excluded by \Dzero \cite{PLB2008D0Limits}.

For the \ac{mSUGRA} grid,
five different colors mark five different regions excluded by earlier experiments.
The legend of the plot lists the name of the corresponding experiments
and in brackets the channel of the search from which these limits were obtained.
The first two exclusion regions are set by the \acs{LEP} experiments
from a search for signatures of two scalar leptons or two charginos, respectively.
The following paragraph explains where the limits included in the official exclusion plots come from.
The \LEP slepton exclusion region shown in \Fig{fig:analysis_exclusion_limit_mSUGRA_CLs}
matches the exclusion limit which is cited in \cite{PLB2008D0Limits} as the result of the LEP2 slepton searches for \tanbeta{3}, $A_0=0$ and $\signmu < 0$
with reference to \cite{LEPSUSYWG/02-06.2},
and in \cite{PLB2009D0Limits} as the result of the LEP slepton searches for \tanbeta{3}, $A_0=0$ and $\signmu > 0$
with reference to \cite{LEPSUSYWG/01-03.1} and \cite{LEPSUSYWG/04-01.1}.
Similar limits are shown as result of a slepton search by the \propername{ALEPH} collaboration,
interpreted in a minimal supergravity setting with \tanbeta{5} and 10, $A_0=0$ and both $\signmu < 0$ and $\signmu > 0$ in \cite{PLB2001AlephLimits}, %
the exclusion limits from the slepton search on \moh for $\signmu < 0$ being slightly weaker than for $\signmu > 0$.
In the same plot in \cite{PLB2001AlephLimits}, an exclusion region from chargino searches is presented,
which extends over the whole \mzero range in the \mzero-\moh plane,
again for \tanbeta{5} and 10, $A_0=0$ and both signs of $\mu$,
and again the limits are weaker in \moh for $\signmu < 0$.
For the chargino search, the border of the exclusion region differs by about \GeV{30}.
An update on these exclusion limits is given in \cite{ALEPH2002,PLB2004AlephLimits}
for \tanbeta{10} and 30, $A_0=0$ and both signs of $\mu$,
showing small improvements for the chargino limits at \tanbeta{10}.
\cite{PRL101CDFLimits} cites an exclusion of the chargino masses below \GeV{103.5} as \LEP direct limit,
with reference to \cite{LEPSUSYWG/01-03.1},
and shows this in relation with limits on \moh up to \GeV{170} for \tanbeta{3}, $A_0=0$ and $\signmu > 0$ and small \mzero. %
\cite{PLB2008D0Limits} also includes such limits under the label LEP2 chargino limits,
with reference to \cite{LEPSUSYWG/02-06.2},
for \tanbeta{3}, $A_0=0$ and $\signmu < 0$
but with significantly smaller exclusion reach only up to \GeV{130} in \moh.
\cite{PLB2009D0Limits} includes LEP chargino limits up to \GeV{170} in \moh for \tanbeta{3}, $A_0=0$ and $\signmu > 0$,
with reference to \cite{LEPSUSYWG/01-03.1} and \cite{LEPSUSYWG/04-01.1}.
In conclusion, the chargino limits cited in the exclusion plots from the \LEP experiments are appropriate for $\signmu > 0$,
but if $\signmu < 0$ as in the scenario assumed in the mSUGRA grid used for this analysis,
the limits from LEP for \moh seem to be weaker than indicated by the official contours.

The other three exclusion regions included in the plots are from Tevatron experiments.
The exclusion regions from the \Dzero experiment are from a search of \ifb{2.3} of data
for production of a chargino \charginon{1} and the second-lightest neutralino \neutralinon{2},
with final states containing three charged leptons and missing transverse energy,
interpreted in the \ac{mSUGRA} framework with \tanbeta{3}, $A_0 = 0$ and $\signmu > 0$ \cite{PLB2009D0Limits},
and from a search of \ifb{2.1} of data for squarks and gluinos
in final states with jets and missing transverse energy,
interpreted under the same assumptions but with $\signmu < 0$ \cite{PLB2008D0Limits}.
The exclusion region from \CDF is the result of a search for gluinos and squarks in \ifb{2.0} of \Tevatron data,
assuming minimal supergravity with \tanbeta{5}, $A_0 = 0$ and negative sign of $\mu$ \cite{CDFNote9229},
which has been published as \cite{PRL102CDFLimit}
but without the plot of the \mzero-\moh plane.

The thick dashed-dotted black line is the result of the official zero-lepton analysis of 2010 data
by the \ATLAS Supersymmetry group \cite{SUSYAnalyse0lepton2010}.
The slim dashed black line is the corresponding result from \CMS on the same amount of data from 2010 \cite{CMSLimits2010}.
The dashed grey lines labelled with particle type (\gluino or \squark) and mass in GeV
indicate the gluino and squark masses for \tanbeta{10}.
As the \acs{mSUGRA} and \acs{MSSM} models have completely different mass spectra,
the limits on the squark and gluino masses read off from this plot need not correspond to the limits shown below.

\subsubsection{Exclusion Plots for the MSSM Grid}

\begin{figure}
  \centering
    \incgraphics{width=\widthwideplot}{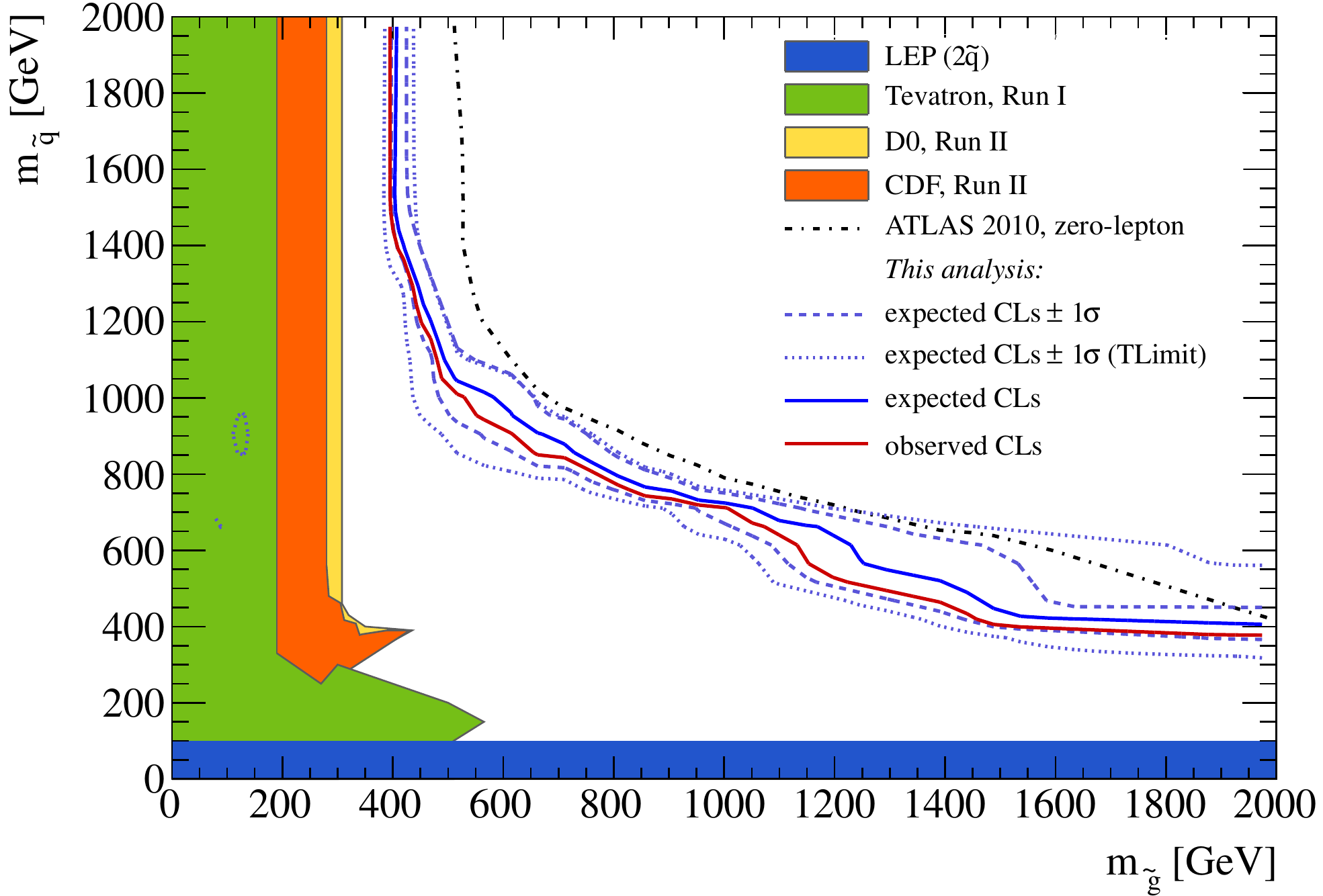}
  \caption{
    \percent{95} \CL exclusion limits in the MSSM plane spanned by $m_\gluino$ and $m_\squark$.
    The solid red line marks the observed exclusion limits obtained from the \cls method in this analysis,
    the solid blue line gives the expected exclusion limit and
    the dashed and dotted blue lines mark the $\pm1\sigma$ bands from two different methods as explained in the text.
    Shaded areas indicate exclusion results from other experiments.
  }
  \label{fig:analysis_exclusion_limit_MSSM_CLs}
\end{figure}

\begin{figure}
  \centering
    \incgraphics{width=\widthwideplot}{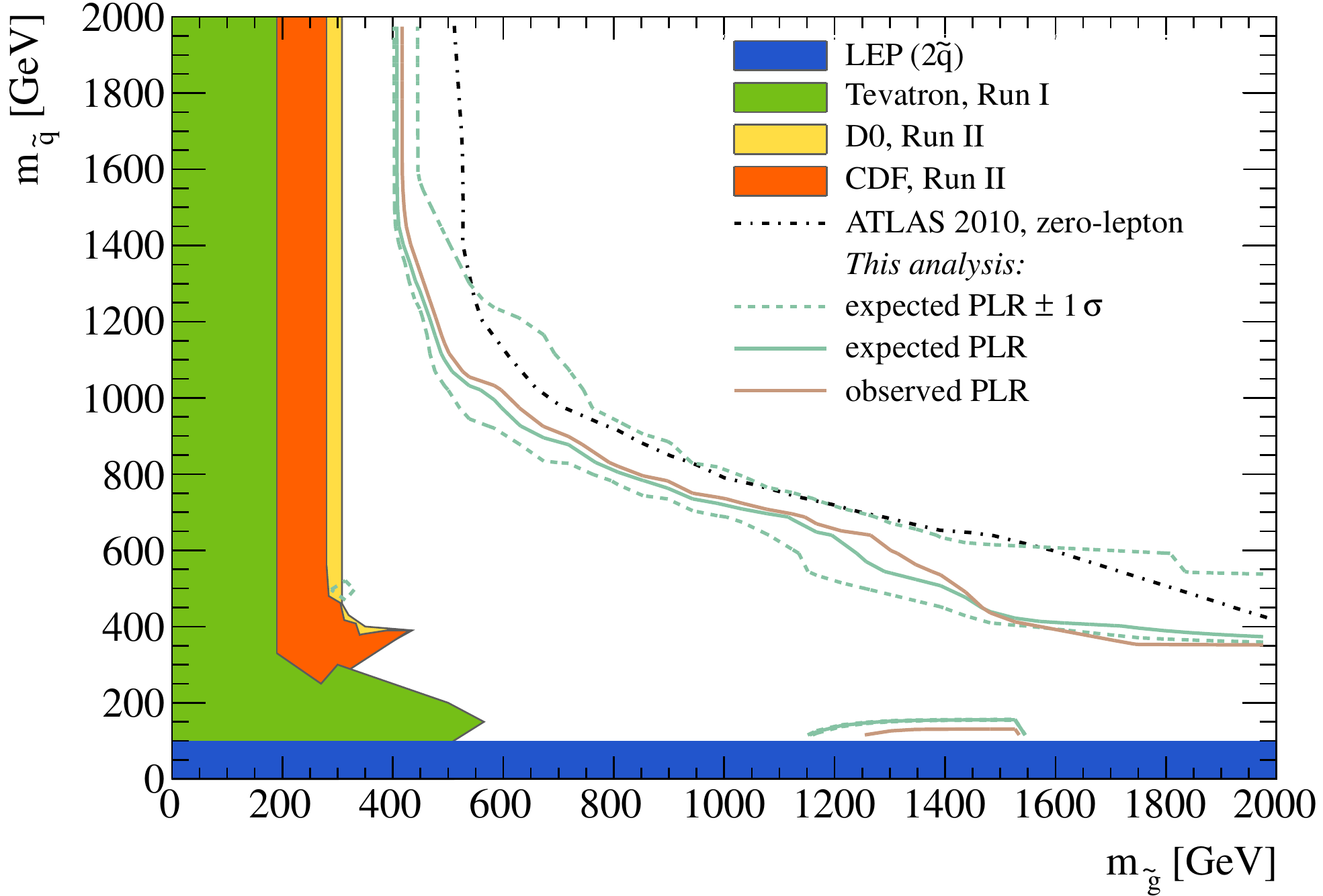}
  \caption{
    \percent{95} \CL exclusion limits in the MSSM plane spanned by $m_\gluino$ and $m_\squark$.
    The brown line marks the observed exclusion limits obtained from the PLR method in this analysis,
    the solid green line gives the expected exclusion limit and
    the dashed green lines mark the $\pm1\sigma$ band.
    Shaded areas mark exclusion results from other experiments.
  }
  \label{fig:analysis_exclusion_limit_MSSM_PLR}
\end{figure}

\begin{figure}
  \centering
    \incgraphics{width=\widthwideplot}{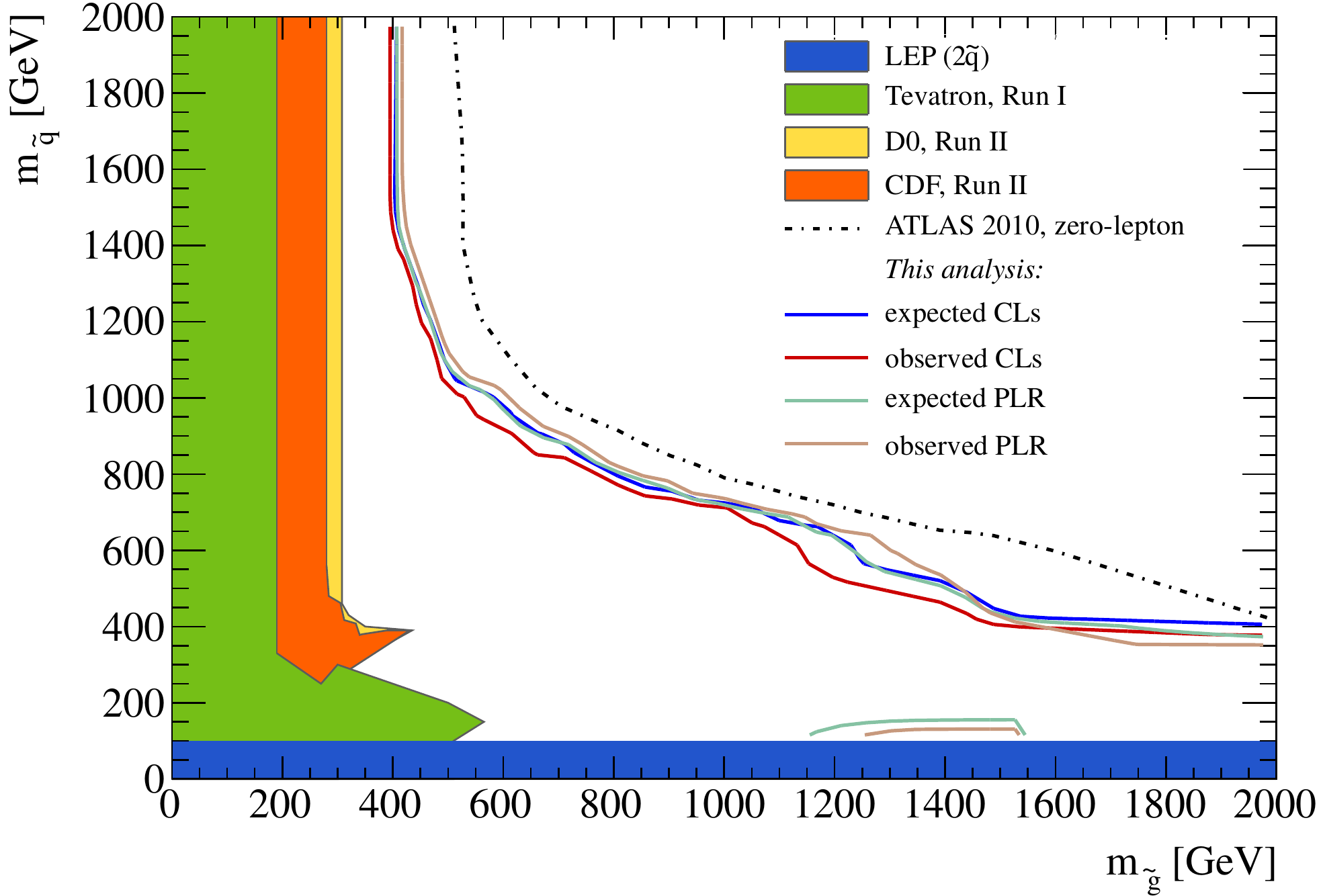}
  \caption{
    \percent{95} \CL exclusion limits in the MSSM plane spanned by $m_\gluino$ and $m_\squark$,
    comparing in particular the results obtained in this analysis from the \cls method with those from PLR.
    The line styles and color coding are the same as in
    \Figs{fig:analysis_exclusion_limit_MSSM_CLs} and \ref{fig:analysis_exclusion_limit_MSSM_PLR}.
  }
  \label{fig:analysis_exclusion_limit_MSSM_Compare}
\end{figure}

\Fig{fig:analysis_exclusion_limit_MSSM_CLs} shows the \percent{95} \CL exclusion limits in the \ac{MSSM} plane,
again in terms of the expected and observed exclusion limits,
as they are obtained from the \cls method in this analysis.
The $\pm1\sigma$ bands are shown as colored dashed lines.
\Figs{fig:analysis_exclusion_limit_MSSM_PLR} and \ref{fig:analysis_exclusion_limit_MSSM_Compare}
show the limits which have been computed using the \ac{PLR} method
and a comparison of the expected and observed limits from \cls and PLR for the MSSM grid in one plot.

Four exclusion limits from earlier experiments have been included in the plots.
At \LEP, the L3 experiment has set a lower bound on squark masses of about \GeV{100} from \ipb{450} of data in the MSSM framework.
It extends over the gluino mass range up to \GeV{550}, %
where the squark mass becomes smaller than the neutralino mass in their setting \cite{PLB2004L3Limits}. %
The other contours are from Run~I and Run~II of the \Tevatron experiments.
The contour from Run~I is governed by the \CDF result,
interpreting \ipb{84} of data in a general MSSM framework with \tanbeta{3} \cite{CDFGluinosSquarks2002}, %
looking for final states including three or more jets and missing transverse energy from the two \acp{LSP}.
The Run~II results from \Dzero are based on \ifb{2.1} of data,
setting limits on the squark und gluino masses 
in an mSUGRA interpretation with \tanbeta{3}, $A_0 = 0$ and $\mu < 0$ \cite{PLB2008D0Limits}.
The results from \CDF use \ifb{2} and \tanbeta{5}, $A_0 = 0$ and $\mu < 0$ \cite{PRL102CDFLimit}. %
The lower part of the boundary of the Run~II exclusion limits close to the diagonal 
is due to the fact that there is no mSUGRA solution for that part of parameter space in their setting.

The limit setting in the MSSM grid using the PLR method suffers from low Monte Carlo statistics
in the region of low signal efficiency %
and large cross sections.
This enhances fluctuations in the $p$-values,
which in the end lead to problems in the contour finding.
When using only the signal regions which contain at least $10$ raw signal Monte Carlo events,
corresponding to roughly one per mille of the available Monte Carlo statistics,
also PLR gives stable results, thus this approach is used here.
\Figs{fig:analysis_exclusion_limit_MSSM_PLR} and \ref{fig:analysis_exclusion_limit_MSSM_Compare}
cannot exclude the region below $m_\squark < \GeV{100}$ or $m_\gluino < \GeV{200}$,
but all of this region is excluded by the \cls method,
for which the effect was not observed, %
as can be seen in \Fig{fig:analysis_exclusion_limit_MSSM_CLs}.

\section{Summary and Conclusion}

The Supersymmetry analysis presented in this chapter has evaluated data taken with the \ATLAS detector in 2010,
corresponding to an integrated luminosity of $\Lint^\text{2010} = \unit[33.4]{pb^{-1}}$
of proton-proton collisions at \seventev.
It validates the official analysis,
which has been done in the \ATLAS Supersymmetry group,
and is optimized for supersymmetric final states with no leptons and high-energetic jets and large missing transverse energy.
The event counts in four signal regions are used
to set limits on mass parameters in two different supersymmetric settings, mSUGRA and MSSM.
Limits are derived with two different methods, \cls and PLR, which give consistent results.
The limits are expressed as exclusion contours at a \percent{95} confidence level
in \Figs{fig:analysis_exclusion_limit_mSUGRA_CLs} through \ref{fig:analysis_exclusion_limit_MSSM_Compare},
where for comparison results from other experiments and the official analyses of 2010 data from \ATLAS and \CMS are included.

The comparison to the exclusion contour from the official \ATLAS analysis in these plots
shows that the limits set in the analysis presented here are weaker.
The excluded mass ranges are smaller by roughly 10 to \percent{15} for the mSUGRA grid and \percent{20} for the MSSM grid.
This is expected and mainly due to the inclusion of pile-up effects,
which have not been considered in the official analysis.
The pile-up effects are taken into account here in a conservative approach
in terms of an additional uncertainty on all backgrounds
for which corresponding Monte Carlo samples are available.
A reweighting of the 2010 Monte Carlo according to the pile-up distribution in data is not possible
because no suitable Monte Carlo samples are available.
Furthermore, an updated estimate of the luminosity has been used here.
This has two consequences:
The uncertainty on the luminosity is smaller, thus improving the limit.
As the uncertainty on the luminosity is small compared to most of the other uncertainties,
this has only a negligible impact though.
On the other hand, the new luminosity estimate gives a \percent{4} smaller integrated luminosity, %
so that the estimation of the number of background events from Monte Carlo is smaller,
whereas the number of observed events is the same as in the official analysis.
This also leads to a weaker limit.

For the continuation of the analysis with 2011 data, a number of changes are due.
Besides using updated object definitions %
and calibrations, %
the most important changes are made necessary by the increase in the instantaneous luminosity,
which has two major consequences.
Firstly, higher rates will make triggers with too low thresholds unaffordable so that these will become prescaled.
In order not to lose sensitivity due to the prescales,
triggers with higher thresholds will have to be used instead.
As the analysis strategy currently assumes the trigger to be fully efficient,
the offline cuts need to follow the online requirements,
which means that harsher offline cuts need to be applied,
depending on the onset of the plateau for the new triggers.
\Sec{sec:results_trigger_performance_measurements_data_2011} discusses the updates in the trigger strategy for the zero-lepton analysis for 2011
and the efficiency of the triggers which have been employed in the analyses of 2011 data.
Reweighting of Monte Carlo events to account for the trigger efficiency
may become important for the zero-lepton analysis
when increasing the statistics by lowering the offline cuts into the turn-on region of the primary trigger becomes attractive,
or, put the other way round, when the trigger thresholds become so high,
that the offline selection can no longer follow without inducing a significant loss of sensitivity in the search.
Note that a steep turn-on curve will be of disadvantage here
because the systematic uncertainties due to the trigger efficiency become large close to the turn-on region.
Secondly, the increase in luminosity also entails an increase in the level of in-time pile-up,
as discussed \eg in the context of the \met model in \Sec{sec:results_met_model_predictions}.
Taking into account these pile-up effects in an appropriate way will then be necessary,
in order not to deteriorate the analysis performance.
For these higher levels of pile-up,
a reweighting of the Monte Carlo events
according to the distribution of the number of concurrent proton-proton interactions in the recorded events needs to be done,
and in fact Monte Carlo samples with a broad spectrum of additional pile-up vertices
have been produced and are in use for the on-going analyses of 2011 data in \ATLAS.
For the 2010 analysis,
in particular for the \ac{MSSM} grid,
the spacing of the grid points was relatively large for intermediate and high values of the squark and gluino masses.
This will also be improved in the 2011 Monte Carlo which has a finer granularity. %

In conclusion, it is interesting to see that the relatively small amount of data collected in 2010 in terms of integrated luminosity
already allows to set limits which are competitive with the existing limits from earlier experiments and often even go far beyond these,
although these are based on a many times larger integrated luminosity but at a lower center-of-mass energy.
Considering the fact that in 2011 already more than a hundred times more data has been collected by \ATLAS,
this demonstrates the impact that the physics results from the LHC will have.

\appendix
\chapter{Appendix}

\section{Trigger Emulation}
\label{sec:tdaq_explain_trigger_emulation}

Writing code that is able to reproduce the decision of the online trigger system has a number of benefits and use cases outlined below.
This kind of simulation of the trigger behavior will be called trigger emulation in the following,
to make clear the distinction from the official trigger simulation that it used to simulate the trigger decision in Monte Carlo.
The main difference between the trigger emulation
which is used in this thesis for different purposes
and the trigger simulation
is that the trigger emulation only emulates the hypothesis algorithms,
but it does not emulate the feature-extraction algorithms.
This would be far more complicated,
because it would require the reconstruction of physics objects from low-level detector data.
Instead, the emulation uses the trigger objects reconstructed online by the feature-extraction algorithms of the trigger system.
This brings some limitations because it is not possible to emulate trigger decisions in events
for which the feature-extraction algorithms, which construct the required trigger objects,
have not been run online so that these objects are not available.
The official trigger simulation used in Monte Carlo covers the full trigger system,
and thus also the feature-extraction algorithms.
It does not only make use of the same code base which is also run online in the \ac{HLT},
but for Level~1 also needs to simulate the trigger algorithms,
which for performance reasions are implemented in hardware.
This means that there may be more or less subtle differences
between the performance of the actual trigger and its counterpart in Monte Carlo.

Although it only emulates a part of the trigger system,
the trigger emulation can still be of use in a number of cases.
First of all, it helps to understand in detail,
how and based on what quantities the decision of the trigger is made.
In particular, a validation of the kind which will be presented below would show
whether the trigger emulation correctly captures all relevant features of the trigger hypothesis algorithms.
The main use with respect to studies of the trigger performance
is that the trigger emulation allows to study trigger definitions
which do not exist in the online trigger menu,
within the limitations \revised{outlined} above.

What is always possible within these limitations
is to emulate triggers with higher thresholds or additional cuts on existing trigger objects.
The main pitfall which has to be taken care of is the principle of early rejection,
which means that if none of the chains within a certain slice, \eg the jet slice,
has issued an accept for a given event
at one of the trigger levels below the highest (Level~1 and Level~2 of the \ATLAS trigger system),
the \revised{feature extraction at} the subsequent levels will not be executed.
This means that the corresponding trigger objects will not be reconstructed.
The trigger emulation will then always reject this event,
although, if the feature-extraction had been run,
they might have created objects which would have fulfilled the trigger selection.
In this case, the decision of the trigger emulation gives the wrong result for this event.
This will in particular cause problems for studies which select event samples based on an orthogonal trigger,
because it is not guaranteed that the trigger objects needed for the trigger performance study are available at all trigger levels,
but only those belonging to the orthogonal trigger which has accepted all of the events.
This means that for this type of trigger,
the feature-extraction must have been executed at all trigger levels.
On the other hand, by selecting a sample of events with the same type of trigger that is under study,
but using a lower threshold in the selection,
it can be ensured that the needed objects are available on all levels.
A special case are the \met triggers,
for which the feature-extraction algorithms are run for all events at all trigger levels
such that the emulation of these triggers in possible in all events that are recorded.

\subsection{Implementation}
In this thesis, the computation of trigger efficiencies is in general based on the emulation of triggers.
Therefore, a validation of the emulation is advisible.
Its results are presented in this \revised{subsection.}
First, a description is given of which triggers are emulated and how this is done,
followed then by plots of the validation results,
which show full agreement between the ``official'' trigger decision stored in the event record and the emulated trigger decision.

The triggers which are studied in this thesis,
and thus have been implemented in the emulation code,
are \met triggers, jet triggers, multi-jet triggers, and combined \jetmet triggers.
The emulation will be briefly sketched in the following.
The \met triggers, being based on an event-global quantity, are the easiest type of trigger to emulate.
The measurements of \met at Level~1, Level~2 and Event Filter can be compared directly against the respective thresholds,
where it has to be taken into account
whether the muon contributions to \met at Level~2 and Event Filter are included or not.
In its current implementation at the time of writing,
the \met computation in the \ac{HLT} does not make use of the muon correction (cf. \Sec{sec:tdaq_met}).
Cuts can be applied at all levels for the full chain \met triggers,
or only at Event Filter for the \met part of the jet-seeded combined \jetmet triggers.
The latter type is referred to as \trigger{EF_EFonly_xe25_noMu} (for a threshold of \GeV{25})
in this thesis (cf. \Sec{sec:analyis_description_cutflow}).
An event will pass Level~1 of the \met trigger chains,
regardless of the configured thresholds,
if the flag signaling an overflow in the computation of \mex or \mey at Level~1 is \revised{raised},
and \revised{also pass} Level~2 if an overflow of \met or \sumet at Level~1 occured,
because Level~2 does not recompute \met, but uses the Level~1 result (cf. \Sec{sec:tdaq_met}).

The emulation of jet triggers is more sophisticated (cf. \Sec{sec:tdaq_jet}),
because there can be an arbitrary number of jet objects at each trigger level,
and the trigger objects constructed by the feature-extraction algorithms have to be matched between the different trigger levels\footnote{
  For the same reason, the computation of jet trigger efficiencies is technically more involved
  than the computation of \met trigger efficiencies, cf. \Sec{sec:results_trigger_performance_measurements_introduction}.
}.
This matching associates Level~1 \acp{RoI}, Level~2 proto-jets and jets reconstructed by the Event Filter,
yielding a jet object which spans all three levels.
Only \acp{RoI} within $|\eta| < 3.2$ are considered.
A jet trigger only accepts an event if the same jet object exceeds the thresholds at all required trigger levels.
If one trigger jet exceeds the trigger threshold at Level~1, but not at Level~2,
and another trigger jet vice versa,
this is not sufficient to issue a trigger signal for this event.
Different types of matching have been implemented.
The matching based on \ac{RoI} words makes use of a unique identifier assigned to every jet \acp{RoI} at Level~1,
which is copied to jet objects seeded by this \ac{RoI} and allows an unambiguous matching.
Problems may be encountered if the \ac{RoI} words are missing in the data-format being used,
in which case a matching based on $\Delta R$ can be employed instead,
or if several jet algorithms are run in parallel in the \ac{HLT},
so that the same jet appears in several variants in the trigger jet collection.
To resolve this, additional flags are necessary which provide information about which jet has been constructed by which algorithm.
In the end, there has never been \revised{a} need for matching of jet trigger objects between Level~2 and Event Filter
because in 2010 jet trigger objects reconstructed at Event Filter level does not affect the trigger decision,
and in 2011, after the introduction of the full-scan algorithm at Event Filter \revised{level},
simply no matching is done between Level~2 and Event Filter.
Based on the existing implementation of jet and \met triggers,
the emulation of multi-jet and \jetmet triggers is straightforward.
The \jetmet triggers are a simple logical AND of the corresponding jet and \met triggers,
although the thresholds may differ from the single object trigger chains,
and there is the special type of jet-seeded \jetmet triggers in 2010,
which does not use cuts on \met at Level~1 and Level~2.
The multi-jet triggers which are vital for the multi-jet Supersymmetry searches,
but have not been employed for the zero-lepton analysis presented in this thesis,
extend on the jet triggers by requiring more than one matched jet at each level satisfying the respective threshold ladder.
In addition to the above,
on data the trigger emulation always checks
that the event carries a valid bunch group code, %
which roughly speaking means that proton-proton collisions are expected in this bunch crossing
and corresponds to the flags BGRP0 and BGRP1 being set at Level~1,
a requirement which should be included in the definition of any trigger on non-empty bunch crossings.

\subsection{Validation}

The validation of the trigger emulation is done on both Monte Carlo and data.
If the emulation is correct, the main difference between Monte Carlo and real data comes from prescales.
The fact that in Monte Carlo all prescales are set to \revised{unity} at all trigger levels
makes a full validation possible,
where no false positives (type I errors) and no false negatives (type II errors) should occur.
A falsely positive decision means that the trigger emulation yields a positive decision for an event
where the actual trigger decision read from the event record is negative.
In data, this may either be due to an error in the emulation or, for prescaled triggers, due to prescales.
Although the raw trigger decision before applying the prescales is also available in the event record,
this cannot be used for validating the full trigger chain for triggers which are prescaled at Level~1 or Level~2,
because if the prescales lead to a veto of an otherwise positive trigger decision at one of the lower levels,
the subsequent levels are not executed and thus no trigger decision is available for these levels.
This ambiguity makes the validation on data less informative.
A falsely negative decision means that the trigger emulation yields a negative decision for an event
where the actual trigger decision read from the event record is positive.
This would always indicate a problem in the emulation.

\begin{figure}
  \centering
  \incgraphics{width=\widthwideplot}{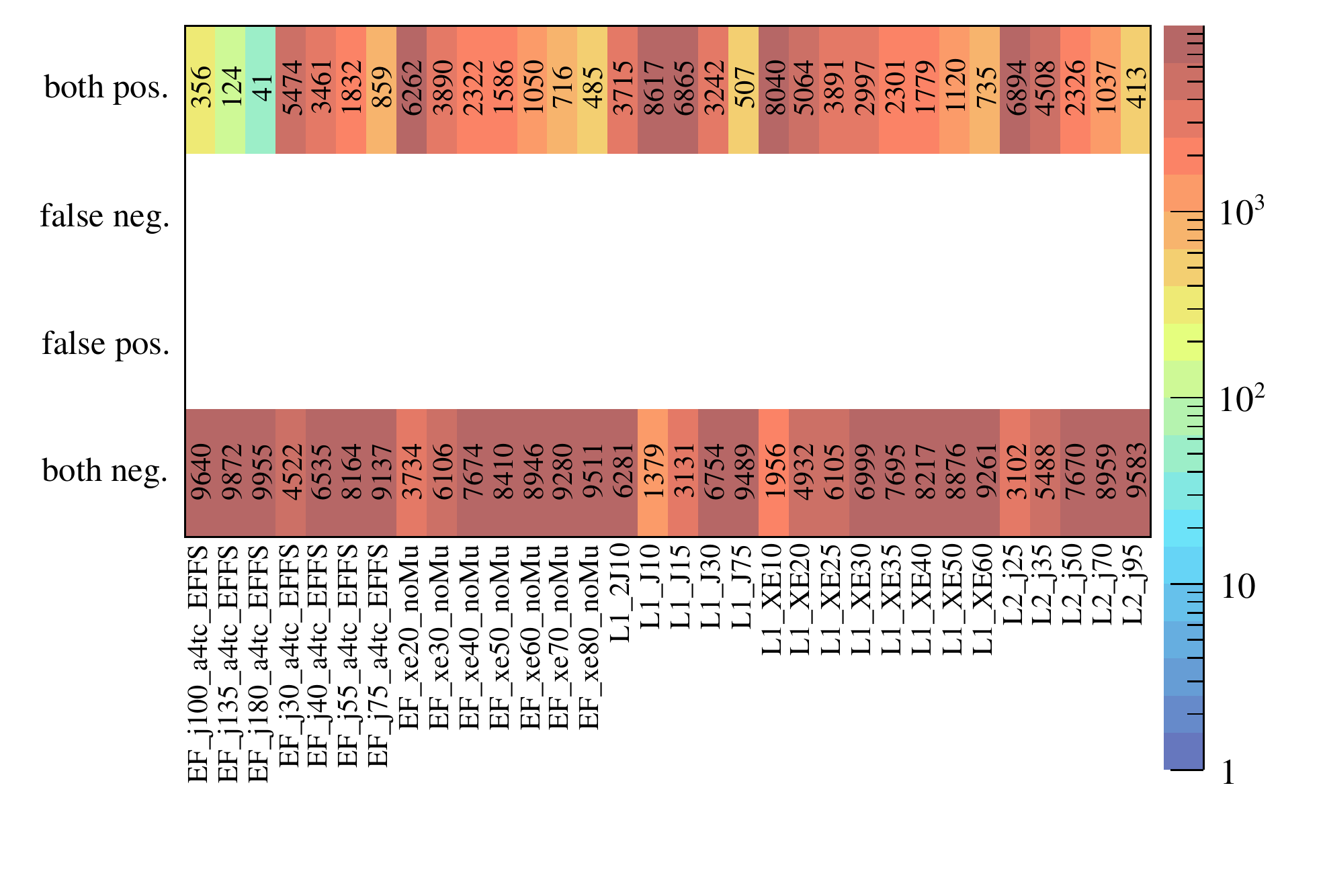}
  \caption{
    Validation of the trigger emulation in Monte Carlo from 2011 (\tagname{mc10}),
    using an event sample with $Z\to\nu\nu$ plus additional jets.
    The four bins on the vertical axis give
    the number of false negatives and positives
    and the number of events for which the emulation and stored trigger decision are both positive or both negative,
    \ie the emulation result is correct.
  }
  \label{fig:results_trigger_performance_validation_MC2011}
\end{figure}

The validation results of the trigger emulation on Monte Carlo (\tagname{mc10}) are shown in \Fig{fig:results_trigger_performance_validation_MC2011}.
On the horizontal axis, all tested trigger chains are listed with their official names.
They include jet triggers with the full-scan algorithm at Event Filter introduced in 2011 (prefix \trigger{EF_j})
and \met triggers (prefix \trigger{EF_xe}).
Some results for Level~1 and Level~2 are also shown.
The vertical axis has four bins,
which give the four possible combinations of true and false positive and negative results of the emulated trigger decision.
The event sample used for this plot has MC~ID 107712 and contains $Z\to\nu\nu$ events plus additional jets.
No discrepancy between the emulation and the actual trigger decision can be seen,
all bins with false decisions are empty.
(No number is printed for bins with zero entries.)

\begin{figure}
  \centering
  \incgraphics{width=\widthwideplot}{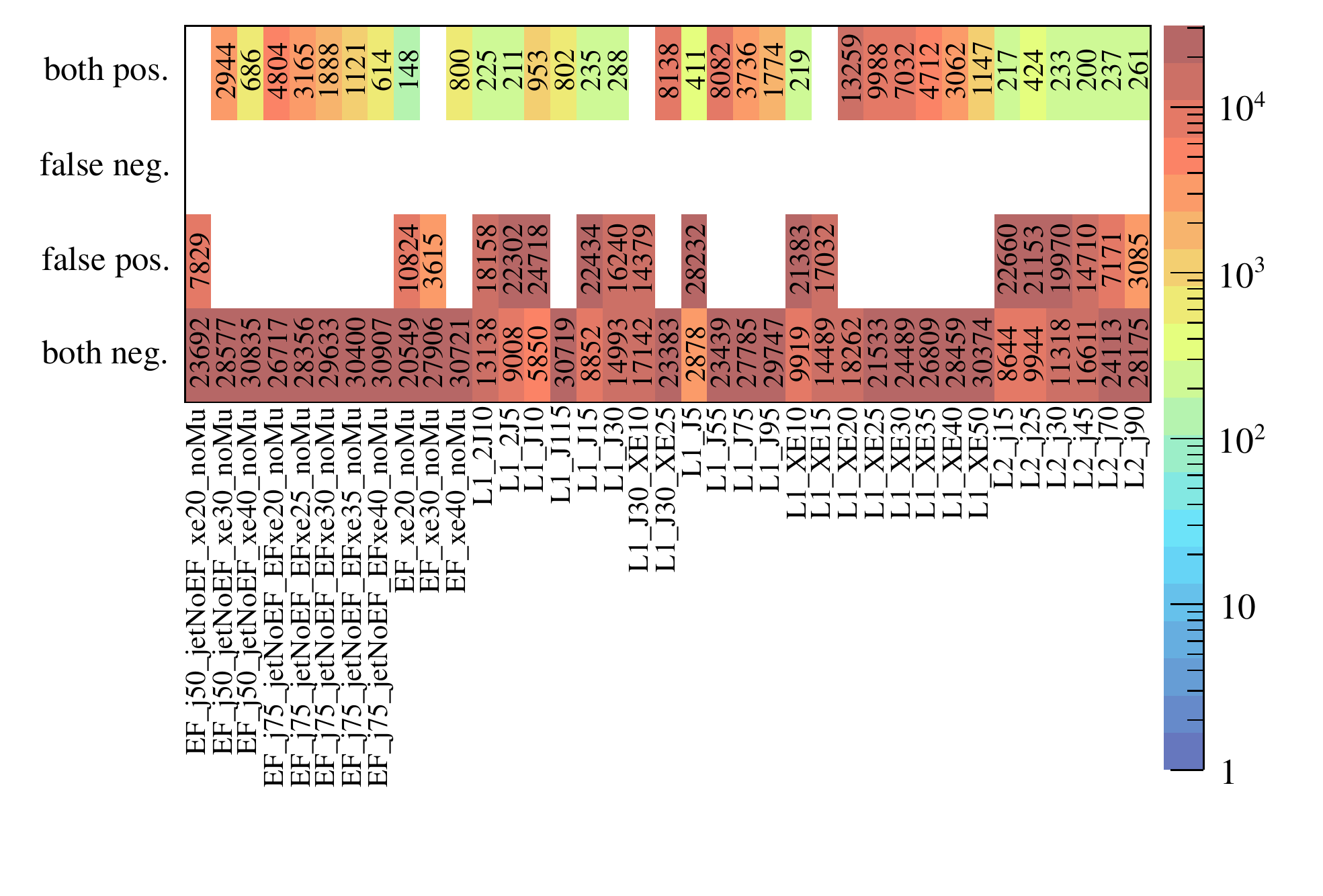}
  \caption{
    Validation of the trigger emulation in data from a run which was taken in 2010.
    The four bins on the vertical axis give
    the number of false negatives and positives
    and the number of events for which the emulation and stored trigger decision are both positive or both negative.
  }
  \label{fig:results_trigger_performance_validation_data2010}
\end{figure}

The plot in \Fig{fig:results_trigger_performance_validation_data2010} is structured \revised{analogously}, %
but this time showing the results of the validation of the trigger emulation on data from 2010.
A number of important differences with respect to \Fig{fig:results_trigger_performance_validation_MC2011} can be seen:
No Event Filter jet chains are listed in this plot
because the Event Filter has not been used to reject events in the jet slice in 2010.
It was therefore sufficient to simulate the jet chains up to Level~2.
Several \jetmet triggers are included in this plot,
whereas no \jetmet triggers are listed in the previous plot for 2011.
The validation for the updated \jetmet triggers has not been implemented for 2011,
because they are a simple logical AND of \met and jet triggers and considered unproblematic.
A striking difference are the many false positives in this plot,
which are, of course, coming from the prescales as explained above.
The false positives only occur for the lowest thresholds within the groups of different types of triggers,
because these are the ones which are prescaled first\footnote{
  Prescales are usually applied at the lowest possible trigger level.
  It has been verified that, for a given trigger chain which has prescales applied at Level~1 or 2 only,
  the false positives vanish if that respective trigger from this chain is used as sample trigger,
  which is the trigger at the highest level of this chain with prescales other than one.
  Another possibility to avoid false positives would be to not count events
  in which the execution of a trigger chain was stopped due to prescales,
  but the trigger decision before prescales is not included in the %
  data format used as input for the validation.
}.
The same check has also been done using the whole data from period 2010~H,
showing agreement between emulation and the actual trigger decision,
with the usual exceptions of false positives due to prescales,
for a large number (about $6\ten{8}$ in total) of tested trigger decisions. %

As was said at the beginning of this section,
the same code that has been used to produce the validation plots
is used for the computation of trigger efficiencies which are presented in this thesis.
The emulation of the trigger decision in the computation of the efficiencies has the advantage
that the efficiencies can be computed
independently \revised{of} changes \revised{in} the trigger menu which \revised{was} run online.
Moreover, efficiencies for newly introduced trigger chains can be computed on data
which was taken before the new chains were defined
to maximize the available statistics,
bearing in mind the above mentioned limitations of the emulation.
Finally, note that the trigger emulation must never be used for sample triggers,
because events that have been rejected online cannot be recovered by emulating the sample trigger,
at least not without the risk of introducing a bias from the collection of triggers which constitute the trigger menu:
If a trigger used as \revised{a} sample trigger were emulated,
the composition of the resulting event sample would depend on the triggers which have been running in parallel,
rather than the sample trigger itself,
because only those events have been recorded and can be recovered which have fired another trigger,
and therefore they represent only a particular (biased) subsample of all possible events.

\section{Production of Trigger Rate Plots from COOL}
\label{sec:results_trigger_rates_describe_COOL_plots}
All data needed to produce plots of the trigger rates can be retrieved from the \COOL da\-ta\-base. %
In this database, the data is organized in folders and is usually indexed by run and luminosity block.
It can be accessed by specifying a range of \ac{IOV} timestamps.
In general, an \ac{IOV} timestamp is a 63-bit unsigned integer used as unique identifier,
obtained by summing up the run number multiplied by $2^{32}$ and the event number.
Here however, for most folders the \ac{IOV} timestamp consists of the run and luminosity-block number. %
Each folder may contain several channels,
with each channel having the same structure \ie the same fields and type of data.

The folder \tagname{TRIGGER/LVL1/Menu}, for example, has four fields:
the IOV timestamp of the run, the channel number, the item name and a version tag. %
Each channel corresponds to one of the 256 possible triggers (trigger items) of the \ac{L1} Trigger.
This folder is used to map the channel identifier to the name of the L1 trigger item,
which is needed for the look-ups in other folders.
The rates at Level~1 are computed from the ratio of two trigger counters which are filled by the \ac{CTP}.
The denominator is the number of LHC turns, $n_\text{LHC turns}$,
within the given luminosity block from the folder \tagname{TRIGGER/LVL1/CTPCORELBDATA}.
This is the maximum number of times a Level-1 trigger could have fired.
The enumerator is the \ac{L1} counter $n_\text{L1}^A$ for the respective trigger $A$,
which counts how often this \ac{L1} trigger actually fires.
Three different counters are available for each trigger,
counting the trigger signals before and after application of the prescales
(but before any dead time or busy veto or the trigger-mask is applied)
and the final number of L1 accepts.
The L1 rate $\nu_\text{L1}^A$ for a trigger $A$ is thus given by
\begin{equation}
  \nu_\text{L1}^A = \frac{n_\text{L1}^A} {n_\text{LHC turns}} \cdot \nu_\text{orbit},
\end{equation}
where $\nu_\text{orbit} = \unit[11.245]{kHz}$ %
is the revolution frequency of the LHC,
\ie the number of times each bunch travels around the LHC ring in one second (cf. \Tab{tab:experimental_setup_lhc_parameters}).
The impact of the reduced center-of-mass energy on this value is calculated in \Sec{sec:appendix_lhc_parameters_revolution_frequency}.

The prescales for Level~1 are taken from the \tagname{TRIGGER/LVL1/Prescales} folder,
which holds the prescale values per \ac{L1} trigger item.
The raw rate of a L1 trigger before prescaling can either be computed
by using the corresponding counter before prescale,
or by multiplying the counter with L1 accepts after prescale by the prescale value.
Using the L1 accept counter will include the dead time in the rate,
which will result in a lower value,
but for normal data taking this difference has been found to only be of the order of \revised{one to two percent.} %

The \ac{HLT} menu is needed to obtain the association between the different chains in the computation of the aggregate prescale for L2 and EF~chains
and for the mapping between trigger name and an integer used as index which in case of the HLT is called the chain counter.
It is stored in the folder \tagname{TRIGGER/HLT/Menu},
from which the chain name, the chain counter and the name of the lower chain are the most important information.
The lower chain is, for L2~chains and EF~chains,
the name of the L1~item or L2~chain,
which seeds the L2~chain or EF~chain, respectively.
The prescales for the HLT chains themselves are stored in the folder \tagname{TRIGGER/HLT/Prescales},
which is indexed by the chain counter. %
The HLT rates are stored in the folders \tagname{TRIGGER/LUMI/LVL2COUNTERS} and \tagname{EFCOUNTERS}.
As they are stored as \acfp{BLOB} they need to be decoded with the \tagname{TrigStatsFromCool} module from \Athena, %
which directly returns the rates and counts.

The instantaneous luminosity per run and luminosity block is stored in two separate folders,
one of which contains luminosity information from online sources (\tagname{TRIGGER/{\allowbreak}LUMI/{\allowbreak}LBLESTONL}),
the other from offline sources (\tagname{LBLESTOFL}).
Each folder has several channels,
corresponding to different sources or algorithms for the luminosity estimate.
For the trigger rate plots always the value from channel $0$ is used,
which is the default instantaneous luminosity in units of $\unit[10^{30}]{Hz / cm^2}$,
taken from the currently preferred luminosity algorithm at any given time.

The average number of interactions per bunch crossing \avgmu can also be obtained from these two folders.
It is computed from the instantaneous luminosity \Linst according to
\begin{equation}
  \avgmu = \frac{\Linst \cdot \sigma_\text{MB}} {\nu_\text{orbit} \cdot n_\text{coll}},
\end{equation}
where $\sigma_\text{MB}$ is the inclusive minimum-bias cross section
which is assumed to be $\sigma_\text{MB} = \unit[71.5]{mb}$ for the LHC at the current center-of-mass energy of \TeV{7}.
$n_\text{coll}$ is the number of colliding bunches in the respective run.
In this computation, the value of \avgmu is averaged over all bunch crossings, %
\ie it assumes that the size and spread of the bunches and thus the luminosity is the same for all bunches.

It should be mentioned that the rates in the \COOL database are known to be incorrect in some rare cases,
due to errors which may occur when collecting the input for writing the rates to \COOL.
The trigger rates measured online are not only stored in the \COOL database, but also in another archive.
Plots of the trigger rates on a daily basis can be accessed
using an online tool called \propername{Web Trigger Rate Presenter} \cite{URL_WTRPA}.
The rates obtained from \COOL have been cross-checked against these plots in several random checks and have been found to agree.
In any case, in the trigger rate plots created from \COOL not single runs are used but whole periods,
and thus the impact from these rare errors is not expected to render the results unusable.
\revised{The trigger rates are available from the COOL database starting with data-taking period 2010~G.}

\section{\texorpdfstring{Implementation of the \cls and PLR Methods} {Implementation of the CLs and PLR Methods}}

\label{sec:analysis_implementation}

This section describes technical details of the implementation of the \cls and the PLR methods,
which are used to evaluate the results of the analysis in \Sec{sec:analysis_susysearch}.
The mathematical background of the methods is presented in \Sec{sec:analysis_methods_description}.
For all computations version \tagname{5.30/00} of \ROOT is used.

\subsection{\texorpdfstring{\cls Method} {CLs Method}}
\label{sec:analysis_implementation_cls}

The computation of exclusion limits with the \cls method
makes use of the \propername{TLimit} class,
which is part of the \propername{ROOT} framework \cite{ROOTUsersGuide}.
Internally, a \acf{MC} method is used to perform pseudoexperiments, %
fluctuating the signal and background expectations within the systematic errors
and computing the probability of having that particular outcome.

The prototype of the relevant member function %
of the \propername{TLimit} class is 
\code{TConfidenceLevel* ComputeLimit(TLi{\allowbreak}\-mit\-DataSource* data}[, options]\code{)}. %
The most important option is the number of \ac{MC} steps or pseudoexperiments.
The \code{TLimit\-Data\-Source} object holds, for each channel,
the hypothesized number of signal and background events as well as the number of events observed in data,
plus the relative systematic uncertainties,
which are all modeled by Gaussian distributions.
Correlations between systematic uncertainties can be specified by assigning the same name to a set of uncertainties,
which then in each \ac{MC} step share a common random number drawn from a standard normal distribution,
\ie with a mean of zero and a variance of one,
multiplied by the value of the relative uncertainty.
The \code{TConfidenceLevel} object stores the results of the \ac{MC} pseudoexperiments and allows to retrieve several derived quantities,
most importantly \cls,
but also \clsb and the expected values of \cls and \clsb under the background-only hypothesis.

The methods of \code{TConfidenceLevel} which return the observed \cls and \clsb values
provide an option called \code{use\_sMC} that is not documented.
This option is deactivated (\ie set to \code{false}) by default.
Nonetheless, in the way the \code{TLimit} class is used here,
it has been found that the results are only stable
if this option is set to \code{true}.
This is also in agreement with comparisons to a private implementation %
of the \cls method,
which does not account for uncertainties.
It is therefore assumed that this option needs to be set to \code{true}.
The methods of \code{TConfidenceLevel} which return the expected \cls and \clsb values,
such as \code{TConfidenceLevel::GetExpectedCLs\_b},
which returns the expected \cls under the back\-ground-only hypothesis,
do not have the \code{use\_sMC} option,
but accept an optional argument called \code{sigma}.
This option is also not documented, but is assumed to return the expected \cls and \clsb values
when letting the expected background fluctuate up or down by a given number of standard deviations
according to the distribution of the test statistic. %
This has also been validated by testing against a private implementation. %

\subsection{PLR Method}
\label{sec:analysis_implementation_plr}
The implementation of the profile likelihood method is done using the \code{ProfileLike\-lihood\-Calculator} class
from the \propername{RooStats} framework \cite{Moneta2010}.
It builds upon \propername{RooFit} and \propername{ROOT},
and itself relies on \propername{Minuit} \cite{Minuit1975} as minimizer.
The \code{ProfileLikelihoodCalculator} is fed with a \code{RooDataSet} object holding the observed number of events
and a \code{Model\-Config} object describing the data model in terms of
the nuisance parameters, the parameter of interest and the \ac{PDF}.
Again, as the signal is only expected to increase the number of observed events,
the signal strength $\mu$ as parameter in the likelihood fit is limited by zero from below.

The model \ac{PDF} is basically a product of a Poisson distribution %
as shown in \Eq{eq:PLR_likelihood_Poisson} for the special case
of only one search channel, $n=1$, and several standard normal distributions for the nuisance parameters:
\begin{equation}
  L = \Poisson{n_\text{obs}}{\mu f_s(s, \vec\nu) + f_b(b, \vec\nu)} \cdot \prod_{u \in {\cal N}} \Normal{\nu_u}.
\end{equation}
In this equation,
${\cal N}$ is the set of statistical and systematic uncertainties listed below,
which correspond to the uncertainties given in \Tab{tab:analysis_overview_systematic_uncertainties}.
All of these are modeled by a standard normal distribution.
\begin{align}
  {\cal N}_c &= \{\text{luminosity, JER, JES, pile-up}\} \\
  {\cal N}_s &= \{\text{statistical (signal), theoretical (signal)}\} \\
  {\cal N}_b &= \{\text{statistical (background), SUSY (background)}\} \\ %
  {\cal N}   &= {\cal N}_c \cup {\cal N}_s \cup {\cal N}_b
\end{align}
Again, ``SUSY'' stands for the additional uncertainties on the background normalization.
The theoretical uncertainties for the signal concern the signal cross section.
${\cal N}_c$ contains the correlated uncertainties,
the other two sets the uncorrelated uncertainties for signal and background.
The expected number of signal and background events is modified by the nuisance parameters according to
\begin{align}
  \label{eq:PLR_fs}
  f_s(s, \vec\nu) &= s \cdot (1 + \sum_{u \in {\cal N}_c} \sigma_u^s \nu_u + \sum_{u \in {\cal N}_s} \sigma_u^s \nu_u) \\
  \label{eq:PLR_fb}
  f_b(b, \vec\nu) &= b \cdot (1 + \sum_{u \in {\cal N}_c} \sigma_u^b \nu_u + \sum_{u \in {\cal N}_b} \sigma_u^b \nu_u)
\end{align}
Here, $\sigma_u^s$ and $\sigma_u^b$ specify the relative uncertainty for signal and background, respectively.
By using standard normal distributions, which are then multiplied by the relative uncertainties,
it is possible to implement uncertainties that are correlated,
but of different size for signal and background.
In the above equations, the \ac{JES} uncertainty is written as a symmetric uncertainty.
The actual implementation allows for an asymmetric \ac{JES} uncertainty,
and the addend $\sigma_\text{JES}^s \nu_\text{JES}$ in \Eqs{eq:PLR_fs} needs to be replaced by
\begin{equation}
  \sigma_\text{JES}^s \nu_\text{JES} \to
  \begin{cases}
    \sigma_\text{JES, up}^s  \nu_\text{JES}   & \text{ for } \nu_\text{JES} \geq 0, \\
    \sigma_\text{JES, down}^s \nu_\text{JES}  & \text{ otherwise},
  \end{cases}
\end{equation}
and analogously for the background in \Eq{eq:PLR_fb}. %
This means that an upward fluctuation is assumed to be equally likely as a downward fluctuation,
but the size of the uncertainty can be scaled independently.

The \code{ProfileLikelihoodCalculator::GetInterval} method returns a \code{LikelihoodIn\-ter\-val} object,
from which, for a given confidence level, lower and upper limits on the parameter of interest $\mu$ can be queried, %
which are computed using the approximation of $-2\ln \lambda(\mu)$ as a $\chi^2$~distribution introduced above. %
Just opposite to the procedure described above for the \cls method,
where a confidence interval was to be derived from the binary decision
whether a parameter value is excluded or not,
here, a $p$-value for the signal-plus-background hypothesis is sought.
To compute this $p$-value,
the value of $-2\ln \lambda(\mu)$ for the nominal signal strength $\mu = 1$ is converted
into a $p$-value using the above approximation.
If this value is smaller than~$\alpha$,
the signal-plus-background hypothesis is reported to be excluded at a certainty level of $1-\alpha$.
The expected limits from the \ac{PLR} method under the assumption of the background-only hypothesis \hypob
are obtained by setting the number of observed events equal to the background expectation from Monte Carlo.
The expected $p$-value for this hypothesis is then computed in the same way as described above.

\section{ \texorpdfstring{Computation of \LHC Parameters} {Computation of LHC Parameters} }
\label{sec:appendix_lhc_parameters}

\subsection{Required Magnetic Field Strength}
\label{sec:appendix_lhc_parameters_magnetic_field_strength}

The following calculation shows that the magnetic field
needed to keep the protons on a track with the radius of the LHC tunnel
is proportional to the beam energy with a proportionality constant of approximately \unit[1]{Tesla/TeV}.
The calculation is very simple and can be done classically by requiring the Lorentz force $\vec F_L$ and centripetal force $\vec F_C$ to be equal:
\begin{align}
  \vec F_L = q\vec v \times \vec B
  \mustbe - m \vec \omega \times \vec r \times \vec \omega %
  = \vec F_C.
\end{align}
$\vec v$ is the velocity, $\vec \omega$ the vectorial angular velocity, $\vec r$ the position vector and
$q$ the charge of the protons in a homogeneous magnetic field $\vec B$.
In the high-energy approximation with $E\approx |\vec p| c$ and assuming $\vec B$ to be perpendicular to $\vec v$,
it follows that
\begin{align}
  \folgt B &= \frac{E}{qrc} \\
    &= \frac{ E \cdot 1.6\ten{-7} \unit[]{J/TeV} }{\unit[2.0\ten{-7}]{m^2/s}} \\
  \folgt B\unit[]{[T]} &= 0.8\,E\unit[]{[TeV]}.
  \label{eq:appendix_bending_magnets}
\end{align}

\subsection{Impact of the Center-of-Mass Energy on the Revolution Frequency}
\label{sec:appendix_lhc_parameters_revolution_frequency}

From \Tab{tab:experimental_setup_lhc_parameters},
it can be seen that the revolution frequency of the proton bunches
around the LHC ring is \unit[11.245]{kHz} at an energy $E$ of \TeV{7} per beam.
At the time of writing, the energy is only \TeV{3.5} per beam (cf. \Sec{sec:experimental_setup_LHC}),
only half the value of the design energy.
As a consequence, the protons are slower and the revolution frequency is lower.
The frequency of \unit[11.245]{kHz} is an input to the computation of the trigger rates (cf. \Sec{sec:results_trigger_rates_describe_COOL_plots}),
thus the question is what the impact of the lower proton energy on this computation is.

The relativistic factor of the protons at $\sqrt{s} = \TeV{14}$ is
\begin{equation}
  \gamma_{14} = E/m_p = 7461, %
\end{equation}
and the relative velocity of the protons with respect to the speed of light $c$ is thus
\begin{equation}
  \frac {v_{14}}c = \sqrt{1 - 1/\gamma_{14}^2} = 1-9\ten{-9}. %
\end{equation}
For $\sqrt{s} = \TeV{7}$, one obtains
\begin{equation}
  \gamma_{7} = E/m_p = 3730,
\end{equation}
and the relative velocity of the protons with respect to the speed of light is
\begin{equation}
  \frac {v_{7}}c = \sqrt{1 - 1/\gamma_{7}^2} = 1-3.6\ten{-8}. %
\end{equation}
The absolute smallness of the change in the velocity makes clear, already at this step,
that reducing the center-of-mass energy to half its design value
has no significant impact on the revolution frequency.
Using \unit[11.245]{kHz} is therefore correct within the precision at which the trigger rates are stored and evaluated.
The exact value is \unit[11.2455]{kHz}, %
given by $C_\text{LHC} \, c / v_{7}$,
where $C_\text{LHC}$ is the circumference of the LHC ring.

\section{\texorpdfstring{Analytic Forms of \sumet and \met} {Analytic Forms of SumET and MET}}
\label{sec:appendix_analytic_forms}

In this section,
the mathematical proofs for the analytic forms
of the probability density functions
for \sumet and \met will be derived,
which have been used in the model presented in \Sec{sec:results_met_model}.

\subsection{\texorpdfstring{Distribution of \sumet for Several Concurrent Interactions} {Distribution of SumET for Several Concurrent Interactions}}
\label{sec:appendix_analytic_form_sumet}

The analytic form for \sumet is motivated by the fact that there are $\mu$ interactions taking place at the same time,
so that their energy depositions are attributed to the same bunch crossing and therefore add up.
Each of the interactions is assumed to be independent of the others and contribute an energy according to the same probability density function (PDF).
This PDF is assumed to be an exponential distribution with rate parameter $\lambda$, defined as
\begin{equation}
  f_{\exp}(x|\lambda) \definedas
  \begin{cases}
    \lambda \exp(-\lambda x) & \text{ for } x \geq 0, \\
    0                        & \text{ for } x < 0.
  \end{cases}
\end{equation}

\paragraph{Theorem.}

Let $S_n$ be the sum of $n > 0$ independent random variables $X_i$, $i \in \{1,\dots,n\}$,
\begin{equation}
  S_n \definedas \sum_{i=1}^{n} X_i,
\end{equation}
each of which is distributed according to $f_{X_i}(x|\lambda) = f_{\exp}(x|\lambda)$ with a common value for $\lambda$.
Then, the PDF $f_{S_n}$ of the random variable $S_n$ is
\begin{equation}
  f_{S_n}(x|\lambda) \definedas
  \begin{cases}
    \frac{ \lambda^n }{(n-1)!} \, x^{n-1} \exp(-\lambda x) & \text{ for } x \geq 0, \\
    0                                                      & \text{ for } x < 0.
  \end{cases}
  \label{eq:appendix_sumet_tobeproven}
\end{equation}

\paragraph{Proof. \textnormal{(By Mathematical Induction.)}}

\noindent Initial Step:
When $n = 1$, then $S_n = X_1$, and thus by definition
\begin{equation}
  f_{S_1}(x|\lambda) = f_{X_1}(x|\lambda) = f_{\exp}(x|\lambda).
\end{equation}
\Eq{eq:appendix_sumet_tobeproven} thus holds for $n = 1$.

\noindent Inductive Step:
The inductive assumption is that \Eq{eq:appendix_sumet_tobeproven} holds for a given $n\geq1$. %
It is to be shown that from this it follows that \Eq{eq:appendix_sumet_tobeproven} also holds for $n+1$.

In general, the addition rule for two independent random variables $U$ and $V$ with probability density functions $f_U$ and $f_V$ is 
\begin{equation}
  \begin{aligned}
    f_W(x) &= f_{U+V}(x) \\
           &= \left( f_{U} \ast f_{V} \right) (x) \definedas \int_{-\infty}^\infty f_U(y) \, f_V(x - y) \,\intd y,
    \label{eq:appendix_pdf_addition_rule}
  \end{aligned}
\end{equation}
where $f_W$ is the probability density function of the random variable $W = U+V$.
Applying the addition rule \eqref{eq:appendix_pdf_addition_rule} to
\begin{equation}
  S_{n+1} = \left( \sum_{i=1}^n X_i \right) + X_{n+1},
\end{equation}
it follows that %
\newcommand{\nolambda}{|\lambda}
\begin{align}
  f_{S_{n+1}}(x\nolambda) &= \left( f_{S_n} \ast f_{X_{n+1}} \right) (x\nolambda) \nonumber\\
    &= \int_{-\infty}^\infty f_{S_n}(y\nolambda) \, \ast f_{\exp} (x-y\nolambda)  \intd y \nonumber\\
    &= \int_0^x \frac{ \lambda^n }{(n-1)!} \, y^{n-1} \exp(-\lambda y)  \, \lambda \exp\left(-\lambda (x-y)\right) \intd y \nonumber\\
    &= \frac{ \lambda^{n+1} }{(n-1)!} \, \exp(-\lambda x) \int_0^x  y^{n-1}  \intd y \nonumber\\
    &= \frac{ \lambda^{n+1} }  {n!} \, {x^n} \exp(-\lambda x),
\end{align}
where the integration range in the third line is due to the probability density functions $f_{S_n}$ and $f_{\exp}$ being zero for negative arguments.

\qed\vspace*{1mm}

Under the model assumptions outlined above,
the distribution of \sumet in events with $\mu$ concurrent interactions
is therefore given by the probability density function $f_{S_\mu}$,
\ie in the notation of \Sec{sec:results_met_model},
\begin{equation}
  P\left(\sumet = x | \mu\right) \equiv f_{S_\mu}(x|\lambda).
\end{equation}

\subsection{\texorpdfstring{Distribution of \met from Resolution Effects} {Distribution of MET from Resolution Effects}}
\label{sec:appendix_analytic_form_met}

In the following, the analytic form for \met under the assumption of only fake \met from resolution effects will be derived.
The assumption of the presence of only this type of fake \met translates into
the distribution of the $x$- and $y$-component of the missing energy in the transverse plane, \mex and \mey,
following a Gaussian distribution of same width $\sigma_x = \sigma_y = \sigma$.
The distribution of \met then follows from the geometrical relation $\met = \sqrt{(\mex)^2 + (\mey)^2}$.
As was done above, the probability distribution will be derived generically and then applied to the physical question.

\paragraph{Theorem.}

Let $X$ and $Y$ be independent random variables following a Gaussian distribution with mean zero and width $\sigma^2$,
\begin{equation}
  f_X(x|\sigma) = f_Y(x|\sigma) = \frac{1}{\sqrt{2\pi\sigma^2}} \, %
    \exp\left(-\frac{x^2}{2\sigma^2} \right).
  \label{eq:appendix_met_gaussians_for_XY}
\end{equation}
The probability density function of the random variable $U = \sqrt{X^2 + Y^2}$ is then given by
\begin{equation}
  f_U(x|\sigma) \definedas
  \begin{cases}
    \frac{x}{\sigma^2} \, \exp\left(-\frac{x^2}{2\sigma^2} \right) & \text{ for } x \geq 0, \\
    0                                                              & \text{ for } x < 0.
  \end{cases}
  \label{eq:appendix_met_tobeproven}
\end{equation}

\paragraph{Proof.}

The general rule for transforming a bivariate probability density function $f_{X,Y}$ is
\begin{equation}
  f_{U,V}(u,v) = \sum_{i=1}^k f_{X,Y} \left( h_{1i}(u,v) , h_{2i}(u,v) \right) |J_i|,
  \label{eq:appendix_multivariate_transformation_rule}
\end{equation}
where $h_{1i}$ and $h_{2i}$ are the inverse functions of the variable transformation $U = g_1(X,Y)$, $V = g_2(X,Y)$
on the $i$-th set of a partition $A_i$ of the support of $f_{X,Y}$.
The partition is chosen such that the inverse of $g_{1,2}$ on each $A_i$ exists,
\ie $\left. g_1\right|_{A_i}$ and $\left. g_2\right|_{A_i}$ are injective functions,
mapping $A_i$ onto the support $B$ of $f_{U,V}$.
$J_i$ is the Jacobian determinant of the inverse transformation on $A_i$ \cite{CasellaBerger2001}.
The transformation chosen here is
\begin{equation}
  (X,Y) \mapsto (U,V) = (\sqrt{X^2+Y^2}, Y).
\end{equation}
As $(X,Y)$ and $(-X,Y)$ are mapped onto the same point by this transformation,
the partition 
\begin{equation}
  \begin{aligned}
    A_1 &= \{(X,Y)\in\setR^2|X<0\} \\
    A_2 &= \{(X,Y)\in\setR^2|X\geq0\}
  \end{aligned}
\end{equation}
of the support of $f_{X,Y}$ is used. %
Hence, on $A_1$,
\begin{align}
         &h_{11}(u,v) = -\sqrt{u^2-v^2} \\
         &h_{21}(u,v) = v \\
  \folgt &J_1 =
    \begin{vmatrix}
      \frac{2u}{2\sqrt{u^2-v^2}} & \frac{-2v}{2\sqrt{u^2-v^2}} \\ %
      0 & 1 %
    \end{vmatrix}
    = \frac{u}{\sqrt{u^2-v^2}}.
\end{align}
On $A_2$, evidently $h_{12} = -h_{11}$ and $h_{22} = h_{21}$,
thus $J_2 = -J_1$.
Using \Eq{eq:appendix_met_gaussians_for_XY} and the independence of $X$ and $Y$,
so that $f_{X,Y} = f_X f_Y$\footnote{
  Here, it is exploited that the means of $f_X(x|\sigma)$ and $f_Y(y|\sigma)$, $x_0$ and $y_0$, respectively,
  are zero by assumption, %
  such that in the enumerator of the exponential function in \Eq{eq:appendix_use_assumption_of_zero_mean}:
  $$(x-x_0)^2 + (y-y_0)^2 = x^2 + y^2 = u^2.$$
  The reason to assume a common width parameter $\sigma$ for both $f_X$ and $f_Y$ is obvious.
},
it follows from \Eq{eq:appendix_multivariate_transformation_rule} that
\begin{equation}
  f_{U,V}(u,v) = \frac{1}{2\pi\sigma^2} \frac{2u}{\sqrt{u^2-v^2}} \, \exp\left(-\frac{u^2}{2\sigma^2}\right).
  \label{eq:appendix_use_assumption_of_zero_mean}
\end{equation}
The desired probability density function of $U = \sqrt{X^2+Y^2}$ is obtained by marginalization,
\begin{align}
  f_{U}(u) &= \int_{\operatorname{supp}V} f_{U,V}(u,v) \intd v  \nonumber\\ 
    &= \frac{u}{\pi\sigma^2} \, \exp\left(-\frac{u^2}{2\sigma^2}\right) \int_{-u}^u \frac 1 {\sqrt{u^2-v^2}} \intd v \nonumber\\ %
    &= \frac u {\sigma^2} \, \exp\left(-\frac{u^2}{2\sigma^2}\right),
  \label{eq:appendix_met_final_eq_of_proof}
\end{align}
where in the second line $v \leq u$ has been used, and $f_U$ is defined on $u\geq 0$.

\qed\vspace*{1mm}

\Eq{eq:appendix_met_tobeproven} can be applied directly to state the distribution of \met in events with only fake \met,
for which, as stated above, the $x$ and $y$ component are assumed,
from the central limit theorem, to be given by a Gaussian distribution
because they are a sum over a large number of independent measurements of the energy deposited in the cells of the calorimeter.
The distribution of \met in the notation of \Sec{sec:results_met_model} is
\begin{equation}
  P(\met = x|\sigma) \equiv f_U(x|\sigma).
\end{equation}

\section{\texorpdfstring{Supplementary Plots: \met Model} {Supplementary Plots: MET Model}}

In the description of the model for the contribution of detector resolution effects to fake \met in \Sec{sec:results_met_model},
some of the plots are included in the main part of the thesis only for either 2010 or 2011.
To confirm that the conclusions from the data from these two data-taking periods are consistent,
here, the two corresponding plots for the respective other data-taking period are shown,
highlighting similarities and differences where applicable.
Moreover, additional plots provide more details.

\begin{figure}
  \centering
  \incgraphics{width=\widthsingleplot}{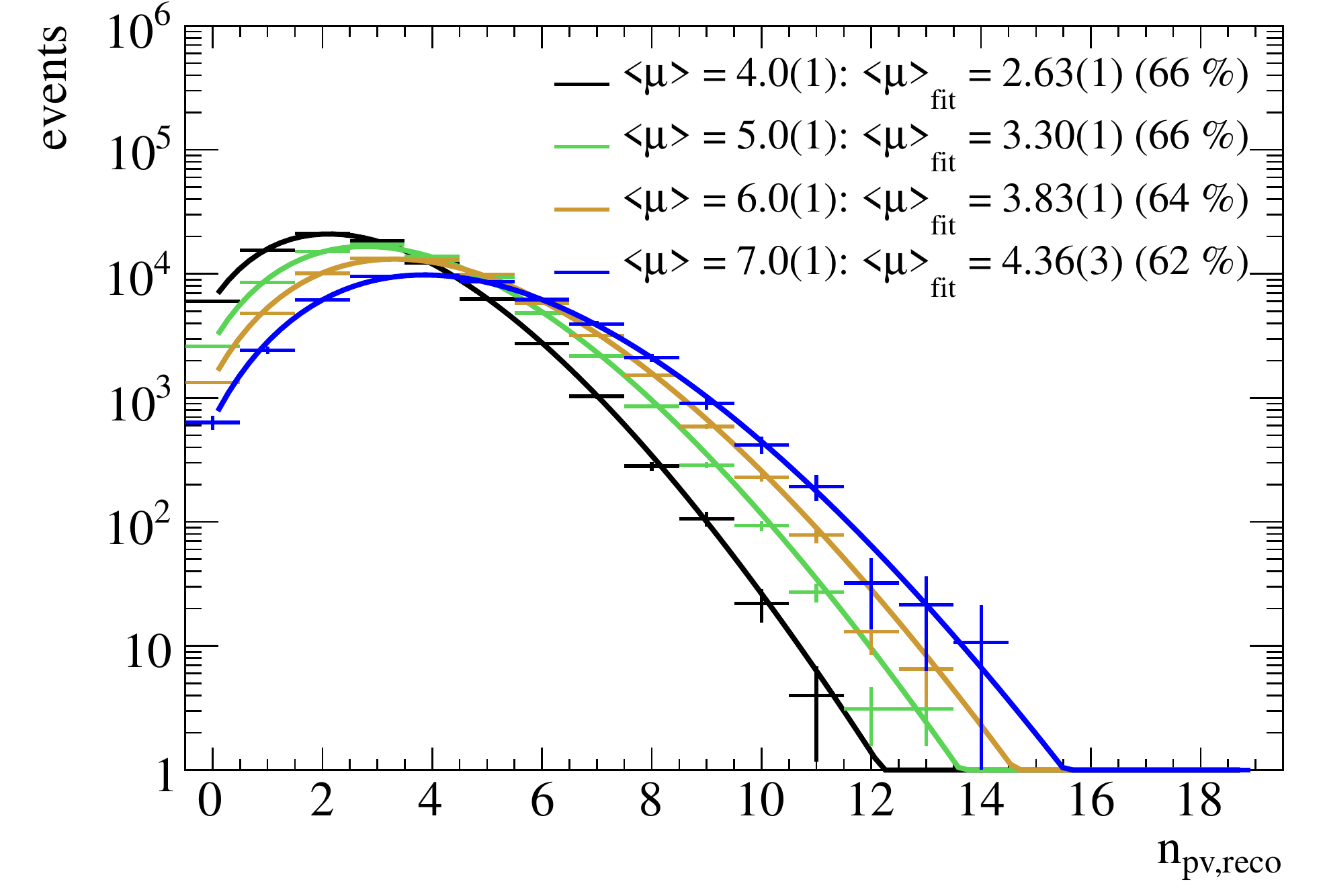}
  \caption{
    Distribution of the number of reconstructed primary vertices \nreco in data from 2011
    for four small event samples with a narrow distribution of \avgmu.
    The data points are fitted with a Poisson distribution (solid lines) with free rate parameter $\mu$
    and the fit result $\avgmu_\text{fit}$ is given in the legend
    together with the ratio of $\avgmu_\text{fit}$ over the nominal \avgmu expressed as a percentage value.
  }
  \label{fig:results_met_model_data_reco_PV_fit_2011}
\end{figure}

\Fig{fig:results_met_model_data_reco_PV_fit_2011} shows
the distribution of the number of reconstructed primary vertices \nreco
for four small event samples with a narrow distribution of \avgmu.
The spread of \avgmu is $0.1$ here.
This is the same plot as in \Fig{fig:results_met_model_data_reco_PV_fit_2010} but for 2011.
The data points are again fitted with a Poisson with free rate parameter $\mu$
and the fit result $\avgmu_\text{fit}$ is given in the legend,
together with the ratio of $\avgmu_\text{fit}$ and the nominal \avgmu taken from the \COOL database expressed as a percentage value in brackets.
The plot mainly differs in the values of \avgmu being higher in 2011 than in 2010.
Another interesting observation is the continuation of the trend
that with higher \avgmu the ratio of $\avgmu_\text{fit} / \avgmu$ gets smaller,
which is an indication for the vertex reconstruction performance becoming worse for busier events,
where the probability of accidentally merging vertices becomes higher.

\begin{figure}
  \centering
  \incgraphics{width=\widthsingleplot}{{{plot_MET_model_11b_2010_MB_mu2.0_Arbeit}}}
  \caption{
    Comparison of fits to the Event Filter \sumet spectrum in events from 2010 data with $\avgmu \approx 2.0$.
    Shown are the distributions in subsamples
    selected by the number of reconstructed vertices \nreco between $1$ (red) and 2 through 6 (gray)
    as well as the full sample (blue).
    The data points represent the distributions in data,
    the curves are fits of the same type as explained for \Fig{fig:results_met_model_sumet_slope_fit_to_mu5}.
  }
  \label{fig:results_met_model_sumet_slope_fit_to_MB_2010}
\end{figure}

\Fig{fig:results_met_model_sumet_slope_fit_to_MB_2010} shows the fit
which is used to extract the slope of \sumet as input parameter of the model for 2010 data.
This plot is analogous to \Fig{fig:results_met_model_sumet_slope_fit_to_mu5},
which is for 2011,
with the difference that for 2010 the \avgmu in data in general is smaller,
so that here events with $\avgmu \approx 2.0$ are used. %
The data in the plot is taken with a minimum-bias trigger.
The fit shown in the plot gives a slope of $\lambda_\text{fit} = 0.0405(9)$ for 2010,
which is in very good agreement with the value of $\lambda_\text{fit} = 0.0410(4)$ obtained from 2011 data.
While for 2011 the fit yields a value of \revised{$\avgmu_\text{fit} = 4.96(5)$} %
for the average number of concurrent interactions,
which is compatible with the nominal value $\avgmu \approx 5$, %
for 2010 the fit converges to \revised{$\avgmu_\text{fit} = 1.52(8)$}, %
which is significantly lower than the nominal value $\avgmu \approx 2.0$.
This may indicate a better description of the data by the model for higher values of \avgmu,
but as well be due to the different trigger selections.
The latter possibility cannot be ruled out
because the number of events taken with random triggers in 2010 or minimum-bias triggers in 2011 is too small.
\Fig{fig:results_met_model_correlation_mu} confirms that there is a general difference between 2010 and 2011.
In this plot, the vertical axis is the value for \avgmu obtained from a fit of the \sumet spectrum above \GeV{50}
in the same way as was done in \Fig{fig:results_met_model_sumet_slope_fit_to_MB_2010}.
This is plotted as function of the nominal \avgmu,
which is taken from the \COOL database and is computed from the instantaneous luminosity.
Every point corresponds to a sample of events with a narrow distribution of \avgmu ($|\avgmu - \avgmu_0| < 0.1$,
where $\avgmu_0$ is the target value of the nominal \avgmu from COOL).
The error bars show the size of the uncertainty from the fit of \sumet.
The five points for $\avgmu<4$ are from 2010, the other four from 2011.
It can be seen that the correlation of the fit result with the nominal value of \avgmu
is better in 2011 than in 2010,
although for $\avgmu \approx 7$ the fit to 2011 data also gives too small a value.

\begin{figure}
  \centering
  \incgraphics{width=\widthsingleplot}{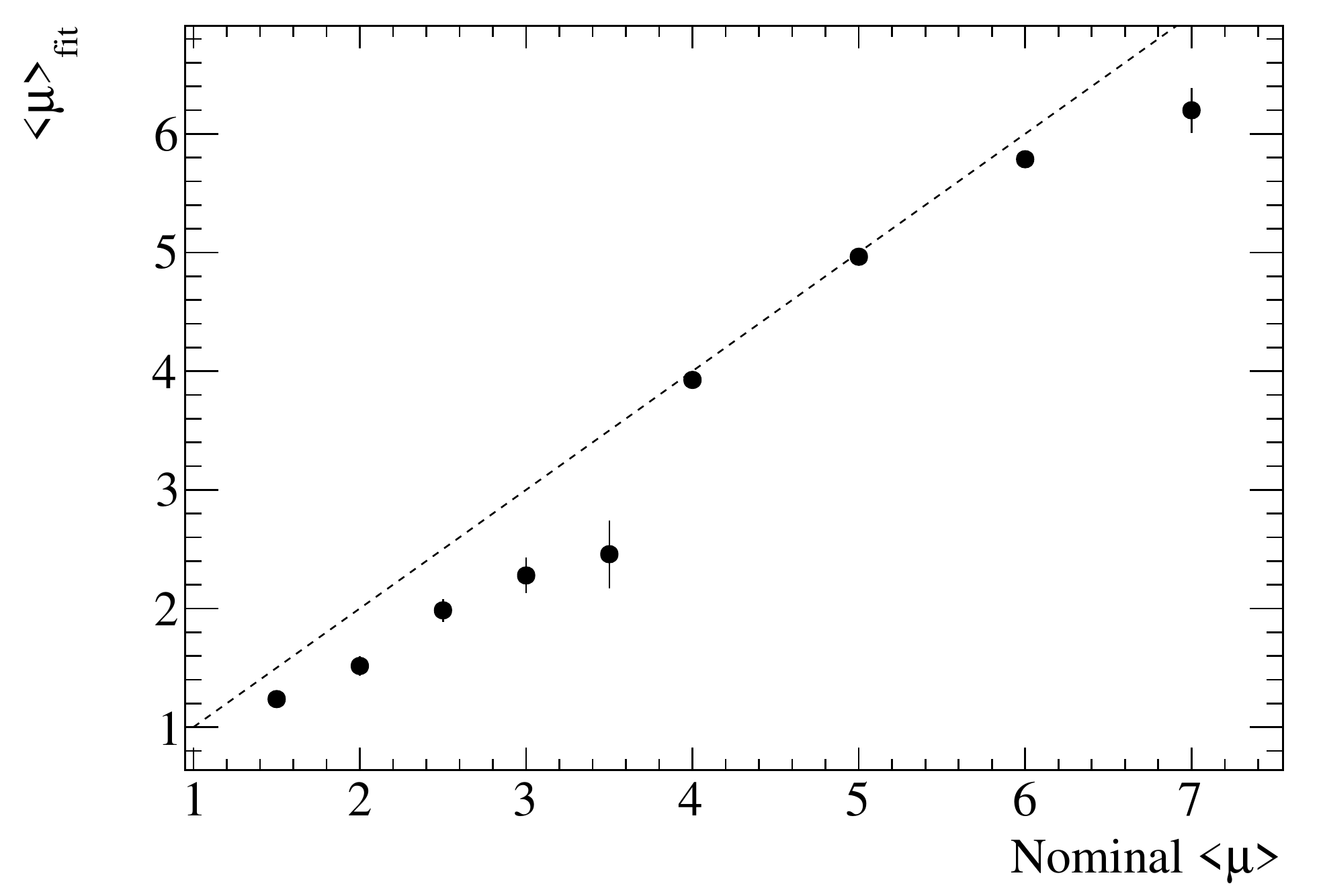}
  \caption{
    Correlation of the value for $\avgmu_\text{fit}$ obtained from a fit of the full \sumet spectrum
    and the nominal \avgmu taken from the \COOL database. %
    The error bars indicate the uncertainties from the fit.
    The dashed line marks the diagonal.
  }
  \label{fig:results_met_model_correlation_mu}
\end{figure}

\begin{figure}
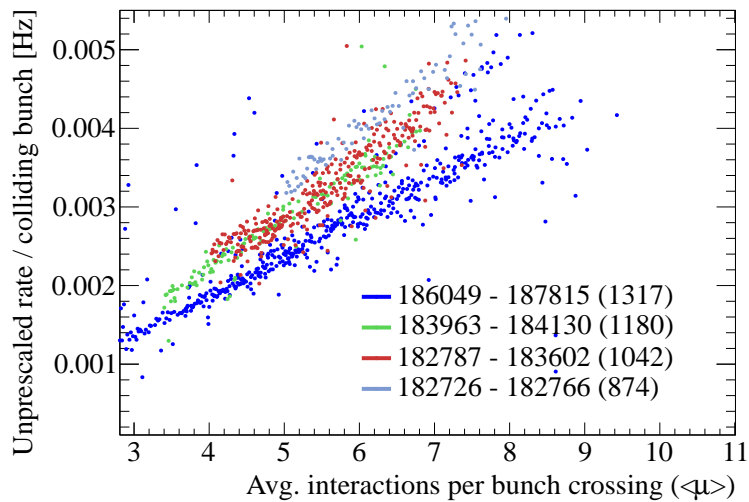

  \centering
  \incgraphics{width=\widthsingleplot}{{{COOL.plot_rates_EF_xe60_noMu_2011K,2011J,2011I,2011H,2011G_withGRL_Xofflmu_cbnbc_3}}}
  \caption{
    Rates of one high-threshold \met trigger, \trigger{EF_xe60_noMu}, in periods 2011 G~--~K,
    as function of the average number of interactions per bunch crossing \avgmu.
    The runs are grouped and colored by the number of colliding bunches which are given in brackets in the legend.
  }
  \label{fig:results_met_model_rate_high_EF_triggers_more_periods}
\end{figure}

Periods 2011 I~--~K have been chosen in \Fig{fig:results_met_model_rate_high_EF_triggers}
to display the rates of high threshold \met triggers
because in these periods the number of colliding bunches is approximately the same value ($1317$ or $1333$) for all runs. %
In \Fig{fig:results_met_model_rate_high_EF_triggers_more_periods},
the rates of one of the triggers, \trigger{EF_xe60_noMu},
are shown for a larger dataset with runs from periods 2011 G~--~K, %
again coloring groups of runs with different numbers of colliding bunches differently.
It can be seen that there are again discontinuities in the rate whenever the number of colliding bunches changes,
although the rates are plotted as function of \avgmu,
for which at the beginning of the discussion of the \met model
it has been shown that the low threshold triggers do not exhibit discontinuities (cf. \Fig{fig:results_met_model_rate_L1_XE20_avgmu}).
In fact, the rates of the high threshold triggers exhibit discontinuities
when plotted as function of \avgmu as well as when plotted as function of the instantaneous luminosity.
This can be explained by the rate being composed of two contributions,
one of which depends in a continuous manner on the instantaneous luminosity per bunch,
\ie the event activity expressed by the average number of interactions,
and the other continuously on the instantaneous luminosity,
independently of the number of colliding bunches.
As discussed before, %
the first part would be attributed to fake \met from noise as well as from mismeasurements of jets,
the second part would be attributed to real \met from invisible particles produced in the hard \revised{scattering} process.
For the low threshold triggers, the rates are completely dominated by fake \met,
so that the discontinuities due to the small fraction of real \met events are not visible.

\section{Supplementary Plots: Trigger Efficiencies in Monte Carlo}
\label{sec:appendix_trigger_efficiency_plots}
This section collects plots of trigger efficiencies that have not been included in the main part of the thesis.
They are shown here for completeness.

\label{sec:appendix_trigger_efficiency_plots_Monte_Carlo}

\begin{figure}
  \centering
  \incgraphics{width=\widthtwoplots}{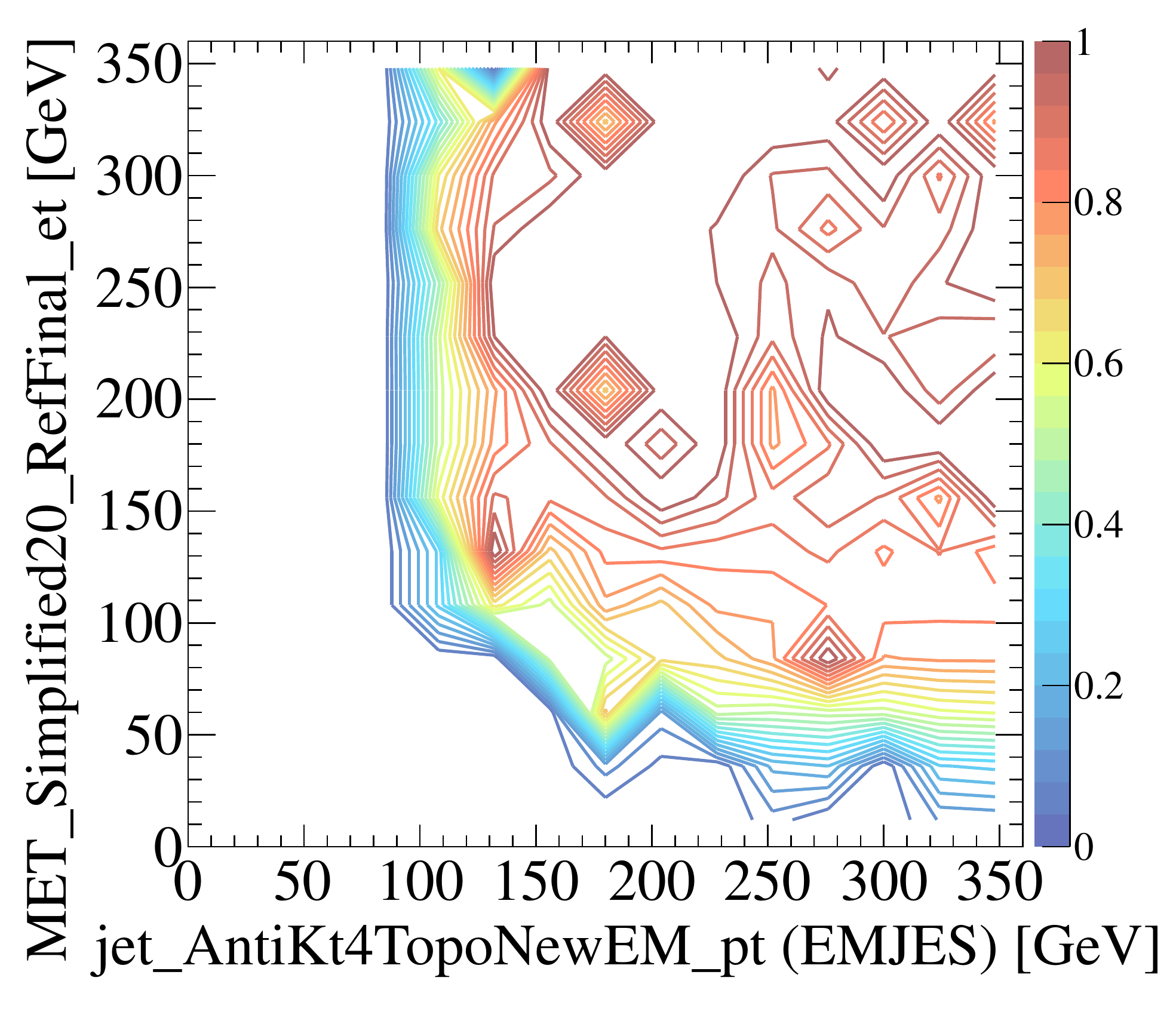}
  \hfill
  \incgraphics{width=\widthtwoplots}{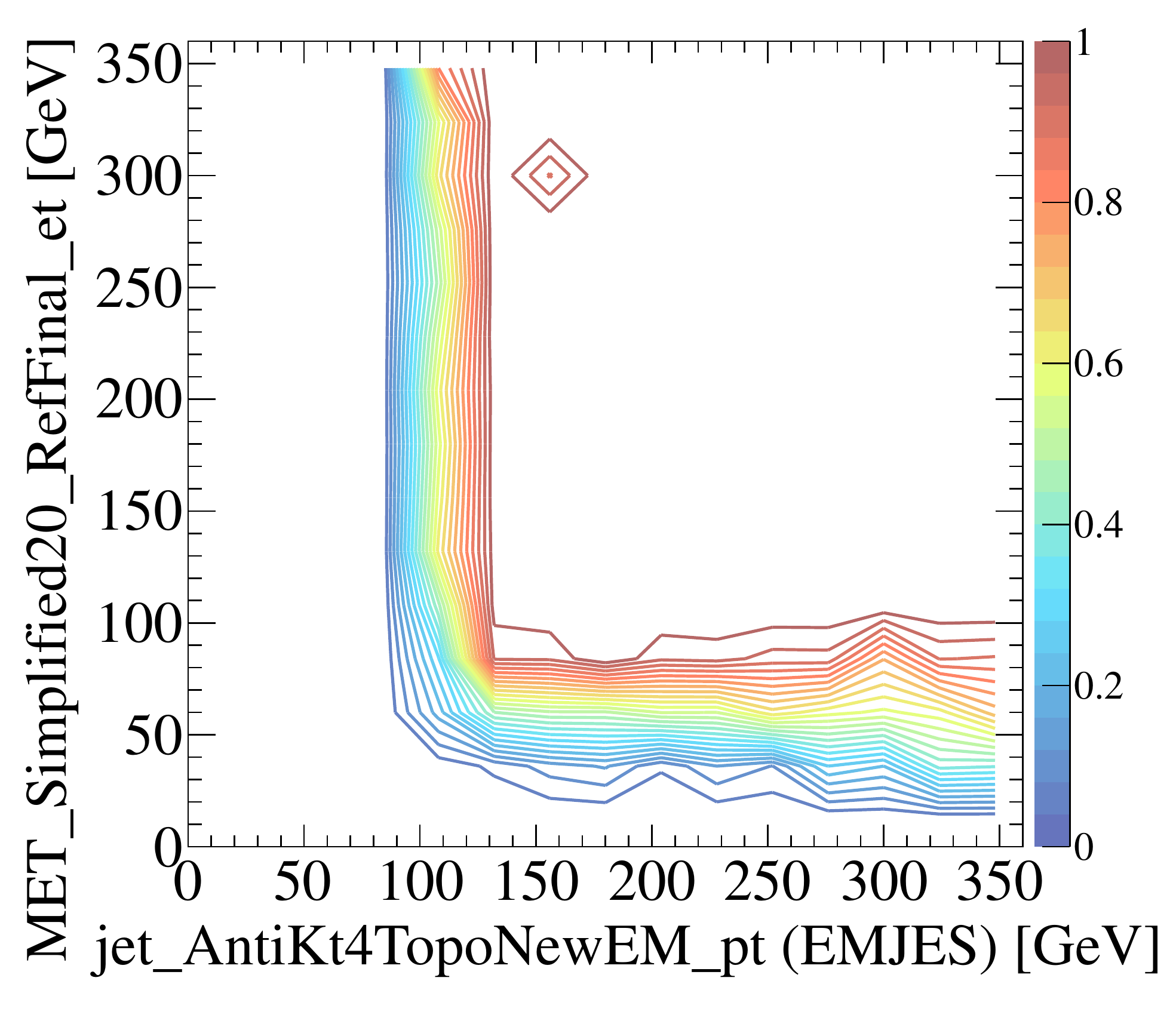}
  \\
  \incgraphics{width=\widthtwoplots}{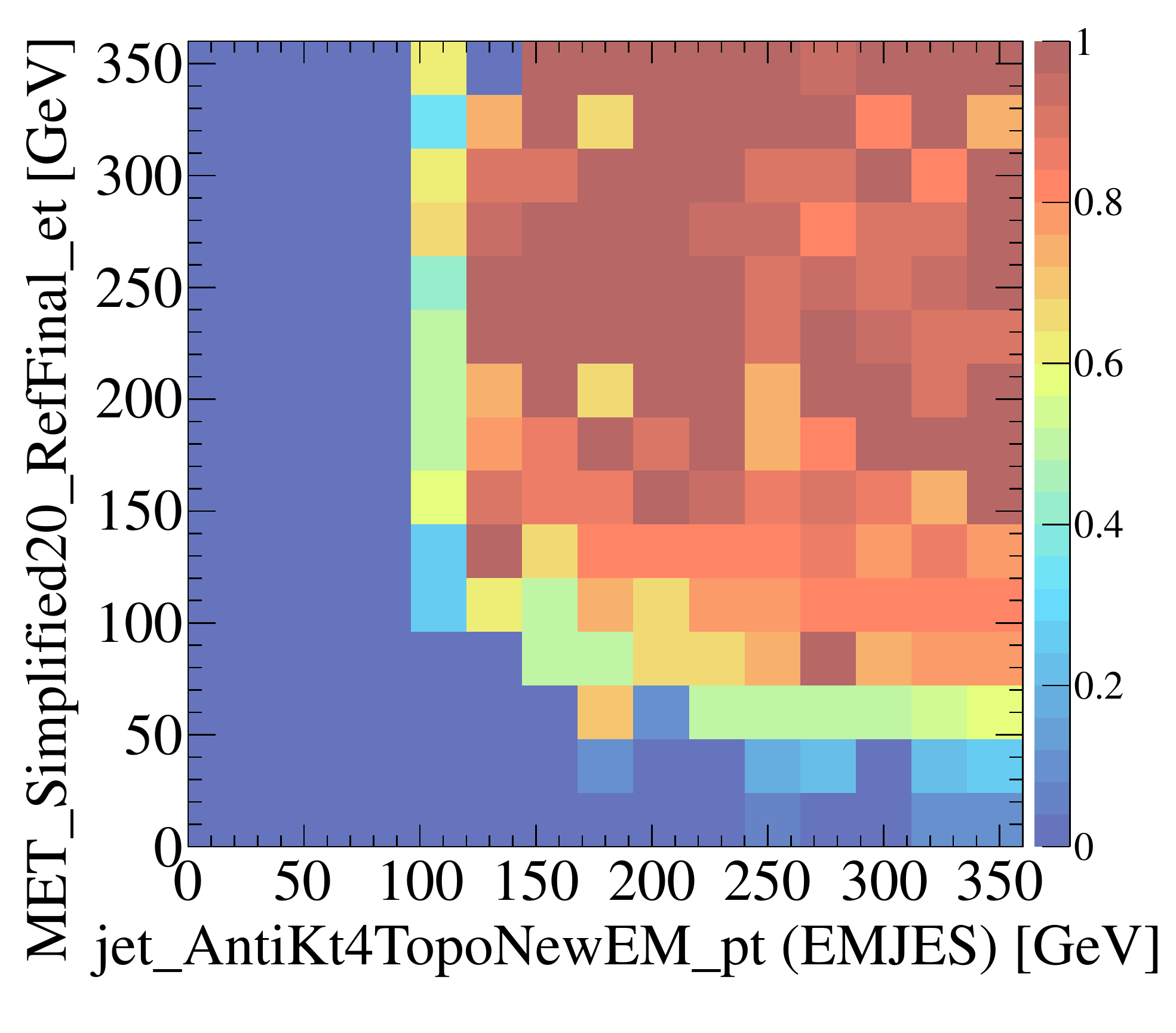}
  \hfill
  \incgraphics{width=\widthtwoplots}{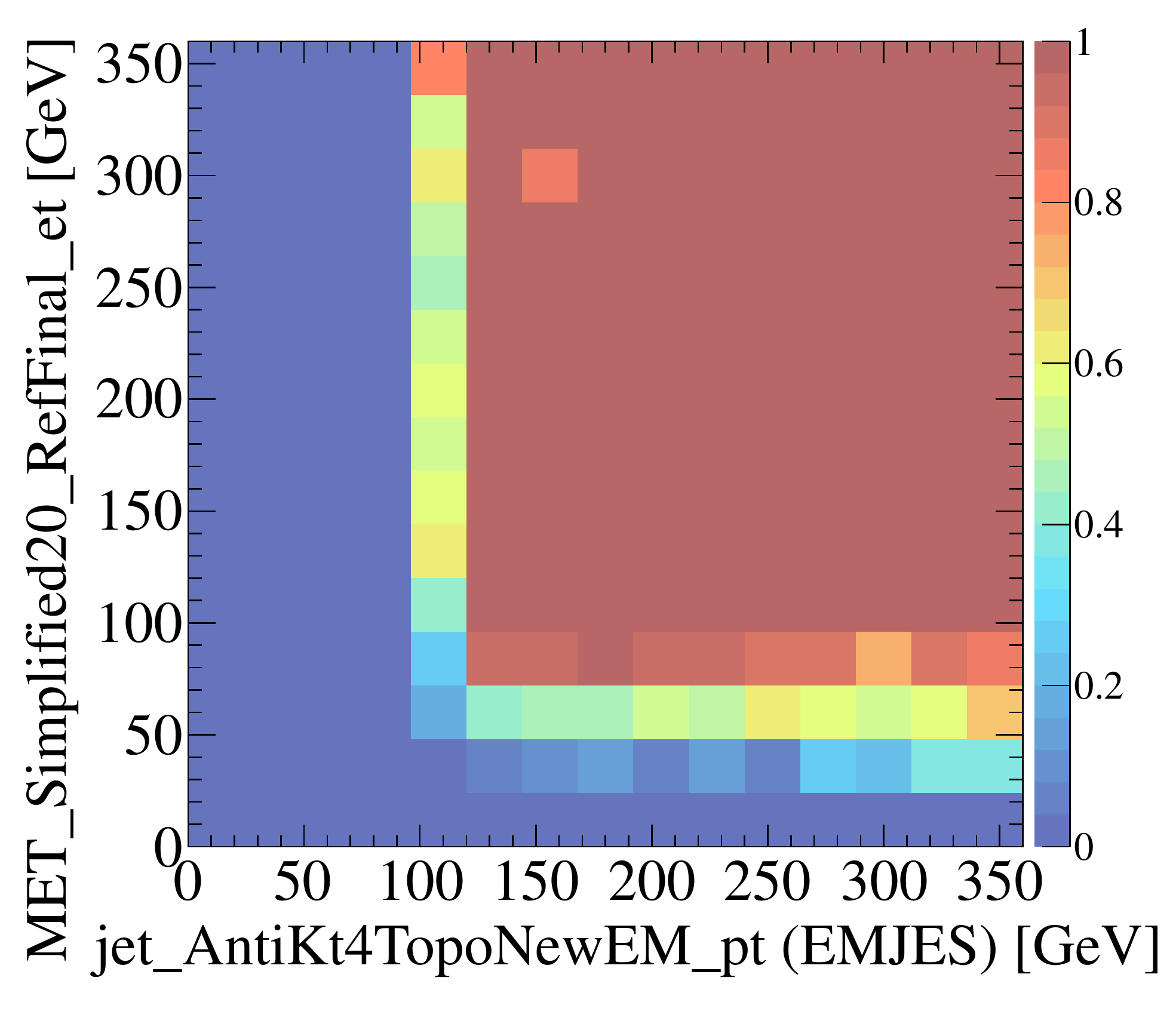}
  \caption{
    Two-dimensional efficiency plots of the combined \jetmet trigger \trigger{EF_j75_a4tc_EFFS_xe45_loose_noMu} in Monte Carlo events,
    for a QCD sample (left) and for a $Z\to\nu\nu$ + jets sample (right).
    The two upper plots show the efficiencies as contours,
    which highlights details in the regions where the efficiencies change,
    in particular in the turn-on regions.
    The two lower plots use a color-coding of the efficiencies,
    thereby making it easier to distinguish regions with low and high efficiencies.
    The SUSY definition of \met is used as offline reference on the vertical axis
    (cf. \Fig{fig:results_trigger_performance_measurements_2d_efficiency_MC_LocHadTopo_2d}).
  }
  \label{fig:results_trigger_performance_measurements_2d_efficiency_MC_SUSYMET_2d}
\end{figure}
\Fig{fig:results_trigger_performance_measurements_2d_efficiency_MC_SUSYMET_2d}
is the analog of \Fig{fig:results_trigger_performance_measurements_2d_efficiency_MC_LocHadTopo_2d},
but has as offline reference the \met variant which is used by the \ATLAS Supersymmetry group for the analysis of 2011 data.
This variant is called \tagname{MET\_Simplified20\_RefFinal} (cf. \Sec{sec:software_reconstruction_met}).
In the same way,
\Figs{fig:results_trigger_performance_measurements_2d_efficiency_MC_SUSYMET_projections} and \ref{fig:results_trigger_performance_measurements_2d_efficiency_MC_SUSYMET_projections_per_subsample}
are the analogs of \Figs{fig:results_trigger_performance_measurements_2d_efficiency_MC_LocHadTopo_projections} and
\ref{fig:results_trigger_performance_measurements_2d_efficiency_MC_LocHadTopo_projections_per_subsample}
with the SUSY definition of missing transverse energy instead of the \tagname{LocHadTopo} variant.

\begin{figure}
  \centering
  \incgraphics{width=\widthtwoplots}{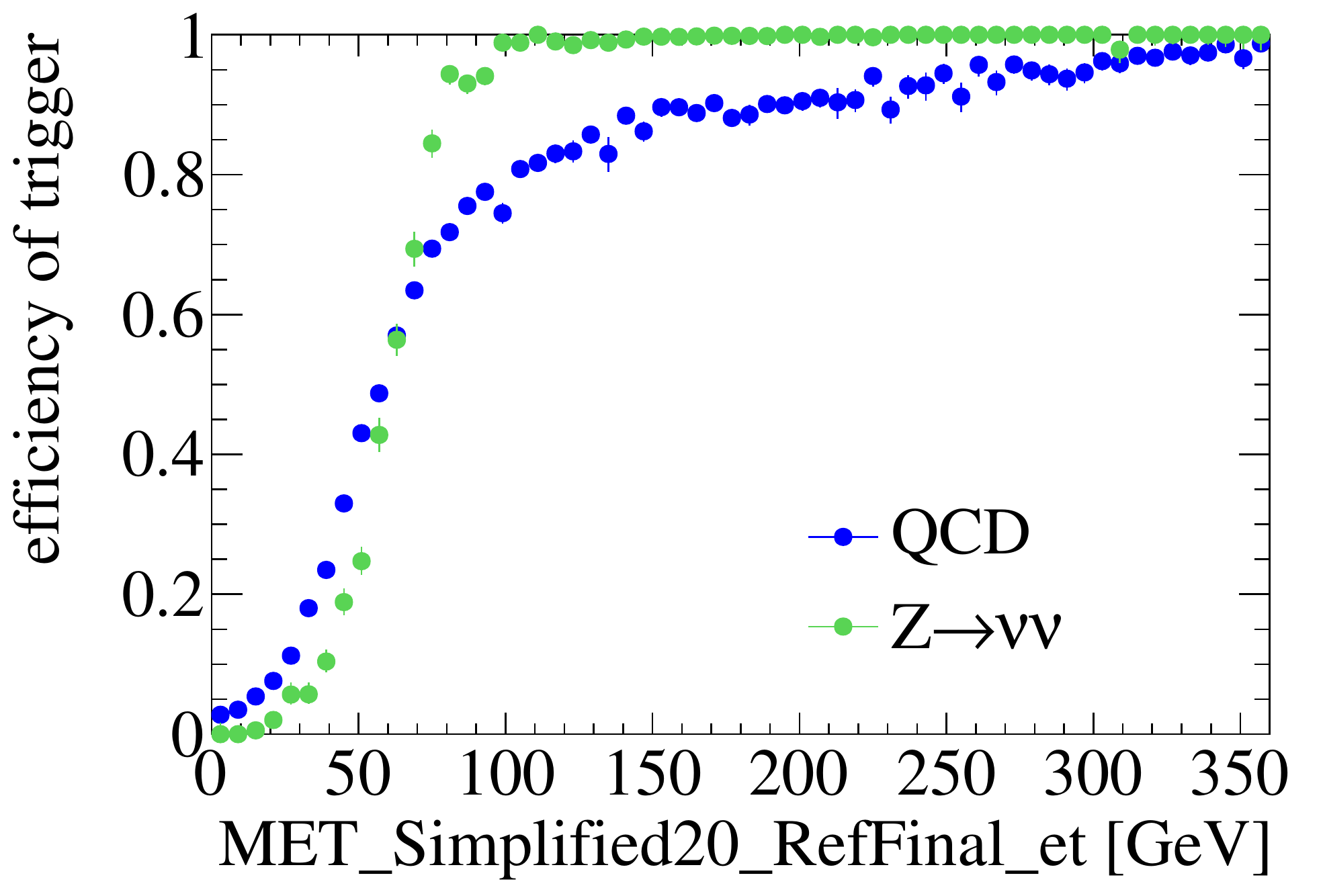}
  \hfill
  \incgraphics{width=\widthtwoplots}{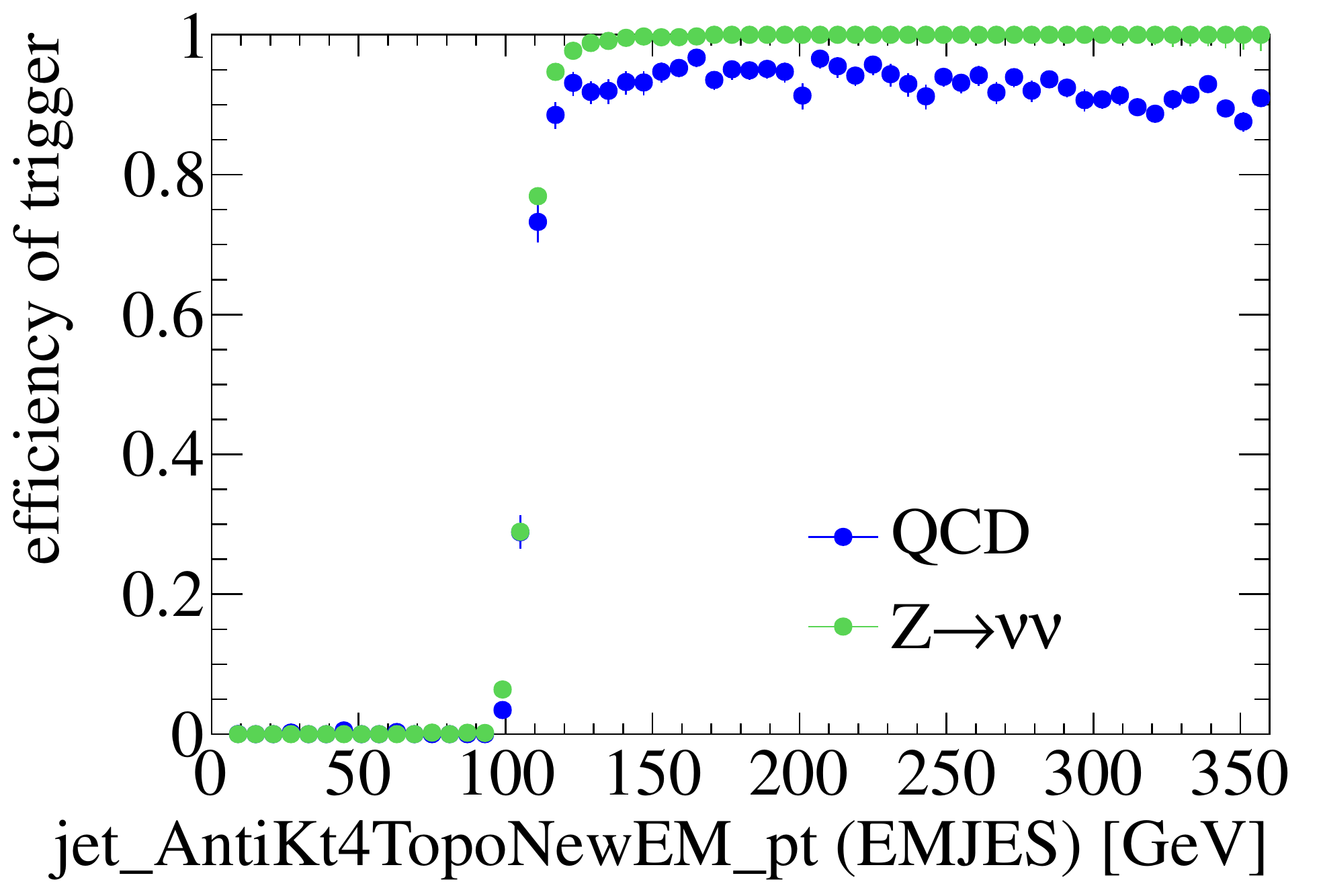}
  \caption{
    Projections of the efficiency of the combined \jetmet trigger \trigger{EF_j75_a4tc_EFFS_xe45_loose_noMu} in Monte Carlo events
    onto the \met (left) and jet \pt (right) axis,
    after cuts on the respective orthogonal variable at \GeV{130},
    for the combined QCD sample and for the combined $Z\to\nu\nu$ + jets sample.
    The SUSY definition of \met is used as offline reference here
    (cf. \Fig{fig:results_trigger_performance_measurements_2d_efficiency_MC_LocHadTopo_projections}).
  }
  \label{fig:results_trigger_performance_measurements_2d_efficiency_MC_SUSYMET_projections}
\end{figure}

\begin{figure}
  \centering
  \incgraphics{width=\widthtwoplots}{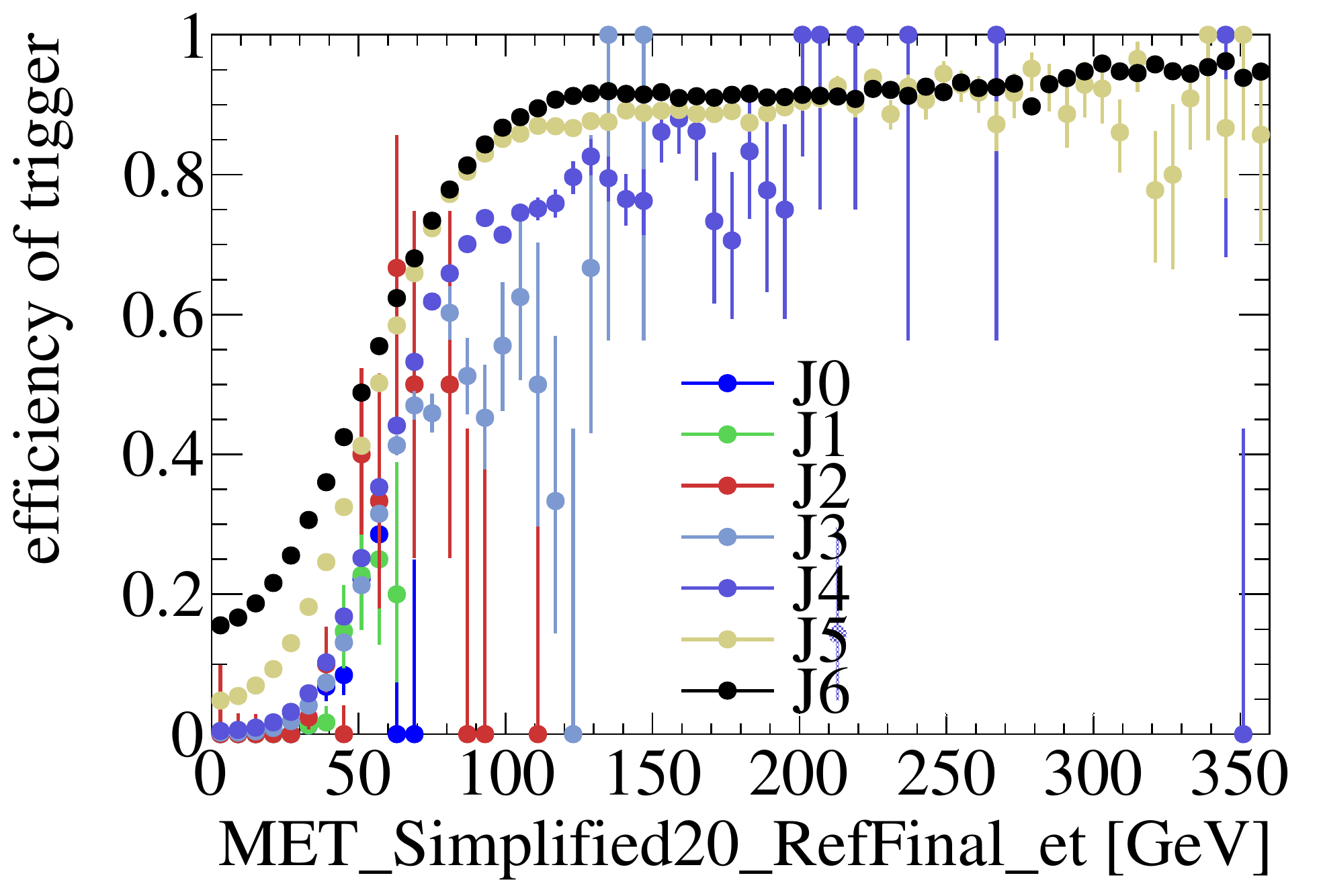}
  \hfill
  \incgraphics{width=\widthtwoplots}{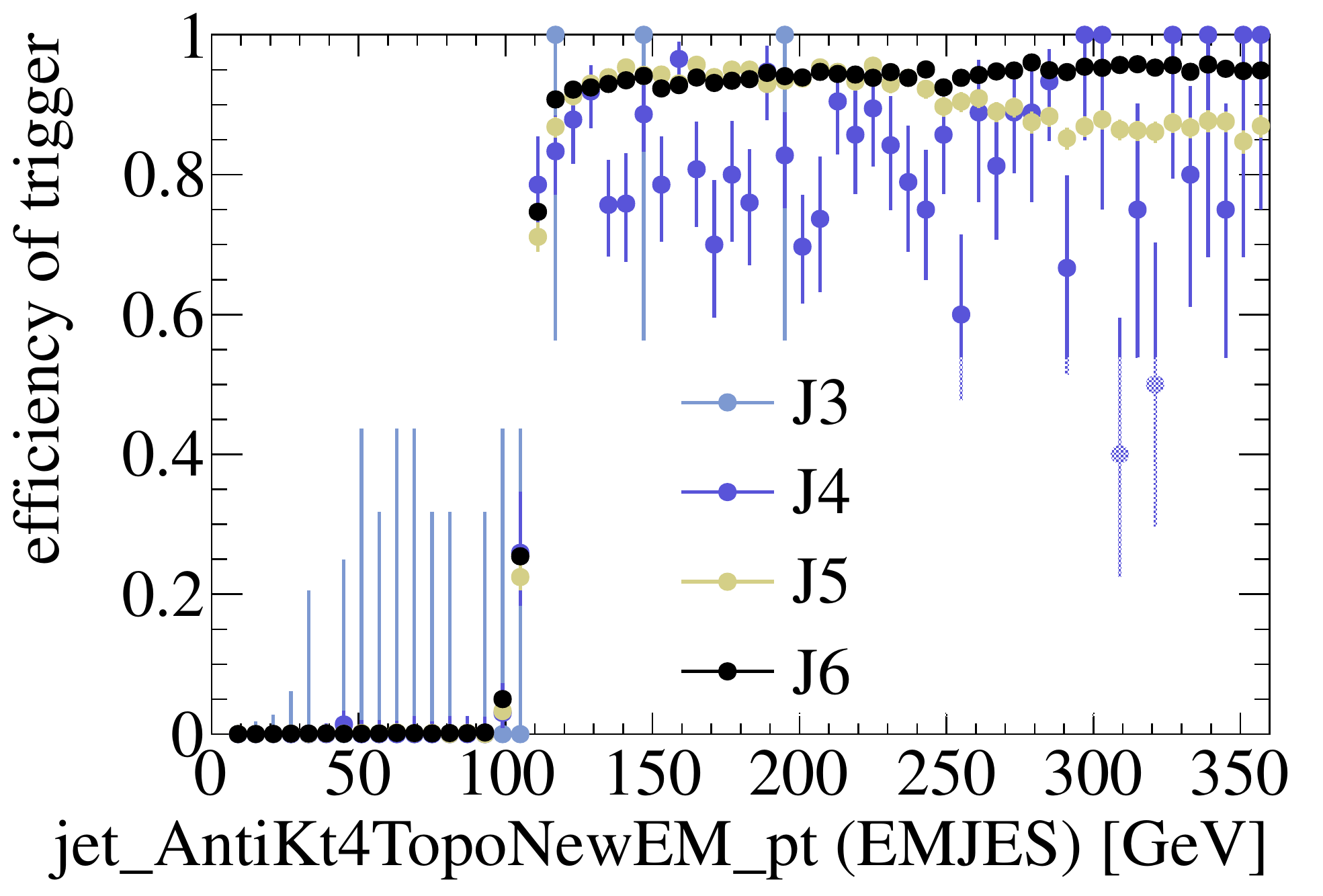}
  \\
  \incgraphics{width=\widthtwoplots}{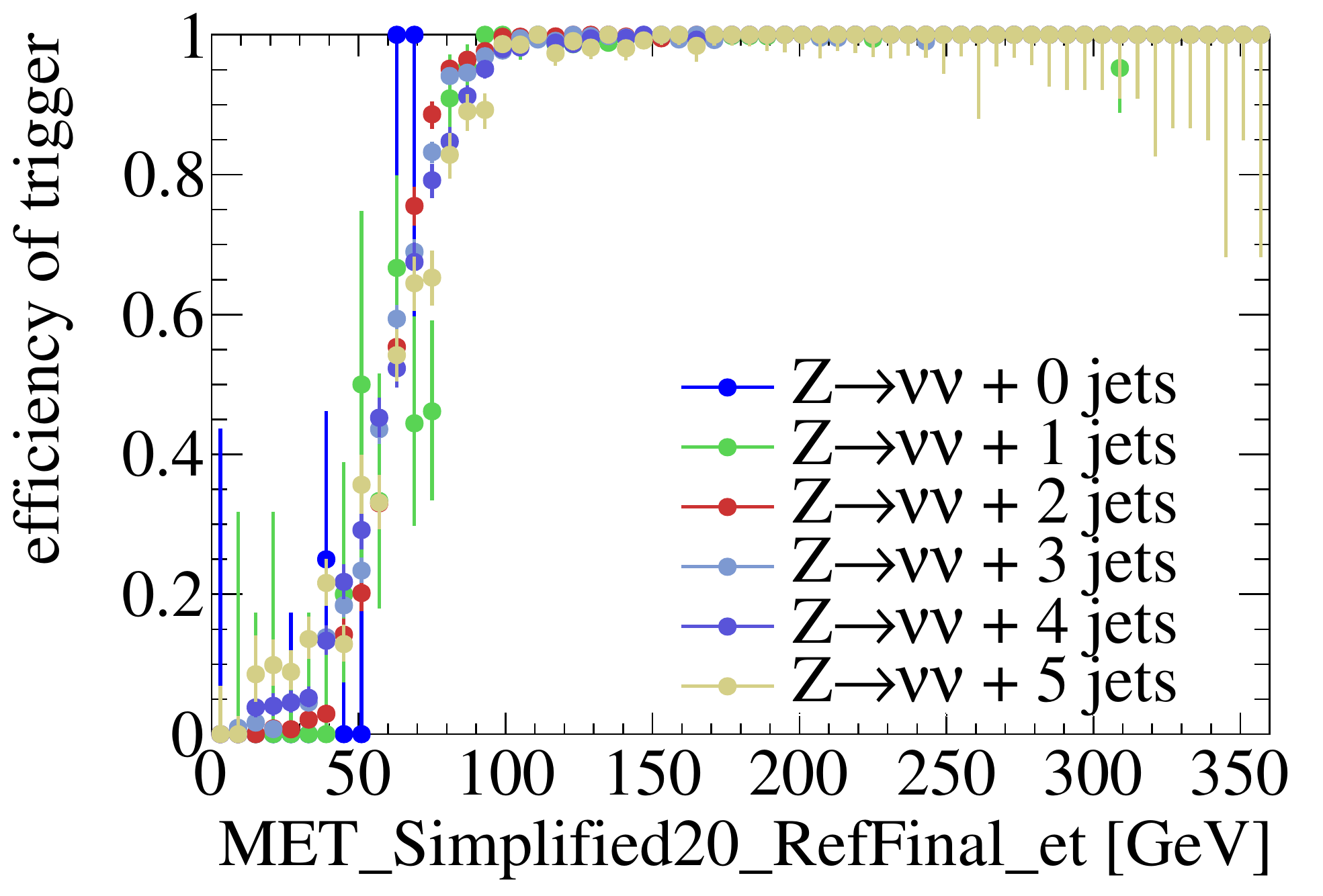}
  \hfill
  \incgraphics{width=\widthtwoplots}{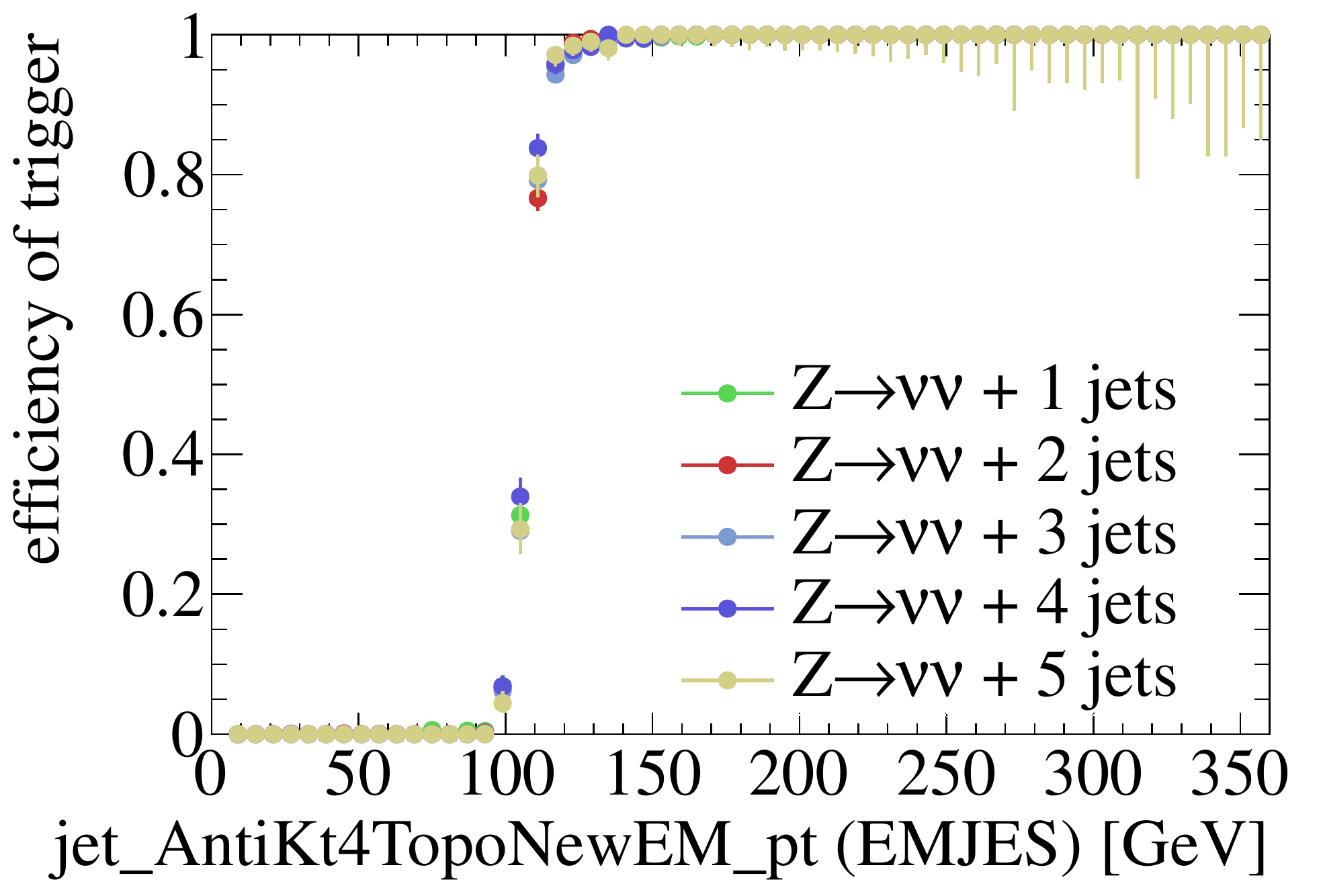}
  \caption{
    Projections of the efficiency of the combined \jetmet trigger \trigger{EF_j75_a4tc_EFFS_xe45_loose_noMu} in Monte Carlo events
    for each subsample of the QCD sample (top) and the $Z\to\nu\nu$ + jets sample (bottom) individually.
    Left: onto offline \met, right: onto jet \pt,
    after cuts on the respective orthogonal variable at \GeV{130}.
    The SUSY definition of \met is used as offline reference here
    (cf. \Fig{fig:results_trigger_performance_measurements_2d_efficiency_MC_LocHadTopo_projections_per_subsample}).
  }
  \label{fig:results_trigger_performance_measurements_2d_efficiency_MC_SUSYMET_projections_per_subsample}
\end{figure}

For the $Z\to\nu\nu$ + jets sample,
the turn-on curves for the two offline references look very similar.
In particular, the projection onto \met shows some small differences for the QCD samples.
The turn-ons for J5 and J6 are sharper,
but do not reach a plateau efficiency of \percent{100}.
The combination in \Fig{fig:results_trigger_performance_measurements_2d_efficiency_MC_SUSYMET_projections}
shows a lower efficiency of the trigger at the beginning of the plateau
when plotted as function of the SUSY \met definition.

\FloatBarrier

\section{Overview of Monte Carlo Datasets}
\label{sec:appendix_MC_datasets}

\afterpage{%
\begin{landscape}
  \centering
  \small
  \begin{longtable}{llld{2.2}d{2.2}rrrrl}
    \toprule
    Process & Subprocess & MC set ID & \multicolumn{1}{c}{$\sigma$ [pb]} & \multicolumn{1}{c}{$\sigma_\text{AMI}$ [pb]}
         & \multicolumn{1}{c}{$\epsilon\sub{gen}$} & \multicolumn{1}{c}{$k_\text{eff}$}
         & \multicolumn{1}{c}{$N_\text{sample}$}   & ${\cal L}_\text{MC}$ & Dataset tag\\
    \midrule
    \endhead
    $Z\to e^+e^-$ + jets  & $N_p = 0$ & 107650 & 830.13 & 664.10 & 1.000 & 1.25 & \numprint{304216} & 366 &  e529\_s765\_s767\_r1302\_r1306\_p305\\
     & $N_p = 1$ & 107651 & 166.24 & 132.99 & 1.000 & 1.25 & \numprint{63440} & 382 &  e529\_s765\_s767\_r1302\_r1306\_p305\\
     & $N_p = 2$ & 107652 & 50.28 & 40.23 & 1.000 & 1.25 & \numprint{19497} & 388 &  e529\_s765\_s767\_r1302\_r1306\_p305\\
     & $N_p = 3$ & 107653 & 13.92 & 11.14 & 1.000 & 1.25 & \numprint{5499} & 395 &  e529\_s765\_s767\_r1302\_r1306\_p305\\
     & $N_p = 4$ & 107654 & 3.62 & 2.89 & 1.000 & 1.25 & \numprint{1499} & 414 &  e529\_s765\_s767\_r1302\_r1306\_p305\\
     & $N_p \geq 5$ & 107655 & 0.94 & 0.75 & 1.000 & 1.25 & 500 & 532 &  e529\_s765\_s767\_r1302\_r1306\_p305\\
    $Z\to \mu^+\mu^-$ + jets  & $N_p = 0$ & 107660 & 830.13 & 663.79 & 1.000 & 1.25 & \numprint{303947} & 366 &  e529\_s765\_s767\_r1302\_r1306\_p305\\
     & $N_p = 1$ & 107661 & 166.24 & 132.95 & 1.000 & 1.25 & \numprint{62996} & 379 &  e529\_s765\_s767\_r1302\_r1306\_p305\\
     & $N_p = 2$ & 107662 & 50.28 & 40.38 & 1.000 & 1.25 & \numprint{18993} & 378 &  e529\_s765\_s767\_r1302\_r1306\_p305\\
     & $N_p = 3$ & 107663 & 13.92 & 11.16 & 1.000 & 1.25 & \numprint{5497} & 395 &  e529\_s765\_s767\_r1302\_r1306\_p305\\
     & $N_p = 4$ & 107664 & 3.62 & 2.90 & 1.000 & 1.25 & \numprint{1499} & 414 &  e529\_s765\_s767\_r1302\_r1306\_p305\\
     & $N_p \geq 5$ & 107665 & 0.94 & 0.76 & 1.000 & 1.24 & 499 & 531 &  e529\_s765\_s767\_r1302\_r1306\_p305\\
    $Z\to \tau^+\tau^-$ + jets  & $N_p = 0$ & 107670 & 830.13 & 662.50 & 1.000 & 1.25 & \numprint{303359} & 365 &  e529\_s765\_s767\_r1302\_r1306\_p305\\
     & $N_p = 1$ & 107671 & 166.24 & 133.94 & 1.000 & 1.24 & \numprint{63481} & 382 &  e529\_s765\_s767\_r1302\_r1306\_p305\\
     & $N_p = 2$ & 107672 & 50.28 & 40.30 & 1.000 & 1.25 & \numprint{19492} & 388 &  e529\_s765\_s767\_r1302\_r1306\_p305\\
     & $N_p = 3$ & 107673 & 13.92 & 11.03 & 1.000 & 1.26 & \numprint{5497} & 395 &  e529\_s765\_s767\_r1302\_r1306\_p305\\
     & $N_p = 4$ & 107674 & 3.62 & 2.80 & 1.000 & 1.29 & \numprint{1499} & 414 &  e529\_s765\_s767\_r1302\_r1306\_p305\\
     & $N_p \geq 5$ & 107675 & 0.94 & 0.78 & 1.000 & 1.20 & 499 & 531 &  e529\_s765\_s767\_r1302\_r1306\_p305\\
    $Z\to \nu\bar{\nu}$ + jets  & $N_p = 0$ & 107710 & 35.96 & 3538.80 & 0.008 & 1.32 & \numprint{2999} & 83 &  e530\_s765\_s767\_r1302\_r1306\_p305\\
     & $N_p = 1$ & 107711 & 549.69 & 731.35 & 0.607 & 1.24 & \numprint{44487} & 81 &  e530\_s765\_s767\_r1302\_r1306\_p305\\
     & $N_p = 2$ & 107712 & 247.42 & 222.50 & 0.878 & 1.27 & \numprint{39491} & 160 &  e530\_s765\_s767\_r1302\_r1306\_p305\\
     & $N_p = 3$ & 107713 & 74.71 & 62.00 & 0.966 & 1.25 & \numprint{11995} & 161 &  e530\_s765\_s767\_r1302\_r1306\_p305\\
     & $N_p = 4$ & 107714 & 18.58 & 15.87 & 0.990 & 1.18 & \numprint{7993} & 430 &  e530\_s765\_s767\_r1302\_r1306\_p305\\
     & $N_p \geq 5$ & 107715 & 6.02 & 4.38 & 0.999 & 1.38 & \numprint{2500} & 415 &  e530\_s765\_s767\_r1302\_r1306\_p305\\
\pagebreak
    $W\to e\nu_e$ + jets  & $N_p = 0$ & 107680 & 8288.15 & 6870.50 & 1.000 & 1.21 & \numprint{1381931} & 167 &  e511\_s765\_s767\_r1302\_r1306\_p305\\
     & $N_p = 1$ & 107681 & 1550.14 & 1293.80 & 1.000 & 1.20 & \numprint{258408} & 167 &  e511\_s765\_s767\_r1302\_r1306\_p305\\
     & $N_p = 2$ & 107682 & 452.09 & 376.60 & 1.000 & 1.20 & \numprint{188896} & 418 &  e511\_s765\_s767\_r1302\_r1306\_p305\\
     & $N_p = 3$ & 107683 & 120.97 & 101.29 & 1.000 & 1.19 & \numprint{50477} & 417 &  e511\_s765\_s767\_r1302\_r1306\_p305\\
     & $N_p = 4$ & 107684 & 30.33 & 25.25 & 1.000 & 1.20 & \numprint{12991} & 428 &  e511\_s765\_s767\_r1302\_r1306\_p305\\
     & $N_p \geq 5$ & 107685 & 8.27 & 7.12 & 1.000 & 1.16 & \numprint{3449} & 417 &  e511\_s765\_s767\_r1302\_r1306\_p305\\
    $W\to\mu\nu_\mu$ + jets  & $N_p = 0$ & 107690 & 8288.15 & 6871.10 & 1.000 & 1.21 & \numprint{1386038} & 167 &  e511\_s765\_s767\_r1302\_r1306\_p305\\
     & $N_p = 1$ & 107691 & 1550.14 & 1294.70 & 1.000 & 1.20 & \numprint{255909} & 165 &  e511\_s765\_s767\_r1302\_r1306\_p305\\
     & $N_p = 2$ & 107692 & 452.09 & 376.08 & 1.000 & 1.20 & \numprint{187860} & 416 &  e511\_s765\_s767\_r1302\_r1306\_p305\\
     & $N_p = 3$ & 107693 & 120.97 & 100.72 & 1.000 & 1.20 & \numprint{50887} & 421 &  e511\_s765\_s767\_r1302\_r1306\_p305\\
     & $N_p = 4$ & 107694 & 30.33 & 25.99 & 1.000 & 1.17 & \numprint{12991} & 428 &  e511\_s765\_s767\_r1302\_r1306\_p305\\
     & $N_p \geq 5$ & 107695 & 8.27 & 7.13 & 1.000 & 1.16 & \numprint{3498} & 423 &  e511\_s765\_s767\_r1302\_r1306\_p305\\
    $W\to\tau\nu_\tau$ + jets  & $N_p = 0$ & 107700 & 8288.15 & 6873.30 & 1.000 & 1.21 & \numprint{1365491} & 165 &  e511\_s765\_s767\_r1302\_r1306\_p305\\
     & $N_p = 1$ & 107701 & 1550.14 & 1295.40 & 1.000 & 1.20 & \numprint{254753} & 164 &  e511\_s765\_s767\_r1302\_r1306\_p305\\
     & $N_p = 2$ & 107702 & 452.09 & 375.07 & 1.000 & 1.21 & \numprint{188446} & 417 &  e511\_s765\_s767\_r1302\_r1306\_p305\\
     & $N_p = 3$ & 107703 & 120.97 & 101.77 & 1.000 & 1.19 & \numprint{50472} & 417 &  e511\_s765\_s767\_r1302\_r1306\_p305\\
     & $N_p = 4$ & 107704 & 30.33 & 25.76 & 1.000 & 1.18 & \numprint{12996} & 428 &  e511\_s765\_s767\_r1302\_r1306\_p305\\
     & $N_p \geq 5$ & 107705 & 8.27 & 7.00 & 1.000 & 1.18 & \numprint{3998} & 483 &  e511\_s765\_s767\_r1302\_r1306\_p305\\
    $W\to b\bar{b}$ + jets  & $N_p = 0$ & 106280 & 3.90 & 3.31 & 1.000 & 1.18 & \numprint{6499} & \numprint{1665} &  e524\_s765\_s767\_r1302\_r1306\_p305\\
     & $N_p = 1$ & 106281 & 3.17 & 2.68 & 1.000 & 1.19 & \numprint{5500} & \numprint{1734} &  e524\_s765\_s767\_r1302\_r1306\_p305\\
     & $N_p = 2$ & 106282 & 1.71 & 1.38 & 1.000 & 1.24 & \numprint{2997} & \numprint{1755} &  e524\_s765\_s767\_r1302\_r1306\_p305\\
     & $N_p \geq 3$ %
      & 106283 & 0.73 & 0.66 & 1.000 & 1.11 & \numprint{1500} & \numprint{2049} &  e524\_s765\_s767\_r1302\_r1306\_p305\\
    \bottomrule
    \caption{
      \normalsize
      Detailed list of the Monte Carlo samples for the \Wjets and \Zjets background.
    }
    \label{tab:analysis_MC_datasets_details_WZ}
  \end{longtable}

\newpage
  \begin{table}
    \centering
    \small
    \hspace*{-5mm}
    \begin{tabular}{p{15mm}p{33mm}ld{2.2}d{2.2}rrrrl}
    \toprule
    Process & Subprocess & MC set ID & \multicolumn{1}{c}{$\sigma$ [pb]} & \multicolumn{1}{c}{$\sigma_\text{AMI}$ [pb]}
         & \multicolumn{1}{c}{$\epsilon\sub{gen}$} & \multicolumn{1}{c}{$k_\text{eff}$}
         & \multicolumn{1}{c}{$N_\text{sample}$}   & ${\cal L}_\text{MC}$ & Dataset tag\\
    \midrule
    $t\bar{t}$ & semi- / fully leptonic & 105200 & 89.40 & 144.12 & 0.556 & 1.12 & \numprint{999387} & \numprint{8648} &  e510\_s765\_s767\_r1302\_r1306\_p305\\
    & fully hadronic & 105204 & 71.39 & 144.28 & 0.444 & 1.11 & \numprint{149899} & \numprint{1626} &  e540\_s765\_s767\_r1302\_r1306\_p305\\
    \multirow{2}{15mm}{Single top} & $t$-channel $W\to e\nu_e$ & 108340 & 7.15 & 7.15 & 1.000 & 1.00 & \numprint{9993} & 831 &  e508\_s765\_s767\_r1302\_r1306\_p305\\
    & $t$-channel $W\to \mu\nu_\mu$ & 108341 & 7.15 & 7.18 & 1.000 & 1.00 & \numprint{9997} & 837 &  e508\_s765\_s767\_r1302\_r1306\_p305\\
    & $t$-channel $W\to \tau\nu_\tau$ & 108342 & 7.15 & 7.13 & 1.000 & 1.00 & \numprint{10000} & 829 &  e508\_s765\_s767\_r1302\_r1306\_p305\\
    & $s$-channel $W\to e\nu_e$ & 108343 & 0.47 & 0.47 & 1.000 & 1.00 & \numprint{9950} & \numprint{17962} &  e534\_s765\_s767\_r1302\_r1306\_p305\\
    & $s$-channel $W\to \mu\nu_\mu$ & 108344 & 0.47 & 0.47 & 1.000 & 1.00 & \numprint{9996} & \numprint{17979} &  e534\_s765\_s767\_r1302\_r1306\_p305\\
    & $s$-channel $W\to \tau\nu_\tau$ & 108345 & 0.47 & 0.47 & 1.000 & 1.00 & \numprint{9996} & \numprint{18081} &  e534\_s765\_s767\_r1302\_r1306\_p305\\
    & $Wt$ channel, all $W$ decays & 108346 & 14.58 & 14.58 & 1.000 & 1.00 & \numprint{14995} & 911 &  e508\_s765\_s767\_r1302\_r1306\_p305\\
    \multirow{3}{15mm}{Pythia QCD dijets} & J0 & 105009 & \multicolumn{1}{c}{$1.262\cdot10^{10}$} & \multicolumn{1}{c}{$9.857\cdot10^{9}$} & 1.000 & 1.28 & \numprint{1379184} & <1 &  e468\_s766\_s767\_r1303\_r1306\_p305\\
    & J1 & 105010 & \multicolumn{1}{c}{$8.679\cdot10^{8}$} & \multicolumn{1}{c}{$6.781\cdot10^{8}$} & 1.000 & 1.28 & \numprint{1395383} & <1 &  e468\_s766\_s767\_r1303\_r1306\_p305\\
    & J2 & 105011 & \multicolumn{1}{c}{$5.247\cdot10^{7}$} & \multicolumn{1}{c}{$4.099\cdot10^{7}$} & 1.000 & 1.28 & \numprint{1398078} & <1 &  e468\_s766\_s767\_r1303\_r1306\_p305\\
    & J3 & 105012 & \multicolumn{1}{c}{$2.808\cdot10^{6}$} & \multicolumn{1}{c}{$2.194\cdot10^{6}$} & 1.000 & 1.28 & \numprint{1397430} & <1 &  e468\_s766\_s767\_r1303\_r1306\_p305\\
    & J4 & 105013 & \numprint{112261} & \numprint{87704} & 1.000 & 1.28 & \numprint{1397401} & 12 &  e468\_s766\_s767\_r1303\_r1306\_p305\\
    & J5 & 105014 & \numprint{3008} & \numprint{2350} & 1.000 & 1.28 & \numprint{1391612} & 463 &  e468\_s766\_s767\_r1303\_r1306\_p305\\
    & J6 & 105015 & 43.0 & 33.6 & 1.000 & 1.28 & \numprint{1347654} & \numprint{31316} &  e468\_s766\_s767\_r1303\_r1306\_p305\\
    & J7 & 105016 & 0.18 & 0.14 & 1.000 & 1.28 & \numprint{1125428} & \numprint{6280290} &  e468\_s766\_s767\_r1303\_r1306\_p305\\
    \bottomrule
    \end{tabular}
    \caption{
      Detailed list of the Monte Carlo samples for the top quark and QCD background.
      For the MC@NLO samples, \ie all top quark samples in this table, %
      the unweighted total event count is given.
    }
    \label{tab:analysis_MC_datasets_details_top_dijet}
  \end{table}
\end{landscape}
}

\Tabs{tab:analysis_MC_datasets_details_WZ} and \ref{tab:analysis_MC_datasets_details_top_dijet}
list all Monte Carlo samples that were used in the analysis in \Sec{sec:analysis_susysearch}
to model the backgrounds from Standard Model processes.
The symbolic %
description of the processes follows the common convention that
in case that no charge state, generation or flavor is given explicitly,
all possibilities are understood to be included which are in agreement with the conservation laws from the Standard Model.
In addition to the process name, the subprocess description and the ID of the Monte Carlo sample,
the tables include the final cross section $\sigma$ used in the analysis for the normalization of the Monte Carlo to the data,
the cross section $\sigma_\text{AMI}$ stored in the \ac{AMI} database,
and the \ac{MC} generator efficiency $\epsilon\sub{gen}$.
The factor $k_\text{eff}$ is the ratio of the final cross section after application of all normalization factors and the cross section from the \ac{AMI} database,
given by $k_\text{eff} = \sigma / (\sigma_\text{AMI} \cdot \epsilon_\text{gen})$.
$N_\text{sample}$ is the number of simulated events in the subsample.
To get a feeling whether the number of events in Monte Carlo is sufficient,
the integrated luminosity ${\cal L}_\text{MC}$ that corresponds to the size of the samples is given,
where ${\cal L}_\text{MC} = N_\text{sample} / \sigma$.
Only for the dijet samples with extremely large cross sections,
the sample luminosity ${\cal L}_\text{MC}$ is smaller than the integrated luminosity $\Lint^{2010} = \unit[33.4]{pb^{-1}}$
(cf. \Sec{sec:analysis_luminosity_calculation}) of the dataset to be evaluated.
The dataset tags of the MC samples are also given (cf. \Sec{sec:definition_nomenclature}). %

All top quark samples in \Tab{tab:analysis_MC_datasets_details_top_dijet} are generated with MC@NLO.
A characteristic feature of MC@NLO is the appearance of negative event weights arising from the subtraction method \cite{MCatNLO2002}.
This is a consequence of the way the events are generated in the integration over phase space.
For the MC@NLO samples, the unweighted total event count is given.
Due to the negative-weight events,
the value of ${\cal L}_\text{MC}$ given in the \Tab{tab:analysis_MC_datasets_details_top_dijet} is smaller %
than the one obtained from the unweighted total event count.
No errors are available on the cross sections given in \Tabs{tab:analysis_MC_datasets_details_WZ} and \ref{tab:analysis_MC_datasets_details_top_dijet}.
It is therefore difficult to estimate the number of significant digits,
but they will probably be less than the number of digits included in the tables.
Nevertheless, it is common to present these numbers cutting them at the same position,
rather than introducing some arbitrary rounding, so this habit is adopted here.
The number of events in the table is exact.

\removesection{

Other Monte Carlo samples with similar processes which have not been used for the background estimation are:
\begin{itemize}
  \item $W\to b\bar{b}$ without $\Delta R$ cuts, 107280\dots3: These samples are similar to 106280\dots3, %
    but generated without phase space cuts. They cannot be used in conjunction with the $W$ + light jets samples, %
    because they have significant overlap. %
    Besides, in the \tagname{NTUP\_SUSY} format they are not available without overlaid pile-up events.
  \item Drell-Yan: The hadronic production of lepton pairs in the process $q\bar{q}\to\gamma^* / Z\to l^+l^-$ is known as the Drell-Yan process \cite{DrellYan1970}. %
    In order to avoid overlap with the $Z$ + jets samples these samples are not used.
    (The $Z$+jets samples include the Drell-Yan contribution and take interference between $Z$ and $\gamma^*$ into account.)
    Still their contribution has been evaluated and is found to be a negligible background anyway
    as from Monte Carlo no events are expected to reach the signal region. %
  \item Pair production of weak bosons ($WW$, $ZZ$, $WZ$), 105985\dots7: These samples are filtered for events with at least 1 lepton with $\pt > \GeV{10}$ and $|\eta| < 2.8$.
    Their contribution is estimated to be below one percent.
\end{itemize}

}

\begin{landscape}
  \begin{table}
    \centering
    \tiny
    \setlength{\tabcolsep}{4pt}
    \begin{tabular}{r|rrrrrrrrrrrrrrrrrrrrrrrrrrrrrrrrr|l}
    & \Vert{-2000006} & \Vert{-2000005} & \Vert{-2000004} & \Vert{-2000003} & \Vert{-2000002} & \Vert{-2000001} & \Vert{-1000037} & \Vert{-1000024} & \Vert{-1000006} & \Vert{-1000005} & \Vert{-1000004} & \Vert{-1000003} & \Vert{-1000002} & \Vert{-1000001} & \Vert{1000001} & \Vert{1000002} & \Vert{1000003} & \Vert{1000004} & \Vert{1000005} & \Vert{1000006} & \Vert{1000021} & \Vert{1000022} & \Vert{1000023} & \Vert{1000024} & \Vert{1000025} & \Vert{1000035} & \Vert{1000037} & \Vert{2000001} & \Vert{2000002} & \Vert{2000003} & \Vert{2000004} & \Vert{2000005} & \Vert{2000006} & \\
    \hline
    -2000006 &  &  &  &  &  &  &  &  &  &  &  &  &  &  &  &  &  &  &  &  &  &  &  &  &  &  &  &  &  &  &  &  &  &  $\bar{t}_R$ \\
    -2000005 &  &  &  &  &  &  &  &  &  &  &  &  &  &  &  &  &  &  &  &  &  &  &  &  &  &  &  &  &  &  &  &  &  &  $\bar{b}_R$ \\
    -2000004 &  &  & 2.0 & 2.7 & 2.9 & 2.9 &  &  &  &  & 1.3 & 1.7 & 1.7 & 1.8 & 3.5 & 3.9 & 2.7 & 2.4 &  &  & 3.4 & 3.4 &  &  &  &  &  & 2.5 & 3.1 & 1.6 & 5.2 &  &  &  $\bar{c}_R$ \\
    -2000003 &  &  & 2.7 & 2.6 & 3.2 & 3.2 &  &  &  &  & 1.6 & 2.0 & 2.1 & 2.1 & 3.8 & 4.2 & 3.0 & 2.7 &  &  & 3.8 & 3.1 &  &  &  &  &  & 2.8 & 3.3 & 5.2 & 1.6 &  &  &  $\bar{s}_R$ \\
    -2000002 &  &  & 2.9 & 3.2 & 2.9 & 3.3 &  &  &  &  & 1.8 & 2.1 & 2.3 & 2.3 & 3.9 & 4.3 & 3.2 & 2.9 &  &  & 3.9 & 3.8 &  &  & 0.0 &  &  & 3.0 & 5.2 & 2.1 & 2.0 &  &  &  $\bar{u}_R$ \\
    -2000001 &  &  & 2.9 & 3.2 & 3.3 & 3.0 &  &  &  &  & 1.8 & 2.2 & 2.3 & 2.3 & 4.0 & 4.4 & 3.2 & 2.8 &  &  & 3.9 & 3.3 &  &  &  &  &  & 5.2 & 3.5 & 2.2 & 1.9 &  &  &  $\bar{d}_R$ \\
    -1000037 &  &  &  &  &  &  &  &  &  &  &  &  &  &  &  &  &  &  &  &  &  &  &  &  &  &  &  &  &  &  &  &  &  &  $\chi_2^-$ \\
    -1000024 &  &  &  &  &  &  &  &  &  &  &  &  &  &  &  & 0.5 &  &  &  &  &  &  &  &  &  &  &  &  &  &  &  &  &  &  $\chi_1^-$ \\
    -1000006 &  &  &  &  &  &  &  &  &  &  &  &  &  &  &  &  &  &  &  &  &  &  &  &  &  &  &  &  &  &  &  &  &  &  $\bar{t}_1$ \\
    -1000005 &  &  &  &  &  &  &  &  &  &  &  &  &  &  &  &  &  &  &  &  &  &  &  &  &  &  &  &  &  &  &  &  &  &  $\bar{b}_1$ \\
    -1000004 &  &  & 1.3 & 1.6 & 1.8 & 1.8 &  &  &  &  & 2.0 & 2.7 & 2.8 & 2.9 & 2.5 & 2.9 & 2.8 & 5.2 &  &  & 3.4 & 2.2 &  &  &  &  &  & 3.5 & 3.9 & 2.7 & 2.4 &  &  &  $\bar{c}_L$ \\
    -1000003 &  &  & 1.7 & 2.0 & 2.1 & 2.2 &  &  &  &  & 2.7 & 2.6 & 3.2 & 3.2 & 2.8 & 3.2 & 5.2 & 3.2 &  &  & 3.8 & 2.5 &  &  &  &  &  & 3.8 & 4.2 & 3.0 & 2.7 &  &  &  $\bar{s}_L$ \\
    -1000002 &  &  & 1.7 & 2.1 & 2.3 & 2.3 &  &  &  &  & 2.8 & 3.2 & 2.9 & 3.3 & 3.0 & 5.2 & 2.1 & 1.8 &  &  & 3.9 & 2.6 &  &  &  &  &  & 3.9 & 4.3 & 3.1 & 2.8 &  &  &  $\bar{u}_L$ \\
    -1000001 &  &  & 1.8 & 2.1 & 2.3 & 2.3 &  &  &  &  & 2.9 & 3.2 & 3.3 & 2.9 & 5.2 & 3.4 & 2.1 & 1.7 &  &  & 3.9 & 2.7 &  &  &  &  &  & 4.0 & 4.4 & 3.2 & 2.9 &  &  &  $\bar{d}_L$ \\
    1000001 &  &  & 3.5 & 3.8 & 3.9 & 4.0 &  &  &  &  & 2.5 & 2.8 & 3.0 & 5.2 & 4.2 & 5.1 & 3.8 & 3.6 &  &  & 4.7 & 3.2 &  & 0.3 &  &  &  & 3.6 & 4.1 & 2.8 & 2.5 &  &  &  $d_L$ \\
    1000002 &  &  & 3.9 & 4.2 & 4.3 & 4.4 &  & 0.5 &  &  & 2.9 & 3.2 & 5.2 & 3.4 & 5.1 & 5.0 & 4.2 & 3.9 &  &  & 5.1 & 3.6 &  &  & 0.3 &  &  & 4.0 & 4.5 & 3.2 & 2.9 &  &  &  $u_L$ \\
    1000003 &  &  & 2.7 & 3.0 & 3.2 & 3.2 &  &  &  &  & 2.8 & 5.2 & 2.1 & 2.1 & 3.8 & 4.2 & 2.6 & 2.8 &  &  & 3.8 & 2.5 &  &  &  &  &  & 2.8 & 3.2 & 1.9 & 1.7 &  &  &  $s_L$ \\
    1000004 &  &  & 2.4 & 2.7 & 2.9 & 2.8 &  &  &  &  & 5.2 & 3.2 & 1.8 & 1.7 & 3.6 & 3.9 & 2.8 & 2.0 &  &  & 3.4 & 2.1 &  &  &  &  &  & 2.5 & 2.9 & 1.6 & 1.4 &  &  &  $c_L$ \\
    1000005 &  &  &  &  &  &  &  &  &  &  &  &  &  &  &  &  &  &  &  &  &  &  &  &  &  &  &  &  &  &  &  &  &  &  $b_1$ \\
    1000006 &  &  &  &  &  &  &  &  &  &  &  &  &  &  &  &  &  &  &  &  &  &  &  &  &  &  &  &  &  &  &  &  &  &  $t_1$ \\
    1000021 &  &  & 3.4 & 3.8 & 3.9 & 3.9 &  &  &  &  & 3.4 & 3.8 & 3.9 & 3.9 & 4.7 & 5.1 & 3.8 & 3.4 &  &  & 6.3 & 4.3 &  &  &  &  &  & 4.7 & 5.1 & 3.8 & 3.4 &  &  &  ${g}$ \\
    1000022 &  &  & 3.4 & 3.1 & 3.8 & 3.3 &  &  &  &  & 2.2 & 2.5 & 2.6 & 2.7 & 3.2 & 3.6 & 2.5 & 2.1 &  &  & 4.3 &  &  &  &  &  &  & 3.8 & 4.8 & 3.1 & 3.4 &  &  &  $\chi_1^0$ \\
    1000023 &  &  &  &  &  &  &  &  &  &  &  &  &  &  &  &  &  &  &  &  &  &  &  &  &  &  &  &  &  &  &  &  &  &  $\chi_2^0$ \\
    1000024 &  &  &  &  &  &  &  &  &  &  &  &  &  &  & 0.3 &  &  &  &  &  &  &  &  &  &  &  &  &  &  &  &  &  &  &  $\chi_1^+$ \\
    1000025 &  &  &  &  & 0.0 &  &  &  &  &  &  &  &  &  &  & 0.3 &  &  &  &  &  &  &  &  &  &  &  &  & 0.7 &  &  &  &  &  $\chi_3^0$ \\
    1000035 &  &  &  &  &  &  &  &  &  &  &  &  &  &  &  &  &  &  &  &  &  &  &  &  &  &  &  &  &  &  &  &  &  &  $\chi_4^0$ \\
    1000037 &  &  &  &  &  &  &  &  &  &  &  &  &  &  &  &  &  &  &  &  &  &  &  &  &  &  &  &  &  &  &  &  &  &  $\chi_2^+$ \\
    2000001 &  &  & 2.5 & 2.8 & 3.0 & 5.2 &  &  &  &  & 3.5 & 3.8 & 3.9 & 4.0 & 3.6 & 4.0 & 2.8 & 2.5 &  &  & 4.7 & 3.8 &  &  &  &  &  & 4.1 & 4.9 & 3.8 & 3.5 &  &  &  $d_R$ \\
    2000002 &  &  & 3.1 & 3.3 & 5.2 & 3.5 &  &  &  &  & 3.9 & 4.2 & 4.3 & 4.4 & 4.1 & 4.5 & 3.2 & 2.9 &  &  & 5.1 & 4.8 &  &  & 0.7 &  &  & 4.9 & 5.0 & 4.2 & 3.9 &  &  &  $u_R$ \\
    2000003 &  &  & 1.6 & 5.2 & 2.1 & 2.2 &  &  &  &  & 2.7 & 3.0 & 3.1 & 3.2 & 2.8 & 3.2 & 1.9 & 1.6 &  &  & 3.8 & 3.1 &  &  &  &  &  & 3.8 & 4.2 & 2.6 & 2.7 &  &  &  $s_R$ \\
    2000004 &  &  & 5.2 & 1.6 & 2.0 & 1.9 &  &  &  &  & 2.4 & 2.7 & 2.8 & 2.9 & 2.5 & 2.9 & 1.7 & 1.4 &  &  & 3.4 & 3.4 &  &  &  &  &  & 3.5 & 3.9 & 2.7 & 2.0 &  &  &  $c_R$ \\
    2000005 &  &  &  &  &  &  &  &  &  &  &  &  &  &  &  &  &  &  &  &  &  &  &  &  &  &  &  &  &  &  &  &  &  &  $b_R$ \\
    2000006 &  &  &  &  &  &  &  &  &  &  &  &  &  &  &  &  &  &  &  &  &  &  &  &  &  &  &  &  &  &  &  &  &  &  $t_R$ \\
    \hline
     &  $\bar{t}_R$  &  $\bar{b}_R$  &  $\bar{c}_R$  &  $\bar{s}_R$  &  $\bar{u}_R$  &  $\bar{d}_R$  &  $\chi_2^-$  &  $\chi_1^-$  &  $\bar{t}_1$  &  $\bar{b}_1$  &  $\bar{c}_L$  &  $\bar{s}_L$  &  $\bar{u}_L$  &  $\bar{d}_L$  &  $d_L$  &  $u_L$  &  $s_L$  &  $c_L$  &  $b_1$  &  $t_1$  &  ${g}$  &  $\chi_1^0$  &  $\chi_2^0$  &  $\chi_1^+$  &  $\chi_3^0$  &  $\chi_4^0$  &  $\chi_2^+$  &  $d_R$  &  $u_R$  &  $s_R$  &  $c_R$  &  $b_R$  &  $t_R$  & \\
    \end{tabular}
    \caption{
      Overview of all pairs of supersymmetric particles found in the final states in the Monte Carlo samples comprising the MSSM grid.
      The numbers give the decadic logarithm of the number of events in which the two supersymmetric particles,
      given by the respective row and column,
      have been found.
      The tilde commonly used to mark supersymmetric partners of Standard Model particles has been suppressed here.
      All final states fall into one of the classes defined in \Tab{tab:analysis_subprocesses_ids}.
    }
    \label{tab:susygrid_missing_states_mssm}
  \end{table}
  \setlength{\tabcolsep}{6pt}
  \newcommand{\hl}[1]{\cellcolor{lightblue}#1}
  \begin{table}
    \centering
    \tiny
    \setlength{\tabcolsep}{4pt}
    \begin{tabular}{r|rrrrrrrrrrrrrrrrrrrrrrrrrrrrrrrrr|l}
    & \Vert{-2000006} & \Vert{-2000005} & \Vert{-2000004} & \Vert{-2000003} & \Vert{-2000002} & \Vert{-2000001} & \Vert{-1000037} & \Vert{-1000024} & \Vert{-1000006} & \Vert{-1000005} & \Vert{-1000004} & \Vert{-1000003} & \Vert{-1000002} & \Vert{-1000001} & \Vert{1000001} & \Vert{1000002} & \Vert{1000003} & \Vert{1000004} & \Vert{1000005} & \Vert{1000006} & \Vert{1000021} & \Vert{1000022} & \Vert{1000023} & \Vert{1000024} & \Vert{1000025} & \Vert{1000035} & \Vert{1000037} & \Vert{2000001} & \Vert{2000002} & \Vert{2000003} & \Vert{2000004} & \Vert{2000005} & \Vert{2000006} & \\
    \hline
    -2000006 &  &  &  &  &  &  &  &  &  &  &  &  &  & 0.0 &  & \hl{-2.1} &  & \hl{-1.3} & \hl{-1.8} & \hl{-1.5} &  & \hl{-2.6} &  & \hl{-2.9} &  &  & \hl{-1.5} &  &  &  &  & 0.0 & 3.8 &  $\bar{t}_R$ \\
    -2000005 &  & \hl{-1.2} & \hl{-1.9} & \hl{-2.2} & \hl{-2.4} & \hl{-2.4} &  &  &  & \hl{-1.2} & \hl{-1.5} & \hl{-1.8} & \hl{-2.0} & \hl{-2.0} & 3.2 & 3.6 & 2.2 & 1.7 & 1.8 & \hl{-0.8} & \hl{-3.9} & \hl{-2.1} & \hl{-1.3} & 0.0 &  & 0.0 &  & 2.8 & 3.3 & 1.8 & 1.4 & 4.1 & \hl{-0.3} &  $\bar{b}_R$ \\
    -2000004 &  & \hl{-1.9} & 1.6 & 2.5 & 2.7 & 2.6 &  &  &  & \hl{-1.4} & 1.5 & 1.8 & 2.1 & 2.1 & 3.4 & 3.8 & 2.4 & 2.0 & 1.9 &  & 4.1 & 2.7 & 1.9 &  &  & 0.3 &  & 3.0 & 3.4 & 2.0 & 4.1 & 1.3 &  &  $\bar{c}_R$ \\
    -2000003 &  & \hl{-2.2} & 2.5 & 2.4 & 3.0 & 3.0 &  &  &  & \hl{-1.9} & 1.9 & 2.4 & 2.5 & 2.5 & 3.7 & 4.2 & 2.8 & 2.5 & 2.4 &  & 4.5 & 2.5 & 1.6 &  &  & 0.0 &  & 3.3 & 3.8 & 4.1 & 1.9 & 1.9 &  &  $\bar{s}_R$ \\
    -2000002 &  & \hl{-2.4} & 2.7 & 3.0 & 2.8 & 3.2 &  &  &  & \hl{-2.1} & 2.2 & 2.5 & 2.7 & 2.7 & 3.9 & 4.3 & 3.0 & 2.6 & 2.6 &  & 4.7 & 3.2 & 2.3 & 0.0 & 0.3 & 1.0 &  & 3.5 & 4.3 & 2.6 & 2.2 & 2.0 &  &  $\bar{u}_R$ \\
    -2000001 &  & \hl{-2.4} & 2.6 & 3.0 & 3.2 & 2.8 &  &  &  & \hl{-2.0} & 2.1 & 2.6 & 2.7 & 2.7 & 3.9 & 4.3 & 3.0 & 2.7 & 2.6 &  & 4.7 & 2.7 & 1.6 &  &  & 0.3 &  & 4.2 & 3.9 & 2.6 & 2.1 & 2.1 &  &  $\bar{d}_R$ \\
    -1000037 &  &  &  &  &  &  &  &  &  &  &  & 2.0 &  & 2.4 &  & 2.8 &  & 2.0 &  & \hl{-3.4} & 2.8 & \hl{-1.5} &  &  &  &  &  &  &  &  &  &  & \hl{-1.5} &  $\chi_2^-$ \\
    -1000024 &  &  &  &  &  &  &  &  &  &  &  & 3.0 &  & 3.5 &  & 4.3 &  & 3.5 &  & \hl{-2.4} & 4.4 & 0.0 &  &  &  &  &  &  &  &  &  &  & \hl{-2.9} &  $\chi_1^-$ \\
    -1000006 &  &  &  &  &  &  &  &  &  &  &  & \hl{-1.2} &  & \hl{-1.7} &  & \hl{-2.0} &  & \hl{-1.0} & \hl{-1.8} & 5.2 &  & \hl{-2.1} &  & \hl{-2.4} &  &  & \hl{-3.4} &  &  &  &  & \hl{-0.5} & \hl{-1.4} &  $\bar{t}_1$ \\
    -1000005 &  & \hl{-1.2} & \hl{-1.4} & \hl{-1.9} & \hl{-2.1} & \hl{-2.0} &  &  &  & \hl{-1.4} & \hl{-1.9} & \hl{-2.4} & \hl{-2.5} & \hl{-2.5} & 2.9 & 3.4 & 1.9 & 1.4 & 4.3 & \hl{-2.2} & \hl{-4.1} & \hl{-2.5} & \hl{-2.8} & 0.0 & \hl{-0.8} & \hl{-1.5} &  & 3.3 & 3.8 & 2.3 & 2.0 & 1.9 & \hl{-2.2} &  $\bar{b}_1$ \\
    -1000004 &  & \hl{-1.5} & 1.5 & 1.9 & 2.2 & 2.1 &  &  &  & \hl{-1.9} & 1.6 & 2.4 & 2.6 & 2.6 & 2.9 & 3.5 & 2.0 & 4.0 & 1.4 & \hl{-1.2} & 4.1 & 2.0 & 2.8 & 3.5 & 0.3 & 1.4 & 1.9 & 3.4 & 3.8 & 2.4 & 2.0 & 1.8 & \hl{-1.2} &  $\bar{c}_L$ \\
    -1000003 &  & \hl{-1.8} & 1.8 & 2.4 & 2.5 & 2.6 & 2.0 & 3.0 & \hl{-1.2} & \hl{-2.4} & 2.4 & 2.4 & 2.9 & 3.0 & 3.4 & 3.7 & 4.0 & 2.3 & 1.9 &  & 4.5 & 2.8 & 3.0 & 0.5 & 1.0 & 1.8 &  & 3.7 & 4.1 & 2.8 & 2.4 & 2.1 &  &  $\bar{s}_L$ \\
    -1000002 &  & \hl{-2.0} & 2.1 & 2.5 & 2.7 & 2.7 &  &  &  & \hl{-2.5} & 2.6 & 2.9 & 2.8 & 3.1 & 3.4 & 4.3 & 2.5 & 2.4 & 2.2 & \hl{-1.3} & 4.7 & 2.3 & 3.3 & 3.7 & 0.6 & 1.9 & 2.1 & 3.9 & 4.3 & 3.0 & 2.7 & 2.4 & \hl{-1.2} &  $\bar{u}_L$ \\
    -1000001 & 0.0 & \hl{-2.0} & 2.1 & 2.5 & 2.7 & 2.7 & 2.4 & 3.5 & \hl{-1.7} & \hl{-2.5} & 2.6 & 3.0 & 3.1 & 2.7 & 4.1 & 3.8 & 2.5 & 2.2 & 2.2 &  & 4.7 & 2.9 & 3.2 & 1.0 & 1.0 & 2.0 &  & 3.9 & 4.3 & 3.0 & 2.6 & 2.4 &  &  $\bar{d}_L$ \\
    1000001 &  & 3.2 & 3.4 & 3.7 & 3.9 & 3.9 &  &  &  & 2.9 & 2.9 & 3.4 & 3.4 & 4.1 & 4.2 & 5.0 & 3.7 & 3.5 & \hl{-3.3} & \hl{-2.8} & 5.4 & 3.5 & 3.9 & 4.6 & 1.7 & 2.7 & 3.6 & 4.2 & 4.6 & 3.3 & 2.9 & \hl{-2.8} & \hl{-1.8} &  $d_L$ \\
    1000002 & \hl{-2.1} & 3.6 & 3.8 & 4.2 & 4.3 & 4.3 & 2.8 & 4.3 & \hl{-2.0} & 3.4 & 3.5 & 3.7 & 4.3 & 3.8 & 5.0 & 5.0 & 4.1 & 3.8 & \hl{-3.8} &  & 5.9 & 3.2 & 4.4 & 0.9 & 1.9 & 3.0 &  & 4.6 & 5.0 & 3.7 & 3.4 & \hl{-3.2} &  &  $u_L$ \\
    1000003 &  & 2.2 & 2.4 & 2.8 & 3.0 & 3.0 &  &  &  & 1.9 & 2.0 & 4.0 & 2.5 & 2.5 & 3.7 & 4.1 & 2.4 & 2.5 & \hl{-2.4} & \hl{-1.1} & 4.5 & 2.6 & 3.0 & 3.0 & 0.8 & 1.7 & 1.9 & 3.3 & 3.7 & 2.3 & 1.9 & \hl{-1.8} &  &  $s_L$ \\
    1000004 & \hl{-1.3} & 1.7 & 2.0 & 2.5 & 2.6 & 2.7 & 2.0 & 3.5 & \hl{-1.0} & 1.4 & 4.0 & 2.3 & 2.4 & 2.2 & 3.5 & 3.8 & 2.5 & 1.7 & \hl{-1.9} &  & 4.1 & 1.7 & 2.8 & 0.0 & 0.0 & 1.5 &  & 2.9 & 3.4 & 1.9 & 1.4 & \hl{-1.3} &  &  $c_L$ \\
    1000005 & \hl{-1.8} & 1.8 & 1.9 & 2.4 & 2.6 & 2.6 &  &  & \hl{-1.8} & 4.3 & 1.4 & 1.9 & 2.2 & 2.2 & \hl{-3.3} & \hl{-3.8} & \hl{-2.4} & \hl{-1.9} & \hl{-1.4} &  & \hl{-4.1} & \hl{-2.4} & \hl{-2.7} &  & \hl{-0.9} & \hl{-1.6} &  & \hl{-2.9} & \hl{-3.4} & \hl{-1.9} & \hl{-1.4} & \hl{-1.2} &  &  $b_1$ \\
    1000006 & \hl{-1.5} & \hl{-0.8} &  &  &  &  & \hl{-3.4} & \hl{-2.4} & 5.2 & \hl{-2.2} & \hl{-1.2} &  & \hl{-1.3} &  & \hl{-2.8} &  & \hl{-1.1} &  &  &  &  & \hl{-2.1} &  &  &  &  &  &  &  &  &  &  &  &  $t_1$ \\
    1000021 &  & \hl{-3.9} & 4.1 & 4.5 & 4.7 & 4.7 & 2.8 & 4.4 &  & \hl{-4.1} & 4.1 & 4.5 & 4.7 & 4.7 & 5.4 & 5.9 & 4.5 & 4.1 & \hl{-4.1} &  & 6.4 & 4.5 & 4.6 & 4.8 & 2.7 & 2.9 & 3.3 & 5.5 & 5.9 & 4.5 & 4.1 & \hl{-3.9} &  &  ${g}$ \\
    1000022 & \hl{-2.6} & \hl{-2.1} & 2.7 & 2.5 & 3.2 & 2.7 & \hl{-1.5} & 0.0 & \hl{-2.1} & \hl{-2.5} & 2.0 & 2.8 & 2.3 & 2.9 & 3.5 & 3.2 & 2.6 & 1.7 & \hl{-2.4} & \hl{-2.1} & 4.5 & \hl{-0.5} & \hl{-0.3} & \hl{-0.5} & \hl{-1.1} & \hl{-1.3} & \hl{-1.5} & 3.3 & 4.3 & 2.4 & 2.6 & \hl{-1.9} & \hl{-2.1} &  $\chi_1^0$ \\
    1000023 &  & \hl{-1.3} & 1.9 & 1.6 & 2.3 & 1.6 &  &  &  & \hl{-2.8} & 2.8 & 3.0 & 3.3 & 3.2 & 3.9 & 4.4 & 3.0 & 2.8 & \hl{-2.7} &  & 4.6 & \hl{-0.3} & 0.0 & 0.0 &  &  &  & 2.3 & 3.3 & 1.5 & 1.9 & \hl{-1.4} &  &  $\chi_2^0$ \\
    1000024 & \hl{-2.9} & 0.0 &  &  & 0.0 &  &  &  & \hl{-2.4} & 0.0 & 3.5 & 0.5 & 3.7 & 1.0 & 4.6 & 0.9 & 3.0 & 0.0 &  &  & 4.8 & \hl{-0.5} & 0.0 & 0.0 &  &  &  &  & 0.6 &  &  &  & \hl{-0.3} &  $\chi_1^+$ \\
    1000025 &  &  &  &  & 0.3 &  &  &  &  & \hl{-0.8} & 0.3 & 1.0 & 0.6 & 1.0 & 1.7 & 1.9 & 0.8 & 0.0 & \hl{-0.9} &  & 2.7 & \hl{-1.1} &  &  &  &  &  & 0.0 & 1.5 &  &  &  &  &  $\chi_3^0$ \\
    1000035 &  & 0.0 & 0.3 & 0.0 & 1.0 & 0.3 &  &  &  & \hl{-1.5} & 1.4 & 1.8 & 1.9 & 2.0 & 2.7 & 3.0 & 1.7 & 1.5 & \hl{-1.6} &  & 2.9 & \hl{-1.3} &  &  &  &  &  & 1.1 & 2.0 & 0.3 & 0.0 &  &  &  $\chi_4^0$ \\
    1000037 & \hl{-1.5} &  &  &  &  &  &  &  & \hl{-3.4} &  & 1.9 &  & 2.1 &  & 3.6 &  & 1.9 &  &  &  & 3.3 & \hl{-1.5} &  &  &  &  &  &  &  &  &  &  &  &  $\chi_2^+$ \\
    2000001 &  & 2.8 & 3.0 & 3.3 & 3.5 & 4.2 &  &  &  & 3.3 & 3.4 & 3.7 & 3.9 & 3.9 & 4.2 & 4.6 & 3.3 & 2.9 & \hl{-2.9} &  & 5.5 & 3.3 & 2.3 &  & 0.0 & 1.1 &  & 4.2 & 5.0 & 3.8 & 3.4 & \hl{-3.2} &  &  $d_R$ \\
    2000002 &  & 3.3 & 3.4 & 3.8 & 4.3 & 3.9 &  &  &  & 3.8 & 3.8 & 4.1 & 4.3 & 4.3 & 4.6 & 5.0 & 3.7 & 3.4 & \hl{-3.4} &  & 5.9 & 4.3 & 3.3 & 0.6 & 1.5 & 2.0 &  & 5.0 & 5.0 & 4.2 & 3.9 & \hl{-3.7} &  &  $u_R$ \\
    2000003 &  & 1.8 & 2.0 & 4.1 & 2.6 & 2.6 &  &  &  & 2.3 & 2.4 & 2.8 & 3.0 & 3.0 & 3.3 & 3.7 & 2.3 & 1.9 & \hl{-1.9} &  & 4.5 & 2.4 & 1.5 &  &  & 0.3 &  & 3.8 & 4.2 & 2.5 & 2.5 & \hl{-2.3} &  &  $s_R$ \\
    2000004 &  & 1.4 & 4.1 & 1.9 & 2.2 & 2.1 &  &  &  & 2.0 & 2.0 & 2.4 & 2.7 & 2.6 & 2.9 & 3.4 & 1.9 & 1.4 & \hl{-1.4} &  & 4.1 & 2.6 & 1.9 &  &  & 0.0 &  & 3.4 & 3.9 & 2.5 & 1.6 & \hl{-1.8} &  &  $c_R$ \\
    2000005 & 0.0 & 4.1 & 1.3 & 1.9 & 2.0 & 2.1 &  &  & \hl{-0.5} & 1.9 & 1.8 & 2.1 & 2.4 & 2.4 & \hl{-2.8} & \hl{-3.2} & \hl{-1.8} & \hl{-1.3} & \hl{-1.2} &  & \hl{-3.9} & \hl{-1.9} & \hl{-1.4} &  &  &  &  & \hl{-3.2} & \hl{-3.7} & \hl{-2.3} & \hl{-1.8} & \hl{-1.3} &  &  $b_R$ \\
    2000006 & 3.8 & \hl{-0.3} &  &  &  &  & \hl{-1.5} & \hl{-2.9} & \hl{-1.4} & \hl{-2.2} & \hl{-1.2} &  & \hl{-1.2} &  & \hl{-1.8} &  &  &  &  &  &  & \hl{-2.1} &  & \hl{-0.3} &  &  &  &  &  &  &  &  &  &  $t_R$ \\
    \hline
     &  $\bar{t}_R$  &  $\bar{b}_R$  &  $\bar{c}_R$  &  $\bar{s}_R$  &  $\bar{u}_R$  &  $\bar{d}_R$  &  $\chi_2^-$  &  $\chi_1^-$  &  $\bar{t}_1$  &  $\bar{b}_1$  &  $\bar{c}_L$  &  $\bar{s}_L$  &  $\bar{u}_L$  &  $\bar{d}_L$  &  $d_L$  &  $u_L$  &  $s_L$  &  $c_L$  &  $b_1$  &  $t_1$  &  ${g}$  &  $\chi_1^0$  &  $\chi_2^0$  &  $\chi_1^+$  &  $\chi_3^0$  &  $\chi_4^0$  &  $\chi_2^+$  &  $d_R$  &  $u_R$  &  $s_R$  &  $c_R$  &  $b_R$  &  $t_R$  & \\
    \end{tabular}
    \caption{
      Overview of all pairs of supersymmetric particles found in the final states in the Monte Carlo samples comprising the mSUGRA grid.
      The numbers give the decadic logarithm of the number of events in which the two supersymmetric particles,
      given by the respective row and column,
      have been found.
      Negative signs and the background color
      indicate states which do not fall into any of the classes defined in \Tab{tab:analysis_subprocesses_ids}.
      The tilde commonly used to mark supersymmetric partners of Standard Model particles has been suppressed here.
    }
    \label{tab:susygrid_missing_states_tanbeta3}
  \end{table}
  \setlength{\tabcolsep}{6pt}
\end{landscape}

\newpage
\subsection{Overview of Final States in the Monte Carlo Signal Grids}
\label{sec:appendix_overview_final_state_counts}

\Tabs{tab:susygrid_missing_states_mssm} and \ref{tab:susygrid_missing_states_tanbeta3} give an overview
in which the frequency is listed at which given pairs of supersymmetric particles
are found in the final states from all the Monte Carlo samples comprising the two signal grids
discussed in \Sec{sec:results_0lanalysis_final_state_dependent_xsec}.
The rows and columns are labelled both with the particle numbers
from the Monte Carlo particle numbering scheme of the Particle Data Group \cite{PDB2010}
and the usual particle symbols.
Note that the scalar quarks, apart from the scalar top and bottom quark,
are written using indices indicating the chirality. %
In the particle numbering scheme from the Particle Data Group,
the particle numbers $\pm1000005$, $\pm2000005$, $\pm1000006$ and $\pm2000006$
stand for the mixed states $\tilde{b}_{1,2}$ and $\tilde{t}_{1,2}$, respectively,
the lighter mixed state having the smaller number.

From \Tab{tab:susygrid_missing_states_mssm},
it can be seen that almost all final states fall into the classes with IDs~1 through~4
($\squark\gluino$, $\gluino\gluino$, $\squark\squark$ and $\squark\bar{\squark}$),
for which the cross sections are known.
Thus, only a negligibly small number of events has to be discarded due to missing cross section information.
This is different for the mSUGRA grid,
for which the final state counts are shown in \Tab{tab:susygrid_missing_states_tanbeta3}.
Here, for a larger number of final states cross sections have been computed,
but still there are combinations which do not fall into any of the classes defined in \Tab{tab:analysis_subprocesses_ids}.
They are marked in the table with a negative sign.
The impact of these states is discussed in \Secs{sec:results_0lanalysis_final_state_dependent_xsec} and \ref{sec:appendix_impact_unidentified_events}.

\subsection{Impact of Unidentified Events}
\label{sec:appendix_impact_unidentified_events}
\begin{figure}
  \centering
  \incgraphics{width=\widthtwoplots}{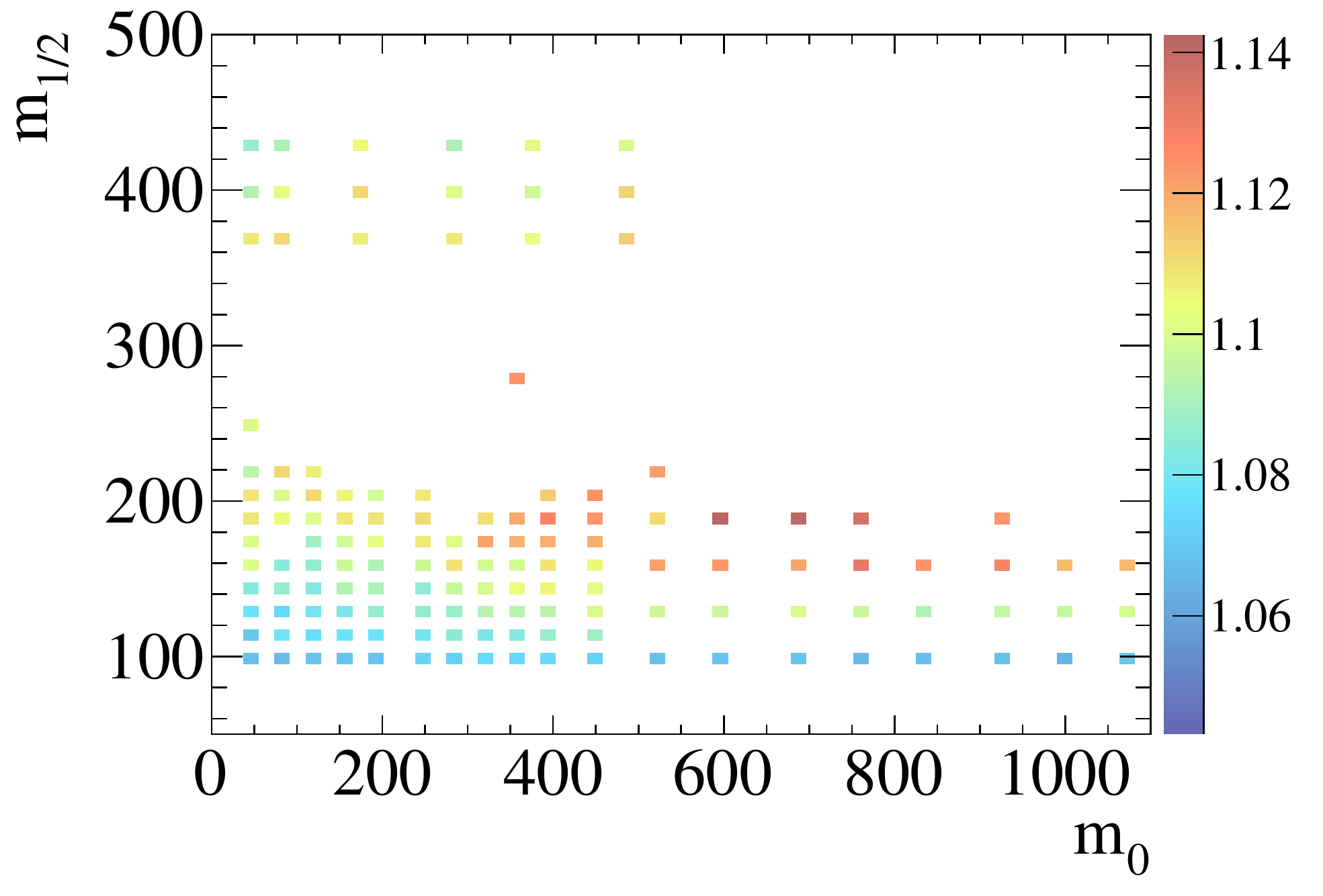}
  \hfill
  \incgraphics{width=\widthtwoplots}{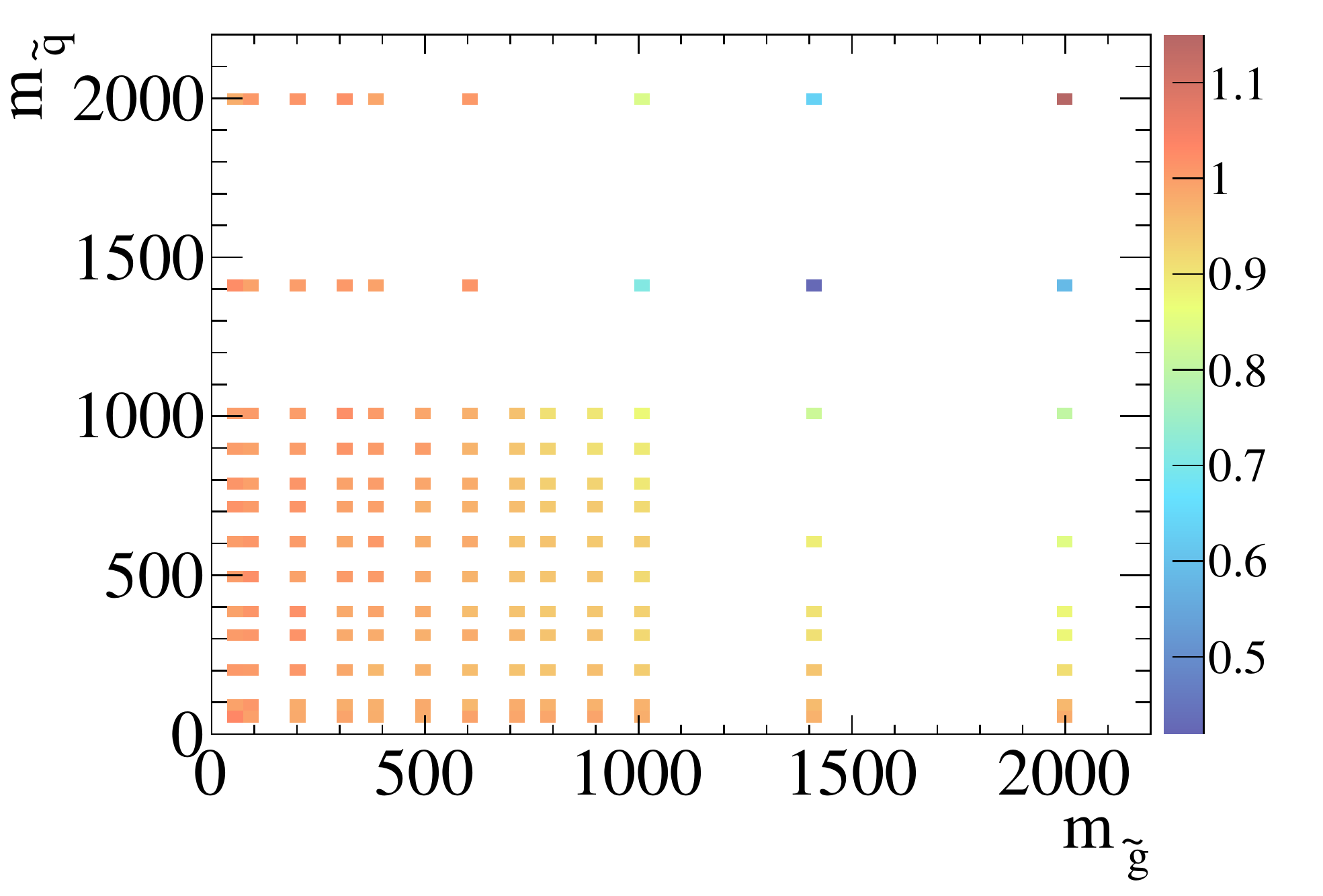}
  \caption{
    Ratio of the leading-order cross section for all identified final states over the generator cross section.
    Left: mSUGRA grid, right: MSSM grid.
  }
  \label{fig:analysis_impact_neglected_events_xsec_ratio}
\end{figure}

The approach in \Sec{sec:analysis_SUSY_GRIDs} does not take the cross section into account,
and it is conceivable that if the ignored subprocesses have a comparably large cross section,
the fraction of neglected processes is much larger in terms of cross section than in terms of event counts.
Therefore, as a second cross-check one can try to estimate the cross section of the neglected processes.
The \ac{NLO} cross section of the full Monte Carlo sample cannot be used for this,
because if it were known, the problem would not have arised in the beginning.
But a computation of the cross section of the full sample is available
at \ac{LO} from the \ac{MC} generator \propername{Herwig++} via the \propername{AMI}.
Of course, the LO generator cross section also only includes
the processes that have been included in the event generation.

\Fig{fig:analysis_impact_neglected_events_xsec_ratio} shows the ratio 
of the LO cross section for all identified final states over the generator cross section.
The LO cross section for all identified final states is computed using
the \kfactor between the \propername{Herwig++} LO cross section and \propername{Prospino} NLO cross section for the MSSM grid %
and the \propername{Prospino} LO-NLO \kfactor for the mSUGRA grid,
for which the \kfactor with respect to \propername{Herwig++}
is not provided in the cross section file. %

The gap in the plot for the mSUGRA grid is due to the fact that for these samples the generator cross section
for some reason has not been stored in the \propername{AMI} database.
This plot shows only values above 1, which may be due to the fact
that here the \propername{Prospino} LO-NLO \kfactor are used to compute the LO cross section from the NLO \propername{Prospino} cross section
so that the LO cross sections of two different generators, \propername{Herwig++} and \propername{Prospino}, are compared.
One could try to renormalize the ratio to restore a mean of one,
but being meant only as a second cross-check of the magnitude of an effect that in any case cannot lead to wrong results,
the spread of the values shall suffice to convey the basic message of the plot,
\ie that there are no outliers which would indicate that a large fraction of the cross section were missing.
In the right plot, which is for the MSSM grid,
the normalization seems to be more reasonable,
and the plot shows in general the expected behavior.
The ratio decreases from the left to the right and exhibits the same tendency
as the corresponding plot in \Fig{fig:analysis_impact_neglected_events},
although the magnitude of the effect appears to be larger.
In conclusion, the plots have to be interpreted with some care,
and their informative value is unfortunately not too high,
but at least they do not hint that the cross sections of the unidentified states comprise an surprisingly large fraction.

\removesection{ %

\subsection{Signal Grid Overview}
\todo{Mehr Grid-Plots? (siehe expert0lepton\_plot.py)}

\todo{Plots giving overview of basic quantities in SUSY grids, \eg xsec etc. (xsec is here now, what else? raw number of events in MC?)}
\todo{Plots: Signaleffizienzen}

\subsection{Overview of Data Periods}
\todo{List of periods for 2010 and 2011, Periods2011.ods}

\section{Implementation-Related Details}

\subsection{Cutflow per Data Period}
\todo{Tabelle mit einzelnen cutflows für Perioden A-I getrennt, erlaubt zu sehen wieviel Daten in welcher Periode genommen wurden und ggf. wie die Schnitteffizienzen jeweils sind}

\subsection{Analysis flow}
\todo{technical implementation details; beschreibe: data10 wird im GRID auf NTUP\_SUSY behandelt, Python-Skripte zur Verwaltung (Expert2010),  mc09 wird von NTUP\_SUSY geslimmed (Reduktion der Daten in Prozent?), dann lokal weiterverarbeitet, ermöglicht verschiedene Prozessierungen und einfachere Neuberechnung / andere Variationen (JER, JES etc.);  Beschreibe Verwendung von SFrame statt Athena}

\subsection{List of D3PD variables needed in the analysis}
\todo{einfache Tabelle mit Liste der Ausgabe von SUSYCommonSelection}

} %

\section{Jet Spectrum in Pythia QCD Samples}
\label{sec:appendix_MC_study_QCD_jets}

\begin{figure}
  \centering
  \incgraphics{width=\widthsingleplot}{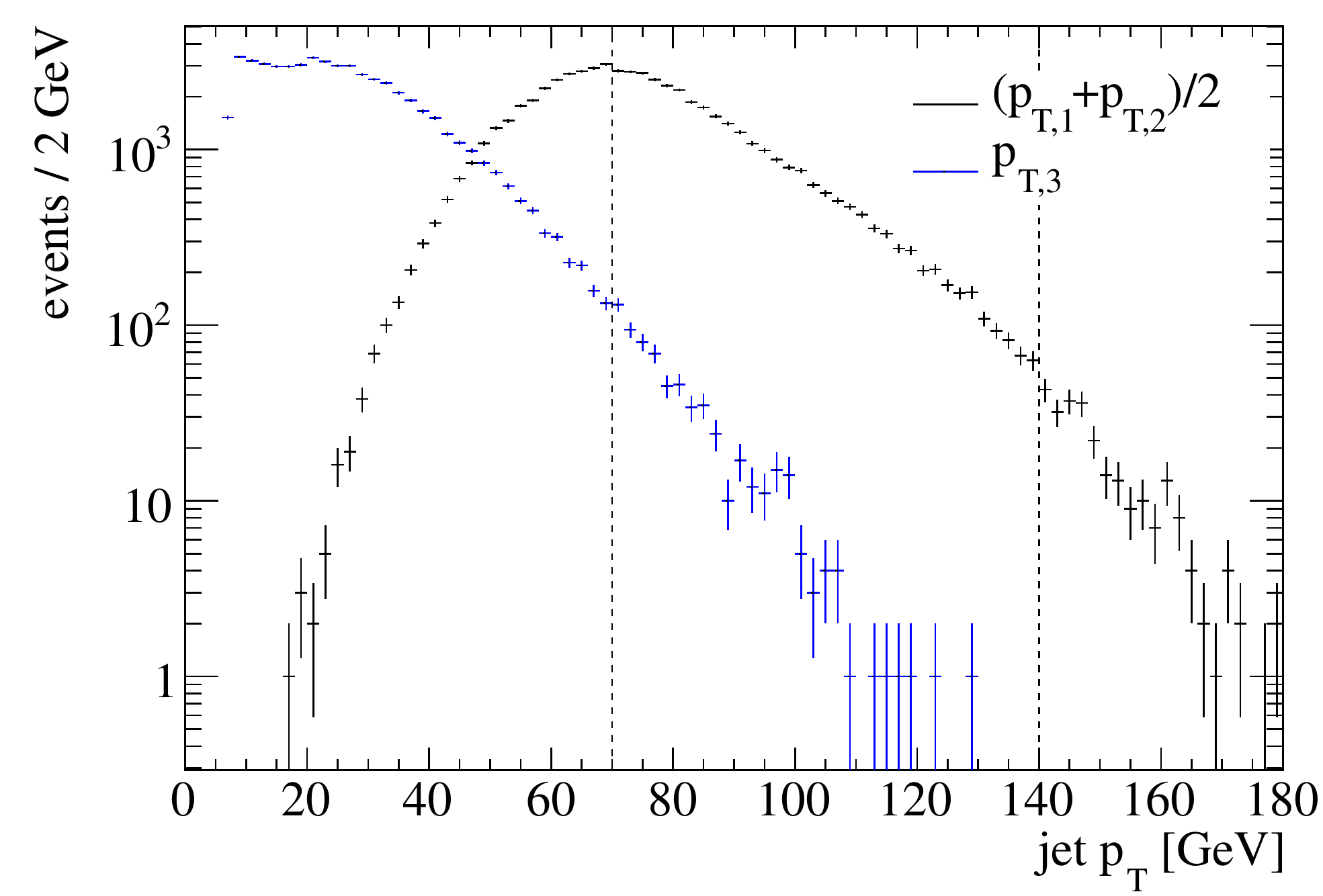}
  \caption{
    Distribution of jet \pt in the subsample J3 containing \propername{Pythia} QCD Monte Carlo.
    The solid histogram line gives the average of the first and second leading jet,
    the dashed histogram line gives the third leading jet (if present).
    The two vertical dashed lines mark the lower and upper bound of the $\hat{p}_T$ interval that is covered by J3.
  }
  \label{fig:QCD_J3_1}
\end{figure}

\Fig{fig:QCD_J3_1} shows the distribution of \pt for the leading jets in one of the subsamples of \propername{Pythia} QCD Monte Carlo.
The subsample chosen is J3 so that according to \Tab{tab:analyis_Pythia_dijet_ptslices},
it includes events with $\unit[70]{GeV} < \hat{p}_T < \unit[140]{GeV}$.
What this means in practice can be seen from the plot, together with some additional interesting observations.
The solid line in the plot is the average of the \pt of the first and second leading jet from all events of this sample.
The \pt is reconstructed offline at \EMJES scale (the \tagname{AntiKt4TopoEMJES} collection is used, cf. \Sec{sec:software_jet_reconstruction})
and thus smeared by detector effects,
some of which will lead to an underestimation of the momentum,
others to an overestimation.
If parts of the jet go into an uninstrumented region or are otherwise out of acceptance of the detector
the corresponding part of the energy will be lost and lead to a decrease of the visible momentum.
The same tendency is caused by invisible particles such as neutrinos created within the jet
that carry away momentum but cannot be detected and particles which are created at such an angle
that they lie outside the radius of the cone which is used for the reconstruction.
On the other hand, overlaps with other jets or particles which happen to fall into the same region of the calorimeter
that is summed up to give the energy of the jet
will add energy to the jet on top of the energy of the original parton,
and in addition to this, there are statistical fluctuations from the energy measurements in the calorimeter.
Another important contribution will come from initial- and final-state radiation
that smears out the effects of the original kinematics selection
already at parton level \cite{Pythia6.4}. %
All of these effects will lead to a broadening of the jet resolution,
but the overall normalization of the scale is corrected by the \EMJES factor,
so that the averaged peak correctly appears at \GeV{70},
marked by the left vertical line as the lower bound of the $\hat{p}_T$ interval.
The sorting of the reconstructed jets will lead to a systematic bias which pushes the peak
for the leading jets to higher and the peak for the second leading jets to lower values,
therefore the average of the two leading jets is used in \Fig{fig:QCD_J3_1}.
\pt values outside $\unit[70]{GeV} < \pt < \unit[140]{GeV}$ are due to mismeasurements
and the limited resolution.
This can also be seen from the fact that within this interval,
the slope is roughly constant,
but for higher \pt,
the distribution falls off much faster.
The dashed line shows that the third leading jet is already much softer,
peaking close to zero.
There is a cut-off at low \pt values,
because only jets reconstructed with a $\pt > \GeV{7}$ %
are written into the jet collections.

\section{Computing}
\label{sec:computing}

\removesection{ %
\begin{quote}
 There are two ways of constructing a software design:\\
 One way is to make it so simple that there are obviously no deficiencies,\\
 the other way is to make it so complicated that there are no obvious deficiencies.\\
 The first method is far more difficult.
 \begin{flushright}
    \textit{C.A.R. Hoare.}
 \end{flushright}
\end{quote}
} %

This section documents technical aspects of the software
which has been employed in the context of this thesis,
in particular the \ROOT framework, which is widely used in high-energy physics,
and \Athena, the framework developed and used within the \ATLAS collaboration.

\subsection{ROOT} %
\inindex{ROOT@\ROOT}
One of the most widely used general frameworks in high-energy physics in the \ROOT framework,
which was developed starting in 1994 in the context of the NA49 experiment at CERN.
It is designed to be able to handle the large amounts of data \cite{ROOT1997,ROOTUsersGuide}
which are produced in modern collider experiments.
\ROOT is written in C++ and relies on the CINT interpreter \cite{CINT} to provide a command-line interface
and script processor that accepts C++-style commands.
In addition to this interactive environment,
the core features of \ROOT include a proprietary file format
that allows to store and mine large amounts of data using vertical data storage techniques (see below),
a graphical user-interface for the visualization of data mainly in terms of histograms and projections,
and libraries that provide a wide spectrum of methods from basic mathematical functions to advanced tools for statistical analyses.
Coming from the physics community, its main area of application still lies therein.

The native ROOT file format is a compressed binary format
which is machine-independent and optimized for quick data access in large files.
ROOT files contain both the data and a description of the type of data.
An expression that has been coined in this context is to \define{persistify} data.
In object-oriented programming languages like C++,
complex data is represented as objects which are instantiated from classes.
These objects cannot only store data,
but also provide methods to process and access the data and reflect complex relations between data and objects.
The problem is that C++ does not have built-in means to save these objects from memory to file.
Of course, it is possible to store primitive data types and to dump the binary representation of a object as it is stored in memory,
but all references that are stored as pointers,
and therefore depend on the absolute position of the object in memory,
are then rendered useless.
This binary representation is also highly dependent on the platform, the compiler and revision of the code
and cannot be interpreted without access to the underlying C++ code in which the object is defined.
Objects therefore normally only exist in a transient state,
as long as the process that created them is running,
but of course the objects created as results of the processing of physical data need to be saved
to be able to distribute them and access them again later.

This is made possible by the reflection capabilities\footnote{
  In computer science, reflection means that a program can access and modify its own structure,
  in order to be able, for example, to retrieve information about the type of an object or the relative position of data members at run-time.
  The reflection database of ROOT is Reflex which provides a type description layer on top of C++ \cite{URL_Reflex}.
}
of \ROOT and the underlying CINT interpreter, respectively.
They allow to create \define{dictionaries} describing the data,
from which C++ classes can automatically be produced
which can be used to write transient objects to a file and read them back again.
Backwards-compatibility is a feature that is very important in experiments which are supposed to collect data over decades,
and is ensured by ROOT's support of schema evolution.
Another feature of the ROOT file format is that it allows to organize data in a nested folder structure within the ROOT file itself.

The data that needs to be stored in collider experiments can neatly be split into independent \define{\index{event}s},
each of which corresponds to the collection of all data attributed to particles that have been produced in the same bunch-crossing.
In practice, the association of reconstructed\footnote{
  The reconstruction algorithms are described in \Sec{sec:software_reconstruction_event_reconstruction}.
}
objects to a certain bunch crossing is not trivial at all,
due to the high frequency of collisions within the detector.
Particles which are produced in the collision
are usually highly relativistic,
but even in the short time a particle with a velocity near the speed of light needs to cross the \ATLAS detector volume,
several more interactions will have taken place
because of the targeted collision frequency of \unit[40]{MHz}.
This means that at any point in time,
the remnants of collisions from several subsequent bunch-crossings
will leave traces in the different detector systems concurrently.
Every collision event is described by the reconstructed objects that have been associated with it
and by event-wise quantities such as the missing transverse energy, %
status flags for detector components \etc.
The \index{event record} is the collection of all recorded data which belong to a single event.

\inindex{TTree@\code{TTree}}
Instead of writing objects into ROOT files directly,
these are normally organized in \code{TTree}s.
A \code{TTree} is a collection of branches,
each of which can be thought of as a vector of objects of a given type. %
All entries in all of these vectors with the same index are understood to belong to the same event.
The data of each event is therefore distributed over all vectors.
Of course, the number of reconstructed particles of a certain type may be different for every event,
and therefore the entries of these vectors are not objects, but vectors of different sizes themselves.
The advantage of storing data in this vertical structure is that branches can be switched off selectively
and the processing event by event can be considerably sped up by this.

\subsection{The Athena Framework} %

Every experiment that produces raw data in such amounts that they can only be handled with computers
needs a software framework for the evaluation of the data.
\ROOT is a general framework that is independent of the experiment.
In addition to and often on top of \ROOT,
every larger experiment in high-energy physics has its own proprietary framework
that is adapted to its specific requirements.

\inindex{Athena framework@\Athena framework}
In \ATLAS,
the main software framework is called {\Athena}\footnote{ %
  Not to be confused with one of the experiments at the Antiproton Decelerator (AD) at CERN bearing the same name,
  which successfully produced and detected cold antihydrogen atoms \cite{Athena2002}.
}.
\Athena is an enhanced version of the Gaudi framework \cite{Gaudi2001},
originally developed by the \LHCb experiment \cite{ATLASComputingTDR2005}
with the envisioned long lifetime of the software and maintainability in mind.
Being the main software of an active experiment,
\Athena is under constant development.
Roughly once a year a new major revision is released.
The releases relevant for this thesis are number 15 and 16 \cite{URL_AthenaReleases}.
Most of \Athena is coded in C++, and the configuration is done through Python scripts,
which allow changes in configuration parameters without recompiling.
This could also be done using text files, but scripts offer much more flexibility.

\Athena is written as a general framework for the wide range of applications in high-energy physics.
It is used to run detector simulations and create the Monte Carlo in the central production system,
it is used in the online system to steer the detector, in particular in the data-taking system and \ac{HLT}, %
it incorporates the event data model and the reconstruction algorithms used to evaluate the raw detector output,
is used for the reprocessing of data on the basis of improved calibrations,
and is also intended to be suitable for end-user analysis.

\begin{figure}
  \centering
  \incgraphics{width=\widthsingleplot}{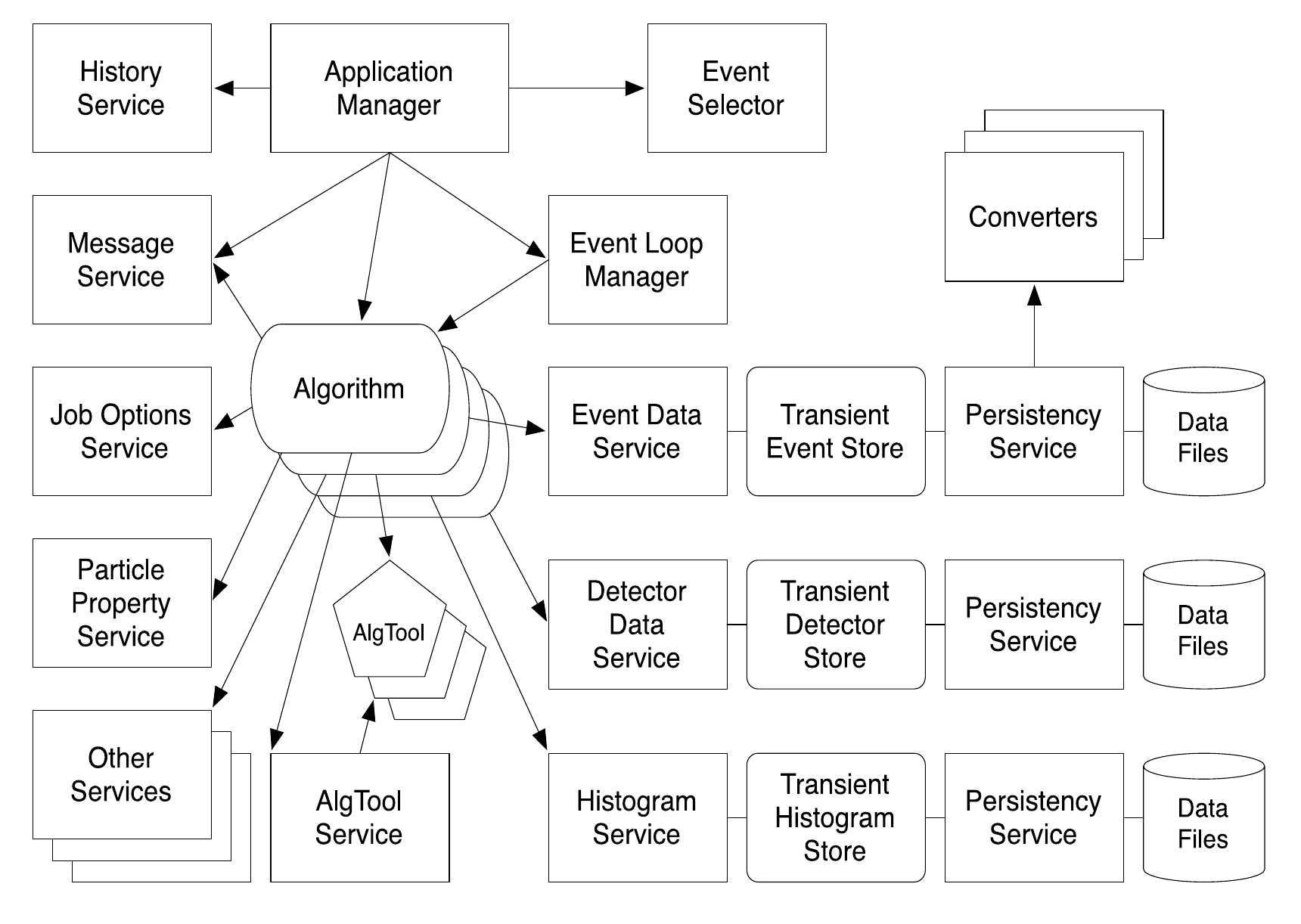}
  \caption{
    Overview of the major components of the \Athena framework and their relationships in terms of navigability and usage.
  }
  \label{fig:athena_components} %
\end{figure}
An overview of the structure and the major components of the \Athena framework is given in \Fig{fig:athena_components},
and used here to exemplify the purpose of the central parts an event-based analysis framework in general consists of.
The application manager is the central algorithm that manages and coordinates the execution of all other components.
The second important steering unit is the event loop manager,
which coordinates the processing of all events present in the input data one after the other.
All kinds of debug and log messages are fed into the message service,
which, depending on its configuration, decides which messages to present to the user.
Every event is passed through a number of algorithms,
which retrieve their configuration from the job options service
and access the data within the event through the event data service which is called \code{StoreGate} in \Athena. %
Additional event-independent information about the detector (its geometry, alignment or calibrations) 
and static data such as particle properties are provided by the detector data service, particle property and similar services.
Although \Athena is not framework used for the analysis presented in this analysis,
but instead the much lighter framework \propername{SFrame} \cite{URL_SFrame} is used,
many of these concepts can also be found there, too,
and can be generalized to most event-based analyses.

Special methods are implemented for classes which are defined in the \Athena framework,
allowing to persistify objects that were instantiated from these class through the Persistency Service (cf. \Fig{fig:athena_components}).
These persistifiers can alleviate the need to use exactly the same code revision the data was stored with,
when reading the data back from a file.
The downside of this is that most of the data formats used in \ATLAS
cannot be read using \ROOT only, \ie without the \Athena framework%
\removesection{
  \footnote{
    A compromise is \propername{AthenaROOTAccess} which allows to access files
    containing data in \Athena's native format directly from \ROOT,
    without using the (full) \Athena framework.
    However, a complete independence is not possible in general,
    as the transient classes and converters from \Athena are needed to decode the data.
  }
}.
An overview of the data formats in \ATLAS is given in the following.

\subsection{Data formats}
\label{sec:definition_dataformats}

\inindex{Data formats}
Data taken with the \ATLAS detector is made available for analysis in several formats,
which mainly differ in volume, the level of detail %
and the level of processing of the detector output.
Three main classes of formats can be distinguished.
The raw output from the detector is stored in raw data (RAW) and \ac{RDO} files,
before any reconstruction of physics objects is done.
These files contains low-level output from the detectors
which for most analyses is unnecessarily intricate to use.
From this output, physics objects are created in the reconstruction step of data processing (cf. \Sec{sec:software_reconstruction_event_reconstruction}).
These objects are stored in the \ac{ESD} and \ac{AOD} data format,
which is able to hold the description of physics objects like electrons or tracks of charged particles and the relation between these objects,
such that \eg for a muon,
the tracks associated to it in the process of reconstruction,
and in turn for the track,
the position of hits in different subdetectors can be stored and accessed in the analysis.
The \acp{AOD} are a format derived from the \acp{ESD} and contain a subset of the information.
Due to the flexibility of these formats that requires special converters for persistifying the information,
they cannot be accessed by \ROOT directly, but only in combination with the \Athena libraries.
Many analyses therefore use the simple \ac{NTUP} data format,
which obviates the need to use \Athena
because these files, also called (flat) \define{\index{n-tuple}s},
contain only simple vectors and objects that can be processed by \ROOT directly.
This comes at the price of reduced flexibility and often an again increased data volume,
because instead of being able to make links between objects,
part of the information has to be stored redundantly in different places.
Many versions of ntuples, differing by their variable content,
are automatically produced in the central production,
from the higher data formats such as AOD and ESD.
One of these ntuple variants are the SUSY ntuples, NTUP\_SUSY,
which are adapted to the requirements of the \ATLAS Supersymmetry group.
Note that this distinction is done from the point-of-view of analysis,
\ie from the objects contained within the files rather than the underlying physical file format,
which is the \index{ByteStream format} for RAW,
a format which reflects the format in which the data are delivered from the detector read-out systems,
and the \index{POOL} ROOT format for RDO, AOD and ESD files, which are object-oriented representations,
where RDO is an object-oriented representation of the ByteStream data, \ie not in terms of physics objects. %

The data formats described above, besides technical aspects in terms of accessibility,
are different compromises between data size and level of detail.
Even when using the NTUP format,
which is supposed to be the smallest representation with regard to event size,
reducing the amount of data that needs to be processed in subsequent analysis steps may be desired or required.
For the reduction of the data volume,
three different approaches are available:
\begin{itemize}
  \item Events that fail very basic cuts like the trigger requirement or (for data) the \acf*{GRL} selection
    can be discarded from the dataset.
    This is called \textit{skimming}.
    The efficiency of this reduction depends strongly on the cut and, on Monte Carlo, on the sample.
  \item Objects that do not satisfy the criteria to be identified as a certain physics object can be removed from object container.
    This is called \textit{thinning}.
    For example, jets that do not have a certain minimum \pt will not be stored in the jet object collections,
    so this kind of thinning is done right after the object reconstruction.
    From a technical point of view, thinning is more difficult to implement in flat ntuples
    because the entries connected to an object need to be removed from all branches.
  \item
    Variables, \ie object or event properties, that are not needed in the analysis can be removed (\textit{slimming} of the dataset).
    This is easily possible in flat ntuples thanks to the structure of this data format,
    where data is not stored in event objects, %
    but every variable is a vector and entries at the same offset in all vectors are assumed to belong to the same event.
    It is thus possible to remove variables by deleting the corresponding vector without having to loop over all events.
\end{itemize}
\subsection{Nomenclature of Dataset Names in ATLAS}
\label{sec:definition_nomenclature}

\inindex{Dataset nomenclature}
The general format of the name of an \ATLAS dataset is \cite{ATL-COM-GEN-2007-003}
\begin{center}
  \textit{Project.[OtherFields.]DataType.Version[/]}
\end{center}
If the dataset name ends with a slash, then the dataset is a container holding several datasets.
The project part of the dataset name gives a general idea of what is contained in the dataset.
The following project names are of special importance in the context of this thesis:
\begin{itemize}
  \item \tagname{mc$vv$}: Datasets which contain Monte Carlo events begin with \tagname{mc$vv$},
    where %
    $vv$ gives the release version of the Monte Carlo,
    which roughly corresponds to the two last digits of the year in which the Monte Carlo production has been in started.
  \item \tagname{data$yy$}: This is the project name of datasets containing data taken with the \ATLAS detector,
    where $yy$ specifies the year in the which the data has been taken.
  \item \tagname{user}:
    Datasets which are not centrally produced, but uploaded or created by users,
    start with \tagname{user} to make them easily distinguishable from official productions.
\end{itemize}

\noindent The full name of a Monte Carlo dataset will have the form
\begin{center}
  \textit{mcNN\_subProject.datasetNumber.physicsShort.prodStep.dataType.Version}
\end{center}
The project part of the name typically ends with a specification of the center-of-mass energy used in the simulation,
which can be \tagname{900GeV}, \tagname{7TeV}, \tagname{10TeV}, or \tagname{2TeV} for \TeV{2.36}.
\textit{datasetNumber} is a 6 digit decimal number used as unique identifier of Monte Carlo datasets.
It is associated with a particular type of physics process,
which is also described in a slightly easier to remember textual form in \textit{physicsShort}.
The \textit{datasetNumber} uniquely identifies the class of events simulated in this sample across different release versions of the Monte Carlo.
\textit{prodStep} gives the production step used to create the data,
which is also reflected in the \textit{Version} field.
\textit{dataType} specifies the type of data, \eg RAW, ESD, AOD, or NTUP, and is explained below, as is the \textit{Version} field.

The full name of a dataset with real data will have the form
\begin{center}
  \textit{dataNN\_subProject.runNumber.streamName.prodStep.dataType.Version}
\end{center}
The fields are again mostly self-explanatory:
\textit{runNumber} is the number of the run the data belongs to
and \textit{streamName} is the name of the data stream.
\textit{subProject}, \textit{prodStep}, \textit{dataType} and \textit{Version} are the same as for Monte Carlo datasets.

The \textit{dataType} field is important to select the data format (see above) best suited for the analysis. %
The \textit{Version} field encodes the configuration of each of the processing steps that were passed by this dataset.
It consists of several entries separated by underscores, each consisting of a letter and a number
the meaning of which can be resolved via the \ac{AMI}.
For Monte Carlo datasets, the \textit{Version} field typically contains a combination of \tagname{e}, \tagname{s} and \tagname{r},
describing the event generation, detector simulation and reconstruction, respectively.
For datasets containing real data, the first two steps do not apply, of course.

\subsection{Databases in ATLAS}
\label{sec:software_ATLAS_databases}

Apart from storing the detector output,
from which the events and physics objects are reconstructed,
running a large machine like the \ATLAS detector requires a lot of additional bookkeeping.
Part of the information stored in these databases is metadata,
\ie second-level information about other data stored elsewhere.
For example, this can be information about the distribution of the datasets over the various grid sites.
Another part is primary information like the status of the detector,
summaries of the measurements or the configuration of the trigger system.
The \acf{AMI} is a catalogue,
not only of the datasets with Monte Carlo simulated and real events.
It also stores the software version tags that have been used in the simulation of Monte Carlo and processing of data.
The \ATLAS trigger database (\propername{TriggerDB}) is a relational database
that holds the information that is needed to configure the trigger system in online data taking \cite{TriggerConfig2008}. %
This information needs to be deployed to the \acf*{CTP} and all worker nodes at Level~2 and Event Filter,
and needs to be changeable during data taking in realtime.
It is also needed in the simulation of Monte Carlo events,
and the information is replicated to the \COOL database (see below),
from which it is picked up in the reprocessing of the data and stored also in the event records.

The \acl{COOL}, called \COOL for short after the name of the underlying application programming interface,
records information about the state of the detector at the time when events are collected \cite{ATLASComputingTDR2005},
such as the alignment of a tracking device, temperatures measured by a sensor or the trigger configuration.
Rather than being an event-based record of detector data,
the data in \COOL database describes the state of the detector during given intervals of time called \acl{IOV}.
The \COOL database is versioned,
which means that for some objects in the database there may exist several different versions for the same \ac{IOV}.
This may be the case, for example, when calibration or alignment parameters are updated,
or an updated luminosity computation becomes available,
which supersedes the old calibration valid for the same time.
\COOL is a common project of both \ATLAS and the \LHCb experiment at the \LHC.
Metadata about the detector configuration and conditions are also stored in the \ac{COMA} database \cite{URL_COMA},
which offers a compact overview of subsets of important information from other databases.

\addchap{Image Credits}

The following list gives external sources of figures which have been used in this thesis:

\renewcommand{\copyright}{\textcopyright~}
\newcommand*\oldurlbreaks{}
\let\oldurlbreaks=\UrlBreaks
\renewcommand{\UrlBreaks}{\oldurlbreaks\do\?\do\-\do\&\do\=\do\_\do\/\do\.}

\begin{itemize}
  \item
    \Fig{fig:theory_gut_unification}:
    Plot taken from \cite{LangackerPolonsky1993}.
    The plots have been rearranged within the figure.
    \copyright 1993 by The American Physical Society.
  \item
    \begin{sloppypar}
    \Fig{fig:example_instant_lumi}:
    Available from \url{https://twiki.cern.ch/twiki/bin/view/AtlasPublic/LuminosityPublicResults}, version of 16.12.2011. %
    Original title: ``\ATLAS instantaneous luminosity''.
    Original description: ``\ATLAS instantaneous luminosity profiles as measured online for representative \LHC fills with 7 TeV centre-of-mass energy in 2010.
    The gray shaded curves give the delivered luminosity, the green shaded curves show the delivered luminosity during stable beam conditions allowing \ATLAS to turn on their tracking devices, and the yellow shaded curves give the recorded luminosity with the entire detector available. [...] The luminosity values shown have been calibrated with van-der-Meer beam-separation scan data. The error on the luminosity is estimated to be \percent{11}, dominated by the uncertainty in the beam intensities.''
    \end{sloppypar}
  \item
    \begin{sloppypar}
    \Fig{fig:example_pulse_shape}:
    Downloaded from %
    \url{https://lar-elec-automatic-validation.web.cern.ch/lar-elec-automatic-validation/cgi-bin/PulseShape.sh?sub_sub_partition=EMBA&ft=0&sl=7&ch=124&campaign=185&gain=M} on 28.09.2011.
    \end{sloppypar}
  \item
    \Fig{fig:experimental_setup_lhc_layout}:
    Downloaded from CERN Document Server, record LHC-PHO-1993-005, \url{http://cdsweb.cern.ch/record/841542} on 10.06.2011.
    Original title: ``Layout of the LEP tunnel including future LHC infrastructures. L'ensemble du tunnel LEP avec les futures infractrustures LHC.''
    \copyright 1993 CERN.
  \item
    \begin{sloppypar}
    \Fig{fig:example_total_lumi}:
    Available from \url{https://twiki.cern.ch/twiki/bin/view/AtlasPublic/LuminosityPublicResults}, version of 16.12.2011. %
    Original title: ``Total Integrated Luminosity in 2011''.
    Original description: ``Cumulative luminosity versus day delivered to (green), and recorded by \ATLAS (yellow) during stable beams and for pp collisions at 7 TeV centre-of-mass energy in 2011. The delivered luminosity accounts for the luminosity delivered from the start of stable beams until the \LHC requests \ATLAS to turn the sensitive detector off to allow a beam dump or beam studies. Given is the luminosity as determined from counting rates measured by the luminosity detectors. These detectors have been calibrated with the use of the van-der-Meer beam-separation method, where the two beams are scanned against each other in the horizontal and vertical planes to measure their overlap function.''
    \end{sloppypar}
  \item
    \Fig{fig:lhc_cross_sections}:
    Plot taken from \cite{ATLASHLTTDR2003}, page 34.
    This is a modified version of the plot in \cite{Campbell2007}, page 95.
  \item
    \Fig{fig:experimental_setup_atlas_full_detector_view}:
    Downloaded from the CERN Document Server, record CERN-GE-0803012, \url{http://cdsweb.cern.ch/record/1095924} on 02.11.2011.
    Original title: ``Computer generated image of the whole ATLAS detector''.
    \ATLAS Experiment \copyright 2008 CERN.
  \item
    \Figs{fig:experimental_setup_atlas_inner_detector_view}: %
    Downloaded from the CERN Document Server, record CERN-GE-0803014, \url{http://cdsweb.cern.ch/record/1095926} on 02.11.2011.
    Original title: ``Computer generated image of the ATLAS inner detector''.
    \ATLAS Experiment \copyright 2008 CERN.
  \item
    \Fig{fig:experimental_setup_atlas_calorimeters}: %
    Downloaded from the CERN Document Server, record CERN-GE-0803015, \url{http://cdsweb.cern.ch/record/1095927} on 04.11.2011.
    Original title: ``Computer Generated image of the ATLAS calorimeter''.
    \ATLAS Experiment \copyright 2008 CERN.
  \item
    \Fig{fig:experimental_setup_atlas_muon_systems}: %
    Downloaded from the CERN Document Server, record CERN-GE-0803017, \url{http://cdsweb.cern.ch/record/1095929} on 04.11.2011.
    Original title: ``Computer generated image of the ATLAS Muons subsystem''.
    \ATLAS Experiment \copyright 2008 CERN.
  \item
    \Fig{fig:experimental_setup_atlas_muon_systems_schematic}: %
    Plot taken from \cite{CSCNotes}, page 163.
  \item
    \Fig{fig:tdaq_system_overview_schematic}:
    Figure taken from \cite{ATL-DAQ-PROC-2010-017}, page 2.
  \item
    \Fig{fig:athena_components}: %
    Figure taken from \cite{ATLASComputingTDR2005}, page 28.
\end{itemize}

\bibliographystyle{unsrt85} %

\begin{btSect}{doktor_noweb}

\addchap{Bibliography}

\btPrintCited
\end{btSect}
\begin{btSect}{doktor_web}

\addchap{Web Citations}

Internet resources are a vital part of contemporary research.
However, web citations are naturally volatile and subject to change.
They have thus been avoided whenever possible.
For each reference listed here, the date of the information retrieval is specified.
Some of the resources are password protected.
\btPrintCited
\end{btSect}
\addchap{Acronyms}

\label{sec:appendix_acronyms}

\begin{acronym}[XRHIANX]
\acro{ALFA} {Absolute Luminosity For ATLAS} %
\acro{AMI}  {Atlas Metadata Interface}
\acro{AMSB} {Anomaly-Mediated Supersymmetry Breaking}
\acro{AOD}  {Analysis Object Data}
\acro{ASCII}{American Standard Code for Information Interchange}
\acro{BCID} {Bunch-Crossing Identifier}
\acro{BLOB} {Binary Large Object}
\acro{BSM}  {Beyond Standard Model}
\acro{CDM}  {Cold Dark Matter}
\acro{CERN} {Conseil Européen pour la Recherche Nucléaire}
\acro{CLIC} {Compact Linear Collider}
\acro{CMS}  {Compact Muon Solenoid}
\acro{COMA} {Conditions Metadata}
\acro{COOL} {\ATLAS Conditions Database}
\acro{CPU}  {Central Processing Unit}
\acro{CSC}  {Cathode Strip Chamber}
\acro{CTP}  {Central Trigger Processor}
\acro{DDM}  {Distributed Data Management}
\acro{DQ2}  {Don Quixote 2}
\acro{EDM}  {Event Data Model}
\acro{EF}   {Event Filter}
\acro{EM}   {electromagnetic}
\acro{ESD}  {Event Summary Data}
\acro{FEx}  {Feature Extraction}
\acro{GMSB} {Gauge-Mediated Supersymmetry Breaking}
\acro{GRL}  {Good Runs List} %
\acro{GUT}  {Grand Unified Theory}\acrodefplural{GUT}{Grand Unified Theories}
\acro{HEC}  {Hadronic End-cap Calorimeter}
\acro{HLT}  {High-Level Trigger}
\acro{ILC}  {International Linear Collider}
\acro{IOV}  {Interval Of Validity}
\acro{JER}  {Jet-Energy Resolution}
\acro{JES}  {Jet-Energy Scale}
\acro{L1}   {Level~1}
\acro{L2}   {Level~2}
\acro{LAr}  {Liquid Argon Gas}
\acro{LB}   {Luminosity Block}
\acro{LEP}  {Large Electron-Positron Collider}
\acro{LHC}  {Large Hadron Collider}
\acro{LLR}  {Log-Likelihood Ratio}
\acro{LO}   {Leading Order}
\acro{LSP}  {Lightest Supersymmetric Particle}
\acro{LUCID}{Luminosity measurement Using a \cerenkov Integrating Detector}
\acro{MB}   {Minimum Bias}
\acro{MBTS} {Minimum-Bias Trigger Scintillators}
\acro{MC}   {Monte Carlo}
\acro{MDT}  {Monitored Drift Tube}
\acro{MET}  {Missing Transverse Energy}
\acro{MSSM} {Minimal Supersymmetric Standard Model}
\acro{mSUGRA}{Minimal Supergravity}
\acro{NLO}  {Next-to-Leading Order}
\acro{NLSP} {Next-Lightest Supersymmetric Particle}
\acro{NMSSM}{Next-to-Minimal Supersymmetric Standard Model}
\acro{NNLO} {next-to-next-to-leading order}
\acro{NTUP} {ROOT-readable ntuple}
\acro{PCL}  {Power-Constrained Limits}
\acro{PDF}  {Parton Distribution Function}
\acro{PLR}  {Profile Likelihood Ratio}
\acro{QCD}  {Quantum Chromodynamics}
\acro{QED}  {Quantum Electrodynamics}
\acro{QFT}  {Quantum Field Theory}
\acro{RDO}  {Raw Data Objects}
\acro{RGE}  {Renormalization Group Equation}
\acro{RMS}  {Root Mean Square}
\acro{RoI}  {Region of Interest}  \acrodefplural{RoI}{Re\-gions of Interest}
\acro{ROB}  {Read-Out Buffer}
\acro{RPC}  {Resistive Plate Chamber}
\acro{RPV}  {$R$-Parity Violating}
\acro{SCT}  {Silicon Microstrip Tracker}
\acro{SLHA} {Supersymmetry Les Houches Accord}
\acro{SUSY} {Supersymmetry}
\acro{TDAQ} {Trigger and Data-Acquisition System}
\acro{TDR}  {Technical Design Report}
\acro{TGC}  {Thin Gap Chamber}
\acro{TRT}  {Transition Radiation Tracker}
\acro{UE}   {Underlying Event}
\acro{UED}  {Universal Extra Dimensions}
\acro{VOMS} {Virtual Organization Management System}
\acro{WIMP} {Weakly Interacting Massive Particle}
\acro{XML}  {Extensible Markup Language}
\acro{ZDC}  {Zero-Degree Calorimeter}
\end{acronym}

Further explanations of acronyms can be found in the appendix of \cite{ATLASHLTTDR2003,ATLASComputingTDR2005}.
\addcontentsline{toc}{chapter}{Index}
\printindex

\end{document}